\documentclass{ucdthesis}

\usepackage{pdfpages}

\usepackage{amssymb}

\usepackage{amsmath}

\usepackage{array}

\usepackage{color}

\usepackage[colorlinks=true,urlcolor=black,menucolor=black,linkcolor=black,filecolor=black,citecolor=black]{hyperref} 

\newcommand{\xxp}{\left( x,x' \right)}
\newcommand{\beq}{\begin{equation}}
\newcommand{\eeq}{\end{equation}}

\newcommand{\fld}{\Phi}
\newcommand{\Gret}{G_{\text{ret}}}
\newcommand{\Grad}{\tilde{g}}
\newcommand{\fldtail}{\fld_\mu^{\text{tail}}}
\newcommand{\mass}{m}

\newcommand{\alp}{\alpha}

\newcommand{\gam}{\gamma}
\newcommand{\eps}{\epsilon}
\newcommand{\lam}{\lambda}
\newcommand{\sig}{\sigma}

\newcommand{\rstar}{r_\ast}

\newcommand{\phif}{\Delta \phi}

\newcommand{\Bef}{\mathcal{B}}
\newcommand{\unorm}{\tilde{u}}

\newcommand{\Aout}{A^{\text{(out)}}}
\newcommand{\Ain}{A^{\text{(in)}}}

\newcommand{\uin}{u^{\text{(in)}}_{l \omega}}
\newcommand{\uup}{u^{\text{(up)}}_{l \omega}}

\newcommand{\sech}{\text{sech}}
\newcommand{\nn}{\nonumber}

\newcommand{\lmax}{l_{\text{cut}}}
\newcommand{\nmax}{n_{\text{max}}}
\newcommand{\Ztt}{{\sigma^{t'}}_{t'}}
\newcommand{\Zyy}{{\sigma^{\rho'}}_{\rho'}}

\newcommand{\Phipartial}{\Phi_{\text{partial}}}
\newcommand{\GQNM}{\Gret^{\text{QNM}}}
\newcommand{\II}{\mathcal{I}}
\newcommand{\Agam}{\mathcal{A}(\gam)}

\newcommand{\R}{\rho}
\newcommand{\Rs}{\rho_\ast}

\title{Green Functions and Radiation Reaction From a Spacetime Perspective}
\author{Barry Wardell}

\degree{Philosophiae Doctor}

\college{Engineering, Mathematical and Physical Sciences}
\school{Complex and Adaptive Systems Laboratory\\
	and\\
	School of Mathematical Sciences}

\advisor{Professor Adrian C.~Ottewill}
\head{Dr. M\'ich\'eal \'O Searc\'oid}

\thesisdate{\today}

\begin{document}

\pagestyle{plain}

\maketitle

\begin{abstract}
Accurate modelling of gravitational wave emission by extreme mass ratio inspirals is essential for their detection by LISA, the proposed space-based gravitational wave detector. A leading perturbative approach involves the calculation of the self-force acting upon the smaller orbital body. In this thesis, we present methods for calculating the self-force which are motivated by a desire to gain a deep understanding of the self-force and the effect that the geometry of spacetime has on it.

The basis of this work is the first full application of the Poisson-Wiseman-Anderson method of `matched expansions' to compute the self-force acting on a point particle moving in a curved spacetime. The method employs two expansions for the Green function which are respectively valid in the `quasilocal' and `distant past' regimes, and which are matched together within the normal neighborhood.

Building on a fundamental insight due to Avramidi, we provide a system of transport equations for determining key fundamental bi-tensors, including the tail-term, $V(x,x')$, appearing in the Hadamard form of the Green function. These bitensors are central to a broad range of problems from radiation reaction and the self-force to quantum field theory in curved spacetime and quantum gravity. Using their transport equations, we show how the quasilocal Green function may be computed throughout the normal neighborhood both numerically and as a covariant Taylor series expansion. These calculations are carried out for several black hole spacetimes.

Finally, we present a complete application of the method of matched expansions. The calculation is performed in a static region of the spherically symmetric Nariai spacetime ($dS_2 \times  \mathbb{S}^2$), where the matched expansion method is applied to compute the scalar self-force acting on a static particle. We find that the matched expansion method provides insight into the non-local properties of the self-force. The Green function in Schwarzschild spacetime is expected to share certain key features with Nariai. In this way, the Nariai spacetime provides a fertile testing ground for developing insight into the non-local part of the self-force on black hole spacetimes.
\end{abstract}
\chapter*{Acknowledgments}

I would like to thank my supervisor, Adrian Ottewill, for his endless support and encouragement. I would also like to thank the other members of staff and students of the School of Mathematical Sciences and CASL. In particular, Marc Casals and Sam Dolan deserve a special mention for all the help they have given me with my research, as does Kirill Ignatiev for helping with the derivation of some tricky formul\ae.

Several researchers have generously given me their time through the years. Paul Anderson, Ardeshir Eftekharzadeh and Warren Anderson gave valuable time and effort to resolving some issues that arose during this work. Correspondences with Antoine Folacci have always been helpful and informative. Brien Nolan has given key insights, in particular with regard to analytic calculations in the Nariai spacetime. Also my appreciation to the many others from the Capra meetings for numerous interesting conversations. 

Finally I would like to thank my family, all my friends and my girlfriend Sin\'ead for all their support. 

This research was financially supported by the Irish Research Council for Science, Engineering and Technology: funded by the National Development Plan.

\tableofcontents
\listoffigures
\listoftables

\clearpage
\pagenumbering{arabic}
\pagestyle{fancy}

\chapter*{Overview} \label{ch:overview}
\addcontentsline{toc}{chapter}{Overview}
\markboth{}{Overview}

The last decade has seen a surge of interest in the nascent field of gravitational wave astronomy.  Gravitational waves -- propagating ripples in spacetime -- are generated by some of the most violent processes in the known universe, such as supernovae, black hole mergers and galaxy collisions. These powerful processes are often hidden from the view of `traditional' electromagnetic-wave telescopes behind shrouds of dust and radiation. On the other hand, gravitational waves are not strongly absorbed or scattered by intervening matter, and carry information about the dynamics at the heart of such processes. The prospects seem good for direct detection of gravitational waves in the near future. The interest in gravitational wave astronomy has been increasingly spurred on by the recent and upcoming construction of both ground and space based gravitational wave observatories. A number of ground-based detectors (such as LIGO \cite{LIGO}, VIRGO \cite{VIRGO} and GEO600 \cite{GEO}) are now in the data collection phase. 

Gravitational wave astronomy will enter a new era with the launch of the first space-based observatory: the Laser Interferometer Space Antenna (LISA) \cite{LISA}. It is hoped that this joint NASA/ESA mission, presently in the design and planning phase, will be launched within a decade. It will be preceded by a pathfinder mission, due for launch in 2010 \cite{Bell:2008}. The primary difference between ground and space based detectors is the frequency band in which they operate. Environmental effects such as seismic noise mean that ground based detectors are only capable of making observations above $1Hz$. Space based detectors have the benefit of not being affected by this large amount of low frequency noise, so are free to explore much lower frequency bands, where a considerable number of interesting gravitational wave sources are expected to be found. The LISA mission will detect gravitational waves in the frequency band 0.1-0.0001Hz.

Data analysis methods such as matched filtering may be applied to improve the sensitivity of detectors by separating the weak GW signals from a noisy background \cite{Vallisneri:2009}. To improve the effectiveness of this data analysis, it is necessary to have accurate predictions of the waveforms of the gravitational radiation emitted by candidate sources. As an essential prerequisite to the creation of accurate templates for the gravitational wave emission of such sources, it is necessary to develop a theoretical model of the system.

Black hole binary systems are a key target for gravitational wave (GW) observatories worldwide. Breakthroughs in numerical relativity \cite{Pretorius:2005} in the last five years have led to a rapid advance in the theoretical modelling of comparable-mass binaries, where the partners are of similar mass. Progress in numerical relativity continues apace \cite{Gonzalez-Sperhake-Bruegman-2008}.

One of the primary targets for the LISA mission are the so-called \emph{Extreme Mass Ratio Inspirals} (EMRIs): compact binaries in which one partner (mass $M$) is significantly more massive than the other (mass $m$).
EMRIs with mass ratios of $\mu \equiv m/M \gtrsim 10^{-9}$ are possible, for example for a solar-mass black hole orbiting a supermassive black hole \cite{Hughes:Drasco:2005}. Those with $\mu \sim 10^{-7} - 10^{-5}$ are expected to be detectable by LISA out to Gpc distances \cite{Gair:2004}. Mass ratios of up to $m/M\sim 1/10$ have been studied by numerical relativists \cite{Gonzalez-Sperhake-Bruegman-2008}; smaller ratios are presently beyond the scope of numerical relativity due to the existence of two distinct and dissimilar length scales in the system. Perturbative approaches seem more likely to succeed in the extreme-mass regime.

EMRI systems may be perturbatively modelled by making the approximation that the smaller mass is travelling in the background spacetime of the larger mass. The compact nature of the smaller mass means that it will distort the curvature of the spacetime in which it is moving to a non-negligible extent. As a result, it does not exactly follow a geodesic of the background spacetime generated by the larger mass, $M$, but rather, it follows a geodesic of a total \emph{effective} spacetime generated by both the larger mass and a regularized perturbation from the smaller mass \cite{Detweiler-Whiting-2003}. However, if the mass ratio is extreme, the deviation of the smaller body's motion from the background geodesic will be (locally) small and may be interpreted as arising from a \emph{self-force}, created by the smaller mass $m$ interacting with its own gravitational field. To leading order, the self-force acceleration is proportional to $m$. The computation of this self-force is of fundamental importance to the accurate calculation of the orbital evolution of such EMRI systems and hence to the prediction of the gravitational radiation waveform. Unfortunately, finding the instantaneous self-force in a curved spacetime is not at all straightforward; it turns out to depend on the \emph{entire past history} of the
smaller mass, $m$.

In this thesis, we focus primarily not on the gravitational self-force described thus far, but rather on the analogous scalar self-force. Instead of considering a smaller mass, $m$, to be generating a gravitational field, we consider a point particle with scalar charge, $q$. This charge couples to a massless scalar field which is then the cause of the self-force. Otherwise, we leave the problem exactly as posed above.

The idea of a self-force has a long history in physics. In the late 19th century it was well-known that a charge undergoing an acceleration 
in flat spacetime will generate electromagnetic radiation, and will feel a corresponding \emph{radiation reaction}. The self-acceleration of a charged point particle in flat spacetime is given by the well-known Abraham-Lorentz-Dirac formula \cite{Dirac-1938}. Radiation reaction implies that the `classical' model of the atom (a point-particle electron orbiting a compact nucleus) is unstable. The observed stability of the atom remained a puzzle for many years, and provided a key motivation for the development of quantum mechanics.
In the 1960s, DeWitt and Brehme \cite{DeWitt:1960} derived for the first time a formula for the self-force acting on an electrically-charged point particle in a curved background, and a correction was later provided by Hobbs \cite{Hobbs:1968a}. More recently, this expression was recovered by Quinn and Wald \cite{Quinn:Wald:1997} and in a rigorous derivation by Gralla, Harte and Wald \cite{Gralla:Harte:Wald:2009}. The gravitational self-force acting on a point mass was found in 1997 by two groups working concurrently and independently: Mino, Sasaki and Tanaka \cite{Mino:Sasaki:Tanaka:1996} and Quinn and Wald \cite{Quinn:Wald:1997}. A subsequent derivation using matched asymptotic expansions was given by Poisson \cite{Poisson:2003}. Recent developments have put the gravitational self-force on a thoroughly firm footing through a rigorous treatment by Gralla and Wald \cite{Gralla:Wald:2008}. For the case of a minimally-coupled scalar charge, Quinn used a adaptation of the axiomatic approach of Ref.~\cite{Quinn:Wald:1997} to derive an expression for the scalar self-force \cite{Quinn:2000}. Many of these developments are summarized in 2004/05 reviews by Poisson \cite{Poisson:2003} (where the scalar case was extended
to cater for non-minimal coupling) and Detweiler \cite{Detweiler:2005}. In the subsequent period, a range of complementary approaches to the self-force problem have been developed \cite{Galley:Hu:Lin:2006, Harte:2008, Futamase:Hogan:Itoh:2008, Flanagan:Hinderer:2008, Norton:2009}. 

The organization of this thesis is as follows. In Chapter~\ref{ch:intro} we introduce the concept of a bi-tensor and define some useful quantities such as the world function, the Van Vleck determinant and the bi-tensor of parallel transport. We also introduce the scalar Green function and its expression within the normal neighborhood in the Hadamard form. Finally, we introduce several black hole spacetimes which are of particular interest for self-force calculations.

In Chapter~\ref{ch:sf}, we discuss the scalar self-force and the equations of motion of a scalar charge moving in a curved background spacetime. We also introduce the MiSaTaQuWa equation for the self-force and present the method of matched expansions as a way of evaluating it. This leads to a derivation of the quasilocal self-force, the contribution to the self-force from the recent past of the particle. We consider cases of both geodesic and non-geodesic motion.

In Chapter~\ref{ch:coordex} we describe a coordinate expansion approach to the calculation of the scalar Green function within the quasilocal region. Using a high-order coordinate Taylor series representation, valid in spherically symmetric spacetimes, we show that the expansion is valid throughout a significant portion of the normal neighborhood. Applying the method of Pad\'e approximants, we show that the domain of the series representation may be extended beyond its radius of convergence to within a short distance of the normal neighborhood boundary.

The coordinate approach of Chapter~\ref{ch:coordex} is limited to spherically symmetric spacetimes. This precludes its application to the most astrophysically relevant spacetime, the Kerr black hole. We present a solution to this problem in Chapter~\ref{ch:covex}, where we derive a system of transport equations allowing us to compute the Green function for a general spacetime in a totally covariant way. These transport equations may be solved either numerically or as a covariant series. We investigate both methods and present some results for specific spacetimes.

These results are applied, in Chapter~\ref{ch:qlsf}, to the calculation of the quasilocal self-force. We study a range of examples of both geodesic and non-geodesic motion in several spacetimes. We compare our results  with other results achieved using alternative methods.

Finally, in Chapter~\ref{ch:nariai}, we combine the quasilocal results with calculations of the contribution to the self-force from the `distant past' to produce an example application of the method of matched expansions. We conduct our calculation in the Nariai spacetime and consider as a specific example the case of a static particle - a particle with fixed spatial coordinates. This toy model yields considerable insight into the self-force and the effect that the geometry of spacetime has on it.

Some portions of this thesis were done in collaboration with Marc Casals, Sam Dolan and Adrian Ottewill. For the sake of clarity and a coherent presentation of the matched expansions method, that work is included here  in full, with sections for which I was not a primary contributor indicated by an asterisk ($*$). Additionally, it should be noted that many of the results presented here have previously appeared in journal articles by the authors \cite{Ottewill:Wardell:2008, Ottewill:Wardell:2009, Casals:Dolan:Ottewill:Wardell:2009, QL, Transport}.

Throughout this thesis, we use units in which $G=c=1$ and adopt curvature conventions of \cite{Misner:Thorne:Wheeler:1974}, including the metric signature convention $\{-+++\}$. We denote symmetrization of indices using brackets (e.g. $(a b)$), anti-symmetrization using square brackets (eg. $[a b]$) and exclude indices from symmetrization by surrounding them by vertical bars (e.g. $(a | b | c)$). Roman letters are used for free indices and Greek letters for indices summed over all spacetime dimensions. The Roman letters $i, j, k, l$ are used for indices over spatial dimensions only.

\chapter{Introduction} \label{ch:intro}

\section{A Brief Review of Green functions, Bitensors and Covariant Expansions}
\label{sec:review}

\subsection{Classical Green functions}\label{sec:classical-green}

We take an arbitrary field $\varphi^{A}(x)$ where ${}^{A}$ denotes the spinorial/tensorial index appropriate to the field,
and consider wave operators 
which are second order partial differential operators of the form~\cite{Avramidi:2000}
\begin{equation}
\label{eq:Wave-operator}
\mathcal{D}^{A}{}_B  = \delta^{A}{}_B (\square - m^2) + P^{A}{}_B
\end{equation}
where $\square = g^{\alpha\beta}\nabla_{\alpha}\nabla_{\beta}$, $g^{\alpha\beta}$ is the (contravariant) metric tensor and $\nabla_{\alpha}$ is the covariant derivative defined by a connection $\mathcal{A}^{A}{}_{B\alpha}$:
$\nabla_{\alpha}\varphi^{A}= \partial_{\alpha}\varphi^{A}+  \mathcal{A}^{A}{}_{B\alpha} \varphi^{B}$, $m$ is the mass of the field and $P^{A}{}_B(x)$ is a
possible potential term.

In the classical theory of wave propagation in curved spacetime, a fundamental object is the retarded Green function, $G_{\mathrm{ret}}{}^{B}{}_{C'} \xxp$, where $x'$ is a spacetime point in the causal past of the spacetime field point, $x$. The retarded Green function is a solution of the inhomogeneous wave equation,
\begin{equation}
\label{eq:Wave}
\mathcal{D}^{A}{}_B G_{\mathrm{ret}}{}^{B}{}_{C'} \xxp = - 4\pi \delta^{A}{}_{C'}\delta \xxp ,
\end{equation}
 with support on and within the past light-cone of the field point. (The factor of $4\pi$ is a matter of convention, our choice here is consistent with 
 Ref.~\cite{Poisson:2003}.) 
Finding the retarded Green function globally can be extremely hard. However, provided $x$ and $x'$ are sufficiently close (within a normal neighborhood\footnote{More precisely, the Hadamard parametrix used in the quasilocal region requires that $x$ and $x'$ lie within a \emph{causal domain} -- a \emph{convex normal neighborhood} with causality condition attached. This effectively requires that $x$ and $x'$ be connected by a unique non-spacelike geodesic which stays within the causal domain. However, as we expect the term \emph{normal neighborhood} to be more familiar to the reader, we will use it throughout this thesis, with implied assumptions of convexity and a causality condition.\label{def:causal domain}}), we can use the Hadamard form for the retarded Green function solution \cite{Hadamard,Friedlander},
which in 4 spacetime dimensions takes the form
\begin{equation}
\label{eq:Hadamard}
G_{\mathrm{ret}}{}^{A}{}_{B'}\xxp = \theta_{-} \xxp \left\lbrace U^{A}{}_{B'} \xxp \delta \left( \sigma \xxp \right) - V^{A}{}_{B'} \xxp \theta \left( - \sigma \xxp \right) \right\rbrace ,
\end{equation}
where $\theta_{-} \xxp$ is analogous to the Heaviside step-function, being $1$ when $x'$ is in the causal past of $x$, and $0$ otherwise, $\delta \xxp$ is the covariant form of the Dirac delta function, $U^{AB'}\xxp$ and $V^{AB'}\xxp$ are symmetric bi-spinors/tensors and  are regular for $x' \rightarrow x$. The bi-scalar $\sigma \xxp$ is the Synge~\cite{Poisson:2003} world function, which  is equal to one half of the squared geodesic distance between $x$ and $x'$.
The first term, involving $U^{A}{}_{B'} \xxp$, in Eq.~(\ref{eq:Hadamard}) represents the \emph{direct} part of the Green function 
while the second term, involving $V^{A}{}_{B'} \xxp$,  is known as the \emph{tail} part of the Green function. This tail term represents back-scattering off the spacetime geometry and is, for example, responsible for the quasilocal contribution to the self-force.

Within the Hadamard approach, the symmetric bi-scalar $V^{AB'}\xxp$ is expressed in terms of a formal expansion in increasing powers of $\sigma$ \cite{Decanini:Folacci:2005a}:
\begin{equation}
\label{eq:V}
V^{AB'}\xxp = \sum_{r=0}^{\infty} V_{r}{}^{AB'}\xxp \sigma ^{r}\xxp
\end{equation}
The coefficients $U^{AB'}$ and  $V_{r}{}^{AB'}$ are determined by imposing the wave equation, using the identity $\sigma_{\alpha} \sigma^{\alpha}= 2 \sigma =\sigma_{\alpha'} \sigma^{\alpha'}$,
and setting the coefficient of each \textsl{manifest} power of $\sigma$ equal to zero. Since $V^{A}{}_{B'}$ is symmetric for self-adjoint wave operators we are free to apply the wave 
equation either at $x$ or at $x'$; here we choose to apply it at $x'$. We find that 
$U^{AB'} \xxp = \Delta^{1/2} \xxp g^{AB'}\xxp$,  where $\Delta \xxp$ is the Van Vleck-Morette determinant defined as~\cite{Poisson:2003}
\begin{align}
\label{eq:vv-def}
\Delta \xxp =& - \left[ -g \left( x \right) \right] ^{-1/2} \det \left( -\sigma_{\alpha \beta '} \xxp \right) \left[ -g \left( x' \right) \right] ^{-1/2} \nonumber \\
 =& \det \left( -g^{\alpha'}{}_\alpha \xxp \sigma^{\alpha}{}_{ \nu '} \xxp \right)
\end{align}
with $g^{\alpha'}{}_\alpha\xxp$ being the bi-vector of parallel transport (defined fully below) and where $g^{AB'}$ is the bi-tensor of parallel transport appropriate to the tensorial nature of the field, eg.
\begin{equation}
 g^{A B'} = \begin{cases}
             1 & \text{(scalar)}\\
             g^{a b'} & \text{(electromagnetic)} \\
             g^{a' (a} g^{b) b'} & \text{(gravitational)}.
            \end{cases}
\end{equation}
In making the identification \eqref{eq:vv-def}, we have used the transport equation for the Van Vleck-Morette determinant:
\begin{equation}
\label{eq:VVtransport}
\sigma^{\alpha} \nabla_{\alpha} \ln \Delta = (4 - \square \sigma) .
\end{equation}
The coefficients $V_{r}^{AB'}\xxp$ satisfy the recursion relations
\begin{subequations}
\label{eq:RecursionV}
\begin{align}
\label{eq:recursionVn}
  \sigma^{\alpha'} (\Delta ^{-1/2} V^{AB'}_{r})_{;\alpha'}  
 + \left( r+1 \right)  \Delta ^{-1/2}  V_{r}^{AB'} + {\frac{1}{2r}} \Delta ^{-1/2}  \mathcal{D}^{B'}{}_{C'} V^{AC'}_{r-1} = 0 
\end{align}
for $r \in \mathbb{N}$ along with the `initial condition'
\begin{eqnarray}
\label{eq:recursionV0}
\sigma^{\alpha'} (\Delta ^{-1/2} V_0^{AB'}){}_{;\alpha'} 
+ \Delta ^{-1/2} V^{AB'}_0 + {\frac{1}{2}}\Delta ^{-1/2} \mathcal{D}^{B'}{}_{C'} ( \Delta ^{1/2} g^{AC'}) &=& 0 .
\end{eqnarray}
\end{subequations}
These are transport equations which may be solved in principle within a normal neighborhood by direct integration along the geodesic from $x$ to $x'$. The complication is that the calculation of $ V^{AB'}_{r}$ requires the calculation of second derivatives of $ V^{AB'}_{r-1}$ in directions off the geodesic;
we address this issue below.

Finally we note that the Hadamard expansion is an ansatz not a Taylor series. For example, in deSitter spacetime for a conformally invariant scalar theory all  the $V_r$'s are non-zero while $V\equiv0$.

\subsection{The quantum theory}\label{sec:quantum-green}
In curved spacetime a fundamental object of interest is the Feynman Green 
function defined for a quantum field $\hat \varphi^A(x)$ in the state $| \Psi \rangle$ by
\[
G_\mathrm{f}^{AB'}(x,x')= i \mathrm{T}\left[ \langle \Psi | \hat \varphi^A(x) \hat \varphi^{B'}(x')| \Psi \rangle \right] .
\]
where $\mathrm{T}$ denotes time-ordering.
The Feynman Green function may be related to the advanced and retarded Green functions of the classical theory by
the covariant commutation relations~\cite{DeWitt:1965}
\begin{multline}
G_\mathrm{f}{}^{AB'}(x,x')= \\\frac{1}{8\pi} \left(G_\mathrm{adv}^{AB'}(x,x') + G_\mathrm{ret}^{AB'}(x,x')\right)
+  \frac{i}{2}  \langle \Psi | \hat \varphi^A(x) \hat \varphi^{B'}(x')+ \hat \varphi^{B'}(x')\hat \varphi^A(x)| \Psi \rangle .
\end{multline}
The anticommutator function $\langle \Psi | \hat \varphi^A(x) \hat \varphi^{B'}(x')+ \hat \varphi^{B'}(x')\hat \varphi^A(x)| \Psi \rangle$
clearly satisfies the homogeneous wave equation so that the Feynman Green function satisfies the equation
\[
 \mathcal{D}^{A}{}_B G_{\mathrm{f}}{}^{B}{}_{C'} \xxp = - \delta^A{}_{C'} \delta(x,x') .
\]

Using the proper-time formalism~\cite{DeWitt:1965}, the identity 
\[
i \int\limits_0^\infty \mathrm{d}s\> e^{- \epsilon s} \exp({ i s x}) = - \frac{1}{x + i \epsilon}, \qquad (\epsilon > 0),
\]
allows the causal properties of the Feynman function to be encapsulated in the formal expression
\[
G_\mathrm{f}{}^{A}{}_{C'} \xxp
 = i \int\limits_0^\infty \mathrm{d}s\> e^{- \epsilon s} \exp(i s \mathcal{D})^{A}{}_{B} \delta^B{}_{C'} \delta(x,x')
\] 
where the limit $\epsilon \rightarrow 0+$ is understood.
The integrand
\begin{align}
    K^{A}{}_{C'}(x,x';s) = \exp(i s \mathcal{D})^{A}{}_{B} \delta^B{}_{C'} \delta(x,x')
\end{align}
clearly satisfies the Schr\"odinger/heat equation
\begin{align}
    \frac {1}{i} \frac{\partial K^{A}{}_{C'}}{\partial s} (x,x';s)= \mathcal{D}^A{}_{B}K^{B}{}_{C'}(x,x';s)
\end{align}
together with the initial condition 
$
    K^{A}{}_{B'}(x,x';0) = \delta^A{}_{B'}(x,x') 
$.
The trivial way in which the mass $m$ enters these equations allows it to be eliminated through the
prescription
\begin{align}
    K^{A}{}_{C'}(x,x';s) = e^{ - i m^2 s }K_0{}^{A}{}_{C'}(x,x';s),
\end{align}
with the massless heat kernel satisfying the equation
\begin{align}
\label{eq:K0heateqn}
    \frac {1}{i} \frac{\partial K_0{}^{A}{}_{C'}}{\partial s} (x,x';s)= (\delta^{A}{}_B \square  + P^{A}{}_B)K_0{}^{B}{}_{C'}(x,x';s)
\end{align}
together with the `initial condition' 
$  K_0{}^{A}{}_{B'}(x,x';0) = \delta^A{}_{B'}\delta(x,x') $.

In $4$-dimensional Minkowski spacetime without potential, the massless heat kernel is readily obtained as
\begin{align}
    K_0{}^{A}{}_{B'}(x,x';s) =  \frac{1}{(4 \pi s)^2} \exp \left(-\frac{\sigma}{2 i s}\right) \delta^{A}{}_{B'} \qquad\textrm{(flat spacetime)} .
\end{align}
This motivates the ansatz~\cite{DeWitt:1965} that in general the massless heat kernel allows the representation
\begin{align}
    K_0{}^{A}{}_{B'}(x,x';s) \sim  \frac{1}{(4 \pi s)^2}\exp \left(-\frac{\sigma}{2 i s}\right)  \Delta^{1/2}\xxp \Omega{}^{A}{}_{B'}(x,x';s)\ ,
\end{align}
where $\Omega{}^{A}{}_{B'}(x,x';s)$ possesses the following asymptotic expansion as $s \rightarrow 0+$:
\begin{align}
\label{eq:dewittseries}
    \Omega{}^{A}{}_{B'}(x,x';s) \sim   \sum\limits_{r=0}^\infty a_r^{A}{}_{B'}(x,x') (i s)^r\ ,
\end{align}
with $a_0{}^{A}{}_{B'}(x,x)= \delta^{A}{}_{B'}$ and $a_r{}^{A}{}_{B'}(x,x')$ has dimension $(\textrm{length})^{-2 r}$.
The inclusion of the explicit factor of $\Delta^{1/2}$ is simply a matter of convention; by including it we are following DeWitt, but many authors, including D\'ecanini and Folacci \cite{Decanini:Folacci:2005a}, choose instead to include it in the
series coefficients 
\begin{align}
A_r{}^{A}{}_{B'}(x,x')=\Delta^{1/2}a_r{}^{A}{}_{B'}(x,x').
\end{align}
It is clearly trivial to convert between the two conventions and, in any case, the coincidence limits agree.

Now, requiring our expansion to satisfy Eq.~(\ref{eq:K0heateqn}) and using the symmetry of $\Omega{}^{A}{}_{B'}(x,x';s)$ to allow operators to act at $x'$, we find that
$\Omega{}^{A}{}_{B'}(x,x';s)$ must satisfy
\[
\frac{1}{i}\frac{\partial \Omega{}^{AB'}}{\partial s} + \frac{1}{i s} \sigma^{\alpha'} \Omega^{AB'}{}_{;\alpha'} 
 = \Delta^{-1/2} (\delta^{B'}{}_{C'} \square  + P^{B'}{}_{C'})\left(\Delta^{1/2}  \Omega{}^{AC'}(x,x';s)\right) .
\]
Inserting the expansion Eq.~(\ref{eq:dewittseries}), the coefficients $a_{n}^{AB'}\xxp$ satisfy the recursion relations
\begin{subequations}
\label{eq:Recursiona}
\begin{align}
 \sigma^{\alpha'} a_{r+1}^{\phantom{n}AB'}{}_{;\alpha'}
 + \left( r+1 \right)  a_{r+1}^{\phantom{n}AB'} -
 \Delta^{-1/2} (\delta^{B'}{}_{C'} \square  + P^{B'}{}_{C'})\left(\Delta^{1/2}  a_r{}^{AC'}\right) = 0 
\end{align}
for $n \in \mathbb{N}$ along with the `initial condition'
\begin{eqnarray}
\sigma^{\alpha'} a_0{}^{AB'}{}_{;\alpha'} &=& 0 ,
\end{eqnarray}
\end{subequations}
with the implicit requirement that they be regular as $x' \to x$.

Comparing \eqref{eq:RecursionV} and \eqref{eq:Recursiona}, one can see that the Hadamard and (mass independent) DeWitt\footnote{The Hadamard and DeWitt coefficients also appear in the literature under several other guises. They may be called DeWitt, Gilkey, Heat Kernel, Minakshisundaram, Schwinger or Seeley coefficients, or any combination thereof (yielding acronyms such as DWSC, DWSG and HDMS). In the coincidence limit, it has been proposed that they be called Hadamard-Minakshisundaram-DeWitt (HaMiDeW) \cite{Gibbons} coefficients. For the remainder of this thesis, we will refer to them as either DeWitt (for the coefficients $a_k{}^A{}_{B'}$) or Hadamard (for the coefficients $V_r{}^A{}_{B'}$) coefficients.} coefficients are related for a theory of mass $m$ by
\begin{equation}
\label{eq:v-a-relation}
V_r{}^{A}{}_{B'}(x,x') = \frac{\Delta^{1/2}\xxp}{2^{r+1} r!} \sum \limits_{k=0}^{r+1}
(-1)^k \frac{(m^2)^{r-k+1}}{(r-k+1)!} a_k{}^{A}{}_{B'}(x,x')
\end{equation}
with inverse
\begin{equation}
a_{r+1}{}^{A}{}_{B'}(x,x') = \Delta^{-1/2} \sum \limits_{k=0}^{r}
(-2)^{k+1} \frac{k!}{(r-k)!} (m^2)^{r-k} V_k{}^{A}{}_{B'}(x,x') +  \frac{(m^2)^{r+1}}{(r+1)!} .
\end{equation}
In particular,
\begin{equation}
\label{eq:v-massless-a-relation}
V_r^{(m^2=0)}{}^{A}{}_{B'}(x,x') = \frac{\Delta^{1/2}\xxp}{2^{r+1} r!} 
(-1)^{r+1} a_{r+1}{}^{A}{}_{B'}(x,x')  .
\end{equation}

These relations enable us to relate the `tail term' of the
massive theory to that of  massless theory by
\begin{multline}
V(x,x'){}^{A}{}_{B'} =\\ \sum\limits_{r=0}^\infty   V_r^{(m^2=0)}{}^{A}{}_{B'} (x,x')  
\frac{\left(2\sigma\right)^r r! J_r\left( (-2 m^2 \sigma)^{1/2} \right)}{ (-2 m^2 \sigma)^{r/2}} 
+ m^2 \Delta^{1/2} \frac{J_1\left( (-2 m^2 \sigma)^{1/2} \right)}{ (-2 m^2 \sigma)^{1/2}} \delta{}^{A}{}_{B'},
\end{multline}
where $J_r(x)$ are Bessel functions of the first kind. This last expression is obtained by using \eqref{eq:v-massless-a-relation} in \eqref{eq:v-a-relation}, substituting the result into \eqref{eq:V} and interchanging the order of summation (upon doing so, the sum over $k$ yields the Bessel functions).

\subsection{Bi-tensors and Covariant Expansions}\label{sec:bitensors}

\begin{figure}
 \begin{center}
 \includegraphics[width=4.5cm]{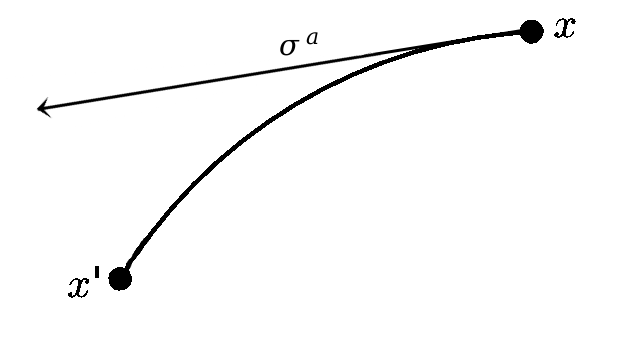}
 \includegraphics[width=4.5cm]{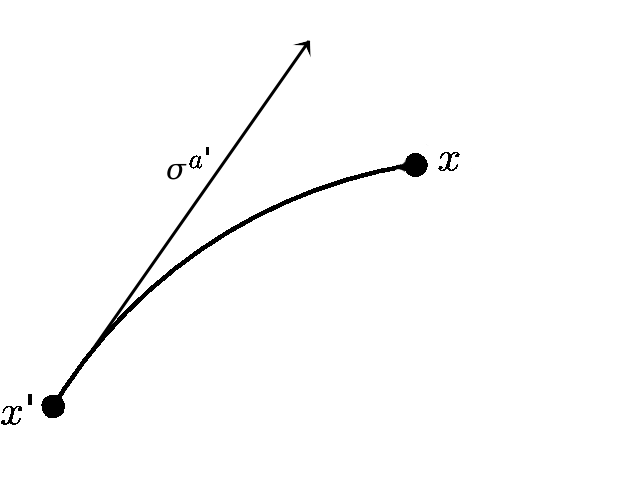}
\end{center}
\caption[Derivative of $\sigma(x,x')$]{Derivatives of $\sigma(x,x')$ at $x$ (left) and at $x'$ (right).}
\label{fig:sigma}
\end{figure}

The Synge world-function, $\sigma (x,x')$, is a bi-scalar, i.e. a scalar at $x$ and at $x'$ defined to be equal to half the square of the geodesic distance between the two points.
The world-function may be defined by the fundamental identity
\begin{equation}
\label{eq:SigmaDefiningEq}
 \sigma_\alpha \sigma^\alpha = 2 \sigma = \sigma_{\alpha'} \sigma^{\alpha'},
\end{equation}
together with the boundary condition $\lim\limits_{x'\to x} \sigma (x,x')=0 $
and $\lim\limits_{x'\to x} \sigma_{a b} (x,x')= g_{ab}(x) $.
Here we indicate derivatives at the (un-)primed point by (un-)primed indices:
\begin{align}
 \sigma^a &\equiv \nabla^a \sigma & \sigma_a &\equiv \nabla_a \sigma &
 \sigma^{a'} &\equiv \nabla^{a'} \sigma & \sigma_{a'} &\equiv \nabla_{a'} \sigma .
\end{align}
$\sigma^a $ is a vector at $x$ of length equal to the geodesic distance between $x$ and $x'$, tangent to the geodesic at $x$ and oriented in the direction $x'\to x$ while $\sigma^{a'} $ is a vector at $x'$ of length equal to the geodesic distance between $x$ and $x'$, tangent to the geodesic at $x'$ and oriented in the opposite direction (see Fig.~\ref{fig:sigma}).

The covariant derivatives of $\sigma$ may be written as
\begin{align}
 \sigma^{a}(x,x') &= (s - s') u^{a} & \sigma^{a'}(x,x') &= (s' - s) u^{a'}
\end{align}
where $s$ is an affine parameter and $u^{a}$ is tangent to the geodesic.
For time-like geodesics, $s$ may be taken as the proper time along the geodesic while
$u^{a}$ is the 4-velocity of the particle and
\begin{equation}
\label{eq:sigma-def}
 \sigma (x,x') = -\frac{1}{2} (s - s')^2 .
\end{equation}
For null geodesics, $u^{a}$ is null and  $\sigma (x,x') = 0$.

Another bi-tensor of frequent interest is the  bi-vector of parallel transport, $g_{a b'}$ defined by the equation
\begin{equation}
 g_{a b' ;\alpha} \sigma^{\alpha} = 0 = g_{a b' ;\alpha'} \sigma^{\alpha'}
\end{equation}
with initial condition $\lim\limits_{x'\to x} g_{a b'} (x,x')= g_{ab}(x) $.
From the definition of a geodesic it follows that
\begin{equation}
 g_{a \alpha'} \sigma^{\alpha'} = - \sigma_{a} \qquad \mathrm{and} \qquad g_{\alpha a '} \sigma^{\alpha} = - \sigma_{a'} 
\end{equation}

Given a bi-tensor $T_{a}$ at $x$, the parallel displacement bi-vector allows us write $T_{a}$ as a bi-tensor at $x'$, obtained by parallel transporting $T_{a}$ along the geodesic from $x$ to $x'$ and vice-versa,
\begin{align}
 T_{\alpha} g^{\alpha}_{\phantom{a} a'} &= T_{a'} &
 T_{\alpha'} g^{\alpha'}_{\phantom{a'} a} &= T_{a}.
\end{align}

Any sufficiently smooth bi-tensor $T_{a_1 \cdots a_m a_1' \cdots a_n'}$ may be expanded in a local covariant Taylor series about the point $x$:
\begin{equation}
\label{eq:AxExpansion}
 T_{a_1 \cdots a_m a_1' \cdots a_n'} (x,x')= \sum_{k=0}^\infty \frac{1}{n!} t_{a_1 \cdots a_m a_1' \cdots a_n' ~ \alpha_1 \cdots \alpha_k} (x) \sigma^{\alpha_1} \cdots \sigma^{\alpha_k}
\end{equation}
where the $t_{a_1 \cdots a_m a_1' \cdots a_n' ~ \alpha_1 \cdots \alpha_k}$ are the coefficients of the series and are local tensors at $x$. Similarly, we can also expand about $x'$:
\begin{equation}
 T_{a_1 \cdots a_m a_1' \cdots a_n'} (x,x')= \sum_{k=0}^\infty \frac{1}{n!} t_{a_1 \cdots a_m a_1' \cdots a_n' ~ \alpha_1' \cdots \alpha_k'} (x') \sigma^{\alpha_1'} \cdots \sigma^{\alpha_k'}
\end{equation}

For many fundamental bi-tensors, one would typically use the DeWitt approach \cite{DeWitt:1960} to determine the coefficients in these expansions as follows:
\begin{enumerate}
 \item Take covariant derivatives of the defining equation for the bi-tensor (the number of derivatives required depends on the order of the term to be found).
 \item Replace all known terms with their coincidence limit, $x \rightarrow x'$.
 \item Sort covariant derivatives, introducing Riemann tensor terms in the process.
 \item Take the coincidence limit $x' \rightarrow x$ of the result.
\end{enumerate}
This method allows all coefficients to be determined recursively in terms of lower order coefficients and Riemann tensor polynomials. Although this method proves effective for determining the lowest few order terms by hand \cite{Christensen:1976vb,Christensen:1978yd} and can be readily implemented in software \cite{Christensen:1995}, it does not scale well with the order of the term considered and it is not long before the computation time required to calculate the next term is prohibitively large. This issue can be understood from the fact that the calculation yields extremely large intermediate expressions which simplify tremendously in the end. The fact that the final expressions are so short relative to these intermediate expressions suggests that the algorithm is lacking efficiency. It is therefore desirable to find an alternative approach which is more efficient and better suited to implementation in software. In Chapter~\ref{ch:covex}, we will describe one such approach which proves to be highly efficient.

\section{Black Hole Spacetimes}
From astrophysical considerations, EMRI systems are expected to consist of either static, spherically symmetric black holes, or, more likely, axially symmetric spinning black holes. In this section, we will introduce some black hole spacetimes which are of particular interest for self-force calculations.

\subsection{Spherically Symmetric Spacetimes}
\subsubsection{Schwarzschild} \label{subsubsec:schw}
The Schwarzschild spacetime represents a static, spherically symmetric black hole parametrized by a single quantity, its mass, $M$. It has a line element given by
\beq
\label{eq:schwle}
ds^2 = -\left(1 - \frac{2M}{r}\right) dt^2 + \left(1 - \frac{2M}{r}\right)^{-1} dr^2 + r^2 d \Omega^2_2, \quad d\Omega_2^2=d\theta^2+\sin^2\theta d\phi^2.
\eeq
Schwarzschild is a vacuum spacetime, having vanishing Ricci tensor and scalar.

The equations governing the geodesics in Schwarzschild spacetime can be derived from the Lagrangian \cite{Chandrasekhar,Hartle},
\begin{equation}
 \mathcal{L} = \frac12 \left[-\dot{t}^2 \left(1 - \frac{2M}{r}\right) + \dot{r}^2 \left(1 - \frac{2M}{r}\right)^{-1} + r^2 \dot{\theta}^2 + (r^2 \sin^2\theta) \dot{\phi}^2 \right],
\end{equation}
where the overdot indicates differentiation with respect to $s$ and where $\mathcal{L}=-1$ for timelike geodesics and $\mathcal{L}=0$ for null geodesics. Since the spacetime is spherically symmetric, without loss of generality, we can set $\theta=\pi/2$ and $\dot{\theta}=0$. The geodesics are then parametrized by two constants of the motion
\begin{align}
 e = \dot{t} \left(1 - \frac{2M}{r}\right)
\end{align}
and
\begin{align}
 l = r^2 \dot{\phi},
\end{align}
corresponding to the energy per unit mass and angular momentum per unit mass, respectively. Using this parametrization, the geodesic equations are:
\begin{subequations}
\begin{align}
 \dot{t} =& \frac{e}{\left(1 - \frac{2M}{r}\right)}\\
 \dot{r} =& \pm \sqrt{e^2-\left(1-\frac{2M}{r}\right)\left(\epsilon + \frac{l^2}{r^2}\right)}\\
 \dot{\phi} =& \frac{l}{r^2},
\end{align}
\end{subequations}
where $\epsilon = 1$ for timelike geodesics, $\epsilon=0$ for null geodesics and the sign of the square root in the radial equation is chosen dependent on whether the radial motion is inwards or outwards. For the purposes of numerically solving the geodesic equations, this sign choice is troublesome and it proves better to work with the second order version of the radial equation\footnote{In practise, Eq.~\eqref{eq:rddot-schw} can be numerically solved by writing it as a pair of coupled first order equations for $r$ and $\dot{r}$.},
\begin{equation}
\label{eq:rddot-schw}
 \ddot{r} = \frac{l^2 (r-3M)}{r^4} - \frac{M\epsilon}{r^2},
\end{equation}
which is independent of whether the geodesic is moving inwards or outwards in the radial direction. We also point out that it is highly advantageous in a numerical integration (particularly in the case of elliptical orbits) to make use of an algorithm incorporating adaptive step-sizing. Furthermore, algorithms which make use of the Jacobian (such as Burlisch-Stoer) perform orders of magnitude better than those that do not.

\begin{figure}
 \begin{center}
 \includegraphics[width=14.5cm]{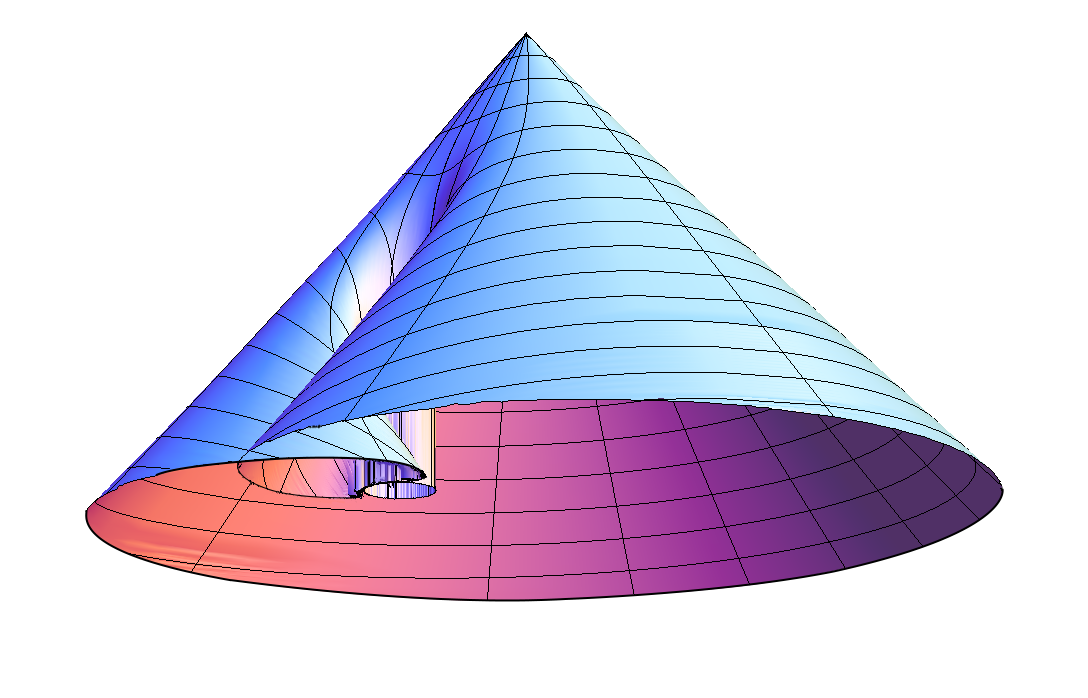}
 \includegraphics[width=14.5cm]{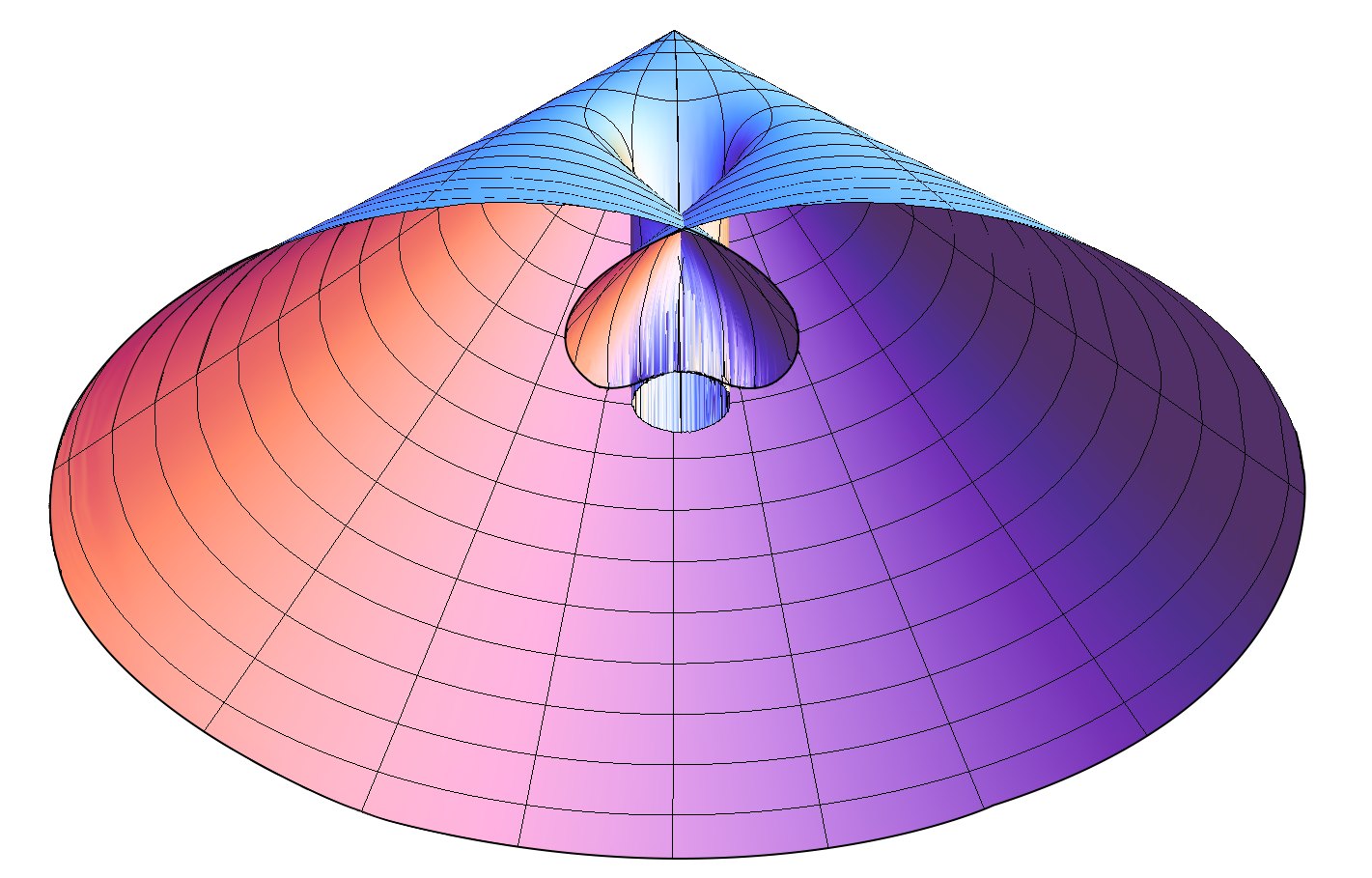}
\end{center}
\caption[Schwarzschild light cone]{\emph{The light cone in Schwarzschild spacetime is self-intersecting.} Geodesics initially moving in different directions will re-intersect at a later time. The spherical symmetry of Schwarzschild means that an infinite number of geodesics intersect at a \emph{caustic}. These caustics appear in the image as the points where the cone intersects itself. The Hadamard form for the Green function is valid for points separated along a geodesic, provided there is no other geodesic also connecting those points (i.e. within a normal neighborhood). In this light-cone diagram, the normal neighborhood for geodesics emanating from the vertex of the cone is the interior of the cone (and including the cone itself), excluding the region in the middle-left bounded by the null geodesics which have already passed through a caustic.}
\label{fig:schw-light-cone}
\end{figure}

There is a null circular orbit at $r=3M$. This means that null geodesics which come close to this orbit may circle the black hole one or more times before going out to infinity. As a result, we find that the light cone in Schwarzschild is self-intersecting (see Fig.~\ref{fig:schw-light-cone}\footnote{This figure is inspired by one  previously produced by Perlick \cite{Perlick}.}), resulting in \emph{caustics}, points where neighboring geodesics converge, at the antipodal points (i.e. at an angle $\pi$, equivalently the opposite side of the black hole).

\subsubsection{Reissner-N\"ordstrom}
The Schwarzschild spacetime may be generalized to allow the black hole to have a charge, $Q$, resulting the Reissner-Nordstr\"{o}m spacetime, with line element
\begin{equation}
\label{eq:RNMetric}
ds^2 = - \left( 1-\frac{2M}{r} + \frac{Q^2}{r^2}\right) dt^2 + \left( 1-\frac{2M}{r} + \frac{Q^2}{r^2}\right)^{-1} dr^2 + r^2 d\Omega_2^2.
\end{equation}
Reissner-N\"ordstrom has a vanishing Ricci scalar, but non-vanishing Ricci tensor. This property will prove useful in calculating the self-force for non-geodesic motion in Chapter~\ref{ch:qlsf}. We will see that it highlights the difference made by the non-geodesicity of the motion at a lower order than a spacetime with vanishing Ricci tensor would.

\subsection{Axially Symmetric Spacetimes}
\subsubsection{Kerr}
The spacetime of a spinning black hole is given by the Kerr metric. In Boyer-Lindquist coordinates, its line-element is
\begin{multline}
\label{eq:KerrMetric}
ds^2 = - \left(1-\frac{2Mr}{\Sigma}\right)dt^2 
			- \frac{4aMr\sin^2\theta}{\Sigma}dtd\phi
			+ \frac{\Sigma}{\Delta}dr^2\\
			+ \Sigma d\theta^2
			+ \left(\Delta+\frac{2Mr(r^2+a^2)} {\Sigma}\right) \sin^2\theta d\phi^2
\end{multline}
where
\begin{eqnarray}
\Sigma &=& r^2+a^2\cos^2\theta\\
\Delta &=& r^2-2Mr+a^2
\end{eqnarray}
and the black hole is parametrized by two quantities: its mass, $M$, and its angular momentum per unit mass, $a$. In analogy to the Schwarzschild case, geodesics of the Kerr spacetime can be parametrized by constants of the motion. In addition to two constants $e$ and $l$, which are analogous to those of Schwarzschild, the geodesics of Kerr have a third constant of the motion, the Carter constant \cite{Carter:1968}. In the cases studied in this thesis, however, for simplicity we will restrict ourselves to motion in the equatorial plane (by the symmetry of the spacetime, this motion will remain in the equatorial plane) and will only be concerned with the constants $e$ and $l$.

The equations governing the timelike geodesics in the equatorial plane may be written in terms of these two constants \cite{Chandrasekhar,Hartle}:
\begin{subequations}
\begin{align}
\dot{t} &= \frac{1}{\Delta} \left[ \left(r^2 + a^2 + \frac{2Ma^2}{r} \right)e - \frac{2Ma}{r}l \right]\\
\dot{\phi} &= \frac{1}{\Delta} \left[ \left(1- \frac{2M}{r} \right)l + \frac{2Ma}{r}e \right]\\
\dot{r}^2 &= e^2-1 - 2 V_{\textrm{eff}} (r, e, l),
\end{align}
\end{subequations}
where $V_{\textrm{eff}} (r, e, l)$ is an effective potential given by
\begin{equation}
V_{\rm{eff}}(r,e,l) = -\frac{M}{r} + \frac{l^2-a^2\left(e^2-1\right)}{2r^2}-\frac{M\left(l-ae\right)^2}{r^3}.
\end{equation}
For null geodesics, the radial equation takes a different form and depends on the impact parameter, $b=|l/e|$, and on whether the orbit is with or against the rotation of the black hole (determined by $\hat{l} \equiv \textrm{sign}(l)$) \cite{Chandrasekhar,Hartle}:
\begin{align}
\dot{r}^2 &= e^2-W_{\textrm{eff}} (r, b, \hat{l}),
\end{align}
where $W_{\textrm{eff}} (r, b, \sigma)$ is an effective potential given by
\begin{equation}
W_{\rm{eff}}(r,b,\sigma) = \frac{1}{r^2}\left[1- \left( \frac{a}{b} \right)^2 -\frac{2M}{r}\left(1-\hat{l} \frac{a}{b}\right)^2\right].
\end{equation}

\subsection{Kerr-Newman}
As was the case with Schwarzschild, the Kerr spacetime may be generalized to allow the black hole to have a charge, giving the Kerr-Newman spacetime. In Boyer-Linquist coordinates, the Kerr-Newman metric is \cite{Boyer-Lindquist}:
\begin{multline}
\label{eq:KerrNewmanMetric}
ds^2 = - \frac{\Delta - a^2 + z^2}{\rho^2} dt^2 
	+ \frac{2(\Delta-r^2-a^2)(a^2-z^2)}{a \rho^2}dtd\phi
	+ \frac{\rho^2}{\Delta}dr^2\\
	+ \rho^2  d\theta^2
	+ \frac{a^2-z^2}{a^2 \rho^2} \left( (a^2+r^2)^2 - \Delta (a^2-z^2) \right) d\phi^2, 
\end{multline}
where $\Delta=r^2-2Mr+a^2+Q^2$, $\rho^2 = r^2 + a^2 \cos^2 \theta$ and $z=a \cos \theta$. Here, $Q$ is the charge per unit mass of the black hole. Like Reissner-N\"ordstrom, the Kerr-Newman spacetime has vanishing Ricci scalar, but non-vanishing Ricci tensor, which again will prove useful in Chapter~\ref{ch:qlsf}.

\section{The Nariai Spacetime} \label{sec:Nariai}
The spacetimes of greatest astrophysical interest are undoubtedly Schwarzschild and Kerr. However, in developing new methods and techniques,
it can also prove useful to consider other, simpler spacetimes. One such example is the product spacetime
\begin{equation}
 dS_2 \times S_2,
\end{equation}
called the Nariai spacetime (the line element will be given later, in Eq.~\eqref{eq:Nariai-le}). The deSitter$-2$ subspace, $dS_2$ is a $2$-hyperboloid, while the $S_2$ subspace is a $2$-sphere. Nariai is not a black hole spacetime and so may appear at first to be of little interest for astrophysical self-force
calculations. On the contrary, as will be discussed in this section, Nariai is very useful as a toy model spacetime, giving fundamental insight into
self-force calculations on the Schwarzschild spacetime. We begin this section with some motivation for this use of Nariai as a toy model for
Schwarzschild.

To evaluate the retarded Green function (\ref{modesum1}) we require solutions to the homogeneous scalar field equation on the appropriate curved background. In the absence of sources, the scalar field equation (\ref{scalar-field-eq1}) is  
\beq
 \frac{1}{\sqrt{-g}} \, \partial_\mu \left( \sqrt{-g} g^{\mu \nu} \partial_{\nu} \fld \right) - \xi R \Phi = 0, \label{scalar-field-eq2}
\eeq
where $\xi$ is the curvature coupling constant.
For the Schwarzschild spacetime, the line element is given by \eqref{eq:schwle}.
We decompose the field in the usual way, 
\beq
\fld(x) = \int^{\infty}_{-\infty} d\omega_S \sum_{l=0}^{+\infty}\sum_{m=-l}^{+l} c_{l m \omega_S} \Phi_{l m \omega_S}(x)
\eeq
where
\beq
 \Phi_{l m \omega_S} (x)=
 \frac{u_{l\omega_S}^{(S)}(r)}{r} Y_{lm}(\theta, \phi) e^{-i \omega_S t_S},
\eeq
and where $Y_{lm}(\theta, \phi)$ are the spherical harmonics, $c_{l m \omega_S}$ are the coefficients in the mode decomposition, and the radial function $u_{l \omega_S}^{(S)}(r)$ satisfies the radial equation
\beq
\left[ \frac{d^2}{d \rstar^2} + \omega_S^2 - V_l^{(S)}(r) \right] u_{l\omega_S}^{(S)}(r) = 0  \label{rad-eq-schw}
\eeq
with an effective potential 
\beq
V_l^{(S)}(r) = f_S(r) \left( \frac{l(l+1)}{r^2} + \frac{2M}{r^3} \right)  . \label{Veff-Schw}
\eeq
where $f_S(r) = 1-2M/r$. Here $\rstar$ is a tortoise (Regge-Wheeler) coordinate, defined by 
\beq
\frac{d\rstar}{dr} = f^{-1}_S(r)
\quad \quad \Rightarrow \quad 
\rstar = r + 2M \ln( r/2M - 1 )  -  (3 M - 2M \ln 2 ) .  \label{rstar-defn}
\eeq
The outer region $r\in (2M,+\infty)$ of the Schwarzschild black hole is now covered by $\rstar\in(-\infty,+\infty)$.
Note that we have chosen the integration constant for our convenience so that, in the high-$l$ limit, the peak of the potential barrier (at $r=3M$) coincides with $\rstar = 0$.

\subsection{P\"{o}schl-Teller Potential and Nariai Spacetime}
\begin{figure}
 \begin{center}
  \includegraphics[width=8cm]{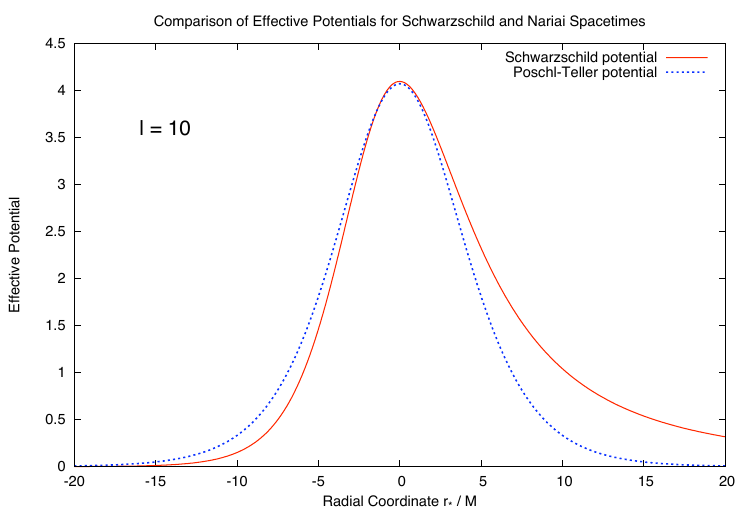}
 \end{center}
 \caption[Effective Potentials for Schwarzschild and P\"{o}schl-Teller radial wave equations]{\emph{Effective Potentials for Schwarzschild (\ref{Veff-Schw}) and P\"{o}schl-Teller (\ref{VPT}) radial wave equations}. Note that here the Schwarzschild tortoise coordinate is defined in (\ref{rstar-defn}) so that the peak is near $\rstar = 0$. The constants in (\ref{VPT}) are $V_0 = l(l+1)$, $\rstar^{(0)} = 0$ and $\alp = 1/(\sqrt{27}M)$.}
 \label{fig:Veff}
\end{figure}

Unfortunately, to the best of our knowledge, closed-form solutions to (\ref{rad-eq-schw}) with potential (\ref{Veff-Schw}) are not known. However, there is a closely-related potential for which exact solutions are available: the so-called P\"{o}schl-Teller potential~\cite{Poschl:Teller:1933},
\beq
V_l^{(PT)}(\rstar) =  \frac{\alp^2 V_0}{\cosh^2(\alp (\rstar-\rstar^{(0)}))}  \label{VPT}
\eeq
where $\alp$, $V_0$ and $\rstar^{(0)}$ are constants ($V_0$ may depend on $l$). Unlike the Schwarzschild potential, the P\"oschl-Teller potential is symmetric about $\rstar^{(0)}$, and decays exponentially in the limit $\rstar \rightarrow \infty$. Yet, like the Schwarzschild potential, it has a single peak, and with appropriate choice of constants, the P\"{o}schl-Teller potential can be made to fit the Schwarzschild potential in the vicinity of this peak (see Fig. \ref{fig:Veff}). 
In the Schwarzschild spacetime, the peak of the potential barrier is associated with the unstable photon orbit at $r = 3M$. This photon orbit may lead to singularities in the Green function. Hence by building a toy model which includes an unstable null orbit, we hope to capture the essential features of the distant past Green function. Authors have found that the P\"{o}schl-Teller potential is a useful model for exploring (some of the) properties of the Schwarzschild solution, for example the quasinormal mode frequency spectrum \cite{Ferrari:Mashhoon:1984, Berti:Cardoso:2006}.

An obvious question follows: is there a spacetime on which the scalar field equation reduces to a radial wave equation with a P\"oschl-Teller potential? The answer turns out to be: yes \cite{Cardoso:Lemos:2003, Zerbini:Vanzo:2004}! The relevant spacetime was first introduced by Nariai in 1950 \cite{Nariai:1950, Nariai:1951}.

To show the correspondence explicitly, let us define the line element
\beq 
 ds^2 = -f(\R) dt_N^2 + f^{-1}(\R) d\R^2 + d\Omega^2_2,   \label{eq:Nariai-le}
\eeq
where $f(\R) = 1 - \R^2$ and $\R \in (-1,+1)$. Line element (\ref{eq:Nariai-le}) describes the central diamond of the Penrose diagram of the Nariai spacetime (Fig. \ref{fig:penrose}), which is described more fully in Sec.~\ref{subsec:nariai}.  Consider the homogeneous wave equation (\ref{scalar-field-eq2}) on this spacetime.
We seek separable solutions of the form $\fld (x)= u^{(N)}_{l\omega_N}(\R) Y_{lm}(\theta, \phi) e^{-i \omega_N t_N}$, where the label `N' denotes `Nariai'. The radial function satisfies the equation
\beq
f(\R) \frac{d}{d \R} \left( f(\R) \frac{d u^{(N)}_{l\omega_N} }{d \R} \right)   +   \left( \omega_N^2 - f(\R) [ l (l+1) + \xi R] \right) u^{(N)}_{l\omega_N} (\R) = 0  \label{Nar-rad-eq1}
\eeq
where $\xi$ is the curvature coupling constant and $R = 4$ is the Ricci scalar. Now let us define a new tortoise coordinate in the usual way,
\beq
\frac{d \Rs}{d \R} = f^{-1}(\R) \quad \Rightarrow \quad \Rs = \tanh^{-1} \R .   \label{eq:rhostar}
\eeq
Note that $f(\R) = \sech^2(\Rs)$ and the tortoise coordinate is in the range
\beq
 \Rs \in (-\infty,+\infty).
\eeq
Hence the radial equation (\ref{Nar-rad-eq1}) may be rewritten in P\"oschl-Teller form,
\beq
\left(   \frac{d^2 }{d \Rs^2} + \omega_N^2 - \frac{U_0}{\cosh^2 \Rs}  \right) u^{(N)}_{l\omega_N}(\Rs) =  0
\label{rad-eq-nar}
\eeq
where
$
U_0 =  l(l+1) + 4\xi   \label{U0-def}.
$
We take the point of view that, as well as being of interest in its own right, the Nariai spacetime can provide insight into the propagation of waves on the Schwarzschild spacetime. The closest analogy between the two spacetimes is found by making the associations
\beq
 \Rs   \rightleftharpoons  \alp \rstar  ,   \quad \quad t_N   \rightleftharpoons    \alp t_S   , \quad \quad \omega_N   \rightleftharpoons \omega_S / \alp \quad \quad \text{where} \quad \alp = 1/(\sqrt{27}M) .
\eeq
Fig.~\ref{fig:Veff} shows the corresponding match between the potential barriers $V_l^{(S)}(\rstar)$ and $V_l^{(PT)}(\Rs)$. In the following sections, we drop the label `N', so that $t \equiv t_N$ and $\omega \equiv\omega_N$. 

The solutions of Eq. (\ref{rad-eq-nar}) are presented in Sec.~\ref{subsec:radial-solns}. First, though, we consider the Nariai spacetime in more detail.

\subsection{Nariai spacetime}\label{subsec:nariai}
\begin{figure}[htb]
\begin{center}
\includegraphics[width=14cm]{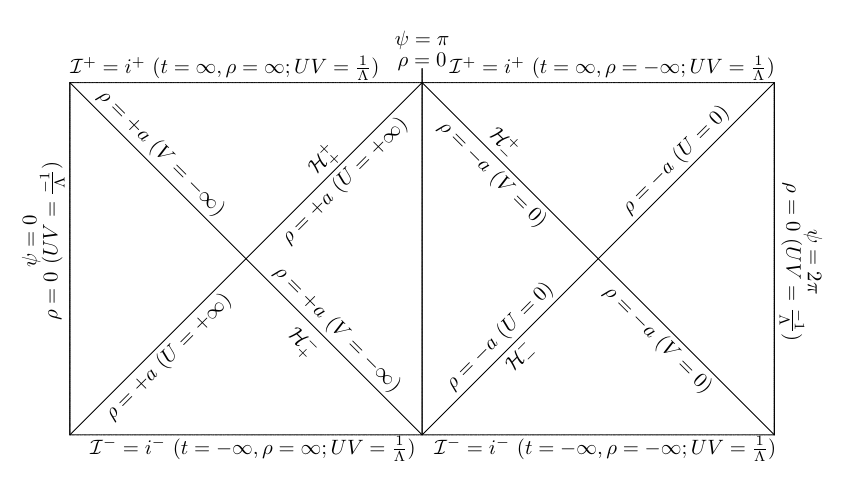}
\caption[Penrose diagram for the Nariai spacetime]{\emph{Penrose diagram for the Nariai spacetime} in coordinates $(\psi,\zeta)$. 
The hypersurfaces $\psi=0$ and $\psi=2\pi$ are identified. 
Past/future timelike infinity $i^{-/+}$ coincides with past/future null infinity 
$\mathcal{I}^{-/+}$, and they are all spacelike hypersurfaces.
Thus, there exist observer-dependent past and future cosmological event-horizons, here marked as $\mathcal{H}^{\pm}_{\pm}$
for an observer along $\psi=\pi$.
}
\label{fig:penrose}
\end{center}
\end{figure}

The Nariai spacetime~\cite{Nariai:1950, Nariai:1951} may be constructed from an embedding in a 6-dimensional Minkowski space
\beq
ds^2=-dZ_0^2+\sum_{i=1}^5dZ_i^2
\eeq
of a 4-D surface determined by the two constraints,
\beq
-Z_0^2+Z_1^2+Z_2^2=a^2, \qquad Z_3^2+Z_4^2+Z_5^2=a^2, \qquad \text{where\ } a>0,
\eeq
corresponding to a hyperboloid and a sphere, respectively.
The entire manifold is covered by the coordinates $\{\mathcal{T},\psi,\theta,\phi\}$ defined via
\begin{align}
&Z_0=a\sinh\left(\frac{\mathcal{T}}{a}\right),\quad &Z_1&=a\cosh\left(\frac{\mathcal{T}}{a}\right)\cos\psi,\quad  &Z_2&=a\cosh\left(\frac{\mathcal{T}}{a}\right)\sin\psi,\quad \\
&Z_3=a\sin\theta\cos\phi,\quad  &Z_4&=a\sin\theta\sin\phi,\quad  & Z_5&=a\cos\theta,
\end{align}
with $\mathcal{T}\in(-\infty,+\infty), \psi\in [0,2\pi), \theta\in [0,\pi], \phi\in [0,2\pi)$. The line-element is given by
\beq
ds^2=-d\mathcal{T}^2+a^2\cosh^2\left(\frac{\mathcal{T}}{a}\right)d\psi^2+a^2d\Omega_2^2.
\eeq
From this line-element one can see that the spacetime has the following features:
(1) it has geometry $dS_2\times \mathbb{S}^2$ and topology $\mathbb{R}\times\mathbb{S}^1\times\mathbb{S}^2$ (the radius of the 1-sphere diminishes with time down
to a value $a$ at $\mathcal{T}=0$ and then increases monotonically with time $\mathcal{T}$,
whereas the 2-spheres have {\it constant} radius $a$),
(2) it is symmetric (ie, $R_{\mu\nu\rho\sigma;\tau}=0$), with $R_{\mu \nu} = \Lambda g_{\mu \nu}$, and constant Ricci scalar, $R = 4\Lambda$, 
where $\Lambda=1/a^2$ is the value of the cosmological constant,
(3) it is spherically symmetric (though not isotropic), homogeneous and locally (not globally) static,
(4) its conformal structure can be obtained by noting the Kruskal-like coordinates defined via 
$U= -(1-\Lambda UV)(Z_0+Z_1)/2,\ V= -(1-\Lambda UV)(Z_0-Z_1)/2$, for which the line-element is then
\beq
ds^2=-\frac{4dUdV}{\left(1-\Lambda UV\right)^2}+d\Omega_2^2.
\eeq
Its two-dimensional conformal {\it Penrose diagram} is shown in Fig.~\ref{fig:penrose} (see, e.g.,~\cite{Ortaggio:2002}),
where we have defined the conformal time $\zeta\equiv 2\exp\left(\mathcal{T}/a\right)\in(0,\pi)$.
Its Penrose diagram differs from that of de Sitter spacetime in that here each point represents a 2-sphere of {\it constant} radius;
note also that the corresponding angular coordinate $\psi$ in de Sitter spacetime has a different range, $\psi\in  [0,\pi)$,
as corresponds to its $\mathbb{R}\times\mathbb{S}^3$ topology.
Past and future timelike infinity $i^{\pm}$ coincide with past and future null infinity $\mathcal{I}^{\pm}$, respectively,
and they are all spacelike hypersurfaces. A consequence of the latter fact is the existence of 
{\it `past/future (cosmological) event horizons'}~\cite{Hawking:Ellis,Gibbons:Hawking:1977,Ortaggio:2002}:
not all events in the spacetime will be influentiable/observable by a geodesic observer;
the boundary of the future/past of the worldline of the observer is its past/future (cosmological) event horizon.

In this thesis, we consider the {\it static} region of the Nariai spacetime which is covered by the coordinates $\{t,\R,\theta,\phi\}$, where 
$\R \equiv a\tanh (\Rs/a)\in(-a,+a),\ \Rs\equiv (v-u)/2\in (-\infty,+\infty),\ t\equiv (v+u)/2 \in (-\infty,+\infty)$ and the null coordinates $\{u,v\}$ are given via $U=ae^{-u/a},\ V=-ae^{v/a}$. 
This coordinate system, $\{t,\R,\theta,\phi\}$, covers the diamond-shaped region in the Penrose diagram (Fig.~\ref{fig:penrose}) around the hypersurface, say, $\psi=\pi$ 
(because of homogeneity we could choose any other $\psi=constant$ hypersurface).
We denote by $\mathcal{H^-_{\pm}}$ the past cosmological event horizon at $\R=\pm a$ of an observer moving along $\psi=\pi$;
similarly, $\mathcal{H^+_{\pm}}$ will denote its future cosmological event horizon at $\R=\pm a$.
Interestingly, Ginsparg and Perry~\cite{Ginsparg:Perry:1983} showed that this static region is obtained from the 
Schwarzschild-deSitter black hole spacetime
as a particular limiting procedure in which the event and cosmological horizons of Schwarzschild-deSitter coincide 
(see also~\cite{Dias:Lemos:2003,Cardoso:Dias:Lemos:2004,Bousso:Hawking:1995,Bousso:Hawking:1996}).

Note that there are three hypersurfaces $\rho=0$, only two of which (those corresponding to $\psi=0$ and $2\pi$) are identified
(the one corresponding to $\psi=\pi$ is not).
Without loss of generality, we will take $\Lambda=1=a$.
The line-element corresponding to this static coordinate system is given in (\ref{eq:Nariai-le}).

\subsection{Geodesics on Nariai spacetime}\label{subsec:Nariai-geodesics}
Let us now consider geodesics on the Nariai spacetime. Our chief motivation is to find the orbiting geodesics, the analogous rays to those shown in Fig.~\ref{fig:circ_orbits} for the Schwarzschild spacetime. We wish to find the coordinate times $t - t'$ for which two angularly-separated points \emph{at the same `radius'}, $\R$, (the quotes indicate the fact that the area of the 2-spheres does not depend on $\R$) may be connected by a null geodesic. We expect the Green function to be singular at these times.

We will assume that particle motion takes place within the central diamond of the Penrose diagram in Fig.~\ref{fig:penrose}; that is, the region $-1 < \R < 1$ (notwithstanding the fact that timelike geodesics may pass through the future horizons $\mathcal{H}^+_+$ and $\mathcal{H}^+_-$ in finite proper time). Without loss of generality, let us consider motion in the equatorial plane $(\theta = \pi / 2)$ described by the world line $z^\mu(\lam) = [t(\lam), \R(\lam), \pi/2, \phi(\lam)]$ with tangent vector $u^\mu = [\dot{t}, \dot{\R}, 0, \dot{\phi}]$, where the overdot denotes differentiation with respect to an affine parameter $\lam$. Symmetry implies two constants of motion, $k = f(\R) \dot{t}$ and $h = \dot{\phi}$. The radial equation is $\dot{\R} = \pm H (\R^2 - \R_0^2)^{1/2}$ where $\R_0 = \sqrt{1 - k^2/H^2}$ is the closest approach point and $H^2 = h^2 + \kappa m_0^2$. Here, $m_0$ is the scaling of the affine parameter (which is equal to the particle's rest mass in the case of a massive particle
following a timelike geodesic) and $\kappa = +1$ for timelike geodesics, $\kappa = 0$ for null geodesics, and $\kappa = -1$ for spacelike geodesics. We still have the freedom to rescale the affine parameter, $\lam$, by choosing a value for $m_0$. It is conventional to rescale so that $\lam$ corresponds to proper time or distance, that is, set $m_0 = 1$. Instead, we will rescale so that $\lam = \phi$, that is, we set $h=1$.

Let us consider a geodesic that starts at $\R = \R_1$, $\phi = 0$ which returns to `radius' $\R = \R_1$ after passing through an angle of $\phif$ (N.B. $\phif$ is unbounded, $\phif\in [0,+\infty)$. In some cases we will refer to the bounded minimal angular distance, $\gamma\in [0,\pi]$). The geodesic distance in this case is $s = -\kappa (H^2 - 1)^{1/2} \phif$. It is straightforward to show that
\beq
\R(\phi) = \R_1 \frac{\cosh \left(H\phi - H\phif/2\right)}{\cosh \left(H \phif / 2 \right)} ,  \label{yphi}
\eeq
hence the radius of closest approach, $\rho_0$, is
\beq
\R_0 = \R_1 \sech(H \phif / 2)   \label{y0} .
\eeq
The coordinate time $\Delta t_1$ it takes to go from $\R=\R_1$, $\phi = 0$ to $\R=\R_1$, $\phi = \phif$ is
\beq
\Delta t_1 = 2 \rho_{*_1} + \ln \left( \frac{\R_1 - \R_0^2 + \sqrt{(1-\R_0^2)(\R_1^2-\R_0^2)}}{\R_1 + \R_0^2 - \sqrt{(1-\R_0^2)(\R_1^2 - \R_0^2)}} \right)
\label{tbar-def1}
\eeq
where $\rho_{*_1}  = \tanh^{-1}(\R_1)$.  Substituting (\ref{y0}) into (\ref{tbar-def1}) yields $\Delta t_1$ as a function of the angle $\phif$,
\beq
\Delta t_1 = 2 \rho_{*_1} + \ln \left( \frac{1-\R_1 \sech^2(H \phif/2) + \tanh(H \phif/2) \sqrt{1 - \R_1^2 \sech^2(H \phif/2)}}{ 1+ \R_1 \sech^2(H \phif/2) - \tanh(H \phif/2) \sqrt{1 - \R_1^2 \sech^2(H \phif/2) }}  \right).  \label{geodesic-time}
\eeq
This takes a particularly simple form as $\R_1 \rightarrow 1$,
\beq
\Delta t_1 \sim 2 \rho_{*_1} + \ln\left( \sinh^2( H \phif / 2 ) \right),\quad\text{for}\quad\ \R_1 \rightarrow 1.  \label{sing-time-inf}
\eeq
As $\phif \rightarrow \infty$, the geodesic coordinate time increases linearly with the orbital angle $\phif$
\beq
\Delta t_1 \sim 2 \rho_{*_1} +  H \phif ,\quad \text{for}\quad\phif \rightarrow \infty, \R_1 \rightarrow 1.
\label{sing-time-periodic}
\eeq
In other words, for fixed spatial points near $\R= 1$, the geodesic coordinate times $\Delta t_1$ are very nearly periodic, with period $2 \pi H$. Results (\ref{yphi}), (\ref{geodesic-time}), (\ref{sing-time-inf}) and (\ref{sing-time-periodic}) will prove useful when we come to consider the singularities of the Green function in Secs. \ref{subsec:Poisson} and \ref{subsec:Hadamard}. 

\subsection{Hadamard Green Function: Exact solution} \label{sec:nariai-exact}
With Nariai being a highly symmetric product spacetime, it is not surprising that analytic expressions for many functions of the affine length can be found. The spacetime can be separated almost entirely into its two constituent subspaces, with each subspace then being treated independently. The world function, $\sigma$, for the total spacetime then separates into a component from the $dS_2$ part, $\lambda$, and a component from the $S_2$ part, $\gamma$, \cite{Nolan:2009}
\begin{equation}
\label{eq:sigma-nariai-separated}
 \sigma = \hat \sigma + \tilde \sigma = -\frac12 \lambda^2 + \frac12 \gamma^2.
\end{equation}
where we use the convention that $\hat{~}$ denotes a quantity in the $dS_2$ part and $\tilde{~}$ denotes a quantity in the $S_2$ part. Each of these components remains symmetric in $x$ and $x'$.
As a result, the connection coefficients, metric and wave operator also separate,
\begin{align}
 \Gamma^{a}_{b c} =&
  \left(\begin{array}{c|c}
    \hat\Gamma^{\hat a}_{\hat b \hat c} & 0 \\ \hline
    0 & \tilde\Gamma^{\tilde a}_{\tilde b \tilde c}
  \end{array}\right), & 
 g_{a b} =&
  \left(\begin{array}{c|c}
    \hat g_{\hat a \hat b} & 0 \\ \hline
    0 & \tilde g_{\tilde a \tilde b}
  \end{array}\right),
 & \Box =& \hat\Box + \tilde\Box.
\end{align}
Furthermore, we have an independent defining equation, \eqref{eq:sigma-def}, for each component of the world function
\begin{align}
 \hat\sigma^\alpha \hat\sigma_\alpha &= 2 \hat \sigma & 
 \tilde\sigma^\alpha \tilde\sigma_\alpha &= 2 \tilde \sigma.
\end{align}
Finally, it is shown in Ref.~\cite{Nolan:2009} that the affine parameter can be written in analytic form for each subspace:
\begin{subequations}
\label{eq:gamma-lambda-nariai}
\begin{align}
 \gamma &= \cos^{-1} \left[ \cos \left(\phi - \phi'\right) \sin \theta \sin \theta' + \cos \theta \cos \theta' \right]\\
 \lambda &= \cosh^{-1} \left[ \sqrt{(1 - \rho^2)(1 - {\rho'}^2)}~ \cosh \left(t - t'\right) + \rho \rho'\right].
\end{align}
\end{subequations}
For null geodesics, given the choice of affine parameter given in Sec.~\ref{subsec:Nariai-geodesics} and since $\sigma = 0$, we find that both affine parameters are equal to $\phi$ (the angle coordinate on $S_2$):
\begin{align}
 \gamma & = \phi & \lambda = \phi,
\end{align}
and the tangent vectors are
\begin{align}
 u_\theta &= 1 & u_\phi &= -1 & u_\rho &= \lambda_{,\rho} & u_t &= \lambda_{,t}~.
\end{align}
In this case, we can view null geodesic motion in the full spacetime as consisting of a spacelike geodesic in the $S_2$ subspace and a timelike geodesic in the $dS_2$ subspace. Without loss of generality, we restrict geodesics to the plane $\theta=\pi/2$. For geodesics outside the $\theta=\pi/2$ plane, one simply rotates the coordinates so that the geodesics are in the plane.

Both subspaces are maximally symmetric, so we may now apply the results of \cite{Allen:Jac:86} to find analytic expressions for $\hat\Box\lambda = -\coth\lambda$, $\tilde\Box\gamma=\cot\lambda$, $g_{a b'}$, $\sigma_{a b}$, $\sigma_{a b'}$ and their derivatives (it is also straightforward to calculate derivatives of $\sigma$ without the use of Ref.~\cite{Allen:Jac:86} by covariantly differentiating Eq.~\eqref{eq:sigma-nariai-separated} and making use of \eqref{eq:gamma-lambda-nariai}). This yields
\begin{subequations}
\allowdisplaybreaks
\begin{align}
 \hat g_{a}{}^{b'} &= - \frac{\sinh \lambda}{\lambda} \hat \sigma_{a b'} - \left(\frac{\sinh \lambda}{\lambda} - 1\right) \lambda_{,\rho} \lambda_{,\rho'} \hat{g}^{\rho \rho'}\\
 \tilde g_{a}{}^{b'} &= \delta_{a}{}^{b'}\\
 \hat g_{a b' ;c'} &= -\tanh{\frac{\lambda}{2}} \left( \hat g_{c' b'} \hat{u}_{a} + \hat g_{c' a} \hat{u}_{b'}\right)\\
 \tilde g_{a b' ;c'} &= -\tan{\frac{\gamma}{2}} \left( \tilde g_{c' b'} \tilde{u}_{a} + \tilde g_{c' a} \tilde{u}_{b'}\right)\\
 \hat \sigma_{a' b'} &= -\left(\lambda_{,a'} \lambda_{,b'} + \lambda \lambda_{,a' b'} - \hat \Gamma^{ \alpha'}_{a' b'} \lambda_{, \alpha'}\right)\\
 \hat \sigma_{a b'} &= -\left(\lambda_{,a} \lambda_{,b'} + \lambda \lambda_{,a b'}\right)\\
 \hat \Box \hat \sigma &= 1 + \lambda \coth \lambda\\
 \tilde \Box \tilde \sigma &= 1 + \gamma \cot \gamma \\
 \tilde \sigma_{a' b'} &= \left(\gamma_{,a'} \gamma_{,b'} + \gamma \gamma_{,a' b'} - \tilde \Gamma^{ \alpha'}_{a' b'} \gamma_{,\alpha'}\right)\\
 \tilde \sigma_{a b'} &= \left(\gamma_{,a} \gamma_{,b'} + \gamma \gamma_{,a b'}\right)\\
 \hat \Delta^{1/2} &= \sqrt{\lambda}~{\sinh{\lambda}}\\
 \tilde \Delta^{1/2} &= \sqrt{\gamma}~{\sin{\gamma}}\\
 \hat \sigma_{a' b' c'} &= \left(\hat \sigma_{a' b'}\right)_{;c'}\\
 \tilde \sigma_{a' b' c'} &= \left(\tilde \sigma_{a' b'}\right)_{;c'}\\
 \hat \sigma_{a' b' c' d'} &= \left(\hat \sigma_{a' b'}\right)_{;c' d'}\\
 \tilde \sigma_{a' b' c' d'} &= \left(\tilde \sigma_{a' b'}\right)_{;c' d'}\\
 \Box \Delta^{1/2} &= \hat \Box \hat\Delta^{1/2} + \tilde \Box \tilde \Delta^{1/2}\nonumber \\
 &= \frac12 \left(4 - \coth^2 \lambda + \cot^2 \gamma \right) \sqrt{\frac{\gamma}{\sin{\gamma}} \frac{\lambda}{\sinh{\lambda}}}
\end{align}
and, for null geodesics,
\begin{align}
 V_0 &= -\frac18 \left(2\lambda + \coth\lambda - \cot\lambda\right)\left(\sinh\lambda \sin\lambda\right)^{-1/2}.
\end{align}
\end{subequations}

These expressions allow us to compute all of these quantities exactly and will prove useful as a check on alternative methods for calculating the Green function. In particular, analytic expressions for these quantities were invaluable as a way of verifying our numerical code described in Sec.~\ref{sec:numerical}.
\chapter{The Scalar Self-force and Equations of Motion} \label{ch:sf}

The idea of a self-force has a long history in physics. Derivations of expressions for the scalar, electromagnetic and scalar self-force have been given by several researchers \cite{DeWitt:1960,Quinn:Wald:1997,Mino:Sasaki:Tanaka:1996,Quinn:2000,Galley:Hu:Lin:2006,Harte:2008,Gralla:Wald:2008,Futamase:Hogan:Itoh:2008,Flanagan:Hinderer:2008,Norton:2009}. In this thesis, we restrict our attention to the simplest case: a point-like scalar charge $q$ of mass $\mass$ coupled to a massless scalar field $\fld(x)$ moving on a curved background geometry. The scalar field $\fld(x)$ satisfies the field equation
\beq
 \left( \square - \xi R \right) \fld(x)  = - 4 \pi \rho(x)  \label{scalar-field-eq1}
\eeq
where $\square$ is the d'Alembertian on the curved background created by, e.g. a black hole of mass $M$, $R$ is the Ricci scalar, and $\xi$ is the coupling to the scalar curvature. The charge density, $\rho$, of the point particle is 
\beq
 \rho (x) = \int_{\gamma} q \, \frac{\delta^4 (x^\mu - z^\mu(\tau) )}{\sqrt{-g}} \, d \tau
\eeq
where $z(\tau)$ describes the worldline $\gamma$ of the particle with proper time $\tau$, 
$g_{\mu \nu}$ is the background metric, $g = \text{det}( g_{\mu \nu} )$, and $\delta^4(\cdot)$ is the four-dimensional Dirac distribution. The scalar field propagates along and within the null cone, while the particle itself will (approximately) follow a time-like geodesic of the background. As a result the particle may `feel' its own field. In this sense, the field is seen to
exert a radiation reaction on the particle, creating a self-force \cite{Quinn:2000}
\beq \label{S-F}
f_\mu^{\text{self}} = q \nabla_\mu \fld_R
\eeq
which leads to the equations of motion for the scalar particle
\beq
\label{eq:ma-field}
ma^\mu = (g^{\mu \nu} + u^\mu u^\nu ) f_\nu^{\text{self}} = q (g^{\mu \nu} + u^\mu u^\nu ) \nabla_\nu \fld_R
\eeq
where $u^\mu$ is the particle's four-velocity, $a^\mu$ is its four-acceleration and  $\fld_R$ is the {\it radiative} part of the field. Identifying the correct radiative field (which is regular at the particle's position) is the essential step in the derivation of the self-force \cite{Poisson:2003}. Note that the projection operator $g^{\mu \nu} + u^\mu u^\nu$ has been applied here to ensure that $u_{\mu} a^\mu = 0$.
The mass, $m$, appearing in (\ref{eq:ma-field}) is the `dynamical' (and renormalized) particle's mass, which
in the scalar case is not necessarily a constant of motion \cite{Quinn:2000}. Rather, it evolves according to 
\beq
\label{eq:dmdtaufield}
\frac{d \mass}{d \tau} = - q u^\mu \nabla_\mu \fld_R .
\eeq
In other words, a scalar particle may radiate away its mass through the emission of monopolar waves.

\section{MiSaTaQuWa Equation for the Self-force} \label{sec:misataquwa}
A leading method for computing the derivative of the radiative field, $\nabla_\nu \fld_R$, and hence the self-force, is based on \emph{mode sum regularization} (MSR). The MSR approach was developed by Barack, Ori and collaborators \cite{Barack:Ori:2000, Barack:2001, Barack:Mino:Nakano:2002, Barack:Sago:2007} and Detweiler and coworkers \cite{Detweiler:Messaritaki:Whiting:2002, Detweiler:2005, Vega:Detweiler:2008}. The method has been applied to the Schwarzschild spacetime to compute, for example, the gravitational self-force for circular orbits \cite{Barack:Sago:2007} and the scalar self-force for eccentric orbits \cite{Haas:2007}. The application to Kerr is in progress \cite{Barack:Golbourn:Sago:2007, Barack:Ori:Sago:2008}. It was recently shown \cite{Sago:Barack:Detweiler:2008} that the gravitational self-force computed in the Lorenz gauge is in agreement with that found in the Regge-Wheeler gauge \cite{Detweiler:Messaritaki:Whiting:2002, Detweiler:2005}. Further gauge-independent comparisons, and comparison with the predictions of Post-Newtonian theory \cite{Blanchet:Grishchuk:Schaefer:2009, Damour:Nagar:2009} are presently under consideration \cite{Detweiler:2008}.

One drawback of the MSR method is that it gives relatively little geometric insight into the physical origin of the self-force. Other methods \cite{Wiseman:2000,Anderson:Eftekharzadeh:Hu:2006,Anderson:Wiseman:2005,Anderson:Flanagan:Ottewill:2004}, based on the MiSaTaQuWa equation first derived by DeWitt and Brehme in the 1960's, rely on an expression for the derivative of the radiative field involving the integral of the retarded Green function over the entire past worldline of the particle\footnote{Note that we include here the field mass, $m_\text{field}$, for the sake of generality. In the self-force calculations considered in this thesis, we will be considering massless fields, i.e. setting $m_\text{field}=0$.},
\begin{multline}
\label{eq:rad-field-deriv}
\nabla_\mu \fld_R \left( z(\tau) \right) = \\ \left(\frac{1}{2}m_{\rm{field}}^2- \frac{1}{12}(1 - 6\xi) R \right) q u_\mu + q ( g_{\mu \nu} + u_{\mu} u_{\nu} ) \left( \frac{1}{3} \dot{a}^\nu + \frac{1}{6} {R^\nu}_\lam u^\lam  \right)
 + \fldtail\left( z(\tau) \right).
\end{multline}
Here, ${R^\nu}_\lam$ is the Ricci tensor of the background metric, $m_{\rm{field}}$ is the field mass and $\dot{a}^\nu$ is the derivative with respect to proper time of the four-acceleration $a^\nu = \tfrac{d u^{\nu}}{d \tau}$. The first two sets of terms are evaluated locally \cite{Quinn:Wald:1997, Quinn:2000, Poisson:2003} without posing any real difficulty. The task is therefore to elucidate the final, non-local term, $\fldtail$, the so-called \emph{tail integral},
\beq
\fldtail \left( z(\tau) \right)= q \lim_{\eps \rightarrow 0^+} \int_{-\infty}^{\tau-\eps} \nabla_\mu \Gret (z(\tau), z(\tau^\prime)) d \tau^\prime  \label{tail-integral}
\eeq
where $\Gret(x, x^\prime)$ is the retarded scalar Green function and the limiting feature at the upper limit of integration avoids the singular nature of the Green function at the point $x=x'$. Note that the tail integral depends on the entire past history of the particle's motion. The calculation of the Green function poses a formidable challenge and is the main obstacle to progress.

The Green function would normally be evaluated using a mode-sum approach. One would calculate a sufficient number of the modes and sum them up to obtain the Green function solution and hence the self-force. However, the limiting feature in the upper limit of integration causes great difficulties for this method. As $\epsilon$ decreases, it is necessary to take an increasing number of modes to obtain an accurate solution. When $\epsilon$ tends towards 0, the number of modes required becomes infinite and it is no longer possible to accurately compute the integral. A novel solution to this problem, based on \emph{matched expansions}, was suggested by Poisson and Wiseman in 1998 \cite{Poisson:Wiseman:1998}. Their idea was to compute the self-force by matching together two independent expansions for the Green function, valid in `quasilocal' and `distant past' regimes. This suggestion was analyzed by Anderson and Wiseman \cite{Anderson:Wiseman:2005}, who concluded in 2005 that ``this approach remains, in our opinion, in the category of `promising but possessing some technical challenges'.'' The present thesis represents the first practical implementation of this method.

\section{The Method of Matched Expansions}\label{sec:matched-expansions}
\begin{figure}
 \begin{center}
  \includegraphics[width=7cm]{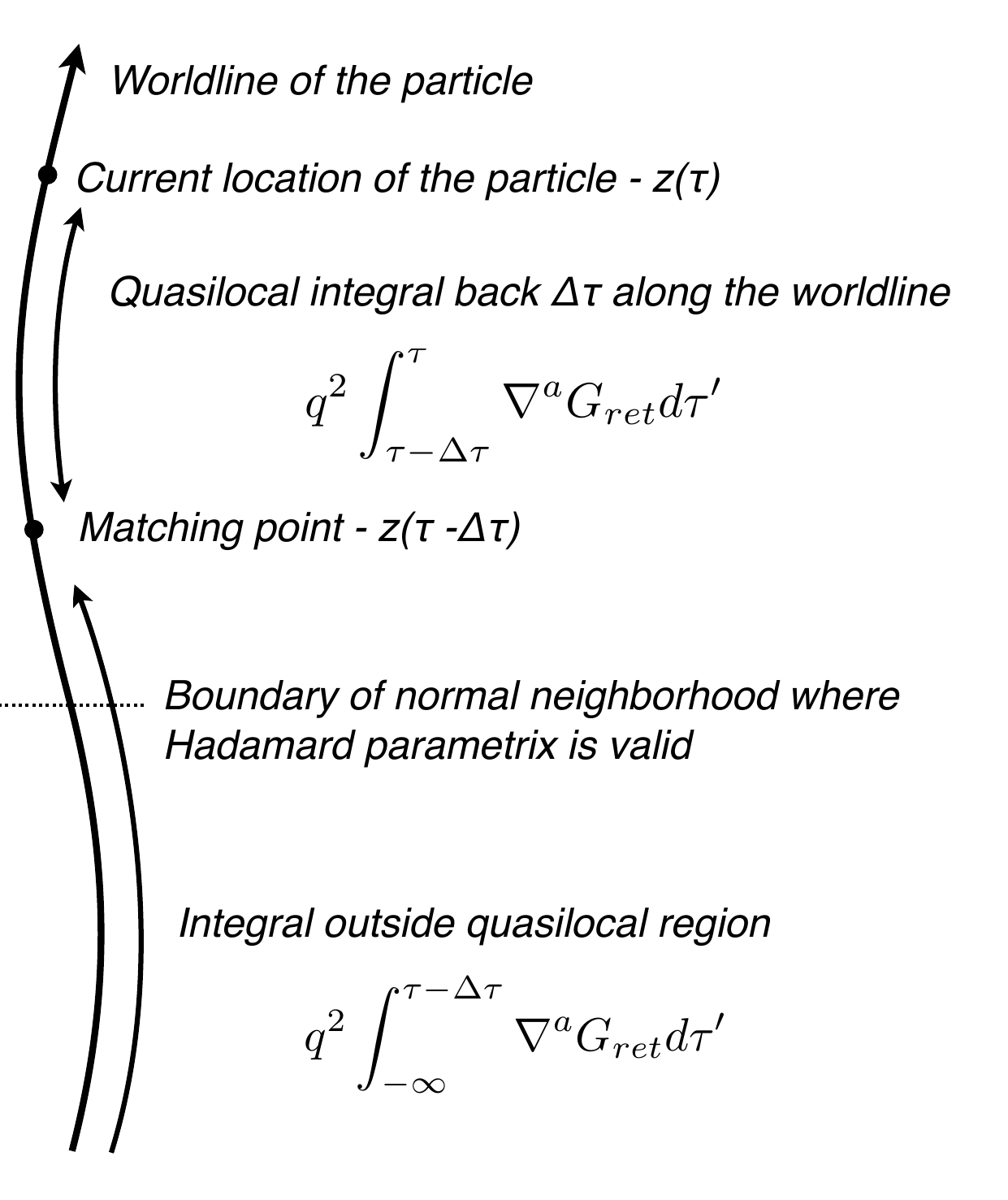}
 \end{center}
 \caption[Method of matched expansions]{\emph{In the method of matched expansions, the tail integral is split into quasilocal (QL) and distant past (DP) parts. }}
 \label{fig:matchedexpansion}
\end{figure}

We now briefly outline the Poisson-Wiseman-Anderson method of `matched expansions' \cite{Poisson:Wiseman:1998, Anderson:Wiseman:2005}.
The tail integral (\ref{tail-integral}) may be split into so-called \emph{quasilocal} (QL) and \emph{distant past} (DP) parts, as shown in Fig.~\ref{fig:matchedexpansion}. That is,
\begin{eqnarray}
\label{eq:SplitForce}
 \fldtail \left( z(\tau) \right) &=&  \fld^{\text{(QL)}}_\mu \left( z(\tau) \right) +  \fld^{\text{(DP)}}_\mu  \left( z(\tau) \right) \label{eq:fld-QL-DP} \nonumber \\
&=& q \lim_{\eps \rightarrow 0^+} \int_{\tau - \Delta \tau}^{\tau - \eps} \nabla_\mu \Gret(z(\tau), z(\tau^\prime)) d\tau^\prime \\
& & \qquad+~ q \int^{\tau - \Delta \tau}_{-\infty} \nabla_\mu \Gret(z(\tau), z(\tau^\prime)) d\tau^\prime
\end{eqnarray}
where $\tau - \Delta \tau$ is the \emph{matching time}, with $\Delta \tau$ being a free parameter in the method (see Fig.~\ref{fig:matchedexpansion}).

The QL and DP parts may be evaluated separately using independent methods. In particular, if we choose $\Delta \tau$  to be sufficiently small so that $z(\tau)$ and $z(\tau - \Delta \tau)$ are within a \emph{normal neighborhood}
(see footnote \ref{def:causal domain}) \cite{Friedlander}, then the QL part may be evaluated by expressing the Green function in the Hadamard parametrix \cite{Hadamard}. In other words, if $z(\tau)$ and $z(\tau - \Delta \tau)$ are connected by a unique non-spacelike geodesic, then the QL integral is simply
\beq
\label{eq:qlsf}
q^{-1} \fld^{\text{(QL)}}_\mu  \left( z(\tau) \right) = - \lim_{\eps \rightarrow 0^+} \int_{\tau - \Delta \tau}^{\tau - \eps} \nabla_\mu V( z(\tau), z(\tau^\prime)) d\tau^\prime~,
\eeq
where $V(x,x^\prime)$ is a smooth, symmetric bi-scalar describing the propagation of radiation within the light cone (see Sec. \ref{subsec:QL} for full details).  
Methods for evaluating the Hadamard Green function will be discussed in the Chapters~\ref{ch:coordex}, \ref{ch:covex} and \ref{ch:qlsf}. The approach ultimately yields an expression for the QL self force in terms of the separation of the points $x$ and $x^\prime$. In Sec. \ref{subsec:QL} we apply these methods to determine the quasilocal Green function, as required for a matched expansion calculation of the self-force in the Nariai spacetime.

Evaluating the contribution to the Green function from the `distant past' is a greater challenge. One possibility is to decompose the Green function into a sum over \emph{angular modes} and an integral over frequency. 
In a spherically symmetric spacetime, the `retarded' Green function may be defined in terms of a Laplace integral transform and an angular $l$-mode decomposition as follows,
\beq
\Gret (x, x^{\prime}) = \frac{1}{2\pi} \int_{-\infty+ic}^{+\infty+ic} d \omega e^{-i \omega (t-t^\prime)} \sum_{l=0}^{+\infty} (2l+1) P_l(\cos \gam) \Grad_{l \omega}(r, r^\prime).
\label{modesum1}
\eeq
Here $c$ is a positive constant, $t$ and $r$ are appropriate time and radial coordinates, and $\cos \gam = \cos \theta \cos \theta^\prime + \sin \theta \sin \theta^\prime \cos( \phi - \phi^\prime )$, where $\gam$ is the angle between the spacetime points $x$ and $x^\prime$. The radial Green function $\Grad_{l \omega}(r, r^\prime)$ may be constructed from two linearly-independent solutions of a radial equation. Since the DP Green function does not need to be extended to coincidence ($\tau^\prime \rightarrow \tau$), the mode sum does not require regularization (though it may still be regularized if desired). However, Anderson and Wiseman \cite{Anderson:Wiseman:2005} found the convergence of the mode sum to be poor, noting that going from 10 modes to 100 increased the accuracy by only a factor of three. 

In Chapter~\ref{ch:nariai}, we explore a new method for evaluating the `\emph{distant past}' contribution, based on an expansion in so-called \emph{quasinormal modes}. 
The integral over frequency in equation (\ref{modesum1}) may be evaluated by deforming the contour in the complex plane \cite{Leaver:1986, Andersson:1997}. This is shown in Fig. \ref{fig:contours}. In the Schwarzschild case there arise three distinct contributions to the Green function, from the three sections of the frequency integral in (\ref{modesum1}):
 \begin{enumerate}
  \item An integral along high-frequency arcs, which leads to a prompt response. Andersson \cite{Andersson:1997} has shown that this is $0$ for black hole spacetimes at sufficiently late times.
  \item A `quasinormal mode sum', arising from the residues of poles in the lower half-plane of complex frequency $\omega$. This contribution leads to a damped ringing, characteristic of the black hole.
  \item An integral along a branch cut, which leads to a power-law tail (among possible other features).
 \end{enumerate}
The three parts (1--3) are commonly supposed to dominate the scattered signal at early, intermediate and late times, respectively \cite{Leaver:1986,Andersson:1997}. (This may be slightly misleading, however; Leaver \cite{Leaver:1986} notes that, in addition, the branch cut integral (part 3) ``contributes heavily to the initial burst of radiation''). In Chapter~\ref{ch:nariai}, we investigate an alternative spacetime, introduced by Nariai in 1950~\cite{Nariai:1950,Nariai:1951} and discussed in Sec.~\ref{sec:Nariai}, in which the power-law tail (part 3) is absent. We demonstrate that, on the Nariai spacetime, at suitably `late times', the distant past Green function may be written as a sum over quasinormal modes (defined in Sec. \ref{subsec:DP}). We use the sum to compute the Green function, the tail field and the self-force for a static particle.

A key question to be addressed in this work is the following: how much of the self-force arises from the quasilocal region, and how much from the distant past? If the Green function falls off fast enough then only the QL integral would be needed, and, since the QL integral is restricted to the normal neighborhood, only the Hadamard parametrix is required. Unfortunately this is not necessarily the case; Anderson and Wiseman \cite{Anderson:Wiseman:2005} note that there are simple situations in which the DP integral in (\ref{modesum1}) gives the dominant contribution to the self-force.

Using the methods presented in Chapter~\ref{ch:nariai}, we are able to compute the retarded Green function and the integrand of Eq.~(\ref{tail-integral}) as a function of time along the past worldline. We will show that the DP contribution cannot be neglected. In particular, we find that the Green function and the integrand of Eq.~(\ref{tail-integral}) is singular whenever the two points $z^{\mu}(\tau)$ and $z^\mu(\tau^\prime)$ are connected by a null geodesic and that the singular form of the Green function changes every time a null geodesic passes through a \emph{caustic}. On a spherically-symmetric spacetime, caustics occur at the antipodal points.

\begin{figure}
 \begin{center}
  \includegraphics[width=15cm]{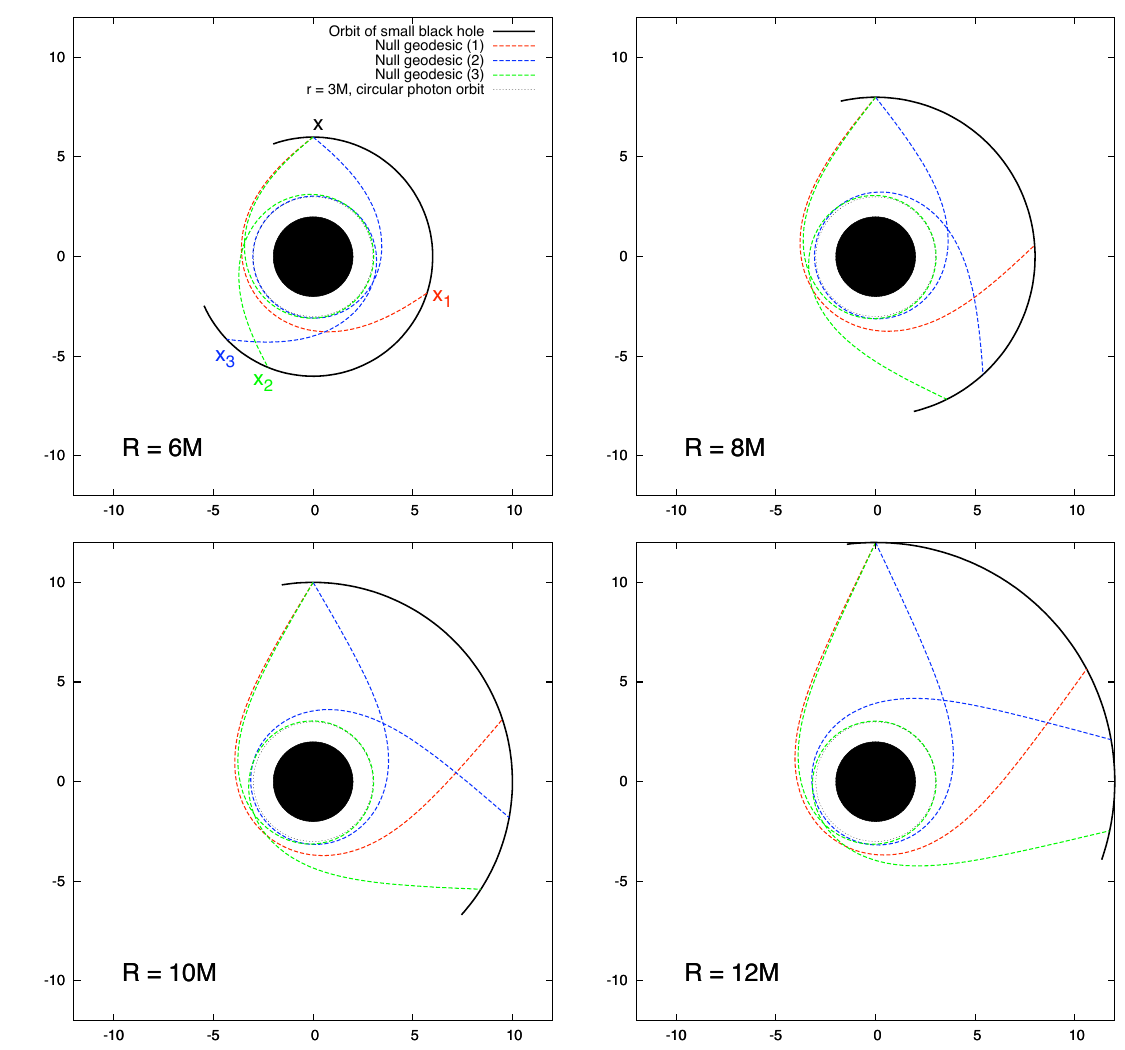}
 \end{center}
 \caption[Orbiting null geodesics on the Schwarzschild spacetime]{\emph{Orbiting null geodesics on the Schwarzschild spacetime that intersect timelike circular orbits of various radii $R = 6M$, $8M$, $10M$ and $12M$}. The null geodesics are shown as dotted (coloured) lines, and the timelike circular orbit is shown as a 
 solid (black) line. The spacetime point $x$ is connected to $x_1, x_2, $ etc.~by null geodesics, as well as by the timelike circular geodesic. The Green function is singular when $x^\prime = x_1, x_2, $ etc. Note that between $R=6M$ and $R=8M$ the ordering of the points $x_2$ and $x_3$ becomes reversed.}
 \label{fig:circ_orbits}
\end{figure}

On Schwarzschild spacetime, the presence of an unstable photon orbit at $r=3M$ implies that a null geodesic originating on a timelike worldline may later re-intersect the timelike worldline, by orbiting around the black hole. Hence the effect of caustics may be significant. For example, Fig. \ref{fig:circ_orbits} shows orbiting null geodesics on the Schwarzschild spacetime which intersect timelike circular orbits of various radii. We believe that understanding the singular behavior of the integrand of Eq.~(\ref{tail-integral}) is a crucial step in understanding the origin of the non-local part of the self-force. As we shall see in Chapter~\ref{ch:nariai}, the Nariai spacetime proves a fertile testing ground.

\subsection{The Static Particle} \label{subsec:static}
Before continuing, we will briefly review the extensive literature that has studied the specific case of the self-force on a static particle. This will be of use in subsequent sections and chapters. 

A static particle -- a particle with constant spatial coordinates -- has been the focus of several scalar self-force calculations, in particular for the Schwarzschild spacetime \cite{SW:1980,Wiseman:2000,Rosenthal:2004,Anderson:Wiseman:2005,Anderson:Eftekharzadeh:Hu:2006,CTW:2007,Ottewill:Wardell:2008}. Although a static particle may not be a particularly physical case, it is frequently chosen because it involves relatively straightforward calculations and has an exact solution for the Schwarzschild spacetime. It therefore provides a good testing ground for new approaches to the calculation of the self-force.

Smith and Will~\cite{SW:1980} calculated the self-force on a static {\it electric} charge in the Schwarzschild background and found
it to be non-zero. In~\cite{Wiseman:2000}, Wiseman considered the analogous case of a static {\it scalar} charge in the case of 
{\it minimal-coupling} (i.e., $\xi=0$) in Schwarzschild. Using isotropic coordinates, he managed to sum the Hadamard series for the Green function in the static case (i.e., the ``Helmholtz"-like equation in Schwarzschild with zero-frequency, $\omega=0$) and thus obtain in closed form the field created by the static charge in the scalar and also electrostatic (already found in~\cite{Copson:1928,Linet:1976} using a different method) cases. He then found the self-force to be zero in the scalar, minimally-coupled case.

In~\cite{CTW:2007}, the calculation of the self-force on a static scalar charge in Schwarzschild is extended to the case of {\it non-minimal coupling} ($\xi\neq 0$) and is found to be zero as well. The fact that the value of the scalar self-force in Schwarzschild is the same (zero) independently of the value of the coupling constant is in agreement with the {\it Quinn-Wald axioms}~\cite{Quinn:Wald:1997,Quinn:2000}:
their method relies only on the field equations, and these are independent of the coupling constant in a Ricci-flat spacetime such as Schwarzschild.
The calculation (without using the Quinn-Wald axioms) is by no means trivial, however, since the effect of the coupling constant might be felt through the stress-energy tensor (in fact,~\cite{CTW:2007} corrected a previous result in~\cite{Zelnikov:Frolov:1982a,Zelnikov:Frolov:1982b}, where the self-force had been incorrectly found to be non-zero). Rosenthal \cite{Rosenthal:2004} has also considered the case of a static particle in Schwarzschild and used it as an example application of the {\it massive field approach} \cite{Rosenthal:2003} to self-force calculations (in Appendix~\ref{app:rosenthal-massive}, we apply this method to the Nariai case).

On the other hand, 
using the conformal invariance of Maxwell's equations,
Hobbs~\cite{Hobbs:1968b} showed that, in a {\it conformally-flat} spacetime, the `tail' contribution (\ref{tail-integral})
to the self-force on an {\it electric} charge (on any motion, static or not) is zero.
The only possible contribution to the self-force might then come from the local Ricci terms, which are zero in cases of physical interest
such as in the de Sitter universe.

One would expect the `tail' contribution to the 
{\it scalar} self-force to also be zero 
for a charge undergoing any motion in a conformally-flat 4-D spacetime with conformal-coupling (i.e., $\xi=1/6$).
In a recent article~\cite{Bezerra:Khus}, it was further argued using Hobbs' result that the self-force should be zero
 for a static charge (where the time-independence effectively reduces the problem to a 3-D spatial one), in a spacetime such that its 3-D spatial section is conformally-flat and with conformal-coupling in 3-D (i.e., $\xi=1/8$).
Indeed, 
they showed
that the scalar self-force on a massless static particle in a {\it wormhole spacetime} (with non-zero Ricci scalar and where the 3-D spatial section is conformally-flat) is equal to zero at $\xi=1/8$ and it actually changes sign at this 3-D conformal value.
[Note that in this case the local contribution to the self-force is exactly zero, so there may only be a `tail' contribution - if any].
It is not completely clear to us, however, that the argument that the `tail' contribution to the self-force is zero due to the conformality of the physical
system carries through intact from a 4-D spacetime to its 3-D spatial section
- Note that, from the Hadamard form in 3-D~\cite{Hadamard,Decanini:Folacci:2005a}, the corresponding
Green function  (the so-called {\it `scalarstatic'} or `electrostatic' Green function)
never (regardless of whether there is conformality or not) possesses a `tail' part $V(x,x')$, and the self-force is calculated from this static
Green function, not via Eqs.(\ref{S-F}) and (\ref{eq:rad-field-deriv}), but rather via, e.g., Eq.(\ref{S-F static}) of Chapter~\ref{ch:nariai}.

In Chapter~\ref{ch:nariai}, we study the self-force on a static particle in Nariai spacetime. The Nariai spacetime, not being Ricci-flat and being conformal to a wormhole spacetime (and so with a conformally-flat 3-D spatial section), suggests a very interesting playground for calculating the self-force: What role does the coupling constant $\xi$ play? Do particular values such as $\xi=1/6$ (4-D conformal-coupling) and
$\xi=1/8$ (3-D conformal-coupling, so a particular value in the case of a static charge) yield `particular' values (namely, zero) for the self-force? Are the Quinn-Wald axioms satisfied?

\section{Quasilocal Contribution: Geodesic Motion}
\label{sec:qlsf-geodesic}
Within the quasilocal region, where $x$ and $x'$ are within a normal neighborhood
\footnote{Anderson and Wiseman \cite{Anderson:Wiseman:2005} have calculated some explicit values for the normal neighborhood boundary for a particle following a circular geodesic in Schwarzschild spacetime. These values place an upper bound estimation on the geodesic distance in the past at which we can place the matching point $\Delta \tau$.} 
, we can use the scalar version of the Hadamard form, \eqref{eq:Hadamard}, for the retarded Green function,
\begin{equation}
\label{eq:Hadamard-scalar}
G_{ret}\left( x,x' \right) = \theta_{-} \left( x,x' \right) \left\lbrace U \left( x,x' \right) \delta \left( \sigma \left( x,x' \right) \right) - V \left( x,x' \right) \theta \left( - \sigma \left( x,x' \right) \right) \right\rbrace 
\end{equation}
where $\theta_{-} \left( x,x' \right)$ is analogous to the Heaviside step-function (i.e. $1$ when $x'$ is in the causal past of $x$, $0$ otherwise), $\delta \left( x,x' \right)$ is the standard Dirac delta function, $U \left( x,x' \right)$ and $V \left( x,x' \right)$ are symmetric bi-scalars having the benefit that they are regular for $x' \rightarrow x$ and $\sigma \left( x,x' \right)$ is the Synge \cite{Synge,Poisson:2003,DeWitt:1965} world function, defined in Sec.~\ref{sec:bitensors}.

The first term (involving $U \left( x,x' \right)$) in Eq. (\ref{eq:Hadamard-scalar}) is the \emph{direct} part of the Green function and does not contribute to the self-force. This can be understood by the fact that the presence of $\delta \left( \sigma \left( x,x' \right) \right)$ in the direct part of $G_{ret}$ means that it is only non-zero on the past light-cone of the particle. However, since the quasilocal integral in Eq. (\ref{eq:SplitForce}) is totally internal to the past light-cone, this term does not contribute to the integral. The second term in Eq. (\ref{eq:Hadamard-scalar}), the \emph{tail} part of the Green function, is responsible for the self-force. Its calculation is therefore the primary focus in any calculation of the quasilocal self-force.

We can now substitute Eq. (\ref{eq:Hadamard-scalar}) into Eq. (\ref{eq:qlsf}) to obtain the final form of the quasilocal self-force:
\begin{equation}
\label{eq:QLSFInt}
f^{a}_\mathrm{QL} (z(\tau))= \lim_{\epsilon \rightarrow 0} q^2 \int_{\tau - \Delta \tau}^{\tau-\epsilon} \nabla^{a} G_{ret} \left( z(\tau),z(\tau') \right) d\tau '.
\end{equation}
Note that it is no longer necessary to take the limit since $V\left( x,x' \right)$ is regular everywhere.

Unlike the cases of electromagnetic and gravitational fields, where the four-acceleration is exactly equal to the self-force divided by the mass, there is one further step required in order to obtain the equations of motion in the presence of a scalar field. In this case, the self-force contains two components, one related to the four-acceleration, the other related to the change in mass:
\begin{equation}
f^{a}_{\rm QL} = \frac{dm}{d\tau} u^{a} + m a^{a}
\end{equation}

The 4-acceleration is always orthogonal to the 4-velocity, so in order to obtain it from the self-force, we take the projection of $f^{a}$ orthogonal to $u^{a}$:
\begin{equation}
\label{eq:ma}
ma^{a} = P^{a}_{\beta} f^{\beta}
\end{equation}
where
\begin{equation}
\label{eq:Projection}
P^{b}_{a} = \delta ^{b}_{a} + u_{a} u^{b}
\end{equation}
is the projection orthogonal to $u^\alpha$.
The remaining component of the self-force (which is tangent to the 4-velocity) is simply the rate of change of the mass:

\begin{equation}
\label{eq:dmdtau}
\frac{dm}{d\tau} = -f^{\alpha} u_{\alpha}
\end{equation}

\subsection{Covariant Calculation}
\label{sec:cov}
The calculation of $V(x,x')$ was previously discussed in Sec.~\ref{sec:classical-green} in the general (not necessarily scalar) case. In this section, we repeat the calculation in the particular case of a scalar field. The symmetric bi-scalar $V\left( x,x' \right)$ is expressed in terms of an expansion in increasing powers of $\sigma$ \cite{Decanini:Folacci:2005a}:
\begin{equation}
\label{eq:V-scalar}
V \left( x,x' \right) = \sum_{n=0}^{\infty} V_{n}\left( x,x' \right) \sigma ^{n}\left( x,x' \right)
\end{equation}
Substituting the expansion (\ref{eq:V-scalar}) into the wave equation, Eq.~(\ref{eq:Wave}), using the identity $\sigma_{\alpha} \sigma^{\alpha} = 2 \sigma$ and explicitly setting the coefficient of each manifest power of $\sigma$ equal to zero, we find (in 4-dimensional spacetime) that the coefficients $V_{n}\left( x,x' \right)$ satisfy the scalar version of the recursion relations, \eqref{eq:RecursionV},
\begin{multline}
\label{eq:RecursionV-scalar}
\left( n+1 \right) \left( 2n +4 \right) V_{n+1} + 2 \left( n+1 \right) V_{n+1;\mu}\sigma^{\mu} \\
- 2 \left( n+1 \right)V_{n+1}\Delta ^{-1/2}\Delta ^{1/2}_{\phantom{1/2} ;\mu} \sigma^{\mu} + \left( \Box _{x} - m^2 - \xi R \right) V_n = 0 \quad \mathrm{for}~ n \in \mathbb{N}
\end{multline}
along with the initial conditions
\begin{subequations}
\label{eq:InitialV}
\begin{eqnarray}
2V_0 + 2V_{0;\mu}\sigma^{\mu} - 2V_0 \Delta ^{-1/2}\Delta ^{1/2}_{\phantom{1/2} ;\mu} \sigma^{\mu} + \left( \Box _{x} - m^2 - \xi R \right) U_0 &=& 0\\
U_0 &=& \Delta ^{1/2}
\end{eqnarray}
\end{subequations}
The quantity $\Delta \left( x,x' \right)$ is the Van Vleck-Morette defined in Sec.~\ref{sec:bitensors}.

It is important to note that the Hadamard expansion is not a Taylor series and is not unique. Indeed, in Appendix~\ref{sec:DeWittVacuum} we show how, under the assumption of a massless field on a vacuum background spacetime, the terms from $V_0$ and $V_1$ conspire to cancel to third order when combined to form $V$. Such cancellation is also seen, for example, in deSitter spacetime, where for a conformally invariant theory all the $V_n$'s are non-zero while $V\equiv0$. 

In order to solve the recursion relations (\ref{eq:RecursionV-scalar}) and (\ref{eq:InitialV}) for the coefficients $V_{n} \left( x,x' \right)$, it is convenient to first write $V_n \left( x,x' \right)$ in terms of its covariant Taylor series expansion about $x = x'$:
\begin{equation}
\label{eq:Vn}
V_n \left( x,x' \right) = \sum _{p=0}^{\infty} \frac{ \left( -1 \right) ^p }{p!} v_{n (p)} \left( x,x' \right),
\end{equation}
where the $v_{n (p)}$ are bi-scalars of the form
\begin{equation}
v_{n (p)} \left( x,x' \right) = v_{n\,\,\alpha_1 \dots \alpha_p} (x) \sigma^{\alpha_1} \left( x,x' \right) \dots \sigma^{\alpha_p} \left( x,x' \right).
\end{equation}

D\'{e}canini and Folacci \cite{Decanini:Folacci:2005a} calculated expressions for these $v_{n (p)}$ up to $O \left( \sigma ^{2} \right)$, or equivalently $O \left( \Delta \tau ^{4} \right)$, where $\Delta \tau$ is the magnitude of the proper time separation between $x$ and $x'$. In Chapter~\ref{ch:covex}, we show how this calculation may be implemented in computer algebra to calculate the expansions to $O \left( \Delta\tau^{20} \right)$ and beyond. In Appendix \ref{sec:cov-5}, we show how, using the symmetry of the Green function in $x$ and $x'$ an even order covariant series expansion may be extended by one further order in the proper time separation of the points, $\Delta \tau$, or equivalently one half order in $\sigma \left( x,x' \right)$.

We now turn to the evaluation of the self-force. We could calculate the self-force given this form for $V_n \left( x,x' \right)$, however, it proves easier to work with the full $V\left( x,x' \right)$ expressed in the form of a covariant Taylor expansion, i.e. an expansion in increasing powers of $\sigma^{a}$:
\begin{equation}
\label{eq:Vt}
V \left( x,x' \right) = \sum _{p=0}^{\infty} \frac{ \left( -1 \right) ^p }{p!} v_{\alpha_1 \dots \alpha_p} (x) \sigma ^{\alpha_1} \left( x,x' \right) \dots \sigma ^ {\alpha_p} \left( x,x' \right)
\end{equation}
Explicit expressions for the $v_{a_1 \dots a_p} (x)$, can then be easily obtained order by order by combining the relevant $v_{n\,\,a_1 \dots a_p}$, as described in Appendix \ref{sec:DeWittVacuum}.

Substituting Eq.~(\ref{eq:Vt}) into Eq.~(\ref{eq:QLSFInt}), taking the covariant derivative and noting that the $v_{a_1 \dots a_p}$ are symmetric under exchange of all indices, $a_1, \dots, a_p$, we obtain an expression for the quasilocal self-force in terms of $v_{a_1 \dots a_p}$, $\sigma$ and their derivatives:
\begin{multline}
\label{eq:SigmaExpandedIntegral}
f^{a}_{\rm QL} = -q^2 \int_{\tau - \Delta \tau}^{\tau} \Big[
	v^{; a}
	- v^{\phantom{\beta} ; a}_{\beta} \sigma^{\beta}
	- v_{\beta} \sigma^{\beta a}
	+ \frac{2}{2!} v_{\beta \gamma} \sigma^{\beta a} \sigma^{\gamma}
	+ \frac{1}{2} v^{\phantom{\beta \gamma} ; a}_{\beta \gamma} \sigma^{\beta} \sigma^{\gamma}
	\\
	- \frac{3}{3!} v_{\beta \gamma \delta} \sigma^{\beta a} \sigma^{\gamma} \sigma^{\delta}
	- \frac{1}{3!} v^{\phantom{\beta \gamma \delta} ; a}_{\beta \gamma \delta} \sigma^{\beta} \sigma^{\gamma} \sigma^{\delta}
	+ \frac{4}{4!} v_{\beta \gamma \delta \epsilon} \sigma^{\beta a} \sigma^{\gamma} \sigma^{\delta} \sigma^{\epsilon}
	\\
	+ \frac{1}{4!} v^{\phantom{\beta \gamma \delta \epsilon} ; a}_{\beta \gamma \delta \epsilon} \sigma^{\beta} \sigma^{\gamma} \sigma^{\delta} \sigma^{\epsilon}
	- \frac{5}{5!} v_{\beta \gamma \delta \epsilon \zeta} \sigma^{\beta a} \sigma^{\gamma} \sigma^{\delta} \sigma^{\epsilon} \sigma^{\zeta}
	+ O\left( \sigma^{5/2} \right)
	\Big] d\tau '
\end{multline}

Now, by definition, for a time-like geodesic, 
\begin{subequations}
\begin{eqnarray}
\label{eq:Synge}
\sigma &=& -\frac{1}{2}\left( \tau -\tau' \right)^2\\
\label{eq:SyngeDeriv}
\sigma^{a} &=& -\left( \tau -\tau' \right) u^{a}
\end{eqnarray}
\end{subequations}
where $u^{a}$ is the 4-velocity of the particle. We may also write $\sigma^{a b}$ in terms of a covariant Taylor series expansion \cite{DeWitt:1960,Poisson:2003,Decanini:Folacci:2005a},
\begin{multline}
\label{eq:Sigma_ab}
\sigma^{a b} = g^{a b}
	- \frac{1}{3}R^{a \phantom{\alpha} b}_{\phantom{a} \alpha \phantom{b} \beta}\sigma^{\alpha} \sigma^{\beta}
	+ \frac{1}{12}R^{a \phantom{\alpha} b}_{\phantom{a} \alpha \phantom{b} \beta ; \gamma} \sigma^{\alpha} \sigma^{\beta} \sigma^{\gamma}
	\\ - \left[ \frac{1}{60} R^{a \phantom{\alpha} b}_{\phantom{a} \alpha \phantom{b} \beta ; \gamma \delta} + \frac{1}{45}R^{a}_{\phantom{a} \alpha \rho \beta}R^{\rho \phantom{\gamma} b}_{\phantom{\rho} \gamma \phantom{b} \delta} \right] \sigma^{\alpha} \sigma^{\beta} \sigma^{\gamma} \sigma^{\delta}
	+ O\left[ \left(\sigma^{a}\right)^{5} \right]
\end{multline}

Substituting Eqs.~(\ref{eq:Synge}), (\ref{eq:SyngeDeriv}) and (\ref{eq:Sigma_ab}) into Eq.~(\ref{eq:SigmaExpandedIntegral}), the integrand becomes an expansion in powers of the proper time, $\tau - \tau'$:
\begin{multline}
\label{eq:TauExpandedIntegral}
f^{a}_{\rm QL} = -q^2 \int_{\tau - \Delta \tau}^{\tau} \Big[
	A^{a} {\left( \tau - \tau ' \right)}^0 
	+ B^{a} {\left( \tau - \tau ' \right)}^1 
	+ C^{a} {\left( \tau - \tau ' \right)}^2 
 \\
	+ D^{a} {\left( \tau - \tau ' \right)}^3 
	+ E^{a} {\left( \tau - \tau ' \right)}^4 
	+ O \left( \left( \tau - \tau ' \right)^5 \right)
	\Big] d\tau ',
\end{multline}
where
\begin{subequations}
\begin{eqnarray}
A^{a} &=& v^{;a} - v_{\beta} g^{\beta a} \label{eq:QLSFgenA} \\
B^{a} &=& v^{\phantom{\beta} ;a}_{\beta} u^{\beta} - v_{\beta \gamma} g^{\beta a} u^{\gamma} \label{eq:QLSFgenB} \\
C^{a} &=& \frac{1}{3} v_{\beta} R^{\beta \phantom{\gamma} a}_{\phantom{\beta} \gamma \phantom{a} \delta} u^{\gamma} u^{\delta}
					+ \frac{1}{2} v^{\phantom{\beta \gamma} ;a}_{\beta \gamma} u^{\beta} u^{\gamma}
					- \frac{1}{2} v_{\beta \gamma \delta} g^{\beta a} u^{\gamma} u^{\delta} \label{eq:QLSFgenC} \\
D^{a} &=& \frac{1}{12} v_{\beta} R^{\beta \phantom{\gamma} a}_{\phantom{\beta} \gamma \phantom{a} \delta ;\epsilon} u^{\gamma} u^{\delta} u^{\epsilon}
					+ \frac{1}{3} v_{\beta \gamma} R^{\beta \phantom{\delta} a}_{\phantom{\beta} \delta \phantom{a} \epsilon} u^{\gamma} u^{\delta} u^{\epsilon}
					+ \frac{1}{6} v^{\phantom{\beta \gamma \delta} ;a}_{\beta \gamma \delta} u^{\beta} u^{\gamma} u^{\delta}
					\nonumber \\ &&- \frac{1}{6} v_{\beta \gamma \delta \epsilon} g^{\beta a} u^{\gamma} u^{\delta} u^{\epsilon} \label{eq:QLSFgenD} \\
E^{a} &=& v_{\beta} \left( \frac{1}{60} R^{\beta \phantom{\gamma} a}_{\phantom{\beta} \gamma \phantom{a} \delta ;\epsilon \zeta} + \frac{1}{45} R^{\beta}_{\phantom{\beta} \gamma \rho \delta} R^{\rho \phantom{\epsilon} a}_{\phantom{\rho} \epsilon \phantom{a} \zeta} \right) u^{\gamma} u^{\delta} u^{\epsilon} u^{\zeta}
					\nonumber \\
					&&+ \frac{1}{12} v_{\beta \gamma} R^{\beta \phantom{\delta} a}_{\phantom{\beta} \delta \phantom{a} \epsilon ;\zeta} u^{\gamma} u^{\delta} u^{\epsilon} u^{\zeta}
					+ \frac{1}{6} v_{\beta \gamma \delta} R^{\beta \phantom{\epsilon} a}_{\phantom{\beta} \epsilon \phantom{a} \zeta} u^{\gamma} u^{\delta} u^{\epsilon} u^{\zeta}
					\nonumber \\
					&&+ \frac{1}{24} v^{\phantom{\beta \gamma \delta \epsilon} ;a}_{\beta \gamma \delta \epsilon} u^{\beta} u^{\gamma} u^{\delta} u^{\epsilon}
					- \frac{1}{24} v_{\beta \gamma \delta \epsilon \zeta} g^{\beta a} u^{\gamma} u^{\delta} u^{\epsilon} u^{\zeta} \label{eq:QLSFgenE}.
\end{eqnarray}
\end{subequations}
The coefficients of the powers of $\tau - \tau '$ are all local quantities at $x$, but the integral is in terms of the primed coordinates, $x'$, so the coefficients may be treated as constants while performing the integration. The result is that we only have to perform a trivial integral of powers of $\tau - \tau'$. Performing the integration gives us our final result, the quasilocal part of the self force in terms of an expansion in powers of the matching point, $\Delta \tau$:
\begin{equation}
\label{eq:QLSFgen}
f^{a}_{\rm QL} = -q^2 \left(
		   A^{a}{\Delta \tau}^1 
		+ \frac{1}{2} B^{a}{\Delta \tau}^2 
		+ \frac{1}{3} C^{a}{\Delta \tau}^3
		+ \frac{1}{4} D^{a}{\Delta \tau}^4
		+ \frac{1}{5} E^{a}{\Delta \tau}^5  + O \left( \Delta \tau^6 \right) \right).
\end{equation}

It is possible to simplify Eqs. (\ref{eq:QLSFgenA}) - (\ref{eq:QLSFgenE}) further by using the results of Appendix \ref{sec:cov-5} to rewrite all the odd order $v_{a_1 \dots a_p}$ terms in terms of the lower order even terms. It is then straightforward to substitute in for the $v_{a_1 \dots a_p}$ to get an expression for the quasilocal self-force in terms of only the scalar charge, field mass, coupling to the scalar background, 4-velocity, and products of Riemann tensor components of the background spacetime. However, we choose not to do so here as that would result in excessively lengthy expressions. Instead, we direct the reader to Chapter~\ref{ch:covex} and Ref.~\cite{Decanini:Folacci:2005a} for general expressions for the $v_{n~ a_1 \dots a_p}$ for a scalar field and to Appendix \ref{sec:DeWittVacuum}, where we give the expressions for the $v_{n~ a_1 \dots a_p}$ for massless scalar fields in vacuum spacetimes up to the orders required.

Once an expression has been obtained for the self-force using this method, it is straightforward to use Eqs. (\ref{eq:ma}) - (\ref{eq:dmdtau}) to obtain the equations of motion.

\subsection{Simplification of the quasilocal self-force for massless fields in vacuum spacetimes}
\label{sec:simplify}
Eq. (\ref{eq:QLSFgen}) is valid for any geodesic motion and any scalar field (massless or massive) in any background spacetime. At first it may appear that this result is quite difficult to work with due to the length of the expressions for the $v_{n~ a_1 \dots a_p}$. However, if we make the two assumptions that:
\begin{enumerate}
\item The scalar field is massless ($m_{\rm field}=0$)\label{item:massless},
\item The background spacetime is a vacuum spacetime ($R_{\alpha \beta}=0=R$)\label{item:vacuum},
\end{enumerate}
then the majority of these terms vanish and we are left with the much more manageable expressions for the $v_{n~ a_1 \dots a_p}$ given in Appendix \ref{sec:DeWittVacuum} for a massless field in a vacuum spacetime up to $O \left( \sigma^{5/2} \right)$. We can see immediately from these expressions that all terms involving $v_{\alpha}$, $v_{\alpha \beta}$ and $v_{\alpha \beta \gamma}$ identically vanish. That leaves expressions involving only $v_{\alpha \beta \gamma \delta}$ and $v_{\alpha \beta \gamma \delta \epsilon}$ and higher order terms. However, Eq. (\ref{eq:vcompsym2-appendix}) gives us a relation which allows us to express any odd order term (in this case, $v_{\alpha \beta \gamma \delta \zeta}$) in terms of the lower even order terms. This, along with the fact that $v_{a_1 \dots a_p}$ is totally symmetric, leads to a vastly simplified expression for the quasilocal self-force on a scalar particle following a geodesic in a vacuum background spacetime:
\begin{multline}
\label{eq:SimpForce}
f^{a}_{\rm QL} = \\ q^2 g^{\beta a}  \left[
		\frac{1}{4!} v_{\beta \gamma \delta \epsilon} u^\gamma u^\delta u^\epsilon {\Delta \tau}^4 - \frac{1}{5!} \left( \frac{1}{2} v_{\gamma \delta \epsilon \zeta ;\beta} - 2 v_{\beta \gamma \delta \epsilon ;\zeta} \right)u^\gamma u^\delta u^\epsilon u^\zeta {\Delta \tau}^5
		+ O \left( \Delta \tau^6 \right) \right]
\end{multline}

We see that a large number of terms in (\ref{eq:QLSFgen}) either identically vanish or cancel each other out. In fact, at the first three orders in $\Delta \tau$, we get no contribution to the quasilocal self-force whatsoever. Additionally, our fourth and fifth order coefficients take on a much simpler form than the general case given in Eqs. (\ref{eq:QLSFgenD}) and (\ref{eq:QLSFgenE}).

\section{Quasi-local Contribution: Non-geodesic Motion}
\label{sec:qlsf-non-geodesic}
In the previous section, we gave an expression for the scalar self-force on a scalar charge travelling on a curved background spacetime. However, this expression was only valid provided the particle's path was that of a geodesic of the background spacetime. With some additional care, this result can be extended relatively easily to allow for non-geodesic particle motion. The calculation remains largely the same as for geodesic motion, so we will focus here here on the differences caused by the non-geodesicity of the motion. As before, we are primarily concerned with the non-local component of the self-force, \eqref{eq:fld-QL-DP}. In the most frequently studied cases of geodesic motion ($a^{a}=0$, $\dot{a}=0$) in Ricci-flat spacetimes ($R_{ab}=0$, $R=0$) and without field mass ($m_{\rm field}=0$), the local terms of Eq.~\eqref{eq:rad-field-deriv} are all identically zero and the full self-force is given by the non-local component. However, for non-geodesic motion, the 4-acceleration will always be non-zero so there will always be some local terms to consider when calculating the self-force.

\begin{figure}
\begin{center}
\includegraphics[width=6cm]{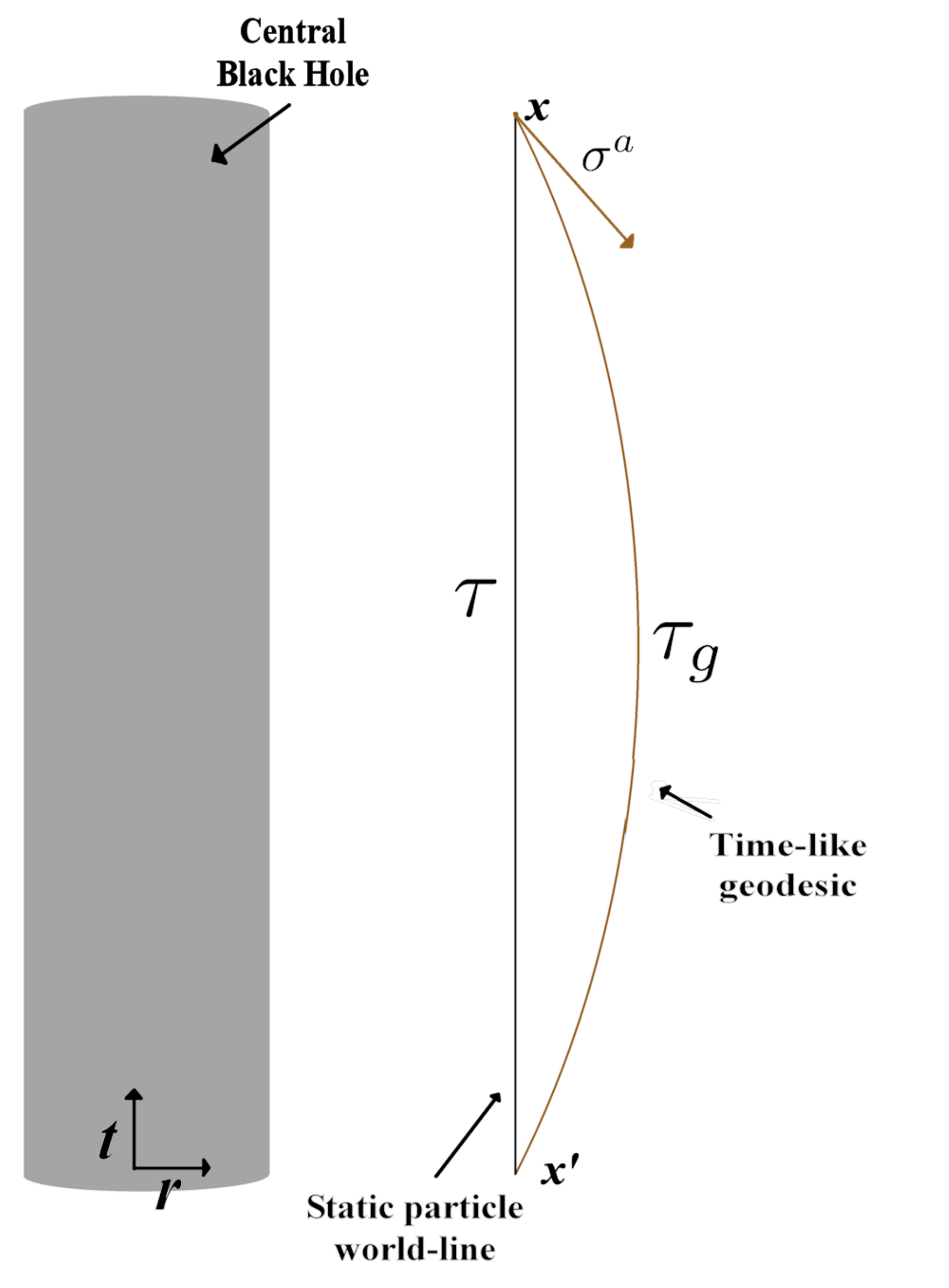}
\end{center}
\caption[Worldline of a static particle]{The worldline of a static particle ($x^i=\mathrm{constant}$) in Schwarzschild spacetime is not a geodesic, so the geodesic proper time, $\tau_g$, and the particle proper time, $\tau$, are not the same. The integral in Eq.~(\ref{eq:QLSFInt}) is along the particle worldline (black), while $\sigma^a$ is along the time-like geodesic (red).}
\label{fig:non-geodesic}
\end{figure}

As before, we focus on the quasilocal contribution, \eqref{eq:qlsf}, using the Hadamard form, \eqref{eq:Hadamard-scalar}, and leave the remaining `distant past' contribution to be computed by other means. We now expand $V(x,x')$ in two ways. First, we express it in the form of a covariant Taylor series expansion as given in Eq.~\eqref{eq:Vt}. In the previous section, we were able to use this expression for $V(x,x')$ to compute the self-force for geodesic motion. Unfortunately, things are less straightforward when non-geodesic motion is allowed for. The problem arises as a result of the presence of $\sigma^a$ in this expression. As demonstrated in Fig.~(\ref{fig:non-geodesic}), it encodes the proper time, $\tau_g$, along a geodesic of the background spacetime, defined through $\sigma^a\sigma_a = - \tau_g{}^2$. However, the integral in Eq.~(\ref{eq:qlsf}) is along the world line of the particle.  In the previous case of geodesic motion, this was not a problem as in that case $\tau_g$ is a natural parameter along the world line. For non-geodesic motion, this is no longer the case. To proceed with the calculation using this expansion of $V(x,x')$ would require us to first express the geodesic proper time $\tau_g$ in terms of the integration variable, i.e. the particle's proper time $\tau$. This is a non-trivial task for general motion in general spacetimes.

An easier resolution of this problem comes from expressing $V(x,x')$ in a second form, as a non-covariant Taylor series expansion in the coordinate separation of the points, $\Delta x^a$:
\begin{equation}
\label{eq:Vc}
V \left( x,x' \right) = \sum _{p=0}^{\infty} \frac{ \left( -1 \right) ^p }{p!} \hat{v}_{\alpha_1 \dots \alpha_p} (x) \Delta x^{\alpha_1} \dots \Delta x^{\alpha_p}
\end{equation}
where $\Delta x^{\alpha} = x^\alpha - x^{\alpha '}$ and the quantities $\hat{v}_{\alpha_1 \dots \alpha_p} (x)$ may be expressed in terms of combinations of the (known) quantities $v_{\alpha_1 \dots \alpha_p} (x) $ as follows.

First, the world function $\sigma \left(x,x'\right)$ is expressed in terms of an expansion in powers of the coordinate separation of $x$ and $x'$, $\Delta x^\alpha$:
\begin{multline}
\label{eq:sigma-coord}
 \sigma = \frac{1}{2} g_{\alpha \beta} \Delta x^\alpha \Delta x^\beta
	 + A_{\alpha \beta \gamma} \Delta x^\alpha \Delta x^\beta \Delta x^\gamma
	 + B_{\alpha \beta \gamma \delta} \Delta x^\alpha \Delta x^\beta \Delta x^\gamma \Delta x^\delta \\
	 + C_{\alpha \beta \gamma \delta \epsilon} \Delta x^\alpha \Delta x^\beta \Delta x^\gamma \Delta x^\delta  \Delta x^\epsilon
	 + \ldots
\end{multline}
where the coefficients $A_{\alpha \beta \gamma}$, $B_{\alpha \beta \gamma \delta}$, $C_{\alpha \beta \gamma \delta \epsilon}$, \dots are to be determined. Differentiating Eq.~(\ref{eq:sigma-coord}), the coordinate expansion of the derivative of $\sigma$ is:
\begin{multline}
\label{eq:sigmaderiv-coord}
 \sigma_a = g_{a \alpha} \Delta x^\alpha
	 + (\frac12 g_{\alpha \beta ,a} + 3 A_{a \alpha \beta} )\Delta x^\alpha \Delta x^\beta
	 + (A_{\alpha \beta \gamma ,a} + 4 B_{a \alpha \beta \gamma}) \Delta x^\alpha \Delta x^\beta \Delta x^\gamma
	 \\+ (B_{\alpha \beta \gamma \delta ,a} + 5 C_{a \alpha \beta \gamma \delta} )\Delta x^\alpha \Delta x^\beta \Delta x^\gamma \Delta x^\delta
	 + \ldots .
\end{multline}
Substituting expansions (\ref{eq:sigma-coord}) and (\ref{eq:sigmaderiv-coord}) into the defining relationship
\begin{equation}
\label{eq:sigma-to-sigmaderiv}
 2 \sigma = \sigma^\alpha \sigma_\alpha
\end{equation}
and equating powers of $\Delta x^\alpha$, we get expressions for each coefficient in terms of the lower order coefficients. The lowest terms are given by:
\begin{subequations}
\begin{eqnarray}
 A_{a b c} 	&=& -\frac{1}{4}g_{(a b ,c)} \\
 B_{a b c d}	&=& -\frac{1}{3} \left[ A_{(a b c ,d)} +  g^{\alpha \beta} \left(\frac{1}{8}g_{(a b ,|\alpha|} g_{c d) ,\beta}
			+ \frac{3}{2} g_{(a b ,|\alpha|} A_{|\beta| c d)}\right. \right.\nonumber \\
& &\left. \left.~~~
			+ \frac{9}{2} A_{\alpha (a b} A_{|\beta| c d)}\right) \right] \\
 C_{a b c d e}	&=& -\frac{1}{4} \left[ B_{(a b c d ,e)} +  g^{\alpha \beta} \left(12 A_{\alpha (a b} B_{|\beta| c d e)}
			+ 3 A_{\alpha (a b} A_{c d e) ,\beta}\right. \right.\nonumber \\
& &\left. \left.~~~
			+ 2 g_{(a b |,\alpha|} B_{|\beta| c d e)}
			+ \frac{1}{2} A_{(a b c |,\alpha|} g_{d e) ,\beta}\right)\right]
\end{eqnarray}
\end{subequations}
This procedure may be easily extended to higher orders using a computer algebra package.

Next, we obtain a relation between the known coefficients of the covariant expansion, Eq.~(\ref{eq:Vt}), and those of the coordinate expansion, Eq.~(\ref{eq:Vc}), by first substituting Eq.~(\ref{eq:sigmaderiv-coord}) into Eq.~(\ref{eq:Vt}) and then equating the two expansions. In this way we find the following expressions for the $\hat{v}_{\alpha_1 \dots \alpha_p} (x)$ in terms of the $v_{\alpha_1 \dots \alpha_p} (x)$:
\begin{subequations}
\begin{eqnarray}
\hat{v} 	&=& v\\
\hat{v}_a 	&=& v_{a}\\
\label{eq:vab-hat}
\hat{v}_{ab} 	&=& v_{ab} + v_{\alpha} \Gamma^{\alpha}_{ab}\\
\hat{v}_{abc}	&=& v_{abc} + 3 v_{(a |\alpha|} \Gamma_{b c)}^{\alpha} 
		+ 6v_{\alpha} \left( A_{(a b c)}^{\phantom{(a b c)},\alpha} + 4 B^{\alpha}_{\phantom{\alpha} a b c}\right)\\
\hat{v}_{abcd}	&=& v_{abcd} + 6 v_{\alpha (a b} \Gamma_{c d)}^{\alpha}
		+ 12 v_{\alpha \beta} \left(\frac{1}{4}\Gamma_{(a b}^{\alpha}\Gamma_{c d)}^{\beta}+2 g_{(a}^{\phantom{(a}\alpha}A_{b c d)}^{\phantom{b c d)},\beta}+ 8 g_{(a}^{\phantom{(a}\alpha} B^{\beta}_{\phantom{\beta} b c d)}\right)
		\nonumber \\
		& & -24 v_{\alpha} \left( B_{(a b c d)}^{\phantom{(a b c d)},\alpha}
			+5C^{\alpha}_{\phantom{\alpha} (a b c d)}\right)\\
\label{eq:vhat5}
\hat{v}_{abcde}	&=& \frac{1}{2} \hat{v}_{,(a b c d e)} - \frac{5}{2} \hat{v}_{(a b , c d e)} + \frac{5}{2} \hat{v}_{(a b c d , e)} 
\end{eqnarray}
\end{subequations}
where the $\Gamma_{a b}^{\alpha}$ are the Christoffel symbols of the second kind. We note here the presence of the Christoffel symbols in the relation between the covariant and coordinate expansions. In the case of a static particle (i.e. purely time separation of the points), these take on a particularly clear form by being directly related to the 4-acceleration of the particle. In particular, the leading order at which this effect appears is given in Eq.~(\ref{eq:vab-hat}), which in this case becomes:
\begin{equation}
\hat{v}_{tt}	= v_{tt} + v_{\alpha} \Gamma_{t t}^{\alpha} = v_{tt} + 3 v_{\alpha} a^{\alpha}
\end{equation}
where $a^\alpha$ is 4-acceleration of the particle.

Although $\hat{v}_{abcde}$ could alternatively be given in terms of the $v_{\alpha_1 \dots \alpha_p} (x)$, we have found that the expression we give proves easier to work with. It is found by taking five symmetrized partial derivatives of the equation
\begin{equation}
\label{eq:vsymmetry}
 V(x,x') = V(x',x)
\end{equation}
and then taking taking the coincidence limit $x' \rightarrow x$.  It is a special case of the general result, Eq.~\eqref{eq:vcompsym2-appendix}, of Appendix~\ref{sec:cov-5}.

Now, we simply substitute expansion (\ref{eq:Vc}) into Eq.~(\ref{eq:QLSFInt}) and, since $V(x,x')$ is a scalar, take a partial rather than covariant derivative to get an easily evaluated expression for the self-force (in this case, it is most natural to work with an expression for the self-force in covariant rather than contravariant form):
\begin{multline}
\label{eq:DxExpandedIntegral}
f_{a}^{\rm QL} = -q^2 \int_{\tau - \Delta \tau}^{\tau} \Big[
	\hat{v}_{,a} - \hat{v}_{a}
	- \left( \hat{v}_{\alpha ,a} - \hat{v}_{\alpha a} \right) \Delta x^\alpha
	+ \frac{1}{2} \left( \hat{v}_{\alpha \beta ,a} - \hat{v}_{\alpha \beta a} \right) \Delta x^\alpha \Delta x^\beta\\
	- \frac{1}{3!} \left( \hat{v}_{\alpha \beta \gamma ,a} - \hat{v}_{\alpha \beta \gamma a} \right) \Delta x^\alpha \Delta x^\beta \Delta x^\gamma
	+ \frac{1}{4!} \left( \hat{v}_{\alpha \beta \gamma \delta ,a} - \hat{v}_{\alpha \beta \gamma \delta a} \right) \Delta x^\alpha \Delta x^\beta \Delta x^\gamma \Delta x^\delta
	\\ + O\left( \Delta x^{5} \right)
	\Big] d\tau '
\end{multline}

Finally, we obtain the equations of motion by projecting orthogonal and perpendicular to the particle 4-velocity, as in Sec.~\ref{sec:qlsf-geodesic}.
This expression may be evaluated for a specific particle path in a specific spacetime by writing the coordinate separations $\Delta x^\alpha$ in terms of the particle proper time separation $\left(\tau - \tau'\right)$. Specific examples of such an evaluation are given in Chapter~\ref{ch:qlsf}.

\section{Static Particle in a stationary spacetime}
\label{sec:stationary}
The expression for the quasilocal contribution to the self-force given in Eq.~(\ref{eq:DxExpandedIntegral}) (and the subsequent equations of motion) take on a significantly simple form if the spacetime is assumed to be stationary and if we assume the spatial coordinate of the particle to be fixed. Stationarity allows us to introduce coordinates $(t,x^i)$ such that the metric tensor components, $g_{a b}$ are all independent of the time coordinate. In addition, for a static particle, $x^i = \mathrm{constant}$, and only the time component of the contravariant 4-velocity, $u^t$, is non-zero. These conditions hold, of course, for the case of a particle held at rest in Schwarzschild spacetime which has already received much attention in the literature \cite{Anderson:Wiseman:2005,Wiseman:2000}.

Since the spatial coordinates of the particle are held fixed, the points $x$ and $x'$ are now only separated in the time direction. Furthermore, in this case it is straightforward to relate the time coordinate to the proper time along the particle's world line. As a result we can rewrite the coordinate separation of the points $x$ and $x'$ in terms of the proper time separation:
\begin{equation}
 \Delta x^t = \Delta t = u^t (\tau - \tau ')  = \frac{1}{\sqrt{-g_{tt}}} (\tau - \tau ')  , \qquad \Delta x^i =0.
\end{equation}
Substituting this into Eq.~(\ref{eq:DxExpandedIntegral}) and performing the straightforward integral of powers of $(\tau-\tau')$ gives
\begin{multline}
\label{eq:force-integrated}
f_{a}^{\rm QL} = -q^2\Big[
	\left( \hat{v}_{,a} - \hat{v}_{a} \right) \Delta \tau
	- \frac{1}{2!} \left( \hat{v}_{t ,a} - \hat{v}_{t a} \right) u^t \Delta \tau ^2
	+ \frac{1}{3!} \left( \hat{v}_{t t ,a} - \hat{v}_{t t a} \right) (u^t)^2 \Delta \tau ^3 \\
	- \frac{1}{4!} \left( \hat{v}_{t t t ,a} - \hat{v}_{t t t a} \right) (u^t)^3 \Delta \tau ^4
	+ \frac{1}{5!} \left( \hat{v}_{t t t t ,a} - \hat{v}_{t t t t a} \right) (u^t)^4 \Delta \tau ^5
	+ O\left( \Delta \tau ^{6} \right)
	\Big]
\end{multline}

We can now use Eq.~(\ref{eq:vhat5}) for $\hat{v}_{a b c d e}$, along with the analogous equations for  $\hat{v}_{a}$ and $\hat{v}_{a b c}$,
\begin{eqnarray}
 \hat{v}_{a} &=& \frac{1}{2} \hat{v}_{,a}\\
 \hat{v}_{a b c} &=& -\frac{1}{4} \hat{v}_{,(a b c)} + \frac{3}{2} \hat{v}_{(a b , c)} 
\end{eqnarray}
to eliminate several terms in this expression. While this did not previously prove particularly beneficial in the general case, the fact that partial derivatives with respect to $t$ of these fundamentally geometric objects vanish in a stationary spacetime means that many of the extra terms introduced by the substitution will also vanish. Indeed it is easy to see from Eq.~(\ref{eq:vcompsym-appendix}) that (in this case of purely time separated points) any term of order $2k+1$ can be related to the order $2k$ term:
\begin{equation}
\hat{v}_{\underbrace{\scriptstyle tt\dots tt}_{2k}b} = \frac{1}{2} \hat{v}_{{\underbrace{\scriptstyle{tt\dots tt}}_{2k}}\, ,b} 
\end{equation}
As a result, Eq.~(\ref{eq:force-integrated}) may be taken to arbitrary order to give:
\begin{align}
\label{eq:force-integrated-simplified}
f_{a}^{\rm QL} =& -q^2 \Big[
	\frac{1}{2} \sum_{k=0}^{\infty} \frac{1}{(2k+1)!} \hat{v}_{\underbrace{\scriptstyle tt\dots tt}_{2k}\, , a} \Delta \tau^{2k+1} (u^t)^{2k}
	\\&\qquad \qquad+ \sum_{k=1}^{\infty} \frac{1}{(2k)!} \hat{v}_{\underbrace{\scriptstyle tt\dots tt}_{2k-1} a} \Delta \tau^{2k} (u^t)^{2k-1}
	\Big]
\end{align}

To proceed further, we benefit from differentiating between the time and spatial components:
\begin{align}
\label{eq:force-integrated-covariant-time}
f_{t}^{\rm QL} =& -q^2 \Big[
	\sum_{k=1}^{\infty} \frac{1}{(2k)!} \hat{v}_{\underbrace{\scriptstyle tt\dots tt}_{2k}} \Delta \tau^{2k} (u^t)^{2k-1}
	\Big]\\
\label{eq:force-integrated-covariant-spatial}
f_{i}^{\rm QL} =& -q^2 \Big[
	\frac{1}{2} \sum_{k=0}^{\infty} \frac{1}{(2k+1)!} \hat{v}_{\underbrace{\scriptstyle tt\dots tt}_{2k}\, , i} \Delta \tau^{2k+1} (u^t)^{2k}
	\\&\qquad \qquad+ \sum_{k=1}^{\infty} \frac{1}{(2k)!} \hat{v}_{\underbrace{\scriptstyle tt\dots tt}_{2k-1} i} \Delta \tau^{2k} (u^t)^{2k-1}
	\Big]
\end{align}
From this, it is clear that the $t$ component of the quasilocal self-force only appears at even orders in $\Delta \tau$, while the spatial components may appear at both even and odd orders (this is consistent with one's physical intuition: the fact that there is an odd power of $u^t$ in $f_t^{\rm{QL}}$ means that it is odd overall, as might be expected given its role in determining energy balance). Note, however, that for a static spacetime - i.e. imposing time reversal invariance in addition to stationarity - the second term in Eq.~(\ref{eq:force-integrated-covariant-spatial}) will vanish since each of the $\hat{v}_{t\dots t i}$ must be zero in that case (as we have an odd number of $t$'s).  As a result, in a static spacetime, the spatial component of the quasilocal self-force will only appear at odd orders in $\Delta \tau$.

By Eq.~(\ref{eq:dmdtau}) and the fact that only the time component of the contravariant 4-velocity is non-zero, we can now rewrite the rate of change of the particle's mass as:
\begin{eqnarray}
\label{eq:dmdtau-stationary-spacetime}
 \frac{dm}{d\tau} &=& -f_{\alpha} u^{\alpha} = - f_t u^t \nonumber \\
	&=& q^2
	\sum_{k=1}^{\infty} \frac{1}{(2k)!} \hat{v}_{\underbrace{\scriptstyle tt\dots tt}_{2k}} \Delta \tau^{2k} (u^t)^{2k}
\end{eqnarray}
Similarly, by Eq.~(\ref{eq:ma}) we can now write the time and spatial components of the mass times 4-acceleration:
\begin{subequations}
\label{eq:ma-stationary-spacetime}
\begin{eqnarray}
ma^{t}_{\rm QL} &=& -\frac{g_{ti}g^{ti}}{g_{tt}} f_t + g^{ti} f_i \\
ma^{i}_{\rm QL} &=& g^{ti} f_{t} + g^{i j} f_{j}
\end{eqnarray}
\end{subequations}
where $f_t$ and $f_i$ are as given in Eqs.~(\ref{eq:force-integrated-covariant-time}) and (\ref{eq:force-integrated-covariant-spatial}).

Again, it is interesting to note the effect of imposing that the spacetime be static. For a static spacetime, the metric components odd in $t$ vanish, so our result simplifies to:
\begin{subequations}
\label{eq:ma-static-spacetime}
\begin{eqnarray}
ma^{t}_{\rm QL} &=& 0 \\
ma^{i}_{\rm QL} &=& g^{i j} f_{j} =  -q^2 g^{i j} \frac{1}{2} \sum_{k=0}^{\infty} \frac{1}{(2k+1)!} \hat{v}_{\underbrace{\scriptstyle tt\dots tt}_{2k}, j} \Delta \tau^{2k+1} (u^t)^{2k}
\end{eqnarray}
\end{subequations}

\chapter{Coordinate Expansion of Hadamard Green Function} \label{ch:coordex}
Quasi-local series expansions -- expansions in the separation of two points $x$ and $x'$ -- are a frequently used tool for calculations of fields on curved spacetimes. Often, as a final step in the calculation, the coincidence limit, $x' \rightarrow x$, is taken. In these cases, the precise convergence properties of the series are of little interest. However, as is clear from Chapter~\ref{ch:sf}, there are cases such as self-force calculations where we would like the points to remain separated.  These calculations require a quasilocal expansion of the retarded Green function, $G_{ret}(x,x')$. As expression \eqref{eq:qlsf} requires the Green function for the points separated up to an amount $\Delta\tau$ along a worldline, it begs the question: how large can the separation of the points be before the series expansion is no longer a valid representation of the Green function?

To the authors knowledge, this question has not yet been quantitatively answered. It is well known that the Hadamard parametrix for the Green function (upon which quasilocal calculations are based) is valid provided $x$ and $x'$ lie within a \emph{normal neighborhood} (see Footnote \ref{def:causal domain}) \cite{Friedlander}. However, this does not necessarily guarantee that a series representation will be convergent everywhere within this normal neighborhood. In fact, we will show that the series is only convergent within a smaller region, the size of which is given by the \emph{circle of convergence} of the series. However, this does not preclude the use of the quasilocal expansion to calculate the Green function \emph{outside} the circle of convergence (but within the normal neighborhood). As we will show, Pad\'{e} resummation techniques, which have been extremely successful in other areas \cite{Damour:Iyer:Sathyaprakash,Porter:Sathyaprakash}, are also effective in extending the series beyond its circle of convergence.

In this chapter, we will focus in particular on calculating the Green function for two spherically symmetric spacetimes: Schwarzschild and Nariai. The Nariai spacetime~\cite{Nariai:1950,Nariai:1951}, discussed in detail in Sec.~\ref{sec:Nariai}, arises naturally from efforts to consider a simplified version of Schwarzschild. It retains some of the key features of Schwarzschild (such as the presence of an unstable photon orbit and a similar effective radial potential which diminishes exponentially on one side), but frequently yields more straightforward calculations. This makes it an ideal testing ground for new methods which are later to be applied to the more complicated Schwarzschild case. In the present work, we will use the line element of the static region of the Nariai spacetime (with cosmological constant $\Lambda=1$ and Ricci scalar $R=4$) in the form \eqref{eq:Nariai-le} where it yields a wave equation with potential which is seen to closely resemble that of the Schwarzschild metric, \eqref{eq:schwle}.

In Sec.~\ref{sec:WKB} we use an adaptation of the Hadamard-WKB method developed by Anderson and Hu \cite{Anderson:2003} to efficiently calculate the coordinate series expansion of $V(x,x')$ to very high order for both Nariai and Schwarzschild spacetimes. This WKB method involves an expansion in a parameter $\chi$, which is large both for large-$l$ and for large-$\omega$. In Sec.~\ref{sec:convergence} we use convergence tests to determine the radius of convergence of our series and show that, as expected, it lies within the convex normal neighborhood. We also estimate the local truncation error arising from truncating the series at a specific order. Using the method of Pad\'{e} approximants, we show in Sec.~\ref{sec:Pade} how the domain of validity of the coordinate series can be extended beyond its radius of convergence to give an accurate representation of $V(x,x')$ to within a small distance of the edge of the normal neighborhood.

\section{Hadamard-WKB Calculation of the\texorpdfstring{\\}{} Green Function}
\label{sec:WKB}

For the present quasilocal calculation, we need to consider the retarded Green function only for the points $x$ and $x'$ lying within a \emph{normal neighborhood}. This allows us to express the retarded Green function in the Hadamard parametrix \cite{Hadamard,Friedlander},
\eqref{eq:Hadamard-scalar}. As the quasilocal self-force calculation which motivates us requires the Green function within the light-cone only, we will only concern ourselves here with the calculation of the function $V(x,x')$.

The fact that $x$ and $x'$ are close together suggests that an expansion of $V(x,x')$ in powers of the coordinate separation of the points,
\begin{equation}
\label{eq:CoordGreen}
V\left( x,x' \right) = \sum_{i,j,k=0}^{\infty} v_{ijk}(r) \left( t - t' \right)^{2i} \left( \cos \gamma - 1 \right)^j (r-r')^k,
\end{equation}
where $\gamma$ is the angular separation of the points, 
may give a good representation of the function within the quasilocal region. Note that, as a result of the spherical symmetry of the spacetimes we will be considering, the expansion coefficients, $v_{ijk}(r)$, are only a function of the radial coordinate, $r$. Anderson and Hu \cite{Anderson:2003} have developed a Hadamard-WKB method for calculating these coefficients. They applied their method to the Schwarzschild case and subsequently found the coefficients to $14^{th}$ order using the \emph{Mathematica} computer algebra system \cite{Anderson:Eftekharzadeh:Hu:2006}. In the present work, we adapt their method to also allow for spacetimes of the Nariai form, (\ref{eq:Nariai-le}). In particular, we consider a class of spacetimes of the general form
\begin{equation}
\label{eq:diagonal-metric}
ds^2 = -f(r) dt^2 + f^{-1}(r) dr^2 + g(r) \left( d\theta^2 + \sin^2\theta d\phi^2 \right),
\end{equation}
where $f(r)$ and $g(r)$ are arbitrary functions of the radial coordinate, $r$, and previously Anderson and Hu had set $g(r) = r^2$, but allowed the metric components $g_{rr}$ and $g_{tt}$ to be independent functions of $r$. The form of the Nariai metric given in Eq.~(\ref{eq:Nariai-le}) falls into the class \eqref{eq:diagonal-metric}, with $f(r) = 1- r^2$ and $g(r) = 1$. For $f(r) = 1-\frac{2M}{r}$ and $g(r)=r^2$, this is the Schwarzschild metric of Eq.~(\ref{eq:schwle}).

The method presented in this section differs from that of Ref.~\cite{Anderson:2003} in the details of the WKB approach used, but otherwise remains very similar. Our alternative WKB approach, based on that of Refs.~\cite{Howard:1985,Winstanley:2007} proves extremely efficient when implemented in a computer algebra package.

Following the prescription of Ref.~\cite{Anderson:2003}, the Hadamard parametrix for the real part of the Euclidean Green function (corresponding to the Euclidean metric arising from the change of coordinate $\tau = i t$) is \footnote{Note that this definition of the Green function differs from that of Ref.~\cite{Anderson:2003} by a factor of $4\pi$ and the definition of $V(x,x')$ differs by a further factor of 2.}
\begin{equation}
\label{eq:HadamardEuclideanGF}
 \text{Re} \left[ G_E (-i \tau,\vec{x};-i\tau',\vec{x}') \right]= \frac{1}{2\pi} \left( \frac{U(x,x')}{\sigma(x,x')} + V(x,x') \ln (|\sigma(x,x')|) + W(x,x') \right) ,
\end{equation}
where $U(x,x')$, $V(x,x')$ and $W(x,x')$ are real-valued symmetric bi-scalars.

Additionally, for the points $x$ and $x'$ separated farther apart in the time direction than in other directions,
\begin{equation}
 \sigma (x,x') = -\frac{1}{2} f(r) (t-t')^2 + O\left[(x-x')^3\right]
\end{equation}
so the logarithmic part of Eq.~(\ref{eq:HadamardEuclideanGF}) is given by
\begin{equation}
\label{eq:VlnTau}
 \frac{1}{\pi} V(x,x') \ln(\tau - \tau').
\end{equation}
Therefore, in order to find $V(x,x')$, it is sufficient to find the coefficient of the logarithmic part of the Euclidean Green function. We do so by considering the fact that the Euclidean Green function also has the exact expression for the spacetimes of the form given in Eq.~(\ref{eq:diagonal-metric}):
\begin{equation}
\label{eq:EuclideanGF}
  G_E (-i \tau,x;-i\tau',x') = \frac{1}{\pi} \int_{0}^{\infty} d\omega \cos \left[ \omega (\tau - \tau') \right] \sum_{l=0}^{\infty} (2l+1) P_l (\cos \gamma) C_{\omega l} p_{\omega l}(r_<) q_{\omega l}(r_>),
\end{equation}
where $p_{\omega l}$ and $q_{\omega l}$ are solutions (normalised by $C_{\omega l}$) to the homogeneous radial equation comint from the scalar wave
equation in the curved background (\ref{eq:diagonal-metric}) (and where $r_< $, $r_>$ are the smaller/larger of $r$ and $r'$, respectively), along with the fact that
\begin{align}
\label{eq:WKB-int-log-relation}
 \int_\lambda^{\infty} d\omega \cos \left[ \omega (\tau - \tau') \right] \frac{1}{\omega^{2n+1}} &= \frac{(-1)^{n+1}}{(2n)!}(\tau-\tau')^{2n}\log\left(\tau-\tau'\right) + \cdots \nonumber\\
	&= \frac{-1}{(2n)!}(t-t')^{2n}\log\left(t-t'\right)+\cdots ,
\end{align}
where $\lambda$ is a low frequency cut-off justified by the fact that we will only need the $\log(\tau-\tau')$ term from the integral. The omitted terms on the right hand side are smooth and do not contribute to the order $\log(t-t')$ part, so are not of interest to the present calculation. We can therefore find $V(x,x')$ as an expansion in powers of the time separation of the points by expressing the sum
\begin{equation}
\label{eq:WKBsum}
\sum_{l=0}^{\infty} (2l+1) P_l (\cos \gamma) C_{\omega l} p_{\omega l}(r_<) q_{\omega l}(r_>)
\end{equation}
of Eq.~(\ref{eq:EuclideanGF}) as an expansion in inverse powers of $\omega$. This is achieved using a WKB-like method based on that of Refs.~\cite{Howard:1985,Winstanley:2007}. Given the form (\ref{eq:EuclideanGF}) for the Euclidean Green function, the radial functions $S(r)=p_{\omega l}(r)$ and $S(r)=q_{\omega l}(r)$ must both satisfy the homogeneous wave equation,
\begin{equation}
\label{eq:WKB-radial-eq}
 f \frac{d^2 S}{dr^2} + \frac{1}{g} \frac{d}{dr}(f g) \frac{dS}{dr} - \left[ \frac{\omega^2}{f} + \frac{l(l+1)}{g} + m_{\text{field}}^2 + \xi R\right] S = 0
\end{equation}
where $m_{\text{field}}$ is the scalar field mass and $\xi$ is the coupling to the scalar curvature, $R$. Next, given the Wronskian $W(r) = C_{\omega l}(p_{\omega l}q_{\omega l}' - q_{\omega l}p_{\omega l}')$, its derivative is
\begin{equation}
 W' = C_{\omega l}(p_{\omega l}q_{\omega l}'' - q_{\omega l}p_{\omega l}'') = - \frac{1}{fg} (fg)' W
\end{equation}
and the Wronskian condition is therefore
\begin{equation}
\label{eq:WKBWronskian}
 C_{\omega l}(p_{\omega l}q'_{\omega l} - q_{\omega l}p'_{\omega l}) = -\frac{1}{fg}.
\end{equation}
We now explicitly assume that $r>r'$ and define the function
\begin{equation}
 B(r,r') = C_{\omega l} p_{\omega l}(r') q_{\omega l}(r).
\end{equation}
Since the sum of Eq.~(\ref{eq:WKBsum}) (and hence $B(r,r')$) is only needed as an expansion in powers of $(r-r')$, we expand $B(r,r')$ about $r'=r$ and (using Eq.~(\ref{eq:WKB-radial-eq}) to replace second order and higher derivatives of $B(r,r')$ with expressions involving $B(r,r')$ and $\partial_{r'}B(r,r')$) find that
\begin{multline}
 B(r,r') = \beta(r) + \alpha(r) (r'-r) + \left\{ \left[\frac{2 (\eta+\chi^2)}{(fg)^2}\right] \beta(r) - \left[\ln(fg) \right]' \alpha(r)  \right\} \frac{(r'-r)^2}{2} + \cdots\\ (\textrm{for\ }r > r')
\end{multline}
where
\begin{align}
\beta(r) \equiv& \left[ B(r,r') \right]_{r' \to r^-} = C_{\omega l} p_{\omega l}(r) q_{\omega l}(r),\\
 \alpha(r) \equiv& \left[\partial_{r'}B(r,r')\right]_{r' \to r^-} =  C_{\omega l} p'_{\omega l}(r) q_{\omega l}(r)
\end{align}
and 
\begin{align}
\label{eq:eta_defn}
 \eta(r) &\equiv -\frac{1}{4}f g+\left(m_{\text{field}}^2+\xi  R\right) f g^2\\
 \label{eq:chi_defn}
 \chi^2(r) &\equiv \omega^2 g^2 +  f g \left(l+\frac{1}{2}\right)^2 .
\end{align}
It will therefore suffice to calculate $\beta(r)$ and $\alpha(r)$. Furthermore, using Eq.~\eqref{eq:WKBWronskian} we can relate $\alpha(r)$ to the derivative of $\beta(r)$,
\begin{equation}
 \alpha (r) = \frac{\beta'(r)}{2} + \frac{1}{2f(r)g(r)},
\end{equation}
so it will, in fact, suffice to find $\beta(r)$ and its derivative, $\beta'(r)$.

Using Eqs.~(\ref{eq:WKB-radial-eq}) and (\ref{eq:WKBWronskian}), it is immediate to see that $\beta(r)$ must satisfy the nonlinear differential equation
\begin{equation} \label{eq:beta-ode}
 f g \frac{d}{dr}\left(f g \frac{d\sqrt{\beta}}{dr}\right)-\left(\eta +\chi ^2\right) \sqrt{\beta}+ \frac{1}{4 \beta^{3/2}}=0.
\end{equation}

The short distance behavior of the Green function is determined by the high-$\omega$ and/or high-$l$ behavior of the
integrand of Eq.~\eqref{eq:EuclideanGF}, so we seek to express $\beta(r)$ as an expansion in inverse powers of $\chi$.  To keep track of this 
expansion we may replace  $\chi$ in Eq.~(\ref{eq:beta-ode}) by $\chi/\epsilon$ where $\epsilon$ is a formal expansion 
parameter which we eventually set to 1.
Then, to balance at leading order we require
\begin{align}
(\chi/\epsilon)^2 \sqrt{\beta} \sim \frac{1}{4 \beta^{3/2}} \quad \implies \beta \sim \frac{\epsilon}{2 \chi} .
\end{align}
We now write 
\begin{align} \label{eq:beta}
\beta(r)&= \epsilon \beta_0(r) + \epsilon^2\beta_1(r) + \ldots
\end{align}
where $\beta_0(r) \equiv 1/(2 \chi(r))$, insert this form for $\beta(r)$ in Eq.~(\ref{eq:beta-ode}), and solve formally order by order in $\epsilon$ to find a recursion relation for the $\beta_n(r)$. 
On doing so and using  Eq.~(\ref{eq:chi_defn}) to eliminate $(l+\frac12)^2$ in favor of $\omega^2$ and $\chi^2$ we find
that we can write
\begin{align} \label{eq:beta-n}
\beta_n(r) = \sum\limits_{m=0}^{2n} \frac{A_{n,m}(r) \omega^{2m} }{\chi^{2n+2m+1}}
\end{align}
so, for example,
\begin{align}   
\beta_1(r) &=\frac{A_{1,0}(r)}{\chi^3}+\frac{A_{1,1}(r) \omega^2}{\chi^5}+\frac{A_{1,2}(r)\omega^4}{\chi^7}\\
\beta_2(r) &=\frac{A_{2,0}(r)}{\chi^5}+\frac{A_{2,1}(r) \omega^2}{\chi^7}+\frac{A_{2,2}(r)\omega^4}{\chi^9}+\frac{A_{2,3}(r) \omega^6}{\chi^{11}}+\frac{A_{2,4}(r)\omega^8}{\chi^{13}}. 
\end{align}
The recursion relations for $\beta_n(r)$ may then be re-expressed to allow us to recursively solve for the $A_{n,m}(r)$. Such a recursive calculation is ideally suited to implementation in a computer algebra system (CAS). Even on a computer, this recursive calculation becomes very long as $n$ becomes large and, in fact, dominates the time required to calculate the series expansion of $V(x,x')$ as a whole. For this reason, we have made available an example implementation in \emph{Mathematica}, including precalculated results for several spacetimes of interest \cite{Hadamard-WKB-Code}. This code calculates analytic results
for $A_{n,m}(r)$ on a moderate Linux workstation up to order $n\sim 25$, corresponding to a separation $|x-x'|^{50}$, in Schwarzschild and Nariai spacetimes in the order of 1~hour of CPU time.

We also note at this point that knowledge of the $A_{n,m}(r)$ (and their $r$ derivative, which is straightforward to calculate) are all that is required to find the series expansion of $\beta'(r)$ and hence $\alpha(r)$. This can be seen by differentiating Eqs.~\eqref{eq:beta} and \eqref{eq:beta-n} with respect to $r$ to get
\begin{align}
 \beta'(r) &= \epsilon \beta'_0 + \epsilon^2 \beta'_1 + \ldots,
\end{align}
with
\begin{align}
\label{eq:betap-n}
 \beta'_n(r) =& \sum_{m=0}^{2n} \left[ \frac{A'_{n,m}(r) \omega^{2m} }{\chi^{2n+2m+1}} - (n+m+\frac12) \frac{2 \chi \chi' A_{n,m}(r)}{\chi^{2n+2m+3}} \right]\nonumber\\
	=&  \sum_{m=0}^{2n+1} \left\{ A'_{n,m}(r)  - (n+m+\frac12) A_{n,m}(r)\frac{(fg)'}{fg} \right.\\
& \qquad \left.- (n+m-\frac12)A_{n,m-1}(r)\left[ (g^2)'-\frac{(fg)'}{fg}g^2\right]\right\} \frac{\omega^{2m} }{\chi^{2n+2m+1}}
\end{align}
where we have used Eq.~\eqref{eq:chi_defn} to write 
\begin{equation}
2 \chi \chi' = \chi^2 (fg)'/(fg) + \omega^2 ((g^2)' -g^2 (fg)'/(fg)),
\end{equation}
and we use the convention $A_{n,-1}=A_{n,2m+1} = 0$.
With the $A_{n,m}(r)$ and their first derivatives calculated, we are faced with the sum over $l$ in Eq.~(\ref{eq:WKBsum}), where Eqs.~\eqref{eq:beta-n} and \eqref{eq:betap-n} yield sums of the form
\begin{equation}
\label{eq:WKB-simplified-sum}
 \sum_{l=0}^{\infty} 2 (l+{\textstyle\frac{1}{2}}) P_l (\cos \gamma) \frac{D_{n,m}(r) \omega^{2m}}{\chi^{2n+2m+1}}.
\end{equation}
with
\begin{equation}
 D_{n,m} = A_{n,m}(r)
\end{equation}
or
\begin{multline}
D_{n,m} = A'_{n,m}(r)  - (n+m+\frac12) A_{n,m}(r)\frac{(fg)'}{fg} \\- (n+m-\frac12)A_{n,m-1}(r)\left[ (g^2)'-\frac{(fg)'}{fg}g^2\right]
\end{multline}

Since we are considering the points $x$ and $x'$ to be close together, we can treat $\gamma$ as a small quantity and expand the Legendre polynomial in a Taylor series about $\gamma=0$, or, more conveniently, in powers of $(\cos \gamma -1)$ about $(\cos \gamma -1)=0$. It is straightforward to express each term in this series as a polynomial in even powers of $(l+\frac{1}{2})$:
\begin{align}
P_l(\cos \gamma) =& {}_2F_1\left(-l,l+1;1;(1-\cos\gamma)/2\right) \\=&\sum_{p=0}^l \frac{\bigl((l+\frac12)^2-(1-\frac12)^2\bigr)\cdots
\bigl((l+\frac12)^2-(p-\frac12)^2\bigr)}{2^p (p!)^2} (\cos\gamma-1)^p
\end{align}
 The calculation of the sum in Eq.~(\ref{eq:EuclideanGF}) therefore reduces to the calculation of sums of the form
\begin{equation}
\label{eq:WKBsum2}
2 D_{n,m}(r) \sum_{l=0}^{\infty}  \frac{(l+{\textstyle\frac{1}{2}})^{2p+1} \omega^{2m}}{\chi^{2n+2m+1}}.
\end{equation}
For fixed $\omega$ and large $l$ the summand behaves as $l^{2(p-n-m)}$, and so only converges if $p<n+m$.
If $p \geq n+m$, we first split the summand as
\begin{align}
&\frac{(l+{\textstyle\frac{1}{2}})^{2p+1} \omega^{2m}}{\chi^{2m+2n+1}} \nonumber\\
&\quad = \frac{(l+{\textstyle\frac{1}{2}})^{2(p-m-n)}\omega^{2m}}{(fg)^{m+n+1/2}}\left(1+\frac{\omega^2 g/f}{   \left(l+\frac{1}{2}\right)^2} \right)^{-m-n-1/2}\\
&\quad = \frac{(l+{\textstyle\frac{1}{2}})^{2(p-m-n)}\omega^{2m}}{(fg)^{m+n+1/2}}\left\{ \sum_{k=0}^{p-m-n}
\frac{(-1)^k(m+n+1/2)_k}{k!} \left(\frac{\omega^2 g/f}{   \left(l+\frac{1}{2}\right)^2} \right)^{k}
+\right.\nonumber \\
&\qquad+ \left. \left[ \left(1+\frac{\omega^2 g/f}{   \left(l+\frac{1}{2}\right)^2} \right)^{-m-n-1/2}-\sum_{k=0}^{p-m-n}
\frac{(-1)^k (m+n+1/2)_k}{k!} \left(\frac{\omega^2 g/f}{   \left(l+\frac{1}{2}\right)^2} \right)^{k}
\right]\right\}
\end{align}
where $(\alpha)_k=\Gamma(\alpha+k)/\Gamma(\alpha)$ is the Pochhammer symbol and the sum correspond to the first $(p-n-m)$ terms in the expansion of $ (1+x)^{-n-m-1/2}$ about $x\equiv(\omega^2 g/f)/(l+\frac12)^2=0$.
The terms outside the square brackets correspond to positive powers of $\omega$ and so contribute to the light cone singularity, not the tail term $V(x,x')$ with which we are concerned in this section. By contrast, the term in square brackets behaves
as $(l+\frac12)^{-2}$ and so converges as $l\to\infty$ and will contribute to the tail term. We denote this term, with its
prefactor as
\begin{multline}
\label{eq:reg_integrand}
\left[\frac{(l+{\textstyle\frac{1}{2}})^{2p+1} \omega^{2m}}{\chi^{2m+2n+1}}\right]_\mathrm{reg} 
=
\frac{\omega^{2(p-n)}}{f^{p+1/2}g^{2(n+m)-p+1/2}} x^{m+n-p}\times \\
\left[  \left(1+ x \right)^{-m-n-1/2} -\sum_{k=0}^{p-m-n}
\frac{(-1)^k (m+n+1/2)_k}{k!} x^{k} \right]
\end{multline}
and adopt the understanding that the sum vanishes if $p<m+n$.
 
To proceed further, we use the Sommerfeld-Watson formula \cite{Watson:1918},
\begin{align}
\sum_{l=0}^{\infty} F(l+{\textstyle\frac12}) =& \text{Re}  \left[ \frac{1}{i} \int_\gamma dz \> F(z) \tan (\pi z) \right] \nonumber \\ =& \int_{0}^{\infty} F\left(\lambda\right) d\lambda - \text{Re} \left( i \int_{0}^{\infty} \frac{2}{1+e^{2\pi \lambda}} F\left(i \lambda \right) d\lambda \right)
\end{align}
which is valid provided we can rotate the contour of integration for $F(z) \tan(\pi z)$ from just above the real axis to the positive imaginary axis.
Defining $z \equiv (f/g)^{1/2} \lambda/\omega = 1/\sqrt{x}$, where $\lambda\equiv (l+1/2)$, the sum \eqref{eq:WKBsum2} can then be written as the contour integral,
\begin{multline}
\label{eq:WKBintegrals}
 \frac{\omega^{2(p-n)+1}}{f^{p+1}g^{2(m+n)-p}}\left[
\int_{0}^{\infty} \mathrm{d}z\>\left[ \frac{ z^{2p+1} }{(1 +z^2)^{m+n+1/2}}\right]_\mathrm{reg}  \right.\\
\left.+ (-1)^p \text{Re} \left( \int_{0}^{\infty} \frac{2\mathrm{d}z}{1+e^{2\pi z \omega \sqrt{g/f}}} \frac{ z^{2p+1}}{(1 - z^2)^{m+n+1/2}} \right)\right]
\ .
\end{multline}
Note that there is no need to include the regularization terms in the second integral as their contribution is manifestly imaginary and so will not contribute to the final answer.

For $p<m+n$ the first integral in Eq.~(\ref{eq:WKBintegrals}) may be performed immediately as
\begin{equation} 
\int_{0}^{\infty} \mathrm{d}z\> \frac{ z^{2p+1} }{(1 +z^2)^{m+n+1/2}} =  \frac{p!}{2 (m+n-p-1/2)_{p+1}} \ .
\end{equation}
For $p\geq m+n$, we use the regularized integrand arising from Eq.~(\ref{eq:reg_integrand}) and temporarily introduce an ultraviolet cutoff $1/\epsilon^2$ to get
\begin{equation}  
 \int_{\epsilon}^{\infty} \mathrm{d}x \> x^{m+n-p-3/2}\left[(1+x)^{-m-n-1/2} -  \sum_{k=0}^{p-m-n}
\frac{(-1)^k (m+n+1/2)_k}{k!} x^{k} \right]\ .
\end{equation}
Integrating by parts $p-m-n+1$ times, the regularization subtraction terms ensure that the boundary term goes to zero in the limit $\epsilon\to 0$ and we are left with
\begin{multline}  
(-1)^{p-m-n+1} \frac{(m+n+1/2)_{p-m-n+1}}{(1/2)_{p-m-n+1}}\int_{\epsilon}^{\infty} \mathrm{d}x \> x^{-1/2} (1+x)^{-p-3/2}\\
=
 \frac{(-1)^{p-m-n+1} \pi p!}{\Gamma(m+n+1/2)\Gamma(p-m-n+3/2)}\>,
\end{multline}
where the last equality reflects that after these integrations by parts are completed, the
remaining integral is finite in the limit $\epsilon\to 0$.
 
\begin{figure}[ht]
 \begin{center}
  \includegraphics[width=8cm]{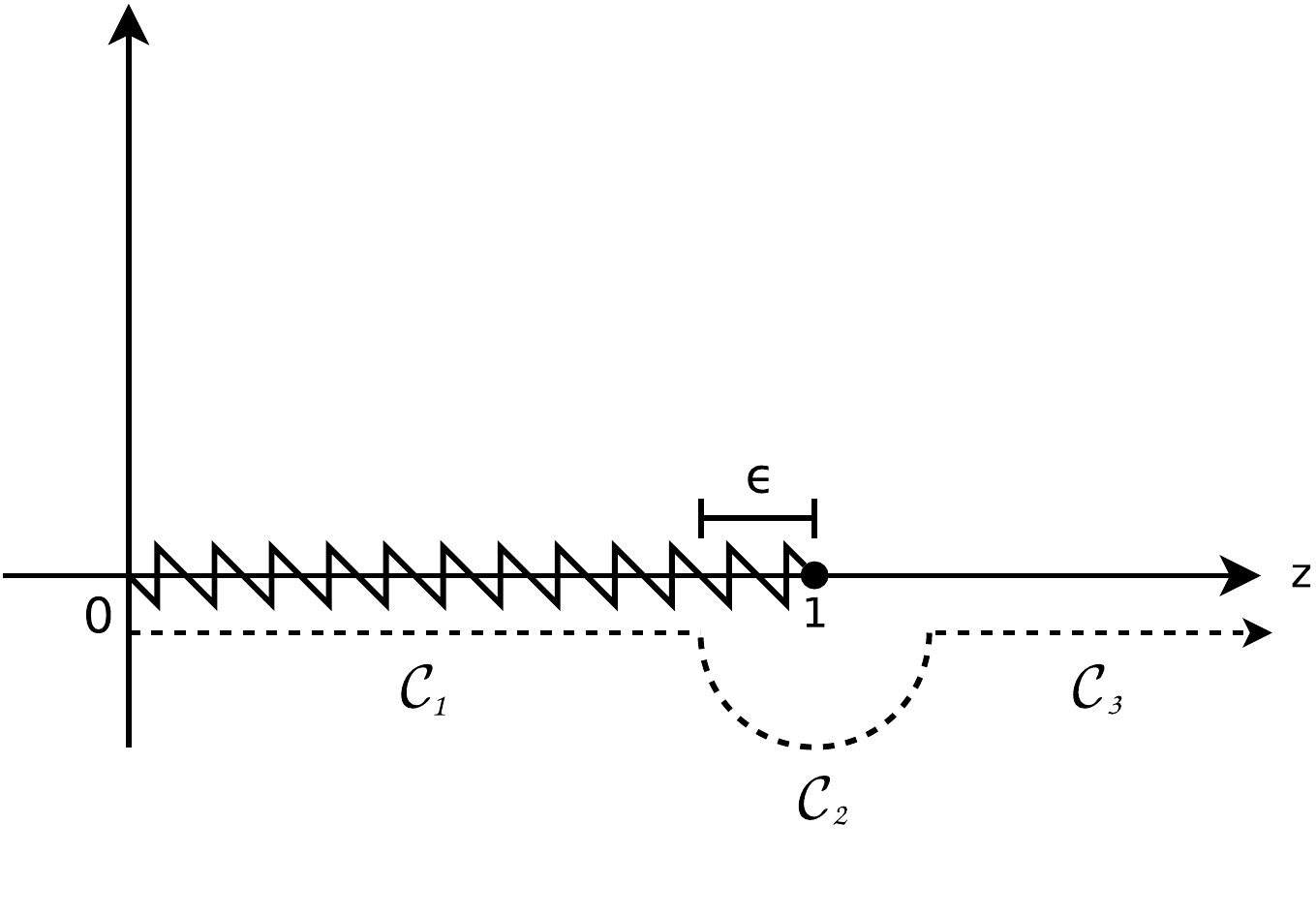}
 \end{center}
 \caption[Contour integral]{The second integral in Eq.~(\ref{eq:WKBintegrals}) has a pole at $z=1$, so we split it into three parts: (1) an integral from $0$ to $1-\epsilon$, (2) an arc of radius $\epsilon$ about $z=1$, and (3) an integral from $z=1+\epsilon$ to $\infty$.}
 \label{fig:wkbcontour}
\end{figure}

The second integral in Eq.~(\ref{eq:WKBintegrals}) is understood as a contour integral as illustrated in Fig.~\ref{fig:wkbcontour}. The integrand is understood to be defined on the complex plane cut from $z=-1$ to $1$
and additionally possesses singularities at $z=-1$ and $z=1$. To handle these in a fashion consistent with the Watson-Sommerfeld prescription we consider the contour in 3 parts: $\mathcal{C}_1$ running just below the cut from $0$ to $1-\epsilon$,
 $\mathcal{C}_2$ a semicircle of radius $\epsilon$ about $z=1$, and $\mathcal{C}_3$ running along the real axis from $z=1+\epsilon$ to $\infty$.

The integrand along $\mathcal{C}_3$ is manifestly imaginary and so gives zero contribution.
Writing the integrand as
\begin{equation}
\frac{ G(z)}{(1 - z)^{m+n+1/2}}
\end{equation}
where
\begin{equation}
\label{eq:Gdef}
 G(z) = \frac{z^{2p+1}}{(1+e^{2\pi z \omega \sqrt{g/f}})(1+z)^{m+n+1/2}}
\end{equation}
it is straightforward to see that
\begin{align}
\text{Re} \int_{\mathcal{C}_2} \frac{ G(z)}{(1 - z)^{m+n+1/2}} \mathrm{d}z = - \sum\limits_{k=0}^\infty \frac{(-1)^{k}}{k!} G^{(k)}(1) 
\frac{\epsilon^{k-m-n+1/2}}{k-m-n+\frac{1}{2}}\ ,
\end{align}
while integrating by parts $m+n$ times
\begin{multline}
\int_{\mathcal{C}_1} \frac{ G(z)}{(1 - z)^{m+n+1/2}} \mathrm{d}z = \sum\limits_{k=0}^{m+n-1} \frac{(-1)^{k}}{k!} G^{(k)}(1) 
\frac{\epsilon^{k-m-n+1/2}-1}{k-m-n+\frac{1}{2}} \\
 + \int\limits_0^{1-\epsilon} \frac{ \mathrm{d} z}{(1-z)^{m+n+1/2}}
  \left[  G(z) - \sum\limits_{k=0}^{m+n-1} \frac{(-1)^{k}}{k!} G^{(k)}(1) (1-z)^k \right]\ .
\end{multline}
Adding the contributions from these components, it is clear that the $\epsilon \to 0$ divergences cancel and we are left with
\begin{multline}
\label{eq:Gresult}
- \sum\limits_{k=0}^{m+n-1} \frac{(-1)^{k}}{k!} \frac{G^{(k)}(1)}{k-m-n+\frac{1}{2}} \\+ \int\limits_0^{1} \frac{ \mathrm{d} z}{(1-z)^{m+n+1/2}}
  \left[  G(z) - \sum\limits_{k=0}^{m+n-1} \frac{(-1)^{k}}{k!} G^{(k)}(1) (1-z)^k \right]\ ,
\end{multline}
where the subtraction terms in the integrand ensure that the integral here is well-defined.

While we can take this analysis further~\cite{Ottewill:Winstanley:Young:2009}, 
for our current purpose we note that we only need the expansion of Eq.~(\ref{eq:Gresult}) in terms of inverse \textit{powers} of $\omega$
as  $\omega\to \infty$.
From Eq.~(\ref{eq:Gdef}), it is immediately apparent that the terms in the sum in (\ref{eq:Gresult}) are exponentially small
and so may be ignored for our purposes here.  Indeed we may simultaneously increase the upper limit
on the two sums in Eq.~(\ref{eq:Gresult}) without changing the result.  Increasing it by 1 (or more), the
integrand increases from $z=0$, peaks and then decreases to $0$ at $z=1$ with the peak approaching $z=0$ as $\omega\to\infty$.
Standard techniques from statistical mechanics then dictate that the $\omega\to\infty$ asymptotic form of
the integral follows from expanding the integrand, aside from the `Planck factor',
about $z=0$ and extending the upper limit to $\infty$. Again, in doing so the contribution from the summation within the integrand gives an exponentially small contribution
so that the powers of $\omega$ are determined simply by
\begin{align}
\int\limits_0^{\infty} \frac{ \mathrm{d} z}{(1+e^{2\pi z \omega \sqrt{g/f}})}
   \mathop{\mathrm{Series}}\limits_{z=0} \left[\frac{z^{2p+1}}{(1-z^2)^{m+n+1/2}}\right],
\end{align}
where $\mathop{\mathrm{Series}}\limits_{z=0}\left[\cdots\right]$ denotes the series expansion about $z=0$. 
The coefficients in the series are known analytically, so to expand in inverse powers of $\omega$ we only need to compute integrals of the form
\begin{equation}
\int_0^{\infty} \frac{z^{2N-1}}{1+e^{2\pi   z \omega \sqrt{g/f}}} \, dz
\end{equation}
which have the exact solutions \cite{GradRyz}
\begin{equation}
\left(1-2^{1-2N}\right)\frac{f^N}{g^N \omega^{2N}}\frac{|B_{2N}|}{4N},
\end{equation}
where $B_N$ is the $N$th Bernoulli number. This expression allows us to calculate the integrals very quickly.

Applying this method for summation over $l$, Eq.~(\ref{eq:EuclideanGF}) takes the form required by Eq.~(\ref{eq:WKB-int-log-relation}) so we now have the logarithmic part of the Euclidean Green function and therefore $V(x,x')$ as the required power series in $(t-t')$, $(\cos \gamma - 1)$ and $(r-r')$. Because of the length of the expressions involved, we have made available online \cite{Hadamard-WKB-Code} a \emph{Mathematica} code implementing this algorithm, along with precalculated results for several spacetimes of interest including Schwarzschild, Nariai and Reissner-Nordstr\"om. These expressions are in total agreement with those calculated using the covariant method of Chapter~\ref{ch:covex}. They also correct a factor of $2$ error in Ref.~\cite{Anderson:2003} (this correction is also made in the errata \cite{Anderson:2003:err1,Anderson:2003:err2} for Ref.~\cite{Anderson:2003}).

\section{Convergence of the Series}
\label{sec:convergence}
We have expressed $V(x,x')$ as a power series in the separation of the points. This series will, in general, not be convergent for all point separations -- the maximum point separation for which the series remains convergent will be given by its \emph{radius of convergence}. In this section, we explore the radius of convergence of the series in the Nariai and Schwarzschild spacetimes and use this as an estimate on the region of validity of our series.

For simplicity, we will consider points separated only in the time direction, so we will have a power series in $(t-t')$,
\begin{equation}
\label{eq:CoordGreenT}
V\left( x,x' \right) = \sum_{n=0}^{\infty} v_{n}(r) \left( t - t' \right)^{2n},
\end{equation}
where $v_{n}(r)$ is a real function of the radial coordinate, $r$ only. We will also consider cases where the points are separated by a fixed amount in the spatial directions, or where the separation in other directions can be re-expressed in terms of a time separation, resulting in a similar power series in $(t-t')$, but with the coefficients, $v_{n}(r)$, being different. This will give us sufficient insight without requiring overly complicated convergence tests.

In the next section, we review some tests that will prove useful. In Secs.~\ref{sec:Nariai-tests} and \ref{sec:Schw-tests} we present the results of applying those tests in Nariai and Schwarzschild spacetimes, respectively.
\subsection{Tests for Estimating the Radius of Convergence} \label{sec:Tests}
\subsubsection{Convergence Tests}
For the power series (\ref{eq:CoordGreenT}), there are two convergence tests which will be useful for estimating the radius of convergence. The first of these, the \emph{ratio test}, gives an estimate of the radius of convergence, $\Delta t_{RC}$,
\begin{equation}
 \Delta t_{RC} = \lim_{n\rightarrow\infty}\sqrt{\left|\frac{v_{n}}{v_{n+1}}\right|}.
\end{equation}
Although strictly speaking, the large $n$ limit must be taken, in practice we can calculate enough terms in the series to get a good estimate of the limit by simply looking at the last two terms. The ratio test runs into difficulties, however, when one of the terms in the series is zero. Unfortunately, this occurs frequently for many cases of interest. It is possible to avoid this issue somewhat by considering non-adjacent terms in the series, i.e. by comparing terms of order $n$ and $n+m$, 
\begin{equation}
\label{eq:ratio-test}
 \Delta t_{RC} = \lim_{n\rightarrow\infty}\left|\frac{v_{n}}{v_{n+m}}\right|^{\frac{1}{2m}}.
\end{equation}
Using the ratio test in this way gives better estimates of the radius of convergence, although the results are still somewhat lacking.

To get around this difficulty, we also use a second test, the \emph{root test},
\begin{equation}
  \Delta t_{RC} = \limsup_{n\rightarrow\infty} \left|\frac{1}{v_n}\right|^{\frac{1}{2n}},
\end{equation}
which is well suited to power series. This gives us another estimate of the radius of convergence. Again, in practice the last calculated term in the series will give a good estimate of the limit.

It may appear that only the (better behaved) root test is necessary for estimating the radius of convergence of the series. However, extra insight can be gained from including both tests. This is because for the power series \eqref{eq:CoordGreenT}, the ratio test typically gives values \emph{increasing} in $n$ while the root test gives values \emph{decreasing} in $n$, effectively giving lower and upper bounds on the radius of convergence.

\subsubsection{Normal Neighborhood} \label{subsubsec:NN}
The Hadamard parametrix for the retarded Green function, (\ref{eq:Hadamard-scalar}), is only guaranteed to be valid provided $x$ and $x'$ are within a \emph{normal neighborhood} (see footnote \ref{def:causal domain}). This arises from the fact that the Hadamard parametrix involves the Synge world function, $\sigma(x,x')$, which is only defined for the points $x$ and $x'$ separated by a \emph{unique geodesic}. It is plausible that the radius of convergence of our series could exactly correspond to the normal neighborhood size, $t_{\text{NN}}$. This turns out to not be the case, although it is still helpful to give consideration to $t_{\text{NN}}$ as we might expect it to place an upper bound on the radius of convergence of the series. This upper bound can be understood from the fact that the series will have a radius of convergence determined by the nearest singularity of $V(x,x')$, along with the fact that we expect $V(x,x')$ to be singular at the boundary of the normal neighborhood\footnote{The expectation that $V(x,x')$ is singular at the boundary of the normal neighborhood may be justified as follows: $V_0(x,x')$ is defined through an equation involving $\Delta^{1/2}(x,x')$, which is singular at a \emph{caustic}. As a result, we would also expect $V_0(x,x')$ to be singular at a caustic (this is supported by our numerical calculations, described in Chapter~\ref{ch:covex}). Since $V(x,x')$ satisfies the homogeneous wave equation with characteristic initial data given by $V_0(x,x')$, we expect this singularity to propagate along characteristics (null geodesics). Therefore, $V(x,x')$ will be singular where the past light-cone of the caustic intersects the time-like geodesics emanating from $x$. This intersection also corresponds to the boundary of the normal neighborhood.}

The normal neighborhood size will be given by the minimum time separation of the spacetime points such that they are connected by two non-spacelike geodesics. For typical cases of interest for self-force calculations there will be a particle following a time-like geodesic, so $t_{\text{NN}}$ will be given by the minimum time taken by a null geodesic intersecting the particle's worldline twice. In typical black hole spacetimes, this geodesic will orbit the black hole once before re-intersecting the particle's worldline.

Another case of interest is that of the points $x$ and $x'$ at constant spatial positions, separated by a constant angle $\Delta\phi$. Initially, (i.e. when $t=t'$), the points will be separated only by spacelike geodesics. After sufficient time has passed for a null geodesic to travel between the points (going through an angle $\Delta\phi$), they will be connected first by a null geodesic and subsequently by a sequence of unique timelike geodesics. Since the geodesics are unique, $t_{\text{NN}}$ will not be given by this \emph{first} null geodesic time. Rather, $t_\text{NN}$ will be given by the time taken by the \emph{second} null geodesic (passing through an angle $2\pi - \Delta\phi$). This subtle, but important, distinction will be clearly evident when we study specific cases in the next sections.

\subsubsection{Relative Truncation Error}
Knowledge of the radius of convergence alone does not give information about the accuracy of the series representation of $V(x,x')$. The series is necessarily truncated after a finite number of terms, introducing a \emph{truncation error}. The local fractional truncation error can be estimated by the ratio between the highest order term in the expansion ($O \left( \Delta \tau ^n \right)$, say) and the sum of all the terms up to that order,
\begin{equation} \label{eq:trunc-error}
\epsilon \equiv \frac{v_n (t-t')^{2n}}{\sum_{i=0}^{n} v_i (t-t')^{2i}}.
\end{equation}
Reference \cite{Anderson:Wiseman:2005} previously calculated these error estimates, but only considered the first two terms in the series. Since we now have a vastly larger number of terms available, it is worthwhile to consider these again to determine the accuracy of the high order series.

\subsection{Nariai Spacetime} \label{sec:Nariai-tests}
\subsubsection{Normal Neighborhood} \label{subsec:Nariai-NN}

Allowing only the time separation of the points to change, we consider two cases of interest for the Nariai spacetime:
\begin{enumerate}
 \item The static particle which has a normal neighborhood determined by the minimum coordinate time taken by a null geodesic circling the origin ($\rho=0$) before returning. This is the time taken by a null geodesic, starting at $\rho=\rho_1$ and returning to $\rho'=\rho_1$, while passing through an angle $\Delta\phi=2\pi$. \label{enum:static-nariai}
 \item Points at fixed radius, $\rho_1$, separated by an angle, say $\pi/2$. In this case, there is a null geodesic which goes through an angle $\Delta\phi=\pi/2$ when travelling between the points. However, there will not yet be any other (non-spacelike) geodesic connecting them, so this will not give $t_{\text{NN}}$. Instead, it is the next null geodesic, which goes through $\Delta\phi=3\pi/2$, that gives the normal neighborhood boundary. \label{enum:static-nariai-angle}
\end{enumerate}

In both cases, the coordinate time taken by the null geodesic to travel between the points is given by \eqref{geodesic-time},
\begin{equation}
 t_{NN} = 2 \tanh^{-1}(\rho_1) + \ln \left( \frac{1-\rho_1 \text{sech}^2(\Delta\phi/2) + \tanh(\Delta\phi/2) \sqrt{1 - \rho_1^2 \text{sech}^2(\Delta\phi/2)}}{ 1+\rho_1 \text{sech}^2(\Delta\phi/2) - \tanh(\Delta\phi/2) \sqrt{1 - \rho_1^2 \text{sech}^2(\Delta\phi/2) }}  \right)  \label{eq:Nariai-geodesic-time}.
\end{equation}

\subsubsection{Results}
For the Nariai spacetime, the ratio test suffers from difficulties arising from zeros of the terms in the series. This is because the $i^{th}$ term of the series (of order $(t-t')^{2i}$) has $2i$ roots in $\rho$. In other words, the higher the order of the terms considered, the more likely one of the coefficients is to be near a zero, and not give a useful estimate of the radius of convergence. We have therefore compared non-adjacent terms to avoid this issue as much as possible, by choosing $m=n/2$ in Eq.~\eqref{eq:ratio-test}. This choice of $m$ only affects the number and location of the points where the test fails. The \emph{envelope} (interpolating over the points where the test fails) remains unaffected. Fortunately, the root test is much less affected by such issues and can be used without any adjustments.

\begin{figure}
  \begin{center}
  \includegraphics[width=7cm]{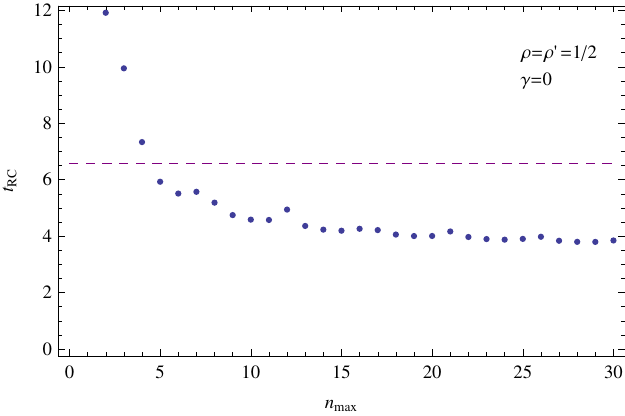}
  \includegraphics[width=7cm]{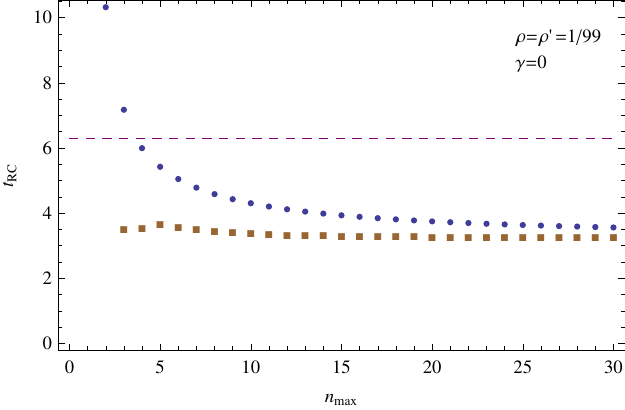}
  \includegraphics[width=7cm]{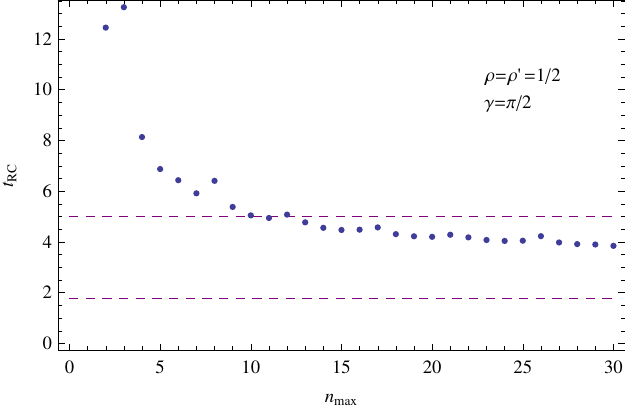}
  \includegraphics[width=7cm]{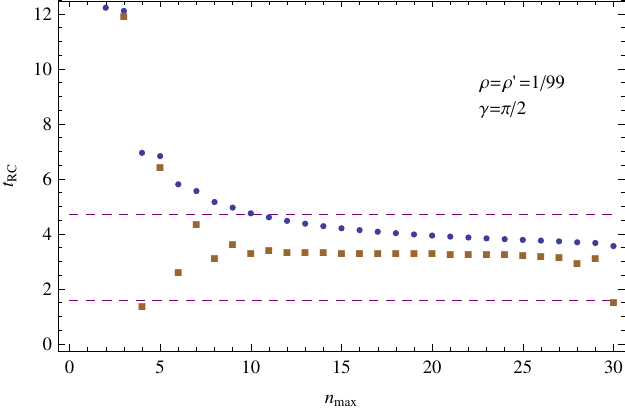}
\end{center}
 \caption[Radius of convergence of Taylor series as a function of the order for the Nariai spacetime]{\emph{Radius of convergence as a function of the number of terms in the series for the Nariai spacetime with curvature coupling $\xi=1/6$.} The limit as $n\rightarrow\infty$ will give the actual radius of convergence, but it appears that just using the terms up to $n=30$ is giving a good estimate of this limit. The radius of convergence is estimated by the root test (blue dots) and ratio test (brown squares) and is compared against the normal neighborhood size (purple dashed line) calculated from considerations on null geodesics (see Sec.~\ref{subsubsec:NN} and Sec.~\ref{subsec:Nariai-NN}). Note that plots for the ratio test were omitted in cases where it did not give meaningful results. 

Also, the lower plots both have two lines estimating the normal neighborhood size. The lower line indicates the time for a null geodesic to go \emph{directly} between the points, passing through an angle $\pi/2$. It is the upper line, corresponding to the time taken by a null geodesic circling the potential (and going through an angle $3\pi/2$) which gives the true value of the normal neighborhood size (see Sec.~\ref{subsec:Nariai-NN}).}
 \label{fig:nariai-roc2}
\end{figure}

In Fig.~\ref{fig:nariai-roc2}, we fix the radial position of the points at $\rho=\rho'=1/2$ (left panels) and $\rho=\rho'=1/99$ (right panels) and plot the results of applying the root test (blue dots) and ratio test (brown squares) for (\ref{enum:static-nariai}) static particle (top panels) and (\ref{enum:static-nariai-angle}) points at fixed spatial points separated by an angle $\gamma=\pi/2$ (bottom panels) as a function of the maximum order, $n_\text{max}$, of the terms of the series considered, $v_{n_\text{max}} \left(t-t^\prime \right)^{2n_\text{max}}$. It seems in both cases that the plot is limiting towards a constant value for the radius of convergence. In case (\ref{enum:static-nariai}), we see that this gives a radius of convergence that is considerably smaller than the normal neighborhood size (purple dashed line). For case (\ref{enum:static-nariai-angle}), we again see that the values from the convergence tests are limiting towards a value for the radius of convergence. However, as discussed in Sec.~\ref{subsubsec:NN}, it is not the first null geodesic (lower dashed purple line), but the second null geodesic (upper dashed purple line) that determines the normal neighborhood and places an upper bound on the radius of convergence of the series.

\begin{figure}
  \begin{center}
  \includegraphics[width=7cm]{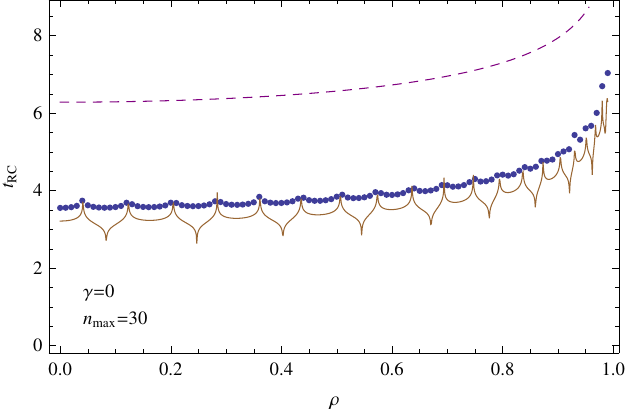}
  \includegraphics[width=7cm]{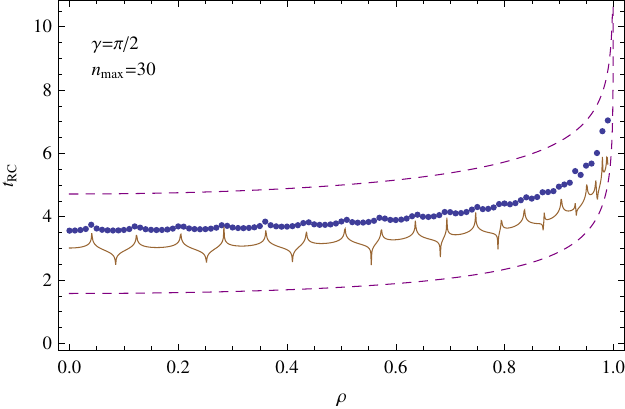}
\end{center}
 \caption[Radius of convergence of Taylor series as a function of radial position for the Nariai spacetime]{\emph{Estimates of the domain of validity (i.e. the radius of convergence) of the series expansion of $V(x,x')$ as a function of radial position in Nariai spacetime.} The root test on $O\left[(t-t')^{60}\right]$ series is given as blue dots, the ratio test is given by the brown line (see footnote \ref{fn:blips}), and the normal neighborhood estimate from null geodesics is given by dashed purple lines. The left plot is for case (\ref{enum:static-nariai}), the static particle, and the right plot is for case (\ref{enum:static-nariai-angle}), points separated by an angle of $\pi/2$. In both cases, the series is clearly divergent before the normal neighborhood boundary. This boundary is sometimes given by the second, rather than the first null geodesic as can be seen in the plot on the right. In particular, this is the case for the points separated by an angle $\gamma=\pi/2$ since they are initially separated by only spacelike geodesics (see Sec.~\ref{subsec:Nariai-NN}).}
 \label{fig:nariai-roc}
\end{figure}

In Fig.~\ref{fig:nariai-roc} we use the root test (blue dots) and ratio test (brown line)\footnote{\label{fn:blips}Note that the `blips' in the ratio test are an artifact of the zeros of the series coefficients used and are not to be taken to have any physical meaning. In fact, the `blips' occur at different times when considering the series at different orders, so they should be ignored altogether. The ratio test plots should therefore only be fully trusted \emph{away} from the `blips'. Near the blips, it is clear that one could interpolate an approximate value, however we have not done so here as the plot is to be taken only as an \emph{indication} of the radius of convergence.} to investigate how the radius of convergence of the series varies as a function of the radial position of the points, $\rho_1$. Again, we look at both cases
(\ref{enum:static-nariai}) (left panel) and (\ref{enum:static-nariai-angle}) (right panel). As a reference, we compare to the normal neighborhood size (purple dashed line). We find that, regardless of the radial position of the points, the radius of convergence of the series is well within the normal neighborhood, by an almost constant amount. As before, in the case (\ref{enum:static-nariai-angle}) of the points separated by an angle, we find that it is the second, not the first null geodesic that gives the normal neighborhood size.

\begin{figure}
  \begin{center}
  \includegraphics[width=7cm]{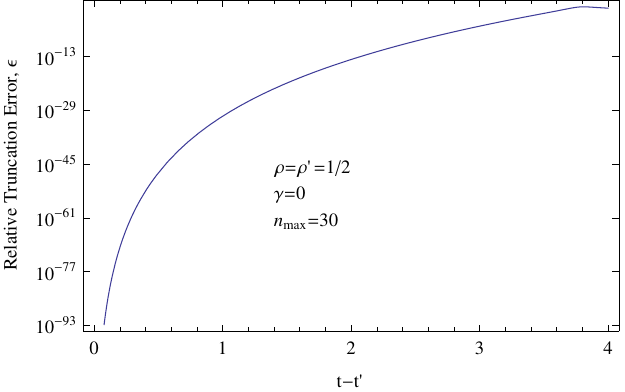}
  \includegraphics[width=7cm]{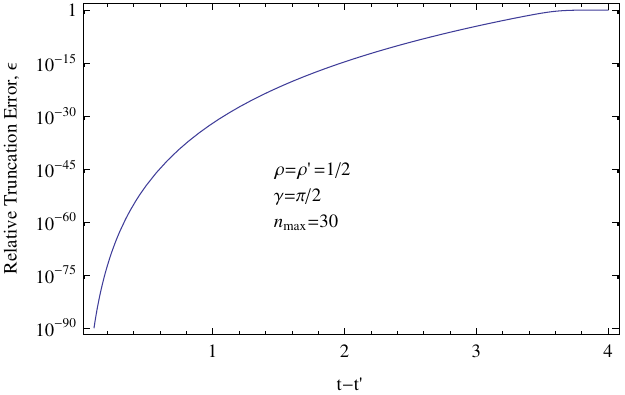}
\end{center}
 \caption[Relative truncation error in Nariai spacetime]{\emph{Relative truncation error in Nariai spacetime} arising from truncating the series expansion for $V(x,x')$ at order $|x-x'|^{60}$ (i.e. $n_\text{max}=30$) for case (\ref{enum:static-nariai}) the static particle (left panel) and case (\ref{enum:static-nariai-angle}) points separated by an angle $\pi/2$ (right panel). In both cases, the radial points are fixed at $\rho=\rho'=1/2$. The series is extremely accurate until we get close to the radius of convergence (see Fig.\ref{fig:nariai-roc2}).}
 \label{fig:nariai-trunc}
\end{figure}

With knowledge of the radius of convergence of the series established, it is also important to estimate the accuracy of the series within that radius. To that end, we plot in Fig.~\ref{fig:nariai-trunc} the relative truncation error, (\ref{eq:trunc-error}), as a function of the time separation of the points at a fixed radius, $\rho=\rho'=1/2$. We find that the $60^{th}$ order series is extremely accurate to within a short distance of the radius of convergence of the series.

\subsection{Schwarzschild Spacetime} \label{sec:Schw-tests}
\subsubsection{Normal Neighborhood}
For a fixed spatial point at radius $r_1$ in the Schwarzschild spacetime, we would like to find the null geodesic that intersects it twice in the shortest time. This geodesic will orbit the black hole once before returning to $r_1$. Clearly, the coordinate time $t_{\text{NN}}$ for this orbit can only depend on $r_1$. The periapsis radius, $r_p$, will be reached half way through the orbit. For the radially inward half of this motion, the geodesic equations can be rearranged to give
\begin{align}
 \pi = \int_0^\pi d\phi &= - \int_{r_1}^{r_p} \frac{dr}{r^2 \sqrt{\frac{1}{r_p^2}\left(1-\frac{2M}{r_p}\right)-\frac{1}{r^2}\left(1-\frac{2M}{r}\right)}}\\
 \frac{t_{NN}}{2} = \int_0^{t_{NN}/2} dt &= - \int_{r_1}^{r_p} \frac{dr}{\left(1-\frac{2M}{r}\right)\sqrt{1-\frac{r_p^2}{r^2}\left(1-\frac{2M}{r}\right)\left(1-\frac{2M}{r_p}\right)^{-1}}}
\end{align}
For a given point $r_1$, we numerically solve the first of these to find the periapsis radius, $r_p$, and then solve the second to give the normal neighborhood size, $t_{\text{NN}}$.

\subsubsection{Results}
The ratio test proves more stable for Schwarzschild spacetime than it was for Nariai spacetime. The series coefficients still have a large number of roots in $r$, but most are within $r=6M$ and therefore do not have an effect for the physically interesting radii, $r\ge6M$.

\begin{figure}
  \begin{center}
  \includegraphics[width=7cm]{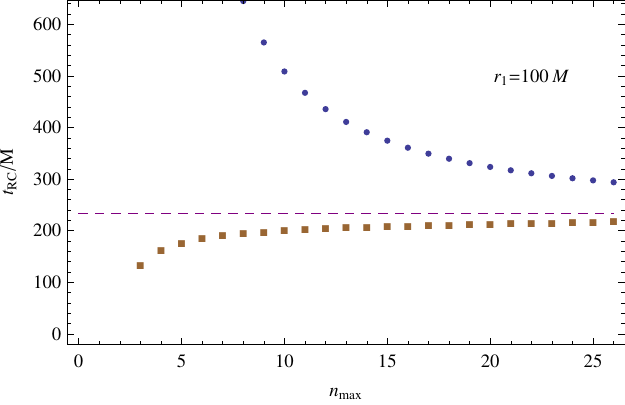}
  \includegraphics[width=7cm]{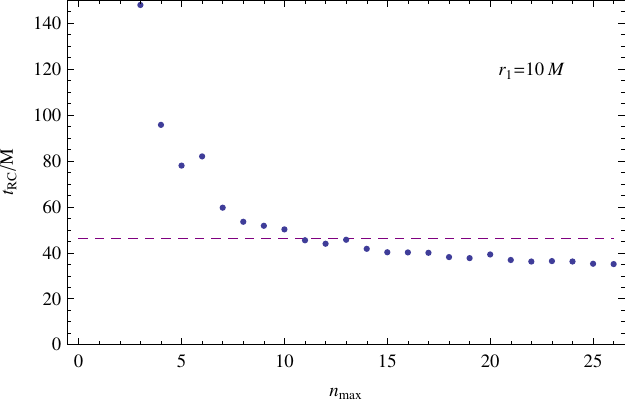}
 \end{center}
 \caption[Radius of convergence of Taylor series as a function of the order for the Schwarzschild spacetime]{\emph{Radius of convergence as a function of the number of terms considered for the Schwarzschild spacetime.} The root test (blue dots), the ratio test (brown squares -- see footnote \ref{fn:blips}), and the first null geodesic (purple dashed line) are given. Left panel: Static particle at $r_1=100M$. Right panel: Static particle at $r_1=10M$. The radius of convergence is given by the limit as the number of terms $n_{\text{max}} \to \infty$. It is apparent that the right plot is asymptoting to a value near, but slightly lower than the normal neighborhood size. This may also be case for the left plot, although higher order terms would be required for a conclusive result. Note that curves for the ratio test were omitted in cases where it did not give meaningful results.}
 \label{fig:schw-static-roc-terms}
\end{figure}

Figure \ref{fig:schw-static-roc-terms} shows that the radius of convergence, $\Delta t_{RC}$, given by the root test is a decreasing function of the order of the term used, while that given by the ratio test is increasing. This effectively gives an upper and lower bound on the radius of convergence of the series. There is some `noise' in the ratio test plot at lower radii (where that test is failing to give meaningful results), but we simply ignore this and omit the ratio test in this case. For the root test, the terms up to order $(t-t')^{52}$ were used, while for the ratio test, adjacent terms in the series up to order $(t-t')^{52}$ were compared.

\begin{figure}
  \begin{center}
  \includegraphics[width=7cm]{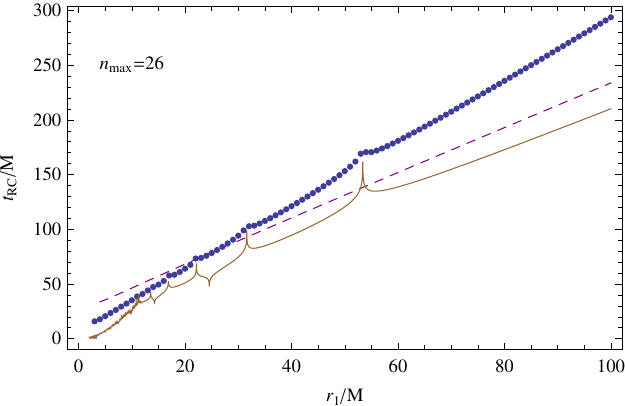}
 \end{center}
 \caption[Radius of convergence of Taylor series as a function of radial position for the Schwarzschild spacetime]{\emph{Radius of convergence as a function of radial position for a static point in Schwarzschild spacetime.} The root test (blue dots), the ratio test (brown line -- see footnote \ref{fn:blips}), and the first null geodesic (purple dashed line) are shown.}
 \label{fig:schw-static-roc-radius}
\end{figure}

In Fig. \ref{fig:schw-static-roc-radius} we apply the root (blue dots) and ratio (brown line -- see footnote \ref{fn:blips}) tests for the case of a static particle at a range of radii in Schwarzschild spacetime. Using the root test as an upper bound and the ratio test as a lower bound, it is clear that the radius of convergence is near, but likely slightly lower than the normal neighborhood size (purple dashed line). In this case, for the root test, the term of order $(t-t')^{52}$ was used, while for the ratio test, terms of order $(t-t')^{52}$ and $(t-t')^{26}$ were compared.

\begin{figure}
  \begin{center}
  \includegraphics[width=7cm]{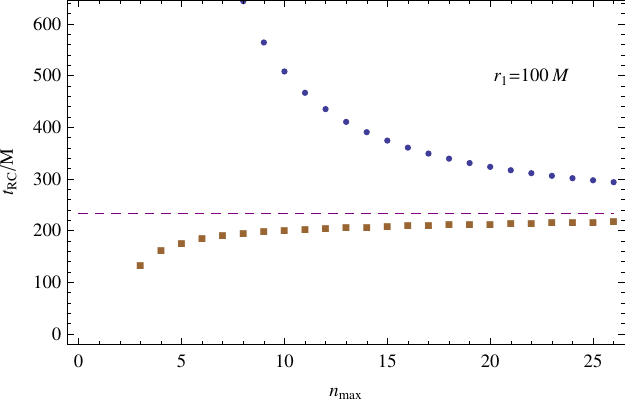}
  \includegraphics[width=7cm]{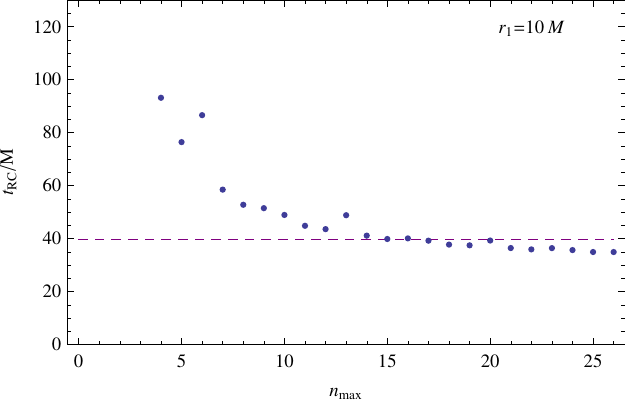}
 \end{center}
 \caption[Radius of convergence of Taylor series as a function of the order for a circular geodesic in Schwarzschild spacetime]{\emph{Radius of convergence as a function of the number of terms considered for points separated along a circular geodesic in Schwarzschild spacetime.} The root test (blue dots), the ratio test (brown squares -- see footnote \ref{fn:blips}), and the first null geodesic (purple dashed line) are shown. Note that curves for the ratio test were omitted in cases where it did not give meaningful results.}
 \label{fig:schw-circ-roc}
\end{figure}

\begin{figure}
  \begin{center}
  \includegraphics[width=7cm]{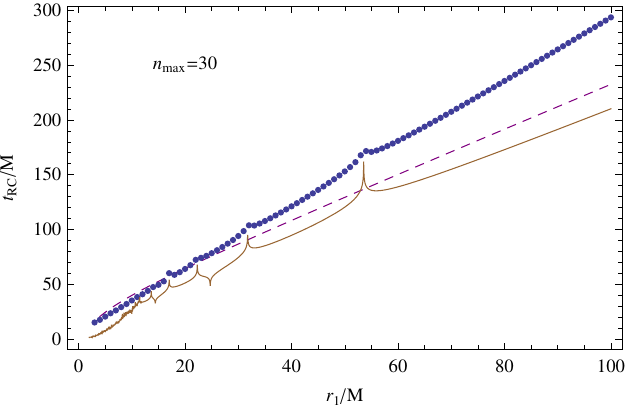}
 \end{center}
 \caption[Radius of convergence of Taylor series as a function of radial position for a circular geodesic in  Schwarzschild spacetime]{\emph{Radius of convergence as a function of radial position for points separated along a circular geodesic in Schwarzschild spacetime.} The root test (blue dots), the ratio test (brown line -- see footnote \ref{fn:blips}), and the first null geodesic (purple dashed line) are shown.}
 \label{fig:schw-circ-roc-radius}
\end{figure}

In Figs.~\ref{fig:schw-circ-roc} and \ref{fig:schw-circ-roc-radius}, we repeat this for the case of the points separated on a circular timelike geodesic. The results are very similar to the static particle case and show the same features.

\section{Extending the Domain of Series Using Pad\'{e} Approximants} \label{sec:Pade}
In the previous section it was shown that the circle of convergence of the series expansion of $V(x,x')$ is smaller than the size of the normal neighborhood. This is not totally unexpected. We would expect the normal neighborhood size to place an upper limit on the radius of convergence, but we can not necessarily expect the radius of convergence to be exactly the normal neighborhood size. However, since the Hadamard parametrix for the Green function is valid everywhere within the normal neighborhood, it is reasonable to hope that it would be possible to find an alternative series representation for $V(x,x')$ which is valid in the region outside the circle of convergence of the original series, while remaining within the normal neighborhood.

The radius of convergence found in the previous section locates the distance (in the complex plane) to the closest singularity of $V(x,x')$. However, that singularity could lie anywhere on the (complex) circle of convergence and will not necessarily be on the real line. In fact, given that the Green function is clearly not singular at the (real-valued) radius of convergence, it is clear that the singularity of $V(x,x')$
does not lie on the real line (see Appendix~\ref{appendix:convergence} for a general discussion of the convergence properties of series).

There are several techniques which can be employed to extend a series beyond its radius of convergence. Provided the circle of convergence does not constitute a \emph{natural boundary} of the function, the method of \emph{analytic continuation} can be used to find another series representation for $V(x,x')$ valid outside the circle of convergence of the original series \cite{M&F,Whittaker:Watson}. This could then be applied iteratively to find series representations covering the entire range of interest of $V(x,x')$. Although this method of analytic continuation should be capable of extending the series expansion of $V(x,x')$, there is an alternative method, the method of \emph{Pad\'{e} approximants} which yields impressive results with little effort.

The method of Pad\'{e} approximants \cite{Bender:Orszag,NumericalRecipes} is frequently used to extend the series representation of a function beyond the radius of convergence of the series. It has been employed in the context of general relativity data analysis with considerable success \cite{Damour:Iyer:Sathyaprakash,Porter:Sathyaprakash}. It is based on the idea of expressing the original series as a rational function (i.e. a ratio of two polynomials $V(x,x')=R(x,x')/S(x,x')$) and is closely related to the continued fraction representation of a function \cite{Bender:Orszag}. This captures the functional form of the singularities of the function on the circle of convergence of the original series.

The Pad\'{e} approximant, $P_M^N (t-t')$ is defined as
\begin{equation}
 P_M^N (t-t') \equiv \frac{\sum_{n=0}^N A_n (t-t')^n}{\sum_{n=0}^M B_n (t-t')^n}
\end{equation}
where $B_0 = 1$ and the other $(M+N+1)$ terms are found by comparing to the first $(M+N+1)$ terms of the original power series. The choice of $M$ and $N$ is arbitrary provided $M+N\le n_{\text{max}}$ where $n_{\text{max}}$ is the highest order term that has been computed for the original series. There are, however, choices for $M$ and $N$ which give the best results. In particular, the diagonal, $P^N_N$, and sub-diagonal, $P^N_{N+1}$, Pad\'e approximants yield optimal results.

\subsection{Nariai spacetime} \label{subsec:padeNariai}
\begin{figure}
  \begin{center}
  \includegraphics[width=7cm]{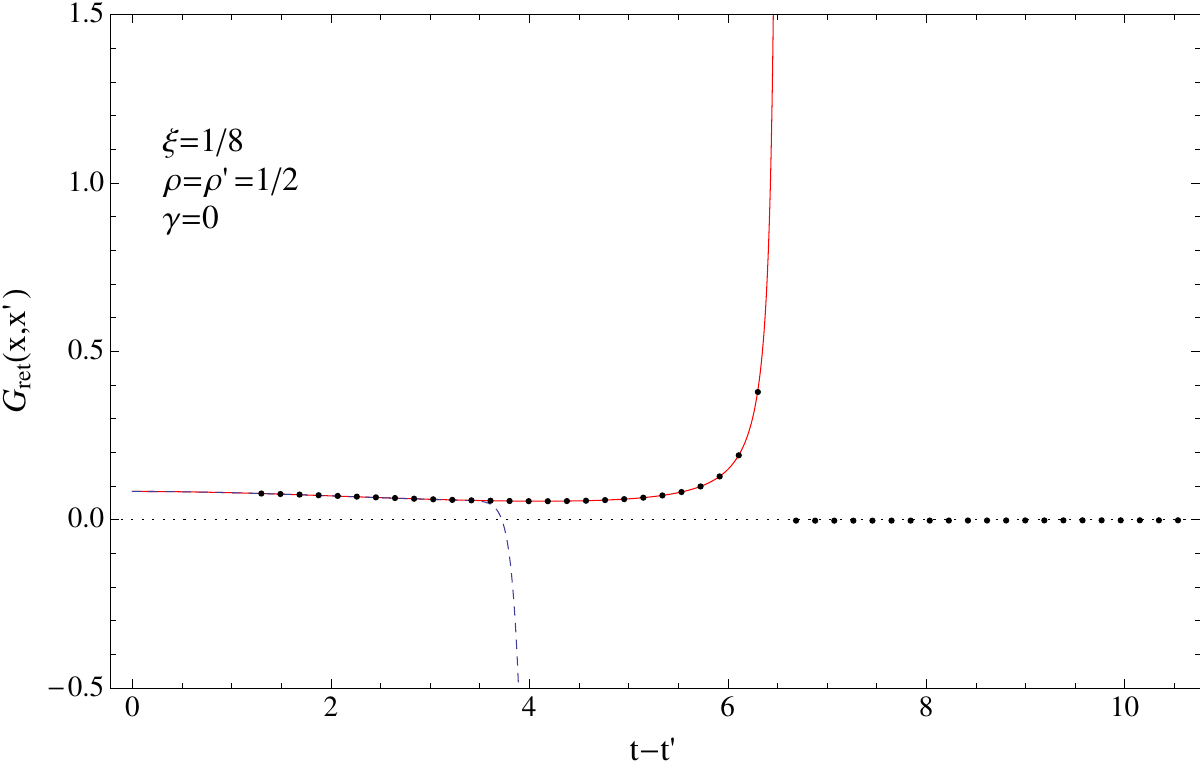}
  \includegraphics[width=7cm]{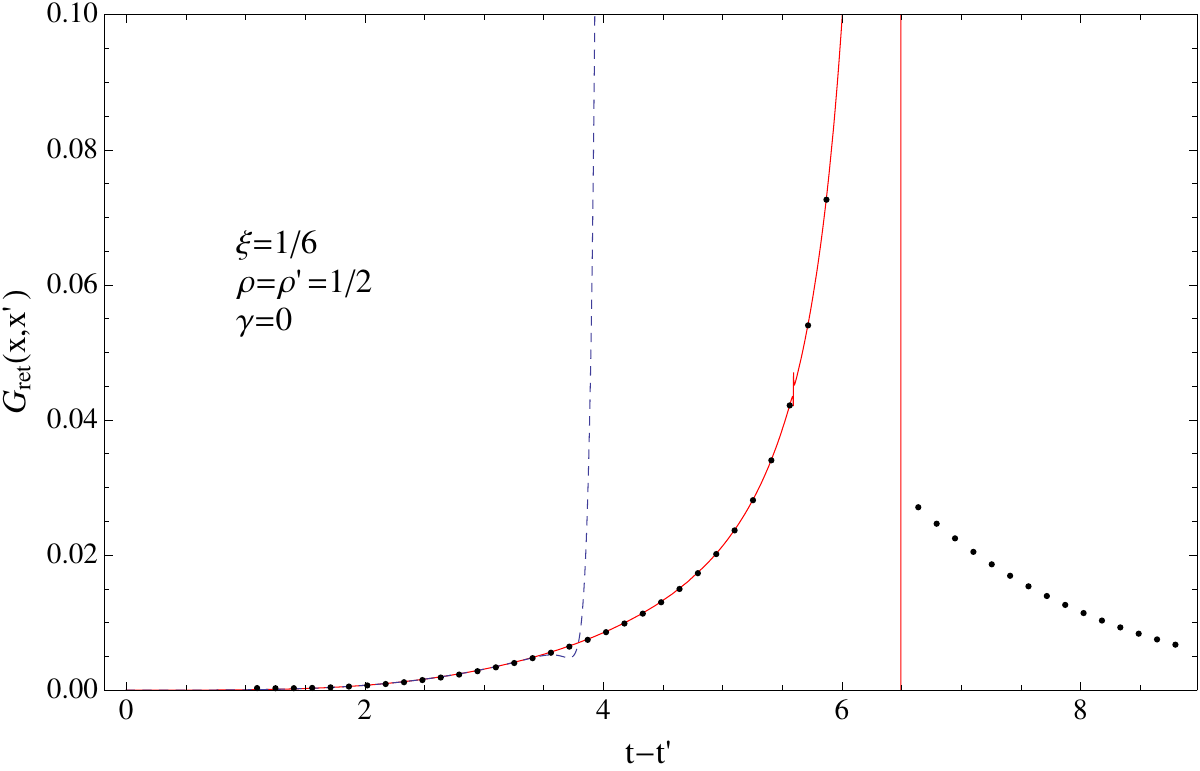}
  \includegraphics[width=7cm]{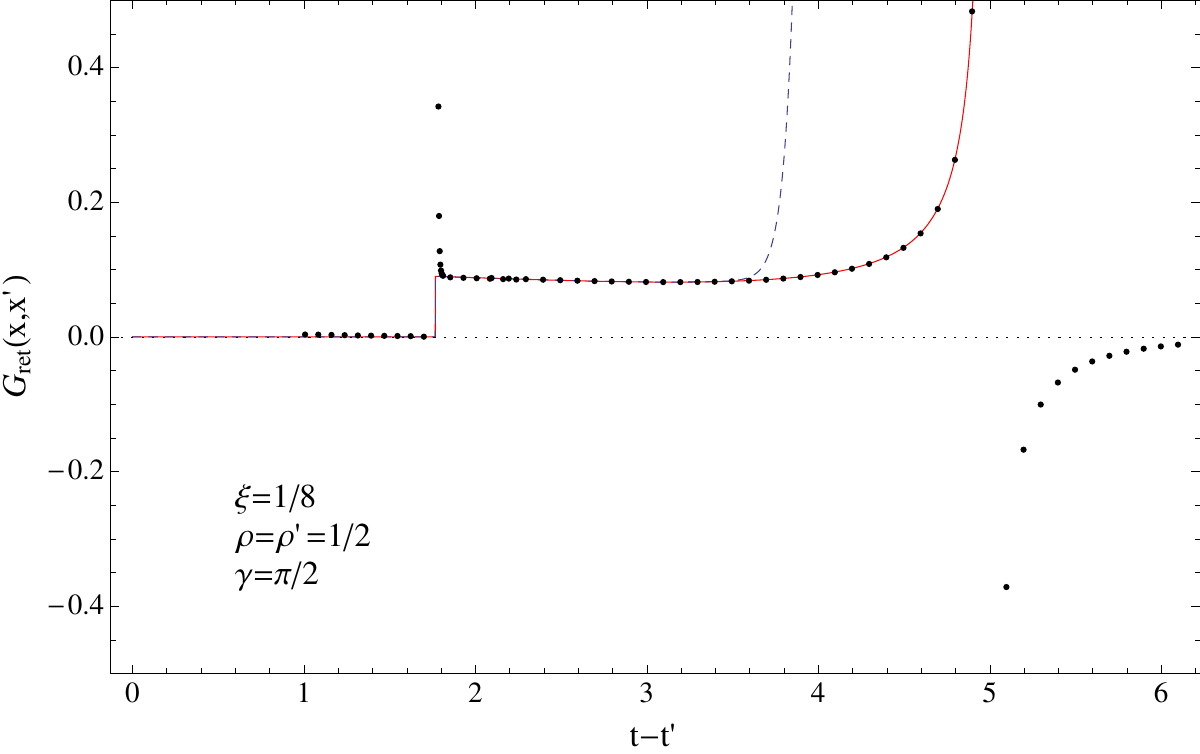}
  \includegraphics[width=7cm]{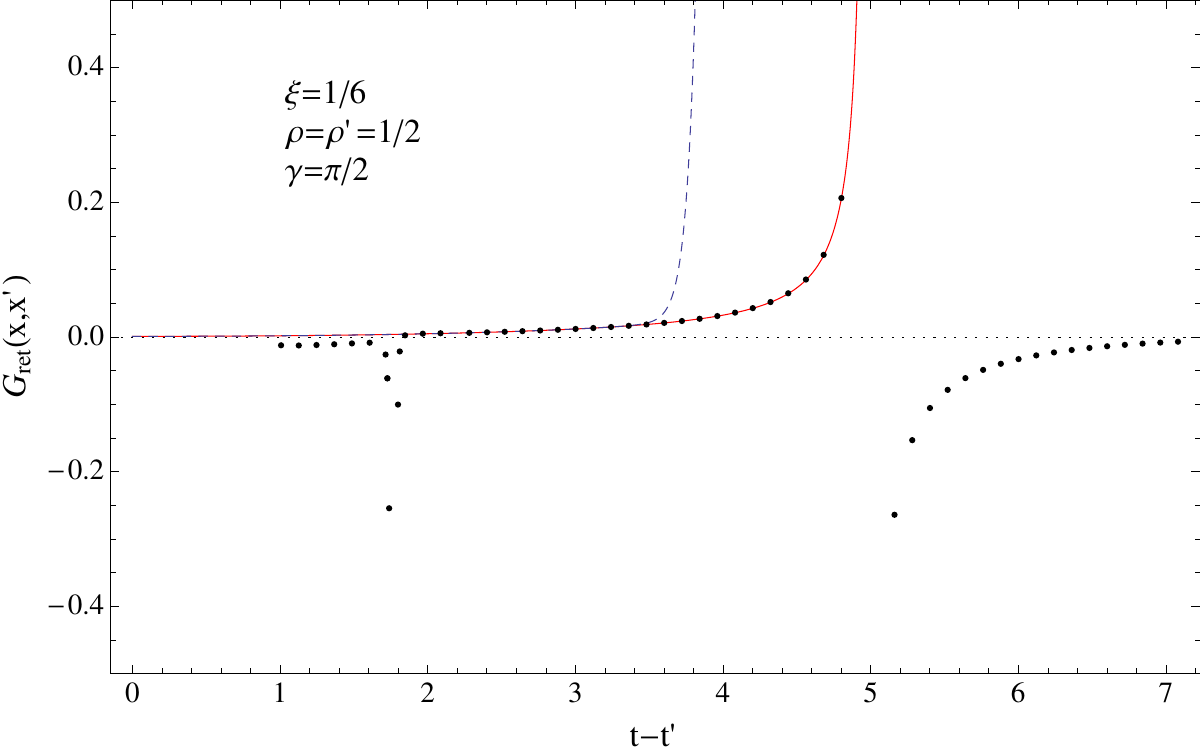}
 \end{center}
 \caption[Comparing Pad\'{e} approximant and coordinate Taylor series]{\emph{Comparison of the  Pad\'{e} approximant and coordinate Taylor series for $\theta( -\sigma(x,x') )  V(x,x')$, with the `exact' Green function calculated from the quasinormal mode sum in Nariai spacetime with curvature coupling $\xi=1/8$ and $\xi=1/6$.} The Pad\'e approximated $V(x,x')$ (red solid line) is in excellent agreement with the quasinormal mode Green function (black dots) up to the normal neighborhood boundary, $t_{NN} \approx 6.56993$ (top panels) and $t_{NN} \approx 4.9956$ (bottom panels). The Taylor series (blue dashed line) diverges outside the radius of convergence of the series.}
 \label{fig:padeCompareSeriesNariai}
\end{figure}

The Green function in Nariai spacetime is known to be given exactly by a quasinormal mode sum \cite{Casals:Dolan:Ottewill:Wardell:2009,Beyer:1999} at sufficiently late times ($(t-t')>2\rho_\star$). We can therefore use this quasinormal mode calculated Green function sum to determine the effectiveness of the Pad\'e resummation. Figure \ref{fig:padeCompareSeriesNariai} compares the Green function calculated from a quasinormal mode sum\footnote{
There are some caveats with how the quasinormal Green function was used. The \emph{fundamental mode} ($n=0$) Green function was used and a \emph{singularity time offset} applied as described in Chapter~\ref{ch:nariai} and Ref.~\cite{Casals:Dolan:Ottewill:Wardell:2009}. In the case where two singularities are present, two sets of fundamental mode Green functions were used, each shifted by an appropriate singularity time offset and matched at an intermediate point.} with both the original Taylor series representation and the Pad\'e resummed series for $V(x,x')$ for a range of cases. We use the Pad\'e approximant $P_{30}^{30}$, computed from the $60^{th}$ order Taylor series. In each case, the Taylor series representation (blue dashed line) diverges near its radius of convergence, long before the normal neighborhood boundary is reached. The Pad\'e resummed series (red line), however, remains valid much further and closely matches the quasinormal mode Green function (black dots) up to the point where the normal neighborhood boundary is reached.

As was shown in Refs.~\cite{Kay:Radzikowski:Wald:1997,Casals:Dolan:Ottewill:Wardell:2009}, the Green function in Nariai spacetime is singular whenever the points are separated by a null geodesic. Furthermore, in Chapter~\ref{ch:nariai} we have derived the functional form of these singularities and shown that they follow a four-fold pattern: $\delta(\sigma)$, $1/\pi \sigma$, $-\delta(\sigma)$, $-1/\pi \sigma$, depending on the number of caustics the null geodesic has passed through (this was also previously shown by Ori \cite{Ori1}). Within the normal neighborhood (where the Hadamard parametrix, \eqref{eq:Hadamard}, is valid), the $\delta(\sigma)$ singularities (i.e. at \emph{exactly} the null geodesic times) will be given by the term involving $U(x,x')$. However, at times other than the \emph{exact} null geodesic times, the Green function will be given fully by $V(x,x')$. For this reason, we expect $V(x,x')$ to reflect the singularities of the Green function near the normal neighborhood boundary.

The Pad\'e approximant attempts to model this singularity of the function $V(x,x')$ (which occurs at the null geodesic time) by representing it as a rational function, i.e. a ratio of two power series. By its nature, this will only faithfully reproduce singularities of integer order. In the Nariai case, however, the asymptotic form of the singularities is known exactly near the singularity times, $t_c$. In cases where the points are separated by an angle $\gamma\in (0,\pi)$ (i.e. away from a caustic), the singularities are expected to have a $1/(t-t'-t_c)$ behavior and it is reasonable to expect the Pad\'e approximant to reproduce the singularity well. When the points are not separated in the angular direction (i.e. at a caustic), however, the singularities behave like $1/(t-t'-t_c)^{3/2}$ and we cannot reasonably expect the Pad\'e approximant to accurately reflect this singularity without including a large number of terms in the denominator.

\begin{figure}
  \begin{center}
  \includegraphics[width=7cm]{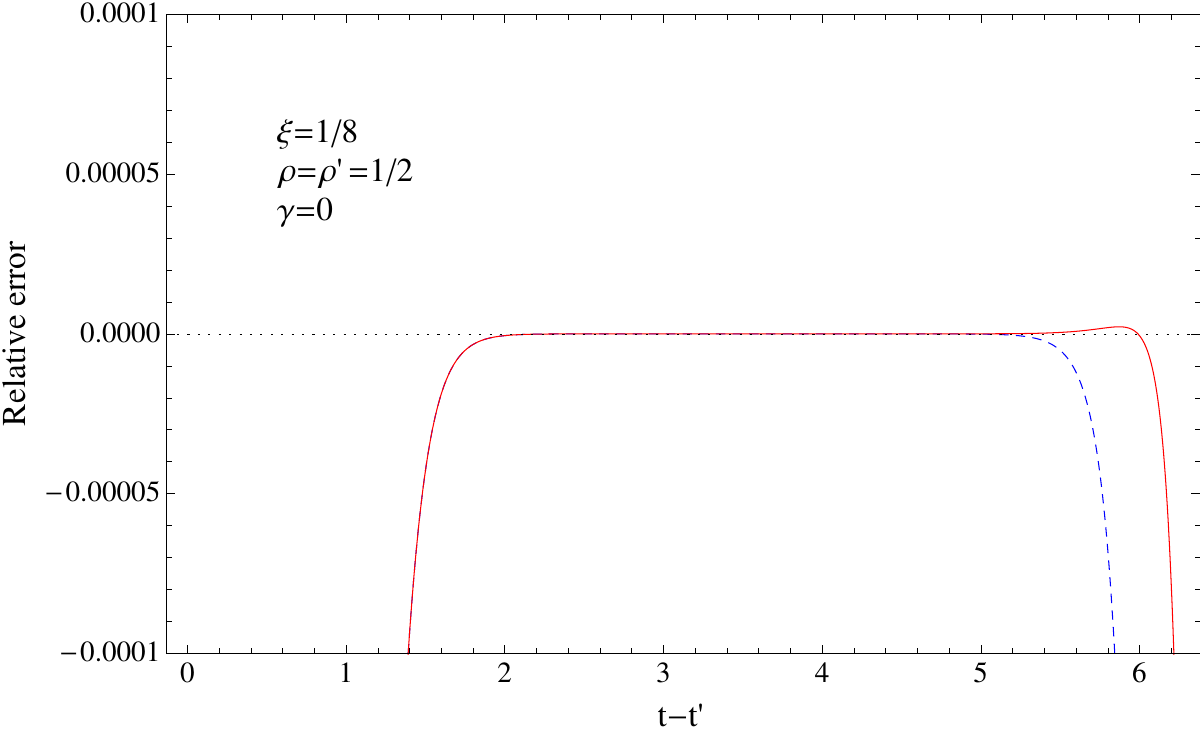}
 \end{center}
 \caption[Relative Error in Improved vs Regular Pad\'{e} Approximant.]{\emph{Relative error in improved vs regular Pad\'{e} approximant.} The relative error in the improved Pad\'e approximant (solid red line) remains small closer to the singularity than the regular Pad\'e approximant (blue dashed line).}
 \label{fig:padeSqrtNariai}
\end{figure}

Given knowledge of the functional form of the singularity, however, it is possible to improve the accuracy of the Pad\'e approximant further. For a singularity of the form $1/S(t)$, we first multiply the Taylor series by $S(t)$.  The result should then have either no singularity, or have a singularity which can be reasonably represented by a power series. The Pad\'e approximant of this new series is then calculated and the result is divided by $S(t)$ to give an \emph{improved Pad\'e approximant}. This yields an approximant which includes the exact form of the singularity and more closely matches the exact Green function near the singularity. In Fig.~\ref{fig:padeSqrtNariai}, we illustrate the improvement with an example case. We consider a static point in Nariai spacetime and compute the error in the Pad\'e approximant relative to the quasinormal mode Green function (with overtone number $n\le8$). The regular Pad\'e approximant (blue dashed line) is compared against the improved Pad\'e approximant (red solid line). The relative error remains small closer to the singularity for the improved Pad\'e approximant case than for the standard Pad\'e approximant. Note that the error for early times arises from the failure of the quasinormal mode sum to converge and does not reflect errors in the series approximations.

\subsection{Schwarzschild spacetime} \label{subsec:padeSchw}
\begin{figure}
  \begin{center}
  \includegraphics[width=7cm]{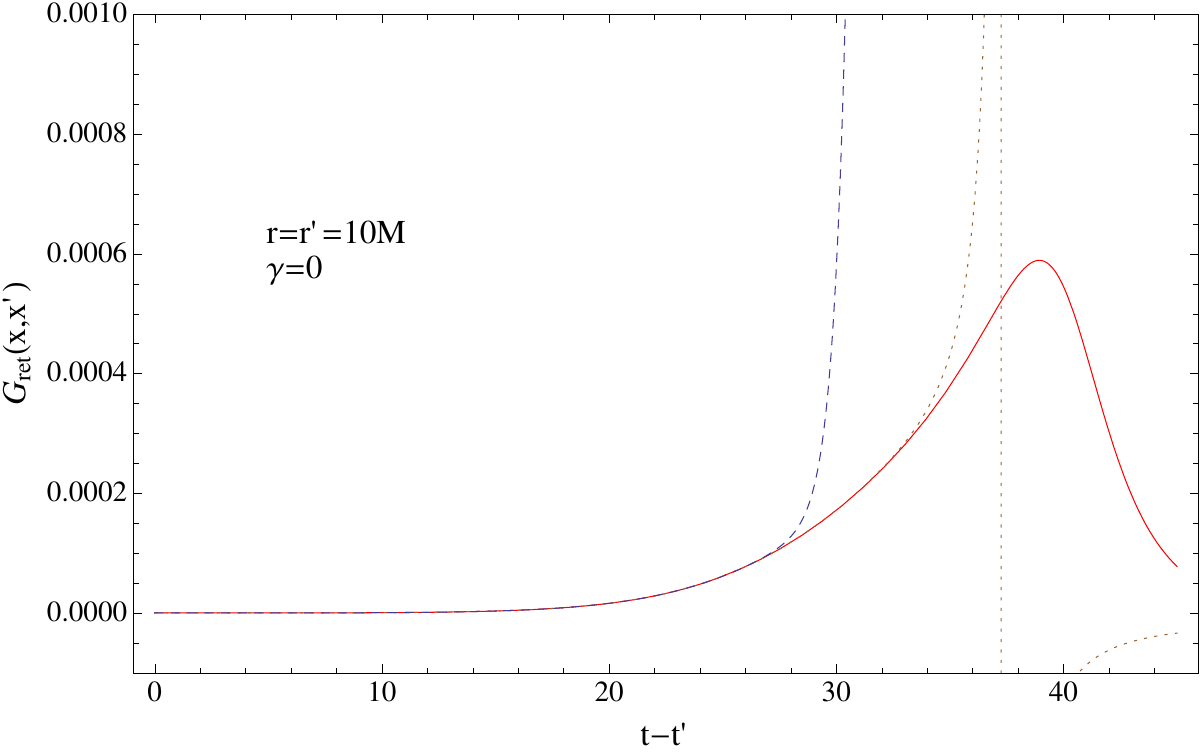}
 \end{center}
 \caption[Comparing Pad\'{e} to Taylor series for Schwarzschild]{\emph{Comparison of the Pad\'{e} approximant to the Taylor series for Schwarzschild spacetime} in the case of a static particle at $r=10M$. The Pad\'e approximants $P^{26}_{26}$ (solid red line) and $P^{24}_{26}$ (dotted brown line) are likely to represent $V(x,x')$ more accurately near the normal neighborhood boundary (at $t-t'\approx46.2471M$) than the regular Taylor series (dashed blue line).}
 \label{fig:padeCompareSeriesSchw}
\end{figure}

For the Schwarzschild case, there is no quasinormal mode sum with which to compare the Pad\'e approximated series\footnote{A quasinormal mode sum could be computed for the Schwarzschild case, but would be augmented by a branch cut integral \cite{Casals:Dolan:Ottewill:Wardell:2009}. This calculation is in progress but has yet to be completed.}. However, given the success in the Nariai case, we remain optimistic that Pad\'e resummation will be successful in the Schwarzschild spacetime. In an effort to estimate the effectiveness of the Pad\'e approximant, we compare in Fig.~\ref{fig:padeCompareSeriesSchw} the series expression for $V(x,x')$ with two different Pad\'e resummations, $P_{26}^{24}$ and $P_{26}^{26}$.  The Pad\'e approximant extends the validity beyond the radius of convergence of the series, but is less successful at reaching the normal neighborhood boundary ($t-t'=t_{NN}\approx 46.2471M$) than in the Nariai case.

The failure of the Pad\'e approximant to reach the normal neighborhood boundary can be understood by the presence of extraneous singularities in the Pad\'e approximant. The zeros of the denominator, $S(t-t')=0$, give rise to singularities which occur at times earlier than the null geodesic time. It is possible that this problem could be reduced to a certain extent using the knowledge of the functional form of the singularities to compute an \emph{improved} Pad\'e approximant (as was successful in the Nariai case). However, to the authors knowledge, the structure of the singularities in Schwarzschild spacetime is not yet known. While it may be possible to adapt the work of Chapter~\ref{ch:nariai} to find the asymptotic form of the singularities in Schwarzschild spacetime, without knowledge of the exact Green function we would not be able to determine whether an improved Pad\'e approximant would truly give an improvement. We therefore leave such considerations for later work.

\subsection{Convergence of the Pad\'e Sequence} \label{subsec:Pade-convergence}
The use of Pad\'e approximants has shown remarkable success in improving the accuracy and domain of the series representation of $V(x,x')$. However, this improvement has not been quantified. There is no general way to determine whether the Pad\'e approximant is truly approximating the correct function, $V(x,x')$ or the domain in which it is valid \cite{NumericalRecipes}. In this subsection, we nonetheless attempt to gain some insight into the validity of the Pad\'e approximants.

The first issue to consider is the presence of extraneous poles in the Pad\'e approximants. In Sec.~\ref{subsec:padeNariai}, the Pad\'e approximant was unable to exactly represent the $1/(t-t'-t_c)^{3/2}$ singularity at $t_c \approx 6.12$ and instead represented it by three (real-valued) simple poles (at $t-t' \approx 6.522, 6.854 \text{ and } 9.488$). This leads to the Pad\'e approximant being a poor representation of the function near the poles. As was shown in Fig.~\ref{fig:padeSqrtNariai}, having exact knowledge of the singularity structure allows for the calculation of an improved Pad\'e approximant without extraneous singularities\footnote{The improved Pad\'e approximant has zeros in its denominator at times $t-t' \approx 2.64, 6.51, 6.59$ and $8.73$. The apparently extraneous singularity within the normal neighborhood (at $t-t'\approx 2.64$) does not cause any difficulty as the numerator also goes to zero at this point.}.

With extraneous poles dealt with, we consider the convergence of the Pad\'e sequence of diagonal and sub-diagonal Pad\'e approximants, 
\begin{equation}
 P = \{P_0^0,~ P_1^0,~ P_1^1,~ P_2^1,~ P_2^2,~ P_3^2,~ P_3^3, \cdots \},
\end{equation}
with $P_N$ being the $N$-th element of the sequence. The convergence of the Pad\'e approximant sequence is determined by the behavior of the denominators, $S_N$ for large $N$ \cite{Bender:Orszag}. Provided $S(x,x')_N$ is not small, the Pad\'e sequence will converge quickly toward the actual value of $V(x,x')$. When the first root of the denominator is at the null geodesic time (i.e. the normal neighborhood boundary), we can, therefore, be optimistic that the Pad\'e sequence will remain convergent until this root is reached and that the Pad\'e approximants will accurately represent the function.

To highlight the improvements made by using Pad\'e approximants over regular Taylor series, we introduce the Taylor sequence (i.e. the sequence of partial sums of the series), $T=\{T_0, T_1, T_2, \cdots\}$, with the $N$th element,
\begin{equation}
 T_N = \sum_{n=0}^{N/2} v_{n} (t-t')^{2n}.
\end{equation}
The two sequences $P_N$ and $T_N$ require approximately the same number of terms in the original Taylor series, so a direct comparison of their convergence will illustrate the improved convergence of the Pad\'e approximants.

\begin{figure}
  \begin{center}
  \includegraphics[width=7cm]{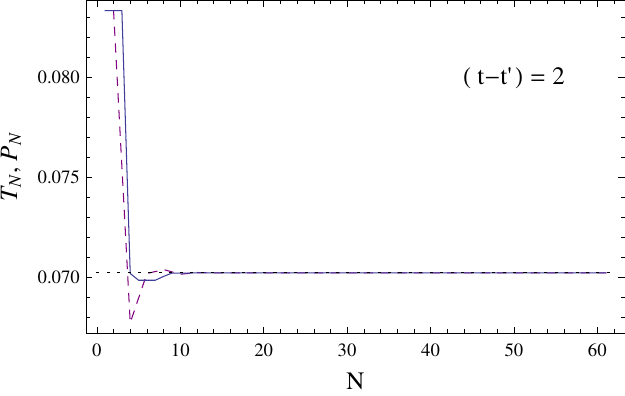}
  \includegraphics[width=7cm]{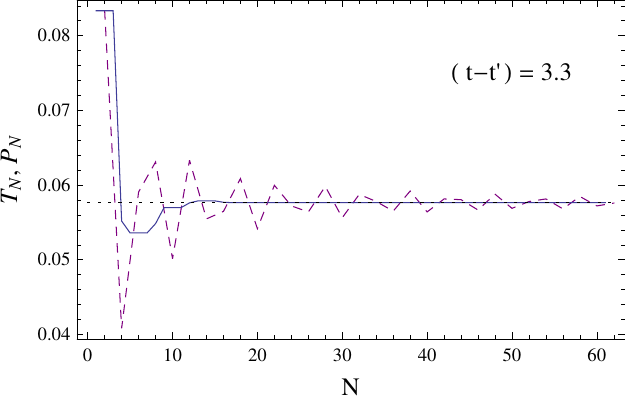}
  \includegraphics[width=7cm]{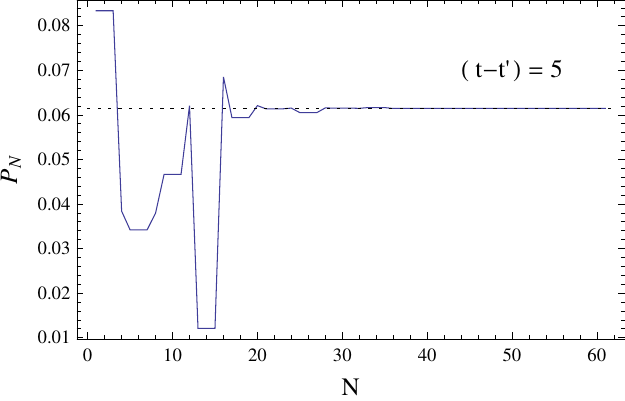}
  \includegraphics[width=7cm]{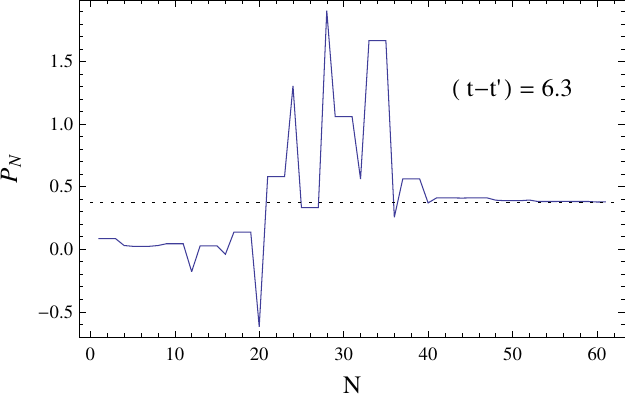}
 \end{center}
 \caption[Convergence of the Taylor and Pad\'{e} Sequences in Nariai]{\emph{Convergence of the Taylor and Pad\'{e} sequences} for the case of static points at $\rho=1/2$ in the Nariai spacetime, with $\xi=1/8$. Within the radius of convergence of the Taylor series, both Pad\'e (blue solid line) and Taylor (purple dashed line) sequences converge to the exact Green function (black dotted line) as calculated from a quasinormal mode sum with overtone number $n\le6$. The Pad\'e sequence converges faster, particularly at later times. Outside the radius of convergence of the Taylor series (in this case, $(t-t')\approx 4$), only the Pad\'e sequence is convergent.}
 \label{fig:padeConvergenceNariai}
\end{figure}

In Fig.~\ref{fig:padeConvergenceNariai}, we plot the Pad\'e sequence (blue line) and Taylor sequence (purple dashed line) for the case of static points at $\rho=1/2$ in the Nariai spacetime, with $\xi=1/8$. For early times (eg. $(t-t')=2$), it is clear that both Pad\'e and Taylor sequences converge very quickly. At somewhat later times (eg. $(t-t')=3.3$), both sequences appear to remain convergent, but the Pad\'e sequence is clearly converging much faster than the Taylor sequence. Outside the radius of convergence of the Taylor series (eg. $(t-t')=5, 6.3$), the Pad\'e sequence is slower to converge, but appears to still do so.

\begin{figure}
  \begin{center}
  \includegraphics[width=7cm]{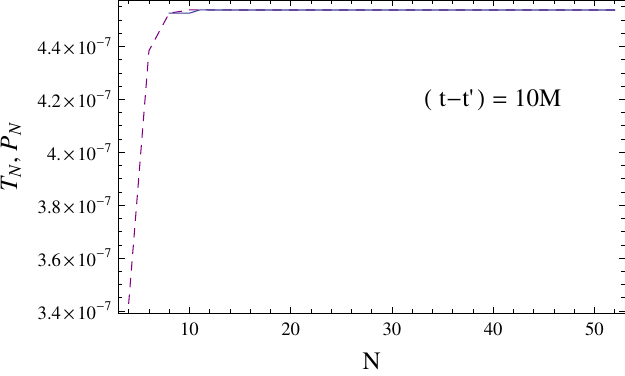}
  \includegraphics[width=7cm]{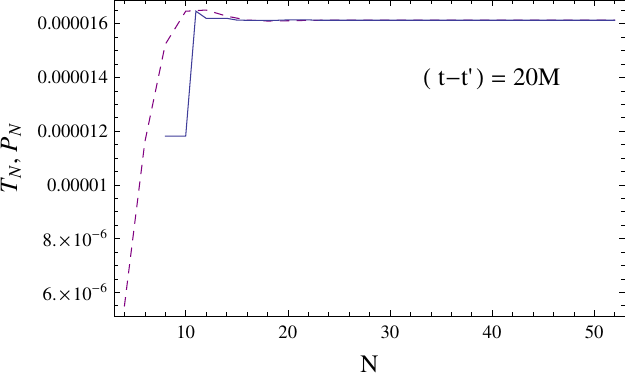}
  \includegraphics[width=7cm]{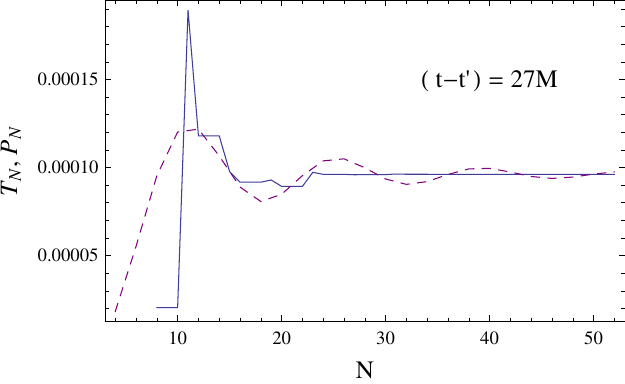}
  \includegraphics[width=7cm]{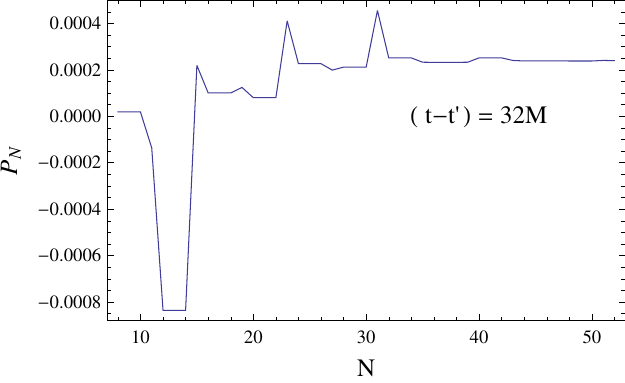}
 \end{center}
 \caption[Convergence of the Taylor and Pad\'{e} sequences in Schwarzschild]{\emph{Convergence of the Taylor and Pad\'{e} Sequences} for the case of static points at $r=10M$ in the Schwarzschild spacetime. As in the Nariai case, the Pad\'e sequence (blue solid line) converges better and at later times than the Taylor sequence (purple dashed line).}
 \label{fig:padeConvergenceSchw}
\end{figure}

In Fig.~\ref{fig:padeConvergenceSchw}, we again plot the Pad\'e sequence, this time for the case of static points at $r=10M$ in the Schwarzschild spacetime. As in the Nariai case, for early times (eg. $(t-t')=10M$), both Pad\'e and Taylor sequences are converging very quickly. At slightly later times (eg. $(t-t')=20M, 27M$) the convergence of the Pad\'e sequence is better than the Taylor sequence. Outside the radius of convergence of the Taylor series, the Pad\'e sequence is slower to converge, but appears to still do so. Unfortunately, the convergence of the series is slower in the Schwarzschild case than in the Nariai case. This is an indication that using more terms may yield a better result\footnote{In the Nariai case (with $\xi=1/8$), the Taylor series for $V(x,x')$ starts at order $(t-t')^0$ and has been calculated to order $(t-t')^{60}$, while for Schwarzschild it starts at $(t-t')^4$ and has been calculated to order $(t-t')^{52}$. Since the Nariai series has several extra orders, it is reasonable to expect it to be a better approximation than the Schwarzschild series.}.

\chapter{Transport Equation Approach to Calculations of Green functions and DeWitt coefficients} \label{ch:covex}

\section{Introduction}
In the previous chapter, we presented methods for obtaining coordinate expansions for
the (tail part of the) retarded Green function in spherically symmetric spacetimes. 
Our ultimate goal in the matched expansion self-force programme is to work in more general spacetimes,
 especially Kerr spacetime. 
As we move away from specific symmetry conditions, we can no longer 
 rely on methods based on a special choice of coordinates in the construction of our quasilocal
 solution and are led instead to consider other techniques such as transport equations and covariant expansion methods. 

Covariant methods for calculating the Green function of the wave operator and the corresponding heat kernel, briefly reviewed in Sec.~\ref{sec:review}, are central to a broad range of problems from radiation reaction to quantum field theory in curved spacetime
and quantum gravity. There is an extremely extensive literature on this topic; here we provide only a very brief overview, referring the reader to the reviews by Vassilevich~\cite{Vassilevich:2003} and Poisson~\cite{Poisson:2003} and references therein for a more complete discussion. These methods have 
evolved from pioneering work by Hadamard~\cite{Hadamard} on the classical theory and
DeWitt~\cite{DeWitt:1960, DeWitt:1965} on the quantum theory.  The central objects in the Hadamard and DeWitt covariant expansions are 
geometrical bi-tensor coefficients $a_n^{AB'}(x,x')$ which are commonly called DeWitt or DeWitt coefficients in the
physics literature. These coefficients are closely related to the short proper-time
asymptotic expansion of the heat kernel of an elliptic operator in a Riemannian space 
and so are commonly called heat kernel coefficients in the mathematics literature. 
Traditionally most attention has focused on the \textit{diagonal value}  of the heat kernel $K^{A}{}_{A}(x,x;s)$,
since the coincidence limits $a_n^{A}{}_{A}(x,x)$ play a central role in the classical theory 
of spectral invariants~\cite{Gilkey} and in the quantum theory of the effective action and trace anomalies~\cite{Birrell:Davies}.  
By contrast, for the quasilocal part of the matched expansion approach to radiation reaction \cite{Ottewill:Wardell:2008, Ottewill:Wardell:2009}, we seek
expansions valid for $x$ and $x'$ as far apart as geometrical methods permit.   

The classical approach to the calculation of these coefficients in the physics literature (briefly discussed in Sec.~\ref{sec:review}) was to use a recursive 
approach developed by DeWitt \cite{DeWitt:1965} in the 1960s. Although these recursive methods work well
 for the first few terms in the expansion \cite{Christensen:1976vb,Christensen:1978yd}, 
 and may be implemented in a tensor software package \cite{Christensen:1995}, the amount of calculation required to compute subsequent terms quickly becomes prohibitively long, even when implemented as a computer program.
  An alternative approach, more common in the
mathematics literature, is to use pseudo-differential operators and invariance theory~\cite{Gilkey}, where a basis of 
curvature invariants of the appropriate
structure is constructed~\cite{Fulling:1992} and then their coefficients determined by explicit evaluation in simple spacetimes.
However, here too, the size of the basis grows rapidly and there seems little prospect of reaching orders 
comparable to those we obtained in the highly symmetric configurations previously studied. 

An extremely elegant, non-recursive approach to the calculation of DeWitt coefficients has been given by Avramidi~\cite{Avramidi:1986,Avramidi:2000}, 
but as his motivation was to study the effective action in quantum gravity he was primarily interested in the coincidence limit of the DeWitt 
coefficients, while in the self-force problem, as noted above, we require point-separated expressions.  In addition, Avramidi introduced his method in the language 
of quantum mechanics; quite distinct from the language of transport equations, such as the Raychaudhuri equation, more familiar to 
discussions of geodesics among relativists.  In this chapter we present 
Avramidi's approach in the language of transport equations and show that it is ideal for numerical and symbolic computation. 
In so doing we are building on the work of D\'ecanini and Folacci~\cite{Decanini:Folacci:2005a} who wrote many of the equations we present (we
indicate below where we deviate from their approach) and 
implemented them explicitly by hand. However, calculations by hand are long and inevitably prone to error, particularly for higher spin and 
for higher order terms in the series and are quite impractical for the very high order expansions we would like for radiation reaction calculations.
Instead, we use the transport equations as the basis for \textsl{Mathematica} code for algebraic calculations and \textsl{C} code for numerical calculations. 
Rather than presenting our results in excessively long equations (our non-canonical expression for $a_7(x,x)$ for a scalar field contains $2\,069\,538$ terms!),
we have provided these codes online \cite{AvramidiCode,TransportCode}. 

In Sec.~\ref{sec:avramidi}, we detail the principles that we consider to encapsulate the key insights of the Avramidi approach
and use these to write down a set of transport equations for the key bi-tensors of the theory. These provide an
adaptation of the Avramidi approach which is ideally suited to implementation on a computer either numerically or symbolically.

In Sec.~\ref{sec:symbolic}, we describe a semi-recursive approach to solving for covariant expansion and briefly describe our \textsl{Mathematica}
implementation of it and its interface to the tensor software package \textsl{xTensor}. 

In Sec.~\ref{sec:numerical}, we present a numerical implementation of the transport equation approach to the calculation of $V(x,x')$ along a null geodesic.

Given our motivation in studying the radiation reaction problem, we shall phrase all the discussions of this chapter in 4-dimensional
spacetime. The reader is referred to Decanini and Folacci~\cite{Decanini:Folacci:2005a} for a discussion of the corresponding
situation in spacetimes of more general (integer) dimension. We do note however that the DeWitt coefficients are purely geometric bi-tensors, formally independent of the spacetime dimension.

\section{Avramidi Approach to Covariant Expansion Calculations}
\label{sec:avramidi}
The traditional approach to the calculation of covariant expansionss of fundamental bi-tensors, due to DeWitt \cite{DeWitt:1960} and described in Sec.~\ref{sec:bitensors}, is to derive a set of recursion relations for the coefficients of the series.  Avramidi \cite{Avramidi:1986} has proposed an alternative, extremely elegant non-recursive method for the calculation of these series coefficients. Translated into the language of transport equations, this approach emphasizes two fundamental principles when doing calculations:
\begin{enumerate}
 \item When expanding about $x$, always try to take derivatives at $x'$. The result is that derivatives only act on the $\sigma^a$'s and not on the coefficients.
 \item Where possible, whenever taking a covariant derivative, $\nabla_{a'}$, contract the derivative with $\sigma^{a'}$
\end{enumerate}
Applying these two principles, Avramidi has derived non-recursive\footnote{Avramidi retains the recurrance relations for the DeWitt coefficients, $a_k$ (and hence the Hadamard coefficients, $V_r$). However, all other relations are non-recursive.} expressions for the coefficients of covariant expansions of several bi-tensors. As Avramidi's derivations use a rather abstract notation, we will now briefly review his technique in a more explicit notation. We will also extend the derivation to include several other bi-tensors and note that equations \eqref{eq:xi-transport}, \eqref{eq:eta-transport}, \eqref{eq:gamma-transport}, \eqref{eq:lambda-eq}, \eqref{eq:vanVleck-transport2}, \eqref{eq:sqrtdelta-zeta}, \eqref{eq:V0-transport} and \eqref{eq:Vr-transport} were previously written down and used by Decanini and Folacci \cite{Decanini:Folacci:2005a}.

Throughout this section, we fix the base point $x$ and allow it to be connected to any other point $x'$ by a geodesic. In all cases, we expand about the fixed point, $x$.

Defining the transport operators $D$ and $D'$ as
\begin{align}
 D & \equiv \sigma^\alpha \nabla_\alpha & D' & \equiv \sigma^{\alpha'} \nabla_{\alpha'},
\end{align}
we can rewrite Eq.~(\ref{eq:SigmaDefiningEq}) as
\begin{align}
 (D-2)\sigma &= 0 & (D'-2)\sigma &= 0.
\end{align}
Differentiating the second of these equations at $x$ and at $x'$, we get
\begin{align}
 \left( D' - 1 \right) \sigma^{a} &= 0 & \left( D' - 1 \right) \sigma^{a'} &= 0,
\end{align}
which, defining
\begin{align}
 \eta^{a}_{\phantom{a} b'} &\equiv \sigma^{a}_{\phantom{b} b'} &
 \xi^{a'}_{\phantom{a'} b'} &\equiv  \sigma^{a'}_{\phantom{a'} b'}
\end{align}
can be rewritten as
\begin{align}
\label{eq:sigma-eta-xi}
 \sigma^{a} &= \eta^{a}_{\phantom{a} \alpha'} \sigma^{\alpha'} &
 \sigma^{a'} &= \xi^{a'}_{\phantom{a'} \alpha'} \sigma^{\alpha'}.
\end{align}
Finally, we define $\gamma^{a'}_{\phantom{a'} b}$, the inverse of $\eta^{a}_{\phantom{a} b'}$,
\begin{equation}
\label{eq:gamma-def}
 \gamma^{a'}_{\phantom{a'} b} \equiv (\eta^{b}_{\phantom{b} a'})^{-1}.
\end{equation}
and also introduce the definition
\begin{equation}
 \lambda^{a}_{\phantom{a} b} \equiv \sigma^{a}_{\phantom{a} b}.
\end{equation}

We will now derive transport equations for each of these newly introduced quantities along with some others which will be defined as required. Many of these derivations involve considerable index manipulations and are most easily (and accurately) done using a tensor software package such as \textsl{xTensor}~\cite{xTensor}.

The transport equations of this section may be derived in a recursive manner, making use of the identities
\begin{multline}
\label{eq:transport-deriving-primed}
D'(\sigma_{a_1'\dots a_n' a_{n+1}'}) = \nabla_{a_{n+1}'} (D' \sigma_{a_1'\dots a_n'}) - \xi^{\alpha'}{}_{a_{n+1}'}  \nabla_{\alpha'}\sigma_{a_1'\dots a_n'} \\
 + \sigma^{\alpha'} R^{c'}{}_{a_1'a_{n+1}'\alpha'}\sigma_{c'\dots a_n'}   + \dots + \sigma^{\alpha'} R^{c'}{}_{a_n'a_{n+1}'\alpha'}\sigma_{a_1'\dots c'}
\end{multline}
and
\begin{align}
\label{eq:transport-deriving-unprimed}
D'(\sigma^{b}{}_{a_1'\dots a_n'}) &= \nabla^{b} (D' \sigma_{a_1'\dots a_n'}) - \eta^{b}{}_{\alpha'}  \nabla^{\alpha'}\sigma_{a_1'\dots a_n'} .
\end{align}
and their generalizations, given below.
This method is particularly algorithmic and well suited to implementation on a computer, thus allowing for the automated derivation of a transport equation for an arbitrary number of derivatives of a bi-tensor.

\subsection{Transport equation for \texorpdfstring{$\xi^{a'}_{~b'}$}{xi\^{ }a'\_b'}}
Taking a primed derivative of the second equation in (\ref{eq:sigma-eta-xi}), we get
\begin{equation}
 \xi^{a'}_{\phantom{a'} b'} = \xi^{a'}_{\phantom{a'} \alpha' b'} \sigma^{\alpha'} + \xi^{a'}_{\phantom{a'} \alpha'}\xi^{\alpha'}_{\phantom{\alpha'} b'}.
\end{equation}

We now commute the last two covariant derivatives in the first term on the right hand side of this equation and rearrange to get:
\begin{equation}
\label{eq:xi-transport}
D' \xi^{a'}_{\phantom{a'} b'} + \xi^{a'}_{\phantom{a'} \alpha'} \xi^{\alpha'}_{\phantom{\alpha'} b'}  - \xi^{a'}_{\phantom{a'} b'} + R^{a'}_{\phantom{a'} \alpha' b' \beta'} \sigma^{\alpha'} \sigma^{\beta'} = 0
\end{equation}

\subsection{Transport equation for \texorpdfstring{$\eta^{a}_{~b'}$}{eta\^{ }a\_b'}}
Taking a primed derivative of the first equation in (\ref{eq:sigma-eta-xi}), we get
\begin{equation}
 \eta^{a}_{\phantom{a} b'} = \eta^{a}_{\phantom{a} \alpha' b'} \sigma^{\alpha'} + \eta^{a}_{\phantom{a} \alpha'}\xi^{\alpha'}_{\phantom{\alpha'} b'}
\end{equation}

In this case, since $\sigma^a$ is a scalar at $x'$, we can commute the two primed covariant derivatives in the first term on the right hand side of this equation without introducing a Riemann term. Rearranging, we get:
\begin{equation}
\label{eq:eta-transport}
D' \eta^{a}_{\phantom{a} b'} + \eta^{a}_{\phantom{a} \alpha'} \xi^{\alpha'}_{\phantom{\alpha'} b'}  - \eta^{a}_{\phantom{a} b'} = 0
\end{equation}

\subsection{Transport equation for \texorpdfstring{$\gamma^{a'}_{~b}$}{gamma\^{ }a'\_b}}
Solving Eq.~(\ref{eq:eta-transport}) for $\xi^{a'}_{\phantom{a'} b'}$ and using \eqref{eq:gamma-def}, we get
\begin{align}
\label{eq:xi-solve-eta-gamma}
 \xi^{a'}_{\phantom{a'} b'} &= \delta^{a'}_{\phantom{a'} b'} - \gamma^{a'}_{\phantom{a'} \alpha'} \left(D' \eta^{\alpha'}_{\phantom{\alpha'} b'}\right)\nonumber \\
	&= \delta^{a'}_{\phantom{a'} b'} + \left( D' \gamma^{a'}_{\phantom{a'} \alpha'} \right) \eta^{\alpha'}_{\phantom{\alpha'} b'},
\end{align}
Next, substituting Eq.~(\ref{eq:xi-solve-eta-gamma}) into Eq.~(\ref{eq:xi-transport}) and rearranging, we get a transport equation for $\gamma^{a'}_{\phantom{a'} b'}$:
\begin{equation}
\label{eq:gamma-transport}
 (D')^2 \gamma^{a'}_{\phantom{a'} b'} + D' \gamma^{a'}_{\phantom{a'} b'} + R^{a'}_{\phantom{a'} \alpha' \gamma' \beta'} \gamma^{\gamma'}_{\phantom{a'} b'} \sigma^{\alpha'} \sigma^{\beta'}
 =0.
\end{equation}

\subsection{Equation for \texorpdfstring{$\lambda^{a}_{~b}$}{lambda\^{ }a\_b}}
Differentiating Eq.~(\ref{eq:SigmaDefiningEq}) at $x$ and $x'$, we get
\begin{equation}
 \eta^{a}_{\phantom{a} b'} = \lambda^{a}_{\phantom{a} \alpha} \eta^{\alpha}_{\phantom{\alpha} b'} + D \eta^{a}_{\phantom{a} b'}
\end{equation}
which is easily rearranged to give an equation for $\lambda^{a}_{\phantom{a} b}$:
\begin{equation}
\label{eq:lambda-eq}
 \lambda^{a}_{\phantom{a} b} = \delta^{a}_{\phantom{a} b} - (D \eta^{a}_{\phantom{a} \alpha'}) \gamma^{\alpha'}_{\phantom{\alpha'} b}.
\end{equation}

\subsection{Transport equation for \texorpdfstring{$\sigma^{a'}_{~~ b' c'}$}{sigma\^{ }a'\_b'c'}}
Applying the identity \eqref{eq:transport-deriving-primed} to \eqref{eq:xi-transport} and simplifying the resulting expression, we get
\begin{align}
\label{eq:transport-sigma-ppp}
(D'-1) \sigma^{a'}_{\phantom{a'} b' c'}
+ \sigma^{\alpha'}_{\phantom{\alpha'} c'} \sigma^{a}_{\phantom{a} \alpha' b'}
+ \sigma^{\alpha'}_{\phantom{\alpha'} b'} \sigma^{a'}_{\phantom{a'} \alpha' c'}
+ \sigma^{a'}_{\phantom{a'} \alpha'} \sigma^{\alpha'}_{\phantom{\alpha'} b' c'}
+ R^{a'}_{\phantom{a'} \alpha' b' \beta' ; c'} \sigma^{\alpha'} \sigma^{\beta'} \nonumber \\
- R^{a'}_{\phantom{a'} \alpha' \beta' b'} \sigma^{\beta'} \sigma^{\alpha'}_{\phantom{\alpha'} c'}
- R^{a'}_{\phantom{a'} \alpha' \beta' c'} \sigma^{\beta'} \sigma^{\alpha'}_{\phantom{\alpha'} b'}
+ R^{\alpha'}_{\phantom{\alpha'} b' \beta' c'} \sigma^{\beta'} \sigma^{a'}_{\phantom{a'} \alpha'}
= 0
\end{align}

\subsection{Transport equation for \texorpdfstring{$\sigma^{a}_{~b' c'}$}{sigma\^{ }a\_b'c'}}
Applying the identity \eqref{eq:transport-deriving-unprimed} to \eqref{eq:xi-transport} and simplifying the resulting expression, we get
\begin{equation}
\label{eq:transport-sigma-upp}
(D'-1) \sigma^{a}_{\phantom{a} b' c'}
+ \sigma^{\alpha'}_{\phantom{\alpha'} b'} \sigma^{a}_{\phantom{a} \alpha' c'}
+ \sigma^{\alpha'}_{\phantom{\alpha'} c'} \sigma^{a}_{\phantom{a} \alpha' b'}
+ \sigma^{a}_{\phantom{a} \alpha'} \sigma^{\alpha'}_{\phantom{\alpha'} b' c'}
+ R^{\alpha'}_{\phantom{\alpha'} b' \beta' c'} \sigma^{a}_{\phantom{a} \alpha'} \sigma^{\beta'}
= 0
\end{equation}

\subsection{Transport equation for \texorpdfstring{$\sigma^{a'}_{~~b' c' d'}$}{sigma\^{ }a'\_b'c'd'}}
Applying the identity \eqref{eq:transport-deriving-primed} to \eqref{eq:transport-sigma-ppp} and simplifying the resulting expression, we get
\begin{multline}
\label{eq:sigma-pppp-transport}
(D'-1)\sigma^{a'}_{\phantom{a'} b' c' d'} 
+ \sigma^{a'}_{\phantom{a'} \alpha' b' c'} \sigma^{\alpha'}_{\phantom{\alpha'} d'} 
+ \sigma^{a'}_{\phantom{a'} \alpha' b' d'} \sigma^{\alpha'}_{\phantom{\alpha'} c'} 
+ \sigma^{a'}_{\phantom{a'} \alpha' c' d'} \sigma^{\alpha'}_{\phantom{\alpha'} b'}
+ \sigma^{a'}_{\phantom{a'} \alpha' b'} \sigma^{\alpha'}_{\phantom{\alpha'} c' d'} 
+ \sigma^{a'}_{\phantom{a'} \alpha' c'} \sigma^{\alpha'}_{\phantom{\alpha'} b' d'} \\
+ \sigma^{a'}_{\phantom{a'} \alpha' d'} \sigma^{\alpha'}_{\phantom{\alpha'} b' c'} 
+ \sigma^{a'}_{\phantom{a'} \alpha'} \sigma^{\alpha'}_{\phantom{\alpha'} b' c' d'}
+ R^{a'}_{\phantom{a'} \alpha' \beta' c'} R^{\alpha'}_{\phantom{\alpha'} d' \gamma' b'} \sigma^{\beta'} \sigma^{\gamma'}
+ R^{a'}_{\phantom{a'} \alpha' \beta' b'} R^{\alpha'}_{\phantom{\alpha'} d' \gamma' c'} \sigma^{\beta'} \sigma^{\gamma'}\\
+ R^{a'}_{\phantom{a'} \alpha' \beta' d'} R^{\alpha'}_{\phantom{\alpha'} c' \gamma' b'} \sigma^{\beta'} \sigma^{\gamma'}
- R^{a'}_{\phantom{a'} \beta' \alpha' d'} R^{\alpha'}_{\phantom{\alpha'} b' \gamma' c'} \sigma^{\beta'} \sigma^{\gamma'}
- R^{a'}_{\phantom{a'} \beta' \alpha' c'} R^{\alpha'}_{\phantom{\alpha'} b' \gamma' d'} \sigma^{\beta'} \sigma^{\gamma'} \\
+ R^{a'}_{\phantom{a'} \beta' b' \gamma' ; c' d'} \sigma^{\beta'} \sigma^{\gamma'}
+ R^{\alpha'}_{\phantom{\alpha'} b' \beta' c' ; d'} \sigma^{\beta'} \sigma^{a'}_{\phantom{a'} \alpha'}
- R^{a'}_{\phantom{a'} \alpha' \beta' c' ; d'} \sigma^{\beta'} \sigma^{\alpha'}_{\phantom{\alpha'} b'}
- R^{a'}_{\phantom{a'} \alpha' \beta' b' ; d'} \sigma^{\beta'} \sigma^{\alpha'}_{\phantom{\alpha'} c'}\\
- R^{a'}_{\phantom{a'} \alpha' \beta' b' ; c'} \sigma^{\beta'} \sigma^{\alpha'}_{\phantom{\alpha'} d'}
+ R^{\alpha'}_{\phantom{\alpha'} c' \beta' d'} \sigma^{\beta'} \sigma^{a'}_{\phantom{a'} b' \alpha'}
+ R^{\alpha'}_{\phantom{\alpha'} b' \beta' d'} \sigma^{\beta'} \sigma^{a'}_{\phantom{a'} c' \alpha'}
+ R^{\alpha'}_{\phantom{\alpha'} b' \beta' c'} \sigma^{\beta'} \sigma^{a'}_{\phantom{a'} d' \alpha'}\\
- R^{a'}_{\phantom{a'} \alpha' \beta' d'} \sigma^{\beta'} \sigma^{\alpha'}_{\phantom{\alpha'} b' c'}
- R^{a'}_{\phantom{a'} \alpha' \beta' c'} \sigma^{\beta'} \sigma^{\alpha'}_{\phantom{\alpha'} b' d'}
- R^{a'}_{\phantom{a'} \alpha' \beta' b'} \sigma^{\beta'} \sigma^{\alpha'}_{\phantom{\alpha'} c' d'} = 0
\end{multline}

\subsection{Transport equation for \texorpdfstring{$\sigma^{a}_{~b' c' d'}$}{sigma\^{ }a\_b'c'd'}}
Applying the identity \eqref{eq:transport-deriving-unprimed} to \eqref{eq:transport-sigma-ppp} and simplifying the resulting expression, we get
\begin{multline}
\label{eq:sigma-uppp-transport}
 (D'-1)\sigma^{a}_{\phantom{a} b' c' d'}  + \sigma^{a}_{\phantom{a} \alpha' b' c'} \sigma^{\alpha'}_{\phantom{\alpha'} d'} + \sigma^{a}_{\phantom{a} \alpha' b' d'} \sigma^{\alpha'}_{\phantom{\alpha'} c'} + \sigma^{a}_{\phantom{a} \alpha' c' d'} \sigma^{\alpha'}_{\phantom{\alpha'} b'} + \sigma^{a}_{\phantom{a} \alpha' b'} \sigma^{\alpha'}_{\phantom{\alpha'} c' d'} \\ + \sigma^{a}_{\phantom{a} \alpha' c'} \sigma^{\alpha'}_{\phantom{\alpha'} b' d'} + \sigma^{a}_{\phantom{a} \alpha' d'} \sigma^{\alpha'}_{\phantom{\alpha'} b' c'} 
+ \sigma^{a}_{\phantom{a} \alpha'} \sigma^{\alpha'}_{\phantom{\alpha'} b' c' d'} 
+ R^{\alpha'}_{\phantom{\alpha'} b' \beta' c'; d'} \sigma^{\beta'} \sigma^{a}_{\phantom{a} \alpha'}\\
+ R^{\alpha'}_{\phantom{\alpha'} b' \beta' c'} \sigma^{\beta'} \sigma^{a}_{\phantom{a} d' \alpha'}
+ R^{\alpha'}_{\phantom{\alpha'} b' \beta' d'} \sigma^{\beta'} \sigma^{a}_{\phantom{a} c' \alpha'}
+ R^{\alpha'}_{\phantom{\alpha'} c' \beta' d'} \sigma^{\beta'} \sigma^{a}_{\phantom{a} b' \alpha'} = 0
\end{multline}

\subsection{Transport equation for \texorpdfstring{$g_{a'}^{~~ b}$}{g\_a'\^{ }b}}
The bi-vector of parallel transport is defined by the transport equation
\begin{equation}
\label{eq:bi-transport}
 D'g_{a'}^{\phantom{a'} b} =  g_{a' \phantom{b} ;\alpha'}^{\phantom{a'} b} \sigma^{\alpha'} \equiv 0.
\end{equation}

\subsection{Transport equation for \texorpdfstring{$g_{a b' ; c'}$}{g\_ab';c'}}
Let
\begin{equation}
 A_{a b c} = g_{b}^{\phantom{b} \alpha'} g_{c}^{\phantom{c} \beta'} g_{a \alpha' ; \beta'}
\end{equation}
Applying $D'$ and commuting covariant derivatives, we get a transport equation for $A_{a b c}$:
\begin{equation}
\label{eq:A-transport}
 D' A_{a b c} + A_{a b \alpha} \xi^{\beta'}_{\phantom{\beta'} \gamma'} g_{\beta'}^{\phantom{\beta'} \alpha} g_{c}^{\phantom{c} \gamma'} + g_{a}^{\phantom{a} \alpha'} g_{b}^{\phantom{b} \beta'} g_{c}^{\phantom{c} \gamma'} R_{\alpha' \beta' \gamma' \delta'} \sigma^{\delta'} = 0
\end{equation}

\subsection{Transport equation for \texorpdfstring{$g_{a b' ; c}$}{g\_ab';c}}
Let
\begin{equation}
 B_{a b c} = g_{b}^{\phantom{b} \beta'} g_{a \beta' ; c}
\end{equation}
Applying $D'$ and rearranging, we get a transport equation for $B_{\alpha \beta \gamma}$:
\begin{equation}
\label{eq:B-transport}
 D' B_{a b c} = - A_{a b \alpha} \eta^{\alpha}_{\phantom{\alpha} \beta'} g_{c}^{\phantom{c} \beta'}
\end{equation}

\subsection{Transport equation for \texorpdfstring{$g_{a ~~ ; c' d'}^{~ b'}$}{g\_a\^{ }b'\_;c'd'}}
Applying $D'$ to $g_{a \phantom{b'} ; c' d'}^{\phantom{a} b'}$, we get
\begin{equation}
 D' g_{a \phantom{b'} ; c' d'}^{\phantom{a} b'} = \sigma^{\alpha'} g_{a \phantom{b'} ; c' d' \alpha'}^{\phantom{a} b'}.
\end{equation}
Commuting covariant derivatives on the right hand side, this becomes
\begin{multline}
 D' g_{a \phantom{b'} ; c' d'}^{\phantom{a} b'} = \sigma^{\beta'} \left(g_{a \phantom{b'} ;\beta' c' d'}^{\phantom{a} b'}  
+ R^{b'}_{\phantom{b'} \alpha' \beta' d'} g^{\phantom{a} \alpha'}_{a \phantom{\alpha'} ;c'}
+ R^{b'}_{\phantom{b'} \alpha' \beta' c'} g^{\phantom{a} \alpha'}_{a \phantom{\alpha'} ;d'}\right. \\ \left.
- R^{\alpha'}_{\phantom{\alpha'} c' \beta' d'} g^{\phantom{a} b'}_{a \phantom{b'} ;\alpha'}
+ R^{b'}_{\phantom{b'} \alpha' \beta' c' ; d'} g^{\phantom{a} \alpha'}_{a \phantom{\alpha'}}
\right).
\end{multline}
Bringing $\sigma^{\beta'}$ inside the derivative in the first time on the right hand side, and noting that $g_{a \phantom{b'} ;\beta'}^{\phantom{a} b'}\sigma^{\beta'} = 0$, this then yields a transport equation for $g_{a \phantom{b'} ;c' d'}^{\phantom{a} b'}$:
\begin{multline}
\label{eq:d2Iinv-transport}
 D' g_{a \phantom{b'} ; c' d'}^{\phantom{a} b'} = 
  -\sigma^{\beta'}_{\phantom{\beta'} c'} g_{a \phantom{b'} ;\beta' d'}^{\phantom{a} b'}  
  -\sigma^{\beta'}_{\phantom{\beta'} d'} g_{a \phantom{b'} ;\beta' c'}^{\phantom{a} b'}  
  -\sigma^{\beta'}_{\phantom{\beta'} c' d'} g_{a \phantom{b'} ;\beta'}^{\phantom{a} b'}  \\
+ R^{b'}_{\phantom{b'} \alpha' \beta' d'} \sigma^{\beta'} g^{\phantom{a} \alpha'}_{a \phantom{\alpha'} ;c'}
+ R^{b'}_{\phantom{b'} \alpha' \beta' c'} \sigma^{\beta'} g^{\phantom{a} \alpha'}_{a \phantom{\alpha'} ;d'}
- R^{\alpha'}_{\phantom{\alpha'} c' \beta' d'} \sigma^{\beta'} g^{\phantom{a} b'}_{a \phantom{b'} ;\alpha'}
+ R^{b'}_{\phantom{b'} \alpha' \beta' c' ; d'} \sigma^{\beta'} g^{\phantom{a} \alpha'}_{a \phantom{\alpha'}}.
\end{multline}

\subsection{Transport equation for \texorpdfstring{$\zeta = \ln \Delta^{1/2}$}{zeta = ln Sqrt(Delta)}}
The Van Vleck determinant, $\Delta$ is a bi-scalar defined by
\begin{equation}
 \Delta \left(x,x'\right) \equiv \det \left[ \Delta^{\alpha'}_{\phantom{\alpha'} \beta'} \right], ~~~ \Delta^{\alpha'}_{\phantom{\alpha'} \beta'} \equiv -g^{\alpha'}_{\phantom{\alpha'} \alpha} \sigma^{\alpha}_{\phantom{\alpha} \beta'} = -g^{\alpha'}_{\phantom{\alpha'} \alpha} \eta^{\alpha}_{\phantom{\alpha} \beta'}
\end{equation}

By Eq.~(\ref{eq:eta-transport}), we can write the second equation here as:

\begin{equation}
 \Delta^{\alpha'}_{\phantom{\alpha'} \beta'} = -g^{\alpha'}_{\phantom{\alpha'} \alpha} \left( D' \eta^{\alpha}_{\phantom{\alpha} \beta'} + \eta^{\alpha}_{\phantom{\alpha} \gamma'} \xi^{\gamma'}_{\phantom{\gamma'} \beta'} \right)
\end{equation}

Since $ D' g^{\alpha'}_{\phantom{\alpha'} \alpha} = g^{\alpha'}_{\phantom{\alpha'} \alpha ; \beta'} \sigma^{\beta'} = 0$, we can rewrite this as

\begin{equation}
 \Delta^{\alpha'}_{\phantom{\alpha'} \beta'} = D' \Delta^{\alpha'}_{\phantom{\alpha'} \beta'} + \Delta^{\alpha'}_{\phantom{\alpha'} \gamma'} \xi^{\gamma'}_{\phantom{\gamma'} \beta'}
\end{equation}

Introducing the inverse $(\Delta^{-1})^{\alpha'}_{\phantom{\alpha'} \beta'}$ and multiplying it by the above, we get
\begin{equation}
\label{eq:vanVleck-transport}
 4 = \xi^{\alpha'}_{\phantom{\alpha'} \alpha'} + D'(\ln \Delta)
\end{equation}
where we have used the matrix identity $\delta \ln \det \mathbf{M} = \rm{Tr} \mathbf{M}^{-1} \delta \mathbf{M}$ to convert the trace to a determinant. This can also be written in terms of $\Delta^{1/2}$:
\begin{equation}
\label{eq:vanVleck-transport2}
 D'(\ln\Delta^{1/2}) =  \frac{1}{2} \left(4 - \xi^{\alpha'}_{\phantom{\alpha'} \alpha'} \right)
\end{equation}

\subsection{Transport equation for the Van Vleck determinant, \texorpdfstring{$\Delta^{1/2}$}{Sqrt(Delta)}}
By the definition of $\zeta$, the Van Vleck determinant is given by
\begin{equation}
\label{eq:sqrtdelta-zeta}
 \Delta^{1/2} = e^{\zeta},
\end{equation}
and so satisfies the transport equation
\begin{equation}
 D' \Delta^{1/2} = \frac12 \Delta^{1/2} \left( 4 - \xi^{\alpha'}_{\phantom{\alpha'} \alpha'} \right).
\end{equation}

\subsection{Equation for \texorpdfstring{$\Delta^{-1/2} D (\Delta^{1/2})$}{1/Sqrt(Delta) D Sqrt(Delta)}}
Defining $\tau = \Delta^{-1/2} D (\Delta^{1/2})$, it is immediately clear that
\begin{align}
\label{eq:tau-eq}
 \tau = \Delta^{-1/2} D (\Delta^{1/2}) = D \zeta
\end{align}

\subsection{Equation for \texorpdfstring{$\Delta^{-1/2} D' (\Delta^{1/2})$}{1/Sqrt(Delta) D' Sqrt(Delta)}}
Defining $\tau' = \Delta^{-1/2} D' (\Delta^{1/2})$, it is immediately clear that
\begin{align}
\label{eq:tau-p-eq}
 \tau' = \Delta^{-1/2} D' (\Delta^{1/2}) = D' \zeta
\end{align}

\subsection{Equation for \texorpdfstring{$\nabla_{a'} \Delta$}{nabla\_a' Delta}}
To derive an equation for $\nabla_{a'} \Delta$, we note that 
\begin{align}
\label{eq:VV-def}
 \Delta &\equiv \det \left[-g^{a'}_{\phantom{a'} \alpha} \eta^{\alpha}_{\phantom{\alpha} b'}\right] = - \det \left[ \eta^{a}_{\phantom{a} b'}\right] \det \left[g_{a}^{\phantom{a} a'} \right],
\end{align}
and make use of Jacobi's matrix identity
\begin{align}
\label{eq:d-det-A}
 \mathrm{d} \left(\det \mathbf{A}\right) &= \mathrm{tr} \left( \mathrm{adj} \left( \mathbf{A} \right) \mathrm{d} \mathbf{A} \right) \nonumber \\
 &= \left(\det \mathbf{A} \right) \mathrm{tr} \left( \mathbf{A}^{-1} \mathrm{d} \mathbf{A} \right)
\end{align}
where the operator $\mathrm{d}$ indicates a derivative. Applying \eqref{eq:d-det-A} to \eqref{eq:VV-def}, we get an equation for $\nabla_{a'}\Delta$:
\begin{equation}
\label{eq:nabla-delta}
 \nabla_{a'} \Delta = - \Delta \left[g_{\alpha'}^{\phantom{\alpha'} \alpha} g_{\alpha \phantom{\alpha'} ;a'}^{\phantom{\alpha} \alpha'} + \gamma^{\alpha'}_{\phantom{\alpha'} \alpha} \sigma^{\alpha}_{\phantom{\alpha} \alpha' a'} \right]
\end{equation}

\subsection{Equation for \texorpdfstring{$\Box' \Delta$}{Box' Delta}}
Applying Jacobi's identity twice, together with $ \mathrm{d} (\mathbf{A}^{-1})=  -\mathbf{A}^{-1} (\mathrm{d} \mathbf{A})\mathbf{A}^{-1}$, we find an identity for the second derivative of a matrix:
\begin{multline}
 \mathrm{d}^2\left(\det \mathbf{A}\right) \\= \left(\det \mathbf{A}\right) \Big[\mathrm{tr}\left(\mathbf{A}^{-1} \mathrm{d}\mathbf{A}\right)\mathrm{tr}\left(\mathbf{A}^{-1} \mathrm{d}\mathbf{A}\right) - \mathrm{tr}\left(\mathbf{A}^{-1} \mathrm{d}\mathbf{A} \mathbf{A}^{-1} \mathrm{d}\mathbf{A}\right) + \mathrm{tr}\left(\mathbf{A}^{-1} \mathrm{d}^2 \mathbf{A}\right)\Big].
\end{multline}
Using this identity in Eq.~\eqref{eq:VV-def}, we get an equation for $\Box'\Delta$,
\begin{multline}
\label{eq:box-delta}
 \Box'\Delta = \Delta \left[ \left(g_{\alpha'}^{\phantom{\alpha'} \alpha} g_{\alpha \phantom{\alpha'} ;\mu'}^{\phantom{\alpha} \alpha'} + \gamma^{\alpha'}_{\phantom{\alpha'} \alpha} \sigma^{\alpha}_{\phantom{\alpha} \alpha' \mu'} \right) \left(g_{\alpha'}^{\phantom{\alpha'} \alpha} g_{\alpha \phantom{\alpha'} }^{\phantom{\alpha} \alpha'; \mu'} + \gamma^{\alpha'}_{\phantom{\alpha'} \alpha} \sigma^{\alpha \phantom{\alpha'} \mu'}_{\phantom{\alpha} \alpha'} \right)\right. \\
 ~\left.
- \left( g_{\alpha'}^{\phantom{\alpha'} \alpha} g_{\alpha \phantom{\beta'} ;\mu'}^{\phantom{\alpha} \beta'} g_{\beta'}^{\phantom{\beta'} \beta} g_{\beta}^{\phantom{\beta} \alpha' ; \mu'} \right) 
 - \left( \gamma^{\alpha'}_{\phantom{\alpha'} \alpha} \sigma^{\alpha}_{\phantom{\alpha} \beta' \mu'} \gamma^{\beta'}_{\phantom{\beta'} \beta} \sigma^{\beta \phantom{\alpha'} \mu'}_{\phantom{\beta} \alpha'}\right)\right. \\
 ~\left. 
 + \left( g_{\alpha'}^{\phantom{\alpha'} \alpha} g_{\alpha \phantom{\alpha'} ;\mu'}^{\phantom{\alpha} \alpha' \phantom{;\mu'} \mu'}\right) 
 + \left( \gamma^{\alpha'}_{\phantom{\alpha'} \alpha} \sigma^{\alpha \phantom{\alpha'} \mu'}_{\phantom{\alpha} \alpha' \phantom{\mu'} \mu'}\right) 
\right]
\end{multline}

\subsection{Equation for \texorpdfstring{$\Box' \Delta^{1/2}$}{Box' Sqrt(Delta)}}
Noting that
\begin{align}
 \Box'\Delta^{1/2} &= \left(\frac12 \Delta^{-1/2}\Delta_{;\mu'}\right)^{;\mu'} = \frac12 \Delta^{-1/2}\Box'\Delta - \frac14 \Delta^{-3/2}\Delta^{;\mu'}\Delta_{;\mu'},
\end{align}
it is straightforward to use Eqs.~\eqref{eq:nabla-delta} and \eqref{eq:box-delta} to find an equation for $\Box'\Delta^{1/2}$:
\begin{multline}
\label{eq:box-sqrt-delta}
\Box' \Delta^{1/2} = \frac12 \Delta^{1/2} \left[\frac12
 \left( g_{\alpha'}^{\phantom{\alpha'} \alpha} g_{\alpha \phantom{\alpha'} ;\mu'}^{\phantom{\alpha} \alpha'} + \gamma^{\alpha'}_{\phantom{\alpha'} \alpha} \sigma^{\alpha}_{\phantom{\alpha} \alpha' \mu'} \right)
 \left( g_{\alpha'}^{\phantom{\alpha'} \alpha} g_{\alpha}^{\phantom{\alpha} \alpha'  ;\mu'} + \gamma^{\alpha'}_{\phantom{\alpha'} \alpha} \sigma^{\alpha \phantom{\alpha'}  ;\mu'}_{\phantom{\alpha} \alpha'}\right)\right. \\
 ~\left.
 - \left( g_{\alpha'}^{\phantom{\alpha'} \alpha} g_{\alpha \phantom{\beta'} ;\mu'}^{\phantom{\alpha} \beta'} g_{\beta'}^{\phantom{\beta'} \beta} g_{\beta}^{\phantom{\beta} \alpha' ; \mu'} \right) 
 - \left( \gamma^{\alpha'}_{\phantom{\alpha'} \alpha} \sigma^{\alpha}_{\phantom{\alpha} \beta' \mu'} \gamma^{\beta'}_{\phantom{\beta'} \beta} \sigma^{\beta \phantom{\alpha'} \mu'}_{\phantom{\beta} \alpha'}\right) \right. \\
 ~\left.
 + \left( g_{\alpha'}^{\phantom{\alpha'} \alpha} g_{\alpha \phantom{\alpha'} ;\mu'}^{\phantom{\alpha} \alpha' \phantom{;\mu'} \mu'}\right) 
 + \left( \gamma^{\alpha'}_{\phantom{\alpha'} \alpha} \sigma^{\alpha \phantom{\alpha'} \mu'}_{\phantom{\alpha} \alpha' \phantom{\mu'} \mu'}\right) 
\right]
\end{multline}

\subsection{Transport equation for \texorpdfstring{$V_0$}{V\_0}}
As is given in Eq.~\eqref{eq:recursionV0}, $V_0$ satisfies the transport equation
\begin{equation}
\label{eq:V0-transport}
 \left(D'+1\right) V^{AB'}_0 + \frac{1}{2} V^{AB'}_0 \left( \xi^{\mu'}_{\phantom{\mu'} \mu'} - 4 \right) + \frac{1}{2} \mathcal{D}^{B'}{}_{C'} (\Delta^{1/2} g^{AC'}) = 0 ,
\end{equation}
or equivalently
\begin{equation}
\label{eq:V0-transport-equiv}
\left(D'+1\right) \left(   \Delta^{-1/2} V^{AB'}_0\right) + \frac{1}{2}  \Delta^{-1/2} \mathcal{D}^{B'}{}_{C'} (\Delta^{1/2} g^{AC'}) = 0.
\end{equation}

In particular, for a scalar field:
\begin{equation}
\label{eq:V0-transport-scalar}
 \left(D'+1\right) V_0 + \frac{1}{2} V_0 \left( \xi^{\mu'}_{\phantom{\mu'} \mu'} - 4 \right) + \frac{1}{2} (\Box'-m^2-\xi R') \Delta^{1/2} = 0.
\end{equation}

\subsection{Transport equations for \texorpdfstring{$V_r$}{V\_r}}
As is given in Eq.~\eqref{eq:recursionVn}, $V_{r}$ satisfies the transport equation
\begin{equation}
\label{eq:Vr-transport}
 \left(D'+r+1\right) V^{AB'}_r + \frac{1}{2} V^{AB'}_r \left( \xi^{\mu'}_{\phantom{\mu'} \mu'} - 4 \right) + \frac{1}{2r} \mathcal{D}^{B'}{}_{C'} V^{AC'}_{r-1} = 0.
\end{equation}
or equivalently
\begin{equation}
\label{eq:Vr-transport-equiv}
\left(D'+r+1\right) \left(  \Delta^{-1/2} V^{AB'}_r\right) + \frac{1}{2r}  \Delta^{-1/2} \mathcal{D}^{B'}{}_{C'} V^{AC'}_{r-1} = 0.
\end{equation}
Comparing with Eq.~(\ref{eq:V0-transport-equiv}), it is clear that Eq.~(\ref{eq:Vr-transport-equiv})  may be taken to include $r=0$ if we replace 
$V^{AC'}_{r-1}/r$ by $\Delta^{1/2} g^{AC'}$.

In particular, for a scalar field:
\begin{equation}
\label{eq:Vr-transport-scalar}
 \left(D'+r+1\right) V_r + \frac{1}{2} V_r \left( \xi^{\mu'}_{\phantom{\mu'} \mu'} - 4 \right) + \frac{1}{2r} (\Box'-m^2-\xi R')  V_{r-1} = 0.
\end{equation}

Together with the earlier equations, the transport equation Eq.~\eqref{eq:V0-transport}  allows us to immediately solve for $V^{AB'}_0$ along a geodesic. 
To obtain the higher order $V_r$ we also need to determine $\square' V_{r-1}^{AB'}$. At first sight this appears to require integrating along a
family of neighboring geodesics but, in fact, again we can write transport equations for it.
First we note the identity
\begin{multline}
\label{eq:recursive-transport}
\nabla_{a'} (D' T^{AB'}{}_{a_1'\dots a_n'}) \\=D'( \nabla_{a'} T^{AB'}{}_{a_1'\dots a_n'})  + \xi^{\alpha'}{}_{a'}  \nabla_{\alpha'}T^{AB'}{}_{a_1'\dots a_n'}
 + \sigma^{\alpha'} \mathcal{R}^{B'}{}_{C'a'\alpha'}T^{AC'}{}_{a_1'\dots a_n'}               
\\
- \sigma^{\alpha'} R^{c'}{}_{a_1'a'\alpha'}T^{AC'}{}_{c'\dots a_n'}   - \dots - \sigma^{\alpha'} R^{c'}{}_{a_n'a'\alpha'}T^{AC'}{}_{a_1'\dots c'}       
\end{multline}
where $\mathcal{R}^{A}{}_{Bcd} = \partial_c \mathcal{A}^A{}_{Bd} -  \partial_c \mathcal{A}^A{}_{Bd} + \mathcal{A}^A{}_{Cd}\mathcal{A}^C{}_{Bc}
-\mathcal{A}^A{}_{Cd}\mathcal{A}^C{}_{Bc}$.
Working with $\tilde V^{AB'}_r=\Delta^{-1/2} V^{AB'}_r$, on differentiating Eq.~(\ref{eq:Vr-transport-equiv}) we obtain
a transport equation for the first derivative of $V^{AB'}_r$
\begin{multline}
(D'+r+1) (\tilde V_r^{AB'}{}_{;a'}) + \xi^{\alpha'}{}_{a'} \tilde V_r^{AB'}{}_{;\alpha'} + \sigma^{\alpha'} \mathcal{R}^{B'}{}_{C'a'\alpha'}\tilde V^{AC'}_r \\+ \frac{1}{2r} \left ( \Delta^{-1/2} \mathcal{D}^{B'}{}_{C'} \left(  \Delta^{1/2}  \tilde V^{AC'}_{r-1}\right)\right)_{;a'} = 0 .
\end{multline}
As noted above, this equation also includes $r=0$ if we replace $\tilde V^{AC'}_{r-1}/r$ in this case by $g^{AC'}$:
\begin{multline}
(D'+1) (\tilde V_0^{AB'}{}_{;a'}) + \xi^{\alpha'}{}_{a'} \tilde V_0^{AB'}{}_{;\alpha'} + \sigma^{\alpha'} \mathcal{R}^{B'}{}_{C'a'\alpha'}\tilde V^{AC'}_0 \\+ 
\frac{1}{2} \left (\Delta^{-1/2} \mathcal{D}^{B'}{}_{C'} \left( \Delta^{1/2}  g^{AC'}\right)\right)_{;a'} = 0 .
\end{multline}
Repeating the process,
\begin{multline}
(D'+r+1) (\tilde V_r^{AB'}{}_{;a'b'}) + \xi^{\alpha'}{}_{b'} \tilde V_r^{AB'}{}_{;a'\alpha'} + \xi^{\alpha'}{}_{a'} \tilde V_r^{AB'}{}_{;\alpha'b'}
+ \sigma^{\alpha'} \mathcal{R}^{B'}{}_{C'b'\alpha'}V_r^{AC'}{}_{;a'} \\
+ \sigma^{\alpha'} \mathcal{R}^{B'}{}_{C'a'\alpha'}\tilde V_r^{AC'}{}_{;b'} + \xi^{\alpha'}{}_{a';b'} \tilde V_r^{AB'}{}_{;\alpha'}
- \sigma^{\alpha'} R^{\beta'}{}_{a'b'\alpha'}\tilde V_r^{AC'}{}_{;\beta'}
+ \xi^{\alpha'}{}_{b'} \mathcal{R}^{B'}{}_{C'a'\alpha'}\tilde V^{AC'}_r \\ + \sigma^{\alpha'} \mathcal{R}^{B'}{}_{C'a'\alpha';b'}\tilde V^{AC'}_r + \frac{1}{2r} \left (\Delta^{-1/2} \mathcal{D}^{B'}{}_{C'} \left(\Delta^{1/2}  \tilde V^{AC'}_{r-1}\right)\right)_{;a'b'} = 0 ,
\end{multline}
with the $\tilde V_0^{AB'}{}_{;a'b'}$ equation given by the same replacement as above.

Clearly this process may be repeated as many times as necessary.  At each stage we require two more derivatives on $\tilde V^{AC'}_{r-1}$ than on 
$\tilde V^{AC'}_{r}$ but this may be obtained by a bootstrap process grounded by the $\tilde V^{AC'}_0$ equation which involves only 
the fundamental objects $\Delta^{1/2}$  and $ g^{AC'}$ which we have explored above.
While this process quickly becomes too tedious to follow by hand it is straightforward to 
programme.

For example, to determine $V_1$ for a scalar field we first need to solve the two transport equations
\begin{align}
(D'+1) (\tilde V_{0;a'}) + \xi^{c'}{}_{a'} \tilde V_{0;c'}  + 
\frac{1}{2} \left(\Delta^{-1/2} \left(\Box'-m^2-\xi R'\right) \Delta^{1/2} \right)_{;a'} = 0 ,
\end{align}
and 
\begin{align}
&(D'+1) (\tilde V_{0;a'b'}) + \xi^{c'}{}_{b'} \tilde V_{0;a'c'} + \xi^{c'}{}_{a'} \tilde V_{0;c'b'}+
\nonumber\\&\qquad
+ \xi^{c'}{}_{a';b'} \tilde V_{0;c'}
- \sigma^{c'} R^{d'}{}_{a'b'c'}\tilde V_{0;d'}+
+ \frac{1}{2} \left (\Delta^{-1/2} \left(\Box'-m^2-\xi R'\right) \Delta^{1/2} \right)_{;a'b'} = 0 .
\end{align}

In the next two sections we show how the above system of transport equations can be solved either as a series expansion or numerically. 
For sufficiently simple spacetimes, it is also be possible to find closed form solutions which provide a 
useful check on our results.

\section{Semi-recursive Approach to Covariant \texorpdfstring{\\}{}Expansions} \label{sec:symbolic}
In this section, we will investigate solutions to the transport equations of Sec.~\ref{sec:avramidi} in the form of covariant series expansions. The goal is to find covariant series expressions for the DeWitt coefficients. Several methods have been previously applied for doing such calculations, both by hand and using computer algebra \cite{Belkov:1996,Fulling:1990,Gusynin:1994,Belkov:1993,Fulling:HeatKernel,Gilkey:1979,Gilkey:1980,Ven:1998,Anselmi:2007,Dowker:1999,Gusynin:1999,Salcedo:2001,Salcedo:2004,Salcedo:2007,Gayral:2006,Booth:1998,Avramidi:Schimming:1996,Barvinsky:1985,Fliegner:1998}. However, this effort has been focused primarily on the calculation of the \emph{diagonal} coefficients. To the our knowledge, only the work of Decanini and Folacci \cite{Decanini:Folacci:2005a}, upon which our method is based, has been concerned with the off-diagonal coefficients.

Before proceeding further, it is helpful to see how covariant expansions behave under certain operations. First, applying the operator $D'$ to the covariant expansion of any bi-tensor $A_{a_1 \cdots a_m a_1' \cdots a_n'}$ about the point $x$, we get
\begin{align}
 D' A_{a_1 \cdots a_m a_1' \cdots a_n'} (x,x') 
	&= \sum_{k=0}^\infty \frac{(-1)^n}{n!} k~a_{a_1 \cdots a_m a_1' \cdots a_n' ~ \beta_1 \cdots \beta_k} (x) \sigma^{\beta_1} \cdots \sigma^{\beta_k}_{\phantom{\beta_k} \alpha'} \sigma^{\alpha'} \nonumber \\
	&= \sum_{k=0}^\infty \frac{(-1)^n}{n!} k~a_{a_1 \cdots a_m a_1' \cdots a_n' ~ \beta_1 \cdots \beta_k} (x) \sigma^{\beta_1} \cdots \sigma^{\beta_k}
\end{align}
where the second line is obtained by applying Eq.~(\ref{eq:sigma-eta-xi}) to the first line. In other words, applying $D'$ to the $k$-th term in the series is equivalent to multiplying that term by $k$.

Next we consider applying the operator $D$ to the covariant expansion of any bi-tensor $A_{a_1 \cdots a_m a_1' \cdots a_n'}$ about the point $x$. In this case, there will also be a term involving the derivative of the series coefficient, giving 
\begin{align}
 &D A_{a_1 \cdots a_m a_1' \cdots a_n'} (x,x')  \nonumber \\
& \qquad = \sum_{k=0}^\infty \frac{(-1)^n}{n!} \Big[ k~a_{a_1 \cdots a_m a_1' \cdots a_n' ~ \beta_1 \cdots \beta_k} (x) \sigma^{\beta_1} \cdots \sigma^{\beta_k}_{\phantom{\beta_k} \alpha} \sigma^{\alpha} \nonumber \\
& \hspace*{5cm} + a_{a_1 \cdots a_m a_1' \cdots a_n' ~ \beta_1 \cdots \beta_k ; \alpha} (x) \sigma^{\beta_1} \cdots \sigma^{\beta_k} \sigma^{\alpha}\Big] \nonumber \\
& \qquad = \sum_{k=0}^\infty \frac{(-1)^n}{n!} \Big[ k~a_{a_1 \cdots a_m a_1' \cdots a_n' ~ \beta_1 \cdots \beta_k} (x) \sigma^{\beta_1} \cdots \sigma^{\beta_k}\nonumber \\
& \hspace*{5cm} + a_{a_1 \cdots a_m a_1' \cdots a_n' ~ \beta_1 \cdots \beta_k ; \alpha} (x) \sigma^{\beta_1} \cdots \sigma^{\beta_k} \sigma^{\alpha}\Big] .
\end{align}

We can also consider multiplication of covariant expansions. For any two tensors, $A^{a}_{\phantom{a} b}$ and $B^{a}_{\phantom{a} b}$, with product $C^{a}_{\phantom{a} \alpha} \equiv A^{\alpha}_{\phantom{\alpha} c}B^{c}_{\phantom{c} b}$, say, we can relate their covariant expansions by
\begin{align}
 \sum_{n=0}^{\infty} &\frac{(-1)^n}{n!}  C^{a}_{\phantom{a} b ~ \beta_1 \cdots \beta_n} \sigma^{\beta_1} \cdots \sigma^{\beta_n} \nonumber \\ &=\left(\sum_{n=0}^{\infty} \frac{(-1)^n}{n!} A^{a}_{\phantom{a} \alpha ~ \beta_1 \cdots \beta_n} \sigma^{\beta_1} \cdots \sigma^{\beta_n}\right) \left(\sum_{n=0}^{\infty} \frac{(-1)^n}{n!} B^{\alpha}_{\phantom{\alpha} b ~ \beta_1 \cdots \beta_n} \sigma^{\beta_1} \cdots \sigma^{\beta_n}\right)\nonumber\\
	&= \sum_{n=0}^{\infty} \frac{(-1)^n}{n!} \sum_{k=0}^{n} \binom{n}{k} A^{a}_{\phantom{a} \alpha ~ \beta_1 \cdots \beta_k} B^{\alpha}_{\phantom{\alpha} b ~ \beta_{k+1} \cdots \beta_n} \sigma^{\beta_1} \cdots \sigma^{\beta_n}.
\end{align}

Finally, many of the equations derived in the previous section contain terms involving the Riemann tensor at $x'$, $R^{a'}_{\phantom{a'} b' c' d'}$. As all other quantities are expanded about $x$ rather than $x'$, we will also need to rewrite these Riemann terms in terms of their expansion about $x$:
\begin{align}
 R^{a'}_{\phantom{a'} \alpha' b' \beta'} \sigma^{\alpha'} \sigma^{\beta'} &= \sum_{k=0}^{\infty} \frac{(-1)^k}{k!} R^{a}_{\phantom{a} (\alpha |b| \beta ; \gamma_1 \cdots \gamma_k)} \sigma^{\alpha} \sigma^{\beta} \sigma^{\gamma_1} \cdots \sigma^{\gamma_k}\nonumber \\
	&= \sum_{k=2}^{\infty} \frac{(-1)^k}{(k-2)!} \mathcal{K}^{a}_{\phantom{a} b ~ (k)},
\end{align}
where we follow Avramidi \cite{Avramidi:1986} in introducing the definition
\begin{align}
 \mathcal{K}^{a}_{\phantom{a} b ~ (n)} &\equiv R^{a}_{\phantom{a} (\alpha_1 |b| \alpha_2 ; \alpha_3 \cdots \alpha_n)} \sigma^{\alpha_1} \cdots \sigma^{\alpha_n}\nonumber \\
  &\equiv \bar{\mathcal{K}}^{a}_{\phantom{a} b ~ (n)}  \sigma^{\alpha_1} \cdots \sigma^{\alpha_n}.
\end{align}

These four considerations will now allow us to rewrite the transport equations of Sec.~\ref{sec:avramidi} as recursion relations for the coefficients of the covariant expansions of the tensors involved.

\subsection{Recursion relation for coefficients of the covariant \texorpdfstring{\\}{}expansion of \texorpdfstring{$\gamma^{a'}_{~b}$}{gamma\^{ }a'\_b}}
Rewriting Eq.~(\ref{eq:gamma-transport}) in terms of covariant expansions, we get
\begin{align}
  \sum_{k=0}^\infty \frac{(-1)^k}{k!} &(k^2+k)~ \gamma^{a'}_{\phantom{a'} b ~ \alpha_1 \cdots \alpha_k} (x) \sigma^{\alpha_1} \cdots \sigma^{\alpha_k} \nonumber \\
&  + \left( \sum_{k=2}^{\infty} \frac{(-1)^k}{(k-2)!} \mathcal{K}^{a'}_{\phantom{a'} b ~ (k)} \right) 
  \left(\sum_{k=0}^\infty \frac{(-1)^k}{k!} \gamma^{a'}_{\phantom{a'} b ~ \alpha_1 \cdots \alpha_k} (x) \sigma^{\alpha_1} \cdots \sigma^{\alpha_k}\right) = 0.
\end{align}
From this, the $n$-th term in the covariant series expansion of $\gamma^{a'}_{\phantom{a} b}$ can be written recursively in terms of products of lower order terms in the series with $\mathcal{K}$:
\begin{gather}
   \gamma^{a'}_{\phantom{a'} b ~(0)} = -\delta^{a'}_{\phantom{a'} b} \quad \gamma^{a'}_{\phantom{b'} \nu ~(1)} = 0 \nonumber \\
\gamma^{a'}_{\phantom{a'} b ~ (n)} = -\left(\frac{n-1}{n+1}\right) \sum_{k=0}^{n-2} \binom{n-2}{k} g^{a'}_{\phantom{a'} \alpha} g^{\beta}_{\phantom{\beta} \beta'} \bar{\mathcal{K}}^{\alpha}_{\phantom{\alpha} \beta ~ (n-k)}  \gamma^{\beta'}_{\phantom{\beta'} b ~ (k)}.
\end{gather}
Many of the following recursion relations will make use of these coefficients.

\subsection{Recursion relation for coefficients of the covariant \texorpdfstring{\\}{}expansion of  \texorpdfstring{$\eta^{a}_{~b'}$}{eta\^{ }a\_b'}}
Since $\gamma^{a'}_{\phantom{a'} b}$ is the inverse of $\eta^{a}_{\phantom{a} b'}$, we have 
\begin{equation}
 \gamma^{a'}_{\phantom{a'} \alpha} \eta^{\alpha}_{\phantom{\alpha} b'} = \delta^{a'}_{\phantom{a'} b'}.
\end{equation}
Substituting in covariant expansion expressions for $\gamma^{a'}_{\phantom{a'} \alpha}$ and $\eta^{\alpha}_{\phantom{\alpha} b'}$, we find that the $n$-th term in the covariant series expansion of $\eta^{a}_{\phantom{a} b'}$ is
\begin{align}
 \eta^{a}_{\phantom{a} b' ~(0)} &= -\delta^{a}_{\phantom{a} b'} & \eta^{a}_{\phantom{a} b' ~(1)} &= 0 &\eta^{a}_{\phantom{a} b' ~(n)} &= \sum_{k=2}^n \binom{n}{k} g^{a}_{\phantom{a} \alpha'} \gamma^{\alpha'}_{\phantom{\alpha} \beta~ (k)} \eta^{\beta}_{\phantom{\beta} b' ~(n-k)}.
\end{align}

\subsection{Recursion relation for coefficients of the covariant \texorpdfstring{\\}{}expansion of \texorpdfstring{$\xi^{a'}_{~b'}$}{xi\^{ }a'\_b'}}
Writing Eq.~(\ref{eq:xi-solve-eta-gamma}) in terms of covariant series, it is immediately apparent that the $n$-th term in the covariant expansion of $\xi^{a'}_{\phantom{a'} b'}$ is 
\begin{gather}
 \xi^{a'}_{\phantom{a'} b' ~(0)} = \delta^{a'}_{\phantom{a'} b'} \quad \xi^{a'}_{\phantom{a'} b' ~(1)} = 0\nonumber \\
 \xi^{a'}_{\phantom{a'} b' ~(n)} = n~g^{a'}_{\phantom{a'} \alpha} \eta^{\alpha}_{\phantom{\alpha} b' ~(n)} - \sum_{k=2}^{n-2} \binom{n}{k} k~ \gamma^{a'}_{\phantom{a'} \alpha~ (n-k)} \eta^{\alpha}_{\phantom{\alpha} b' ~(k)}.
\end{gather}

\subsection{Recursion relation for coefficients of the covariant \texorpdfstring{\\}{}expansion of \texorpdfstring{$\lambda^{a}_{~b}$}{lambda\^{ }a\_b}}
Using Eq.~(\ref{eq:lambda-eq}), we can write an equation for the $n$-th order coefficient of the covariant expansion of $\lambda^{a}_{\phantom{a} b}$. However, the expression involves the operator $D$ acting on the covariant series expansion of $\eta^{a}_{\phantom{a} b'}$, so we will first need to find an expression for that. As discussed in the beginning of this section, the derivative in $D$ will affect both the coefficient and the $\sigma^{a}$'s. When acting on the $\sigma^{a}$'s, it has the effect of multiplying the term by $n$ as was previously the case with $D'$. When acting on the coefficient, it will add a derivative to it and increase the order of the term (since we will then be adding a $\sigma^{a}$). We now appeal to the fact that the terms in the expansion of $\eta^{a}_{\phantom{a} b'}$ consists solely of products of $\mathcal{K}^{a}_{\phantom{a} b ~ (k)}$. This means that applying the following general rules when $D$ acts on $\mathcal{K}^{a}_{\phantom{a} b ~ (k)}$, we will get the desired result:
\begin{itemize}
 \item $D \mathcal{K}^{a}_{\phantom{a} b ~ (k)} = k \mathcal{K}^{a}_{\phantom{a} b ~ (k)} + \mathcal{K}^{a}_{\phantom{a} b ~ (k+1)}$
 \item When encountering compound expressions (i.e. consisting of more than a single $\mathcal{K}^{a}_{\phantom{a} b ~ (k)}$), use the normal rules for differentiation (product rule, distributivity, etc.)
\end{itemize}
Taking this into consideration and letting $D^+$ signify the contribution at one higher order and $D^0$ signify the contribution that keeps the order the same, we can write the general $n$-th term in the covariant series expansion of $D\eta^{a}_{\phantom{a} b'}$ as
\begin{equation}
 (D \eta^{a}_{\phantom{a} b'} )_{(n)} = D^+ \eta^{a}_{\phantom{a} b' ~(n-1)} - D^0 \eta^{a}_{\phantom{a} b' ~(n)}
\end{equation}
It is then straightforward to write an expression for the $n$-th term in the covariant series expansion of $\lambda^{a}_{\phantom{a} b}$:
\begin{equation}
 \lambda^{a}_{\phantom{a} b ~ (n)} = \sum_{k=0}^{n-2} \binom{n}{k} (n-k) \left( D^+ \eta^{a}_{\phantom{a} \alpha' ~(n-k-1)} - D^0 \eta^{a}_{\phantom{a} \alpha' ~(n-k)} \right) \gamma^{\alpha'}_{\phantom{\alpha'} b ~ (k)}
\end{equation}

\subsection{Recursion relation for coefficients of the covariant \texorpdfstring{\\}{}expansion of \texorpdfstring{$A_{a b c}$}{A\_\{a b c\}}}
We can rewrite Eq.~(\ref{eq:A-transport}) as
\begin{equation}
 (D'+1)(A_{a b \alpha} \gamma^{\alpha}_{\phantom{\alpha} c}) + R_{a b \alpha \beta} \sigma^{\alpha} \gamma^{\beta}_{\phantom{\beta} c} = 0,
\end{equation}
which when rewritten in terms of covariant series becomes
\begin{equation}
 A_{a b c ~ (k)} = -\frac{1}{n+1} \sum_{k=0}^{n} \binom{n}{k} k ~\mathcal{R}_{a b \alpha ~ (k)}  \gamma^{\alpha}_{\phantom{\alpha} c ~ (n-k)} + \sum_{k=0}^{n-2} \binom{n}{k} A_{a b \alpha ~ (k)}\gamma^{\alpha}_{\phantom{\alpha} c ~ (n-k)}
\end{equation}
where we follow Avramidi \cite{Avramidi:1986,Avramidi:2000} in defining
\begin{equation}
 \mathcal{R}_{a b c ~ (n)} \equiv R_{a b (\alpha_1 |c| ; \alpha_2 \cdots \alpha_n)} \sigma^{\alpha_1} \cdots \sigma^{\alpha_n}
\end{equation}

Alternatively, writing Eq.~(\ref{eq:A-transport}) directly in terms of covariant series, we get
\begin{equation}
 A_{a b c ~ (k)} = \frac{n}{n+1} \mathcal{R}_{a b c ~ (n)} - \frac{1}{n+1} \sum_{k=0}^{n-2} \binom{n}{k} A_{a b \alpha ~ (k)} \xi^{\alpha}_{\phantom{\alpha} c ~ (n-k)},
\end{equation}
which has the benefit of requiring half as much computation as the previous expression.

\subsection{Recursion relation for coefficients of the covariant \texorpdfstring{\\}{}expansion of \texorpdfstring{$B_{a b c}$}{B\_\{a b c\}}}
By Eq.~(\ref{eq:B-transport}), we can immediately write an equation for the coefficients of the covariant expansion of $B_{a b c}$:
\begin{equation}
 B_{a b c ~ (n)} = \frac{1}{n} \sum_{k=0}^{n} \binom{n}{k} A_{a b \alpha ~ (k)} \eta^{\alpha}_{\phantom{\alpha} c ~ (n-k)}
\end{equation}

\subsection{Covariant expansion of \texorpdfstring{$\zeta$}{zeta}}
From Eq.~(\ref{eq:vanVleck-transport2}) we immediately obtain expressions for the coefficients of the covariant series of $\zeta$:
\begin{align}
 \zeta_{(0)} &=0 & \zeta_{(1)} &=0 & \zeta_{(n)} = -\frac{1}{2n} \xi^{\rho'}_{\phantom{\rho'} \rho' ~(n)}
\end{align}

\subsection{Recursion relation for \texorpdfstring{$\Delta^{1/2}$}{Sqrt(Delta)}}
Since $\zeta = \ln{\Delta^{1/2}}$, we can write
\begin{equation}
 \Delta^{1/2} D' \zeta = D' \Delta^{1/2}.
\end{equation}
This allows us to write down a recursive equation for the coefficients of the covariant series expansion of $\Delta^{1/2}$,
\begin{equation}
 \Delta^{1/2}_{(n)} = \frac{1}{n} \sum_{k=2}^{n} \binom{n}{k} k~ \zeta_{(k)} \Delta^{1/2}_{(n-k)}.
\end{equation}

\subsection{Recursion relation for \texorpdfstring{$\Delta^{-1/2}$}{1/Sqrt(Delta)}}
Similarly, the equation
\begin{equation}
 - \Delta^{-1/2} D' \zeta = D' \Delta^{-1/2}.
\end{equation}
allows us to write down a recursive equation for the coefficients of the covariant series expansion of $\Delta^{-1/2}$,
\begin{equation}
 \Delta^{-1/2}_{(n)} = - \frac{1}{n} \sum_{k=2}^{n} \binom{n}{k} k\, \zeta_{(k)} \Delta^{-1/2}_{(n-k)}.
\end{equation}

\subsection{Covariant expansion of \texorpdfstring{$\tau$}{tau}}
Eq.~(\ref{eq:tau-eq}) may be immediately written as a covariant series equation,
\begin{equation}
 \tau_{(n)} = -n D^+ \zeta_{(n-1)} + D^0 \zeta_{(n)}.
\end{equation}

\subsection{Covariant expansion of \texorpdfstring{$\tau'$}{tau'}}
Eq.~(\ref{eq:tau-p-eq}) may be immediately written as a covariant series equation,
\begin{equation}
 \tau'_{(n)} = n \zeta_{(n)}.
\end{equation}

\subsection{Covariant expansion of covariant derivative at \texorpdfstring{$x'$}{x'} of a bi-scalar}
Let $T(x,x')$ be a general bi-scalar. Writing $T$ as a covariant series,
\begin{equation}
 T = \sum_{n=0}^{\infty} T_{(n)} = \sum_{n=0}^{\infty} T_{\alpha_1 \cdots \alpha_n} (x)\sigma^{\alpha_1} \cdots \sigma^{\alpha_n},
\end{equation}
and applying a covariant derivative at $x'$, we get
\begin{align}
\label{eq:cd-covex}
 T_{;b'} &= \sum_{n=0}^{\infty} T_{(n) ;b'} \nonumber \\
	 &= \sum_{n=0}^{\infty} n~ T_{(\alpha_1 \cdots \alpha_n)} \sigma^{\alpha_1} \cdots \sigma^{\alpha_{n-1}} \sigma^{\alpha_n}_{\phantom{\alpha_n} b'}\nonumber \\
         &= \sum_{n=0}^{\infty} n~ T_{(\alpha_1 \cdots \alpha_{n-1} \rho)} \sigma^{\alpha_1} \cdots \sigma^{\alpha_{n-1}} \eta^{\rho}_{\phantom{\rho} b'} \nonumber \\
	 &= - \sum_{n=0}^{\infty} \sum_{k=0}^{n} \binom{n}{k} (T_{(k+1)})_{(-1)} \eta_{(n-k)}
\end{align}
where we have introduced the notation
\begin{equation}
 (T_{(n)})_{(-k)} \equiv T_{(\alpha_1 \cdots \alpha_{(n-k)} a_{(n-k+1)} \cdots \alpha_{n})} \sigma^{\alpha_1} \cdots \sigma^{\alpha_{n-k}} 
\end{equation}

\subsection{Covariant expansion of d'Alembertian at \texorpdfstring{$x'$}{x'} of a \texorpdfstring{\\}{}bi-scalar}
Let $T(x,x')$ be a general bi-scalar as in the previous section. Applying \eqref{eq:cd-covex} twice and taking care to include the term involving $g_{a}{}^{b'}$, we can then write the d'Alembertian, $\Box' T(x,x')$ at $x'$ in terms of covariant series,
\begin{equation}
\label{eq:box-covex}
 (\Box' T)_{(n)} =  -\sum_{k=0}^{n} \binom{n}{k} (T_{;a (k+1)})_{(-1)} \eta_{(n-k)} -\sum_{k=1}^{n} \binom{n}{k} A_{(k)} T_{;a (n-k)}.
\end{equation}

\subsection{Covariant expansion of \texorpdfstring{$\nabla_a \Delta^{1/2}$}{nabla\_a Sqrt(Delta)}}
Applying Eq.~(\ref{eq:cd-covex}) to the case $T=\Delta^{1/2}$, we get
\begin{equation}
 \Delta^{1/2}_{;a}= - \sum_{n=0}^{\infty} \sum_{k=0}^{n} \binom{n}{k} \left(\Delta^{1/2}_{(k+1)}\right)_{(-1)} \eta_{(n-k)}.
\end{equation}

\subsection{Covariant expansion of \texorpdfstring{$\Box' \Delta^{1/2}$}{Box Sqrt(Delta)}}
Applying Eq.~(\ref{eq:box-covex}) to the case $T=\Delta^{1/2}$, we get
\begin{equation}
 (\Box' \Delta^{1/2})_{(n)} =  -\sum_{k=0}^{n} \binom{n}{k} \left(\Delta^{1/2}_{;a (k+1)}\right)_{(-1)} \eta_{(n-k)} -\sum_{k=1}^{n} \binom{n}{k} A_{(k)} \Delta^{1/2}_{;a (n-k)},
\end{equation}
where $A_{(n)}$ is the $n$-th term in the covariant series of the tensor defined in \eqref{eq:A-transport}.

\subsection{Covariant expansion of \texorpdfstring{$V_0$}{V\_0}}
The transport equation for $V_0$, Eq.~(\ref{eq:V0-transport}), can be written in the alternative form
\begin{equation}
 \left(D'+1\right) V_0 - V_0 \tau' + \frac{1}{2} (\Box'-m^2 - \xi R) \Delta^{1/2} = 0.
\end{equation}
This equation is then easily written in terms of covariant expansion coefficients,
\begin{equation}
 V_{0~(n)} = \frac{1}{n+1}\left( \sum_{k=0}^{n-2} \binom{n}{k} V_{0 (k)} \tau'_{(n-k)} - \frac{1}{2}\left((\Box '\Delta^{1/2})_{(n)} - (m^2 + \xi R) \Delta^{1/2}_{(n)}\right)\right)
\end{equation}

\subsection{Covariant expansion of \texorpdfstring{$V_r$}{V\_r}}
The transport equation for $V_r$, Eq.~(\ref{eq:Vr-transport}) can also be written in the alternative form
\begin{equation}
 \left(D'+r+1\right) V_r - V_r \tau' + \frac{1}{2r} (\Box'-m^2 - \xi R) V_{r-1} = 0.
\end{equation}
Again, this is easily written in terms of covariant expansion coefficients,
\begin{equation}
 V_{r~(n)} = \frac{1}{r+n+1}\left( \sum_{k=0}^{n-2} \binom{n}{k} V_{r~ (k)} \tau'_{(n-k)} - \frac{1}{2r}\left((\Box V_{r-1})_{(n)} - (m^2 + \xi R) V_{r-1~(n)}\right)\right).
\end{equation}

\subsection{Results}
Tables \ref{table:a-times} and \ref{table:v0-times} illustrate the performance of our implementation on a desktop computer (2.4GHz Quad Core Processor with 8GB RAM).

\begin{table}[htb]
\begin{center}
 \begin{tabular}{|c||c|c|c|}
\hline
  \textbf{$a_n$} & \textbf{Time (seconds)} & \textbf{Number of terms} & \textbf{Memory Used (bytes)}\\
\hline
\hline
 $a_0$ & $0$ & $0$ & $16$ \\ \hline
 $a_1$ & $0$ & $2$ & $288$ \\ \hline
 $a_2$ & $0$ & $7$ & $2\, 984$ \\ \hline
 $a_3$ & $0.016$ & $68$ & $36\, 976$ \\ \hline
 $a_4$ & $0.168$ & $787$ & $522\, 952$ \\ \hline
 $a_5$ & $3.012$ & $10\,183$ & $7\, 993\, 792$ \\ \hline
 $a_6$ & $70.196$ & $141\,691$ & $128\, 298\, 192$ \\ \hline
 $a_7$ & $1016.696$ & $2\,069\,538$ & $2\, 123\, 985\, 816$ \\ \hline
 \end{tabular}
 \end{center}
\caption{Calculation performance of our semi-recursive implementation of the Avramidi method for computing the coincident (diagonal) DeWitt coefficients, $a_k$.}
\label{table:a-times}
\end{table}

\begin{table}[htb]
\begin{center}
 \begin{tabular}{|c||c|c|c|}
\hline
  \textbf{Order} & \textbf{Time (seconds)} & \textbf{Number of terms} & \textbf{Memory Used (bytes)}\\
\hline
\hline
 $0$ & $0$ & $2$ & $528$ \\ \hline
 $1$ & $0$ & $2$ & $288$ \\ \hline
 $2$ & $0$ & $8$ & $3\,056$ \\ \hline
 $3$ & $0.004$ & $12$ & $4\,800$ \\ \hline
 $4$ & $0.007$ & $41$ & $18\,400$ \\ \hline
 $5$ & $0.016$ & $72$ & $34\,328$ \\ \hline
 $6$ & $0.024$ & $189$ & $96\,920$ \\ \hline
 $7$ & $0.048$ & $357$ & $193\,120$ \\ \hline
 $8$ & $0.084$ & $810$ & $464\,240$ \\ \hline
 $9$ & $0.132$ & $1\,568$ & $938\,960$ \\ \hline
 $10$ & $0.188$ & $3\,290$ & $2\,067\,512$ \\ \hline
 $11$ & $0.384$ & $6\,350$ & $4\,164\,256$ \\ \hline
 $12$ & $0.692$ & $12\,732$ & $8\,689\,208$ \\ \hline
 $13$ & $1.308$ & $24\,340$ & $17\,230\,944$ \\ \hline
 $14$ & $2.688$ & $47\,291$ & $34\,697\,312$ \\ \hline
 $15$ & $5.156$ & $89\,397$ & $67\,881\,528$ \\ \hline
 $16$ & $9.692$ & $169\,900$ & $133\,241\,688$ \\ \hline
 $17$ & $19.985$ & $317\,417$ & $256\,127\,816$ \\ \hline
 $18$ & $39.582$ & $593\,371$ & $494\,810\,408$ \\ \hline
 $19$ & $76.724$ & $1\,096\,634$ & $937\,590\,072$ \\ \hline
 $20$ & $122.943$ & $2\,023\,297$ & $1\,766\,643\,856$ \\ \hline
 \end{tabular}
\end{center}
\caption{Calculation performance of the Avramidi method for computing the covariant series expansion of $V_0$.}
\label{table:v0-times}
\end{table}

\section{Numerical Solution of Transport Equations} \label{sec:numerical}
 In this section, we describe the implementation of the numerical solution 
of the transport equations of Sec.~\ref{sec:avramidi}.
We use the analytic results for $\sigma$, $\Delta^{1/2}$, $g_a{}^{b'}$ and $V_0$ for a scalar field in Nariai spacetime from Ref.~\cite{Nolan:2009} and Sec.~\ref{sec:nariai-exact} as a check on our numerical code.

For the purposes of numerical calculations, the operator $\mathcal{D}'$ acting on a general bi-tensor $T^{a' \ldots}_{\phantom{a' \ldots} b' \ldots}$ can be written as
\begin{equation}
 \mathcal{D}' T^{a' \ldots}_{\phantom{a' \ldots} b' \ldots} = (s'-s) \left( \frac{d}{ds}T^{a' \ldots}_{\phantom{a' \ldots} b' \ldots} + T^{\alpha' \ldots}_{\phantom{\alpha' \ldots} b' \ldots} \Gamma^{a'}_{\alpha' \beta'} u^{\beta'} + \cdots - T^{a' \ldots}_{\phantom{a' \ldots} \alpha' \ldots} \Gamma^{\alpha'}_{b' \beta'} u^{\beta'} - \cdots\right),
\end{equation}
where $s$ is the affine parameter, $\Gamma^{a'}_{b' c'}$ are the Christoffel symbols at $x'$ and $u^{a'}$ is the four velocity at $x'$. Additionally, we make use of the fact that
\begin{equation}
 \sigma^{a'} = (s'- s) u^{a'}.
\end{equation}
which allows us to write Eqs.~\eqref{eq:xi-transport}, \eqref{eq:eta-transport}, \eqref{eq:transport-sigma-ppp}, \eqref{eq:transport-sigma-upp}, \eqref{eq:sigma-pppp-transport}, \eqref{eq:sigma-uppp-transport}, \eqref{eq:bi-transport}, \eqref{eq:A-transport}, \eqref{eq:d2Iinv-transport}, \eqref{eq:vanVleck-transport2}, \eqref{eq:box-sqrt-delta} and \eqref{eq:V0-transport} as a system of coupled, tensor ordinary differential equations. These equations all have the general form:

\begin{equation}
\label{eq:numeric-transport}
 \frac{d}{ds}T^{a' \ldots}_{\phantom{a' \ldots} b' \ldots} = (s')^{-1} A^{a' \ldots}_{\phantom{\alpha'} b' \ldots} + B^{a' \ldots}_{\phantom{\alpha'} b' \ldots} + s' C^{a' \ldots}_{\phantom{\alpha'} b' \ldots} - T^{\alpha' \ldots}_{\phantom{\alpha'} b' \ldots} \Gamma^{a'}_{\alpha' \beta'} u^{\beta'} - \cdots + T^{a' \ldots}_{\phantom{a' \ldots} \alpha' \ldots} \Gamma^{\alpha'}_{b' \beta'} u^{\beta'} + \cdots,
\end{equation}
where we have set $s=0$ without loss of generality and where $A^{a' \ldots}_{\phantom{\alpha'} b' \ldots} = 0 $ initially (i.e. at $s'=0$). It is not necessarily true, however, that the derivative of $A^{\alpha' \ldots}_{\phantom{\alpha'} b' \ldots}$ is zero initially. This fact is important when considering initial data for the numerical scheme.

Solving this system of equations along with the geodesic equations for the spacetime of interest will then yield a numerical value for $V_0$. Moreover, since $V=V_0$ along a null geodesic, the transport equation for $V_0$ will effectively give the full value of $V$ on the light-cone. We have implemented this numerical integration scheme for geodesics in Nariai and Schwarzschild spacetimes using the Runge-Kutta-Fehlberg method in GSL \cite{GSL}. The source code of our implementation is available online \cite{TransportCode}.

\subsection{Initial Conditions}
Numerical integration of the transport equations requires initial conditions for each of the bi-tensors involved. These initial conditions are readily obtained by considering the covariant series expansions of $V_0$, $\Delta^{1/2}$, $\xi^{a'}_{\phantom{a'} b'}$, $\eta^{a}_{\phantom{a} b'}$ and $g^{\phantom{a} b'}_{a}$ and their covariant derivatives at $x'$. Initial conditions for all bi-tensors used for calculating $V_0$ are given in Table \ref{table:init}, where we list the transport equation for the bi-tensor, the bi-tensor itself and its initial value.

Additionally, as is indicated in Eq.~\eqref{eq:numeric-transport}, many of the transport equations will contain terms involving $(s')^{-1}$. These terms must obviously be treated with care in any numerical implementation. Without loss of generality, we set $s=0$. Then, for the initial time step ($s'=s$), we require analytic expressions for
\begin{equation}
\label{eq:numerical-limit}
 \lim_{s'\to s} (s')^{-1} A^{\alpha' \ldots}_{\phantom{\alpha'} b' \ldots}
\end{equation}
which may then be used to numerically compute an accurate initial value for the derivative. This limit can be computed from the first order term in the covariant series of $A^{\alpha' \ldots}_{\phantom{\alpha'} b' \ldots}$, which is found most easily by considering the covariant series of its constituent bi-tensors. For this reason, we list in Table~\ref{table:init} the limit as $s'\to 0$ of all required bi-tensors multiplied by $(s')^{-1}$. In Table \ref{table:init-transport} we list the terms $(s')^{-1} A^{\alpha' \ldots}_{\phantom{\alpha'} b' \ldots}$ for each transport equation involving $(s')^{-1}$, along with their limit as $s'\to s$.

\begin{table}[htb]
\renewcommand{\arraystretch}{2}
\begin{center}
 \begin{tabular}{|c||c|c|c|}
\hline
  \textbf{Equation} & \textbf{Bitensor} & \textbf{Initial Condition} & \textbf{$(s')^{-1}$ Initial Condition}\\
\hline
\hline
 \eqref{eq:xi-transport} & $\xi^{a'}_{\phantom{a'} b'}$ & $\delta^{a}_{\phantom{a} b}$ & $0$\\ \hline
 \eqref{eq:eta-transport} & $\eta^{a}_{\phantom{a} b'}$ & $-g_{\phantom{a} b}^{a}$ & $0$\\ \hline
 \eqref{eq:transport-sigma-ppp} & $\sigma^{a'}_{\phantom{a'} b' c'}$ & $0$ & $-\frac23 R^{a}_{\phantom{a} (\alpha |b| c)} u^{\alpha}$ \\ \hline 
 \eqref{eq:transport-sigma-upp} & $\sigma^{a}_{\phantom{a} b' c'}$ & $0$ & $\frac12 R^{a}_{\phantom{a} b \alpha c} u^{\alpha} - \frac13 R^{a}_{\phantom{a} (\alpha |b| c)} u^{\alpha}$\\ \hline 
 \eqref{eq:sigma-pppp-transport} & $\sigma^{a'}_{\phantom{a'} b' c' d'}$ & $-\frac23 R^{a}_{\phantom{a}  (c | b | d)}$ & \multicolumn{1}{|c|}{$\frac12 R^{a}_{\phantom{a} (c |b| d ; \alpha)}u^{\alpha} - \frac23 R^{a}_{\phantom{a} (\alpha |b| d) ;c} u^{\alpha}$} \\
  & & & \multicolumn{1}{|c|}{$- \frac23 R^{a}_{\phantom{a} (\alpha |b| c) ;d} u^{\alpha}$ }
 \\ \hline 
 \eqref{eq:sigma-uppp-transport} & $\sigma^{a}_{\phantom{a} b' c' d'}$ & $-\frac13 R^{a}_{\phantom{a}  (c | b | d)} -\frac12  R^{a}_{\phantom{a}  b  c  d}$ & $-\frac12 R^{a}_{\phantom{a} (c |b| d ; \alpha)}u^{\alpha} + \frac23 R^{a}_{\phantom{a} (\alpha |b| d) ;c} u^{\alpha}$ \\ \hline
 \eqref{eq:bi-transport} & $g_{a'}^{\phantom{a'} b}$ & $g_{a}^{\phantom{a} b}$ & 0\\ \hline
 \eqref{eq:A-transport} & $g_{a \phantom{b'} ; c'}^{\phantom{a} b'}$ & $0$ & $\frac12 R^{b}_{\phantom{b} a \alpha c} u^{\alpha}$ \\ \hline
 \eqref{eq:d2Iinv-transport} & $g_{a \phantom{b'} ; c' d'}^{\phantom{a} b'}$ & $-\frac12 R^{b}_{\phantom{b} a c d}$ & $-\frac23 R^{b}_{\phantom{b} a c (d ; \alpha)} u^{\alpha}$ \\ \hline
 \eqref{eq:vanVleck-transport2} & $\Delta^{1/2}$ & $1$ & 0 \\ \hline
 \eqref{eq:V0-transport} & $V_0$ & $\frac12 m^2 + \frac12 \left(\xi-\frac16\right) R$ & $-\frac14 \left(\xi-\frac16\right) R_{;\alpha} u^{\alpha}$\\
\hline
 \end{tabular}
 \end{center}
\caption{Initial conditions for tensors used in the numerical calculation of $V_0$.}
\label{table:init}
\end{table}

\begin{table}[htb]
\begin{center}
\renewcommand{\arraystretch}{2}
 \begin{tabular}{|c||c|c|}
\hline
  \textbf{Equation} & \textbf{Terms involving $(s')^{-1}$} & \textbf{IC for $(s')^{-1}$ terms} \\
\hline
\hline
 \eqref{eq:xi-transport} & $-(s')^{-1}\left(\xi^{a'}_{\phantom{a'} \alpha'}\xi^{\alpha'}_{\phantom{\alpha'} b'} - \xi^{a'}_{\phantom{a'} b'}\right)$ & $0$ \\ \hline
 \eqref{eq:eta-transport} & $-(s')^{-1}\left(\eta^{a}_{\phantom{a} \alpha'}\xi^{\alpha'}_{\phantom{\alpha'} b'} - \eta^{a}_{\phantom{a} b'}\right)$ & $0$ \\ \hline
 \eqref{eq:transport-sigma-ppp} & \multicolumn{1}{|l|}{$(s')^{-1} \big( \sigma^{a'}_{\phantom{a'} b' c'} - \xi^{a'}_{\phantom{a'} \alpha'} \sigma^{\alpha'}_{\phantom{\alpha'} b' c'}$} & $-\frac23 R^{a}_{\phantom{a} (b |\alpha| c)} u^{\alpha}$ \\
 & \multicolumn{1}{|r|}{$ - \xi^{\alpha'}_{\phantom{\alpha'} b'} \sigma^{a'}_{\phantom{a'} \alpha' c'} - \xi^{\alpha'}_{\phantom{\alpha'} c'} \sigma^{a'}_{\phantom{a'} \alpha' b'}\big)$} & \\\hline 
 \eqref{eq:transport-sigma-upp} &  \multicolumn{1}{|l|}{$(s')^{-1} \big( \sigma^{a}_{\phantom{a} b' c'} - \eta^{a}_{\phantom{a} \alpha'} \sigma^{\alpha'}_{\phantom{\alpha'} b' c'}$} & $-\frac12 R^{a}_{\phantom{a} \alpha b c} u^{\alpha} -\frac13 R^{a}_{\phantom{a} (b |\alpha| c)} u^{\alpha}$ \\
 &  \multicolumn{1}{|r|}{$- \xi^{\alpha'}_{\phantom{\alpha'} b'} \sigma^{a}_{\phantom{a} \alpha' c'} - \xi^{\alpha'}_{\phantom{\alpha'} c'} \sigma^{a}_{\phantom{a} \alpha' b'}\big)$} & \\ \hline 
 \eqref{eq:sigma-pppp-transport} & \multicolumn{1}{|l|}{$(s')^{-1}\big( \sigma^{a'}_{\phantom{a'} b' c' d'} - \sigma^{a'}_{\phantom{a'} \alpha' b'}\sigma^{\alpha'}_{\phantom{\alpha'} c' d'}$} & $-\frac32 R^{a}_{\phantom{a} (b |\alpha| c; d)} u^{\alpha}$\\
 & \multicolumn{1}{|c|}{$- \sigma^{a'}_{\phantom{a'} \alpha' c'}\sigma^{\alpha'}_{\phantom{\alpha'} b' d'} - \sigma^{a'}_{\phantom{a'} \alpha' d'}\sigma^{\alpha'}_{\phantom{\alpha'} b' c'}$} & \\
  & \multicolumn{1}{|c|}{$ - \sigma^{a'}_{\phantom{a'} \alpha' b' c'} \xi^{\alpha'}_{\phantom{\alpha'} d'} - \sigma^{a'}_{\phantom{a'} \alpha' b' d'} \xi^{\alpha'}_{\phantom{\alpha'} c'} $} & \\
 & \multicolumn{1}{|r|}{$- \sigma^{a'}_{\phantom{a'} \alpha' c' d'} \xi^{\alpha'}_{\phantom{\alpha'} b'} - \sigma^{\alpha'}_{\phantom{\alpha'} b' c' d'} \xi^{a'}_{\phantom{a'} \alpha'} \big)$} & \\ \hline
 \eqref{eq:sigma-uppp-transport} & \multicolumn{1}{|l|}{$(s')^{-1}\big( \sigma^{a}_{\phantom{a} b' c' d'} - \sigma^{a}_{\phantom{a} \alpha' b'}\sigma^{\alpha'}_{\phantom{\alpha'} c' d'}$} & \multicolumn{1}{|c|}{$\frac12 R^{a}_{\phantom{a} (c |\alpha| d) ;b}u^{\alpha} - \frac56 R^{a}_{\phantom{a} b \alpha (c ; d)}u^{\alpha}$} \\
 & \multicolumn{1}{|c|}{$- \sigma^{a}_{\phantom{a} \alpha' c'}\sigma^{\alpha'}_{\phantom{\alpha'} b' d'} - \sigma^{a}_{\phantom{a} \alpha' d'}\sigma^{\alpha'}_{\phantom{\alpha'} b' c'}$} & \multicolumn{1}{|c|}{$ + \frac76 R^{a}_{\phantom{a} (d |\alpha b| ;c)}u^{\alpha}$}\\
 & \multicolumn{1}{|c|}{$ - \sigma^{a}_{\phantom{a} \alpha' b' c'} \xi^{\alpha'}_{\phantom{\alpha'} d'} - \sigma^{a}_{\phantom{a} \alpha' b' d'} \xi^{\alpha'}_{\phantom{\alpha'} c'} $} & \\
 & \multicolumn{1}{|r|}{$- \sigma^{a}_{\phantom{a} \alpha' c' d'} \xi^{\alpha'}_{\phantom{\alpha'} b'} - \sigma^{\alpha'}_{\phantom{\alpha'} b' c' d'} \eta^{a}_{\phantom{a} \alpha'} \big)$} &  \\ \hline
 \eqref{eq:bi-transport} & $0$ & $0$ \\ \hline
 \eqref{eq:A-transport} & $-(s')^{-1} g_{a \phantom{b'} ; \alpha'}^{\phantom{a} b'} \xi^{\alpha'}_{\phantom{\alpha'} c'}$ & $-\frac12 R^{b}_{\phantom{b} a \alpha c} u^{\alpha}$\\ \hline
 \eqref{eq:d2Iinv-transport} & \multicolumn{1}{|l|}{$-(s')^{-1} \big(g_{a \phantom{b'} ; \alpha' d'}^{\phantom{a} b'} \xi^{\alpha'}_{\phantom{\alpha'} c'} + g_{a \phantom{b'} ; \alpha' c'}^{\phantom{a} b'} \xi^{\alpha'}_{\phantom{\alpha'} d'}$}& $-\frac23 R^{b}_{\phantom{b} a \alpha (c ; d)} u^{\alpha}$ \\
 & \multicolumn{1}{|r|}{$+ g_{a \phantom{b'} ; \alpha'}^{\phantom{a} b'} \sigma^{\alpha'}_{\phantom{\alpha'} c' d'}\big)$ } &\\ \hline
 \eqref{eq:vanVleck-transport2} & $-(s')^{-1} \Delta^{1/2} \left(\xi^{a'}_{\phantom{a'} a'} - \delta^{a'}_{\phantom{a'} a'}\right)$ & $0$ \\ \hline
 \eqref{eq:V0-transport} & \multicolumn{1}{|l|}{$-(s')^{-1} \big[\left(\xi^{a'}_{\phantom{a'} a'} - \delta^{a'}_{\phantom{a'} a'}\right) V_0 + 2 V_0$} & $\frac14 \left(\xi-\frac16\right) R_{;\alpha} u^{\alpha}$ \\
& \multicolumn{1}{|r|}{$ + (\Box' -m^2 - \xi R)\Delta^{1/2} \big]$} &\\
\hline
 \end{tabular}
 \end{center}
\caption{Initial conditions (ICs) for transport equations required for the numerical calculation of $V_0$.}
\label{table:init-transport}
\end{table}

\subsection{Results}
\begin{figure}
\begin{center}
\quad \qquad \includegraphics[width=12.6cm]{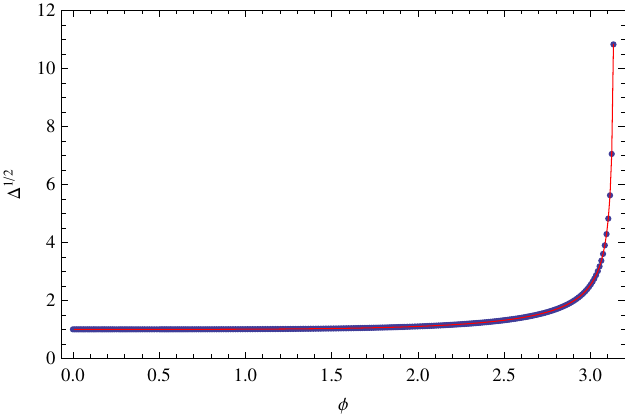}
 \includegraphics[width=14cm]{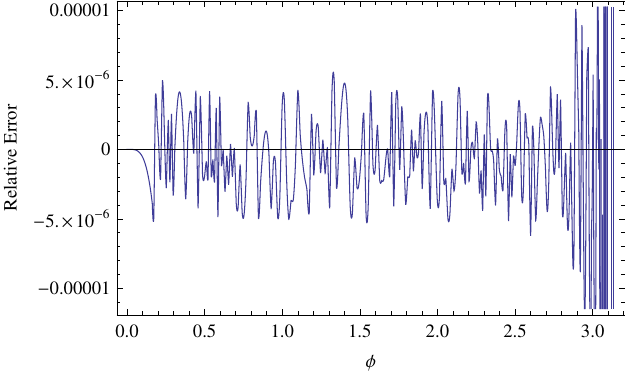}
\end{center}
\caption[Comparison of numerical and exact analytic calculations of $\Delta^{1/2}$]{\emph{Comparison of numerical and exact analytic calculations of $\Delta^{1/2}$ as a function of the angle, $\phi$, along an orbiting null geodesic in Nariai spacetime.} Top: The numerical calculation (blue dots) is a close match with the analytic expression (red line). Bottom: With parameters so that the code completes in $1$ minute, the relative error is within $0.0001\%$ up to the boundary of the normal neighborhood (at $\phi=\pi$).}
\label{fig:nariai-transport-VV-comparison}
\end{figure}

\begin{figure}
\begin{center}
 \includegraphics[width=12.5cm]{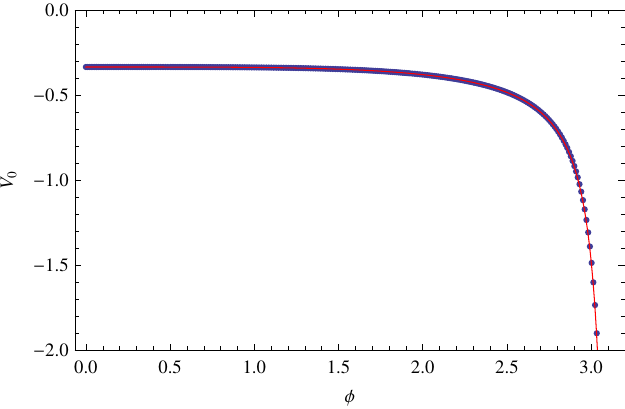}
 \includegraphics[width=12.7cm]{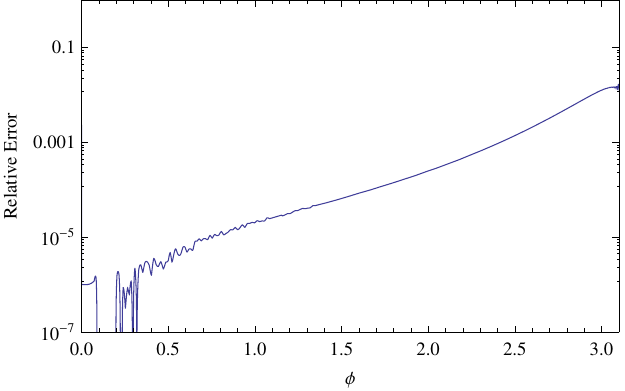}
\end{center}
\caption[Comparison of numerical and exact analytic calculations of $V_0$]{\emph{Comparison of numerical and exact analytic calculations of $V_0$ for a massless, minimally coupled scalar field as a function of the angle, $\phi$, along an orbiting null geodesic in Nariai spacetime.} Top: The numerical calculation (blue dots) is a close match with the analytic expression (red line). The coincidence value is $V(x,x) = \frac12 (\xi-\frac16)R = -\frac13$, as expected. Bottom: With parameters so that the code completes in $1$ minute, the relative error in the numerical calculation is within $1\%$ up to the boundary of the normal neighborhood (at $\phi=\pi$),}
\label{fig:nariai-transport-V0-comparison}
\end{figure}

The accuracy of our numerical code may be verified by comparing with the results of Ref.~\cite{Nolan:2009} and Sec.~\ref{sec:nariai-exact}, which give analytic expressions (for the Nariai spacetime) for all of the bi-tensors used in this chapter. In Figs.~\ref{fig:nariai-transport-VV-comparison} and \ref{fig:nariai-transport-V0-comparison}, we compare analytic and numerical expressions for $\Delta^{1/2}$ and $V_0$, respectively. We consider the null geodesic which starts at $\rho=0.5$ and moves inwards to $\rho=0.25$ before turning around and going out to $\rho=0.789$, where it reaches a caustic (i.e. the edge of the normal neighborhood). The affine parameter, $s$, has been scaled so that it is equal to the angle coordinate, $\phi$. We find that the numerical results faithfully match the analytic solution up the boundary of the normal neighborhood. For the case of $\Delta^{1/2}$ (Fig.~\ref{fig:nariai-transport-VV-comparison}) the error remains less than one part in $10^{-6}$ to within a short distance of the normal neighborhood boundary. The results for $V_0(x,x')$ are less accurate, but nonetheless the relative error remains less $1\%$.
 
\begin{figure}
 \begin{center}
 \includegraphics[width=2cm,angle=90]{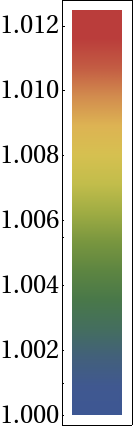}
 \includegraphics[height=15cm,angle=90]{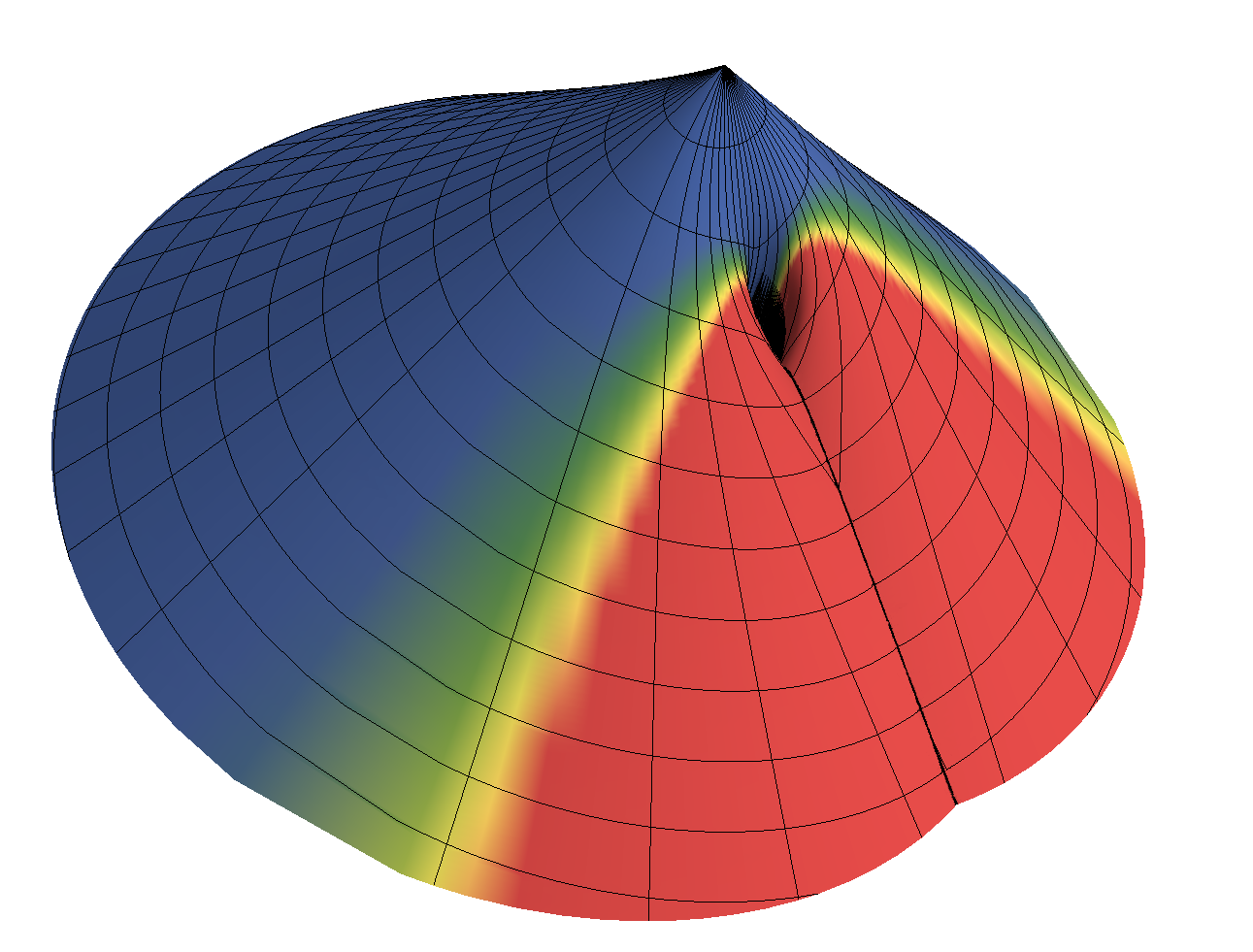}
\end{center}
\caption[$\Delta^{1/2}$ along the light-cone in Schwarzschild spacetime (I)]{\emph{$\Delta^{1/2}$ along the light-cone in Schwarzschild Spacetime.} The point $x$ at the vertex of the cone is fixed at $r=10M$. $\Delta^{1/2}$ increases along any null geodesic up to the caustic where it is singular.}
\label{fig:schw-VV-cone}
\end{figure}

\begin{figure}
 \begin{center}
 \includegraphics[width=2cm,angle=90]{LegendVV.png}
 \includegraphics[height=15cm,angle=90]{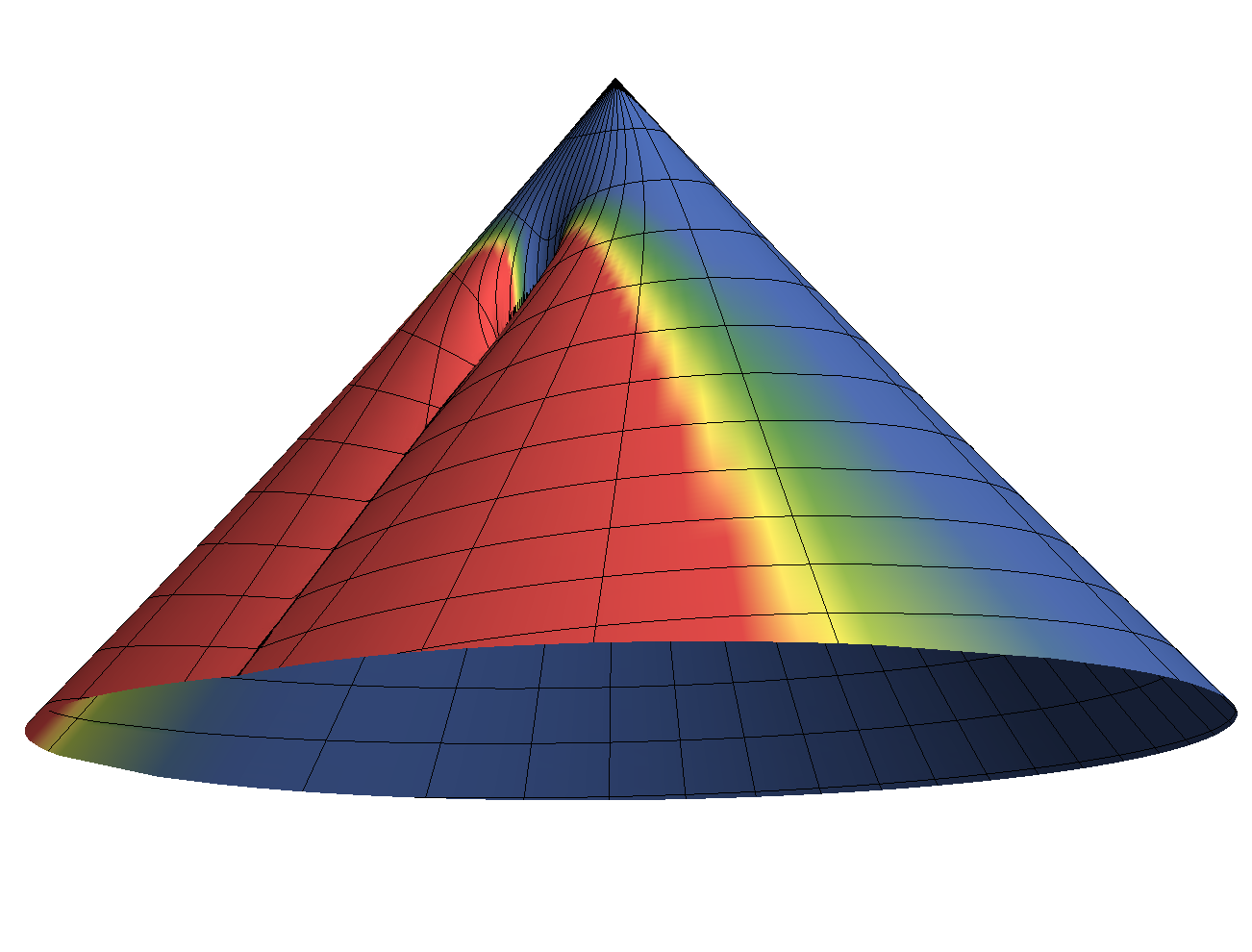}
\end{center}
\caption[$\Delta^{1/2}$ along the light-cone in Schwarzschild spacetime (II)]{\emph{$\Delta^{1/2}$ along the light-cone in Schwarzschild Spacetime.} The point $x$ at the vertex of the cone is fixed at $r=10M$. $\Delta^{1/2}$ increases along any null geodesic up to the caustic where it is singular.}
\label{fig:schw-VV-cone-2}
\end{figure}

In Figs.~\ref{fig:schw-VV-cone} and \ref{fig:schw-VV-cone-2}, we present plots calculated from our numerical code which indicate how $\Delta^{1/2}$ varies over the whole light-cone in Schwarzschild. We find that it remains close to its initial value of $1$ far away from the caustic. As geodesics get close to the caustic, $\Delta^{1/2}$ grows and is eventually singular at the caustic. This is exactly as expected; $\Delta^{1/2}$ is a measure of the strength of focusing of geodesics, where values greater than $1$ correspond to focusing and values less than $1$ correspond to de-focusing. At the caustic, where geodesics are focused to a point, one would expect $\Delta^{1/2}$ to be singular.

\begin{figure}
 \begin{center}
 \includegraphics[width=2cm,angle=90]{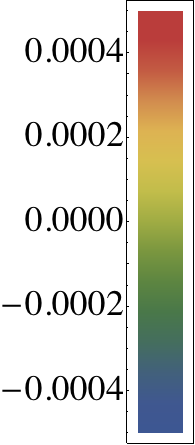}
 \includegraphics[height=14cm,angle=90]{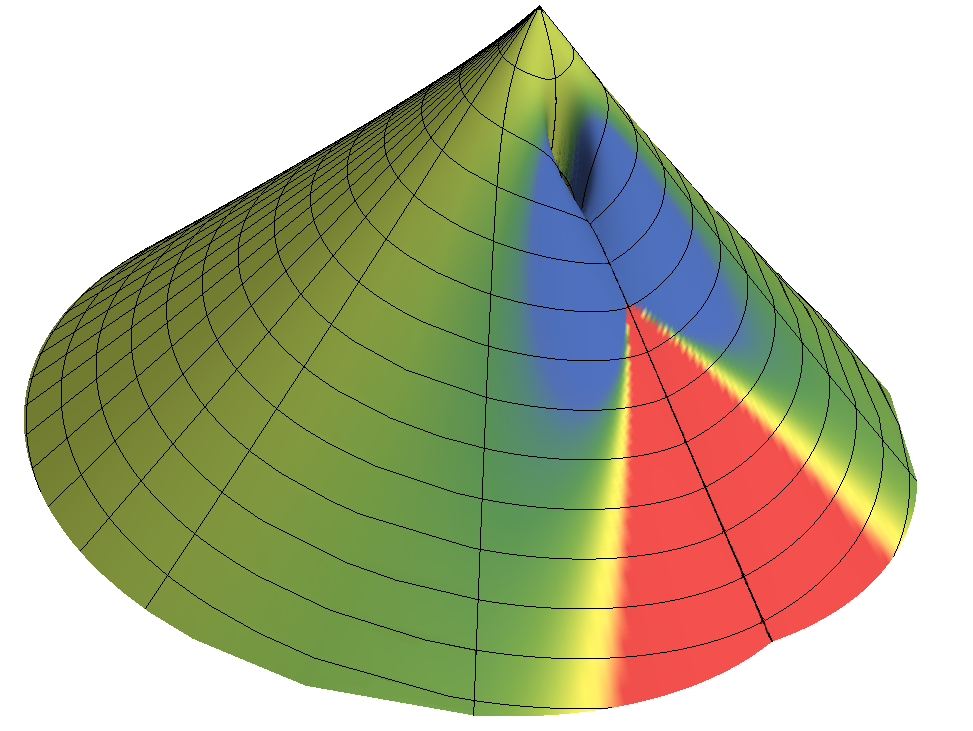}
\end{center}
\caption[$V(x,x')$ along the light-cone in Schwarzschild spacetime (I).]{\emph{$V(x,x')$ for a massless, scalar field along the light-cone in Schwarzschild Spacetime.} The point $x$ (the vertex of the cone) is fixed at $r=6M$. $V(x,x')$ is $0$ initially, then, travelling along a geodesic, it goes negative for a period before turning positive and eventually becoming singular at the caustic.}
\label{fig:schw-V0-cone}
\end{figure}

\begin{figure}
 \begin{center}
 \includegraphics[width=2cm,angle=90]{LegendV0.png}
 \includegraphics[height=12cm,angle=90]{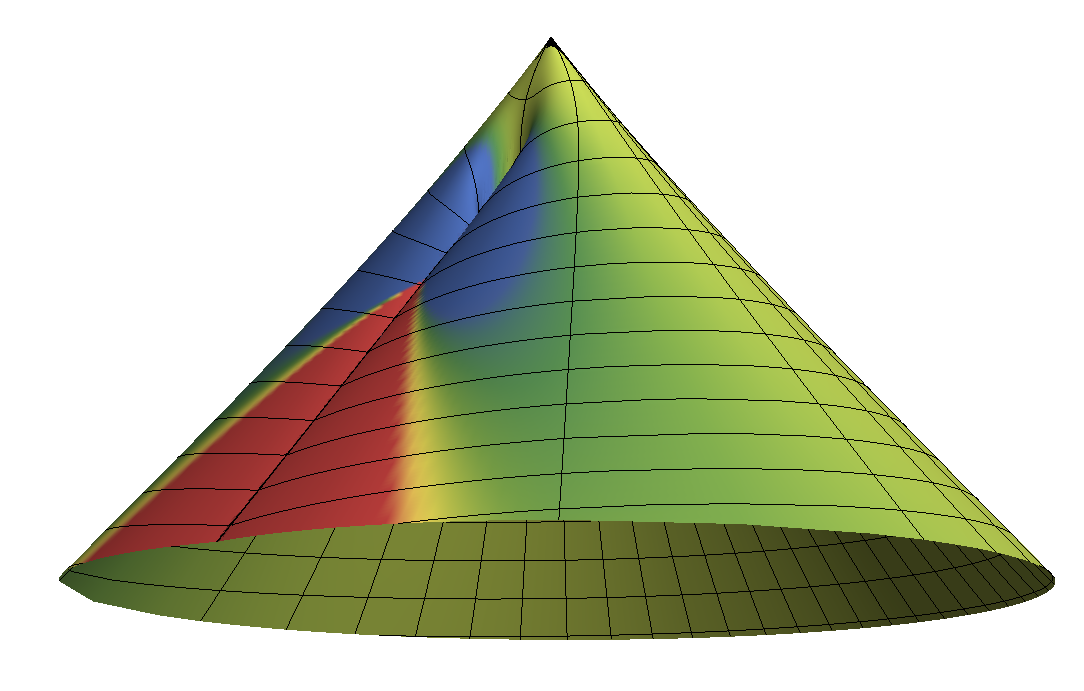}
\end{center}
\caption[$V(x,x')$ along the light-cone in Schwarzschild spacetime (II).]{\emph{$V(x,x')$ for a massless, scalar field along the light-cone in Schwarzschild Spacetime.} The point $x$ (the vertex of the cone) is fixed at $r=6M$. $V(x,x')$ is $0$ initially, then, travelling along a geodesic, it goes negative for a period before turning positive and eventually becoming singular at the caustic.}
\label{fig:schw-V0-cone-2}
\end{figure}

In Figs.~\ref{fig:schw-V0-cone} and \ref{fig:schw-V0-cone-2}, we present similar plots (again calculated from our numerical code) which indicate how $V(x,x')$ varies over the light-cone in Schwarzschild. In this case there is considerably more structure than was previously the case with $\Delta^{1/2}$. There is the expected singularity at the caustic. However, travelling along a geodesic, $V(x,x')$ also becomes negative for a period before turning positive and eventually becoming singular at the caustic.

\begin{figure}
 \begin{center}
 \includegraphics[width=12.5cm]{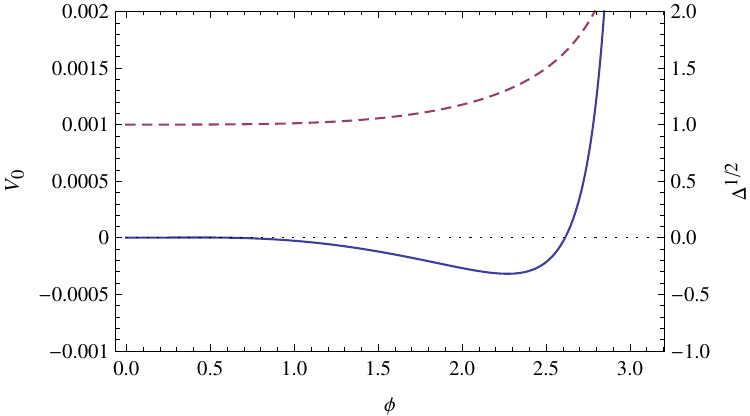}
 \includegraphics[width=12.5cm]{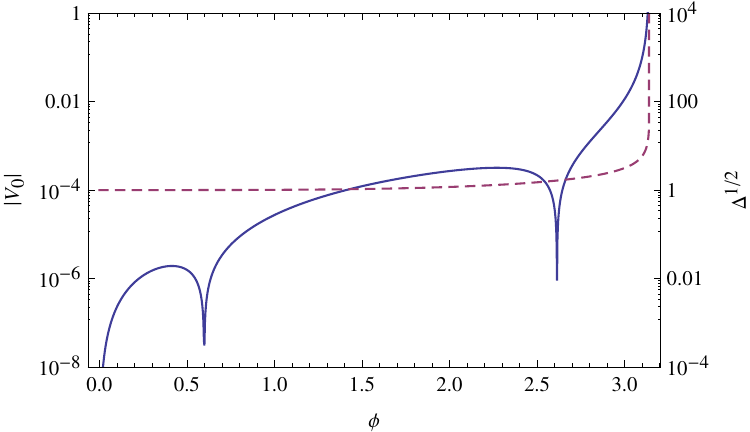}
\end{center}
\caption[$V_0(x,x')$ and $\Delta^{1/2}$ along the timelike circular orbit in Schwarzschild]{\emph{$V_0(x,x')$ (solid blue line) and $\Delta^{1/2}$ (dashed purple line) for a massless, scalar field along the timelike circular orbit at $r=10M$ in Schwarzschild spacetime as a function of the angle, $\phi$ through which the geodesic has passed.} In the logarithmic plot, the absolute value of $V_0(x,x')$ is plotted, since $V_0(x,x')$ is negative between $\phi\sim0.6$ and $\phi \sim 2.6$. The apparent discontinuities in the plot are therefore simply a result of $V_0(x,x')$ passing through $0$ at these points.}
\label{fig:schw-V0-timelike}
\end{figure}

The transport equations may also be applied to calculate $V_r(x,x')$ along a timelike geodesic. In Fig.~\ref{fig:schw-V0-timelike}, we apply our numerical code to the calculation of $V_0(x,x')$ along the timelike circular orbit at $r=10M$ in Schwarzschild.

\chapter{Quasilocal Self-force in Black Hole Spacetimes} \label{ch:qlsf}
In this chapter, we use the results of Chapters~\ref{ch:sf}, \ref{ch:coordex} and \ref{ch:covex} to evaluate the quasilocal portion of the scalar self-force for a variety of geodesics and spacetimes.

In Sec.~\ref{sec:qlsf-geodesic-examples}, we study examples of geodesic motion in both Schwarzschild and Kerr spacetimes. We find that our results are in agreement with those of \cite{Anderson:Wiseman:2005}.

In Sec.~\ref{sec:qlsf-non-geodesic-examples} we consider the case of non-geodesic motion. A strong motivation for doing so is that it allows us to verify our results against those of Refs.~\cite{Anderson:Wiseman:2005} and \cite{Anderson:Eftekharzadeh:Hu:2006}. In fact, the comparison allowed a missing factor of 2 in the results of Ref.~\cite{Anderson:2003} to be discovered \cite{Anderson:2003:err1,Anderson:2003:err2}, having a knock on effect on Refs.~\cite{Anderson:Wiseman:2005} and \cite{Anderson:Eftekharzadeh:Hu:2006}. Also, Wiseman \cite{Wiseman:2000} has shown that the total self-force in this case of a static scalar particle in Schwarzschild spacetime is zero. This example of non-geodesic motion could potentially facilitate a study of the usefulness and accuracy of the matched expansion approach \cite{Ottewill:Wardell:2008,Anderson:Wiseman:2005} to the calculation of the self-force.

\section{Geodesic Motion} \label{sec:qlsf-geodesic-examples}
Expression (\ref{eq:SimpForce}) is a very general result, giving the quasilocal contribution to the self-force on a scalar charge following an arbitrary time-like geodesic in any 4-dimensional vacuum spacetime. While such a general result is of tremendous benefit in terms of flexibility, it is also interesting to examine some specific cases whose gravitational analog are of immediate physical interest. We will consider two well-known black hole spacetimes, Schwarzschild and Kerr, and some astrophysically probable geodesics paths for EMRI systems.

\subsection{Schwarzschild Space-time}
\label{sec:cases-schw}

We begin with the Schwarzschild spacetime as that is the simpler of the two cases. The line-element of this spacetime is given by Eq.~\eqref{eq:schwle}.
From this, we calculate all necessary components of the Riemann tensor and its covariant derivatives and hence values for all the $v_{a_1 \dots a_p}$ of Sec.~\ref{sec:qlsf-geodesic}. While the calculations involved pose no conceptual difficulty, there is a great deal of scope for numerical slips. Fortunately, they are well suited to being done using a computer algebra system such as the \textsl{GRTensorII} \cite{GRTensor} package for the \textsl{Maple} \cite{Maple} computer algebra system.

We consider a particle with a general geodesic $4$-velocity, $u^\alpha$, travelling in this background spacetime. At first glance, it may appear that there are four independent components of this 4-velocity. However, because of the spherical symmetry of the problem, we may arbitrarily choose the orientation of the equatorial axis, $\theta$. By choosing this to line up with the 4-velocity (i.e. $\theta = \frac{\pi}{2}$), we can see immediately that $u^\theta=0$. Furthermore, the particle is following a time-like geodesic, so the normalization condition on the 4-velocity $u_{\alpha} u^\alpha = -1$ holds. This allows us to eliminate one of the three remaining components. In our case, we choose to eliminate $u^t$, which has the equation:

\begin{equation}
\label{eq:ut}
u^t = \sqrt{\left( \frac{r}{r-2M} \right) \left( 1 + \frac{r}{r-2M} \left( u^r \right)^2 + r^2 \left( u^\phi \right)^2 \right)}
\end{equation}

We are now left with just two independent components of the 4-velocity, $u^\phi$ and $u^r$. Substituting in to Eq. (\ref{eq:SimpForce}) the values for the known 4-velocity components $u^\theta$ and $u^t$ along with the Riemann tensor components, we get a general expression for the scalar self-force in terms of the scalar charge, $q$, the matching point, $\Delta \tau$, the mass of the Schwarzschild black hole, $M$, the radial distance of the scalar charge from the black hole, $r$ and the two independent components of the 4-velocity of the scalar charge, $u^r$ and $u^\phi$. We can then use Eqs. (\ref{eq:ma}) - (\ref{eq:dmdtau})  to obtain the quasilocal contribution to the equations of motion:
\begin{subequations}
\allowdisplaybreaks
\begin{align}
\label{eq:QLSF-SchwGen-r}
&ma_{\rm QL}^{r} =
	\frac {3 q^2 M^2}{11200 r^{10}} \Bigg\{ \nonumber \\
& \quad 10 r^2 u^r \left[
	\left(\frac{r-2M}{r}\right) \left(1 + 11\left(ru^\phi\right)^2 + 13\left(ru^\phi\right)^4\right)+\left(u^r\right)^2 \left(1+8\left(ru^\phi\right)^2\right)
	\right]
	{\Delta \tau}^4 \nonumber \\
& \quad	- \Bigg[
	\left(\frac{r-2M}{r}\right) \left(\left(20r-49M\right) + 8\left(11r-28M\right)\left(ru^\phi\right)^2 + 2\left(34r-89M\right)\left(ru^\phi\right)^4\right)
	\nonumber \\
& \quad +\left(u^r\right)^2 \left(\left(  4r-17M\right) - 8\left(31r-59M\right)\left(ru^\phi\right)^2-30\left(14r-29M\right)\left(ru^\phi\right)^4\right)
	\nonumber \\
&\quad	- 16r \left(u^r\right)^4	\left(1+15\left(ru^\phi\right)^2\right)
	\Bigg] {\Delta \tau}^5 + O \left( \Delta \tau ^6 \right) \Bigg\}\\
\label{eq:QLSF-SchwGen-phi}
&ma_{\rm QL}^{\phi} =
	\frac {3 q^2 M^2}{11200 r^{10}}  u^{\phi} \Bigg\{
	10r^2 \left[ 
	\left(\frac{r-2M}{r}\right) \left(7 + 20\left(ru^\phi\right)^2-13\left(ru^\phi\right)^4\right)
	\right. \nonumber \\
& \quad \left. +\left(u^r\right)^2 \left(5+8\left(ru^\phi\right)^2\right)
	\right]{\Delta \tau}^{4}
	+ u^r \Bigg[
	3\left(68r-141M\right) + 48\left(13r-27M\right)\left(ru^\phi\right)^2  \nonumber \\
&\quad  + 30\left(14r-29M\right)\left(ru^\phi\right)^4
 	+ 48 r \left(u^r\right)^2\left(3+5\left(ru^\phi\right)^2\right)
	\Bigg] {\Delta \tau}^{5}
	+ O \left( \Delta \tau ^6 \right) \Bigg\}\\
\label{eq:QLSF-SchwGen-t}
&ma_{\rm QL}^{t} =
	-\frac {3 q^2 M^2}{11200 r^{10}} \sqrt { \frac{r}{r-2M} \left( 1 + \left(ru^\phi\right)^2 + \frac{r}{r-2M} \left(u^r\right)^2\right)}\times  \nonumber \\
	&\quad \Bigg\{
	10 r^2 \left[ 
	\left(\frac{r-2M}{r}\right) \left(7\left(ru^\phi\right)^2+13\left(ru^\phi\right)^4\right)
	+\left(u^r\right)^2 \left(1+8\left(ru^\phi\right)^2\right)
	\right] {\Delta \tau}^4
	\nonumber \\
	&\quad
	 + u^r \left[ 
	\left(20r-49M\right)  - 8\left(17r-31M\right)\left(ru^\phi\right)^2 - 30\left(14r-29M\right)\left(ru^\phi\right)^4\right.
	\nonumber \\
	&
	\quad \left.
	- 16 r \left(u^r\right)^2\left(1+15\left(ru^\phi\right)^2\right)
	\right] {\Delta \tau}^5
	+ O \left( \Delta \tau ^6 \right) \Bigg\}\\
\label{eq:dmdtau-SchwGen}
& \frac{dm}{d\tau} =
	-\frac {3 q^2 M^2}{11200 r^{10}} \Bigg\{
	 5 r^2 \left[
	\left(\frac{r-2M}{r}\right) \left(5+28\left(ru^\phi\right)^2+26\left(ru^\phi\right)^4\right)
	\right.\nonumber \\
	&\left. \quad + ~4 \left(u^r\right)^2 \left(1+4\left(ru^\phi\right)^2\right)
	\right] {\Delta \tau}^4- 3 u^r \Bigg[ 
	\left(20r-41M\right) + 8\left(17r-35M\right)\left(ru^\phi\right)^2 
	\nonumber \\
	& \quad
	+10\left(14r-29M\right)\left(ru^\phi\right)^4
 	+ 16 r \left(u^r\right)^2\left(1+ 5\left(ru^\phi\right)^2\right)
	\Bigg]  {\Delta \tau}^5 + O \left( \Delta \tau ^6 \right) \Bigg\}
\end{align}
\end{subequations}
We would expect that, for motion in the equatorial plane, the $\theta$-component of the 4-acceleration is $0$. This is verified by our calculation: we find $ma_{\rm QL}^{\theta}$ is $0$ up to the order considered. We also find that the other three components, $ma^r_{\rm QL}$, $ma^\phi_{\rm QL}$ and $ma^t_{\rm QL}$ all have their leading order terms at $O\left( \Delta \tau^4 \right)$ and also have terms at order $O\left( \Delta \tau^5 \right)$.

We may also express this result in an alternative way by making use of the constants of motion \cite{Hartle} (discussed in Sec.~\ref{subsubsec:schw}):
\begin{subequations}
\begin{eqnarray}
e &=& \left( \frac{r - 2M}{r} \right) u^t\\
l &=& r^2 u^\phi
\end{eqnarray}
\end{subequations}
and the radial four-velocity,
\begin{eqnarray}
u^r &=& -\sqrt{e^2-1 - 2 V_{\rm{eff}}(r,e,l)}
\end{eqnarray}
where
\begin{equation}
 V_{\rm{eff}}(r,e,l) = -\frac{M}{r} + \frac{l^2}{2r^2} - \frac{Ml^2}{r^3}
\end{equation}
is an effective potential for the motion.

This gives a form for the equations of motion which will prove useful later (as a check on our result for Kerr spacetime):
\begin{subequations}
\allowdisplaybreaks
\begin{eqnarray}
\label{eq:ma-r-schw-el}
ma^r &=& {\frac {3 q^2 M^2}{11200{r}^{17}}}\Bigg[
	\, -\sqrt{e^2-1 - 2 V_{\rm{eff}}}\,( 20\,{l}^{2}{r}^{7}-40\,{l}^{2}M{r}^{6}+50\,{l}^{4}{r}^{5}-100\,{l}^{4}M{r}^{4}\nonumber \\
	& &\quad
	+10\,{e}^{2}{r}^{9}+80\,{e}^{2}{l}^{2}{r}^{7} ) {\Delta \tau}^{4}+\,( -60\,{l}^{2}{r}^{6}+255\,{l}^{2}M{r}^{5}-270\,{l}^{2}{M}^{2}{r}^{4}  \nonumber \\
	& & \quad
	-240\,{l}^{4}{r}^{4}+1008\,{l}^{4}M{r}^{3}-1056\,{l}^{4}{M}^{2}{r}^{2}-180\,{l}^{6}{r}^{2}+750\,{l}^{6}Mr\nonumber \\
	& & \quad
	 -780\,{l}^{6}{M}^{2}-36\,{e}^{2}{r}^{8}+81\,{e}^{2}M{r}^{7}-264\,{e}^{2}{l}^{2}{r}^{6}+552\,{e}^{2}{l}^{2}M{r}^{5}\nonumber \\
	& & \quad
	 -60\,{e}^{2}{l}^{4}{r}^{4}+90\,{e}^{2}{l}^{4}M{r}^{3}+16\,{e}^{4}{r}^{8}+240\,{e}^{4}{l}^{2}{r}^{6} ) {\Delta \tau}^{5} + O \left( \Delta \tau^6 \right)\Bigg]\\
ma^\phi &=& \frac {3 q^2 M^2 l}{11200{r}^{16}}\Bigg[
	( 20\,{r}^{6}-40\,M{r}^{5}+70\,{l}^{2}{r}^{4}-140\,{l}^{2}M{r}^{3}+50\,{l}^{4}{r}^{2}-100\,{l}^{4}Mr \nonumber \\
	& & \quad
	+50\,{e}^{2}{r}^{6}+80\,{e}^{2}{l}^{2}{r}^{4} ) {\Delta \tau}^{4}- \sqrt{e^2-1 - 2 V_{\rm{eff}}}\,( 60\,{r}^{5}-135\,M{r}^{4}+240\,{l}^{2}{r}^{3} \nonumber \\
	& & \quad
	-528\,{l}^{2}M{r}^{2}+180\,{l}^{4}r-390\,{l}^{4}M +144\,{e}^{2}{r}^{5}+240\,{e}^{2}{l}^{2}{r}^{3} ) {\Delta \tau}^{5}
	 \nonumber \\
	& & \quad
	 + O \left(\Delta \tau^6 \right)\Bigg]\\
ma^t &=& {\frac {3 q^2 M^2 e}{11200\left( r-2\,M \right) {r}^{13}}} \Bigg[
	( -10\,{r}^{6}+20\,M{r}^{5}-20\,{l}^{2}{r}^{4} \nonumber \\
	& & \quad+40\,{l}^{2}M{r}^{3}+50\,{l}^{4}{r}^{2}-100\,{l}^{4}Mr
	+10\,{e}^{2}{r}^{6}+80\,{e}^{2}{l}^{2}{r}^{4} ) {\Delta \tau}^{4} \nonumber \\
	& & \quad-\sqrt{e^2-1 - 2 V_{\rm{eff}}}\,( -36\,{r}^{5}+81\,M{r}^{4}-120\,{l}^{2}{r}^{3}
	+264\,{l}^{2}M{r}^{2} \nonumber \\
	& & \quad+180\,{l}^{4}r-390\,{l}^{4}M+16\,{e}^{2}{r}^{5}
	+ 240\,{e}^{2}{l}^{2}{r}^{3} ) {\Delta \tau}^{5}
	 + O \left(\Delta \tau^6 \right)\Bigg]\\
\label{eq:dmdtau-schw-el}
\frac{dm}{d\tau} &=&
	\frac {3 q^2 M^2}{11200{r}^{14}}  \big[
	( -5\,{r}^{6}+10\,M{r}^{5}-40\,{l}^{2}{r}^{4}+80\,{l}^{2}M{r}^{3}-50\,{l}^{4}{r}^{2}+100\,{l}^{4}Mr\nonumber \\
	& &\quad -20\,{e}^{2}{r}^{6}-80\,{e}^{2}{l}^{2}{r}^{4} ) {\Delta \tau}^{4}
	+ \sqrt{e^2-1 - 2 V_{\rm{eff}}} ( 12\,{r}^{5}-27\,M{r}^{4}+120\,{l}^{2}{r}^{3}\nonumber \\
	& &\quad
	-264\,{l}^{2}M{r}^{2}+180\,{l}^{4}r-390\,{l}^{4}M+48\,{e}
^{2}{r}^{5}+240\,{e}^{2}{l}^{2}{r}^{3} ) {\Delta \tau}^{5}\nonumber \\
	& &\qquad
	+ O \left( \Delta \tau^6\right) \big]
\end{eqnarray}
\end{subequations}

These expressions give us the quasilocal contribution to the equations of motion for a particle following any geodesic path in a Schwarzschild background. It is now possible to look at some specific particle paths. In particular, in order to check our results, we consider three cases for which results using other methods are already available in the literature \cite{Anderson:Wiseman:2005,Anderson:Eftekharzadeh:Hu:2006}\footnote{Ref. \cite{Anderson:Eftekharzadeh:Hu:2006} also considers the case of a static particle, but in that case the motion is non-geodesic, so we leave the analysis for Sec.~\ref{sec:qlsf-non-geodesic-examples}.}.
\begin{enumerate}
\item A particle following a circular geodesic
\item A particle under radial in-fall
\item A particle under radial in-fall from rest
\end{enumerate}

\subsubsection{Circular geodesic in Schwarzschild}
For a particle following a circular geodesic in Schwarzschild spacetime, we have the condition $u^r = 0$ along with the previous condition $u^\theta = 0$. Furthermore, for a circular geodesic orbit $u^\phi$ is uniquely determined by the radius, $r$, angle, $\theta$ and mass, $M$ of the Schwarzschild black hole:
\begin{equation}
\label{eq:SchwCirc-u-phi}
u^\phi = \frac{1}{r} \sqrt{\frac{M}{r-3M}}
\end{equation}
Substituting (\ref{eq:SchwCirc-u-phi}) into (\ref{eq:ut}), we also obtain the expression for $u^t$ in terms of the radius, $r$, and mass, $M$ of the Schwarzschild black hole:
\begin{equation}
\label{eq:SchwCirc-u-r}
u^t = \sqrt{\frac{r}{r-3M}}
\end{equation}

We can now use these values for the 4-velocity in Eqs. (\ref{eq:QLSF-SchwGen-r}) - (\ref{eq:dmdtau-SchwGen}). In doing so, the expressions simplify significantly and we obtain a result for the  quasilocal contribution to the equations of motion for a scalar particle following a circular geodesic in terms of $q$, $r$ and $M$ alone:
\begin{subequations}
\label{eq:schw-circ-cov}
\begin{eqnarray}
ma_{\rm QL}^{r} &=& - \frac{3 q^2 M^2 \left( r-2M \right) \left( 20 r^3 - 81 M r^2 + 54 r M^2 + 53 M^3 \right)}{11200 \left( r-3M \right)^2 r^{11}} \Delta \tau^5 \nonumber \\ & & \quad+ O \left( \Delta \tau ^6 \right) \\
ma_{\rm QL}^{\phi} &=& \frac{3 q^2 M^2 \left( r-2M \right)^2 \left( 7r-8M \right)}{1120 \left( r-3M \right)^2r^{10} } \sqrt{\frac{M}{r-3M}} \Delta \tau^4 + O \left( \Delta \tau ^6 \right) \\
ma_{\rm QL}^{t} &=& \frac{3 q^2 M^3 \left( r-2M \right) \left( 7r-8M \right)}{1120 \left( r-3M \right)^2 r^{9}} \sqrt{\frac{r}{r-3M}} \Delta \tau^4 + O \left( \Delta \tau ^6 \right) \\
\frac{dm}{d\tau} &=& \frac { 3 q^2 M^2 \left( r-2M \right) \left( 13M^2 + 2Mr-  5r^2 \right) }{2240 \left( r-3M \right) ^2 r^9} \Delta \tau^{4} + O \left( \Delta \tau ^6 \right) 
\end{eqnarray}
\end{subequations}

Our results exactly match Eqs. (\ref{eq:rCoordCirc}) - (\ref{eq:dmdtauCoordCirc}) containing our corrected version of the results of Anderson and Wiseman \cite{Anderson:Wiseman:2005}.

\subsubsection{Radial geodesic in Schwarzschild}
A particle following a radial geodesic in Schwarzschild spacetime has the conditions on the 4-velocity that $u^\phi = 0$ in addition to the previous condition $u^\theta = 0$. We can therefore write Eq. (\ref{eq:ut}) as:
\begin{equation}
\label{eq:SchwRad-u-t-implicit}
u^t = \sqrt{\frac{r}{r-2M}\left( 1+ \frac{r}{r-2M} \left({u^r}\right)^2 \right) }
\end{equation}
This leaves us with one independent component of the 4-velocity, $u^r$. 
Substituting for $u^\phi$ and $u^t$ into (\ref{eq:QLSF-SchwGen-r}) - (\ref{eq:dmdtau-SchwGen}), the expression simplifies significantly and we obtain a result which is only slightly more complicated than for the case of a particle following a circular geodesic. Our expression for the  quasilocal contribution to the equations of motion is now in terms of the scalar charge, $q$ radius, $r$, the mass, $M$ and the radial 4-velocity of the particle, $u^r$:
\begin{subequations}
\begin{eqnarray}
\label{eq:RadialQLSF-r}
ma_{\rm QL}^{r} &=&
	\frac{3 q^2 M^2 }{11200 r^{11}} \left(r-2M+r \left(u^r\right)^2 \right) \big[ 10 r^2 u^r \Delta \tau^4 \nonumber \\
&&\quad	+ \left(49M - 20r + 16r\left(u^r\right)^{2}\right)  \Delta \tau^5
	+ O \left( \Delta \tau ^6 \right) \big] \\
ma_{\rm QL}^{t} &=&
	\frac{-3 q^2 M^2 u^r }{11200 \left(r-2M\right) r^{9}} \sqrt{\frac{r-2M}{r}+\left({u^r}\right)^2} \big[
	10 r^2 u^r \Delta \tau^4 \nonumber \\
& & \quad + \left( 20r - 49M - 16r \left( u^r \right)^2 \right) \Delta \tau^5
	+ O \left( \Delta \tau ^6 \right)  \big] \\
\label{eq:Radialdmdtau}
\frac{dm}{d\tau} &=&
	\frac{-3 q^2 M^2}{11200 r^{10}} \big[ 5r \left(5r - 10M + 4 r \left(u^r\right)^2 \right) \Delta \tau^4 \nonumber \\
& & \quad
	+u^r \left(60r - 123M + 48r \left(u^r\right)^2\right)\Delta \tau^5
	+ O \left( \Delta \tau ^6 \right) \big]
\end{eqnarray}
\end{subequations}

From angular momentum considerations, we would expect a radially moving particle in a spherically symmetric background to experience no acceleration in the $\theta$ or $\phi$ directions. This is confirmed by our results up to the order considered.

\subsubsection{Radial geodesic in Schwarzschild: Starting from rest}
\label{sec:rad-rest}
In order to calculate the quasilocal contribution to the equations of motion for a particle under radial infall from rest, we begin with expressions (\ref{eq:RadialQLSF-r}) - (\ref{eq:Radialdmdtau}) for a particle under general radial infall. For the sake of simplicity, we will take $\tau=0$ to be the proper time at which the particle is released from rest at a point $r=r_0$. We also assume that at proper time $\tau = \Delta \tau$, the particle will be at the radial point $r$, travelling radially inwards with 4-velocity $u^r$. In this case, expressions (\ref{eq:RadialQLSF-r}) - (\ref{eq:Radialdmdtau}) hold. However, we must now obtain an expression for $u^r = u^r \left(\Delta\tau\right)$ satisfying the initial condition $u^r \left(0\right) = 0$.

We begin with an expression for the radial 4-velocity in radial geodesic motion starting from rest at $r_0$ \cite{Chandrasekhar}:
\begin{equation}
\label{eq:ur-rad-rest}
\left(u^r\right)^2 = \left(\frac{dr}{d\tau}\right)^2 = 2M \left( \frac{1}{r} - \frac{1}{r_0} \right)
\end{equation}

Assuming that the particle has not travelled too far (i.e. $(r-r_0)/r_0 \ll 1$), this equation may be approximately integrated with respect to $r$ from $r_0$ up to $r$, giving an approximate expression for $r_0$:

\begin{equation}
r_0 = r + \frac{M \Delta \tau^2}{2 r_0^2} + O \left(\Delta\tau^3\right) = r + \frac{M \Delta \tau^2}{2 r^2} + O \left(\Delta\tau^3\right) 
\end{equation}

Using this in (\ref{eq:ur-rad-rest}), we get an approximate expression for $u^r$ (we have taken the negative square root since we are considering infalling particles):
\begin{equation}
u^r = -\frac{M\Delta\tau}{r^2} + O \left( \Delta \tau ^2 \right) 
\end{equation}
We can then use this expression in (\ref{eq:RadialQLSF-r}) - (\ref{eq:Radialdmdtau}) to give the quasilocal contribution to the equations of motion for a particle falling radially inward from rest in Schwarzschild spacetime: 
\begin{subequations}
\begin{eqnarray}
\label{eq:RadialRestQLSF}
ma_{\rm QL}^{r} &=&
	- \frac {3 q^2 M^2}{11200 r^{10}}   \left( \frac{r-2M}{r} \right) \left( 20r-39M \right) {\Delta \tau}^5 + O \left( \Delta \tau ^6 \right) \\
ma_{\rm QL}^{t} &=& O \left( \Delta \tau ^6 \right) \\
\label{eq:RadialRestdmdtau}\frac{dm}{d\tau} &=&
	- \frac {3 q^2 M^2}{448 r^8} \left(\frac{r - 2M}{r} \right) {\Delta \tau}^4 + O \left( \Delta \tau ^6 \right) 
\end{eqnarray}
\end{subequations}
As explained in Section \ref{sec:coord-radial}, this result is in agreement with those of Ref. \cite{Anderson:Eftekharzadeh:Hu:2006}.

\subsection{Kerr Space-time}
\label{sec:cases-kerr}
Next, we look at the much more complicated but astrophysically more interesting Kerr spacetime. We work with the line-element in Boyer-Lindquist coordinates, given by Eq.~\eqref{eq:KerrMetric}. 
Although it is possible to use a computer algebra package to obtain a totally general solution for the self-force in Kerr spacetime, the length of the resulting expressions are too unwieldy to be of much benefit in printed form. For this reason, we will only focus here on equatorial motion, so that $\theta = \frac{\pi}{2}$, keeping in mind that more general results may be obtained using the same methodology.

From angular momentum considerations \cite{Hartle}, it is clear that motion in the equatorial plane in Kerr spacetime will always remain in the equatorial plane, i.e. $u^\theta = 0$. Additionally, the motion is parametrized by two quantities analogous to the Schwarzschild case: the conserved energy per unit mass, $e$ and the angular momentum per unit mass along the symmetry axis, $l$. We can therefore write the three non-zero components of the 4-velocity in terms of these conserved quantities \cite{Hartle}:

\begin{subequations}
\begin{eqnarray}
\label{eq:kerr-ur}
u^r = \frac{dr}{d\tau} &=& - \sqrt{e^2-1 - 2 V_{\rm{eff}}(r,e,l)}\\
\label{eq:kerr-uphi}
u^\phi = \frac{d\phi}{d\tau} &=& \frac{1}{\Delta}\left[\left(\frac{r-2M}{r}\right)l + \frac{2Ma}{r}e\right]\\
\label{eq:kerr-ut}
u^t = \frac{dt}{d\tau} &=& \frac{1}{\Delta} \left[\left(r^2 + a^2 + \frac{2Ma^2}{r}\right)e - \frac{2Ma}{r}l\right] 
\end{eqnarray}
\end{subequations}
where
\begin{equation}
V_{\rm{eff}}(r,e,l) = -\frac{M}{r} + \frac{l^2-a^2\left(e^2-1\right)}{2r^2}-\frac{M\left(l-ae\right)^2}{r^3}
\end{equation}
is the effective potential for equatorial motion.

Substituting these values into Eq. (\ref{eq:SimpForce}) and computing the relevant components of the Hadamard coefficients, we arrive at a lengthy but general set of expressions for the equations of motion for a particle following a geodesic in the equatorial plane in Kerr spacetime:
\begin{subequations}
\allowdisplaybreaks
\begin{eqnarray}
ma^r &=& 
	-{\frac {3 q^2 M^2}{11200 r^{14}}} \sqrt{e^2-1 - 2 V_{\rm{eff}}} \Bigg[
	( 20\,{l}^{2}{r}^{4}-40\,{l}^{2}M{r}^{3}+50\,{l}^{4}{r}^{2}-100\,{l}^{4}M{r}^{1} \nonumber \\
	& & \quad
	+10\,{e}^{2}{r}^{6}+80\,{e}^{2}{l}^{2}{r}^{4} )+ a( -120\,el{r}^{4}+80\,elM{r}^{3}-600\,e{l}^{3}{r}^{2} \nonumber \\
	& & \quad
	+400\,e{l}^{3}M{r}^{1}-160\,{e}^{3}l{r}^{4} )+ a^2 ( 150\,{l}^{2}{r}^{2}+500\,{l}^{4}+100\,{e}^{2}{r}^{4} \nonumber \\
	& & \quad
	-40\,{e}^{2}M{r}^{3}+1500\,{e}^{2}{l}^{2}{r}^{2}-600\,{e}^{2}{l}^{2}M{r}^{1}+80\,{e}^{4}{r}^{4} )+ a^3( -300\,el{r}^{2}\nonumber \\
	& & \quad
	-2000\,e{l}^{3}-1400\,{e}^{3}l{r}^{2}+400\,{e}^{3}lM{r}^{1} )+ a^4 ( 150\,{e}^{2}{r}^{2}+3000\,{e}^{2}{l}^{2}\nonumber \\
	& & \quad
	+450\,{e}^{4}{r}^{2}-100\,{e}^{4}M{r}^{1} )- a^5 ( 2000 \,{e}^{3}l )
	+ a^6 ( 500 \,{e}^{4} ) \Bigg]{\Delta \tau}^{4} \nonumber \\
	& & 
	-{\frac {3 q^2 M^2}{11200 r^{18}}}\Bigg[
	\big( 60\,{l}^{2}{r}^{7}-255\,{l}^{2}M{r}^{6}+270\,{l}^{2}{M}^{2}{r}^{5}+240\,{l}^{4}{r}^{5}-1008\,{l}^{4}M{r}^{4} \nonumber \\
	& & \quad
	+1056\,{l}^{4}{M}^{2}{r}^{3}+180\,{l}^{6}{r}^{3} -750\,{l}^{6}M{r}^{2}+780\,{l}^{6}{M}^{2}r+36\,{e}^{2}{r}^{9} \nonumber \\
	& & \quad
	-81\,{e}^{2}M{r}^{8}+264\,{e}^{2}{l}^{2}{r}^{7}-552\,{e}^{2}{l}^{2}M{r}^{6}+60\,{e}^{2}{l}^{4}{r}^{5}-90\,{e}^{2}{l}^{4}M{r}^{4}\nonumber \\
	& & \quad
	-16\,{e}^{4}{r}^{9}-240\,{e}^{4}{l}^{2}{r}^{7} \big)+ a \big( -432\,el{r}^{7}+1176\,elM{r}^{6}-540\,el{M}^{2}{r}^{5} \nonumber \\
	& & \quad
	-2760\,e{l}^{3}{r}^{5}+7728\,e{l}^{3}M{r}^{4}-4224\,e{l}^{3}{M}^{2}{r}^{3}-2160\,e{l}^{5}{r}^{3} \nonumber \\
	& & \quad
	+6600\,e{l}^{5}M{r}^{2}-4680\,e{l}^{5}{M}^{2}r-368\,{e}^{3}l{r}^{7}+1104\,{e}^{3}lM{r}^{6}+1680\,{e}^{3}{l}^{3}{r}^{5} \nonumber \\
	& & \quad
	+360\,{e}^{3}{l}^{3}M{r}^{4}+480\,{e}^{5}l{r}^{7} \big)+ a^2 \big( 660\,{l}^{2}{r}^{5}-1368\,{l}^{2}M{r}^{4}+3000\,{l}^{4}{r}^{3} \nonumber \\
	& & \quad
	-6120\,{l}^{4}M{r}^{2}+2100\,{l}^{6}r-4200\,{l}^{6}M+456\,{e}^{2}{r}^{7}-921\,{e}^{2}M{r}^{6} \nonumber \\
	& & \quad
	+270\,{e}^{2}{M}^{2}{r}^{5}+6720\,{e}^{2}{l}^{2}{r}^{5}-17136\,{e}^{2}{l}^{2}M{r}^{4}+6336\,{e}^{2}{l}^{2}{M}^{2}{r}^{3} \nonumber \\
	& & \quad
	+3120\,{e}^{2}{l}^{4}{r}^{3}-21750\,{e}^{2}{l}^{4}M{r}^{2}+11700\,{e}^{2}{l}^{4}{M}^{2}r+104\,{e}^{4}{r}^{7} \nonumber \\
	& & \quad
	-552\,{e}^{4}M{r}^{6}-5400\,{e}^{4}{l}^{2}{r}^{5}-540\,{e}^{4}{l}^{2}M{r}^{4}-240\,{e}^{6}{r}^{7} \big) \nonumber \\
	& & \quad
	+ a^3 \big( -1800\,el{r}^{5}+2736\,elM{r}^{4}-13600\,e{l}^{3}{r}^{3}+24480\,e{l}^{3}M{r}^{2} \nonumber \\
	& & \quad
	-8400\,e{l}^{5}r+25200\,e{l}^{5}M-6120\,{e}^{3}l{r}^{5}+15120\,{e}^{3}lM{r}^{4} \nonumber \\
	& & \quad
	-4224\,{e}^{3}l{M}^{2}{r}^{3}+5520\,{e}^{3}{l}^{3}{r}^{3}+36000\,{e}^{3}{l}^{3}M{r}^{2}-15600\,{e}^{3}{l}^{3}{M}^{2}r \nonumber \\
	& & \quad
	+5520\,{e}^{5}l{r}^{5}+360\,{e}^{5}lM{r}^{4} \big)+ a^4 \big( 720\,{l}^{2}{r}^{3}+2800\,{l}^{4}r+1140\,{e}^{2}{r}^{5} \nonumber \\
	& & \quad
	-1368\,{e}^{2}M{r}^{4}+22800\,{e}^{2}{l}^{2}{r}^{3}-36720\,{e}^{2}{l}^{2}M{r}^{2}+10500\,{e}^{2}{l}^{4}r \nonumber \\
	& & \quad
	-63000\,{e}^{2}{l}^{4}M+1920\,{e}^{4}{r}^{5}-4704\,{e}^{4}M{r}^{4}+1056\,{e}^{4}{M}^{2}{r}^{3} \nonumber \\
	& & \quad
	-16380\,{e}^{4}{l}^{2}{r}^{3}-32250\,{e}^{4}{l}^{2}M{r}^{2}+11700\,{e}^{4}{l}^{2}{M}^{2}r-1860\,{e}^{6}{r}^{5} \nonumber \\
	& & \quad
	-90\,{e}^{6}M{r}^{4} \big)+ a^5 \big( -1440\,el{r}^{3}-11200\,e{l}^{3}r-16800\,{e}^{3}l{r}^{3}  \nonumber \\
	& & \quad
	+24480\,{e}^{3}lM{r}^{2}+84000\,{e}^{3}{l}^{3}M+13440\,{e}^{5}l{r}^{3}+15000\,{e}^{5}lM{r}^{2} \nonumber \\
	& & \quad
	-4680\,{e}^{5}l{M}^{2}r \big)+ a^6 \big( 720\,{e}^{2}{r}^{3}+16800\,{e}^{2}{l}^{2}r+4600\,{e}^{4}{r}^{3} \nonumber \\
	& & \quad
	-6120\,{e}^{4}M{r}^{2}-10500\,{e}^{4}{l}^{2}r-63000\,{e}^{4}{l}^{2}M-3720\,{e}^{6}{r}^{3} \nonumber \\
	& & \quad
	-2850\,{e}^{6}M{r}^{2}+780\,{e}^{6}{M}^{2}r \big)+ a^7 ( -11200\,{e}^{3}lr+8400\,{e}^{5}lr \nonumber \\
	& & \quad
	+25200\,{e}^{5}lM )+ a^8 ( 2800\,{e}^{4}r-2100\,{e}^{6}r-4200\,{e}^{6}M )
	\Bigg] {\Delta \tau}^{5}\nonumber \\
	& & \quad
	 + O \left( \Delta \tau ^6 \right)\\
ma^\phi &=&
	-\frac{3 q^2 M^2}{11200 ( {r}^{2}-2\,Mr+{a}^{2} ) {r}^{16}} \Bigg[
	\big( -20\,l{r}^{8}+80\,lM{r}^{7}-80\,l{M}^{2}{r}^{6}-70\,{l}^{3}{r}^{6}\nonumber \\
	& & \quad
	+280\,{l}^{3}M{r}^{5}-280\,{l}^{3}{M}^{2}{r}^{4}-50\,{l}^{5}{r}^{4}+200\,{l}^{5}M{r}^{3}-200\,{l}^{5}{M}^{2}{r}^{2}\nonumber \\
	& & \quad
	-50\,{e}^{2}l{r}^{8}+100\,{e}^{2}lM{r}^{7}-80\,{e}^{2}{l}^{3}{r}^{6}+160\,{e}^{2}{l}^{3}M{r}^{5} \big)+ a \big( 60\,e{r}^{8}\nonumber \\
	& & \quad
	-160\,eM{r}^{7}+80\,e{M}^{2}{r}^{6}+570\,e{l}^{2}{r}^{6}-1560\,e{l}^{2}M{r}^{5}+840\,e{l}^{2}{M}^{2}{r}^{4}\nonumber \\
	& & \quad
	+600\,e{l}^{4}{r}^{4}-1700\,e{l}^{4}M{r}^{3}+1000\,e{l}^{4}{M}^{2}{r}^{2}+40\,{e}^{3}{r}^{8}-100\,{e}^{3}M{r}^{7}\nonumber \\
	& & \quad
	+160\,{e}^{3}{l}^{2}{r}^{6}-480\,{e}^{3}{l}^{2}M{r}^{5} \big)+ a^2 \big( -170\,l{r}^{6}+340\,lM{r}^{5}-700\,{l}^{3}{r}^{4} \nonumber \\
	& & \quad
	+1400\,{l}^{3}M{r}^{3}-500\,{l}^{5}{r}^{2}+1000\,{l}^{5}Mr-890\,{e}^{2}l{r}^{6}+2280\,{e}^{2}lM{r}^{5}\nonumber \\
	& & \quad
	-840\,{e}^{2}l{M}^{2}{r}^{4}-1500\,{e}^{2}{l}^{3}{r}^{4}+4800\,{e}^{2}{l}^{3}M{r}^{3}-2000\,{e}^{2}{l}^{3}{M}^{2}{r}^{2} \nonumber \\
	& & \quad
	-80\,{e}^{4}l{r}^{6}+480\,{e}^{4}lM{r}^{5} \big)+ a^3 \big( 210\,e{r}^{6}-340\,eM{r}^{5}+2250\,e{l}^{2}{r}^{4} \nonumber \\
	& & \quad
	-4200\,e{l}^{2}M{r}^{3}+2000\,e{l}^{4}{r}^{2}-5000\,e{l}^{4}Mr+390\,{e}^{3}{r}^{6}-1000\,{e}^{3}M{r}^{5}\nonumber \\
	& & \quad
	+280\,{e}^{3}{M}^{2}{r}^{4}+1400\,{e}^{3}{l}^{2}{r}^{4}-6200\,{e}^{3}{l}^{2}M{r}^{3}+2000\,{e}^{3}{l}^{2}{M}^{2}{r}^{2}\nonumber \\
	& & \quad
	-160\,{e}^{5}M{r}^{5} \big)+ a^4\big( -150\,l{r}^{4}-500\,{l}^{3}{r}^{2}-2400\,{e}^{2}l{r}^{4}+4200\,{e}^{2}lM{r}^{3}\nonumber \\
	& & \quad
	-3000\,{e}^{2}{l}^{3}{r}^{2}+10000\,{e}^{2}{l}^{3}Mr-450\,{e}^{4}l{r}^{4}+3800\,{e}^{4}lM{r}^{3}\nonumber \\
	& & \quad
	-1000\,{e}^{4}l{M}^{2}{r}^{2} \big)+ a^5 \big( 150\,e{r}^{4}+1500\,e{l}^{2}{r}^{2}+850\,{e}^{3}{r}^{4}-1400\,{e}^{3}M{r}^{3} \nonumber \\
	& & \quad
	+2000\,{e}^{3}{l}^{2}{r}^{2}-10000\,{e}^{3}{l}^{2}Mr-900\,{e}^{5}M{r}^{3}+200\,{e}^{5}{M}^{2}{r}^{2} \big) \nonumber \\
	& & \quad
	+ a^6 \big( -1500\,{e}^{2}l{r}^{2}-500\,{e}^{4}l{r}^{2}+5000\,{e}^{4}lMr \big) \nonumber \\
	& & \quad
	+ a^7 \big( 500\,{e}^{3}{r}^{2}-1000\,{e}^{5}Mr \big)
	\Bigg] {\Delta \tau}^{4}  \nonumber \\
	& & 
	-\frac {3 q^2 M^2}{11200( {r}^{2}-2\,Mr+{a}^{2} ) {r}^{16}} \sqrt{e^2-1 - 2 V_{\rm{eff}}} \Bigg[
	\big( 60\,l{r}^{7}-255\,lM{r}^{6}\nonumber \\
	& & \quad
	+270\,l{M}^{2}{r}^{5}+240\,{l}^{3}{r}^{5}-1008\,{l}^{3}M{r}^{4}+1056\,{l}^{3}{M}^{2}{r}^{3}\nonumber \\
	& & \quad
	+180\,{l}^{5}{r}^{3}-750\,{l}^{5}M{r}^{2}+780\,{l}^{5}{M}^{2}r+144\,{e}^{2}l{r}^{7}-288\,{e}^{2}lM{r}^{6}\nonumber \\
	& & \quad
	+240\,{e}^{2}{l}^{3}{r}^{5}-480\,{e}^{2}{l}^{3}M{r}^{4} \big)+a \big( -192\,e{r}^{7} \nonumber \\
	& & \quad 
	+528\,eM{r}^{6}-270\,e{M}^{2}{r}^{5}-2040\,e{l}^{2}{r}^{5}+5688\,e{l}^{2}M{r}^{4} \nonumber \\
	& & \quad 
	-3168\,e{l}^{2}{M}^{2}{r}^{3}-2160\,e{l}^{4}{r}^{3}+6240\,e{l}^{4}M{r}^{2}-3900\,e{l}^{4}{M}^{2}r \nonumber \\
	& & \quad
	-128\,{e}^{3}{r}^{7}+288\,{e}^{3}M{r}^{6}-480\,{e}^{3}{l}^{2}{r}^{5}+1440\,{e}^{3}{l}^{2}M{r}^{4} \big) \nonumber \\
	& & \quad
	+ a^2 \big( 660\,l{r}^{5}-1368\,lM{r}^{4}+3000\,{l}^{3}{r}^{3}-6120\,{l}^{3}M{r}^{2}+2100\,{l}^{5}r \nonumber \\
	& & \quad
	-4200\,{l}^{5}M+3360\,{e}^{2}l{r}^{5}-8352\,{e}^{2}lM{r}^{4}+3168\,{e}^{2}l{M}^{2}{r}^{3} \nonumber \\
	& & \quad
	+5400\,{e}^{2}{l}^{3}{r}^{3}-17460\,{e}^{2}{l}^{3}M{r}^{2}+7800\,{e}^{2}{l}^{3}{M}^{2}r \nonumber \\
	& & \quad
	+240\,{e}^{4}l{r}^{5}-1440\,{e}^{4}lM{r}^{4} \big)+ a^3 \big( -840\,e{r}^{5}+1368\,eM{r}^{4} \nonumber \\
	& & \quad
	-10080\,e{l}^{2}{r}^{3}+18360\,e{l}^{2}M{r}^{2}-8400\,e{l}^{4}r+21000\,e{l}^{4}M-1560\,{e}^{3}{r}^{5} \nonumber \\
	& & \quad
	+3672\,{e}^{3}M{r}^{4}-1056\,{e}^{3}{M}^{2}{r}^{3}-5040\,{e}^{3}{l}^{2}{r}^{3}+22440\,{e}^{3}{l}^{2}M{r}^{2} \nonumber \\
	& & \quad
	-7800\,{e}^{3}{l}^{2}{M}^{2}r+480\,{e}^{5}M{r}^{4} \big)+ a^4 \big( 720\,l{r}^{3}+2800\,{l}^{3}r+11160\,{e}^{2}l{r}^{3} \nonumber \\
	& & \quad
	-18360\,{e}^{2}lM{r}^{2}+12600\,{e}^{2}{l}^{3}r-42000\,{e}^{2}{l}^{3}M+1620\,{e}^{4}l{r}^{3} \nonumber \\
	& & \quad
	-13710\,{e}^{4}lM{r}^{2}+3900\,{e}^{4}l{M}^{2}r \big)+ a^5 \big( -720\,e{r}^{3}-8400\,e{l}^{2}r\nonumber \\
	& & \quad
	-4080\,{e}^{3}{r}^{3}+6120\,{e}^{3}M{r}^{2}-8400\,{e}^{3}{l}^{2}r+42000\,{e}^{3}{l}^{2}M+3240\,{e}^{5}M{r}^{2} \nonumber \\
	& & \quad
	-780\,{e}^{5}{M}^{2}r \big)+ a^6 \big( 8400\,{e}^{2}lr+2100\,{e}^{4}lr-21000\,{e}^{4}lM \big)\nonumber \\
	& & \quad
	+ a^7 \big( -2800\,{e}^{3}r+4200\,{e}^{5}M \big) \Bigg] {\Delta \tau}^{5} + O \left( \Delta \tau ^6 \right)\\
ma^t &=&
	-{\frac {3 q^2 M^2}{11200 ( {r}^{2}-2\,Mr+{a}^{2} ) {r}^{16}}} \Bigg[
	\big( 10\,e{r}^{10}-20\,eM{r}^{9}+20\,e{l}^{2}{r}^{8}-40\,e{l}^{2}M{r}^{7}\nonumber \\
	& & \quad
	-50\,e{l}^{4}{r}^{6}+100\,e{l}^{4}M{r}^{5}-10\,{e}^{3}{r}^{10}-80\,{e}^{3}{l}^{2}{r}^{8} \big) \nonumber \\
	& & \quad
	+ a \big( -60\,l{r}^{8}+160\,lM{r}^{7}-80\,l{M}^{2}{r}^{6}-150\,{l}^{3}{r}^{6}+440\,{l}^{3}M{r}^{5}\nonumber \\
	& & \quad
	-280\,{l}^{3}{M}^{2}{r}^{4}+100\,{l}^{5}M{r}^{3}-200\,{l}^{5}{M}^{2}{r}^{2}+180\,{e}^{2}lM{r}^{7}\nonumber \\
	& & \quad
	+600\,{e}^{2}{l}^{3}{r}^{6}-240\,{e}^{2}{l}^{3}M{r}^{5}+160\,{e}^{4}l{r}^{8} \big)+ a^2 \big( 110\,e{r}^{8} \nonumber \\
	& & \quad
	-240\,eM{r}^{7}+80\,e{M}^{2}{r}^{6}+620\,e{l}^{2}{r}^{6}-2040\,e{l}^{2}M{r}^{5}+840\,e{l}^{2}{M}^{2}{r}^{4} \nonumber \\
	& & \quad
	-550\,e{l}^{4}{r}^{4}-1200\,e{l}^{4}M{r}^{3}+1000\,e{l}^{4}{M}^{2}{r}^{2}-30\,{e}^{3}{r}^{8}\nonumber \\
	& & \quad
	-140\,{e}^{3}M{r}^{7}-1580\,{e}^{3}{l}^{2}{r}^{6}+120\,{e}^{3}{l}^{2}M{r}^{5}-80\,{e}^{5}{r}^{8} \big)\nonumber \\
	& & \quad
	+ a^3 \big( -210\,l{r}^{6}+340\,lM{r}^{5}-650\,{l}^{3}{r}^{4}+1400\,{l}^{3}M{r}^{3}+1000\,{l}^{5}Mr\nonumber \\
	& & \quad
	-750\,{e}^{2}l{r}^{6}+2760\,{e}^{2}lM{r}^{5}-840\,{e}^{2}l{M}^{2}{r}^{4}+2600\,{e}^{2}{l}^{3}{r}^{4}\nonumber \\
	& & \quad
	+3800\,{e}^{2}{l}^{3}M{r}^{3}-2000\,{e}^{2}{l}^{3}{M}^{2}{r}^{2}+1560\,{e}^{4}l{r}^{6}+80\,{e}^{4}lM{r}^{5} \big) \nonumber \\
	& & \quad
	+ a^4 \big( 250\,e{r}^{6}-340\,eM{r}^{5}+2100\,e{l}^{2}{r}^{4}-4200\,e{l}^{2}M{r}^{3}-500\,e{l}^{4}{r}^{2}\nonumber \\
	& & \quad
	-5000\,e{l}^{4}Mr+280\,{e}^{3}{r}^{6}-1160\,{e}^{3}M{r}^{5}+280\,{e}^{3}{M}^{2}{r}^{4}-4500\,{e}^{3}{l}^{2}{r}^{4}\nonumber \\
	& & \quad
	-5200\,{e}^{3}{l}^{2}M{r}^{3}+2000\,{e}^{3}{l}^{2}{M}^{2}{r}^{2}-530\,{e}^{5}{r}^{6}-60\,{e}^{5}M{r}^{5} \big) \nonumber \\
	& & \quad
	+ a^5 \big( -150\,l{r}^{4}-500\,{l}^{3}{r}^{2}-2250\,{e}^{2}l{r}^{4}+4200\,{e}^{2}lM{r}^{3}+2000\,{e}^{2}{l}^{3}{r}^{2}\nonumber \\
	& & \quad
	+10000\,{e}^{2}{l}^{3}Mr+3400\,{e}^{4}l{r}^{4}+3300\,{e}^{4}lM{r}^{3}-1000\,{e}^{4}l{M}^{2}{r}^{2} \big) \nonumber \\
	& & \quad
	+ a^6 \big( 150\,e{r}^{4}+1500\,e{l}^{2}{r}^{2}+800\,{e}^{3}{r}^{4}-1400\,{e}^{3}M{r}^{3}-3000\,{e}^{3}{l}^{2}{r}^{2}\nonumber \\
	& & \quad
	-10000\,{e}^{3}{l}^{2}Mr-950\,{e}^{5}{r}^{4}-800\,{e}^{5}M{r}^{3}+200\,{e}^{5}{M}^{2}{r}^{2} \big) \nonumber \\
	& & \quad
	+ a^7 \big( -1500\,{e}^{2}l{r}^{2}+2000\,{e}^{4}l{r}^{2}+5000\,{e}^{4}lMr \big)+ a^8 \big( 500\,{e}^{3}{r}^{2} \nonumber \\
	& & \quad
	-500\,{e}^{5}{r}^{2}-1000\,{e}^{5}Mr \big)
	\Bigg] {\Delta \tau}^{4} \nonumber \\
	& &
	-\frac {3 q^2 M^2}{11200( {r}^{2}-2\,Mr+{a}^{2} ) {r}^{16}}  \sqrt{e^2-1 - 2 V_{\rm{eff}}} \Bigg[
	\big( -36\,e{r}^{9}+81\,eM{r}^{8} \nonumber \\
	& & \quad
	-120\,e{l}^{2}{r}^{7}+264\,e{l}^{2}M{r}^{6}+180\,e{l}^{4}{r}^{5}-390\,e{l}^{4}M{r}^{4}+16\,{e}^{3}{r}^{9}\nonumber \\
	& & \quad
	+240\,{e}^{3}{l}^{2}{r}^{7} \big)+ \big( 240\,l{r}^{7}-648\,lM{r}^{6}+270\,l{M}^{2}{r}^{5} \nonumber \\
	& & \quad
	+720\,{l}^{3}{r}^{5}-2040\,{l}^{3}M{r}^{4}+1056\,{l}^{3}{M}^{2}{r}^{3}-360\,{l}^{5}M{r}^{2}+780\,{l}^{5}{M}^{2}r\nonumber \\
	& & \quad
	+240\,{e}^{2}l{r}^{7}-816\,{e}^{2}lM{r}^{6}-2160\,{e}^{2}{l}^{3}{r}^{5}+1080\,{e}^{2}{l}^{3}M{r}^{4}-480\,{e}^{4}l{r}^{7} \big) a \nonumber \\
	& & \quad
	+ \big( -456\,e{r}^{7}+921\,eM{r}^{6}-270\,e{M}^{2}{r}^{5}-3360\,e{l}^{2}{r}^{5}+8784\,e{l}^{2}M{r}^{4} \nonumber \\
	& & \quad
	-3168\,e{l}^{2}{M}^{2}{r}^{3}+2280\,e{l}^{4}{r}^{3}+4290\,e{l}^{4}M{r}^{2}-3900\,e{l}^{4}{M}^{2}r-104\,{e}^{3}{r}^{7}\nonumber \\
	& & \quad
	+552\,{e}^{3}M{r}^{6}+5640\,{e}^{3}{l}^{2}{r}^{5}-900\,{e}^{3}{l}^{2}M{r}^{4}+240\,{e}^{5}{r}^{7} \big) {a}^{2} \nonumber \\
	& & \quad
	+ \big( 960\,l{r}^{5}-1368\,lM{r}^{4}+3520\,{l}^{3}{r}^{3}-6120\,{l}^{3}M{r}^{2}-4200\,{l}^{5}M\nonumber \\
	& & \quad
	+4560\,{e}^{2}l{r}^{5}-11448\,{e}^{2}lM{r}^{4}+3168\,{e}^{2}l{M}^{2}{r}^{3}-10560\,{e}^{2}{l}^{3}{r}^{3}\nonumber \\
	& & \quad
	-13560\,{e}^{2}{l}^{3}M{r}^{2}+7800\,{e}^{2}{l}^{3}{M}^{2}r-5520\,{e}^{4}l{r}^{5}+120\,{e}^{4}lM{r}^{4} \big) {a}^{3}\nonumber \\
	& & \quad
	+ \big( -1140\,e{r}^{5}+1368\,eM{r}^{4}-11640\,e{l}^{2}{r}^{3}+18360\,e{l}^{2}M{r}^{2}+2100\,e{l}^{4}r\nonumber \\
	& & \quad
	+21000\,e{l}^{4}M-1920\,{e}^{3}{r}^{5}+4704\,{e}^{3}M{r}^{4}-1056\,{e}^{3}{M}^{2}{r}^{3} \nonumber \\
	& & \quad
	+18000\,{e}^{3}{l}^{2}{r}^{3}+18540\,{e}^{3}{l}^{2}M{r}^{2}-7800\,{e}^{3}{l}^{2}{M}^{2}r+1860\,{e}^{5}{r}^{5} \nonumber \\
	& & \quad
	+90\,{e}^{5}M{r}^{4} \big) {a}^{4}+ \big( 720\,l{r}^{3}+2800\,{l}^{3}r+12720\,{e}^{2}l{r}^{3} \nonumber \\
	& & \quad
	-18360\,{e}^{2}lM{r}^{2}-8400\,{e}^{2}{l}^{3}r-42000\,{e}^{2}{l}^{3}M-13440\,{e}^{4}l{r}^{3} \nonumber \\
	& & \quad
	-11760\,{e}^{4}lM{r}^{2}+3900\,{e}^{4}l{M}^{2}r \big) {a}^{5}+ \big( -720\,e{r}^{3}-8400\,e{l}^{2}r \nonumber \\
	& & \quad
	-4600\,{e}^{3}{r}^{3}+6120\,{e}^{3}M{r}^{2}+12600\,{e}^{3}{l}^{2}r+42000\,{e}^{3}{l}^{2}M+3720\,{e}^{5}{r}^{3}\nonumber \\
	& & \quad
	+2850\,{e}^{5}M{r}^{2}-780\,{e}^{5}{M}^{2}r \big) {a}^{6}+ \big( 8400\,{e}^{2}lr-8400\,{e}^{4}lr \nonumber \\
	& & \quad
	-21000\,{e}^{4}lM \big) {a}^{7}+ \big( -2800\,{e}^{3}r+2100\,{e}^{5}r+4200\,{e}^{5}M \big) {a}^{8}
	\Bigg] {\Delta \tau}^{5} \nonumber \\
	& & \quad
	 + O \left( \Delta \tau ^6 \right)\\
\frac{dm}{d\tau} &=& \frac {3 q^2 M^2}{11200{r}^{15}} \Bigg[
	( -5\,{r}^{7}+10\,M{r}^{6}-40\,{l}^{2}{r}^{5}+80\,{l}^{2}M{r}^{4}-50\,{l}^{4}{r}^{3}+100\,{l}^{4}M{r}^{2} \nonumber \\
	& & \quad
	-20\,{e}^{2}{r}^{7}-80\,{e}^{2}{l}^{2}{r}^{5} )+ a ( 240\,el{r}^{5}-160\,elM{r}^{4}+600\,e{l}^{3}{r}^{3}\nonumber \\
	& & \quad
	-400\,e{l}^{3}M{r}^{2}+160\,{e}^{3}l{r}^{5} )+ a^2 ( -20\,{r}^{5}-300\,{l}^{2}{r}^{3}-500\,{l}^{4}r \nonumber \\
	& & \quad
	-200\,{e}^{2}{r}^{5}+80\,{e}^{2}M{r}^{4}-1500\,{e}^{2}{l}^{2}{r}^{3}+600\,{e}^{2}{l}^{2}M{r}^{2} \nonumber \\
	& & \quad
	-80\,{e}^{4}{r}^{5} )+ a^3 ( 600\,el{r}^{3}+2000\,e{l}^{3}r+1400\,{e}^{3}l{r}^{3}-400\,{e}^{3}lM{r}^{2} ) \nonumber \\
	& & \quad
	+ a^4 ( -300\,{e}^{2}{r}^{3}-3000\,{e}^{2}{l}^{2}r-450\,{e}^{4}{r}^{3}+100\,{e}^{4}M{r}^{2} ) \nonumber \\
	& & \quad
	+ a^5 (2000\,{e}^{3}rl ) - a^6 ( 500 \,{e}^{4}r ) \Bigg] {\Delta \tau}^{4} \nonumber \\
	& &
	+ \frac {3 q^2 M^2}{11200{r}^{15}} \sqrt{e^2-1 - 2 V_{\rm{eff}}} \Bigg[
	\big( 12\,{r}^{6}-27\,M{r}^{5}+120\,{l}^{2}{r}^{4}-264\,{l}^{2}M{r}^{3}\nonumber \\
	& & \quad
	+180\,{l}^{4}{r}^{2}-390\,{l}^{4}Mr+48\,{e}^{2}{r}^{6}+240\,{e}^{2}{l}^{2}{r}^{4} \big)\nonumber \\
	& & \quad
	+ a \big( -720\,el{r}^{4}+528\,elM{r}^{3}-2160\,e{l}^{3}{r}^{2}+1560\,e{l}^{3}Mr-480\,{e}^{3}l{r}^{4} \big) \nonumber \\
	& & \quad
	+ a^2  \big(60\,{r}^{4}+1080\,{l}^{2}{r}^{2}+2100\,{l}^{4}+600\,{e}^{2}{r}^{4}-264\,{e}^{2}M{r}^{3}\nonumber \\
	& & \quad
	+5400\,{e}^{2}{l}^{2}{r}^{2}-2340\,{e}^{2}{l}^{2}Mr+240\,{e}^{4}{r}^{4} \big) \nonumber \\
	& & \quad
	+ a^3 \big( -2160\,el{r}^{2}-8400\,e{l}^{3}-5040\,{e}^{3}l{r}^{2}+1560\,{e}^{3}lMr \big)\nonumber \\
	& & \quad
	+ a^4 \big( 1080\,{e}^{2}{r}^{2}+12600\,{e}^{2}{l}^{2}+1620\,{e}^{4}{r}^{2}-390\,{e}^{4}Mr \big) \nonumber \\
	& & \quad
	- a^5 \big( 8400\,{e}^{3}l\big) + a^6 \big( 2100\,{e}^{4} \big)
	\Bigg] {\Delta \tau}^{5} + O \left( \Delta \tau ^6 \right)
\end{eqnarray}
\end{subequations}

On physical grounds, we expect the $\theta$ component of the four-acceleration to be $0$. This expectation is verified at all orders considered here.
Furtherore, it is straightforward to see that these results exactly match Eqs. (\ref{eq:ma-r-schw-el}) - (\ref{eq:dmdtau-schw-el}) in the limit $a \rightarrow 0$.

\subsubsection{Release from instantaneous rest}
\label{sec:kerr-equ-rest}
We now consider the motion of a particle which is released from rest relative to an observer at spatial infinity. This is analogous to the case of radial infall from rest in Schwarzschild (Section \ref{sec:rad-rest}) and we proceed with the calculation in exactly the same way. Imposing the initial conditions $u^r(r=r_0)=0$ and $u^\phi(r=r_0)=0$, Eqs. (\ref{eq:kerr-ur}) and (\ref{eq:kerr-uphi}) can be solved for the constants of motion:

\begin{subequations}
\begin{eqnarray}
\label{eq:kerr-e}
 e &=& \sqrt{\frac{r_0 -2 M}{r_0}}\\
\label{eq:kerr-l}
 l &=& -\frac{2Ma}{r_0}\sqrt{\frac{r_0}{r_0-2M}}
\end{eqnarray}
\end{subequations}

We then substitute these into Eq. (\ref{eq:kerr-ur}) and integrate to get an approximate expression for $r_0$ in terms of $r$:

\begin{equation}
 r_0 = r + \left(M r_0^2 - \left(l^2 - a^2 \left(e^2-1\right)\right)r_0 + 3M\left(l-ae\right)^2\right) \frac{\Delta \tau^2}{2r_0^4}
\end{equation}

This, along with the equations for $e$ and $l$ then allow us to express the 4-velocity as:
\begin{subequations}
\begin{eqnarray}
 u^r &=& -\left(\frac{r^2-2rM+a^2}{r^3\left(r-2M\right)}\right) M \Delta \tau\\
 u^\theta &=& 0\\
 u^\phi &=& \frac{1}{\Delta}\left[ -\left(\frac{r-2M}{r}\right)\frac{2Ma}{r}\sqrt{\frac{r}{r-2M}} + \frac{2Ma}{r}\sqrt{\frac{r-2 M}{r}}\right]\\
u^t &=& \frac{1}{\Delta} \left[\left(r^2 + a^2 + \frac{2Ma^2}{r}\right)\sqrt{\frac{r -2 M}{r}} + \frac{4M^2a^2}{r^2}\sqrt{\frac{r}{r-2M}}\right] 
\end{eqnarray}
\end{subequations}

We then substitute these values along with the components of the Hadamard coefficients into Eq.~(\ref{eq:SimpForce}) and project orthogonal and tangent to the $4$-velocity to get the equations of motion:

\begin{subequations}
\allowdisplaybreaks
\begin{eqnarray}
ma^r &=& -{\frac {3 {q}^{2}{M}^{2}}{11200 ( {r}^{2}-2\,Mr+{a}^{2} ) ^{3} ( 2\,M-r ) ^{3}{r}^{23}}}\, \Bigg[\nonumber \\
	& & \quad
	- ( 39\,M-20\,r ) ( 2\,M-r ) ^{7}{r}^{15}-4 a^2 \, ( 252\,{M}^{3}-160\,r{M}^{2} \nonumber \\
	& & \quad
	-162\,{r}^{2}M+95\,{r}^{3} )  ( 2\,M-r ) ^{6}{r}^{12}- 12 a^4 \, ( 848\,{M}^{5}-640\,r{M}^{4} \nonumber \\
	& & \quad
	-1404\,{r}^{2}{M}^{3}+960\,{r}^{3}{M}^{2}+291\,{r}^{4}M-185\,{r}^{5} )  ( 2\,M-r ) ^{5}{r}^{9} \nonumber \\
	& & \quad
	-4 a^6 \, \big( 12288\,{M}^{7}-10240\,r{M}^{6}-42720\,{r}^{2}{M}^{5}+32640\,{r}^{3}{M}^{4} \nonumber \\
	& & \quad
	+20844\,{r}^{4}{M}^{3}-14880\,{r}^{5}{M}^{2}-2292\,{r}^{6}M+1545\,{r}^{7} \big)  ( 2\,M-r ) ^{4}{r}^{6} \nonumber \\
	& & \quad
	- a^8 \big( 98304\,{M}^{9}-81920\,r{M}^{8}-811008\,{r}^{2}{M}^{7}+655360\,{r}^{3}{M}^{6}\nonumber \\
	& & \quad
	+758592\,{r}^{4}{M}^{5}-583680\,{r}^{5}{M}^{4}-187920\,{r}^{6}{M}^{3}+138240\,{r}^{7}{M}^{2}\nonumber \\
	& & \quad
	+13347\,{r}^{8}M-9420\,{r}^{9} \big)  ( 2\,M-r ) ^{3}{r}^{3}+12 a^{10} \, ( 8\,{M}^{2}-{r}^{2} )  \big( 15872\,{M}^{7} \nonumber \\
	& & \quad
	-12800\,r{M}^{6}-30272\,{r}^{2}{M}^{5}+24000\,{r}^{3}{M}^{4}+10704\,{r}^{4}{M}^{3}-8200\,{r}^{5}{M}^{2}\nonumber \\
	& & \quad
	-921\,{r}^{6}M+675\,{r}^{7} \big) ( 2\,M-r ) ^{2}{r}^{2}-10 a^{12} \, ( 2\,M-r ) ( 7296\,{M}^{5} \nonumber \\
	& & \quad
	-5760\,r{M}^{4}-4704\,{r}^{2}{M}^{3}+3680\,{r}^{3}{M}^{2}+489\,{r}^{4}M-370\,{r}^{5} )  ( 8\,{M}^{2} \nonumber \\
	& & \quad
	-{r}^{2} ) ^{2}r+100 a^{14} \, ( 9\,M-7\,r )  ( 8\,{M}^{2}-{r}^{2} ) ^{4}
	 \Bigg] {\Delta \tau}^{5} + O\left(\Delta \tau^6 \right)\\
ma^\phi &=&{\frac {3{q}^{2}{M}^{2}a}{112 ( -2\,Mr+{r}^{2}+{a}^{2} ) ^{3} ( 2\,M-r ) ^{3/2}{r}^{{37/2}} }}\, \Bigg[
	{r}^{8} ( 2\,M-r ) ^{4}\nonumber \\
	& & \quad
	+a^2 {r}^{5} ( 16\,{M}^{2}-7\,{r}^{2} )  ( 2\,M-r ) ^{3}\nonumber \\
	& & \quad
	+16\,a^4 {r}^{2} ( 4\,{M}^{4}-6\,{M}^{2}{r}^{2}+{r}^{4} ) ( 2\,M-r ) ^{2}\nonumber \\
	& & \quad
	-5\,a^6 r ( 2\,M-r )  ( -3\,{r}^{2}+8\,{M}^{2} )  ( 8\,{M}^{2}-{r}^{2} )\nonumber \\
	& & \quad
	+5\, a^8 ( 8\,{M}^{2}-{r}^{2} ) ^{2}
	\Bigg] {\Delta \tau}^{4}+ O\left(\Delta \tau^6 \right)\\
ma^t &=&{\frac {3{q}^{2}{M}^{3}{a}^{2}}{56( {r}^{2}-2\,Mr+{a}^{2} ) ^{5} ( 2\,M-r ) ^{5/2}{r}^{{43/2}}}}\, 
	( 2\,M{r}^{3}-{r}^{4}+ ( 8\,{M}^{2}-{r}^{2} ) {a}^{2} )\nonumber \\
	& & \quad
	  \Bigg[
	-{r}^{12} ( 2\,M-r ) ^{7}-a^2 {r}^{9}( 40\,{M}^{3}-16\,r{M}^{2}-30\,{r}^{2}M+9\,{r}^{3} ) ( 2\,M-r ) ^{5}\nonumber \\
	& & \quad
	-a^4 {r}^{6} ( 256\,{M}^{5}-64\,r{M}^{4}-560\,{r}^{2}{M}^{3}+128\,{r}^{3}{M}^{2}+138\,{r}^{4}M\nonumber \\
	& & \quad
	-31\,{r}^{5} )  ( 2\,M-r ) ^{4}-2\,a^6 {r}^{3} ( 256\,{M}^{7}-1664\,{M}^{5}{r}^{2}+224\,{M}^{4}{r}^{3}\nonumber \\
	& & \quad
	+1100\,{M}^{3}{r}^{4}-184\,{M}^{2}{r}^{5}-147\,M{r}^{6}+27\,{r}^{7} )  ( 2\,M-r ) ^{3}\nonumber \\
	& & \quad
	+a^8 {r}^{2} ( 6144\,{M}^{7}-10752\,{M}^{5}{r}^{2}+1024\,{M}^{4}{r}^{3}+3680\,{M}^{3}{r}^{4}\nonumber \\
	& & \quad
	-496\,{M}^{2}{r}^{5}-324\,M{r}^{6}+51\,{r}^{7} )  ( 2\,M-r ) ^{2}- 5\,a^{10}\,r ( 2\,M-r )  ( -5\,{r}^{5}\nonumber \\
	& & \quad
	+36\,{r}^{4}M+24\,{r}^{3}{M}^{2}-272\,{r}^{2}{M}^{3}+384\,{M}^{5} )  ( 8\,{M}^{2}-{r}^{2} )\nonumber \\
	& & \quad
	+5 a^{12}\, ( 32\,{M}^{3}-8\,{r}^{2}M+{r}^{3} )  ( 8\,{M}^{2}-{r}^{2} ) ^{2}
	 \Bigg] {\Delta \tau}^{4} + O\left(\Delta \tau^6 \right)\\
\frac{dm}{d\tau} &=& -{\frac {3 {q}^{2}{M}^{2}}{448{r}^{20} ( {r}^{2}-2\,Mr+{a}^{2} ) ^{4} ( 2\,M-r )^{2}}}\, \Bigg[
	{r}^{15} ( r-2\,M ) ^{7}\nonumber \\
	& & \quad
	+16\, a^2 {r}^{12} ( {r}^{2}-2\,{M}^{2} )  ( r-2\,M ) ^{6}\nonumber \\
	& & \quad
	+12\, a^4 {r}^{9} ( 7\,{r}^{4}-40\,{r}^{2}{M}^{2}+32\,{M}^{4} )  ( r-2\,M ) ^{5}\nonumber \\
	& & \quad
	+8\, a^6 {r}^{6} ( 27\,{r}^{6}-276\,{M}^{2}{r}^{4}+672\,{M}^{4}{r}^{2}-256\,{M}^{6} )  ( r-2\,M ) ^{4}\nonumber \\
	& & \quad
	+ a^8 {r}^{3} ( {r}^{2}-8\,{M}^{2} )  ( -512\,{M}^{6}+3264\,{M}^{4}{r}^{2}-2232\,{M}^{2}{r}^{4}\nonumber \\
	& & \quad
	+309\,{r}^{6} )  ( r-2\,M ) ^{3}+12\, a^{10} {r}^{2} ( 21\,{r}^{4}-96\,{r}^{2}{M}^{2}\nonumber \\
	& & \quad
	+64\,{M}^{4} ) ( r-2\,M ) ^{2} ( {r}^{2}-8\,{M}^{2} ) ^{2}+10\, a^{12} r ( r-2\,M )  ( -24\,{M}^{2}\nonumber \\
	& & \quad
	+11\,{r}^{2} )  ( {r}^{2}-8\,{M}^{2} ) ^{3}+20\, a^{14} ( {r}^{2}-8\,{M}^{2} ) ^{4}
	\Bigg] {\Delta \tau}^{4}\nonumber \\
	& & \quad
	+ O\left(\Delta \tau^6 \right) 
\end{eqnarray}
\end{subequations}
Again, it is straightforward to see that these results exactly match Eqs. (\ref{eq:RadialRestQLSF}) - (\ref{eq:RadialRestdmdtau}) in the limit $a \rightarrow 0$.

\subsection{Coordinate Calculation}
\label{sec:coord}
In the previous subsections, a covariant expansion approach was used to calculate the quasilocal contribution to the self-force. An alternative approach to the same calculation is to work with coordinate, rather than covariant expansions. To do so, we follow the method prescribed by Anderson and Wiseman \cite{Anderson:Wiseman:2005} and begin with the same expression we had for the covariant calculation, \eqref{eq:qlsf}.

Next, instead of using a covariant expansion for $V \left( x,x' \right)$ as we did previously, we now use the coordinate expansion calculated in Chapter~\ref{ch:coordex}:
\begin{equation}
\label{eq:WKBGreen}
V\left( x,x' \right) = \sum_{i,j,k=0}^{\infty} v_{ijk} \left( t - t' \right)^{2i} \left( \cos \left( \gamma \right) - 1 \right)^j \left(r - r'\right)^k 
\end{equation}
We recall that the coordinate $\gamma$ here is the angle on the 2-sphere between $x$ and $x'$. For the spherically symmetric Schwarzschild spacetime, we can set $\theta=\frac{\pi}{2}$ without loss of generality, so that $\gamma=\phi$. We next take the partial derivative with respect to the coordinates $r$, $\phi$, and $t$ and convert from those coordinates to proper time $\tau$. Next, we integrate over $\tau$ as we did previously in the covariant calculation and project orthogonal and tangent to the 4-velocity to obtain a final expression for the quasilocal contribution to the equations of motion. The coordinate transformation is dependent on the particle's path so it is in this way that the motion of the particle affects the self-force and equations of motion. We now work through this calculation for two of the particle paths investigated previously in the covariant calculation.

\subsubsection{Circular Geodesic}
\label{sec:coord-circ}
For a circular geodesic, the geodesic equations can be integrated to give
\begin{equation}
	r-r'=0,~ ~ ~\theta - \theta'=0,~ ~ ~\phi - \phi'=\frac{1}{r}\sqrt{\frac{M}{r-3M}}({\tau}-{\tau}'),~ ~ ~
	t-t'=\sqrt{\frac{r}{r-3M}}({\tau}-{\tau}')
\end{equation}
We now substitute Eq.~(\ref{eq:WKBGreen}) into (\ref{eq:qlsf}), take the partial derivative, then use the above to express the coordinate separations in terms of proper time separations. Finally, we do the easy integration over $\tau'$ and project orthogonal and parallel to the 4-velocity to obtain an expression for the quasilocal contribution to the equations of motion for a particle following a circular geodesic in Schwarzschild spacetime. The radial component of the quasilocal mass times 4-acceleration is
\begin{equation}
\label{eq:rCoordCirc}
ma_{\rm QL}^{r} = ma_{\rm QL}^{r} \left[ 5 \right] {\Delta}{\tau}^5 + ma_{\rm QL}^{r} \left[ 7 \right] {\Delta}{\tau}^7 + O\left({\Delta}{\tau}^9\right)
\end{equation}
where
\begin{subequations}
\begin{eqnarray}
ma_{\rm QL}^{r} \left[ 5 \right] &=& -\frac{3 q^2 M^2 \left( r-2M \right) \left( 20 r^3 - 81 M r^2 + 54 M^2 r + 53 M^3 \right)}{11200 \left( r-3M \right) ^2 r^{11}} \\
ma_{\rm QL}^{r} \left[ 7 \right] &=& -\frac{q^2 M^2 \left( r-2M \right) }{188160 \left( r-3M \right)^3 r^{14} } \left( 560 r^5 - 6020 r^4 M + 21465 r^3 M^2\right. \nonumber \\
	& & \qquad
\left.	 - 26084 r^2 M^3 - 5281 rM^4 + 21510 M^5 \right)
\end{eqnarray}
\end{subequations}
The $\theta$ component of the quasilocal mass times $4$-acceleration is $0$ (up to the orders considered) as would be expected.
The $\phi$ component of the quasilocal mass times $4$-acceleration is:
\begin{equation}
\label{eq:phiCoordCirc}
ma_{\rm QL}^{\phi} = ma_{\rm QL}^{\phi} \left[ 4 \right] {\Delta}{\tau}^4 + ma_{\rm QL}^{\phi} \left[ 6 \right] {\Delta}{\tau}^6 + O\left({\Delta}{\tau}^8\right)
\end{equation}
where
\begin{subequations}
\begin{eqnarray}
ma_{\rm QL}^{\phi} \left[ 4 \right] &=& \frac{3 q^2 M^2 \left( r-2M \right)^2 \left( 7r-8M \right)}{1120 \left( r-3M \right) ^2 r^{10}} \sqrt{\frac{M}{r-3M}}\\
ma_{\rm QL}^{\phi} \left[ 6 \right] &=& \frac{q^2 M^2 \left( r-2M \right) }{13440 r^{13} \left( r-3M \right) ^3} \sqrt{\frac{M}{r-3M}}\times\\ & & \left( 126 r^4 - 1129 r^3 M + 3447 M^2 r^2 -4193 M^3 r  + 1623 M^4 \right)\nonumber 
\end{eqnarray}
\end{subequations}
The $t$ component of the quasilocal mass times 4-acceleration is:
\begin{equation}
\label{eq:tCoordCirc}
ma_{\rm QL}^{t} = ma_{\rm QL}^{t} \left[ 4 \right] {\Delta}{\tau}^4 + ma_{\rm QL}^{t} \left[ 6 \right] {\Delta}{\tau}^6 + O\left({\Delta}{\tau}^8\right)
\end{equation}
where
\begin{subequations}
\begin{eqnarray}
ma_{\rm QL}^{t} \left[ 4 \right] &=& \frac{3 q^2 M^3 \left( r-2M \right) \left( 7r-8M \right)}{1120 \left( r-3M \right)^2 r^{9}} \sqrt{\frac{r}{r-3M}}\\
ma_{\rm QL}^{t} \left[ 6 \right] &=& \frac{q^2 M^3}{13440 \left( r-3M \right)^3 r^{12}}  \sqrt{\frac{r}{r-3M}} \times\\ & &\left( 126 r^4 - 1129 r^3 M + 3447 r^2 M^2 - 4193 r M^3 + 1623 M^4 \right) \nonumber 
\end{eqnarray}
\end{subequations}
Finally, the quasilocal contribution to the mass change is:
\begin{equation}
\frac{dm}{d\tau} = \frac{dm}{d\tau} \left[ 4 \right] {\Delta}{\tau}^4 + \frac{dm}{d\tau} \left[ 6 \right] {\Delta}{\tau}^6 + O\left({\Delta}{\tau}^8\right)
\end{equation}
where
\begin{subequations}
\begin{eqnarray}
\frac{dm}{d\tau} \left[ 4 \right] &=& \frac { 3 q^2 M^2 \left( r-2M \right) \left( 13M^2 + 2Mr-  5r^2 \right) }{2240 \left( r-3M \right) ^2 r^9} \\
\label{eq:dmdtauCoordCirc}
\frac{dm}{d\tau} \left[ 6 \right] &=& \frac { q^2 M^2  \left( r-2M\right) }{ 6720 \left( r-3M \right)^3 r^{12}}\times \\ &&\left( 651\,{M}^{4}-344\,{M}^{3}r-261\,{M}^{2}{r}^{2}+189\,M{r}^{3}-28\,{r}^{4} \right)\nonumber
\end{eqnarray}
\end{subequations}

Direct comparison between these results for the equations of motion and those of Ref.~\cite{Anderson:Wiseman:2005} for the self-force are not immediately possible. However, it is straightforward to project the results of Ref.~\cite{Anderson:Wiseman:2005} orthogonal to and tangent to the 4-velocity. In doing so, we find that the results of Ref.~\cite{Anderson:Wiseman:2005} agree in their general structure, but have incorrect numerical coefficients. Furthermore, we note that these results are in agreement with Eqs.~\eqref{eq:schw-circ-cov}.

\subsection{Radial Geodesic: Infall from rest}
\label{sec:coord-radial}
The case of a particle following a radial geodesic has been investigated by Anderson, Eftekharzadeh and Hu \cite{Anderson:Eftekharzadeh:Hu:2006}. They consider a particle held at rest until a time $t=0$. The particle is then allowed to freely fall radially inward. In such a case they have shown that, provided the particle falls for a sufficiently short amount of time, it is possible to obtain an expression for the entire self-force, rather than just the quasilocal part. Unfortunately, direct comparison with our result is not immediately possible as their results give the full self-force, rather than just the quasilocal contribution from the radial infall phase. However, as it is straightforward to calculate the quasilocal contribution using the method they prescribe, we have done so and found agreement with our results once the corrections to Ref. \cite{Anderson:2003} are taken into account.

As we are only interested in the leading order behavior for comparison to our own results, we take a slightly more straightforward approach to the calculation. From the arguments in Section \ref{sec:rad-rest}, it is clear that
\begin{equation}
\label{eq:coord-tau-radrest}
	r-r'\approx-\frac{M}{2 r^2}\left(\tau^2 - \tau'^2\right),~ ~ ~\theta - \theta'=0,~ ~ ~\phi - \phi'=0 ,~ ~ ~
	t-t'\approx\sqrt{\frac{r}{r-2M}}({\tau}-{\tau}')
\end{equation}
As in the circular geodesic case, we now substitute Eq. (\ref{eq:WKBGreen}) into (\ref{eq:qlsf}), take the partial derivative, then use Eq. (\ref{eq:coord-tau-radrest}) to express the coordinate separations in terms of proper time separations. Finally, we do the integration and project orthogonal and parallel to the 4-velocity to obtain an expression for the quasilocal contribution to the equations of motion for a particle starting at rest at $r=r_0, ~\tau = 0$ and subsequently falling radially inwards for a proper time $\Delta\tau$ to a radius $r$:
\begin{subequations}
\begin{eqnarray}
\label{eq:RadialRest-coord}
ma_{\rm QL}^{r} &=&
	- \frac {3 q^2 M^2}{11200 r^{10}}   \left( \frac{r-2M}{r} \right) \left( 20r-39M \right) {\Delta \tau}^5 + O \left( \Delta \tau ^6 \right)\\
ma_{\rm QL}^{t} &=& O \left( \Delta \tau ^6 \right) \\
\label{eq:RadialRestdmdtau-coord}\frac{dm}{d\tau} &=&
	- \frac {3 q^2 M^2}{448 r^8} \left(\frac{r - 2M}{r} \right) {\Delta \tau}^4 + O \left( \Delta \tau ^6 \right) 
\end{eqnarray}
\end{subequations}
This is in exact agreement with our covariant results in Eqs. (\ref{eq:RadialRestQLSF}) - (\ref{eq:RadialRestdmdtau}).

\section{Quasi-local Self-force: Non-Geodesic Motion} \label{sec:qlsf-non-geodesic-examples}
In this section, we consider non-geodesic motion in two black hole spacetimes, Reissner-Nordstr\"om and Kerr-Newman. These spacetimes were chosen as, while through their spherical and axial symmetry they retain much of the features of Schwarzschild and Kerr, their non-vanishing Ricci tensor means that the effects of non-geodesic motion become apparent at lower order than would otherwise be the case.

\subsection{Static Particle in Reissner-Nordstr\"{o}m\texorpdfstring{\\}{} spacetime}
\label{sec:rn}
We now calculate the quasilocal contribution to the scalar self-force on a static particle in Reissner-Nordstr\"{o}m spacetime, with line element given by \eqref{eq:RNMetric}. Since this is an example of a particle at rest in a static spacetime, we may use Eqs.~(\ref{eq:dmdtau-stationary-spacetime}) and (\ref{eq:ma-static-spacetime}) in order to calculate the equations of motion. Computing and substituting in the expressions for the relevant $\hat{v}_{t \dots t} (x)$ in Reissner-Nordstr\"{o}m spacetime, along with the time component of the 4-velocity, 
\begin{equation}
u^t = \left(1-\frac{2M}{r}+\frac{Q^2}{r^2}\right)^{-1/2}
\end{equation}
we arrive at our result for the quasilocal contribution to the equations of motion of a scalar particle held at rest in Reissner-Nordstr\"{o}m spacetime:
\begin{subequations}
\label{eq:staticresult}
\begin{eqnarray}
ma_{\rm QL}^{r} &=& -q^2\Big[\frac{Q^2 (Q^2-2 M r+r^2)}{60 r^{11}} (5 Q^2-8 M r+3 r^2) \Delta\tau^3
	+\frac{1}{6720 r^{15}}\big(1568 Q^8\nonumber \\
	& & \quad
	-7154 M Q^6 r +10647 M^2 Q^4 r^2 +2604 Q^6 r^2-5265 M^3 Q^2 r^3\nonumber \\
	& & \quad
	-7424 M Q^4 r^3+198 M^4 r^4+5147 M^2 Q^2 r^4+1236 Q^4 r^4-171 M^3 r^5\nonumber \\
	& & \quad
	 -1566 M Q^2 r^5+36 M^2 r^6+144 Q^2 r^6\big) \Delta\tau^5 + O(\Delta\tau^7)\Big]\\
\frac{dm}{d\tau} &=& -q^2\Big[\frac{Q^2 (Q^2-2 M r+r^2) }{20 r^8}\Delta\tau^2
	+\frac{1}{1344 r^{12}}\big(196 Q^6-588 M Q^4 r\nonumber \\
	& & \quad
	+429 M^2 Q^2 r^2+204 Q^4 r^2-18 M^3 r^3-268 M Q^2 r^3+9 M^2 r^4\nonumber \\
	& & \quad
	+36 Q^2 r^4\big) + O(\Delta\tau^6)\Big].
\end{eqnarray}
\end{subequations}
The $\theta$ and $\phi$ components of the four-acceleration are expected to be $0$ as a result of spherical symmetry. The $t$ component is also expected to be $0$ since we have a static particle in a static spacetime. These expectations are confirmed up to the orders calculated.

In the limit $Q \rightarrow 0$, this result reduces to the Schwarzschild case, which may be compared to Ref.~\cite{Anderson:Wiseman:2005}. However, Anderson and Wiseman use an incorrect expression to relate $\Delta t$ to $\left(\tau-\tau'\right)$ and also use the results of Ref.~\cite{Anderson:2003} prior to the corrections given in the subsequent errata \cite{Anderson:2003:err1,Anderson:2003:err2}. Once these two issues are corrected, we find our results are in exact agreement.

\subsection{Static Particle in Kerr-Newman spacetime}
\label{sec:kn}
To illustrate the flexibility of the covariant series approach presented in Chapters~\ref{ch:sf} and \ref{ch:coordex}, we will now investigate the case of a particle in Kerr-Newman spacetime at rest relative to an observer at spatial infinity. While it does not appear that the WKB approach of Ref.~\cite{Anderson:2003} could not be extended to such a spacetime, the covariant approach is easily applied to any spacetime, including Kerr-Newman.

We work with the Kerr-Newman metric in Boyer-Lindquist coordinates given by the line element of Eq.~\eqref{eq:KerrNewmanMetric}. For the case of a particle in such a spacetime at rest relative to an observer at spatial infinity, the condition on the motion is that the spatial components of the 4-velocity vanish, i.e. $u^i=0$. Additionally, the time component of the 4-velocity may be written as:
\begin{equation}
\label{eq:coord-to-proper-kn}
	u^t =\sqrt{\frac{\rho^2}{\Delta - a^2 + z^2}}
\end{equation}

Now, since Kerr-Newman is an example of a stationary spacetime, we may use Eqs.~(\ref{eq:dmdtau-stationary-spacetime}) and(\ref{eq:ma-stationary-spacetime}) in order to calculate the equations of motion:
\begin{subequations}
\allowdisplaybreaks
\begin{align}
&\frac{dm}{d\tau} = -\frac{q^2 Q^2}{20 \Delta  \rho ^8}(6 a^4+\{Q^2+r [-2 M+r]\}^2-4 \{Q^2+r [-2 M+r]\} z^2+z^4\nonumber \\
&\quad+6 a^2 \{Q^2-2 M r+r^2-z^2\}) \Delta\tau^2- \frac{q^2}{1344 \Delta ^2 \rho ^{14}}\boldsymbol{\Big(} 28 Q^{10} \{7 r^2+z^2\}\nonumber \\
&\quad-4 Q^8 \{[343 M-149 r] r^3+r [7 M+234 r] z^2+131 z^4\} +Q^6 \{r^4 [3565 M^2\nonumber  \\
&\quad	 - 3044 M r+640 r^2]-2 r^2 [215 M^2-2654 M r+1076 r^2] z^2+[37 M^2+2304 M r\nonumber  \\
&\quad	-1100 r^2] z^4+684 z^6\}- Q^4 \{[2 M-r] r^5 [2043 M^2-1532 M r+276 r^2]\nonumber  \\
&\quad	+r^3 [-1124 M^3+9971 M^2 r-7516 M r^2+1440 r^3] z^2+ r [166 M^3+3119 M^2 r\nonumber  \\
&\quad	-3140 M r^2+864 r^3] z^4+[137 M^2+2428 M r-1008 r^2] z^6-156 z^8\}\nonumber \\
&\quad	+ Q^2 \{r^6 [-2 M+r]^2 [447 M^2-268 M r+36 r^2]-4 [2 M-r] r^4 [85 M^3\nonumber \\
&\quad	-739 M^2 r+447 M r^2-63 r^3] z^2+ 4 r^2 [55 M^4+371 M^3 r-545 M^2 r^2\nonumber  \\
&\quad	+327 M r^3-72 r^4] z^4+4 r [109 M^3+516 M^2 r-427 M r^2+72 r^3] z^6 \nonumber \\
&\quad	+ 3 [27 M^2-104 M r+84 r^2] z^8-36 z^{10}\}-9 M^2 \{[2 M-r]^3 r^3\nonumber  \\
&\quad	+9 r^2 [-2 M+r]^2 z^2+9 [2 M-r] r z^4+z^6\} \rho ^4+ 12 a^6 \rho ^2 \{154 Q^4\nonumber \\
&\quad	+15 M^2 \rho ^2+30 Q^2 [-7 M r+2 \rho ^2]\} +6 a^4 \{42 Q^6 [13 r^2+9 z^2] -8 Q^4 [(224 M\nonumber \\
&\quad	-89 r) r^3 + 5 (28 M-9 r) r z^2+44 z^4]-45 M^2 [2 M r-r^2+z^2] \rho ^4 \nonumber \\
&\quad	+Q^2 [M^2 (1459 r^4+846 r^2 z^2+59 z^4)- 20 M r (53 r^2-24 z^2) \rho ^2\nonumber  \\
&\quad	+180 (r^2-z^2) \rho ^4]\} +4 a^2 \{7 Q^8 [59 r^2+23 z^2] -Q^6 [(2135 M-906 r) r^3\nonumber \\
&\quad	+ r (623 M+288 r) z^2+690 z^4] + Q^4 [r^4 (3639 M^2-2994 M r+601 r^2)  \nonumber \\
&\quad	+r^2 (670 M^2+1236 M r-543 r^2) z^2+ (55 M^2+2214 M r-1083 r^2) z^4+61 z^6]  \nonumber \\
&\quad	+27 M^2 [r^2 (-2 M+r)^2-3 r (-2 M+r) z^2+z^4] \rho ^4- Q^2 [4 M^3 r (524 r^4\nonumber \\
&\quad	+61 r^2 z^2+41 z^4)+M^2 (-2446 r^6+1251 r^4 z^2+1776 r^2 z^4+95 z^6)  \nonumber \\
&\quad	+ 15 M r (61 r^4-118 r^2 z^2+3 z^4) \rho ^2-108 (r^4-3 r^2 z^2+z^4) \rho ^4]\}\boldsymbol{\Big)} \Delta\tau^4 + O(\Delta\tau^6)\\
&ma^t=-\frac{q^2 (Q^2-2 M r) (a^2-z^2) }{75600 \Delta ^{5/2} \rho ^{17}} \boldsymbol{\Bigg[} 
	11340 Q^2 \rho^8 \Delta (2 a^2+Q^2-2 M r+r^2-z^2)\Delta\tau^2 \nonumber \\
&\quad	+ \boldsymbol{\Big(}Q^8 \{23807 r^4+36895 r^2 z^2+9632 z^4\}
	-Q^6 \{[119701 M-49004 r] r^5  \nonumber\\
&\quad	 +8 [21670 M-4053 r] r^3 z^2+r [32923 M+60726 r] z^4+44146 z^6\}   \nonumber\\
&\quad	 +2025 M^2 \{r^2 [-2 M+r]^2-3 r [-2 M+r] z^2+z^4\} \rho ^6\nonumber\\
&\quad	+3 Q^4 \{2 M^2 [32985 r^6+44155 r^4 z^2+5033 r^2 z^4+775 z^6]-6 M r [8618 r^4  \nonumber\\
&\quad	 -3695 r^2 z^2-7549 z^4] \rho ^2+[9749 r^4-20300 r^2 z^2+2999 z^4] \rho ^4\} \nonumber\\
&\quad	 -3 Q^2 \{M^3 [37174 r^7+47910 r^5 z^2+5970 r^3 z^4+4450 r z^6]\nonumber\\
&\quad	-M^2 [40349 r^6-20415 r^4 z^2-34905 r^2 z^4-2725 z^6] \rho ^2 +9 M r [1509 r^4\nonumber\\
&\quad	 -3090 r^2 z^2+295 z^4] \rho ^4-1350 [r^4-3 r^2 z^2+z^4] \rho ^6\} +675 a^4 \rho ^4 \{124 Q^4\nonumber\\
&\quad	 +15 M^2 \rho ^2+30 Q^2 [-5 M r+\rho ^2]\}+675 a^2 \rho ^2 \{Q^6 [145 r^2\nonumber\\
&\quad	+103 z^2]-2 Q^4 [(227 M-84 r) r^3+(143 M-30 r) r z^2+54 z^4] -15 M^2 [2 M r\nonumber\\
&\quad	-r^2+z^2] \rho ^4 +2 Q^2 [M^2 (7 r^2+z^2) (25 r^2+11 z^2)-2 M r (56 r^2-33 z^2) \rho ^2\nonumber\\
&\quad	+15 (r^2-z^2) \rho ^4]\}\boldsymbol{\Big)} \Delta\tau^4 +O(\Delta\tau^6)\boldsymbol{\Bigg]}\\
&ma^r =
	-\frac{q^2 (a^2+Q^2+r (-2 M+r))}{6720 \Delta^2  \rho ^{18}} \boldsymbol{\Bigg[}112 Q^2 \rho^6 \Delta  \boldsymbol{\Big(}30 a^4 r+r \{Q^2+r [-2 M\nonumber \\
&\quad+r]\} \{5 Q^2+r [-8 M+3 r]\} +2 \{-2 M^2 r -11 Q^2 r-9 r^3+M [Q^2+21 r^2]\} z^2 \nonumber \\
&\quad	+\{-4 M+9 r\} z^4+6 a^2 \{5 Q^2 r+4 r^3-6 r z^2+M [-9 r^2+z^2]\}\boldsymbol{\Big)}\Delta\tau^3 \nonumber \\
&\quad	+\boldsymbol{\Big(}56 Q^{10} r \{28 r^2+z^2\}+14 Q^8 \{r^4 [-735 M+298 r]+10 [13 M-62 r] r^2 z^2 \nonumber \\
&\quad	+[M-270 r] z^4\}+Q^6 \{r^5 [24955 M^2-19786 M r+3840 r^2]-2 r^3 [5285 M^2\nonumber\\
&\quad	-23710 M r+8492 r^2] z^2+r [763 M^2+11622 M r-4496 r^2] z^4+8 [-144 M\nonumber\\
&\quad	+907 r] z^6\}+Q^4 \{r^6 [-26559 M^3+30642 M^2 r-11462 M r^2+1380 r^3]\nonumber \\
&\quad	+r^4 [18645 M^3-85118 M^2 r+56148 M r^2-9744 r^3] z^2-r^2 [3097 M^3+5010 M^2 r \nonumber \\
&\quad	-4760 M r^2+1728 r^3] z^+[83 M^34+1886 M^2 r-25348 M r^2+9792 r^3] z^6\nonumber \\
&\quad	+2 [607 M+198 r] z^8\}-9 M^2 \{[11 M-4 r] r^4 [-2 M+r]^2+3 r^2 [-4 M^3 \nonumber\\
&\quad	+80 M^2 r-71 M r^2+16 r^3] z^2-9 r [4 M^2-19 M r+8 r^2] z^4+[-9 M+16 r] z^6\} \rho ^4\nonumber\\
&\quad	+Q^2 \{4 M^4 r [2682 r^6-2531 r^4 z^2+780 r^2 z^4-55 z^6]+M^2 r [8315 r^8\nonumber\\
&\quad	-45844 r^6 z^2+4336 r^4 z^4+20872 r^2 z^6-1335 z^8]-2 M^3 [7865 r^8-25324 r^6 z^2\nonumber\\
&\quad	+2250 r^4 z^4-740 r^2 z^6+109 z^8]-6 M [309 r^8-2719 r^6 z^2+2639 r^4 z^4+41 r^2 z^6\nonumber\\
&\quad	-26 z^8] \rho ^2+144 r [r^6-12 r^4 z^2+18 r^2 z^4-4 z^6] \rho ^4\}\nonumber\\
&\quad	+84 a^6 \rho ^2 \{176 Q^4 r+15 M^2 r \rho ^2+15 Q^2 [M (-15 r^2+z^2)\nonumber\\
&\quad	+4 r \rho ^2]\}+2 a^2 \{56 Q^8 r [118 r^2+37 z^2]-7 Q^6 [3 (1525 M-604 r) r^4\nonumber\\
&\quad	+2 r^2 (299 M+588 r) z^2+(-89 M+1692 r) z^4]+2 Q^4 [3 r^5 (8491 M^2\nonumber\\
&\quad	-6487 M r+1202 r^2)+r^3 (-1918 M^2+16755 M r-5604 r^2) z^2+r (-175 M^2\nonumber\\
&\quad	+16965 M r-7578 r^2) z^4+3 (-369 M+544 r) z^6]+54 M^2 [r^3 (24 M^2-22 M r\nonumber\\
&\quad	+5 r^2)+r (-4 M^2+45 M r-20 r^2) z^2+(-3 M+10 r) z^4] \rho ^4\nonumber\\
&\quad	+Q^2 [4 M^3 (-6812 r^6+1705 r^4 z^2-514 r^2 z^4+41 z^6)\nonumber\\
&\quad	+6 M^2 r (4892 r^6-5365 r^4 z^2-3902 r^2 z^4+307 z^6)\nonumber\\
&\quad	+1080 (r^9-2 r^7 z^2-5 r^5 z^4+2 r z^8)-15 M (671 r^6-1839 r^4 z^2\nonumber\\
&\quad	+399 r^2 z^4-3 z^6) \rho ^2]\}+6 a^4 \{168 Q^6 r [26 r^2+17 z^2]\nonumber\\
&\quad	+56 Q^4 [r^4 (-240 M+89 r)+2 r^2 (-61 M+13 r) z^2+(10 M-63 r) z^4]\nonumber\\
&\quad	-45 M^2 [(13 M-6 r) r^2-(M-8 r) z^2] \rho ^4+Q^2 [7 M^2 r (1459 r^4\nonumber\\
&\quad	+550 r^2 z^2-45 z^4)-10 M (689 r^4-519 r^2 z^2+24 z^4) \rho ^2\nonumber\\
&\quad	+360 r (3 r^2-4 z^2) \rho ^4]\}\boldsymbol{\Big)} \Delta\tau^5 + O(\Delta\tau^7) \boldsymbol{\Bigg]}\\
&ma^\theta =
	\frac{a z \sin \theta }{6720 \Delta^2  \rho ^{18}} \boldsymbol{\Bigg[} 112 Q^2 \rho^6 \Delta \boldsymbol{\Big(}30 a^4+5 Q^4+r^2 \{20 M^2-28 M r+9 r^2\}\nonumber\\
&\quad	+2 \{16 M-9 r\} r z^2+3 z^4+6 a^2 \{5 Q^2-10 M r+6 r^2-4 z^2\}-2 Q^2 \{10 M r\nonumber\\
&\quad	-7 r^2+8 z^2\}\boldsymbol{\Big)} \Delta\tau^3+ \boldsymbol{\Big(} 56 Q^{10} \{31 r^2\nonumber\\
&\quad	+4 z^2\}-28 Q^8 \{5 [88 M-45 r] r^3+2 r [4 M+115 r] z^2+131 z^4\}\nonumber\\
&\quad	+Q^6 \{r^4 [32515 M^2-32704 M r+7912 r^2]-2 r^2 [1757 M^2-18928 M r\nonumber\\
&\quad	+7508 r^2] z^2+[259 M^2+16128 M r-9752 r^2] z^4+4104 z^6\}\nonumber\\
&\quad	-2 Q^4 \{r^5 [18949 M^3-27967 M^2 r+13136 M r^2-1962 r^3]+r^3 [-4662 M^3\nonumber\\
&\quad	+36765 M^2 r-26924 M r^2+4896 r^3] z^2+r [581 M^3+10711 M^2 r\nonumber\\
&\quad	-14632 M r^2+4536 r^3] z^4+3 [137 M^2+2428 M r-904 r^2] z^6-390 z^8\}\nonumber\\
&\quad	-18 M^2 \{[7 M-8 r] r^3 [-2 M+r]^2+18 r^2 [6 M^2-7 M r+2 r^2] z^2\nonumber\\
&\quad	+3 [15 M-8 r] r z^4+2 z^6\} \rho ^4+Q^2 \{28 M^4 r^2 [599 r^4\nonumber\\
&\quad	-210 r^2 z^2+55 z^4]+8 M^3 r [-3999 r^6+5881 r^4 z^2+1135 r^2 z^4\nonumber\\
&\quad	+327 z^6]+M^2 [21499 r^8-47896 r^6 z^2-21452 r^4 z^4+12060 r^2 z^6\nonumber\\
&\quad	+405 z^8]-120 M r [50 r^6-181 r^4 z^2+62 r^2 z^4+13 z^6] \rho ^2+144 [4 r^6-18 r^4 z^2\nonumber\\
&\quad	+12 r^2 z^4-z^6] \rho ^4\}+84 a^6 \rho ^2 \{176 Q^4+15 M^2 \rho ^2+60 Q^2 [-4 M r+\rho ^2]\}\nonumber\\
&\quad	+4 a^2 \{28 Q^8 [127 r^2+46 z^2]14 Q^6 [(1328 M- 603 r) r^3+2 r (178 M+33 r) z^2\nonumber\\
&\quad	-+345 z^4]+27 M^2 [2 (7 M-5 r) (2 M-r) r^2+4 (9 M-5 r) r z^2\nonumber\\
&\quad	+5 z^4] \rho ^4-Q^2 [28 M^3 r (665 r^4+58 r^2 z^2+41 z^4)+3 M^2 (-7755 r^6+2152 r^4 z^2\nonumber\\
&\quad	-+4049 r^2 z^4+190 z^6)-540 (2 r^8-5 r^4 z^4-2 r^2 z^6+z^8)+30 M r (303 r^4\nonumber\\
&\quad	-416 r^2 z^2+9 z^4) \rho ^2]+Q^4 [7 M^2 (4583 r^4+750 r^2 z^2+55 z^4)\nonumber\\
&\quad	+42 M r (-671 r^4+130 r^2 z^2+369 z^4)+6 (992 r^4-1355 r^2 z^2+61 z^4) \rho ^2]\}\nonumber\\
&\quad	+6 a^4 \{1512 Q^6 [3 r^2+2 z^2]-224 Q^4 [(67 M-27 r) r^3+8 (5 M-2 r) r z^2\nonumber\\
&\quad	+11 z^4]-90 M^2 [7 M r-4 r^2+3 z^2] \rho ^4+Q^2 [-1120 M r (8 r^4+5 r^2 z^2-3 z^4)\nonumber\\
&\quad	+7 M^2 (1755 r^4+950 r^2 z^2+59 z^4)+360 (4 r^2-3 z^2) \rho ^4]\}
\boldsymbol{\Big)}\Delta\tau^5 + O(\Delta\tau^7)	\boldsymbol{\Bigg]}\\
&ma^\phi = 
	-\frac{q^2 a}{75600 \Delta ^{3/2} \rho ^{17}} \boldsymbol{\Bigg[} 
	11340 Q^2 \rho^8 \Delta (2 a^2+Q^2-2 M r+r^2-z^2) \Delta\tau^2 \nonumber \\
&\quad	+\boldsymbol{\Big(} Q^8 \{23807 r^4+36895 r^2 z^2+9632 z^4\}-Q^6 \{[119701 M-49004 r] r^5 \nonumber \\
&\quad	+8 [21670 M-4053 r] r^3 z^2+r [32923 M +60726 r] z^4+44146 z^6\} \nonumber \\
&\quad	+2025 M^2 \{r^2 [-2 M+r]^2-3 r [-2 M+r] z^2+z^4\} \rho ^6 \nonumber \\
&\quad	+3 Q^4 \{2 M^2 [32985 r^6+44155 r^4 z^2+5033 r^2 z^4+775 z^6] \nonumber \\
&\quad	-6 M r [8618 r^4-3695 r^2 z^2-7549 z^4] \rho ^2+[9749 r^4-20300 r^2 z^2 \nonumber \\
&\quad	+2999 z^4] \rho ^4\}-3 Q^2 \{M^3 [37174 r^7+47910 r^5 z^2+5970 r^3 z^4 \nonumber \\
&\quad	+4450 r z^6]-M^2 [40349 r^6-20415 r^4 z^2-34905 r^2 z^4-2725 z^6] \rho ^2 \nonumber \\
&\quad	+9 M r [1509 r^4-3090 r^2 z^2+295 z^4] \rho ^4-1350 [r^4-3 r^2 z^2+z^4] \rho ^6\} \nonumber \\
&\quad	+675 a^4 \rho ^4 \{124 Q^4+15 M^2 \rho ^2 + 30 Q^2 [-5 M r \nonumber \\
&\quad	+\rho ^2]\}+675 a^2 \rho ^2 \{Q^6 [145 r^2+103 z^2]-2 Q^4 [(227 M-84 r) r^3 \nonumber \\
&\quad	+(143 M-30 r) r z^2+54 z^4] -15 M^2 [2 M r-r^2+z^2] \rho ^4+2 Q^2 [M^2 (7 r^2 \nonumber \\
&\quad	+z^2) (25 r^2+11 z^2)-2 M r (56 r^2-33 z^2) \rho ^2+15 (r^2-z^2) \rho ^4]\} \boldsymbol{\Big)}\Delta\tau^4+ O(\Delta\tau^6)\boldsymbol{\Bigg]}
\end{align}
\end{subequations}
As expected, in the limit $a \rightarrow 0$, these reduce to Eqs.~(\ref{eq:staticresult}).


\subsection{Coordinate Calculation}
An alternative approach in the special case of a static, spherically symmetric spacetime is to use the Hadamard-WKB approach of Chapter~\ref{ch:coordex} to calculate the coordinate expansion of the retarded Green function. Upon doing so for Reissner-Nordstr\"{o}m spacetime, we find that the coefficients of the expansion of $V(x,x')$ relevant to the current static particle calculation are:
\begin{subequations}
\begin{equation}
 \tilde v_{000}=\tilde v_{001}=0
\end{equation}
and
\begin{align}
 \tilde v_{100}=&-\frac{Q^2 \left(Q^2-2 M r+r^2\right)^2 t^2}{20 r^{10}}\\
 \tilde v_{101}=&-\frac{Q^2 \left(Q^2-2 M r+r^2\right) \left(5 Q^2-8 M r+3 r^2\right)}{20 r^{11}}\\
 \tilde v_{200}=&-\frac{\left(Q^2-2 M r+r^2\right)^2}{1344 r^{16}} \Big[196 Q^6+12 r (17 r-49 M) Q^4\\
 & \qquad +r^2
   \left(429 M^2-268 r M+36 r^2\right) Q^2 +9 M^2 r^3 (r-2 M)\Big]\nonumber \\
 \tilde v_{201}=&-\frac{\left(Q^2-2 M r+r^2\right)}{1344 r^{17}} \Big[1568 Q^8+14 Q^6 r (-511 M+186 r)\\ &\qquad+9 M^2 r^4 \left(22 M^2-19 M r+4 r^2\right) +Q^4 r^2 \left(10647 M^2-7424 M r+1236 r^2\right)\nonumber\\
&\qquad+Q^2 r^3 \left(-5265 M^3+5147 M^2 r-1566 M r^2+144 r^3\right)\Big]\nonumber 
\end{align}
\end{subequations}
Additionally, for a particle held at rest in Reissner-Nordstr\"{o}m, we can immediately relate the coordinate point separations $\Delta x^\alpha$ to $\tau - \tau'$,
\begin{equation}
\label{eq:coord-to-proper}
	\Delta r = 0,~ ~ ~\Delta \theta =0,~ ~ ~\Delta \phi =0,~ ~ ~
	\Delta t =\left(1-\frac{2M}{r}+\frac{Q^2}{r^2}\right)^{-1/2} \left({\tau}-{\tau}'\right)
\end{equation}

Taking the partial derivative of Eq.~(\ref{eq:WKBGreen}), relating the coordinate separations to proper time separations using Eqs.~(\ref{eq:coord-to-proper}) and performing the integral over $\tau'$, we arrive at a result which is in exact agreement with Eqs.~(\ref{eq:staticresult}).

\chapter{The Scalar Self-force in Nariai Spacetime} \label{ch:nariai}
In the following sections we demonstrate that accurate self-force calculations via matched expansions are indeed feasible. We apply the method to compute the self-force for a scalar charge at fixed position in the Nariai spacetime (introduced in Sec.~\ref{sec:Nariai}).  We introduce a method for calculating the `distant past` Green function using an expansion in quasinormal modes. The effect of \emph{caustics} upon wave propagation is examined. This work is intended to lay a foundation for future studies of self-force in black hole spacetimes through matched expansions. The prospects for extending the calculation to the Schwarzschild spacetime appear good, although the work remains to be conducted.

This chapter is organised as follows. Section \ref{sec:scalarGF} is concerned with the scalar Green function on the Nariai spacetime. We begin in Sec.~\ref{subsec:FG} by expressing the Green function as a sum over angular modes and integral over frequency. We show in Sec.~\ref{subsec:DP} that performing the integral over frequency leaves a sum of residues: a so-called `quasinormal mode sum', which may be matched onto a `quasilocal' Green function, described in previous chapters.

In Sec.~\ref{sec:singularities} we consider the singular structure of the Green function. In Sec.~\ref{subsec:large-l} we demonstrate that the Green function is singular on the null surface, even beyond the boundary of the normal neighborhood and through caustics. We show that the singular behavior arises from the large-$l$ asymptotics of the quasinormal mode sums. To investigate further, we employ two closely-related methods for converting sums into integrals, namely, the Watson transform and Poisson sum (Sec.~\ref{subsec:watson-and-poisson}). The form of the Green function close to the null cone is studied in detail in Secs.~\ref{subsec:Poisson} and \ref{subsec:Hadamard}, and asymptotic expressions are derived. 

In Sec.~\ref{sec:summary} we take a short breath and rewrite the main equations obtained until then for the Green function.

Section \ref{sec:sf} describes the calculation of the self-force for the specific case of the static particle. We show in Sec.~\ref{subsec:static full Green} that the massive-field approach of Rosenthal \cite{Rosenthal:2003,Rosenthal:2004} may be adapted to the Nariai spacetime. This provides an independent check on the matched expansion calculation which is described in Sec.~\ref{subsec:matched-static}. Relevant numerical methods are outlined in Sec.~\ref{subsec:nummeth}.

In Sec.~\ref{sec:results} we present a selection of significant numerical results. We start in Sec.~\ref{subsec:results:GF} by examining the properties of the quasinormal mode Green function. In Sec.~\ref{subsec:results-asymptotics} we test the asymptotic expressions describing the singularity structure. In Sec.~\ref{subsec:results:matched} we show that the `quasilocal' and `distant past' Green functions match in an appropriate regime. In Sec.~\ref{subsec:results:SF} we present results for the self-force on a static particle.

\section{The Scalar Green Function}\label{sec:scalarGF}



\subsection{Retarded Green function as a Mode Sum} \label{subsec:FG}

The retarded Green function for a scalar field on the Nariai spacetime is defined by Eq.~(\ref{eq:Wave}), together with appropriate causality conditions. As mentioned in Sec.~\ref{sec:matched-expansions} the retarded Green function may be obtained through a Laplace integral transform and an angular $l$-mode sum, 
\beq
\Gret(t, \Rs; t^\prime, \Rs^\prime; \gam) = 
\frac{1}{2\pi}\int_{-\infty+ic}^{+\infty+ic} d\omega\sum_{l=0}^{+\infty} \tilde{g}_{l\omega}(\Rs, \Rs')
(2l+1)P_l(\cos\gamma)e^{-i\omega (t-t')} ,  \label{eq:Gret:mode-sum}
\eeq
where $\Rs$ and $t$ are the `tortoise' and `time' coordinates in the line element (\ref{eq:Nariai-le}), $c$ is a positive real constant, $t - t^\prime$ is the coordinate time difference, and $\gam$ is the spatial angle separating the points. 
The remaining ingredient in this formulation is the one-dimensional (radial) Green function $\Grad_{l\omega}(\Rs, \Rs^\prime)$ which satisfies
\beq
\left[ \frac{d^2}{d \Rs^2} + \omega^2 - \frac{U_0}{\cosh^2 \Rs} \right] \Grad_{l\omega}(\Rs, \Rs^\prime) = - \delta( \Rs - \Rs^\prime )  \label{rad-gf}
\eeq
The radial Green function may be constructed from two linearly-independent solutions of the radial equation (\ref{rad-eq-nar}).
To ensure a \emph{retarded} Green function we apply causal boundary conditions: no flux may emerge from the past horizons $\mathcal{H}^-_-$ and $\mathcal{H}^-_+$ (see Fig.~\ref{fig:penrose}). To this end, we will employ a pair of solutions denoted $u_{l \omega}^{\text{in}}$ and $u_{l \omega}^{\text{up}}$, in analogy with the Schwarzschild case. These solutions are defined in the next subsection.

\subsubsection{Radial Solutions}\label{subsec:radial-solns}
The homogeneous radial equation (\ref{rad-eq-nar}) may be rewritten as the Legendre differential equation
\beq
\frac{d}{d\R} \left( (1-\R^2) \frac{d u_{l\omega} }{d \R}  \right) +  \left( \beta (\beta + 1) - \frac{\mu^2}{1 - \R^2} \right) u_{l\omega}= 0  \label{legendre-de}
\eeq
where
\begin{eqnarray}
\mu = \pm i \omega, \quad \quad \beta = -1/2 + i \lam ,  \label{mu-nu-def} \\
 \lambda = \pm \sqrt{(l+1/2)^2 + d},   \quad \quad d = 4\xi - 1/2  .  \label{lam-def}
\end{eqnarray}
We choose $\mu = i \omega$, $\lam = \sqrt{(l+1/2)^2 + d}$ and note that the choice of signs will not have a bearing on the result.
The value of the constant $\xi$ in the conformally-coupled case in a $D$-dimensional spacetime is: $(D-2)/(4(D-1))$. 
Note that for conformal coupling in 4-D ($\xi = 1/6$) the constant is $d = 1/6$, and for minimal coupling ($\xi = 0$) we have $d = -1/2$. For the special value $\xi = 1/8$ we have $d = 0$. The possible significance of the value $\xi = 1/8$, the conformal coupling factor in three dimensions, was recently noted in a study of the self-force on wormhole spacetimes \cite{Bezerra:Khus}. 

\begin{figure}
\begin{center}
\includegraphics[width=15cm]{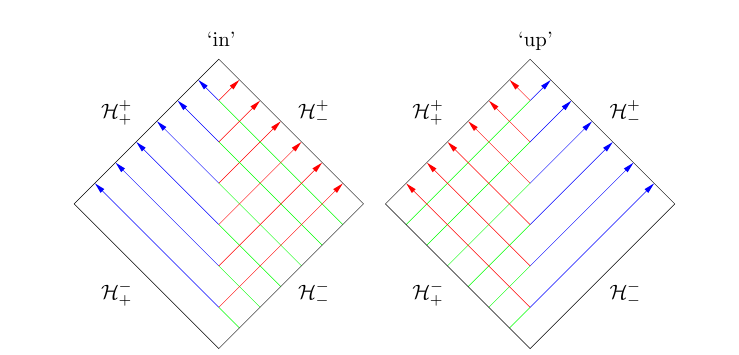}
\caption[Penrose diagrams for IN and UP radial solutions.]{\emph{Penrose diagrams for IN and UP radial solutions \cite{Frolov}}. The IN solution satisfies the boundary condition that no waves should emerge from the black hole (left). The UP solution satisfies boundary conditions such that it gives outgoing waves at infinity (right).  }
 \label{fig:inup}
\end{center}
\end{figure}

The solutions of Eq.~(\ref{legendre-de}) are associated Legendre functions of complex order, which are defined in terms of hypergeometric functions as follows (Ref.~\cite{GradRyz} Eq.~(8.771)),
\beq
P^{\mu}_\beta(\R) = \frac{1}{\Gamma(1-\mu)} \left( \frac{1+\R}{1-\R} \right)^{\mu/2} \, {}_2 F_1 \left(-\beta, \beta+1; 1 - \mu ; \frac{1-\R}{2} \right)   .  \label{LegP-def}
\eeq
Note that $P^{\mu}_{-1/2+i\lambda}(\R)=P^{\mu}_{-1/2-i\lambda}(\R)$, so clearly any results will be independent
of the choice of sign for $\lambda$ in (\ref{lam-def}).
In the particular case $\mu = 0$, the solutions belong to the class of conical functions (Eq.~(8.840) of Ref.~\cite{GradRyz}). We define the pair of linearly-independent solutions to be
\begin{eqnarray}
\uin(\R) &=& \Gamma(1 - \mu) P_{\beta}^\mu (-\R),  \label{uin-def} \\
\uup(\R) &=& \Gamma(1 - \mu) P_{\beta}^\mu (\R).   \label{uup-def}
\end{eqnarray}
These solutions are labelled ``in'' and ``up'' because they obey analogous boundary conditions (see Penrose diagram in Fig.~\ref{fig:inup}) to the ``ingoing at horizon'' and ``outgoing at infinity'' solutions that are causally appropriate in the Schwarzschild case \cite{Andersson:1997}. 
It is straightforward to verify that the ``in'' and ``up'' solutions obey
\begin{eqnarray}
\uin  &\sim& e^{-i \omega \Rs}  \quad \quad \text{as} \quad \Rs \rightarrow -\infty \\ 
\uup &\sim& e^{+i \omega \Rs}  \quad \quad \text{as} \quad \Rs \rightarrow +\infty
\end{eqnarray}
To find the asymptotes of $\uin$ near $\R = 1$, we may employ the series expansion
\begin{eqnarray}
{}_2 F_1 (a, b; c; z) &=& \frac{\Gamma(c) \Gamma(a+b-c)}{\Gamma(a) \Gamma(b)} (1-z)^{c-a-b} \left[ \sum_{k=0}^\infty \frac{(c-a)_k(c-b)_k}{(c+1-a-b)_k} \frac{(1-z)^k}{k!} \right]  \nn \\
 && + \, \frac{\Gamma(c) \Gamma(c-a-b)}{\Gamma(c-a) \Gamma(c - b)} \left[ \sum_{k=0}^\infty \frac{a_k b_k}{(1+a+b-c)_k} \frac{(1-z)^k}{k!}  \right]  \label{2F1-power-series}
\end{eqnarray}
where $(z)_k \equiv \Gamma(z + k) / \Gamma(z)$ is the Pochhammer symbol. In our case $a = -\beta$, $b = \beta + 1$, $c = 1 - \mu$ and $1 - z = (1 - \R) / 2$. It is straightforward to show that
\beq
\uin (\Rs) \sim \left \{  \begin{array}{ll} 
e^{-i \omega \Rs}, & \Rs \rightarrow -\infty ,  \\
\Aout_{l \omega} e^{i \omega \Rs} + \Ain_{l \omega} e^{-i \omega \Rs},
 \quad & \Rs \rightarrow + \infty ,
 \end{array} \right.    \label{bc-in}
\eeq
where
\begin{eqnarray}
 \Ain_{l\omega}  &=&  \frac{ \Gamma(1 - i \omega ) \Gamma(- i \omega) }{ \Gamma(1 + \beta - i\omega) \Gamma( - \beta - i \omega) } , \label{Ain-defn} \\
 \Aout_{l \omega}  &=& \frac{ \Gamma(1 - i\omega) \Gamma(i\omega)}{ \Gamma(1 + \beta) \Gamma(- \beta) } ,  \label{Aout-defn}
\end{eqnarray}
with $\beta$ as defined in (\ref{mu-nu-def}).
The ``up'' solution is found from the ``in'' solution via spatial inversion $\R \rightarrow - \R$; hence
\beq
\uup (\Rs) \sim \left \{  \begin{array}{ll} 
\Aout_{l \omega} e^{- i \omega \Rs} + \Ain_{l \omega} e^{i \omega \Rs}, \quad \quad & \Rs \rightarrow -\infty,  \\
 \quad  e^{i \omega \Rs}, & \Rs \rightarrow + \infty .
 \end{array} \right.    \label{bc-up}
\eeq


The Wronskian $W$ of the two linearly-independent solutions $u_{l \omega}^{\text{in}}(\Rs)$ and $u_{l \omega}^{\text{up}}(\Rs)$ can be easily obtained:
\beq
W = u_{l \omega}^{\text{in}}(\Rs) \frac{d u_{l \omega}^{\text{up}}}{d \Rs} - u_{l \omega}^{\text{up}}(\Rs) \frac{d u_{l \omega}^{\text{in}}}{d \Rs} = 2 i \omega A_{l \omega}^{\text{(in)}} 
 \label{Wronskian}
\eeq
The one-dimensional Green function $\Grad_{l \omega}(\Rs, \Rs^\prime)$ is then given by
 \begin{eqnarray}
\Grad_{l \omega}(\Rs, \Rs^\prime) &=& - \frac{1}{W} \left\{   \begin{array}{ll} u_{l \omega}^{\text{in}}(\Rs)   u_{l \omega}^{\text{up}}(\Rs^\prime) , \quad \quad & \Rs < \Rs^\prime , \\  u_{l \omega}^{\text{up}}(\Rs) u_{l \omega}^{\text{in}}(\Rs^\prime),  & \Rs > \Rs^\prime , \end{array} \right. \nonumber \\
 &=&
 \tfrac{1}{2} \Gamma(1+\beta-\mu) \Gamma(-\beta - \mu) P_{\beta}^{\mu}(-\R_<)P_{\beta}^{\mu}(\R_>)
, \label{eq:g-radial} \quad 
\end{eqnarray}
where $\R_<\equiv \text{min}(\R,\R')$ and $\R_>\equiv \text{max}(\R, \R')$.
The four-dimensional retarded Green function can thus be written as 
\begin{multline}
\label{full-ret-Green-mode-sum}
\Gret (x,x')=
\frac{1}{4 \pi }
\sum_{l=0}^{+\infty}  (2l+1)P_l(\cos\gamma)\int_{-\infty + ic}^{+\infty + ic}d\omega e^{-i\omega (t-t')} \\
\times 
\Gamma\left(\frac{1}{2}+ i \lambda -i\omega \right) \Gamma\left(\frac{1}{2} - i \lambda - i \omega \right)
P_{ -1/2 + i \lam}^{i \omega}(-\R_<)P_{ -1/2 + i \lam}^{i \omega}(\R_>).
\end{multline}
Unfortunately, we have not been able to find anywhere in the literature where this retarded Green function has been obtained in closed form.
In this chapter, we will calculate it and extract interesting information about it (such as its singularity structure) by using a combination
of analytic and numerical techniques.

 \subsection{Distant Past Green Function: The Quasinormal Mode Sum *} \label{subsec:DP}

\begin{figure}
 \begin{center}
  \includegraphics[width=7cm]{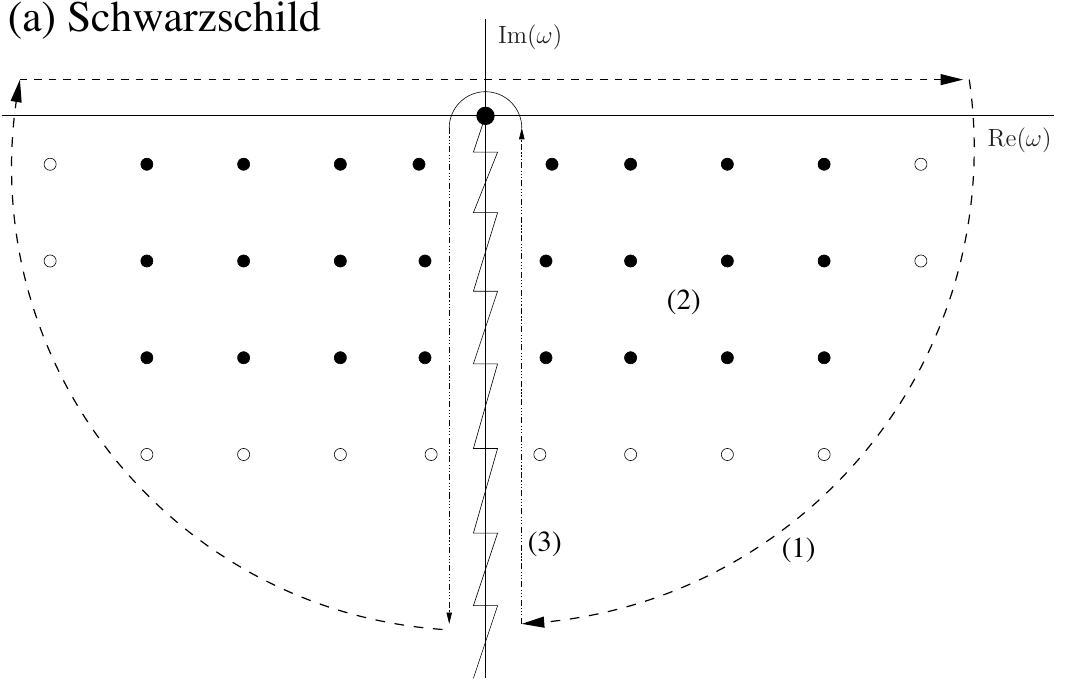}
  \includegraphics[width=7cm]{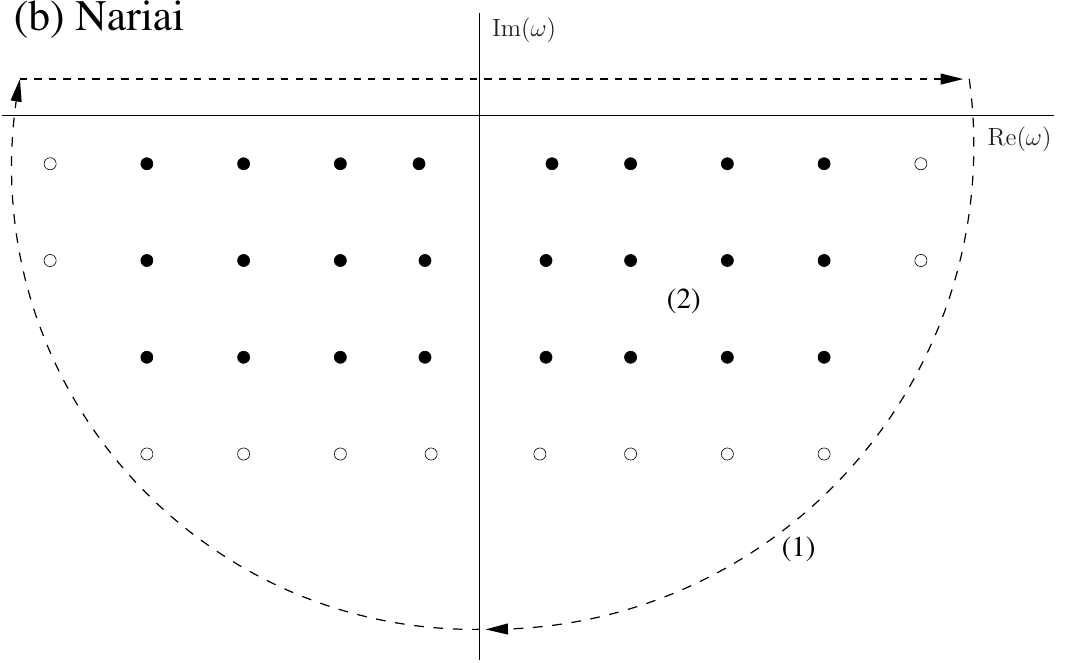}
 \end{center}
 \caption[Contour Integrals]{\emph{Contour Integrals}. These plots show the deformation of the integral over frequency in the complex plane to include the poles of the Green function (quasinormal modes), for two spacetimes: (a) Schwarzschild [left] (including a branch point at $\omega=0$ and the corresponding branch cut) and (b) Nariai [right] (for $\xi>0$).  
}
 \label{fig:contours}
\end{figure}

As discussed in Sec.~\ref{sec:matched-expansions}, the integral over frequency in Eq.~(\ref{eq:Gret:mode-sum}) may be evaluated by deforming the contour in the complex plane \cite{Leaver:1986, Andersson:1997}. The deformation is shown in Fig.~\ref{fig:contours}. The left plot (a) shows the Schwarzschild case, and the right plot (b) shows the Nariai case. 

On the Schwarzschild spacetime, it is well-known that a `power-law tail' arises from the frequency integral along a branch cut along the (negative) imaginary axis (Fig. \ref{fig:contours}, part (3)). In the Schwarzschild case, the branch cut is necessary due to a branch point in $\Grad_{l \omega}(r,r^\prime)$ at $\omega = 0$ \cite{Hartle:Wilkins:1974, Leaver:1986}. In contrast, for the Nariai case with $\xi > 0$,
$\omega=0$ is a regular point of $\Grad_{l \omega}(r,r^\prime)$
(the Wronskian (\ref{Wronskian}) is well-defined and non-zero in the limit $\omega \rightarrow 0$ and the 
Legendre functions $P_{\beta}^{\mu}(\R)$ are regular for $\R\in (-1,+1)$). 
Interestingly, for minimal coupling ($\xi = 0$), we find that $\omega = 0$ is a simple pole of $\Grad_{l \omega}(r,r^\prime)$
when $l=0$, and yet $A^{\text{(in)}}_{l=0,\omega=0}\neq 0$, so this mode does not obey the QNM boundary conditions (see below);
we will not consider the minimal coupling case here, however.
In either case ($\xi>0$ and $\xi=0$), $\omega =0$ is not a branch point and hence power law decay does not arise on the Nariai spacetime.

The simple poles of the Green function (shown as dots in Fig.~\ref{fig:contours}) occur in the lower half of the complex frequency plane. Given that the associated Legendre functions $P^{\mu}_{\beta}(\R)$ have no poles for
fixed $\R\in (-1,+1)$, and the poles of the Green function correspond to the zeros of the Wronskian, (\ref{Wronskian}). The Wronskian is zero when the ``in'' and ``up'' solutions are linearly-dependent. This occurs at a discrete set of (complex) \emph{Quasinormal Mode} (QNM) frequencies $\omega_q$. Beyer \cite{Beyer:1999} has shown that, for the P\"oschl-Teller potential, the corresponding QNM radial solutions form a \emph{complete basis} at sufficiently late times ($t > t_c$, to be defined below). Completeness means that \emph{any} wavefunction obeying the correct boundary conditions at $\Rs \to \pm \infty$ can be represented as a sum over quasinormal modes, to arbitrary precision. Intuitively, we may expect this to mean that, at sufficiently late times, the Green function itself can be written as a sum over the residues of the poles.

\subsubsection{Quasinormal Modes}
Quasinormal Modes (QNMs) are solutions to the radial wave equation, Eq.~(\ref{rad-eq-nar}),
which are left-going ($e^{-i \omega \Rs}$) at $\Rs \rightarrow -\infty$ and right-going ($e^{+ i \omega \Rs}$) as $\Rs \rightarrow +\infty$ . QNMs occur at discrete complex frequencies $\omega = \omega_q$ for which $\Ain_{l \omega_q} = 0$. At QNM frequencies, the ``in'' and ``up'' solutions ($u_{l \omega_q}^{\text{in}}$ and $u_{l \omega_q}^{\text{up}}$) are linearly-dependent and the Wronskian 
,(\ref{Wronskian}), is zero. 

The QNMs of the Schwarzschild black hole have been studied in much detail \cite{Leaver:1985, Nollert:1999, Kokkotas:Schmidt:1999, Cardoso:Witek:2008}. QNM frequencies are complex, with the real part corresponding to oscillation frequency, and the (negative) imaginary part corresponding to damping rate. QNM frequencies $ \omega_{ln}$ are labelled by two integers: $l$, the angular momentum, and $n = 0,1, \ldots \infty$, the \emph{overtone number}. For every multipole $l$, there are an infinite number of overtones. In the asymptotic limit $l \gg n$, the Schwarzschild QNM frequencies approach \cite{Ferrari:Mashhoon:1984, Iyer:1987, Konoplya:2003}
\beq
M \omega_{ln}^{(S)} \approx \frac{1}{\sqrt{27}} \left[ \pm(l+1/2) - i (n+1/2) \right].
\eeq
In general, the damping increases with $n$. The $n=0$ (`fundamental') modes are the least-damped.

The quasinormal modes of the Nariai spacetime are found from the condition $\Ain_{l \omega_{ln}} = 0$. Using (\ref{Ain-defn}), we find
\begin{equation}
1 + \beta - i \omega_{ln} = -n \quad \quad \text{or} \quad \quad -\beta - i\omega_{ln} = -n
\end{equation}
where $n$ is a non-negative integer. These conditions lead to the QNM frequencies
\begin{equation}
\omega_{ln} =  - \lambda - i (n+1/2) ,
\label{QNM freq}
\end{equation}
where $\lam$ is defined in (\ref{lam-def}). (Note we have chosen the sign of the real part of frequency here for consistency with previous studies \cite{Ferrari:Mashhoon:1984, Berti:Cardoso:2006} which use $\sig = -\omega$ as the frequency variable.)

\subsubsection{The Quasinormal Mode Sum}
The quasinormal mode sum is constructed from (\ref{modesum1}) by taking the sum over the residues of the poles of $\Grad_{l\omega}(\R,\R^\prime)$ in the complex-$\omega$ plane. Applying Leaver's analysis \cite{Leaver:1986} to (\ref{modesum1}), with the radial Green function (\ref{eq:g-radial}), we obtain 
\beq
\GQNM(t, \R; t^\prime, \R^\prime; \gam)  = 2 \, \text{Re} \sum_{n=0}^{+\infty} \sum_{l=0}^{+\infty} (2l+1) P_l(\cos \gam) \Bef_{ln} \tilde{u}_{ln}(\R) \tilde{u}_{ln}(\R^\prime) e^{ - i \omega_{ln} 
T},
\label{G-QNM}
\eeq
where
\beq
T \equiv t - t^\prime - \Rs - \Rs^\prime,   \label{T-def}
\eeq
and
the sum is taken over \emph{either} the third or fourth quadrant of the frequency plane only. Here, $\R$ and $t$ are the coordinates in line element (\ref{eq:Nariai-le}), $\Rs$ is defined in (\ref{eq:rhostar}), $\omega_{ln}$ are the QNM frequencies and $\Bef_{ln}$ are the \emph{excitation factors}, defined as
\beq
\Bef_{ln} \equiv \frac{\Aout_{l \omega_{ln}}}{2 \omega_{ln} \left.\frac{d \Ain_{l \omega}}{d \omega}\right|_{\omega_{ln}} } , \label{excitation-factors}
\eeq
and $\unorm_{ln}(\R)$ are the QNM radial functions, defined by
\beq
\unorm_{ln}(\Rs) = \frac{u^{\text{in}}_{l \omega_{ln}} (\Rs)}{\Aout_{l \omega_{ln}} e^{i \omega_{ln} \Rs}} .   \label{unorm-def}
\eeq
The QNM radial functions are normalised so that $\unorm_{ln}(\R) \rightarrow 1$ as $\R \rightarrow 1$ ($\Rs \rightarrow \infty$).
The excitation factors, defined in (\ref{excitation-factors}), may be shown to be
\begin{eqnarray}
\mathcal{B}_{ln} &=&
\frac{1}{2 \, n !}  \frac{ \Gamma(n + 1 - 2i\omega_{ln} ) }{ [ \Gamma(1 - i \omega_{ln} ) ]^2} , \nonumber \\
&=& \frac{1}{2 \, n !} \frac{\Gamma(-n + 2 i \lam ) }{[\Gamma(-n+1/2 + i \lam)]^2} . \label{EF2}
\end{eqnarray}
The steps in the derivation are given in Appendix A.3 of \cite{Berti:Cardoso:2006}.


The Green function (\ref{G-QNM}) now takes the form of a double infinite
series, taken over both angular momentum $l$ and overtone number $n$. Let us
briefly consider the convergence properties of these sums.

The infinite series over $n$ taken at fixed $l$ is convergent if $t - t' > t_c$, where $t_c \equiv \Rs + \Rs^\prime$ (i.e., at `late times' $T > 0$), and
divergent if  $t - t' < t_c$ (i.e., at `early times' $T < 0$). Note that the QNM frequencies have a negative imaginary part. 
The dominant quantity in the series is the
exponential $\exp(-(n+1/2) T)$, which decays or diverges more
powerfully with $n$ than any other factor in the sum
[this can be seen from the exact analytic forms shown below for the excitation factors
and the radial functions and their large-$n$ asymptotic behavior].
More details on
the convergence-with-$n$ in the Schwarzschild case are given in Sec.
IIIB of \cite{Andersson:1997}, and their arguments follow through to
our case.
Beyer \cite{Beyer:1999} has shown in the Nariai case that, for `late times' $T > 0$, QNMs form a \emph{complete basis}. Physically, for the QNM $n$-sum to be appropriate, sufficient coordinate time must elapse for a light ray to propagate inwards from $\R^\prime$, reflect off the potential barrier near $\R = 0$, and propagate outwards again to $\R$. 

The infinite series over $l$, taken at fixed $n$, is divergent.
For large $l$, magnitudes of the factors $(l+1/2) \Bef_{ln}$, $P_l(\cos\gamma)$ and $\tilde{u}_{ln}(\R)$ asymptote as, respectively,
$(l+1/2)^{n+1/2}$, $(l+1/2)^{-1/2}$ (except at $\gamma=0,\pi$, where it is equal to $+1,(-1)^l$, respectively; see Eq.~(\ref{eq:large-l P})) and 
a bounded term. 
Therefore, the limit $l=+\infty$ of the summand in (\ref{G-QNM}) is not zero and so the $l$-series (for fixed $n$) is divergent.
Despite the
series being divergent, we show in Sec.~\ref{subsec:watson} by applying a Watson
transform that well-defined and meaningful values can be extracted
from the series over $l$. This procedure (the Watson transform) is
\emph{not} possible globally, as the series also contain `physical'
divergences, associated with the null wavefront. We show in Sec.~\ref{subsec:large-l} that
the breakdown in validity of the Watson transform is related to a
`coherent phase condition' for terms in the series.

\subsubsection{Green Function near Spatial Infinity, $\R, \R^\prime \rightarrow +1$}
In the limit that both radial coordinates $\Rs$ and $\Rs^\prime$ tend to infinity ($\R,\R' \rightarrow +1$), the QNM sum (\ref{G-QNM}) may be rewritten
\beq
\lim_{\Rs,\Rs^\prime \rightarrow +\infty} \GQNM(T, \gam) = 2 \, \text{Re} \sum_{l=0}^\infty (2l+1) P_l(\cos \gam)  e^{ i \lam T} \sum_{n=0}^{\infty} \Bef_{ln} e^{- (n+1/2) T}
\eeq
where 
$\lam$ was defined in (\ref{lam-def}). Using the expression for the excitation factors (\ref{EF2}), the sum over $n$ can be evaluated explicitly, as follows,
\begin{eqnarray}
\sum_{n=0}^{\infty} \Bef_{ln} e^{- (n+1/2) T} &=& \frac{z^{1/2}}{2}  \sum_{n=0}^\infty \frac{\Gamma(-n + 2 i \lam)}{[\Gamma(-n + 1/2 + i \lam)]^2} \frac{z^n}{n!} \nonumber \\
&=&  \frac{z^{1/2}}{2} \frac{\Gamma(2i \lam)}{[\Gamma(1/2 + i\lam)]^2} \sum_{n=0}^\infty \frac{(2 i \lam)_{-n}}{[(1/2 + i \lam)_{-n}]^2} \frac{z^n}{n!}
\end{eqnarray}
where $z = e^{- T}$ and $(\cdot)_k$ is the Pochhammer symbol. Using the identity $(x)_{-n} = (-1)^n / (1-x)_n$ and the duplication formula $\Gamma(z) \Gamma(z+1/2) = 2^{1-2z} \sqrt{\pi} \, \Gamma(2z)$ we find
\begin{eqnarray}
\sum_{n=0}^{\infty} \Bef_{ln} e^{- (n+1/2) T} &=&  \frac{e^{- T / 2}}{4 \sqrt{\pi}}  \frac{2^{2 i \lam} \Gamma(i \lam)}{\Gamma(1/2 + i\lam)} \, {}_2F_1(-\beta, -\beta; -2\beta; -e^{-T})
\end{eqnarray}
where ${}_2F_1$ is the hypergeometric function, and $\beta$ was defined in (\ref{mu-nu-def}).
Hence the Green function near spatial infinity ($\R, \R^\prime \rightarrow 1$) is 
\begin{multline}
\GQNM(T, \gam) \sim
\frac{ e^{-T / 2} }{\sqrt{\pi}} \, \text{Re} \sum_{l=0}^{+\infty} \frac{(l+1/2) \Gamma(i \lam)}{\Gamma(1/2 + i\lam)} P_l(\cos \gam) e^{i \lam (T + 2 \ln 2)}\\ \times{}_2F_1(-\beta, -\beta; -2\beta; -e^{-T}),\quad \Rs,\Rs^\prime \rightarrow +\infty.  \label{GF-inf}
\end{multline}

\subsubsection{Green Function at Arbitrary `Radii'}
Unfortunately, it is not straightforward to perform the sum over $n$ explicitly for general values of $\R$ and $\R'$. Instead we must include the QNM radial functions $\unorm_{ln}(\Rs)$, defined by (\ref{uin-def}) and (\ref{unorm-def}). We note that the Legendre function appearing in (\ref{uin-def}) can be expressed in terms of a hypergeometric function (see Eq.~(\ref{LegP-def})), and 
the hypergeometric function can be written as power series about $\R = 1$ (Eq.~\ref{2F1-power-series}). 
At quasinormal mode frequencies, the second term in Eq. (\ref{2F1-power-series}) is zero, and hence the wavefunction is purely outgoing at infinity, as expected. Combining results (\ref{LegP-def}), (\ref{uin-def}), (\ref{2F1-power-series}) and (\ref{unorm-def}) we find the normalised wavefunctions to be 
\beq
\unorm_{ln}(\R) = \left( \frac{2}{1+\R} \right)^{i \omega_{ln}} \mathcal{S}_{ln}(\R) ,   \label{unorm-ser}
\eeq
where $\R$ is the radial coordinate in line element (\ref{eq:Nariai-le}) and $\mathcal{S}_{ln}$ is a finite series with $n$ terms,
\begin{align} \label{series Sln}
\mathcal{S}_{ln}(\R) =&  \sum_{k=0}^n \frac{1}{k!} \frac{ (- n + 2i\lambda)_k (-n)_k }{ (-n+1/2 + i\lambda)_k } \left( \frac{1-\R}{2} \right)^{k} \\ =&  {}_2F_{1}\left(-n+2 i \lambda, -n; -n + \frac{1}{2} + i \lambda; \frac{1-\R}{2} \right)  ,
\end{align}
where we have adopted the sign convention $\omega_{ln} =  -\lam - i(n+1/2)$ of (\ref{QNM freq}).
Hence the Green function at arbitrary `radii' may be written as the double sum,
\begin{multline}
\GQNM(x,x^{\prime}) = 2 \, \text{Re} \sum_{l=0}^\infty (2l+1)  P_l(\cos \gam)  e^{ i \lam \left[ T - \ln(2/(1+\R)) - \ln(2/(1+\R^\prime)) \right] }  \\ \times \sum_{n=0}^{\infty} \Bef_{ln} \left[ \frac{4 e^{-T} }{ (1+\R)(1+\R^\prime) } \right]^{n+1/2} \mathcal{S}_{ln}(\R) \mathcal{S}_{ln}(\R^\prime)   \label{G-arbitrary}
\end{multline}

\subsubsection{Green Function Approximation from Fundamental Modes}\label{subsec:fundamental-modes}
Expression (\ref{G-arbitrary}) is complicated and difficult to analyze as it involves a double infinite sum. It would be useful to have a simple approximate expression, with only a single sum, which captures the essence of the physics. At late times, we might expect that the Green function is dominated by the least-damped modes, that is, the $n=0$ fundamental quasinormal modes. If we discard the higher modes $n > 0$, we are left with an approximation to the Green function which does indeed seem to capture the essential features and singularity structure. However, as we show in section \ref{subsec:Poisson}, it does not correctly predict the singularity times. 

The $n=0$ approximation to the Green function is
\begin{multline}
\Gret^{(n=0)}(x, x^{\prime}) =  \left( \frac{e^{-T} }{\pi(1+\R)(1+\R^\prime)} \right)^{1/2} \\ \times \text{Re}\sum_{l=0}^{+\infty} \frac{ (2l+1) P_l(\cos \gam) \Gamma(i \lam)}{\Gamma(1/2 + i\lam)} e^{i \lam \left[ T + \ln(1+\R)(1+\R^\prime) \right]},  \label{G-n0}
\end{multline}
since $\mathcal{S}_{l0}(\R)\equiv 1$.
Note that the sum over $l$ is approximately periodic  in $T$ (exactly periodic in the case $\xi = 1/8$), with period $4 \pi$.

\subsection{Quasilocal Green Function: Hadamard-WKB Expansion} \label{subsec:QL}
To calculate the Green function within the quasilocal region, we may employ either the coordinate, covariant or numerical methods described in previous chapters. In this case, since Nariai is a spherically symmetric spacetime, it proves particularly straightforward to apply the coordinate method of Chapter~\ref{ch:coordex}. We express the retarded Green function in the Hadamard parametrix \cite{Hadamard,Friedlander}, Eq.~\eqref{eq:Hadamard}, as usual and write $V(x,x')$ as an expansion in powers of the coordinate separation of the points, as in Eq.~\eqref{eq:CoordGreen}.
However, for the non-radial motion considered in the present work, we will only need the terms of order  $O\left[(\R-\R')^0\right]$ and $O\left[(\R-\R')^1\right]$. We note that the $O\left[(\R-\R')^1\right]$ terms are easily calculated from the $O\left[(\R-\R')^0\right]$ terms using the identity described in Appendix~\ref{sec:WKB-7},
\begin{equation}
v_{ij1}(\R) = - \frac{1}{2} v_{ij0,\R}(\R).
\end{equation}
We can therefore use the expansion of $V(x,x')$ in
in powers of $(t-t')$ and $(\cos \gamma - 1)$ only,
\begin{equation}
\label{eq:CoordGreen-t-gamma}
V\left( x,x' \right) = \sum_{i,j=0}^{+\infty}  v_{ij}(\R) ~ \left( t - t' \right)^{2i} \left( \cos \gamma - 1 \right)^j,
\end{equation}
to give the quasilocal contribution to the retarded Green function as required in the present context.

\section{Singular Structure of the Green Function}\label{sec:singularities}
In this section we investigate the singular structure of the Green function.
We note that one expects the Green function to be singular when its two argument points are connected by a null geodesic, on account of the `Propagation of Singularities' theorems of Duistermaat and H\"ormander \cite{Duistermaat:Hormander:1972,Hormander:1985} and their application to the Hadamard elementary function 
(which is, except for a constant factor, the imaginary part of the Feynman propagator defined below in Eq.(\ref{eq:Hadamard G_F}))
for the Klein-Gordon equation by, e.g., Kay, Radzikowski and Wald~\cite{Kay:Radzikowski:Wald:1997}:
``if such a distributional bi-solution is singular for sufficiently nearby pairs of
points on a given null geodesic, then it will necessarily remain singular for
all pairs of points on that null geodesic."

We begin in Sec.~\ref{subsec:large-l} by exploring the large-$l$ asymptotics of the quasinormal mode sum expressions (\ref{GF-inf}), (\ref{G-arbitrary}) and (\ref{G-n0}). The large-$l$ asymptotics of the mode sums are responsible for the singularities in the Green function. We argue that the Green function is singular whenever a `coherent phase' condition is satisfied. The coherent phase condition is applied to find the times at which the Green function is singular. We show that the `singularity times' are exactly those predicted by the geodesic analysis of Sec.~\ref{subsec:Nariai-geodesics}. In Sec.~\ref{subsec:watson-and-poisson} we introduce two methods for turning the sum over $l$ into an integral. We show in Sec.~\ref{subsec:watson} that the Watson transform can be applied to extract meaningful values from the QNM sums away from singularities. We show in Sec.~\ref{subsec:Poisson} that the Poisson sum formula may be applied to study the behavior of the Green function near the singularities. We show that there is a four-fold repeating pattern in the singular structure of the Green function, and use uniform asymptotics to improve our estimates. In Sec.~\ref{subsec:Hadamard} we re-derive the same effects by computing the Van Vleck determinant along orbiting geodesics, to find the `direct' part of the Green function arising from the Hadamard form. The two approaches are shown to be consistent.



\subsection{Singularities of the Green function: Large-\texorpdfstring{$l$}{l} Asymptotics *}\label{subsec:large-l}
We expect the Green function $\Gret(x,x^\prime)$ 
to be `singular' if the spacetime points $x$ and $x^\prime$ are connected by a null geodesic. By `singular' we mean that $\Gret(x,x^\prime)$ does not take a finite value, although it may be well-defined in a distributional sense. Here we show that the Green function is `singular' in this sense if the large-$l$ asymptotics of the terms in the sum over $l$ satisfy a \emph{coherent phase condition}. 

\subsubsection{Near spatial infinity $\R,\R'\to+1$}
Insight into the occurrence of singularities in the Green function may be obtained by examining the large-$l$ asymptotics of the terms in the series (\ref{GF-inf}). Let us write
\beq
\lim_{\Rs,\Rs^\prime \rightarrow +\infty} \GQNM(x,x^{\prime}) = \text{Re} \sum_{l=0}^{\infty} \mathcal{G}_l(T, \gam)
\eeq
The asymptotic behavior of the gamma function ratio is straightforward: 
\beq
\Gamma(i \lam) / \Gamma(1/2 + i\lam) \sim \lam^{-1/2} e^{- i \pi / 4},\quad \lam\to +\infty.
\eeq 
The large-$l$ asymptotics of the hypergeometric function are explored in Appendix \ref{appendix-hypergeom}. We find
\begin{multline}
{}_2 F_1(-\beta, -\beta; -2\beta; -e^{- T}) \\ \sim \left( 1 + e^{-T} \right)^{-1/4} \exp \left( i \lam \left[ \ln \left\{ \frac{\sqrt{1+e^{-T}} + 1}{\sqrt{1+e^{-T}} - 1} \right\} - 2 \ln 2 - T \right] \right),\quad  \lam\to +\infty .  \label{eq:hypergeom-asymp}
\end{multline}
For simplicity, let us consider the special case of spatial coincidence $\gam = 0$ (near spatial infinity $\R, \R^\prime \to 1$),
\begin{multline} \label{Gl R->1,large l}
\mathcal{G}_l(T, \gam=0) \\ \sim \frac{e^{-T / 2}}{\sqrt{\pi} (1+e^{-T})^{1/4}} \frac{(l+1/2)}{\lam^{1/2}} \exp \left( i \lam \ln \left[ \frac{\sqrt{1+e^{- T}} + 1}{\sqrt{1+e^{-T}} - 1} \right] - i \pi / 4 \right),\quad \lam \rightarrow \infty  . 
\end{multline}
Asymptotically, the magnitude of the terms in this series grows with $(l+1/2)^{1/2}$. Hence the series is divergent. Nevertheless, due to the oscillatory nature of the series, well-defined values can be extracted (see Sec.~\ref{subsec:watson}), provided that the \emph{coherent phase condition},
\beq
\lim_{l \rightarrow +\infty} \arg \left( \mathcal{G}_{l+1} / \mathcal{G}_{l} \right) = 2 \pi N, \quad \quad N \in \mathbb{Z} , \label{coherent-phase}
\eeq
is {\it not} satisfied (for a related, but different, coherent phase condition or `resonance' see, for example,~\cite{Decanini:Folacci:Jensen:2003}).
In other words, the Green function is `singular' in our sense if Eq.~(\ref{coherent-phase}) is satisfied. 
In this case, 
\beq
 \ln \left( \frac{\sqrt{1+e^{-T}} + 1}{\sqrt{1+e^{-T}} - 1} \right) =  2 \pi N, \quad \quad N \in \mathbb{Z}  . \label{coherent-phase-1}
\eeq
Rearranging, we see that the Green function (\ref{GF-inf}) with $\gam = 0$ is `singular' 
when the `QNM time' $T$ is equal to
\beq
T^{(\R\sim 1,\gamma=0)} =  t -t' - \Rs - \Rs^\prime  = \ln \left[ \sinh^2 ( \pi N ) \right] .
\eeq
Note that the coherent phase condition (\ref{coherent-phase}) implies that the Green function is singular at precisely the null geodesic times (\ref{sing-time-inf}) (with $H=1$ and $\Delta\phi=2\pi N$), derived in Sec.~\ref{subsec:Poisson}. 

For the more general case where the spacetime points $x$, $x^\prime$ are separated by an angle $\gam$ on the sphere, it is straightforward to use the asymptotics of the Legendre polynomials to show that the Green function (\ref{GF-inf}) is singular when
the `QNM time' $T$ is equal to
\beq \label{sing-time-inf,bis}
T^{(\R\sim 1)} = t-t' - \Rs - \Rs^\prime  = \ln \left[ \sinh^2 ( \phif / 2 ) \right] ,  \quad \quad \text{where} \quad \phif = 2\pi N \pm \gam
\eeq
again in concordance with (\ref{sing-time-inf}).

\subsubsection{`Fundamental mode' $n=0$}
In Section \ref{subsec:fundamental-modes} we suggested that a reasonable approximation to the Green function may be found by neglecting the higher overtones $n > 0$. The `fundamental mode' series (\ref{G-n0}) also has singularities arising from the coherent phase condition (\ref{coherent-phase}), but they occur at slightly different times; we find that these times are
\beq
T^{(n=0)} =  \Delta \phi  - \ln[(1+\R)(1+\R^\prime)]    \quad \quad \text{where} \quad \Delta \phi = 2 \pi N \pm \gam .  \label{T-periodic}
\eeq
Towards spatial infinity, (\ref{T-periodic}) simplifies to $T^{(n=0)} = \Delta \phi - 2 \ln 2$, which should be compared with the `null geodesic time' given in (\ref{sing-time-inf}) (with $H=1$, as it corresponds to null geodesics). 
In Sec.~\ref{subsec:Poisson} we compare the singularities of the approximation (\ref{G-n0}) with the singularities of the exact solution (\ref{GF-inf}) at spatial infinity. 
Note that the singularity times $T^{(n=0)}$ are periodic, with period $2\pi$.
Clearly, the periodic times $T^{(n=0)}$, for any $\R=\R'\in (-1,+1)$, are not quite equal to the null geodesic times. Nevertheless, the latter approaches the former as $N \rightarrow \infty$.
These last properties are not surprising:
$n=0$ corresponds to the least-damped modes, which are the dominating ones in the QNM Green function at late times.
Therefore, one would expect that the singularity times, at late times, of the $n=0$ Green function correspond to those 
null geodesics that come in from $\R'$, orbit very near the 2-sphere at $\R=0$ 
(the unstable photon orbit) a very large number of times ($N \rightarrow \infty$) and then they come back out to $\R$.
Since it takes a null geodesic a coordinate time of $2\pi$ (recall that all 2-spheres in Nariai have radius $a=1$) to orbit around 
the 2-sphere at $\R=0$,
this time of $2\pi$ should be (at least at late times) precisely the time difference 
between singularities for the $n=0$ QNM Green function.

\vspace{0.5cm}

To investigate the form of the Green function we now introduce two methods for converting a sum over $l$ into an integral in the complex $l$-plane.
One method -- the Watson transformation -- is used to investigate the Green function away from the null cone, and the other method -- the Poisson sum -- is used near the null cone.

 \subsection{Watson Transform and Poisson Sum *}\label{subsec:watson-and-poisson}
In Sec.~\ref{subsec:DP}, the distant--past Green function was expressed via a sum over $l$ of the form
\beq
I\equiv \text{Re}\sum\limits_{l=0}^{+\infty} \mathcal{F} (l+\textstyle{\frac12}) P_l(\cos \gamma).    \label{eq:lsum}
\eeq
Here, $\mathcal{F} (l +\textstyle{\frac12})$ may be immediately read off from (\ref{GF-inf}), (\ref{G-arbitrary}) and (\ref{G-n0}).   
The so-called \emph{Watson Transform} \cite{Watson:1918} and \emph{Poisson Sum Formula} \cite{M&F,Aki:Richards} 
(see Eq.(\ref{Poisson sum}))
provide two closely-related ways of transforming a sum over $l$ into an integral in the complex $l$-plane. 
The two methods provide complementary advantages in understanding the sum over $l$. 

\begin{figure}
 \begin{center}
  \includegraphics[width=8cm]{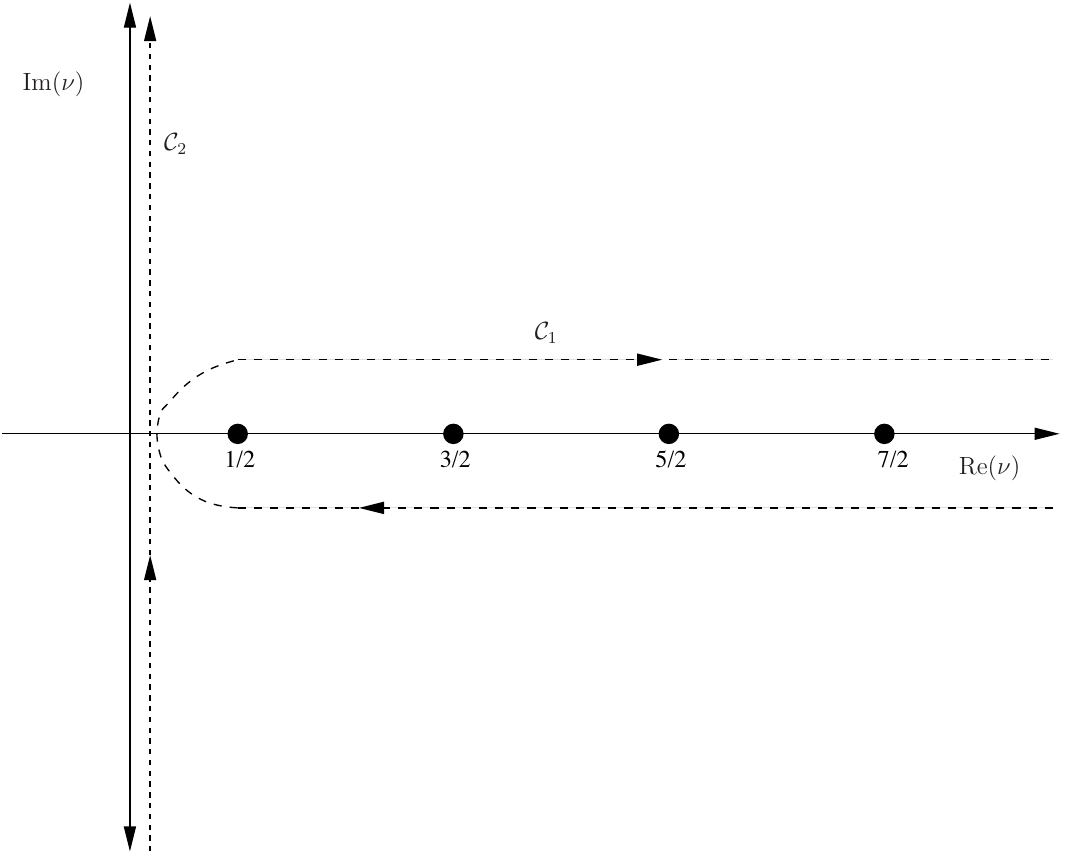}
 \end{center}
 \caption[The Watson Transform]{\emph{The Watson Transform}. The plot shows the contour $\mathcal{C}_1$ that defines the Watson transform in Eq.~(\ref{eq:watson-integral}). Provided the integrand is convergent in both quadrants $I$ and $IV$, the contour may be deformed onto $\mathcal{C}_2$ .
}
 \label{fig:contours2}
\end{figure}

A key element of the Watson transform is that, when extending the Legendre polynomial to non-integer $l$, the function with the appropriate behavior is $P_l(-\cos \gamma)$. This is obscured by the fact that
 $P_l(-\cos \gamma)= (-1)^l P_l(\cos \gamma)$ when $l$ is an integer. Our first step is then to rewrite the sum (\ref{eq:lsum}) as
 \beq
I= \text{Re}\sum\limits_{l=0}^{+\infty} e^{i(2N+1)\pi l} \mathcal{F} (l+\textstyle{\frac12}) P_l(- \cos \gamma), 
\eeq
where we have also introduced an integer $N$ for later convenience.
Using the Watson transform, we may now express the sum (\ref{eq:lsum}) as a contour integral
 \beq
I= \text{Re}\frac{(-1)^N}{2 i } \int_{\mathcal{C}_1}  \, e^{i 2N \pi \nu} \mathcal{F} (\nu) P_{\nu-1/2}(- \cos \gamma)\frac{d\nu}{\cos(\pi \nu)},   \label{eq:watson-integral}
\eeq
where $\nu = l+1/2$. The contour $\mathcal{C}_1$ starts just below the real axis at $\infty$, encloses the points $\nu=\frac12, 1+\frac12, 2+\frac12, \dots $ which are
poles of the integrand and returns to just above the real axis at $\infty$. The contour $\mathcal{C}_1$ is shown in Fig.~\ref{fig:contours2}. If the integrand is exponentially convergent in both quadrants $I$ and $IV$, the contour may be deformed in the complex-$l$ plane onto a contour $\mathcal{C}_2$ parallel to the imaginary axis (see Fig.~\ref{fig:contours2}). Note that there are no poles present inside quadrants $I$ and $IV$ 
(the so-called  `Regge'  poles - see, e.g., \cite{Decanini:Folacci:Jensen:2003}) in this case.
  
To study the asymptotic behavior of the Green function near singularities it is convenient to use the alternative representation of the sum 
obtained by writing
\begin{align}
\frac{1}{\cos(\pi \nu)}&=
\begin{cases}
\displaystyle 2i\sum_{l=0}^{\infty}e^{i\pi (2l+1)(\nu-1/2)}, & \text{Im}(\nu)>0,\\
\displaystyle -2i\sum_{l=0}^{\infty}e^{-i\pi (2l+1)(\nu-1/2)} = -2i\sum_{l=-\infty}^{-1}e^{i\pi (2l+1)(\nu-1/2)}, & \text{Im}(\nu)<0 .
\end{cases} \label{eq:cospi-rep}
\end{align} 
Inserting representation (\ref{eq:cospi-rep}) into (\ref{eq:watson-integral}) leads to the Poisson sum formula,
\beq
I=\sum_{s = -\infty}^{+\infty} (-1)^s \text{Re} \int_0^\infty d \nu e^{2\pi i s \nu}  \mathcal{F} (\nu) P_{\nu-1/2}(\cos \gam) .  \label{eq:poisson-sum}
\eeq
The Poisson sum formula is applied in Sec.~\ref{subsec:Poisson} to study the form of the singularities. 

\subsection{Watson Transform: Computing the Series *}\label{subsec:watson}

Let us now show how the Watson transform may be applied to extract well-defined values from 
series over $l$, even if these series are divergent.
We will illustrate the approach by considering the $n$th-overtone QNM contribution to the Green function, Eq.~(\ref{G-arbitrary}), for which case
\begin{multline}
\mathcal{F}(x,x^{\prime};\nu) = \>  \frac{1}{ n !} \left[ \frac{4 e^{-T} }{ (1+\R)(1+\R^\prime) } \right]^{n+1/2} 2 \nu e^{ i \lam \left[ T - \ln(2/(1+\R)) - \ln(2/(1+\R^\prime)) \right] }  \\ \times  \frac{\Gamma(-n + 2 i \lam ) }{[\Gamma(-n+1/2 + i \lam )]^2}  {}_2F_1\left(-n-2 i \lam ,-n,-n+{\textstyle\frac12}+ i \lam; (1-\R)/2\right) \\\times{}_2F_1\left(-n-2 i \lam,-n,-n+{\textstyle\frac12}+ i \lam; (1-\R^\prime)/2\right), \nn
\end{multline}
where $\lambda$ was defined in (\ref{lam-def}). 
We will choose the integer $N$ so that this contour  may be deformed into the complex plane (as shown in Fig.~\ref{fig:contours2}) to a contour on which the integral converges more rapidly. First, we note that the Legendre function may be written as the sum of waves propagating clockwise and counterclockwise,
\beq
P_{\nu-1/2}(\cos \gam) = \mathcal{Q}^{(+)}_{\nu-1/2}(\cos \gam) + \mathcal{Q}^{(-)}_{\nu-1/2}(\cos \gam),
\eeq
where 
\beq
\mathcal{Q}^{(\pm)}_{\mu} (z) = \frac{1}{2} \left[ P_{\mu}(z) \pm \frac{2i}{\pi} Q_{\mu}(z) \right]  \label{Qpm-def}
\eeq
and here $Q_{\mu}(z)$ is a Legendre function of the second kind. The functions $\mathcal{Q}^{(\pm)}_{\nu-1/2}$ have exponential asymptotics in the limit  $\nu \gam \gg 1$,
\begin{eqnarray}
\mathcal{Q}^{(\pm)}_{\nu-1/2} (\cos \gam) 
&\sim& \left( \frac{1}{2 \pi \nu \sin \gam} \right)^{1/2} e^{\pm i \pi / 4} e^{\mp i \nu\gam}.   \label{exp-approx}
\end{eqnarray}
With these asymptotics, and with the large-$l$ asymptotics (\ref{eq:hypergeom-asymp}) of the hypergeometric functions, one finds that the contour may be rotated to run, for example, along a line $\text{Re}(\nu) = c$ with $c$ a constant between  $0$ and $\frac{1}{2}$ \emph{provided that we choose}
\beq
N=  \begin{cases}
     \left[(T+\log \left((\R+1) (\R^\prime+1)\right)+\gamma )/(2 \pi )\right] & \text{for} \ \mathcal{Q}^{(+)}\\
     \left[(T+\log \left((\R+1) (\R^\prime+1)\right)+2 \pi - \gamma )/(2 \pi )\right]& \text{for}\  \mathcal{Q}^{(-)}\\
     \end{cases}
     \label{eq:critical}
\eeq
where here $[x]$ denotes the greatest integer less than or equal to $x$.

We performed the integrals numerically along $\text{Re}(\nu) = \frac14$ and found rapid convergence except near the critical times defining the jumps in $N$ given by Eq.~(\ref{eq:critical}), when the integrands fall to 0 increasingly slowly. An alternative method for extracting meaningful values from 
divergent series
is described in Sec.~\ref{subsec:nummeth}.

 \subsection{The Poisson sum formula: Singularities and \texorpdfstring{\\}{}Asymptotics *}\label{subsec:Poisson}
In this section we study the singularity structure of the Green function by applying the Poisson sum formula (\ref{eq:poisson-sum}). The first step is to group the terms together so that
\begin{equation}
I= \sum_{N=0}^\infty \II_N \quad \quad \text{where} \quad \quad \II_N \equiv \text{Re} \int_0^{+\infty} d \nu \mathcal{F}(\nu) R_N(\nu, \gam)    \label{InRN}
\end{equation}
and
\begin{equation} \label{eq:RN}
R_N = 
\begin{cases}
 \displaystyle  (-1)^{N/2} \left[ \mathcal{Q}_{\nu-1/2}^{(-)}(\cos \gam) e^{iN\pi\nu} + \mathcal{Q}_{\nu-1/2}^{(+)}(\cos \gam) e^{-iN\pi\nu}  \right] ,  & N \; \text{even} , \\
 \displaystyle (-1)^{(N+1)/2} \left[ \mathcal{Q}_{\nu-1/2}^{(+)}(\cos \gam) e^{i(N+1)\pi\nu} + \mathcal{Q}_{\nu-1/2}^{(-)}(\cos \gam) e^{-i(N+1)\pi\nu}  \right] ,  & N \; \text{odd} .
  \end{cases}
\end{equation}
and $\mathcal{Q}_{\nu - 1/2}^{(\pm)}$ were defined in Eq.~(\ref{Qpm-def}). We can now use the exponential approximations for $\mathcal{Q}_{\nu - 1/2}^{(\pm)}$ given in (\ref{exp-approx}) to establish
\beq
R_N \sim \frac{1}{(2\pi \nu \sin \gam)^{1/2}}  \begin{cases}
\displaystyle (-1)^{N/2} \left[ e^{-i \pi / 4} e^{ i \nu ( N \pi + \gam ) } + \text{c.c.} \right] ,  & \quad N \; \text{even}   , \\
\displaystyle (-1)^{(N+1)/2}  \left[ e^{i \pi / 4} e^{ i \nu ( (N+1) \pi - \gam ) } + \text{c.c.} \right] ,  & \quad N \; \text{odd}  . 
  \end{cases} \label{RN-asymp}
\eeq
It should be borne in mind that the exponential approximations (\ref{exp-approx}) are valid in the limit $\gam \nu \gg 1$. Hence the approximations are not suitable in the limit $\gam \rightarrow 0$ case. Below, we use alternative asymptotics (\ref{Olver-approx}) to investigate this case. 

\subsubsection{`Fundamental mode' $n=0$}
Let us apply the method to the `fundamental mode' QNM series (\ref{G-n0}). In this case we have
\beq
\mathcal{F} (\nu) = \left( \frac{4 e^{- T} }{\pi(1+\R)(1+\R^\prime)} \right)^{1/2} \frac{ \nu \, \Gamma(i \lam)}{\Gamma(1/2 + i\lam)} e^{i \lam \chi} 
\label{F n=0}
\eeq 
where $T$ was defined in (\ref{T-def}), $\lam$ was defined in (\ref{lam-def}), and
\beq
\chi = T + \ln[(1+\R)(1+\R^\prime)]  .  \label{eq:chi}
\eeq
Taking the asymptotic limit $\nu \rightarrow \infty$ we find
\beq \label{F n=0,large l}
\mathcal{F} (\nu) \sim \left( \frac{4 \nu \, e^{-T} }{i \pi(1+\R)(1+\R^\prime)} \right)^{1/2} \, e^{i \nu \chi}, \quad \nu \rightarrow \infty.
\eeq
Now let us combine this with the `exponential approximations' (\ref{RN-asymp}) for $R_N$,
\begin{multline}
\mathcal{F}(\nu) R_N(\nu, \gam) \sim \left( \frac{2 \, e^{-T} }{\pi^2 \sin\gam (1+\R)(1+\R^\prime)} \right)^{1/2} \\\times
\begin{cases}
\displaystyle (-1)^{N/2} \left[-i e^{i \nu (\chi + N\pi + \gam)} + e^{i  \nu (\chi - N\pi - \gam)} \right] , \quad & N \; \text{even} , \\
\displaystyle (-1)^{(N+1)/2} \left[ e^{i \nu (\chi + (N+1)\pi - \gam)} - i e^{i \nu (\chi - (N+1)\pi + \gam)} \right], \quad & N \; \text{odd} , 
\end{cases} \label{Poisson1}
\end{multline}
for $\nu \rightarrow \infty$.
It is clear that the integral in (\ref{InRN}) will be singular if the phase factor in either term in (\ref{Poisson1}) is zero. In other words, each wave $R_N$ gives rise to two singularities, occurring at particular `singularity times'. We are only interested in the singularities for $T > 0$; hence we may neglect the former terms in (\ref{Poisson1}). Now let us note that
\beq
\lim_{\eps \rightarrow 0^+} \int_{0}^{\infty} e^{i \nu (\zeta + i \eps)} d \nu = \lim_{\eps \rightarrow 0^+} \left( \frac{i}{\zeta + i\eps} \right) = i / \zeta + \pi \delta( \zeta )
\eeq
Upon substituting (\ref{Poisson1}) into (\ref{InRN}) and performing the integral, we find
\beq
\II_N^{(n=0)} \sim  \left( \frac{2 \, e^{-T } }{\sin\gam (1+\R)(1+\R^\prime)} \right)^{1/2} 
\begin{cases}
\displaystyle (-1)^{N/2} \delta( t-t' - t^{(n=0)}_{N} )  , \quad & N \; \text{even} , \\
\displaystyle \frac{(-1)^{(N+1)/2}}{\pi \left(t-t' - t^{(n=0)}_{N} \right)}  , \quad & N \; \text{odd} ,
\end{cases}
\label{sing-Poisson}
\eeq
where $\II_N^{(n=0)}$ is $\II_N$ in (\ref{InRN}) with $\mathcal{F} (\nu)$ given by (\ref{F n=0}), and 
\beq \label{sing-time-n=0}
t^{(n=0)}_N = \Rs + \Rs^\prime - \ln \left( (1+\R)(1+\R^\prime) \right) +
\begin{cases}
\displaystyle N  \pi + \gam   , \quad & N \; \text{even} , \\
\displaystyle (N+1) \pi - \gam   , \quad & N \; \text{odd} .
\end{cases}
\eeq
These times $t_{N}^{(n=0)}$ are equivalent to the `periodic' times identified in Sec.~\ref{subsec:large-l}  (Eq.~\ref{T-periodic}) and Sec.~\ref{subsec:Nariai-geodesics} (Eq.~\ref{sing-time-periodic}). 

Let us consider the implications of Eq.~(\ref{sing-Poisson}) carefully. Let us fix the spatial coordinates $\R$, $\R^\prime$ and $\gam$ and consider variations in $t - t'$ only. Each term $\II_N$ corresponds to a particular singularity in the mode sum expression (\ref{G-n0}) for the
($n=0$)-Green function. The $N$th singularity  occurs at $t-t' = t_{N}^{(n=0)}$. For times close to $t_{N}^{(n=0)}$, we expect the term $\II_N$ to give the dominant contribution to the ($n=0$)-Green function. Eq.~(\ref{sing-Poisson}) suggests that the ($n=0$)-Green function has a repeating four-fold singularity structure. The `shape' of the singularity alternates between a a delta-distribution ($\pm \delta( t-t' - t_N^{(n=0)} )$, $N$ even) and a singularity with antisymmetric `wings' ($\pm 1/(t- t' - t_N^{(n=0)})$,  $N$ odd).
 
The $N$th wave may be associated with the $N$th orbiting null geodesic shown in Fig.~\ref{fig:circ_orbits}. Note that `even N' and `odd N' geodesics pass in opposite senses around $\R=0$. Now, $N$ has a clear geometrical interpretation: it is the number of \emph{caustics} through which the corresponding geodesic has passed. Caustics are points where neighboring geodesics are focused, and in a spherically-symmetric spacetime caustics occur whenever a geodesic passes through angles $\Delta \phi = \pi$, $2\pi$, $3\pi, $ etc. Equation (\ref{sing-Poisson}) implies that the singularity structure of the Green function changes each time the wavefront passes through a caustic \cite{Ori1}. 

More accurate approximations to the singularity structure may be found by using the uniform asymptotics established by Olver \cite{Olver:1974}
(as an improvement on the `exponential asymptotics' (\ref{exp-approx})),
\begin{eqnarray}
\mathcal{Q}^{(\pm)}_{\nu-1/2} (\cos \gam) &\sim& \frac{1}{2} \left( \frac{\gam}{\sin \gam} \right)^{1/2} H_0^{(\mp)} (\nu \gam),  \label{Olver-approx}  
\end{eqnarray}
where $H_0^{(\pm)}(\cdot) = J_0(\cdot) \pm i Y_0(\cdot)$ are Hankel functions of the first $(+)$ and second $(-)$ kinds. This approximation (\ref{Olver-approx}) is valid in the large-$\nu$ limit for angles in the range $0 \le \gam < \pi$. With these asymptotics, we replace (\ref{RN-asymp}) with
\begin{multline}
R_N \sim \frac{1}{2} \left( \frac{\gam}{\sin \gam} \right)^{1/2} \\\times
\begin{cases}
\displaystyle (-1)^{N/2} \left[ H_0^{(+)}(\nu \gam) e^{iN\pi\nu} + H_0^{(-)}(\nu \gam) e^{- iN\pi\nu}  \right] ,  &  N \; \text{even}   , \\
\displaystyle (-1)^{(N+1)/2} \left[  H_0^{(-)}(\nu \gam) e^{i(N+1)\pi\nu} + H_0^{(+)}(\nu \gam) e^{- i(N+1)\pi\nu}  \right] ,  & N \; \text{odd}  . 
  \end{cases}  \label{RN-asymp-Hankel}
\end{multline}
In Appendix \ref{appendix:poisson-sum} we derive the following asymptotics for the `fundamental mode'  ($n=0$) Green function,
\begin{multline}
 \II^{(n=0)}_1 \\\sim 
 \begin{cases}
 \displaystyle \frac{2\Agam \, }{\sqrt{\pi} [(2\pi - \gam) - \chi] \, [(2\pi + \gam) - \chi]^{1/2} }  E\left(2\gam/[(2\pi + \gam) - \chi]\right) , & \chi < 2 \pi - \gam , \\
 \displaystyle \frac{-\Agam \sqrt{\pi}}{2 [\chi - (2\pi - \gam)]^{3/2}} \, {}_2 F_1\left( 3/2, 1/2; 2; \frac{\chi - 2\pi - \gam}{\chi - 2\pi + \gam} \right) , & \chi > 2 \pi - \gam  ,
 \end{cases} \label{eq:I1-asymp}
\end{multline}
\begin{multline}
\II^{(n=0)}_2 \\\sim \begin{cases}
\displaystyle  - \sqrt{\frac{2 \pi}{\gam}} \Agam \delta \left( \chi - (2\pi + \gam) \right) ,  & \chi \le 2 \pi + \gam  , \\
\displaystyle   \frac{\Agam \sqrt{\pi}}{2 [\chi - (2\pi - \gam)]^{3/2}} \, {}_2 F_1\left( 3/2, 1/2; 2; \frac{\chi - 2\pi - \gam}{\chi - 2\pi + \gam} \right)  , & \chi > 2 \pi + \gam  ,
   \end{cases}   \label{eq:I2-asymp}
\end{multline}
where $E$ is the complete elliptic integral of the second kind, $\chi$ was defined in (\ref{eq:chi}) and
\beq
\Agam = \left( \frac{\gam}{\sin \gam} \right)^{1/2} \left( \frac{e^{-T}}{\pi (1+\R)(1+\R^\prime)} \right)^{1/2}.  \label{eq:A-gam}
\eeq

\begin{figure}
 \begin{center}
  \includegraphics[width=8cm]{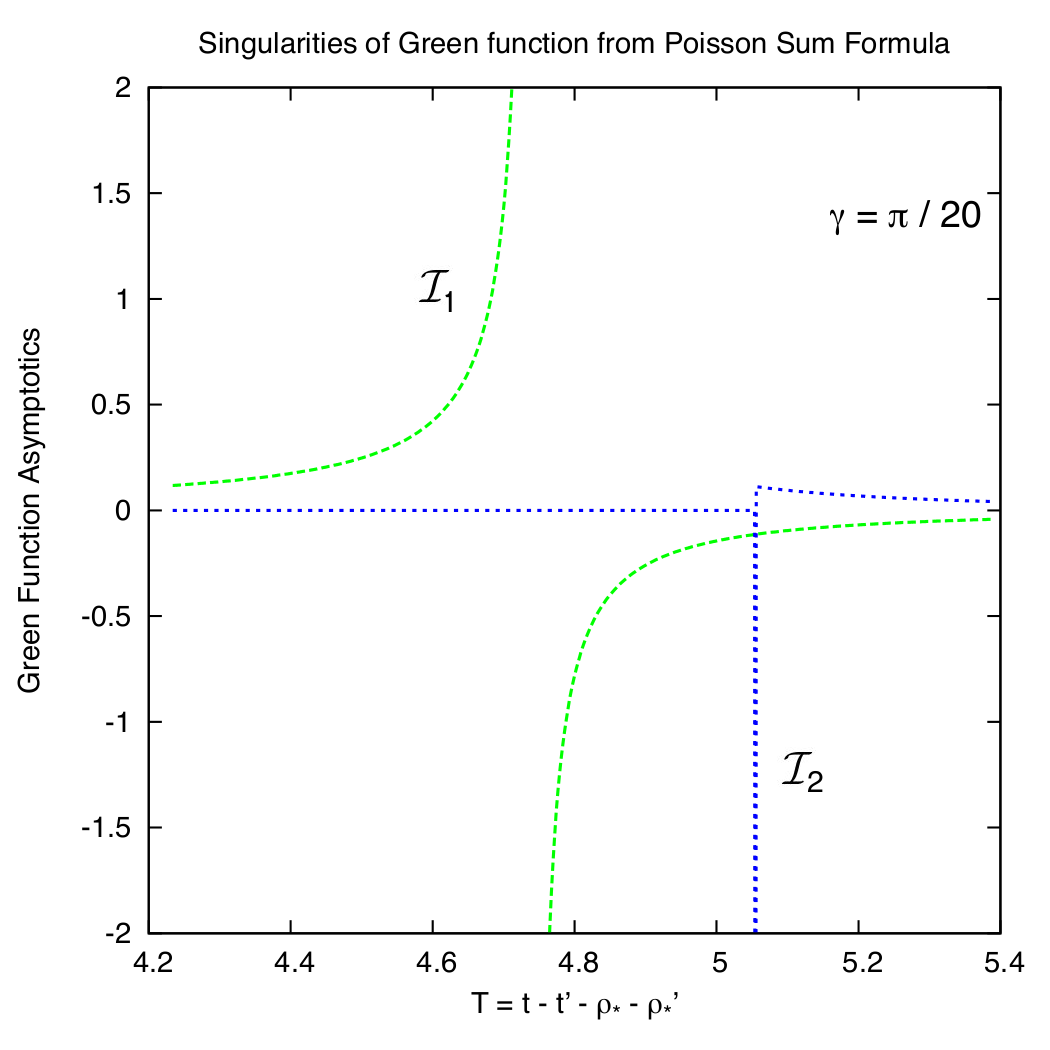}
  \includegraphics[width=8cm]{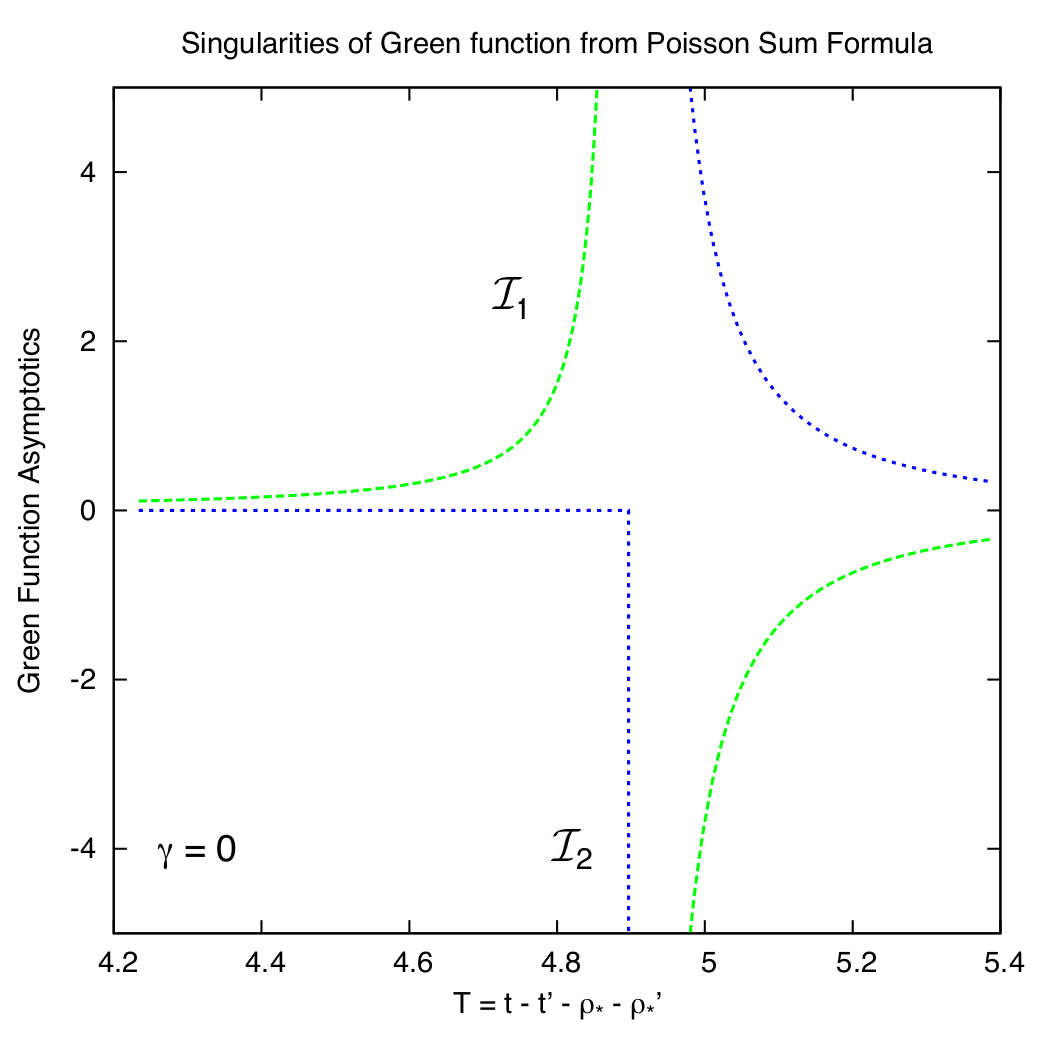}
 \end{center}
 \caption[Singularities of the `fundamental mode' Green function (\ref{G-n0}) near the caustic at $2\pi$ and $\R=\R'\to +1$]{\emph{Singularities of the `fundamental mode' Green function (\ref{G-n0}) near the caustic at $2\pi$ and $\R=\R'\to +1$}. These plots show the $\II_1$  (dashed) and $\II_2$ (dotted) contributions to the Poisson sum, given explicitly by (\ref{eq:I1-asymp}) and (\ref{eq:I2-asymp}). The top plot shows an angular separation $\gam = \pi/20$ and the bottom plot shows angular coincidence $\gam =0$. Note that, for $T > 2 \pi + \gam - 2\ln 2$, the $\II_1$ and $\II_2$ integrals are equal and opposite and will exactly cancel out (see text).}
 \label{fig:I1I2}
\end{figure}

The asymptotics (\ref{eq:I1-asymp}) and (\ref{eq:I2-asymp}) provide insight into the singularity structure near the caustic at $\Delta\phi = 2 \pi$. Figure \ref{fig:I1I2} shows the asymptotics (\ref{eq:I1-asymp}) and (\ref{eq:I2-asymp}) at $\R=\R'\to +1$ for two cases: (i) $\gam = \pi / 20$ (left) and (ii) $\gam = 0$ (right). In the left plot, the $\II_1^{(n=0)}$ integral has a (nearly) antisymmetric form. The $\II_2^{(n=0)}$ integral is a delta function with a `tail'. However, the `tail' is exactly canceled by the $\II_1^{(n=0)}$ integral in the regime $\chi > 2\pi + \gam$. The cancellation creates a step discontinuity in the Green function at $\chi = 2\pi + \gam$.
The form of the divergence shown in the right plot ($\gam = 0$) may be understood by substituting $\gam = 0$ into (\ref{eq:I1-asymp}) to obtain
\beq
\II_1^{(n=0)}(\gam = 0) \sim \left( \frac{e^{-T}}{(1+\R) (1+\R^\prime)} \right)^{1/2} ( 2 \pi - \chi )^{-3/2} .
\eeq

\subsubsection{Near spatial infinity, $\R,\R^\prime\to +1$}

It is straightforward to repeat the steps in the above analysis for the closed-form Green function (\ref{GF-inf}), valid for $\R,\R^\prime\to +1$. We reach a result of the same form as (\ref{sing-Poisson}), but with modified singularity times, 
\beq   \label{sing-time-inf,bis2}
t^{(\R\sim 1)}_{N} \sim \Rs + \Rs^\prime + 
\begin{cases}
\displaystyle \ln \left( \sinh^2 \left([N \pi + \gam] / 2 \right) \right) , \quad & N \text{ even} , \\
\displaystyle \ln \left( \sinh^2 \left([(N+1) \pi - \gam] / 2 \right) \right) , \quad & N \text{ odd} ,
\end{cases}
\eeq
corresponding to the geodesic times (\ref{sing-time-inf}).  
For instance, with the exponential asymptotics (\ref{RN-asymp}) applied to (\ref{GF-inf}) we obtain
\beq
\II^{(\R\sim 1)}_N \sim  \left( \frac{ e^{-T} }{ 2 \sin \gam \sqrt{1 + e^{-T}} } \right)^{1/2} 
\begin{cases}
\displaystyle (-1)^{N/2} \, \delta\left( t-t'- t^{(\R\sim 1)}_{N} \right)  , \quad & N \; \text{even} , \\
\displaystyle \frac{(-1)^{(N+1)/2}}{\pi \left(t-t' - t^{(\R\sim 1)}_{N} \right)}  , \quad & N \; \text{odd} .
\end{cases}
\label{sing-Poisson-2}
\eeq
In Sec.~\ref{sec:results} the asymptotic expressions derived here are compared against numerical results from the mode sums.

\vspace{0.5cm}

We believe that the 4-fold cycle in the singularity structure of the Green function which we have just unearthed using tricks we picked up from
seismology~\cite{Aki:Richards} is
characteristic of the $ \mathbb{S}^2$ topology (different types of cycle arising in different cases).
This cycle may thus also appear in
the more astrophysically interesting case of the Schwarzschild spacetime. Indeed, Ori \cite{Ori1} has suggested that this truly is the case for Schwarzschild. However, since this cycle does not seem to be
widely known in the field of General Relativity (with the notable exception of Ori's work),
in Appendix \ref{appendix-Green on S2} we apply the large-$l$
asymptotic analysis of this section to the simplest case with $ \mathbb{S}^2$
topology: the spacetime of $T\times  \mathbb{S}^2$, where the same cycle
blossoms in a clear manner.

\subsection{Hadamard Approximation and the Van Vleck \texorpdfstring{\\}{}Determinant}\label{subsec:Hadamard}
In this section, we re-derive the singularity structure found in (\ref{sing-Poisson}) and (\ref{sing-Poisson-2}) using a `geometrical' argument based on the Hadamard form of the Green function. 
In Sec.~\ref{subsec:QL} we used the Hadamard parametrix of the Green function to find the quasilocal contribution to the self-force. Strictly speaking, the Hadamard parametrix of Eq.~(\ref{eq:Hadamard}), is only valid when $x$ and $x^\prime$ are within a normal neighborhood \cite{Friedlander} (see footnote \ref{def:causal domain}). Nevertheless, it is plausible (particularly in light of the previous sections) that the Green function near the singularities may be adequately described by a Hadamard-like form, but with contributions from \emph{all appropriate orbiting geodesics} (rather than just the unique timelike geodesic joining $x$ and $x^\prime$). 

We first introduce 
the Feynman propagator $G_F(x,x')$
(see, e.g.,~\cite{DeWitt:1960,Birrell:Davies})
which satisfies
the inhomogeneous scalar wave equation (\ref{eq:Wave}). 
The Hadamard form, which in principle is only valid for points
$x'$ within the normal neighborhood of $x$, for the Feynman propagator in 4-D is~\cite{DeWitt:1960,Decanini:Folacci:2005b}
\beq\label{eq:Hadamard G_F}
G_F(x,x')=\frac{i}{2\pi}\left[\frac{U(x,x')}{\sigma+i\epsilon}+V(x,x')\ln(\sigma+i\epsilon)+W(x,x')\right],
\eeq
where $U(x,x')$, $V(x,x')$ (already introduced in Sec.~\ref{sec:classical-green}) and $W(x,x')$ are bi-tensors which are regular at coincidence ($x\to x'$) and
$\sig(x,x^\prime)$ is the Synge world function: half the square of the geodesic distance \emph{along a specific geodesic} joining $x$ and $x^\prime$.
Note the {\it Feynman prescription} `$\sigma\to\sigma+i\epsilon$' in (\ref{eq:Hadamard G_F}),
where $\epsilon$ is an infinitesimally small positive value, which is necessary so that the Feynman propagator
has the appropriate analytical properties; this implies that Eq.~(\ref{Gret from GF}) below is then satisfied.

The retarded Green function is readily obtained from 
the Feynman propagator by
\begin{equation} \label{Gret from GF}
G_{\text{ret}}(x,x')=
2\theta_-(x,x')\ \text{Re}\left(G_F(x,x')\right),
\end{equation}
which yields (\ref{eq:Hadamard}) inside the normal neighborhood, since $U(x,x')$, $V(x,x')$ and $W(x,x')$ 
are real-valued there.
We posit that the `direct' part of the Green function 
remains in Hadamard form,
\beq
\Gret^{\text{dir.}}(x, x^\prime) = \lim_{\eps \rightarrow 0^+} \frac{1}{\pi} \text{Re} \left[ i \frac{U(x,x^\prime)}{\sig + i \eps} \right] = \text{Re} \left[ U(x,x^\prime) \left( \delta(\sig)+\frac{i}{\pi \sig}\right) \right],  \label{Gdir1}
\eeq
even {\it outside} the normal neighborhood [note that outside the normal neighborhood we still use the term `direct' part to refer to the contribution from the $U(x,x')$ term, even if its support may not be restricted to the null cone anymore]. It is plausible that the Green function near the $N$th singularity (see previous section) is dominated by the `direct' Green function (\ref{Gdir1}) calculated along geodesics near the $N$th orbiting null geodesic. To test this assertion, we will calculate the structure and magnitude of the singularities and compare with (\ref{sing-Poisson-2}).

In four dimensional spacetimes, the symmetric bi-tensor $U(x,x^\prime)$ is given by
\beq
U(x,x^\prime) = \Delta^{1/2}(x, x^\prime),
\eeq
where $\Delta(x,x^\prime)$ is the Van Vleck determinant \cite{VanVleck:1928, Morette:1951, Visser:1993}. In Sec.~\ref{sec:Nariai}, we derived an analytic expression for the Van Vleck determinant,
\beq
\Delta (x,x')= \left( \frac{\gam}{\sin \gam} \right) \left( \frac{ \eta }{ \sinh \eta}  \right).  \label{eq:vanVleckDet}
\eeq
where $\gam \in [0, \pi)$ is the geodesic distance traversed on the two-sphere and $\eta$ is the geodesic distance traversed in the two-dimensional de Sitter subspace. Hence the Van Vleck determinant is singular at the angle $\gam = \pi$. 


This may be seen another way. Using the spherical symmetry, let us assume without loss of generality that the motion is in the plane $\theta = \pi/2$, from which it follows that the equation for ${\sig^\theta}_\theta$ in (\ref{eq:xi-transport}) decouples from the remainder; it is
\beq
\phi' \frac{d {\sig^{\theta'}}_{\theta'}}{d \phi'} + \phi'^2 - {\sig^{\theta'}}_{\theta'} (1 - {\sig^{\theta'}}_{\theta'} ) = 0
\label{eq:V-V transport eq}
\eeq
Here, as before, we have rescaled the affine parameter $s$ to be equal to the angle $\phi$ subtended by the geodesic. Note that here we let $\phi$ take values greater than $\pi$. 
It is straightforward to show that the solution of Eq.~(\ref{eq:V-V transport eq}) is ${\sig^{\theta'}}_{\theta'} = \phi' \cot \phi'$. Hence ${\sig^{\theta'}}_{\theta'}$ is singular at the angles $\phi = \pi$, $2\pi$, $3\pi$, etc.  In other words, the Van Vleck determinant is singular at the antipodal points, where neighboring geodesics are focused: the \emph{caustics}. The 
Van Vleck determinant may be separated in the following manner: $\Delta = \Delta_{\theta} \Delta_{t\rho}$, where
\begin{eqnarray}
\phi' \frac{d \ln \Delta_{\theta}}{d \phi'} &=& 1 - {\sig^{\theta'}}_{\theta'},   \label{Delta_pp}  \\
\phi' \frac{d \ln \Delta_{t\rho} }{d \phi'} &=& 2 - \Ztt - \Zyy  \label{Delta_ty} .
\end{eqnarray}
Eq.~(\ref{Delta_pp}) yields
\beq \label{eq:V-V sing}
\ln  \Delta_{\theta}=\ln \left(\frac{\phi}{\phi_0}\right)
-\int_{\phi_0}^{\phi}d\phi'\cot\phi'
\eeq
which can be integrated analytically by following a Landau contour in the complex $\phi'$-plane around 
the (simple) poles of the integrand (located at $\phi'=k \pi$, $k\in \mathbb{Z}$), which are the caustic points.
Following the Feynman prescription `$\sigma\to\sigma+i\epsilon$', we choose 
the Landau contour so that the poles lie {\it below} the contour.
We then obtain (setting $\phi_0=0$ without loss of generality)
\beq \label{eq:Delta_phi}
\Delta_{\theta} = \left| \frac{\phi}{\sin{\phi}} \right| e^{-i N \pi} .
\eeq
Here, $N$ is the number of caustic points the geodesic has passed through. The phase factor, obtained by continuing the contour of integration past the singularities at $\phi = \pi, 2\pi, $ etc., is crucial. Inserting the phase factor $e^{-i N \pi / 2}$ in (\ref{Gdir1}) leads to exactly the four-fold singularity structure predicted by the large-$l$ asymptotics of the mode sum (\ref{sing-Poisson-2}). That is:
\beq
 G_{N}^{dir} \sim  \left( \frac{\eta}{\sinh \eta } \right)^{1/2} 
 \left( \frac{\phi}{\sin \phi } \right)^{1/2}
 \begin{cases}
 \displaystyle (-1)^{N/2} \delta(\sig), \quad \quad &  N \text{ even}, \\ 
 \displaystyle \frac{(-1)^{(N-1)/2}}{\pi \sig}, \quad & N \text{ odd}.
 \end{cases} 
 \label{GNdir-Hadamard}
\eeq

The accumulation of a phase of `$-i$' on passing through a caustic, and the alternating singularity structure which results, is well-known to researchers in other fields involving wave propagation -- for example, in acoustics~\cite{Kravtsov:1968},
seismology~\cite{Aki:Richards},
symplectic geometry \cite{Arnold} and quantum mechanics~\cite{B&M}  the integer $N$ is known as the Maslov index~\cite{Maslov'65,MaslovWKB}.

We would expect to find an analogous effect in, for example, the Schwarzschild spacetime. The four-fold structure has been noted before by at least one researcher \cite{Ori1}. 
Nevertheless, the effect of caustics on wave propagation in four-dimensional spacetimes does not seem to have received much attention in the gravitational literature (see \cite{Friedrich:Stewart:1983, Ehlers:Newman:2000} for exceptions). 

To compare the singularities in the mode-sum expression (\ref{sing-Poisson-2}) with the singularities in the Hadamard form (\ref{GNdir-Hadamard}), let us consider the `odd-$n$' singularities of $1/\sig$ form. We will rearrange (\ref{sing-Poisson-2}) into analogous form by expanding $\sigma$ to first-order in $t - t_N^{(\R\sim 1)}$, where $t_N^{(\R\sim 1)}$ is the $N$th singularity time for orbiting geodesics starting and finishing at $\R\to 1$.
For the orbiting geodesics described in Sec. \ref{subsec:Nariai-geodesics} we have $\sig = - \tfrac{1}{2} (H^2 - 1) \phi^2$. 
At $\R\to 1$, expanding to first order and using (\ref{sing-time-inf}) yields
\beq
 \sig \sim  - (H \phi) \tanh( H \phi / 2) \left(t - t_N^{(\R\sim 1)}\right) .
\eeq
The mode-sum expression (\ref{sing-Poisson-2}) may then be rewritten in analogous form to the `$N$ odd' expression in (\ref{GNdir-Hadamard}),
\beq
\GQNM \sim (-1)^{(N-1)/2} \frac{\left| \Delta_{\text{(QNM)}} \right|^{1/2}}{\pi \sig} 
\eeq
where
\beq
 \left| \Delta_{\text{(QNM)}} \right|^{1/2} =  \left( \frac{H \eta \sinh(\eta/2)}{2 \cosh^3(\eta/2)} \right)^{1/2}
 \left| \frac{\phi}{\sin \phi} \right|^{1/2} .
\label{Delta-QNM}
\eeq
Here $\eta = H \phi$, where $H$ is the constant of motion introduced in Sec.~\ref{subsec:Nariai-geodesics}. We find very good agreement between (\ref{eq:vanVleckDet}) and (\ref{Delta-QNM}) in the $\phi \gtrsim \pi$ regime. The disagreement at small angles is not unexpected as the QNM sum is invalid at early times (or equivalently, for orbiting geodesics which have passed through small angles $\phi$).



\section{Summary} \label{sec:summary}
After a series of equations and results, for the benefit of the reader we will now take a moment to recap and (literally) rewrite the main
equations obtained thus far.
We recall that 
$\beta=-1/2+i\lambda$, $\lambda= \sqrt{(l+1/2)^2 + 4\xi - 1/2}$, $\Rs = \tanh^{-1}\R$,
that we tend to use
$T=t-t' - \Rs - \Rs^\prime $ as the independent `time' variable, and that the unbounded orbital angle is given by
$ \Delta \phi=
N  \pi + \gam$
for $N$ even and by
$ \Delta \phi=
(N+1) \pi - \gam$
for $N$ odd,
where $N$ is the number of caustics the null geodesic has gone through.

The main equations for the `retarded' Green function we have obtained so far are:
\begin{itemize}

\item QNM `retarded' Green function (valid $\forall x,x'$ where $T>0$), Eq.~(\ref{G-arbitrary}):
\begin{multline} \label{G-arbitrary,summary}
\GQNM(x,x^{\prime}) = 2 \, \text{Re} \sum_{l=0}^\infty (2l+1)  P_l(\cos \gam)  e^{ i \lam \left[ T - \ln(2/(1+\R)) - \ln(2/(1+\R^\prime)) \right] } \\  \quad \quad \quad \times \sum_{n=0}^{\infty} \Bef_{ln} \left[ \frac{4 e^{-T} }{ (1+\R)(1+\R^\prime) } \right]^{n+1/2} \mathcal{S}_{ln}(\R) \mathcal{S}_{ln}(\R^\prime).
\end{multline}
where $\Bef_{ln}$ are the excitation factors (\ref{EF2}) and $\mathcal{S}_{ln}(\R)$ is the finite series (\ref{series Sln}) essentially based on the
radial functions evaluated at the QNM frequencies.

\item QNM `retarded' Green function (after summing over all overtone numbers $n$) at $ \Rs,\Rs^\prime \rightarrow +\infty$ (and $T>0$),
Eq.~(\ref{GF-inf}):
\begin{multline}\label{GF-inf,summary}
\GQNM(T, \gam) \sim
\frac{ e^{-T / 2} }{\sqrt{\pi}} \, \text{Re} \sum_{l=0}^{+\infty} \frac{(l+1/2) \Gamma(i \lam)}{\Gamma(1/2 + i\lam)} \\ \times P_l(\cos \gam) e^{i \lam (T + 2 \ln 2)} {}_2F_1(-\beta, -\beta; -2\beta; -e^{-T}),\quad \Rs,\Rs^\prime \rightarrow +\infty,
\end{multline}
with singularities at times
given by Eq.~(\ref{sing-time-inf,bis}) (also Eq.~(\ref{sing-time-inf}) with $H=1$ and Eq.~(\ref{sing-time-inf,bis2})),
$T^{(\R\sim 1)} =  \ln \left[ \sinh^2 ( \phif / 2 ) \right]$.
The corresponding singularity structure is given by Eq.~(\ref{sing-Poisson-2}):
\beq \label{sing-Poisson-2,summary}
\II^{(\R\sim 1)}_N \sim  \left( \frac{ e^{-T} }{ 2 \sin \gam \sqrt{1 + e^{-T}} } \right)^{1/2} 
\begin{cases}
\displaystyle (-1)^{N/2} \, \delta\left( t-t'- t^{(\R\sim 1)}_{N} \right)  , \quad & N \; \text{even} , \\
\displaystyle \frac{(-1)^{(N+1)/2}}{\pi \left(t-t' - t^{(\R\sim 1)}_{N} \right)}  , \quad & N \; \text{odd},
\end{cases}
\eeq
where $\II^{(\R\sim 1)}_N$ is the contribution $\II_N$ to the `retarded' Green function associated with the $N$th orbiting null geodesic in the case where
the points are located at spatial infinity.

\item Contribution of the `fundamental mode' $n=0$ to the QNM `retarded' Green function (valid $\forall x,x'$ where $T>0$),
Eq.~(\ref{G-n0}):
\begin{multline} \label{G-n0,summary}
\Gret^{(n=0)}(x, x^{\prime}) =  \left( \frac{e^{-T} }{\pi(1+\R)(1+\R^\prime)} \right)^{1/2}\\ \times \text{Re}\sum_{l=0}^{+\infty} \frac{ (2l+1) P_l(\cos \gam) \Gamma(i \lam)}{\Gamma(1/2 + i\lam)} e^{i \lam \left[ T + \ln(1+\R)(1+\R^\prime) \right]},
\end{multline}
with singularities at times
given by Eq.~(\ref{T-periodic}) (also Eq.~(\ref{sing-time-n=0})),
$T^{(n=0)} =  \Delta \phi  - \ln[(1+\R)(1+\R^\prime)]$,
which coincide with $T^{(\R\sim 1)}$ (and with Eq.~(\ref{sing-time-inf}) with $H=1$) for $\R,\R'\to 1$ and $N\to +\infty$.
The corresponding singularity structure is given by Eq.~(\ref{sing-Poisson}):
\beq \label{sing-Poisson,summary}
\II_N^{(n=0)} \sim  \left( \frac{2 \, e^{-T } }{\sin\gam (1+\R)(1+\R^\prime)} \right)^{1/2} 
\begin{cases}
\displaystyle (-1)^{N/2} \delta\left( t-t' - t^{(n=0)}_{N} \right)  , \quad & N \; \text{even} , \\
\displaystyle \frac{(-1)^{(N+1)/2}}{\pi \left(t-t' - t^{(n=0)}_{N} \right)}  , \quad & N \; \text{odd} ,
\end{cases}
\eeq

(for more accurate asymptotics, which include the `tail' in the singularity, see Eqs.~(\ref{eq:I1-asymp}),~(\ref{eq:I2-asymp}))
where $\II_N^{(n=0)}$ is $\II_N$ for $n=0$ only.

\item The extension beyond the normal neighborhood of the `direct part' in the Hadamard form of the `retarded' Green function
has a contribution from the $N$th orbiting null geodesic given by (\ref{GNdir-Hadamard}):
\beq \label{GNdir-Hadamard,summary}
 G_{N}^{dir} \sim  \left( \frac{\eta}{\sinh \eta } \right)^{1/2} 
 \left( \frac{\theta}{\sin \theta } \right)^{1/2}
 \begin{cases}
 \displaystyle (-1)^{N/2} \delta(\sig), \quad \quad &  N \text{ even}, \\ 
 \displaystyle \frac{(-1)^{(N-1)/2}}{\pi \sig}, \quad & N \text{ odd},
 \end{cases} 
\eeq
which, for $\R\to 1$ and $\theta \gtrsim \pi$, agrees well with $\II^{(\R\sim 1)}_N$ with $\eta = H \theta$ (where $H>1$ for timelike geodesics, $H=1$ for null geodesics and $H<1$ for spacelike geodesics).

\end{itemize}

\section{Self-Force on the Static Particle}
\label{sec:sf}


In this section we turn our attention to a simple case: the self-force acting on a static scalar particle in the Nariai spacetime. By `static' we mean a particle with constant spatial coordinates. It is not necessarily at rest, since its worldline may not be a geodesic, and it may require an external force to keep it static.  This is analogous to the case of a static particle in Schwarzschild. In Sec.~\ref{subsec:static full Green} we explore an analytic method for computing the static self-force in Nariai spacetime.  This method, based on the massive field approach of Rosenthal \cite{Rosenthal:2003,Rosenthal:2004} provides an independent check on the matched-expansion approach. In Sec.~\ref{subsec:matched-static} we describe how the method of matched expansions may be applied to the static case. To compute the self-force, we require robust numerical methods for evaluating the quasinormal mode sums such as (\ref{G-arbitrary,summary}); two such methods are outlined in Sec. \ref{subsec:nummeth}. The results of all methods are validated and compared in Sec. \ref{sec:results}. 

\subsection{Static Green Function Approach} \label{subsec:static full Green}
The conventional approach to calculating the self-force on a static particle due to Wiseman \cite{Wiseman:2000} uses
the `scalarstatic' Green function. 
Following Copson \cite{Copson:1928}, Wiseman was able to obtain this Green function by summing the Hadamard series. 
Only by performing the full sum was he able to verify that his Green function satisfied the appropriate boundary conditions.
 Linet \cite{Linet:2005} has classified all spacetimes in which the scalarstatic equation is solvable by the Copson ansatz \cite{Copson:1928}
and, unfortunately, the Nariai metric does not fall into any of the classes given. Therefore, instead we work with the mode form 
for the static Green function. This corresponds to the integrand at $\omega=0$ of Eq.~(\ref{full-ret-Green-mode-sum}) (with integral measure $\frac{d\omega}{2\pi}$), 
 \beq \label{Green static}
G_{static}(\R,\Omega; \R',\Omega') =
\sum_{l=0}^{\infty}
(2l+1)P_l(\cos\gamma)
\frac{\pi P_{ -1/2 + i \lam}(-\R_<)P_{ -1/2 + i \lam}(\R_>)} {2 \cosh (\pi \lambda )}
\eeq
where, as before, $\lam = \sqrt{(l+\frac12)^2 +d}$. This equation, having only one infinite series, is amenable to numerical computation.

To regularize the self-force we follow the method of Rosenthal \cite{Rosenthal:2004}, who used a massive field approach to calculate 
the static self-force in Schwarzschild.  Following his prescription, we calculate the derivative of the scalar field and of a massive scalar field. 
In the limit of the field mass going to infinity, we obtain the derivative of the radiative field which is regular.
This method can be carried through to Nariai spacetime (for details see Appendix~\ref{app:rosenthal-massive}), where it  yields the expression
\beq \label{S-F static}
m a^\R = q^2(1-\R^2)^\frac32  \lim_{\R'\to \R^-} \left [ \partial_\R  G_{static}(\R,\Omega;\R',\Omega) +   \frac{1}{(\R-\R')^2} - \frac{ (\xi R- \frac23)}{2(1-\R^2)} \right],
\eeq
where $R=4$ here.
Note that the massive field approach was developed in~\cite{Rosenthal:2004} for the case $\xi=0$, but it can be seen that it equally
applies to $\xi\neq 0$ too.
The limiting process $\R'\to \R^-$ does not affect the term with $1/(1-\R^2)$ in (\ref{S-F static}). 
We will therefore omit it from the following calculation and we will trivially subtract it from the self-force in the calculations we carry out 
later in Sec.~\ref{subsec:results:SF}. 
The singular (as $\R'\to \R^-$) subtraction term may be expressed in a convenient form using the identity~\cite{Erdelyi:1953,Candelas:Jensen:1986},
\beq \label{reg terms}
\int_0^{+\infty}d\lambda 
\lambda \tanh (\pi\lambda)
\frac{\pi P_{ -1/2 + i \lam}(-\R_<)P_{ -1/2 + i \lam}(\R_>)}{\cosh (\pi\lambda)}=\frac{1}{\R_>-\R_<}
\eeq
Subtracting  (\ref{reg terms}) from (\ref{Green static}), we can express the regularized Green function as 
a sum of two well-defined and easily calculated sums/integrals
\beq \label{Gstatic reg}
G_{static}(\R,\Omega; \R',\Omega) - \frac{1}{\R_>-\R_<} = \mathcal{I} (\R, \R') + \mathcal{J} (\R, \R') 
\eeq
 where
 \beq
\mathcal{I} (\R, \R') = 
   \int_0^{+\infty}d\lambda ~
\lambda \left(1-\tanh (\pi\lambda)\right)
\frac{\pi P_{ -1/2 + i \lam}(-\R_<)P_{ -1/2 + i \lam}(\R_>)}{\cosh (\pi\lambda)}
\eeq
and
\begin{multline}
\mathcal{J} (\R,\R')  = \sum_{l=0}^{+\infty}
(l+{\textstyle{\frac12}})
\frac{\pi P_{ -1/2 + i \lam}(-\R_<)P_{ -1/2 + i \lam}(\R_>)} {\cosh (\pi \lambda )} \\- 
 \int_0^{+\infty}d\lambda \> \lambda
\frac{\pi P_{ -1/2 + i \lam}(-\R_<)P_{ -1/2 + i \lam}(\R_>)}{\cosh (\pi\lambda)}\>.
\end{multline}
The term $\mathcal{J} (\R,\R') $  may either be calculated directly as a sum or by 
using the Watson-Sommerfeld transform to write
\beq
\sum\limits_{l=0}^\infty g\left(l+\frac12\right) =  \text{Re}  \left[ \frac{1}{i} \int_\gamma dz \> \tan (\pi z) g(z) \right]  ,
\eeq
where $\gamma$ runs from $0$ to $\infty$ just above the real axis, and for us
\beq
g(z) = z
\frac{\pi P_{ -1/2 + i \sqrt{z^2 +d}}(-\R_<)P_{ -1/2 + i  \sqrt{z^2 +d}}(\R_>)} {\cosh (\pi  \sqrt{z^2 +d} )} .
\eeq
Writing 
\beq
 \tan (\pi z) = i - \frac{2 i} {1 + e^{- 2 \pi i z}},
\eeq
the first term yields 
\beq
\int_{0}^\infty  dx\>g(x) = \int_{\sqrt{d}}^{\infty}d\lambda \> \lambda
\frac{\pi P_{ -1/2 + i \lam}(-\R_<)P_{ -1/2 + i \lam}(\R_>)}{\cosh (\pi\lambda)}.
\eeq
The contribution from the second term can be best evaluated by rotating the original contour  to a contour
 $\gamma'$, running from $0$ to $i \infty$ just to the right of the imaginary axis.
 This is permitted since the Legendre functions are analytic functions of their parameter
and the contribution from the arc at infinity vanishes for our choice of $g(z)$.
From the form of $g(z)$ it is clear that it possesses poles along the contour $\gamma'$ but these give a 
purely imaginary contribution to the integral.  We conclude that
\begin{align}
   \mathcal{J} (\R, \R') &= - \int_0^{\sqrt{d}}   t dt\>  \tanh(\pi \sqrt{d-t^2})\frac{\pi P_{ -1/2 +i t }(-\R_<)P_{ -1/2 +i t }(\R_>)}{\cosh(\pi t)} 
   + \nonumber \\ & \qquad + \mathcal{P} \int_0^{+\infty} \frac{2\pi  t dt} { (1+e^{2\pi \sqrt{d+t^2}})\cos(\pi t)} P_{ -1/2 -t }(-\R_<)P_{ -1/2 -t }(\R_>)
\end{align}
where $\mathcal{P}$ denotes the Principal Value. These integrals and that defining $\mathcal{I} (\R,\R')$ and their derivatives 
with respect to $\R$ are very rapidly convergent and easily calculated.


\subsection{Matched Expansions for Static Particle} \label{subsec:matched-static}
In Sec.~\ref{sec:matched-expansions}, we outlined the method of matched expansions. In this subsection, we show how to apply the method to a specific case: the computation of the self-force on a static particle in the Nariai spacetime. 

The four velocity of the static particle is simply 
\beq
u^\R = u^\theta = u^\phi = 0, \quad u^t = (1-\R^2)^{-1/2}
\eeq
and hence $d \tau^\prime = (1-\R^2)^{1/2} dt^\prime$. 
We find from Eqs.~(\ref{eq:ma}), (\ref{eq:dmdtau}) and (\ref{eq:rad-field-deriv}) that $ma^t = ma^\theta = ma^\phi = 0$ and 
\begin{align}
 ma^\R &= q^2\left( \frac{1}{3} \dot{a}^\R + \lim_{\eps \rightarrow 0^+} \int_{-\infty}^{\tau-\eps}  g^{\R\R} \partial_\R \Gret (z(\tau), z(\tau^\prime)) d \tau^\prime\right) \label{rad-acc-eq} \\
 \frac{dm}{d\tau} &= -q^2 \left( \frac{1}{12}(1 - 6\xi) R + \left( 1-\R^2 \right)^{-1/2} \lim_{\eps \rightarrow 0^+} \int_{-\infty}^{\tau-\eps} \partial_t \Gret (z(\tau), z(\tau^\prime)) d \tau^\prime\right)  \label{mass-loss-eq}
\end{align}
We note that in the tail integral of the mass loss equation, (\ref{mass-loss-eq}),  the time derivative $\partial_t$ may be replaced with $-\partial_{t^\prime}$ since the retarded Green function is a function of $(t-t')$. Hence we obtain a total integral,
\begin{align}
\left( 1-\R^2 \right)^{-1/2} \lim_{\eps \rightarrow 0^+} \int_{-\infty}^{\tau-\eps} \partial_t \Gret (z(\tau), z(\tau^\prime)) d \tau^\prime 
&= - \lim_{\eps \rightarrow 0^+} \int_{-\infty}^{t - \eps} \partial_{t^\prime} \Gret (z(\tau), z(\tau^\prime)) d t^\prime \nonumber \\
&= - \lim_{\eps \rightarrow 0^+}\left[ \Gret (x,x') \right]_{t'\to-\infty}^{t'=t-\eps}
\end{align}
The total integral depends only on the values of the Green function at the present time and in the infinite past ($t'\rightarrow \infty$). The QNM sum expressions for the Green function (e.g. Eq.~(\ref{GF-inf,summary})) are zero in the infinite past, as the quasinormal modes decay exponentially. The value of the Green function at coincidence ($t^\prime \rightarrow t$) is found from the coincidence limit of the function $-V(x,x')$ in the Hadamard form (\ref{eq:Hadamard}). It is
$
\frac{1}{12}(1 - 6\xi) R
$, 
which exactly cancels the local contribution in the mass loss equation (\ref{mass-loss-eq}). It is no surprise to find that this cancellation occurs -- the local terms were originally derived from the coincidence limit of the Green function. In fact, because $\frac{d \Phi_R}{d \tau} = 0$ due to time-translation invariance, we can see directly from the original equation (\ref{eq:dmdtau}) that the mass loss is zero in the static case. 

Now let us consider the radial acceleration (\ref{rad-acc-eq}). The acceleration keeping the particle in a static position is constant ($\dot{a}^\R = 0$). The remaining tail integral may be split into two parts,
\begin{multline}
ma^\R = 
 q^2  (1-\R^2)^{3/2}  \Bigg[ - \lim_{\eps \rightarrow 0^+} \int_{t-\Delta t}^{t - \eps} \partial_{\R} V  (z(t), z(t^\prime)) d t^\prime \\ + \int_{-\infty}^{t- \Delta t} \partial_{\R} \Gret  (z(t), z(t^\prime)) d t^\prime \Bigg],
 \label{eq:sf-matched}
\end{multline}
where $\Delta t$ is a free parameter corresponding to $\Delta\tau$ introduced in (\ref{eq:fld-QL-DP})
that determines the matching time in the matched expansions method.
For the first part of (\ref{eq:sf-matched}), we use the quasilocal calculation of $V(x,x')$ from Sec.~\ref{subsec:QL}. As $V(x,x'$) is given as a power series in $(\R-\R')$ and $(t-t')$, the derivatives and integrals can be done termwise and are straightforward. The quasilocal integral contribution is therefore simply
\beq
\lim_{\eps \rightarrow 0} \int_{t-\Delta t}^{t - \eps} \partial_{\R} V (z(t), z(t^\prime)) d t^\prime = \frac{1}{2} \sum_{k=0}^{+\infty} \frac{1}{(2k+1)} \partial_\R v_{k0} ( \Delta t )^{2k+1}.
\label{eq:sf-ql}
\eeq

The second part of (\ref{eq:sf-matched}) can be computed using the QNM sum (\ref{G-arbitrary,summary}). To illustrate the approach, let us rewrite (\ref{G-arbitrary,summary}) as 
\beq
\GQNM (\R,t;\R',t;\gamma)= \text{Re} \sum_{l,n} \mathcal{G}_{ln}(\R^\prime,\gamma) e^{- i \omega_{ln} (t - t^\prime - \Rs - \Rs^\prime)} \unorm_{ln}(\R).
\eeq
Applying the derivative with respect to $\R$ and taking the integral with respect to $t^\prime$ leads to
\begin{align}
&\int_{-\infty}^{t - \Delta t} \partial_{\R} \Gret  (z(t), z(t^\prime)) d t^\prime \nonumber \\
&\quad=
\left( \frac{d\R}{d\Rs} \right)^{-1}  \int_{-\infty}^{t - \Delta t} \partial_{t^\prime} \GQNM dt^\prime
+ \text{Re} \sum_{l,n} \int_{-\infty}^{t-\Delta t} \mathcal{G}_{ln} e^{- i \omega_{ln} (t - t^\prime - \Rs - \Rs^\prime)}  \frac{d \unorm_{ln}}{d \R} dt^\prime \nn
 \\
&\quad= \left( 1-\R^2 \right)^{-1} \left[ \Gret \right]^{t'=t - \Delta t} + \sum_{l,n} \frac{  \mathcal{G}_{ln} }{ i \omega_{ln} } e^{- i \omega_{ln} (\Delta t - \Rs - \Rs^\prime)} \frac{d \unorm_{ln}}{d \R}
\label{eq:sf-dp}
\end{align}
It is straightforward to find the derivative of the radial wavefunction from the definition (\ref{unorm-ser}). 
In Sec.~\ref{subsec:nummeth} we outline two methods for numerically computing mode sums such as (\ref{eq:sf-dp}).

The self-force computed via (\ref{eq:sf-matched}), (\ref{eq:sf-ql}) and (\ref{eq:sf-dp}) should be independent of the choice of the matching time (we verify this in Sec.~\ref{subsec:results:SF}). This invariance provides a useful test of the validity of our matched expansions. Additionally, through varying $\Delta \tau$ we may estimate the numerical error in the self-force result. 

\subsection{Numerical Methods for Computing Mode Sums *}\label{subsec:nummeth}
The static-self-force calculation requires the numerical calculation of mode sums like (\ref{eq:sf-dp}). 
We used two methods for robust numerical calculations: (1) `smoothed sum', and (2) Watson transform (described previously in Sec.~\ref{subsec:watson}). We see in Sec. \ref{sec:results} that the results of the two methods are consistent.

The `smoothed sum' method is straightforward to describe and implement.  Let us suppose that we wish to extract a meaningful numerical value from an infinite series
\beq
 \sum_{l=0}^{+\infty} a_l    \label{sum1}
\eeq
such that
$\lim_{l \rightarrow \infty} a_{l}\neq 0$, and so the series is divergent. One may expect it to not be possible for a divergent series to be summed to yield a finite value. However, the summation of divergent series is on a firm footing, having been extensively studied with significant rigour (for example, see Hardy~\cite{Hardy}). For the case of mode sums such as \eqref{eq:sf-dp}, we employ the `$\phi$' method (Sec.~4.7 and Theorem~25 of Ref.~\cite{Hardy}) for summing the divergent series. In this method, the terms, $a_l$, of the series are multiplied by a function $\phi_l(x)$ such that $\phi_l(x) \to 1$ for each $l$ as $x \to 0$. In particular, we choose\footnote{A similar `regulator' function was proposed by Ching, et al. \cite{Ching:1995b}.}
\begin{equation}
 \phi_l(x) = e^{-l^2 x^2/2}
\end{equation}
and define $l_{\rm cut} \equiv \frac{1}{x}$. The `$\phi$' method is regular\footnote{A method for summing a divergent sum is said to be \emph{regular} if it sums every convergent series to its ordinary sum.} provided the condition 
\begin{equation}
 0 \le \phi_{n+1}(x) \le \phi_{n}(x)
\end{equation}


Thus instead of computing the sum \eqref{sum1}, we may compute the sum
\beq
\phi\left(\frac{1}{\lmax}\right) = \sum_{l=0}^{\infty} a_l e^{-l^2 / 2 \lmax^2},  \label{num-meth1}
\eeq
where $\lim_{\lmax\to\infty} \phi\left(\frac{1}{\lmax}\right) = s$ and $s$ is the finite value we extract from the sum \eqref{sum1}.
In practise, for our numerical approximation it is sufficient to compute the finite sum
\beq
S\left(l_\infty\right)\equiv\sum_{l=0}^{l_\infty} a_l e^{-l^2 / 2 \lmax^2}.
\eeq
The choice of $l_\infty$ must be large enough to suppress any high-$l$ oscillations in the result (typically $l_\infty > 4\lmax$). We find that (\ref{num-meth1}) is a good approximation to (\ref{sum1}) provided we are not within $\delta t \sim 1/\lmax$ of a singularity of the Green function. Increasing the cutoff $\lmax$ therefore improves the resolution of the singularities.
The introduction of such a cutoff is  -- although of different magnitude -- in the same spirit as the Feynman prescription of Sec.~\ref{subsec:Hadamard}.



\section{Results}\label{sec:results}
We now present a selection of results from our numerical calculations. 
In Sec. \ref{subsec:results:GF} the distant past Green function is examined. We plot the Green function as a function of coordinate time $t - t'$ for fixed spatial points. A four-fold singularity structure is observed. In Sec.~\ref{subsec:results-asymptotics} we test the asymptotic approximations of the singular structure, derived in Secs. \ref{subsec:Hadamard} and \ref{subsec:Poisson} (Eqs. \ref{sing-Poisson-2,summary} and \ref{GNdir-Hadamard,summary}). We show that the `fundamental mode' ($n=0$) series (\ref{G-n0,summary}) is a good approximation of the exact result (\ref{GF-inf,summary}), if a `time-offset' correction is applied. In Sec. \ref{subsec:results:matched} the quasilocal and distant past expansions for the Green function are compared and matched. We show that the two methods for finding the Green function are in excellent agreement for a range of matching times $\Delta \tau$. In Sec. \ref{subsec:results:SF} we consider the special case of the static particle. We present the Green function, the tail field and the self-force in turn. The radial self-force acting on the static particle is computed via the matched expansion method (described in Secs. \ref{sec:matched-expansions} and \ref{subsec:matched-static}), and plotted as a function of the radial coordinate $\R$, and compared with the result derived in Sec.~\ref{subsec:static full Green}.

 \subsection{The Green Function Near Infinity from Quasinormal Mode Sums *}\label{subsec:results:GF}
 


Let us begin by looking at the Green function for fixed points near spatial infinity, $\R = \R^\prime \rightarrow 1$ (i.e. $\Rs = \Rs^\prime \rightarrow +\infty$). The Green function may be computed numerically by applying either the Watson transform (Sec.~\ref{subsec:watson}) or the `smoothed sum' method (Sec.~\ref{subsec:nummeth}) to the QNM sum (\ref{GF-inf,summary}).

\begin{figure}
 \begin{center}
  \includegraphics[width=10cm]{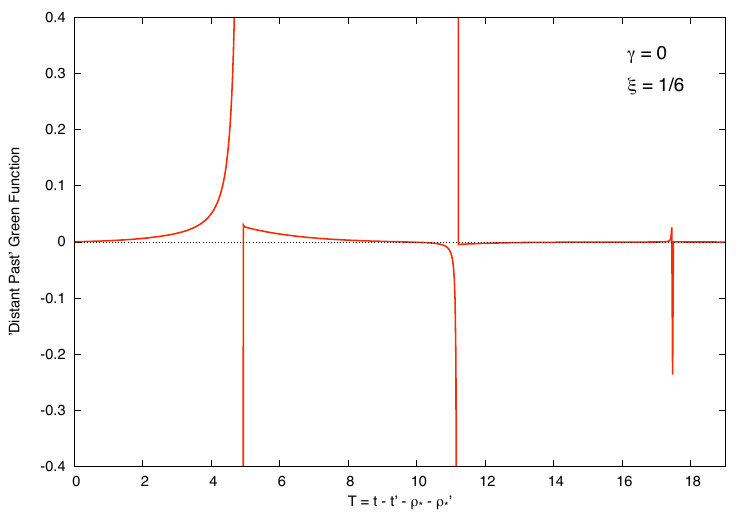}
 \end{center}
 \caption[Distant Past Green Function for spatially-coincident points near infinity ($\R=\R^\prime \rightarrow 1$, $\gam = 0$)]{\emph{Distant Past Green Function for spatially-coincident points near infinity ($\R=\R^\prime \rightarrow 1$, $\gam = 0$)}. The Green function was calculated from mode sum (\ref{GF-inf,summary}) numerically using the smoothed sum method (\ref{num-meth1}) with $\lmax = 200$ and curvature coupling $\xi = 1/6$.}
 \label{gf-inf-gam0}
\end{figure}

Figure \ref{gf-inf-gam0} shows the Green function for fixed spatially-coincident points near infinity ($\R = \R^\prime \rightarrow 1$, $\gam = 0$)
and $\xi=1/6$. The Green function has been calculated from series (\ref{GF-inf,summary}) using the `smoothed sum' method. It is plotted as a function of QNM time, $T =t- t^\prime - (\Rs + \Rs^\prime)$. We see that singularities occur at the times (\ref{sing-time-inf}) predicted by the geodesic analysis of Sec. \ref{subsec:Nariai-geodesics}. In this case, $T_{C} =  \ln[\sinh^2 (N \pi)] \approx 4.893$, $11.180$, $17.463$, etc. At times prior to the first singularity at $T \approx 4.893$, the Green function shows a smooth power-law rise. At the singularity itself, there is a feature resembling a delta-distribution, with a negative sign. Immediately after the singularity the Green function falls close to zero (although there does appear a small `tail'). This behavior is even more marked in the case $\xi = 1/8$ (not shown). A similar pattern is found close to the second singularity at $T \approx 11.180$, but here the Green function takes the opposite sign, and its amplitude is smaller.

\begin{figure}
 \begin{center}
  \includegraphics[width=10cm]{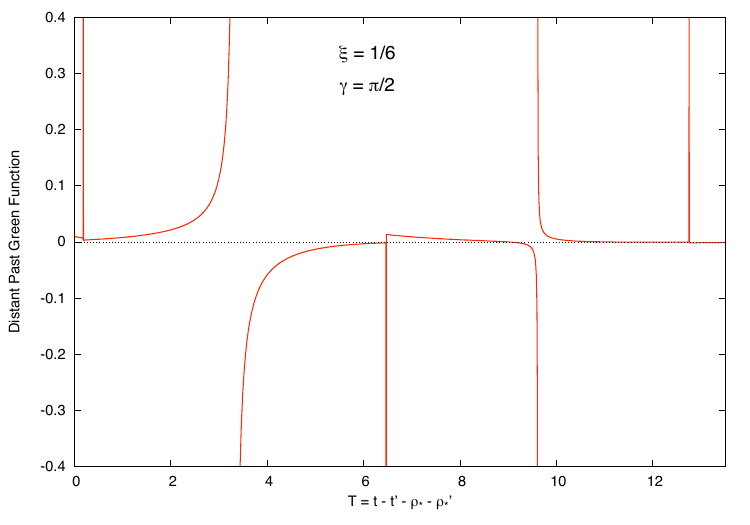}
 \end{center}
 \caption[Distant Past Green Function near spatial infinity ($\R=\R^\prime \rightarrow 1$) for points separated by angle $\gam = \pi /2 $]{\emph{Distant Past Green Function near spatial infinity ($\R=\R^\prime \rightarrow 1$) for points separated by angle $\gam = \pi /2 $}
and $\xi=1/6$. The Green function was calculated from the `fundamental mode' approximation (\ref{G-n0,summary}) numerically using the smoothed sum method with $\lmax = 1000$. Note the four-fold singularity structure (see text). }
 \label{gf-inf-piby2}
\end{figure} 

Figure \ref{gf-inf-piby2} shows the Green function for points near infinity ($\R=\R^\prime \rightarrow1$) separated by an angle of $\gam = \pi / 2$
and $\xi=1/6$. The Green function shown here is computed from the `fundamental mode' approximation (\ref{G-n0,summary}), again using the `smoothed sum' method. In this case, the singularities occur at `periodic' times (\ref{sing-time-periodic}), given by $T = \Delta \phi - 2 \ln 2 \approx 0.1845, 3.326, 6.468,$ etc., where $\Delta \phi = \pi / 2, 3 \pi /2, 5 \pi /2, \ldots$. As discussed, there is a one-to-one correspondence between singularities and orbiting null geodesics, and the four-fold singularity pattern predicted in Sec. \ref{subsec:Poisson} (\ref{GNdir-Hadamard,summary}) and Sec.~\ref{subsec:Hadamard} (\ref{sing-Poisson-2,summary}) is clearly visible. Every `even' singularity takes the form of a delta distribution. Numerically, the delta distribution is manifest as a Gaussian-like spike whose width (height) decreases (increases) as $\lmax$ is increased. By contrast (for $\gam \neq 0, \pi$), every `odd' singularity diverges as $1/(T-T_c)$; it has  antisymmetric wings on either side. The singularity amplitude diminishes as $T$ increases.

\subsection{Asymptotics and Singular Structure *}\label{subsec:results-asymptotics}
\begin{figure}
 \begin{center}
  \includegraphics[width=14cm]{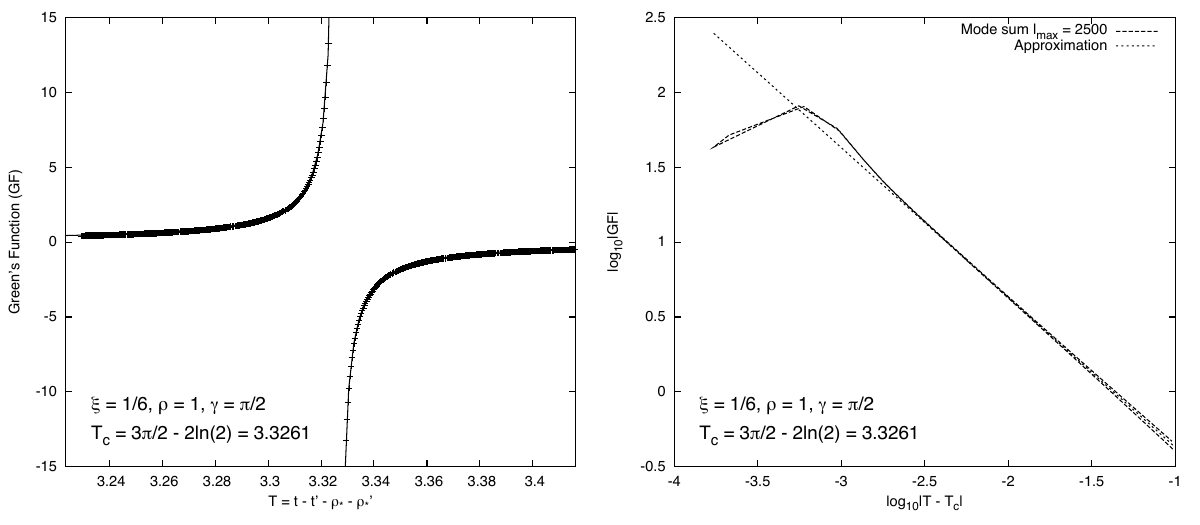}
 \end{center}
 \caption[Green function near the singularity arising from a null geodesic passing through an angle $\Delta \phi = 3\pi/2$ and with $\R = \R^\prime \rightarrow 1$]{\emph{Green function near the singularity arising from a null geodesic passing through an angle $\Delta \phi = 3\pi/2$ and with $\R = \R^\prime \rightarrow 1$}. The `fundamental mode' ($n=0$) Green function (\ref{G-n0,summary}) (with $\lmax = 2500$) is compared with approximations (\ref{sing-Poisson-2,summary}) and (\ref{GNdir-Hadamard,summary}) from considering high-$l$ asymptotics. The approximations give $\Gret \sim -0.04268 / (T - T_c)$ and $\Gret \sim -0.04344 / (T-T_c)$, respectively.  The left panel shows the Green function in the vicinity of the (`periodic') singularity at $T_c  = 3\pi/2 - 2 \ln 2 \approx 3.3261$. The right panel shows the same data on a log-log scale, and compares the mode sum (dashed) with the approximation (dotted). The discrepancy close to the singularity may be improved by increasing $\lmax$.}
 \label{gf-sing-3piby2}
\end{figure}

The analyses of Sec.~\ref{subsec:Poisson} and Sec.~\ref{subsec:Hadamard} yielded approximations for the singularity structure of the Green function. In particular, Eq.~(\ref{sing-Poisson-2,summary}) gives an estimate for the amplitude of the `odd' singularities as $\R ,\R^\prime \rightarrow 1$. We tested our numerical computations against these predictions. Figure \ref{gf-sing-3piby2} shows the Green function near the singularity associated with the null geodesic passing through an angle $\Delta \phi = 3\pi / 2$. The left plot compares the numerically-determined Green function (\ref{G-n0,summary}) with the asymptotic prediction (\ref{sing-Poisson-2,summary}). The right plot shows the same data on a log-log plot. The asymptotic prediction (\ref{sing-Poisson-2,summary}) is a straight line with gradient $-1$, and it is clear that the numerical data is in excellent agreement.

\begin{figure}
 \begin{center}
  \includegraphics[width=10cm]{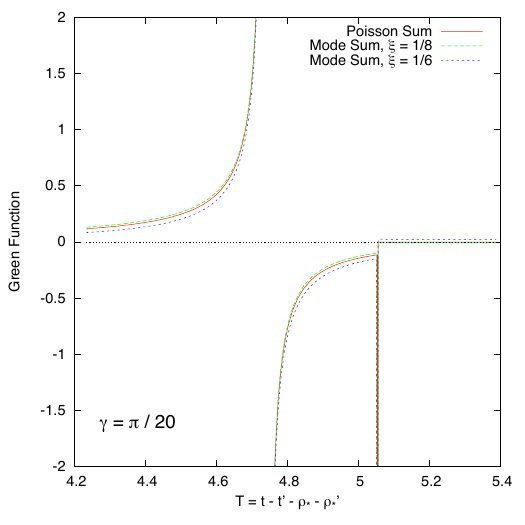}
  \includegraphics[width=10cm]{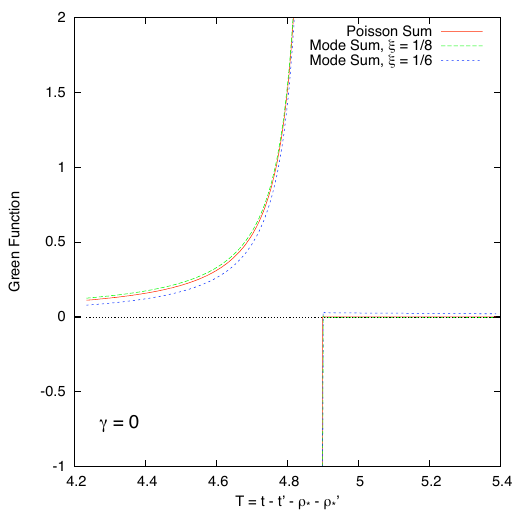}
 \end{center}
 \caption[Singularities of the `Fundamental Mode' Green function compared with asymptotics from the Poisson sum]{\emph{Singularities of the `Fundamental Mode' Green function (\ref{G-n0,summary}) compared with asymptotics from the Poisson sum (\ref{eq:I1-asymp}) and (\ref{eq:I2-asymp})}. The upper plot shows a small angular separation $\gam = \pi / 20$, and the lower plot shows coincidence $\gam = 0$, for $\rho,  \rho' \rightarrow 1$.}
 \label{fig:I1I2-numerical}
\end{figure}

Improved asymptotic expressions for the singular structure of the fundamental mode Green function were given in (\ref{eq:I1-asymp}) and (\ref{eq:I2-asymp}). These asymptotics are valid all the way up to $\gam = 0$. Figure \ref{fig:I1I2-numerical} compares the asymptotic expressions (\ref{eq:I1-asymp}) and (\ref{eq:I2-asymp})(solid line) with numerical computations (broken lines) from the mode sum (\ref{G-n0,summary}). It is clear that the asymptotics  (\ref{eq:I1-asymp}) and (\ref{eq:I2-asymp}) are in excellent agreement with the numerically-determined Green function. Closest agreement is found near the singular times, but the asymptotics provide a remarkably good fit over a range of $t$. 

\begin{figure}
 \begin{center}
  \includegraphics[width=15cm]{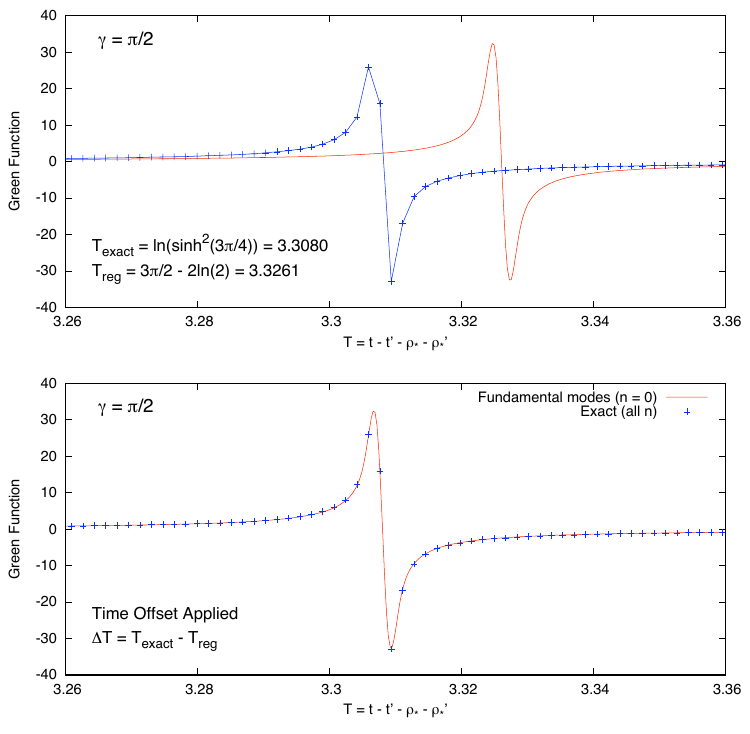}
 \end{center}
 \caption[Singularities of `Fundamental Mode' approximation]{\emph{Singularities of `Fundamental Mode' approximation}.  The top plot shows the Green function near spatial infinity ($\R = \R^\prime\rightarrow 1$, $\Rs = \Rs^\prime \rightarrow \infty$ ), for an angular separation $\gam = \pi / 2$. It compares the exact solution (\ref{GF-inf,summary}) [plain solid (red) line] and the approximation from the fundamental modes [crossed (blue) line], as a function of time $T = t - t^\prime - \Rs - \Rs^\prime$. The singularities occur at two distinct times, marked $T_{exact}$ and $T_{reg}$ respectively. If a time offset is applied to the `Fundamental Mode' approximation (see text), then we find the singularities look remarkably similar (lower plot). }
 \label{fig:singularities}
\end{figure}

In Fig.~\ref{fig:singularities}, the `fundamental mode' ($n=0$) approximation (\ref{G-n0,summary}) is compared with the exact QNM Green function (\ref{GF-inf,summary}) near spatial infinity. Away from singularities, the former is found to be a good approximation to the latter. However, close to singularities this is not the case. The singularities of the `fundamental mode' approximation (\ref{G-n0,summary}) occur at slightly different times to the singularities of the exact solution (\ref{GF-inf,summary}), as discussed in Sec. \ref{subsec:large-l}. For the fundamental mode series (\ref{G-n0,summary}), the singularity times $T_{reg}^N$ given in Eq.~(\ref{T-periodic}) are periodic. For the exact solution (\ref{GF-inf,summary}) near spatial infinity, the singularity times $T_{exact}^{N}$ are precisely the `null geodesic times' given in Eq.~(\ref{sing-time-inf}). Remarkably, if we apply a \emph{singularity time offset} to the `fundamental mode' approximation ($T \rightarrow T + \Delta T$  where $\Delta T = T_{exact}^{N} - T_{reg}^{N}$) we find that the `fundamental mode' Green function is an almost perfect match to the exact Green function. This is clearly shown in the lower plot of Fig.~\ref{fig:singularities}. Comparing the series (\ref{G-n0,summary}) and (\ref{GF-inf,summary}) we see that, in both cases, the magnitude of the terms in the series increases as $(l+1/2)^{1/2}$ in the large-$l$ limit
(see Eqs.(\ref{Gl R->1,large l}) and (\ref{F n=0,large l})). This observation raises the possibility that the $n=0$ modes may give the essential features of the full solution; if true, this would certainly aid the analysis of the Schwarzschild case, where it is probably not feasible to perform a sum over $n$ analytically.

 \subsection{Matched Expansions: Quasilocal and Distant Past}\label{subsec:results:matched}

Let us now turn our attention to the match between quasilocal and distant past Green functions. The 
quasilocal expansion (\ref{eq:CoordGreen}) is valid within the convergence radius of the series, $t-t^\prime<  t_{QL}$, while the QNM $n$-series is convergent at `late' times, $t-t' > \Rs + \Rs'$. 
Hence, a matched expansion method will only be practical if the quasilocal and distant past Green functions overlap in an intermediate regime $\Rs+\Rs^\prime < t-t' < t_{QL}$. It is expected that the convergence radius of the quasilocal series, $ t_{QL}$, will lie within the normal neighborhood, $t_{NN}$, of spacetime point $x$. The size of the normal neighborhood is limited by the earliest time at which spacetime points $x$ and $x'$ may be connected by more than one non-spacelike geodesic. Typically this will happen when a null geodesic has orbited once, taking a time $t_{NN} > \Rs + \Rs^\prime$, so we can be optimistic that an intermediate regime will exist. To test this idea, we computed the quasilocal Green function using (\ref{eq:CoordGreen}), and the distant past Green function (\ref{G-arbitrary,summary})
(or its approximation (\ref{G-n0,summary}) for $n=0$) for a range of situations.

\begin{figure}
 \begin{center}
  \includegraphics[width=14cm]{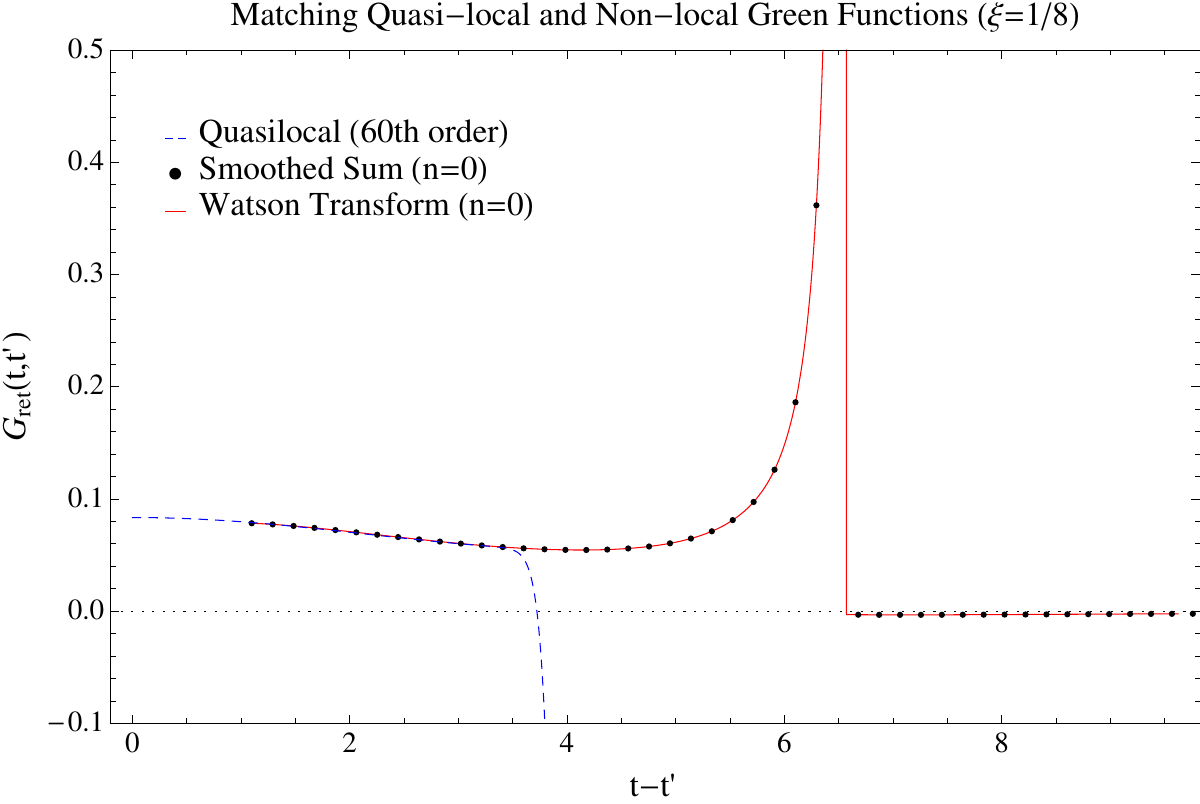}
 \end{center}
 \caption[Matching of the quasilocal and distant past Green functions (I)]{\emph{Matching of the Quasilocal and Distant Past Green Functions} for $\xi=1/8$ and for a static particle at $\R = \R^\prime = 0.5$. Here, the Green functions are plotted as functions of coordinate time, $t - t^\prime$ (\emph{not} `QNM time' as in most other plots). The (black) dots and solid (red) line show the results of the `smoothed sum' and `Watson transform' methods applied to compute the $n=0$ QNM sum (\ref{G-n0,summary}) (see text). The dashed (blue) line shows the quasilocal series expansion taken to order $(t- t^\prime)^{60}$. The distant past Green function cannot be computed for early times $t-t^\prime< \Rs + \Rs^\prime = 2 \tanh^{-1}(1/2) = 1.099$, whereas the quasilocal series diverges at large $t-t^\prime \gtrsim 3.46$. In the intermediate regime, we find excellent agreement (see also Figs. \ref{matching} and \ref{matcherr}). Note that the quasilocal Green function tends to $\frac{1}{12}(1 - 6 \xi)R = \frac{1}{12}$ in the limit $t^\prime \rightarrow t$.}
 \label{matching2}
\end{figure}

Figure \ref{matching2} shows the $n=0$ retarded Green function (\ref{G-n0,summary}) as a function of coordinate time $t - t'$ for a static particle at $\R = \R^\prime = 0.5$. At early times, the quasilocal Green function is well-defined, but the distant past Green function is not. Conversely, at late times the quasilocal series is not convergent. At intermediate times $1.099 < \Delta t \lesssim 3.45$, we find an excellent match. Figure \ref{matching2} also shows that the results of the two numerical methods for evaluating QNM sums are equivalent. That is, the Green function found from the Watson transform (Sec.~\ref{subsec:watson}, red line) coincides with the Green function calculated by the method of smoothed sums (Sec.~\ref{subsec:nummeth}, black dots).

\begin{figure}
 \begin{center}
  \includegraphics[width=10cm]{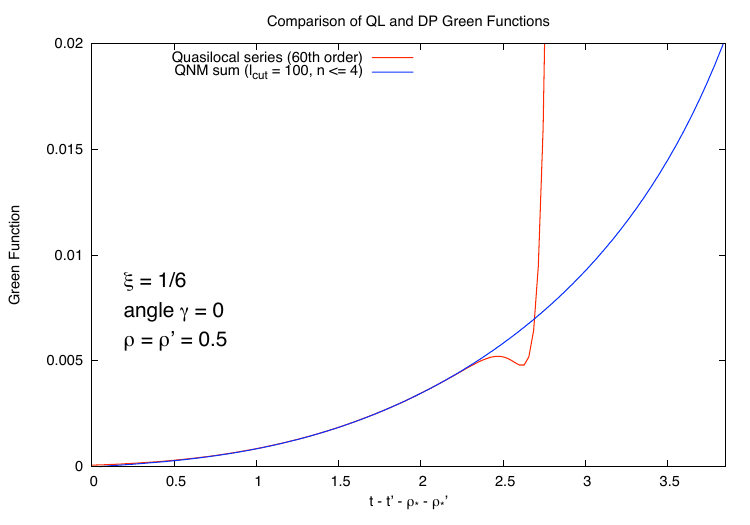}
  \includegraphics[width=10cm]{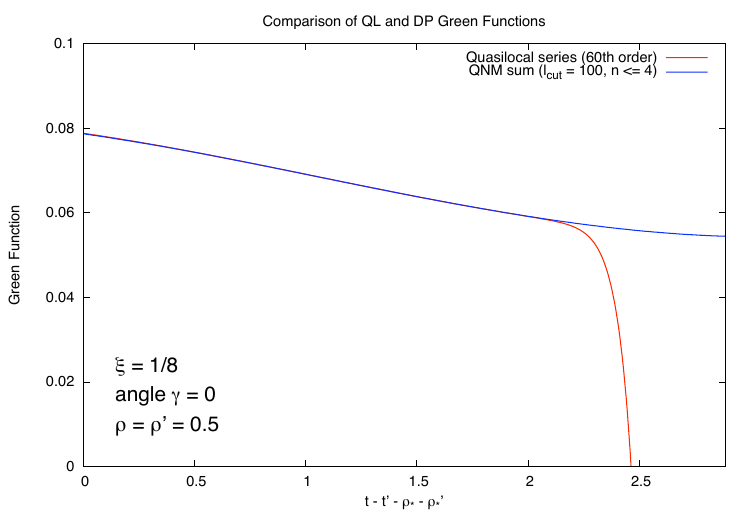}
 \end{center}
 \caption[Matching of the quasilocal and distant past Green functions (II)]{\emph{Matching of the Quasilocal and Distant Past Green Functions.} The upper plot shows curvature coupling $\xi = 1/6$ and the lower plot shows $\xi = 1/8$, for a static particle at $\R = \R^\prime = 0.5$. Note the timescale on the horizontal axis, $T = t - t^\prime - \Rs - \Rs^\prime$. The distant past Green function
 [(blue) lines reaching up to the last value for $T$]
  cannot be computed for $T < 0$, whereas the quasilocal series 
  [(red) lines {\it not} reaching up to the last value for $T$]
  clearly diverges at large $T$. In the intermediate regime, we find excellent agreement.}
 \label{matching}
\end{figure}

Let us now examine the matching procedure in more detail. Figure \ref{matching} shows the match between the distant past and quasilocal Green functions, computed from (\ref{G-arbitrary,summary}) and (\ref{eq:CoordGreen}), in the case $\R = \R^\prime = 0.5$. 
The double sum in (\ref{G-arbitrary,summary}) has been calculated by applying the `smoothed sum' method (\ref{num-meth1})
to the $l$-sum and by summing over $n$ up to the value $n_\text{max}=4$.
In Fig. \ref{matching}, the upper plot shows the case for conformal coupling $\xi = 1/6$ and the lower plot shows the case for $\xi = 1/8$. Note that Green function tends to the constant value $\frac{1}{12}(1 - 6 \xi)R$ in the limit $\Delta t \rightarrow 0^+$. In both cases, we find that the fit between `quasilocal' and `distant past' Green functions is good up to nearly the radius of convergence of the quasilocal series. 

\begin{figure}
 \begin{center}
  \includegraphics[width=14cm]{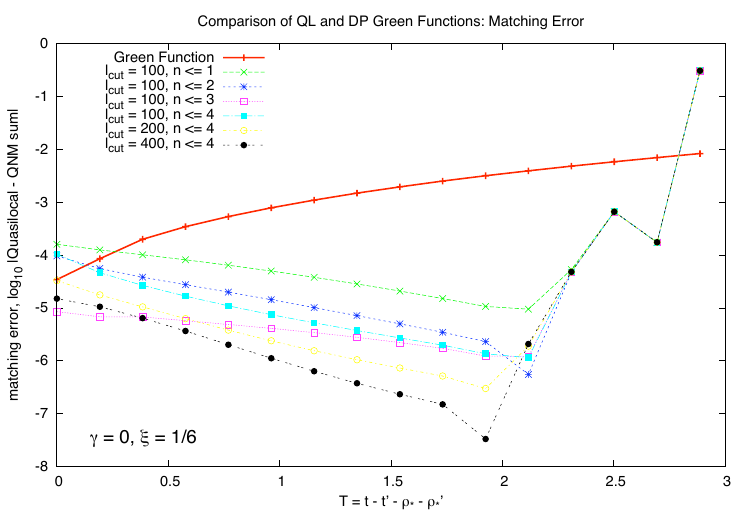}
 \end{center}
 \caption[Error in Matching the Quasilocal and Distant Past Green Functions.]{\emph{Error in Matching the Quasilocal and Distant Past Green Functions.} This plot shows the difference between the quasilocal and distant past Green functions in the matching regime. The magnitude of the Green function as a function of time $T = t - t^\prime - \Rs - \Rs^\prime$ is shown as a solid red line. The broken lines show the `matching error': the difference between the quasilocal and QNM sum Green functions, for various $\lmax$ in (\ref{num-meth1}) (see Sec.~\ref{subsec:nummeth}) and $\nmax$. Note the logarithmic scale on the vertical axis. It is clear that the matching error is reduced by increasing $\nmax$ and $\lmax$, and that the best agreement is found close to the radius of convergence of the quasilocal series (at $T \sim 1.9$). The plot shows that matching accuracy of above one part in $10^4$ is achievable.}
 \label{matcherr}
\end{figure}

Figure \ref{matcherr} quantifies the accuracy of the match between quasilocal and distant past Green functions. Here, we have used the `smoothed sum' method (Sec.~\ref{subsec:nummeth}) to compute the distant past Green function from (\ref{G-arbitrary,summary}). To apply this method, we must choose appropriate upper limits for $l$ (angular momentum) and $n$ (overtone number).  We have experimented with various cutoffs $\lmax$ and $\nmax$. As expected, better accuracy is obtained by increasing $\lmax$ and $\nmax$, although the run time for the code increases commensurately. With care, a relative accuracy of one part in $10^4$ to $10^5$ is possible. This accuracy is sufficient for confidence in the self-force values computed via matched expansions, presented in Sec. \ref{subsec:results:SF}.   

 \subsection{The Self-Force on a Static Particle}\label{subsec:results:SF}
In this section, we present a selection of results for a specific case: a `static' particle at fixed spatial coordinates. Our goal is to compute the self-force as a function of $\R$, to demonstrate the first practical application of the Poisson-Wiseman-Anderson method of matched expansions \cite{Poisson:Wiseman:1998, Anderson:Wiseman:2005}.

\begin{figure}
 \begin{center}
  \includegraphics[width=14cm]{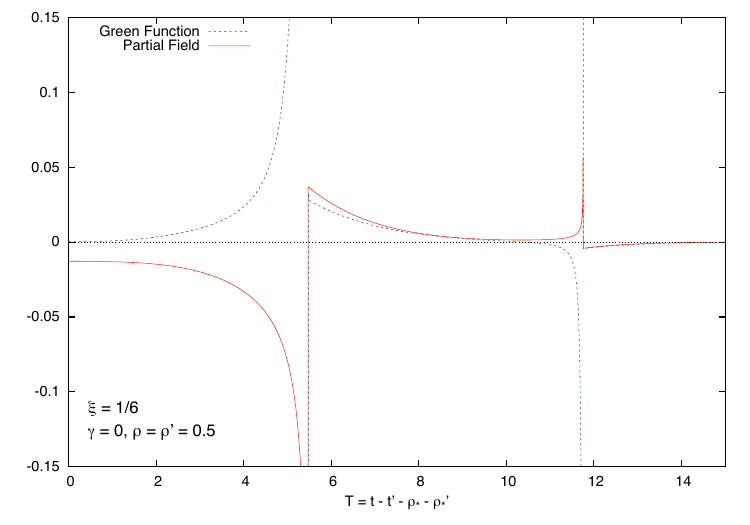}
 \end{center}
 \caption[`Partial Field' Generated by a Static Particle at $\rho=0.5$]{\emph{`Partial Field' Generated by a Static Particle at $\rho=0.5$}. The dotted (blue) 
 line shows the $n=0$ QNM Green function
 (\ref{G-n0,summary}). The solid (red) line shows the partial field $q^{-1} \Phipartial$ defined in Eq.~(\ref{eq:partial-field}). This may be interpreted as the portion of tail field generated by the segment of the static-particle worldline between $t^\prime \to -\infty$ and $t^\prime = t - \Delta t$. Here $\Delta t = T + \Rs + \Rs^\prime $, where $T$ is the time coordinate shown on the horizontal axis.  }
 \label{fig:partial-field}
\end{figure}

The tail field may be found by integrating the Green function with respect to $\tau^\prime$, where $d\tau^\prime = (1-\R^2)^{1/2} dt^\prime$. Integrating a mode sum like (\ref{G-n0,summary}) with respect to $t^\prime$ is straightforward; we simply multiply each term in the sum by a factor $1/(i \omega_{ln})$. Hence it is straightforward to compute a \emph{partial field} defined by
\beq
\Phipartial(\Delta t) = q \left( 1 - \R^2 \right)^{1/2} \int_{-\infty}^{t - \Delta t} \Gret(t-t', \R=\R', \gam=0) d t^\prime .  \label{eq:partial-field}
\eeq
This may be interpreted as ``the field generated by the segment of the static-particle world line lying between $t^\prime \to -\infty$ and $t^\prime = t - \Delta t$''.
In the limit $\Delta t \rightarrow 0$, the partial field $\Phipartial$ will coincide with the tail field. An example of this calculation is shown in Fig.~\ref{fig:partial-field}. Here, $q^{-1} \Phipartial$ is plotted as a function of $T = \Delta t - \Rs - \Rs^\prime$ for a static particle near spatial infinity, $\R \rightarrow 1$. We have used the 
$n=0$ QNM Green function (\ref{G-n0,summary}) together with the 
method of smoothed sums (Sec.~\ref{subsec:nummeth}), with $\lmax = 200$. The `partial field' $\Phipartial$ shares singular points with $\Gret$. Figure \ref{fig:partial-field} shows that a significant amount of the total tail field arises from the segment of the worldline \emph{after} the first singularity. The Green function tends to zero in the limit $\Delta t \rightarrow 0$ (for $\xi = 1/6$). On the other hand, the partial field tends to a constant non-zero value in this limit. The constant value is the tail field.%

\begin{figure}
 \begin{center}
  \includegraphics[width=10cm]{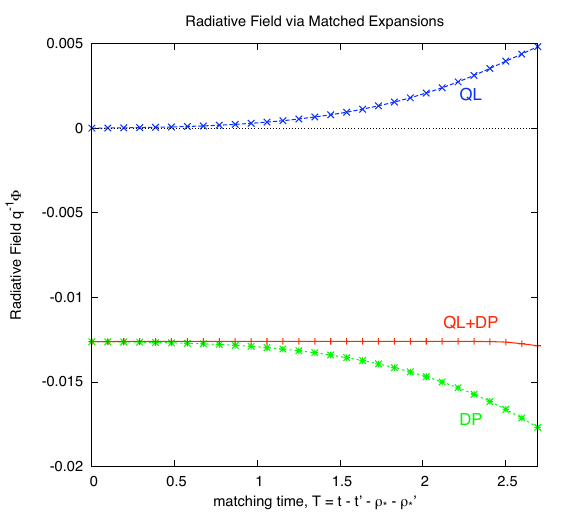}
  \includegraphics[width=10cm]{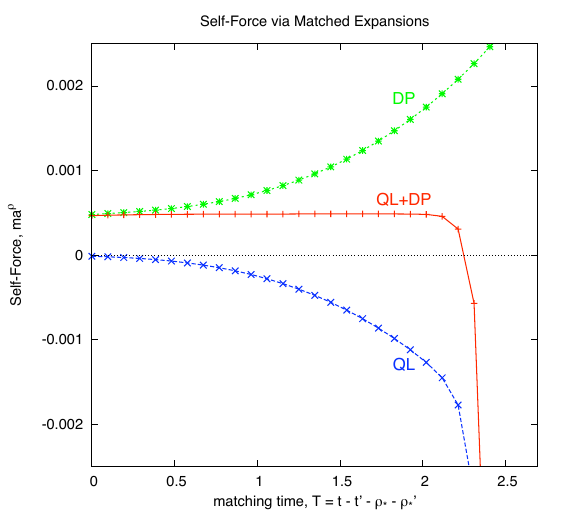}
 \end{center}
 \caption[Illustration of the matching calculation of the self-force for $\xi=1/6$ on a static particle at $\R = \R^\prime = 0.5$]{\emph{Illustration of the matching calculation of the self-force for $\xi=1/6$ on a static particle at $\R = \R^\prime = 0.5$}. The (blue) dashed line shows the quasilocal contribution, integrated from $\tau - \Delta \tau = (1-\R^2)^{1/2} [T+ \Rs + \Rs^\prime]$ to (almost) coincidence. The (green) dotted line shows the distant past contribution, integrated from $-\infty$ to $\tau - \Delta \tau$. The (red) solid line shows the total. In matching region $1 \lesssim T \lesssim 2$ the total approaches a constant, which corresponds to the value of the tail field (upper plot) and radial self-force (lower plot). }
 \label{fig:match}
\end{figure} 

An accurate value for the total tail field is found by using the quasilocal Green function to extend $\Phipartial$ to (almost) coincidence, $\Delta t \rightarrow 0^+$. The method is illustrated in Fig.~\ref{fig:match} (upper plot). The dashed line shows the quasilocal contribution to the tail field, and the dotted line shows the distant past contribution to the tail field, as a function of matching time. The former is the result of integrating from the matching point $\tau - \Delta \tau$ to (almost) coincidence, and the latter from integrating from $-\infty$ to the matching point. Here, $\Delta \tau$ varies linearly with the $x$-axis scale $T$ (see caption). The lower plot illustrates the same calculation for the radial self-force. Here, the dotted line representing the contribution from the distant past is found from the sum (\ref{eq:sf-dp})
using the `smoothed sum' method (\ref{num-meth1}) with $l_{\text{cut}}=400$ for the $l$-sum and summing over $n$
up to the value $\nmax=4$.

\begin{figure}
 \begin{center}
  \includegraphics[width=10cm]{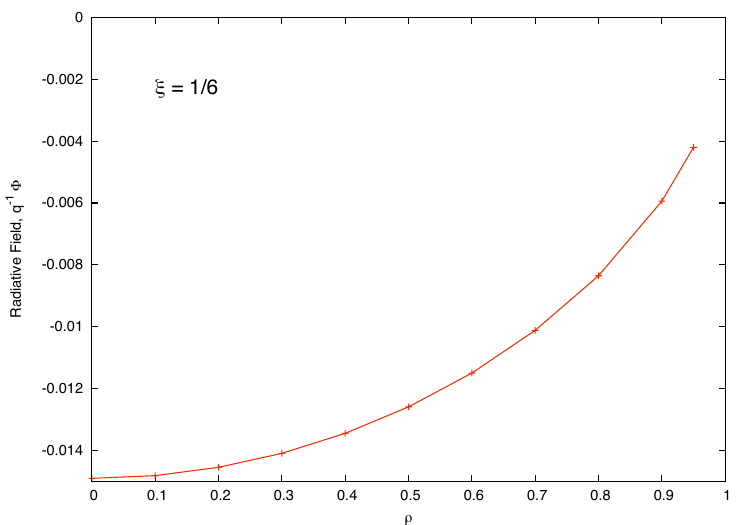}
  \includegraphics[width=10cm]{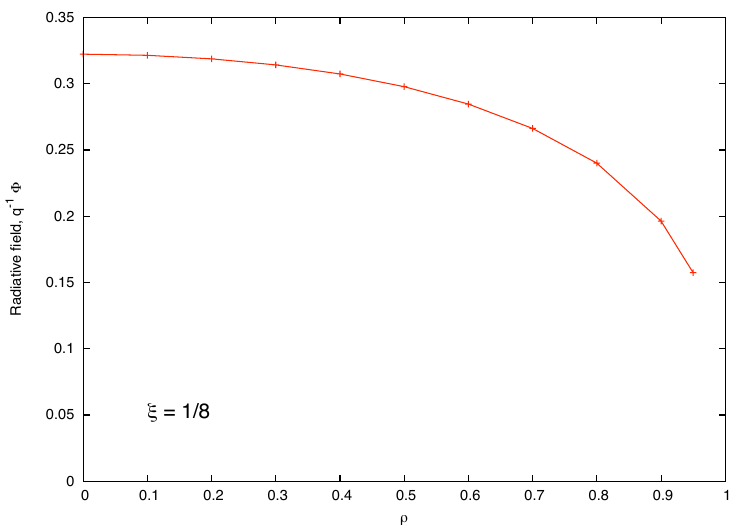}
 \end{center}
 \caption[The Tail Field Generated by the Static Particle]{\emph{The Tail Field Generated by the Static Particle}. The plot shows the tail field for the static particle at $\R=\R^\prime$. For the case $\xi = 1/6$ (top) the field is negative whereas for the case $\xi = 1/8$ it is positive (bottom). Note the differing scales on the vertical axis.}
 \label{fig:field}
\end{figure}

Figure \ref{fig:field} shows the total tail field generated by a static particle in the Nariai spacetime. The field is plotted as a function of $\R$, for two cases: $\xi = 1/6$ and $\xi = 1/8$. In the former case, the field is negative. In the latter case, the field is positive, and about two orders of magnitude greater in amplitude. In both cases, the amplitude of the field is maximal at $\R = 0$ and tends to zero as $\R \rightarrow 1$ as $\Phi_R \sim (1-\R^2)^{1/2}$. 

\begin{figure}
 \begin{center}
  \includegraphics[width=10cm]{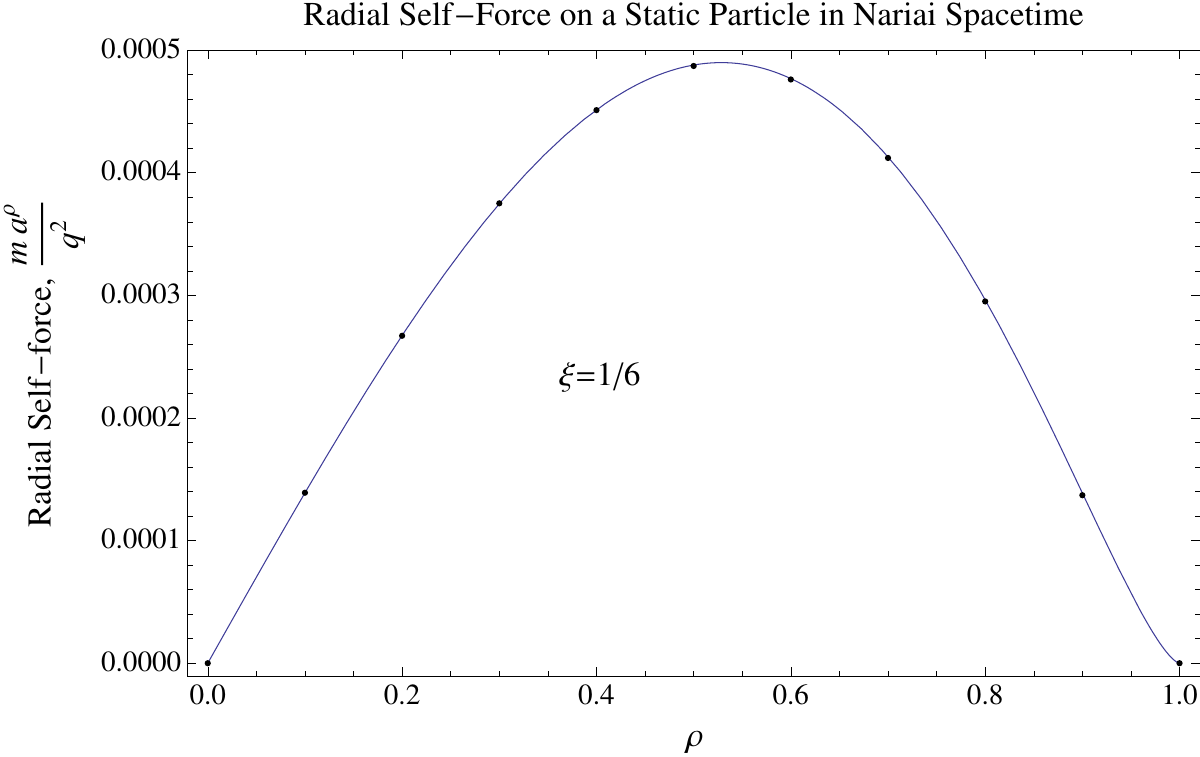}
  \includegraphics[width=10cm]{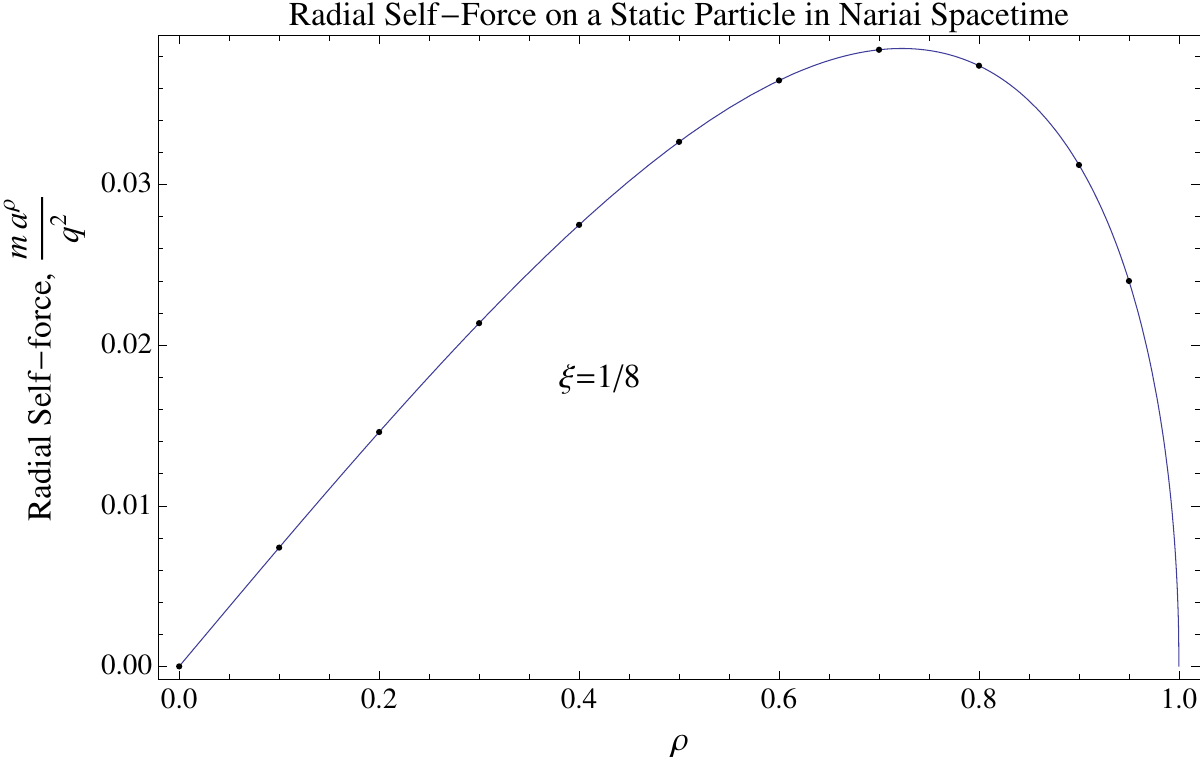}
 \end{center}
 \caption[The Radial Self-Force on the Static Particle]{\emph{The Radial Self-Force on the Static Particle}. The radial component of the self-force, $q^{-2} m a^{\R}$, is plotted as a function of the $\R$ of the static particle. The self-force was calculated by two methods: `matched expansion' [(black) dots] and `massive field regularization' [(blue) curve]. The left plot shows the curvature coupling $\xi = 1/6$ whereas the right plot shows $\xi = 1/8$. Note the difference in scales for the two cases.}
 \label{radial-sf}
\end{figure}

Figure \ref{radial-sf} shows the radial self-force $m a^\R$ acting on a static particle. The results of the matched expansion results are shown as points, and the results of the `massive field regularization method' (described in Sec.~\ref{subsec:static full Green}) are shown as the solid line.  The latter method provides an independent check on the accuracy of the former. We find agreement to approximately six decimal places between the two approaches. We find the self-force at $\R = 0$ to be zero, as expected from the symmetry of the static region of the Nariai spacetime. The self-force also tends to zero as $\R \rightarrow 1$. Between these limits, the self-force rises to a single peak, the magnitude and location of which depends on the curvature coupling $\xi$. We find that the peak of the self-force (in units such that the scalar charge, $q$, is $1$) is approximately $4.9 \times 10^{-4}$ for $\xi = 1/6$ and approximately $3.8 \times 10^{-2}$ for $\xi = 1/8$. 
Note that the argument (see Sec.\ref{subsec:static}) 
in~\cite{Bezerra:Khus} that the scalar self-force with $\xi = 1/8$ should be zero for a static particle
in a spacetime (such as Nariai) possessing a conformally flat 3-D spatial section does not seem to hold here.

\chapter*{Discussion}  \label{ch:conclusions}
\addcontentsline{toc}{chapter}{Discussion}
\markboth{}{Discussion}
The self-force acting on a particle is determined by its entire past history. This is encapsulated in the MiSaTaQuWa equation through an integral of the retarded Green function along a timelike worldline, therefore requiring knowledge of the Green function for points separated along that worldline (other methods negate the need for knowledge of the Green function by computing the regularized field \cite{Barack:Golbourn:2007,Barack:Golbourn:Sago:2007,Barack:Sago:2007} or metric perturbation \cite{Vega:Detweiler:2008} directly). In this thesis, we have presented the first practical demonstration of a self-force calculation using the method of matched expansions, first proposed over a decade ago by Poisson and Wiseman \cite{Poisson:Wiseman:1998}.  The method relies on an expansion of the Green function in two regions: a quasilocal region, corresponding to the recent history of the particle, and a distant past region. In particular, the primary focus of this thesis has been on computing the Green function within the quasilocal region.

The examples of self force calculations presented in this thesis are for the case of a scalar field. From an astrophysical perspective, we are more interested in the case of a gravitational field. The quasilocal contribution to the gravitational self-force has previously been calculated for vacuum spacetimes up to $O(\sigma^{3/2})$ \cite{Anderson:Flanagan:Ottewill:2004}. As is apparent from Chapter~\ref{ch:nariai}, it is necessary to take the calculation to higher order to allow the matching point $\Delta \tau$ to lie far enough into the particle's past to match onto a quasinormal mode sum calculated Green function. Fortunately, the covariant method of Chapter~\ref{ch:covex} is largely independent of the type of field considered, the only issue being the specific form of the transport equations, \eqref{eq:recursionVn} and \eqref{eq:recursionV0} for $V^{AB'}_{r}(x,x')$ and $V^{AB'}_{0}(x,x')$, respectively. The gravitational version of this equation contains additional factors of $g_{a b'}$ (and its derivatives) over its scalar counterpart. However, we have already provided transport equations for these additional bi-tensors and our \textsl{Mathematica} code can calculate their covariant series expansions. It should therefore be possible to calculate the gravitational Green function with little extra effort.

In Chapter~\ref{ch:sf} we introduced the method of matched expansions and derived expressions for the quasilocal contribution to the scalar self-force. These expressions are totally general and may be applied to the computation of the quasilocal scalar self-force for both geodesic and non-geodesic motion in any spacetime, with any coupling constant and any field mass. 

In Chapters \ref{ch:coordex} and \ref{ch:covex} we developed methods for calculating the Green function within the quasilocal region, where the Hadamard form may be used. The coordinate approach of Chapter~\ref{ch:coordex} exploits the spherical symmetry of a spacetime to yield an efficient method for computing the quasilocal Green function. This allowed the coordinate series expansion of $V(x,x')$, appearing in the Hadamard parametrix of the Green function, to be computed to significantly higher order than was done previously \cite{Anderson:2003,Anderson:Eftekharzadeh:Hu:2006}. However, we feel that the method has now been taken to the limit of its potential. It is necessarily restricted to spherically symmetric spacetimes, which are inadequate for a full study of the self-force in the context of EMRIs.

In contrast, the covariant method of Chapter~\ref{ch:covex} is totally general, being valid for any type of field in any spacetime with arbitrary field mass and arbitrary curvature coupling. This method has the flexibility required for a complete study of the quasilocal self-force. Furthermore, it also has applications in other areas such as quantum field theory in curved spacetime and quantum gravity. 

Several of the covariant expansion expressions computed by our code using the Avramidi method have been previously given in Ref.~\cite{Decanini:Folacci:2005a}, albeit to considerably lower order (for example, in Ref.~\cite{Decanini:Folacci:2005a} Decanini and Folacci give $V(x,x')$ to order $\left(\sigma^a\right)^4$ compared to order $\left(\sigma^a\right)^{20}$ here). Comparison between the two results gives exact agreement, providing a reassuring confirmation of the accuracy of both our expressions and those of Ref.~\cite{Decanini:Folacci:2005a} (and confirming the error in Ref.~\cite{Phillips:Hu:2003} found by Decanini and Folacci). For expansions not given by Decanini and Folacci, we have compared our results with those found by Christensen \cite{Christensen:1976vb,Christensen:1978yd}. Again, we have found that our code is in exact agreement.

The expressions for the DeWitt coefficients as computed by our \emph{Mathematica} code are very general. However, they are not necessarily given as a minimal set. For example, the DeWitt coefficient $a_3$ may be written as a sum of four terms, yet our code produces a sum of seven equivalent terms. It is quite probable, however, that a set of transformation rules could be produced to reduce our expression to a canonical form in terms of some basis. Some progress in that direction has already been made. Fulling, et al. \cite{Fulling:1992} (and more recently  D\'{e}canini and Folacci \cite{Decanini:Folacci:2005b}) have establish a linearly independent basis for the Riemann polynomials\footnote{The terms in these polynomials are classified \cite{Fulling:1992} by their rank (number of free indices), order (number of derivatives of the metric - 2 per riemann tensor and 1 per explicit covariant derivative) and degree (number of factors). For example $R_{a \beta \gamma \delta} R_{b}{}^{\beta \gamma \delta}$ is rank 2, order 4 and degree 2, while $R_{(a| \alpha \beta \gamma |} R_{b}{}^{\beta \gamma \delta} R^{\alpha}{}_{c d |\delta| ;e)}{}$ is rank 5, order 7 and degree 3.}, involving product of Riemann and Ricci tensors. The first steps towards the automation of the canonicalization process have been taken with the release of the \emph{Invar} package \cite{Invar1,Invar2} for \emph{Mathematica}, which is so far able to quickly canonicalize the scalar algebraic invariants with up to 7 Riemann tensors and differential invariants up to $12$ metric derivatives. As our code is already written in \textsl{Mathematica} \cite{Mathematica} and has the ability to output into the \textsl{xTensor} \cite{xTensor} notation used by \textsl{Invar}, an extension of the package to allow for the canonicalization of tensor invariants would allow our expressions to be immediately canonicalized with no further effort.

There remains one context in which a coordinate expansion of the Green function is more appropriate: the case of non-geodesic motion. This is not very surprising as the benefit of using a covariant expansion hinges on having a geodesic along which to expand. The coordinate expansion, on the other hand, more naturally characterizes motion which is not along a geodesic. However, as Sec.~\ref{sec:kn} demonstrates, even in this case, beginning with a covariant expansion and computing the coordinate expansion from it is preferable as it allows the calculation to be done for arbitrary spacetimes, rather than only those with spherical symmetry.

The covariant expansion method yields expressions involving Riemann tensor polynomials. In order to compute the Green function in a specific spacetime, it is necessary to explicitly evaluate these polynomials in terms of the spacetime coordinates. In Chapter~\ref{ch:qlsf}, we gave examples of such an evaluation using the GRTensorII \cite{GRTensor} tensor algebra package. Only the leading two orders in the series were considered (up to order $(\sigma^{a})^5$), yet the calculation was already reaching the limit of the ability of GRTensorII given currently available computational power. Improvements in computational performance will be of limited benefit; the complexity of the calculation is exponential in the order of the term considered so that even an order of magnitude increase in performance will make little difference. A better solution would be to develop improved algorithms. Some hope lies in the fact that we are most interested in calculating $V^A{}_{B'}(x,x')$ for massless fields in $4$-dimensional type-D vacuum spacetimes. This suggests \cite{Price:Whiting} that the Geroch-Held-Penrose (GHP) adaptation \cite{GHP} (for a thorough treatment, see Refs.~\cite{Penrose:Rindler:1,Penrose:Rindler:2}) of the Newman-Penrose formalism \cite{NP} may be of use in producing significantly simplified expressions. There is additional hope that the canonicalization process proposed above may yield expressions which are significantly simplified and therefore more easily evaluated for a specific spacetime.

The high order expansions of the quasilocal Green function facilitated an investigation of the convergence properties of the series, described in detail in Chapter~\ref{ch:coordex}. We found that the series was divergent before the normal neighborhood boundary was reached. However, this divergence does not correspond to a singularity in the Green function for real-valued point separation (in general the radius of convergence of a Taylor series extends to the first occurrence in the complex plane of a singularity of the function being approximated). We have found that, using Pad\'e resummation, the domain of validity of the series can be extended beyond the circle of convergence, to within a short distance of the normal neighborhood boundary. For convenience, the Pad\'e resummation was applied to the coordinate expansion of the Green function. We expect it to be equally applicable to the covariant series method of Chapter~\ref{ch:covex}; both expansions are Taylor series and there is no physical singularity in the Green function before the edge of the normal neighborhood is reached. Additionally, our analysis focused on the case of one dimensional series. This was done for reasons of simplicity and clarity. For multi-dimensional series, one could re-express each of the coordinates in terms of a single parameter, as was done in Sec.~\ref{sec:convergence} for the case of a circular geodesic in Schwarzschild. Alternatively, one could employ the generalization of Pad\'e resummation to double (and higher dimensional) power series, as developed by Chisholm \cite{Chisholm:1973,Chisholm:McEwan:1974}.

In Sec.~\ref{sec:numerical}, we discussed a numerical implementation of the transport equation approach to the calculation of $V_0$. This implementation is capable of computing $V_0$ along a geodesic in any spacetime, although we have chosen Nariai and Schwarzschild spacetimes as examples. The choice of Nariai spacetime has the benefit that an expression for $V_0$ is known exactly \cite{Nolan:2009} (the key analytic results are presented in Sec.~\ref{sec:nariai-exact}). This makes it possible to compare our numerical results with the analytic expressions to determine both the validity of the approach and the accuracy of the numerical calculation. Given parameters allowing the code to run in under a minute, we found that the numerical implementation is accurate to less than $1\%$ out as far as the location of the singularity of $V_0$, at the caustic point (i.e. everywhere within the normal neighborhood).

In integrating the transport equations given in Sec.~\ref{sec:avramidi} along a specific geodesic, we are not limited to the normal neighborhood. The only difficulty arises at \emph{caustics}, where some bi-tensors such as $\Delta^{1/2}$ and $V_0$ become singular. However, this is not an insurmountable problem. The singular components may be separated out and methods of complex analysis employed to integrate through the caustics, beyond which the bi-tensors once more become regular (but not necessarily real-valued) \cite{Casals:Dolan:Ottewill:Wardell:2009}. 
This is highlighted in Fig.~\ref{fig:schw-V0-timelike}, where our plot of $\Delta^{1/2}$ and $V_0$ extends outside the normal neighborhood, the boundary of which is at $\phi\approx1.25$, where the first null geodesic re-intersects the orbit. It does not necessarily follow, however, that the Green function outside the normal neighborhood is given by this value for $V(x,x')$. Instead, one might expect to obtain the Green function by considering the sum of the contributions obtained by integrating along \emph{all} geodesics connecting $x$ and $x'$ (other than at caustics, there will be a discrete number of geodesics).


In Chapter~\ref{ch:nariai}, we presented a full application of the method of matched expansions in the Nariai spacetime. 
Through matching the `quasilocal' and `distant past' expansions, the full retarded Green function may be reconstructed. With full knowledge of the Green function, one may accurately compute the `tail' contribution to the self-force. In Sec.~\ref{subsec:matched-static}, we employed the matched expansion method to numerically compute the self-force acting on a static particle in the Nariai spacetime. The resulting value for the self-force is in excellent agreement with the result from an alternative method (Sec.~\ref{subsec:static full Green}), to approximately one part in $10^{6}$. 

This study of the self-force in Nariai spacetime has provided a number of insights into the properties of the Green function, which are of relevance to any future investigation of the Schwarzschild spacetime. Namely,
\begin{itemize}
  \item The Green function $\Gret(x, x^\prime)$ is singular whenever $x$ and $x^\prime$ are connected by a null geodesic. The nature of the singularity depends on the number of caustics that the wave front has passed through. After an even number of caustics, the singularity is a delta-distribution, with support only on the light cone. After an odd number of caustics, the Green function diverges as $1 / \pi \sigma$. A four-fold repeating pattern occurs, i.e. $\delta(\sigma), \, 1/\pi \sig, \, -\delta(\sigma), \, -1/\pi\sig, \, \delta, $ etc.
  \item The four-fold singular structure can be shown to arise from a Hadamard-like ansatz (\ref{Gdir1}) valid even outside the normal neighborhood, if we allow $U(x,x^\prime)$ to pick up a phase of $-i$ upon passing through a caustic. The accumulation of phase may be deduced by analytically continuing the integral for the Van Vleck determinant through the singularities (due to caustics).
  \item The effect of caustics on wave propagation has been well-studied in a number of other fields, such as optics~\cite{Stavroudis}, acoustics~\cite{Kravtsov:1968}, seismology~\cite{Aki:Richards}, symplectic geometry~\cite{Arnold} and quantum mechanics~\cite{B&M}. It may be that mathematical results developed in other fields may be usefully applied to wave propagation in gravitational physics.
  \item The `tail' self-force cannot be calculated from the `quasilocal' contribution alone. For instance, Fig.~\ref{fig:partial-field} would appear to show that a significant part of the tail field is generated by the segment of the world line which lies \emph{outside} of the normal neighborhood. Unlike in flat space, the field generated by an accelerated particle may propagate several times around the black hole before later re-intersecting the worldline of the particle (see Fig.~\ref{fig:circ_orbits}). Radiation from near these orbits will give an important contribution to the self-force which cannot be neglected.
\end{itemize}

The calculation of the `distant past' Green function in Nariai spacetime relies on a `quasinormal mode sum' expansion. The QNM sum is only valid at `late times', $t-t^\prime \ge t_c$, where $t_c$ is approximately the time it takes for a geodesic to reflect from the peak of the potential barrier. One may wonder whether the quasilocal Green  function may be accurately calculated in an overlapping spacetime region at times later than this light reflection time. In the Nariai case, we have demonstrated that this is the case. There is a sufficient regime of overlap in $t-t^\prime$ in which both the `quasilocal' and `QNM sum' expansions are valid for the method to be applied successfully.
There is also a strong indication that this is the case in Schwarzschild. From Sec.~\ref{subsec:Pade-convergence} we  see that the higher order terms of the quasilocal series only have a significant contribution near the radius of convergence (for the Taylor series) or normal neighborhood boundary (for the Pad\'e approximant), while the reflection time  ($t-t'= 2 r_* \approx 12.77 M$, for the case of a static particle at $r=10M$ considered in Sec.~\ref{subsec:Pade-convergence}) is well within the normal neighborhood. We can therefore be optimistic that matched expansion self-force calculations will be possible for Schwarzschild. 

The question of whether this also follows through for the Kerr spacetime is yet to be studied, but there is little reason to expect the answer to be different than in the Schwarzschild and Nariai cases.  The full caustic structure of Kerr spacetime has recently been studied \cite{Bozza:2008} and shown to have additional features compared to the Schwarzschild case. However, if the motion is restricted to the equatorial plane, the caustic structure of Kerr bears a striking resemblance to that of Schwarzschild (see Fig.~\ref{fig:schw-light-cone}).
\pagebreak

 \begin{figure}
  \begin{center}
  \includegraphics[width=15cm]{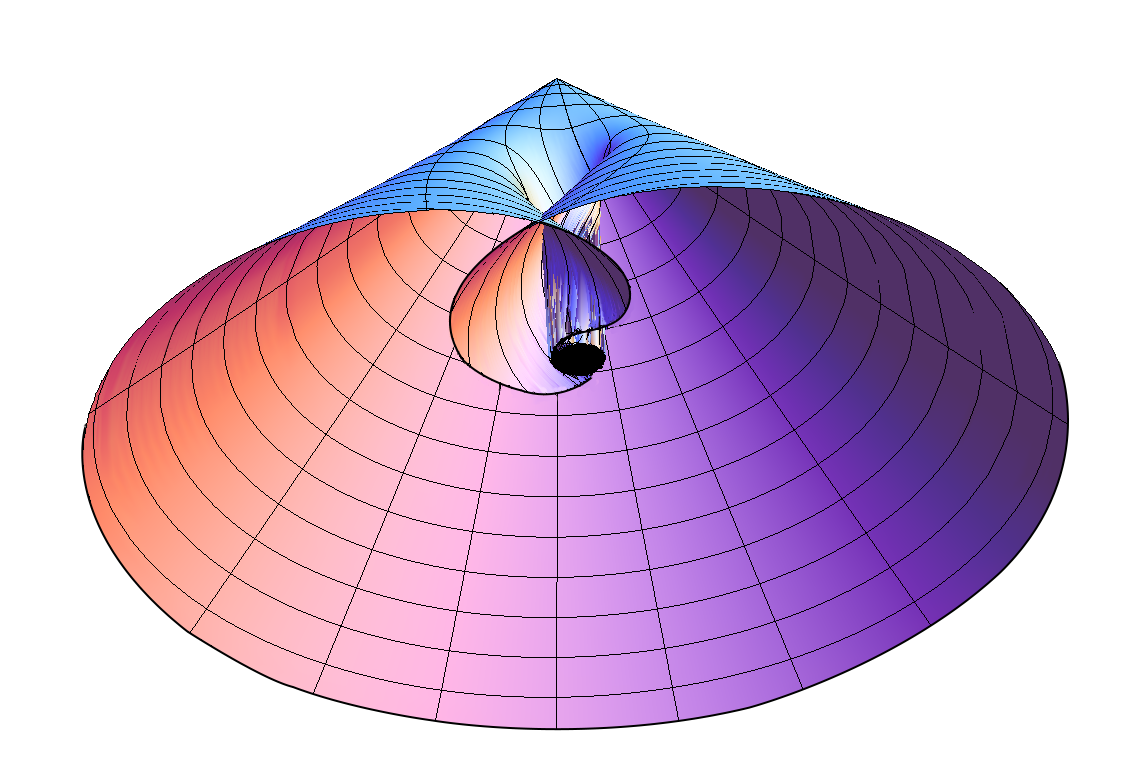}
 \end{center}
{\emph{Light cone for the Kerr spacetime.} The $\theta$ coordinate is suppressed so that only geodesics in the equatorial plane are shown. The caustic structure bears a striking resemblance to that of Schwarzschild (Fig.~\ref{fig:schw-light-cone}).}
\addcontentsline{lof}{figure}{Kerr light cone}
\end{figure}

\renewcommand{\chaptername}{Appendix}
\appendix
\chapter{Odd Order Series Coefficients of Symmetric Bi-scalars}

\section{Covariant Series}
\label{sec:cov-5}
In this appendix, we derive an expression for rewriting odd order coefficients of covariant series expansions of any symmetric bi-scalar in terms of the lower even order coefficients. This not only allows for the simplification of covariant series expressions, but allows the order of an even order covariant expansion to be increased by one order with ease.

Taking, for example, the symmetric bi-scalar $V(x,x')$ appearing in the Hadamard form of the Green function, we have the identity,
\begin{equation}
V\left( x,x' \right) = V\left( x',x \right).
\end{equation}
Repeatedly taking symmetrized covariant derivatives of both sides of this equation and taking coincidence limits, we arrive at an expressions for the odd order terms in terms of all the lower (even) order terms \cite{Brown:Ottewill:1986}:
\begin{equation}
\label{eq:vcompsym-appendix}
V_{a_1a_2\dots a_n}	=  \frac{1}{2}\sum\limits_{k=0}^{n-1} \binom{n}{k} (-1)^k V_{(a_1 a_2 \dots a_{k} ;, a _{k+1} \dots  a_{n})}  \quad \text{for } n \text{ odd}.
\end{equation}
Recursively applying this identity to eliminate all odd order terms, we get expressions for the coefficients of the terms of order $\sigma^\frac{5}{2}$:
\begin{subequations}
\begin{eqnarray}
\label{eq:cov-5}
v_{abcde} &=& \frac{1}{2} v_{;(a b c d e)} - \frac{5}{2} v_{(a b ; c d e)} + \frac{5}{2} v_{(a b c d ; e)} \\
v_{abc} &=& -\frac{1}{4} v_{;(a b c)} + \frac{3}{2} v_{(a b; c)} \\
v_{a} &=& \frac{1}{2} v_{;a}
\end{eqnarray}
\end{subequations}
In general, this can be written as
\begin{equation}
\label{eq:vcompsym2-appendix}
V_{a_1a_2\dots a_n}	=  \sum\limits_{\substack{k=0\\k~\text{even}}}^{n-1}\binom{n}{k} \frac{2(2^{n-k+1}-1)}{n-k+1} B_{n-k+1} V_{(a_1 a_2 \dots a_{k} , a _{k+1} \dots  a_{n})} \quad \text{for } n \text{ odd}
\end{equation}
where the $B_n$ are the Bernoulli numbers \cite{GradRyz}.
These expressions are equally valid for any symmetric bi-tensor, so we may use it for both the calculation of higher order $V_{(n)~a_1 \dots a_n}$ as used in (\ref{eq:V}) and higher order $V_{a_1 \dots a_n}$ as used in (\ref{eq:Vt}).

\section{Odd Order Coordinate Series Coefficients of Symmetric Bi-scalars}
\label{sec:WKB-7}
Similarly to the case of covariant series described in Appendix~\ref{sec:cov-5}, we may express odd order coefficients of \emph{coordinate} series expansions of any symmetric bi-scalar in terms of the lower even order coefficients. Again, taking, for example, the symmetric bi-scalar $V(x,x')$ appearing in the Hadamard form of the Green function, we have the identity,
\begin{equation}
V\left( x,x' \right) = V\left( x',x \right)
\end{equation}
We use expression (\ref{eq:WKBGreen}) for $v \left( x,x' \right)$ in the above to give: 
\begin{multline}
\sum_{i,j,k=0}^{\infty} v_{ijk} \left( r \right) \left( t - t' \right)^{2i} \left( \cos \left( \gamma \right) - 1 \right)^j \left(r - r'\right)^k \\=
	\sum_{i,j,k=0}^{\infty} v_{ijk} \left( r' \right)  \left( t' - t \right)^{2i} \left( \cos \left( \gamma ' \right) - 1 \right)^j \left(r' - r\right)^k \label{eq:SymGreen}
\end{multline}
Now, in a similar vein to Appendix \ref{sec:cov-5}, if we take a sufficient number of symmetrized derivatives of both sides and then take the coincidence limit, it is possible to express odd order coefficients in terms of derivatives of the lower order coefficients.

For the cases of the coordinate series expansion of $V(x,x')$ relevant to the current work, the calculation is made even simpler by spherical symmetry. The coefficients $v_{ijk}$ are functions of $r$ only (i.e. they have no dependence on $t$ or $\gamma$). Additionally, since $(t-t')^{2i}$ and $(\cos(\gamma)-1)^j$ are both even functions, they are invariant under interchange of primed and unprimed coordinates.  This means that we can express (\ref{eq:SymGreen}) in the simpler form:
\begin{equation}
\sum_{k=0}^{\infty} v_{ijk}\left( r \right) \left(r - r'\right)^k = \sum_{k=0}^{\infty} v_{ijk}\left( r' \right) \left(r' - r\right)^k
\end{equation}
Taking $r$-derivatives of both sides of this equation and taking the coincidence limit $r' \rightarrow r$ gives the result
\begin{equation}
v_{ij1} = - \frac{1}{2} v_{ij0,r}
\end{equation}

As a specific example, this can be applied to the results of Ref.~\cite{Anderson:2003} to calculate $v_{301}$, $v_{211}$, $v_{121}$ and $v_{031}$. This gives:
\begin{subequations}
\begin{eqnarray}
v_{301} &=& - \frac{1}{960}   \frac{M^2 \left( r -2M \right) ^2 \left( 598 r M^2 - 195 M r^2 + 20 r^3 - 585 M^3 \right) }{r^{16}} \\
v_{211} &=& \frac{1}{896}   \frac{M^2 \left( r -2M \right)  \left( 3352 r M^2 - 1099 M r^2 + 112 r^3 - 3240 M^3 \right) }{r^{13}} \\
v_{121} &=& - \frac{9}{1120} \frac{M^2 \left( 228 r M^2 - 91 M r^2 + 12 r^3 - 189 M^3 \right) }{r^{10}} \\
v_{031} &=& \frac{1}{3360} \frac{M^2 \left( -155 M r + 56 r^2 + 36 M^2 \right) }{r^7}
\end{eqnarray}
\end{subequations}

\chapter{Vacuum expressions for the DeWitt Coefficients}
\label{sec:DeWittVacuum}
The expressions given in Chapter~\ref{ch:covex} and Ref.~\cite{Decanini:Folacci:2005a} are unnecessarily long and unwieldy for the purposes of the self-force calculations of Chapter~\ref{ch:qlsf}. There, we are only interested in massless fields in vacuum spacetimes ($m_{\rm field}=0$, $R_{\alpha \beta}=0$, $R=0$), so the majority of the terms vanish and we are left with much more manageable expressions for the $v_{n a_1 \dots a_p}$. They are:
\begin{subequations}
\allowdisplaybreaks
\begin{eqnarray}
\label{eq:V0}
v_0 &=& 0\\
v_{0\, a} &=& 0\\
v_{0\, ab} &=& - \frac{1}{720} g_{a b} I\\
v_{0\, abc} &=& - \frac{1}{480} g_{(a b} I_{;c)}\\
v_{0\, abcd} &=& - \frac{2}{525} C^{\rho}_{\phantom{\rho} (a | \sigma | b} \Box C^{\sigma}_{\phantom{\sigma} c | \rho | d)}
	- \frac{2}{105}C^{\rho \sigma \tau}_{\phantom{\rho \sigma \tau} (a} C_{| \rho \sigma \tau | b ; c d)}
	- \frac{1}{280} C^{\rho \phantom{ (a | \sigma | b} ;\tau}_{\phantom{\rho} (a | \sigma | b} C^{\sigma}_{\phantom{\sigma} c | \rho | d) ;\tau}
	\nonumber \\
	& &
	- \frac{1}{56} C^{\rho \sigma \tau}_{\phantom{\rho \sigma \tau} (a ;b} C_{| \rho \sigma \tau | c ; d)} 
	- \frac{2}{1575} C^{\rho \sigma \tau \kappa} C_{\rho (a | \tau | b} C_{| \sigma | c | \kappa | d)}
	\nonumber \\
	& &
	- \frac{2}{525} C^{\rho \kappa \tau}_{\phantom{\rho \kappa \tau} (a} C^{\phantom{| \rho \tau | } \sigma}_{| \rho \tau | \phantom{\sigma} b} C_{| \sigma | c | \kappa | d)}
	- \frac{8}{1575} C^{\rho \kappa \tau}_{\phantom{\rho \kappa \tau} (a} C^{\phantom{| \rho | } \sigma}_{| \rho | \phantom{ \sigma } | \tau | b} C_{| \sigma | c | \kappa | d)}
	\nonumber \\
	& &
	- \frac{4}{1575} C^{\rho \tau \kappa}_{\phantom{\rho \tau \kappa} (a} C^{\phantom{| \rho \tau | } \sigma}_{| \rho \tau | \phantom{\sigma} b} C_{| \sigma | c | \kappa | d)}\\
\label{eq:V1}
v_1 &=& \frac{1}{720} I \\
v_{1\, a} &=& \frac{1}{1440} I_{;a} \\
v_{1\, ab} &=& - \frac{1}{1400} C^{\rho \sigma \tau \kappa} C_{\rho \sigma \tau (a ; b) \kappa}
	+ \frac{1}{1575} C^{\rho \sigma \tau}_{\phantom{\rho \sigma \tau} a} \Box C_{\rho \sigma \tau b}
	+ \frac{29}{25200} C^{\rho \sigma \tau \kappa} C_{\rho \sigma \tau \kappa ; (a b)}
	\nonumber \\
	& &
	+ \frac{1}{1680} C^{\rho \sigma \tau}_{\phantom{\rho \sigma \tau} a ;\kappa} C^{\phantom{\rho \sigma \tau b} ;\kappa}_{\rho \sigma \tau b}
	+\frac{1}{1344} C^{\rho \sigma \tau \kappa}_{\phantom{\rho \sigma \tau \kappa} ;a} C_{\rho \sigma \tau \kappa ;b}
	+ \frac{1}{756} C^{\rho \kappa \sigma \lambda} C^{\tau}_{\phantom{\tau} \rho \sigma a} C_{\tau \kappa \lambda b}
	\nonumber \\
	& &
	- \frac{1}{1800} C^{\rho \kappa \sigma \lambda} C_{\rho \sigma \tau a} C^{\phantom{\kappa \lambda} \tau}_{\kappa \lambda \phantom{\tau} b}
	+ \frac{19}{18900} C^{\rho \sigma \kappa \lambda} C_{\rho \sigma \tau a} C^{\phantom{\kappa \lambda} \tau}_{\kappa \lambda \phantom{\tau} b}\\
\label{eq:V2}
v_2 &=& - \frac{1}{6720} C_{\rho \sigma \tau \kappa} \Box C^{\rho \sigma \tau \kappa}
	- \frac{1}{8960} C_{\rho \sigma \tau \kappa ; \lambda} C^{\rho \sigma \tau \kappa ; \lambda}
	\nonumber \\
	& &
	- \frac{1}{9072} C^{\rho \kappa \sigma \lambda} C_{\rho \alpha \sigma \beta} C^{\phantom{\kappa} \alpha \phantom{\lambda} \beta}_{\kappa \phantom{\alpha} \lambda}
	- \frac{11}{18440} C^{\rho \sigma \kappa \lambda} C_{\rho \sigma \alpha \beta} C^{\phantom{\kappa \lambda} \alpha \beta}_{\kappa \lambda}
\end{eqnarray}
\end{subequations}
where $C_{a b c d}$ is the Weyl tensor and $I = C_{\alpha \beta \gamma \delta}C^{\alpha \beta \gamma \delta}$.

Moreover, we have chosen to work with an expression for $V(x,x')$ as given in (\ref{eq:Vt}), rather than the expression used in Chapter~\ref{ch:covex}, Ref.~\cite{Decanini:Folacci:2005a} and Eq. (\ref{eq:V}). The two sets of coefficients may be related by using 
\begin{equation}
2 \sigma = \sigma^{\alpha} \sigma_{\alpha}
\end{equation}
to rewrite (\ref{eq:V}) in terms of a series expansion in powers of $\sigma^{\alpha}$. We can then equate, power by power, the coefficients of powers of $\sigma^{\alpha}$ to the corresponding terms in (\ref{eq:Vt}) to give:
\begin{subequations}
\begin{eqnarray}
v 		&=& v_0\\
v_a 		&=& v_{0\, a}\\
v_{ab} 		&=& v_{0\, ab} + v_{1}g_{ab}\\
v_{abc}		&=& v_{0\, abc} + 3 v_{1\, (a}g_{bc)}\\
v_{abcd}	&=& v_{0\, abcd} + 6 v_{1\, (ab}g_{cd)} + 6 v_2 g_{(ab}g_{cd)}\\
v_{abcde}	&=& \frac{1}{2} v_{;(a b c d e)} - \frac{5}{2} v_{(a b ; c d e)} + \frac{5}{2} v_{(a b c d ; e)} 
\end{eqnarray}
\end{subequations}
Filling in the values for the $v_{n\, a_1 \dots a_p}$ we see that a large number of terms either cancel or combine to give:
\begin{subequations}
\begin{eqnarray}
v 		&=& 0\\
v_a 		&=& 0\\
v_{ab} 		&=& 0\\
v_{abc}		&=& 0\\
v_{abcd}	&=&-\frac{1}{280}C^{\rho \phantom{(a} \sigma}_{\phantom{\rho} (a \phantom{\sigma} b |;\alpha|}C^{\phantom{|\rho| c |\sigma| d)} ;\alpha}_{|\rho| c |\sigma| d)}
		   -\frac{2}{315}C^{\rho \sigma \tau \kappa}C_{\rho (a |\tau| b}C_{|\sigma| c |\kappa| d)}\nonumber \\
		& &
		   +\frac{1}{105}C^{\rho \phantom{(a} \sigma}_{\phantom{\rho} (a \phantom{\sigma} b}C^{\tau \kappa}_{\phantom{\tau \kappa} |\rho| c}C_{|\tau \kappa \sigma| d)}
		   +\frac{1}{840}C^{\rho \sigma \tau \kappa}C^{\phantom{\rho \sigma} \lambda}_{\rho \sigma \phantom{\lambda} (a}C_{|\tau \kappa \lambda| b}g_{c d)}\nonumber \\
		& &
		   + \frac{1}{8960}\Box I g_{(a b}g_{c d)}
		   -\frac{1}{40320}C^{\rho \sigma \tau \kappa}C^{\phantom{\rho \sigma} u v}_{\rho \sigma}C_{\tau \kappa u v}g_{(a b}g_{c d)} \\
v_{abcde}	&=& \frac{5}{2} v_{(a b c d ;e )}
\end{eqnarray}
\end{subequations}

Here, we have used a slight modification of the basis of Riemann tensor polynomials suggested by Fulling, et al. \cite{Fulling:1992} in order to express the result in its simplest possible canonical form. We have chosen an equivalent basis for Weyl tensor polynomials by ignoring all terms involving $R_{a b}$ and $R$ and replacing $R_{abcd}$ by $C_{abcd}$. We have also ignored all terms which vanish or are not independent under symmetrization of the free indices. Next, we have eliminated terms involving two covariant derivatives of a tensor in favor of terms involving two covariant derivatives of a scalar and single covariant derivatives of a tensor. We have done this as it is computationally faster and easier to calculate the covariant derivative of a scalar than of a tensor. Finally, we have used the identity \cite{Gunther}: 
\begin{equation}
 \nabla_{(r)} C^{\rho \sigma \tau}_{\phantom{\rho \sigma \tau} (a} \nabla_{(s)} C_{|\rho \sigma \tau | b)} = \frac{1}{4} g_{(a b)} \nabla_{(r)} C^{\rho \sigma \tau \kappa} \nabla_{(s)} C_{\rho \sigma \tau \kappa}
\end{equation}
(where $\nabla_{(r)}$ indicates $r$ covariant derivatives) to combine some of the remaining terms. The motivation for working with $V(x,x')$ rather than the $V_n(x,x')$ is immediately apparent from the vast simplification that occurs.

\chapter{Pad\'e Approximants and Convergence Properties of Series}\label{appendix:convergence} 
In this appendix, we will review some of the key features of Pad\'e approximants and of the convergence properties of series. We give two simple examples, which serve to highilight these features for those less familiar with these topics.

In Chapter~\ref{ch:coordex}, we investigated the convergence properties of the series expansion of the function $V(x,x')$ appearing in the Hadamard form of the Green function. We found that the Taylor series was divergent at a point (the radius of convergence) where the actual Green function was behaving normally (i.e. not singular). We asserted that this divergence must be as a result of $V(x,x')$ being singular somewhere (in the complex plane) on the circle of convergence. We also demonstrated how the method of Pad\'e approximants may be used to extend the domain of the series beyond its radius of convergence. Both the divergence of the Taylor series and the effectiveness of the Pad\'e approximants may at first seem mysterious. $V(x,x')$ is a real-valued function with real arguments, so how can complex values possibly have any effect? To aid in the understanding, it is helpful to consider some simple functions and investigate the convergence properties of their series representations and Pad\'e approximants. In this Appendix, we will study two cases which are particularly illuminating. 
Further details, including examples may be found in Refs.~\cite{M&F,Whittaker:Watson,Bender:Orszag,NumericalRecipes}.

\subsubsection*{Example 1}
\begin{figure}
 \begin{center}
  \includegraphics[width=7cm]{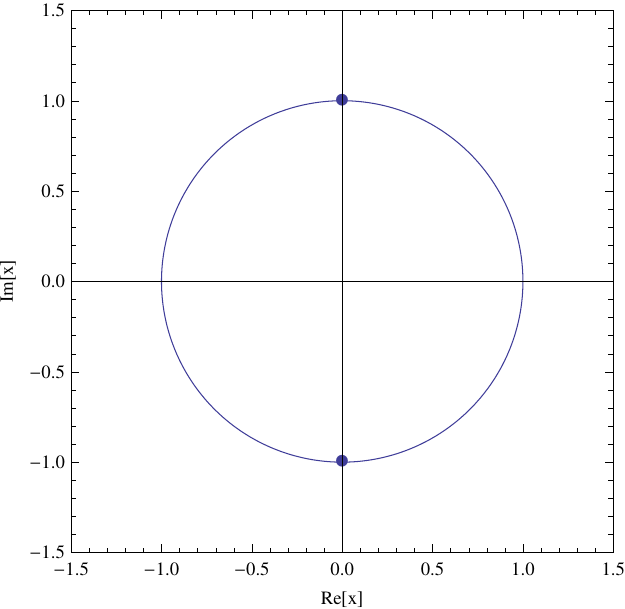}
 \end{center}
 \caption[Circle of convergence of the Taylor series of $1/(1+x^2)$]{\emph{Circle of convergence of the Taylor series of $f(x)=1/(1+x^2)$.} Since $f(x)$ is singular at $x=\pm i$, its Taylor series also diverges everywhere outside its circle of convergence.}
 \label{fig:circle-of-convergence}
\end{figure} 
Consider the real-valued function
\begin{equation}
 f(x) \equiv \frac{1}{1+x^2}
\end{equation}
with real argument $x$. $f(x)$ has the Taylor series expansion
\begin{equation}
 \frac{1}{1+x^2} = \sum_{n=0}^{\infty} (-1)^n x^{2n}
\end{equation}
about $x=0$. This series is only convergent provided $|x|<1$, yet $f(x)$ is clearly finite for all real $x$. So, what is the cause of the failure of the series to diverge for $|x|\ge1$? The answer is found by allowing $x$ to be complex. We then find that $f(x)$ is singular at $x=\pm i$. Somewhat surprisingly, it is this singularity that also causes the Taylor series to diverge for real-valued $|x|\ge1$. In fact, the Taylor series will diverge at \emph{all} points outside the circle of convergence (see Fig.~\ref{fig:circle-of-convergence}).

\begin{figure}
 \begin{center}
  \includegraphics[width=9cm]{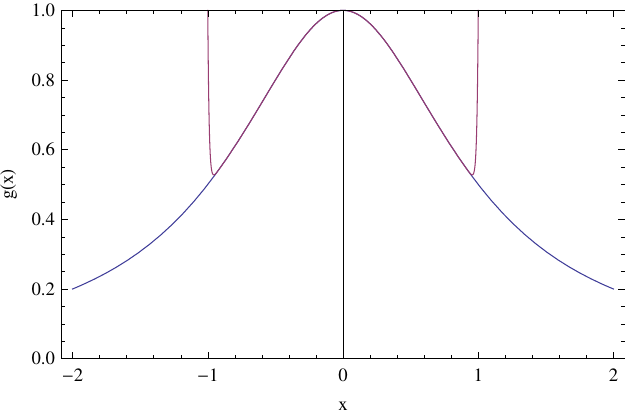}
 \end{center}
 \caption[Series expansion of $1/(1+x^2)$]{\emph{Series expansion of $1/(1+x^2)$.} The Taylor series (purple) diverges at $|x|=1$, whereas $f(x)$ (blue) is finite for all real $x$. The Pad\'e approximant is exactly equal to $f(x)$, so their plots are overlapping.}
 \label{fig:1-1px2}
\end{figure}

We next try to improve the convergence of the Taylor of series using Pad\'e approximants. We find that, provided the Pad\'e approximant is chosen such that it can accurately represent the $1/(1+x^2)$ singularity of $f(x)$, it will successfully extend the domain of the series representation beyond the radius of convergence. This requirement amounts to requiring that the series in the denominator be at least of order $x^2$. In fact, in this case we find that the Pad\'e approximant fully recovers the original function,
\begin{equation}
 P^N_{M} = \frac{1}{1+x^2} \qquad \forall ~M\ge2 .
\end{equation}
This remarkable result is illustrated in Fig.~\ref{fig:1-1px2}.

\subsubsection*{Example 2}
In the previous case, we found that the Pad\'e approximant was able to exactly reproduce the original function from its Taylor series. This success stems from the polynomial nature of the numerator and denominator of $f(x)$. However, Pad\'e resummation may also be extremely effective for functions which may not be represented as a rational function of two polynomials. To illustrate this point, consider the real-valued function
\begin{equation}
 g(x) \equiv \frac{\ln{(1+x)}}{x},
\end{equation}
with real argument $x\in (-1,\infty)$. $g(x)$ has the Taylor series expansion
\begin{equation}
 \frac{\ln{(1+x)}}{x} =  \sum_{n=0}^{\infty} \frac{(-1)^n}{n+1} x^{n}
\end{equation}
about $x=0$. As before, this series is only convergent provided $|x|<1$, yet $g(x)$ is finite for all real $x>-1$. This divergence at $|x|\ge1$ may be attributed to the singularity of $g(x)$ at $x=-1$. 

\begin{figure}
 \begin{center}
  \includegraphics[width=9cm]{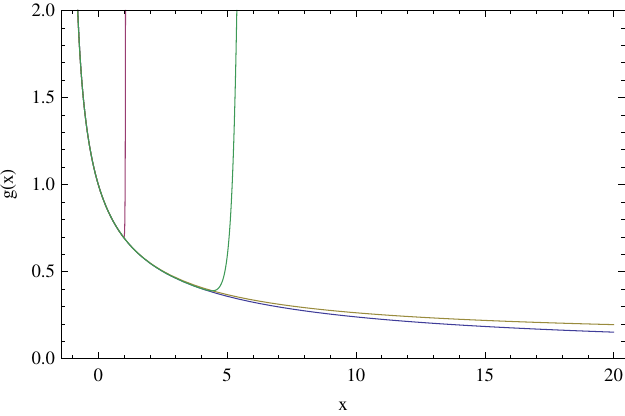}
 \end{center}
 \caption[Series expansion of $\ln(1+x)/x$]{\emph{Series expansion of $g(x)=\ln(1+x)/x$.} The Taylor series (purple) diverges at $|x|=1$, whereas $g(x)$ (blue) is finite for all real $x$. The diagonal Pad\'e approximant $P^2_2$ (brown) dramatically outperforms the off-diagonal Pad\'e approximant $P^{32}_8$ (green) despite needing much fewer terms from the Taylor series.}
 \label{fig:ln1px-x}
\end{figure}

Since $g(x)$ cannot be written as a rational function, one may not expect Pad\'e resummation to be so effective. On the contrary, Fig.~\ref{fig:ln1px-x} shows that Pad\'e resummation may be extremely effetive, particularly when the right choice of $N$ and $M$ is made. In particular, the \emph{diagonal} Pad\'e approximant, $P_2^2$ yields remarkable results with only $4$ terms in the Taylor series required. Not only does its domain extend far beyond the radius of convergence of the Taylor series, it also dramatically outperforms the \emph{off-diagonal} Pad\'e approximant, $P^{32}_8$, while requiring only a tenth as many terms from the original Taylor series. This highlights the importance of choosing the correct Pad\'e approximant for the problem. In general, the choice $N\sim M$ is best.

\chapter{Massive Scalar Field in Nariai}
\label{app:rosenthal-massive}
In Sec.~\ref{subsec:static full Green} we used the massive field approach of Rosenthal \cite{Rosenthal:2003} to compute the regularized self-force acting on a static particle in Nariai spacetime. In this prescription, we must compute the derivative of a massive scalar field in the limit of the field mass going to infinity. In this appendix, we derive an explicit expression for the massive scalar field, closely following the method of Ref.~\cite{Rosenthal:2004}, which considered the Schwarzschild case.

We will solve the massive scalar field equation,
\begin{equation}
\left(\square  - \Lambda ^2- \xi  R\right)\phi^{\Lambda }=-4\pi \rho^{\Lambda} ,
\end{equation}
where $\Lambda$ is the mass of the field, $\xi$ is the coupling to the scalar curvature, $R=4$ is the Ricci scalar in Nariai and $\rho^{\Lambda}$ represents a static point source. Decomposing into spherical harmonics and switching to the `tortoise' coordinate defined by
\begin{equation}
\frac{d\rho_*}{d\rho} = \frac{1}{f},
\end{equation}
where $f(\rho)=1-\rho^2$, the radial part of the wave equation becomes
\begin{equation}
\label{eq:nariai-radial-wave}
\phi^{\Lambda }_{l m}{}_{,\rho_*\rho_*}-\left(f(\rho)l(l+1) +f(\rho) \left(\Lambda ^2+ \xi  R\right)\right)\phi^{\Lambda }_{l m}=-4\pi  \rho^{\Lambda}_{ l m} f(\rho)
\end{equation}
with
\begin{equation}
\rho ^{\Lambda}_{ l m}= q \delta \left(\rho-\rho_0\right)\sqrt{f\left(\rho_0\right)}Y^{l m *}\left(\theta _0,\phi _0\right).
\end{equation}
Assuming two linearly independent solutions $\chi_+$ and $\chi_-$ to the homogeneous equation, both dependent on $l$ , the solution can be expressed as
\begin{equation}
\phi^{\Lambda }_{l m}= -4\pi  q \sqrt{f\left(\rho_0\right)}Y^{l m *}\left(\theta _0,\phi _0\right) \frac{\chi _-\left(\rho_<\right)\chi _+\left(\rho_>\right)}{W\left[\chi _-,\chi _+\right]_{\rho_0}}
\end{equation}
where $W\left[\chi _-,\chi _+\right]_{\rho_0}$ is the Wronskian evaluated at $\rho_0$. Note, that the factor of $f(\rho)$ has been absorbed into the fact that the derivatives of the Wronskian are changed from $\rho$ to $\rho_*$ derivatives.  Now, performing the sum over $m$, we get
\begin{equation}
\label{eq:wkb-phi-l}
\phi^{\Lambda }_l= -2q \left(l+\frac{1}{2}\right)\sqrt{f\left(\rho_0\right)}\frac{\chi _-\left(\rho_<\right)\chi _+\left(\rho_>\right)}{W\left[\chi _-,\chi _+\right]_{\rho_0}}
\end{equation} where there is no Legendre polynomial since we let $\theta -\theta '=0$. We now need to take a $\rho$ derivative and sum over $l$. To do so, we split into a part which is regular as $\rho \to  \rho_0$ and a part which is divergent,
\begin{equation}
\label{eq:wkb-phi-split}
\phi^{\Lambda}_{,\rho}= \sum _{l=0}^{\infty }h_l+ \sum _{l=0}^{\infty }\left(\phi^{\Lambda}_{l,\rho}-h_l\right).
\end{equation}
We use a WKB approximation \cite{Bender:Orszag} to find the large $l$ behavior of $h_l(\rho)$:
\begin{subequations}
\label{eq:wkb-approx}
\begin{align}
\chi_+ &= e^{-S_0/\delta + S_1 - S_2 \delta}\\
\chi_- &= c_1 e^{S_0/\delta + S_1 + S_2 \delta} + c_2 e^{-S_0 / \delta + S_1 - S_2 \delta}.
\end{align}
\end{subequations}
Note that for the outer solution, we want the field to be $0$ at infinity, which fixes the boundary conditions. Also note that $S_0$ and $S_2$ must have the same sign as a result of the differential equations they satisfy. The differential equations satisfied by  $S_0$, $S_1$ and $S_2$ are found by substituting $e^{S_0/\delta + S_1 + S_2 \delta}$ into \eqref{eq:nariai-radial-wave}, and independently setting each power of  $\delta$ equal to zero, noting that the term involving $V(\rho)$ will be large for large $l$ and so balances the dominant $1/\delta^2$ term.
This gives
\begin{subequations}
\begin{align}
 S_0'(\rho) &= \frac{1}{f(\rho)} \sqrt{V(\rho)} \\
 S_1'(\rho) &= - \frac{V'(\rho)}{4 V(\rho)} \\
 S_2'(\rho) &= \frac{1}{f(\rho)}\left(\frac{f(\rho)(f(\rho)V'(\rho))_{,\rho}}{8V(\rho)^{3/2}}-5\frac{(f(\rho)V'(\rho))^2}{32V(\rho)^{5/2}}\right),
\end{align}
\end{subequations}
where the equation for $S_1$ may be solved immediately, giving
\begin{equation}
S_1 = -\frac{1}{4} \ln V(\rho).
\end{equation}
We note that the $S_n$ goes like $L^{1-n}$ and $\Lambda^{1-n}$, where $L\equiv(l+1/2)$. Next, we substitute our two WKB solutions, \eqref{eq:wkb-approx}, into \eqref{eq:wkb-phi-l} and differentiate with respect to $\rho$ to get
\begin{equation}
\phi^{\Lambda}_{l ~,\rho} = \frac{e^{S(\rho)} \left(c_1 + c_2 e^{-2 (S_0(\rho_0) + S_2(\rho_0))}\right) L q S'(\rho)}{c_1 \sqrt{f(\rho_0)} \left(S_0'(\rho_0) + S_2'(\rho_0)\right)},
\end{equation}
where 
\begin{equation}
S(\rho) \equiv S_0(\rho_0) - S_1(\rho_0) + S_2 (\rho_0) - (S_0(\rho) - S_1(\rho) + S_2 (\rho)).
\end{equation}
We can now drop the $c_2$ term since it dies off for large $L$ faster than any negative power of $L$ (since it is like $e^{-L}$), and are left with
\begin{equation}
\label{eq:wkb-phi-l-r}
\phi^{\Lambda}_{l ~,\rho} = \frac{e^{S(\rho)} L q S'(\rho)}{\sqrt{f(\rho_0)} \left(S_0'(\rho_0) + S_2'(\rho_0)\right)}.
\end{equation}
Expanding $S(\rho)$ in a large $L$ asymptotic series, we find that the order $L$ contribution (coming from $S_0$) is given by
\begin{equation}
\label{eq:wkb-alpha}
\alpha \equiv \sin^{-1}(\rho) - \sin^{-1}(\rho_0) \approx \frac{\rho-\rho_0}{\sqrt{1-\rho_0^2}}+\frac{\rho_0 (\rho-\rho_0)^2}{2 \left(1-\rho_0^2\right)^{3/2}}+O\big[(\rho-\rho_0)^3\big].
\end{equation}

We now deal with the essential singularity in \eqref{eq:wkb-phi-l-r} (from the order $L$ term in $S(\rho)$) by rewriting the derivative of the field as
\begin{equation}
\phi^{\Lambda}_{l,~\rho} =
e^{-\alpha L}\left(\frac{e^{S(\rho)+\alpha  L} L q S'(\rho)}{\sqrt{f(\rho_0)}\left(S_0'(\rho_0)+S_2'(\rho_0)\right)}\right)
\end{equation}
and we can find an asymptotic expansion in $L$ for the term in the brackets.
The order $L$ component is
\begin{subequations}
\label{eq:wkb-A-B-C}
\begin{equation}
A_l = -\frac{q (1 - \rho_0^2)^{1/4}}{(1 - \rho^2)^{3/4}}
\end{equation}
The order $1$ component is 
\begin{multline}
B_l = \frac{1}{48} q \left[\frac{-1+\R_0^2}{-1+\R^2}\right]^{1/4}
 \Bigg\{
 \frac{24 \R}{1-\R^2}\\
 +\frac{1}{\sqrt{1-\R^2}}\Bigg[6 \left(-1+4 \Lambda ^2+4 R \xi \right) (\sin^{-1}(\R)-\sin^{-1}(\R_0))\\
 + \frac{12 (-10 + \R) (873649 + 2 \R (-27460 + 779 \R))}{\left(1-\R_0^2\right)^{7/2}}\Bigg]\Bigg\}
\end{multline}
The order $L^{-1}$ component is
\begin{multline}
C_l = \frac{q}{4608 \left(1-\R^2\right)^{7/4} \left(1-\R_0^2\right)^{27/4}}\Bigg\{576 \left(-1+\R_0^2\right)^6 \left(\R^2-\R_0^2\right)\\
+\left(-1+\R^2\right) \Big[-12 (-10+\R) (873649+2 \R (-27460+779 \R))\\-6 \left(1-\R_0^2\right)^{7/2} \left(-1+4 \Lambda ^2+4 R \xi \right) (\sin^{-1}(\R)-\sin^{-1}(\R_0))\Big]^2\\
+\frac{48 \R}{1- \R^2} \left(1-\R^2\right)^{3/2} \left(1-\R_0^2\right)^7 \Big[ -12 (-10+\R) (873649+2 \R (-27460+779 \R))\\-6\left(1-\R_0^2\right)^{7/2}\left(-1+4 \Lambda ^2+4 R \xi \right)( \sin^{-1}(\R)-\sin^{-1}(\R_0))\Big]\Bigg\}.
\end{multline}
\end{subequations}

Next, we must calculate the sum over $L$,
\begin{equation}
\label{eq:wkb-lsum}
\sum_{l=0}^{\infty} h_l = \sum_{l=0}^{\infty} e^{-\alpha L} \left[ A_l L + B_l + C_l L^{-1}  + O(L^{-2})\right].
\end{equation}
where we can make use of the identities
\begin{subequations}
\begin{align}
\sum_{l=0}^{\infty} L e^{-\alpha L} &=  \frac{\cosh(\alpha/2)}{4 \sinh^2(\alpha/2)}\\
\sum_{l=0}^{\infty} e^{-\alpha L} &=  \frac{1}{2 \sinh(\alpha/2)} \\
\sum_{l=0}^{\infty} L^{-1} e^{-\alpha L} &=  2 \tanh^{-1}\left(e^{-\alpha/2}\right)
\end{align}
\end{subequations}
In the end, we only need our result as an expansion in inverse powers of $(\R-\R_0)$, up to order $(\R-\R_0)^0$. Combining \eqref{eq:wkb-alpha}, \eqref{eq:wkb-A-B-C} and \eqref{eq:wkb-lsum} and expanding in $(\R-\R_0)$, we get
\begin{equation}
\label{eq:hl}
\sum_{l=0}^{\infty} h_l = -\frac{q \sqrt{1-\R_0^2}}{(\R-\R_0)^2}+\frac{q \left(-2+3 \Lambda ^2 + 3 R \xi \right)}{6 \sqrt{1-\R_0^2}}+O\left[(\R-\R_0)^1\right].
\end{equation}

With the divergent part of Eq.~\eqref{eq:wkb-phi-split} calculated, we must now calculate the regular part for large $\Lambda$. In this case, we only need the term that is constant in $\R$, so that we have $\R=\R_0$. Letting $L \to \lambda Y$ and using the Riemann integral over Y, we can then compute the sum over $l$ of this regular part to get
\begin{equation}
 -\frac{q \R_0 \Lambda}{2 (1-\R_0^2)}.
\end{equation}
Combining this with Eq.~\eqref{eq:hl} we get the expression for the radial derivative of the scalar field (up to $O\left[ (\R-\R_0)^0 \right]$):
\begin{equation}
 \phi^{\Lambda}_{,\R} =  -\frac{q \sqrt{1-\R_0^2}}{(\R-\R_0)^2}+\left(-\frac{q \R_0 \Lambda}{2 (1-\R_0^2)} + \frac{q \left(-2+3 \Lambda ^2 + 3 R \xi \right)}{6 \sqrt{1-\R_0^2}}\right)+O\left[(\R-\R_0)^1\right].
\end{equation}

Finally, applying the massive field prescription for calculating the self-force:
\begin{equation}
 f_\R = q \lim_{\Lambda \to \infty} \left\{ \lim_{\delta \to 0} \Delta \phi_{,\R} + \frac12 q \left[ \Lambda^2 n_\R + \Lambda a_\R \right] \right\}
\end{equation}
where 
\begin{equation}
 \Delta \phi_{,\R} \equiv \phi_{,\R} - \phi^{\Lambda}_{,\R}
\end{equation}
we see that the $O(\Lambda)$ and $O(\Lambda^2)$ terms exactly cancel and 
we get
\begin{equation}
 f_\R = q~ \phi - \frac{q\sqrt{1-\rho_0^2}}{(\R-\R')^2} + \frac{ (\xi R- \frac23)}{2\sqrt{1-\R^2}}.
\end{equation}
This is exactly the expression, \eqref{S-F static}, used for computing the static self-force in Nariai spacetime in Chapter~\ref{ch:nariai}.

\chapter{Large-\texorpdfstring{$\lam$}{lambda} Asymptotics of the Hypergeometric \texorpdfstring{${}_2F_1$}{2F1} function *}
\label{appendix-hypergeom}
To determine the singularity structure of the Green function in Sec.~\ref{subsec:large-l} we required the large $\lam$ asymptotic behavior of
$
{}_2F_1(-\beta, -\beta; -2 \beta; z)$ where $\beta = -\frac{1}{2} + i \lam$, $z = -e^{-T}$.
The required asymptotics may be found by applying the WKB method \cite{Bender:Orszag} to the hypergeometric differential equation
\beq
z(1-z)\frac{d^2 u}{dz^2} - \left[2\beta(1 - z)  + z \right] \frac{du}{dz} - \beta^2 u = 0
\eeq
which has solutions $u(z) = {}_2F_1(-\beta,-\beta;-2\beta;z)$. 
Inserting the WKB ansatz
\begin{equation}
u(z) \sim e^{ \lam S_0(z) + S_1(z) + \lam^{-1} S_2(z) \dots}   \label{eq:wkb-def}
\end{equation}
immediately yields a quadratic equation for $S_0^\prime$,  
\beq
z(1-z) (S_0^\prime)^2  - 2 i (1-z) S_0^\prime + 1 = 0 .
\eeq
In our case, $z = -e^{- T}$, we require  the root which is finite at $z = 0$. We impose $S_0(0) = 0$ to get
\begin{multline}
S_0^\prime =  \frac{i}{z} \left(1 - (1-z)^{-1/2}  \right)
 \\\Rightarrow  S_0(z)= \ln \left( \frac{(-z)}{4} \, \frac{\left[ \sqrt{1-z} + 1 \right]}{\left[ \sqrt{1 - z} - 1 \right]}\right) = -T - 2\ln 2 + \ln \left( \frac{\sqrt{1+e^{-T}} + 1}{\sqrt{1 + e^{-T}} - 1}  \right) \label{eq:wkb-S2}
\end{multline}
At next order, we obtain the equation
\beq
2 \left[  i (1 - z)  -  z (1 - z) S_0^\prime \right]  S_1^\prime = z(1-z) S_0^{\prime \prime} + (1-2z) S_0^\prime + i .
\eeq
It is straightforward to show that this reduces to
\beq
S_1^\prime = \left( 4(1-z) \right)^{-1} \quad \quad \Rightarrow \quad S_1 = -\frac{1}{4} \ln \left(1 + e^{-T}  \right) .   \label{eq:wkb-S1}
\eeq
Inserting (\ref{eq:wkb-S1}) and (\ref{eq:wkb-S2}) into (\ref{eq:wkb-def}) leads to the quoted result, Eq.~(\ref{eq:hypergeom-asymp}). Of course, the asymptotic approximation may be further refined by taking the WKB method to higher orders. 

\chapter{Poisson Sum Asymptotics *}\label{appendix:poisson-sum}
In this appendix we derive asymptotic approximations for the singular structure of the Green function using the Poisson sum formula (\ref{InRN}). Our starting point is expression (\ref{RN-asymp-Hankel}) for the `$n$=0' fundamental modes, in which the Legendre polynomials $P_l(\cos \gam)$ have been replaced by angular waves $\mathcal{Q}_{\nu-1/2}^{(\pm)}(\cos \gam)$ which are, in turn, approximated by Hankel functions $H^{(\mp)}_0(\nu \gam)$ using (\ref{Olver-approx}). 

Let us consider the $\II_1$ and $\II_2$ integrals arising from substituting (\ref{RN-asymp-Hankel}) into (\ref{InRN}). These integrals are singular at $\chi = 2 \pi - \gam$ and $\chi = 2 \pi + \gam$, respectively. 
First, let us consider $\II_1$ (\ref{InRN}) which can be written
\beq
\II_1 \approx -\Agam \text{Re} \int_0^\infty d\nu (-i \nu)^{1/2} e^{i (\chi - 2\pi) \nu} H_{0}^{(+)}(\nu \gam) 
\label{eq:I1-real}
\eeq
with $\Agam$ as defined in Eq.~(\ref{eq:A-gam}) .

For $\chi > 2 \pi - \gam$, the integral may be computed by rotating the contour onto the \emph{positive} imaginary axis ($\nu = i z$) to obtain
\begin{eqnarray}
\II_1 &\approx& -\frac{2 \Agam}{\pi} \int dz z^{1/2} e^{-(\chi - 2\pi) z} K_0 (\gam z) \nonumber \\
       &\approx& -\frac{\Agam \sqrt{\pi}}{2 [\chi - (2\pi - \gam)]^{3/2}} {}_2F_1 \left( 3/2, 1/2; 2; \frac{\chi - 2\pi - \gam}{\chi - 2\pi + \gam} \right)  \label{eq:I1-after}
\end{eqnarray}
Here we have applied the identity $H_0^{(+)}(ix) = 2 K_0(x) / (i \pi)  $, where $K_0$ is the modified Bessel function of the second kind, and the integral is found from Eq.~6.621(3) of Ref.~ \cite{GradRyz}. 


For $\chi < 2 \pi - \gam$, the integral may be computed by rotating the contour onto the \emph{negative} imaginary axis ($\nu = -i z$). First, we make the replacement 
\begin{equation}
H_0^{(+)}(\nu \gam) = 2 J_0(\nu \gam) - H_0^{(-)}(\nu \gam)
\end{equation}
and note $H_0^{(-)}(- i x) = 2 K_0(x) / (-i \pi)$ to obtain
\begin{eqnarray}
  \II_1 & \approx & \frac{2 \Agam}{\pi} \text{Re} \int dz z^{1/2} e^{-(2\pi - \chi) z} \left[ \pi I_0(\gam z) + i K_0 (\gam z) \right] 
\end{eqnarray}
Here $I_0$ is a modified Bessel function of the first kind. Since we are taking the real part, the $K_0$ term is eliminated, and we obtain
\begin{equation}
  \II_1  \approx  \frac{2 \Agam}{\sqrt{\pi}} \left( 2\pi - \gam - \chi \right)^{-1} \left( 2\pi + \gam - \chi \right)^{-1/2} E\left( \frac{2 \gam}{2 \pi + \gam - \chi} \right)
\end{equation}
where $E$ is the elliptic integral of the second kind defined in, for example, Equation.~$8.111(3)$ of Ref.~\cite{GradRyz}.

The $\II_2$ integral may be calculated in a similar manner. For $\chi < 2\pi + \gam$, we rotate the contour onto the negative imaginary axis,
 \begin{eqnarray}
  \II_2 & \approx & -\Agam \text{Re} \int_0^\infty d \nu (- i \nu)^{1/2} H_0^{(-)}(\nu \gam) e^{i (\chi - 2\pi) \nu} \nonumber \\
   & \approx & \frac{2 \Agam}{\pi} \text{Re} \; i \int_0^\infty dz z^{1/2} e^{-(2\pi - \chi) z} K_0(\gam z)  \quad \quad = 0
\end{eqnarray}
For $\chi > 2\pi + \gam$, we rotate the contour onto the positive imaginary axis after taking the complex conjugate 
 \begin{eqnarray}
  \II_2 & \approx & -\Agam \text{Re} \int_0^\infty d \nu (i \nu)^{1/2} H_0^{(+)}(\nu \gam) e^{i (\chi - 2\pi) \nu} \nonumber \\
   & \approx & \frac{2 \Agam}{\pi} \text{Re} \int_0^\infty dz z^{1/2} e^{-(\chi - \pi) z} \left(i \pi I_0(\gam z) +  K_0(\gam z) \right)  
\end{eqnarray}
The imaginary term does not contribute and hence $\II_2$ is equal and opposite to $\II_1$ defined by Eq.~(\ref{eq:I1-after}) when $\chi > 2 \pi + \gam$.

 \chapter{Green function on  \texorpdfstring{$T\times \mathbb{S}^2$}{T cross S2} *} \label{appendix-Green on S2}
 
 To the best of our knowledge, the four-fold singularity structure for the Green function of Sec.~\ref{sec:singularities} has not been shown before
 in the literature (with the
 exception of~\cite{Ori1}) within the theory of General Relativity.
 We therefore wish to illustrate its derivation and manifestation in the simplest of spacetimes including $ \mathbb{S}^2$-topology:
 \begin{equation} \label{eq:RxS2}
 ds^2=-dt^2+d\Omega^2_2,
 \end{equation}
where the Synge world function is simply given by
$\sigma=\frac{1}{2}\left(-\Delta t^2+\phif^2\right)$, in the case of a conformally-coupled ($\xi=1/8$) scalar field.
Let $x=(t,\theta,\phi)$ denote any point in this spacetime.
In this appendix, $\Delta t\equiv t-t'$ is the time interval between any two spacetime points $x$ and $x'$ (rather than
the free parameter determining the matching time).

We introduce the Wightman function $G_+(x,x')$ (it satisfies the homogeneous scalar wave equation - see, e.g.,~\cite{Birrell:Davies}), from  
which the `retarded' Green function $G_{ret}(x,x')$ is
easily obtained:
\begin{align} \label{eq:G_+ mode sum}
G_+(x,x')&=
\sum_{l=0}^{+\infty}\sum_{m=-l}^{+l}\Phi_{lm}(x)\Phi^*_{lm}(x')=\frac{1}{4\pi}\sum_{l=0}^{+\infty}e^{- i(l+1/2)\Delta t}P_l(\cos\gamma), \\
G_{ret}(x,x')&=-2\theta(\Delta t)\ \text{Im}\left(G_+(x,x')\right),
\end{align}
where
\begin{equation}
 \Phi_{lm}(x)=e^{-i(l+1/2)t}Y_{lm}(\theta,\phi)/\sqrt{(2l+1)}
\end{equation}
 are the Fourier-decomposed
scalar field modes on (\ref{eq:RxS2}) normalized with respect to the scalar product
\beq 
\left(\Phi_{lm},\Phi_{l'm'}\right)=-i\int_{\Sigma}dVn^{\mu}\left[\Phi_{lm}(x)\partial_{\mu}\Phi^*_{l'm'}(x)-
\Phi^*_{l'm'}(x)\partial_{\mu}\Phi_{lm}(x)\right]=\delta_{ll'}\delta_{mm'},
\eeq
where $\Sigma$ is a Cauchy hypersurface with future-directed unit normal vector $n^{\mu}$ and volume element $dV$.

 We now apply exactly the same tricks as in section \ref{subsec:Poisson} in order to derive the four-fold singularity structure in 
 the Green function
 from the large-$l$ asymptotics of the field modes.
 We use the {\it Poisson sum formula}~\cite{M&F,Aki:Richards} 
 \begin{equation} \label{Poisson sum}
 \sum_{l=0}^{+\infty}g(l+1/2)=\sum_{s=-\infty}^{+\infty}(-1)^s\int_0^{+\infty}d\nu g(\nu)e^{2\pi is\nu}
 \end{equation}
 to re-write the mode sum in (\ref{eq:G_+ mode sum}) as
 \begin{align}\label{eq:G_+ mode sum,rewritten}
 4\pi G_+(x,x')&=\sum_{s=-\infty}^{+\infty}(-1)^s\int_0^{+\infty}d\nu e^{-i\nu\Delta t}P_{\nu-1/2}(\cos\gamma)e^{2\pi is\nu}
 =
  \sum_{N=1}^{+\infty}G_+^N, \\
 G_+^N(x,x') &\equiv \int_0^{+\infty}d\nu R_N(\cos\gamma)e^{-i\nu\Delta t}.
 \end{align}

 The Legendre functions $P_{\mu}(\cos\gamma)$ and $Q_{\mu}(\cos\gamma)$, as well as $R_N(\cos\gamma)$,
 are standing waves.
 This is in contrast to $Q_{\mu}^{(\pm)}(\cos\gamma)$, which are travelling waves.
 
 We can now use large-order uniform asymptotics (see \cite{Olver:1974,Jones'01}) for the Legendre functions:
 \begin{align} \label{eq:large-l P}
 P_{\nu-1/2}(\cos\gamma)&\sim \left(\frac{\gamma}{\sin\gamma}\right)^{1/2}J_0(\nu\gamma), \quad |\nu|\to \infty,\\
& \quad \text{``valid in a closed uniform interval containing $\gamma=0$"},\nonumber
 \\
 Q_{\nu-1/2}(\cos\gamma)&\sim -\frac{\pi}{2}\left(\frac{\gamma}{\sin\gamma}\right)^{1/2}Y_0(\nu\gamma), \quad |\nu|\to \infty,
\\& \quad \text{``valid with respect to $\gamma\in (0,\pi/2]$"},\nonumber
 \\ \label{eq:large-l Qpm}
 \mathcal{Q}_{\nu-1/2}^{(\pm)}(\cos\gamma)&\sim \frac{1}{2}\left(\frac{\gamma}{\sin\gamma}\right)^{1/2}H_0^{(\mp)}(\nu\gamma),
 \qquad |\nu|\to\infty,
 \end{align}
 to leading order.
 
 The contribution to the Wightman function from the $N=1$ orbit wave is immediately obtained by using
 the large-order asymptotics of the Legendre function $P_{\nu}(\cos\gamma)$ only, which are ``valid in a closed uniform interval
 containing $\gamma=0$" - this is what we will mean by a result being valid ``near" $\gamma=0$.
 Similarly, we can obtain a result valid ``near" $\gamma=\pi$ by using in (\ref{eq:G_+ mode sum})
 the symmetry $P_l(\cos\gamma)=(-1)^lP_l(\cos(\pi-\gamma))$ for $l\in \mathbb{N}$.
 We then obtain for $N=1$:
 \begin{align} \label{eq:G_+,N=1}
 4\pi G_+^{N=1}(x,x')\sim &\sqrt{\frac{\gamma}{\sin\gamma}} \frac{1}{\sqrt{\gamma^2-\Delta t^2}}, 
 \quad &\text{``near" $\gamma=0$}
 \\
 4\pi G_+^{N=1}(x,x')\sim &\sqrt{\frac{\pi-\gamma}{\sin(\pi-\gamma)}} \frac{-i}{\sqrt{(\pi-\gamma)^2-(\Delta t-\pi)^2}}
 \\=& 
 \sqrt{\frac{\pi-\gamma}{\sin(\pi-\gamma)}} \frac{-i}{\sqrt{-(\Delta t-\gamma)\left[\Delta t-(2\pi-\gamma)\right]}},
 \quad &\text{``near" $\gamma=\pi$}
 \end{align}
 where, for convergence, a small imaginary part was given to $\Delta t$
and/or $\gamma$, in agreement with the Feynman prescription `$\sigma\to \sigma+i\epsilon$'.
 The result for $G_+^{N=1}(x,x')$ valid ``near" $\gamma=0$ is singular at $\Delta t=\pm \gamma$, corresponding
 to $\sigma=0$ before a caustic has been crossed.
It is in accord with the Hadamard form in 3-D~\cite{Decanini:Folacci:2005a} and the 
Van Vleck determinant (\ref{eq:Delta_phi}), before a caustic has been crossed
 (and so without the phase factor).
  The result for $G_+^{N=1}(x,x')$ valid ``near" $\gamma=\pi$ is singular at $\Delta t=\gamma$, corresponding to the
  case where it has not gone through any caustics, and at $\Delta t=2\pi-\gamma$, corresponding to the
  case where it has gone through one caustic; it has thus picked up a factor ``$-i$", as expected.
 Note that these zeros inside the square root in the denominator are simple zeros along the null geodesic, except
 at the caustic point itself, where the two zeros coincide and so it becomes a double zero. 
  
 Similarly to $N=1$, we can use (\ref{eq:G_+ mode sum,rewritten}) and the asymptotics (\ref{eq:large-l Qpm}) together with~\cite{GradRyz}
 \begin{multline}
 I_{\pm}(T,\gamma)\equiv \int_0^{\infty}d\nu e^{-i\nu(T-i\epsilon)}H_0^{(\pm)}(\nu\gamma)\\=
 \frac{1}{\sqrt{\gamma^2-(T-i\epsilon)^2}}\left[1\mp \frac{2i}{\pi}\ln\left(iX+\sqrt{1-X^2}\right)\right],
 \quad T\in \mathbb{R},\ \epsilon>|\text{Im}\gamma|,
 \end{multline}
 where $X\equiv (T-i\epsilon)/\gamma$ (again, a small imaginary part needs to be given for convergence, in accordance with
 the Feynman prescription), 
 in order to obtain for $N>1$:
 \begin{multline} \label{eq:G_+,N>1}
 4\pi G_+^N(x,x')
 \sim \sqrt{\frac{\Xi}{\sin\Xi}} \frac{(-1)^{N/2}}{2}\\ \times
 \begin{cases}
 \left[I_+(\Delta t+N\pi,\Xi)+I_-(\Delta t-N\pi,\Xi)\right],
  &\text{``near" $\gamma=0$} 
 \\
 -i
 \left[I_+(\Delta t+(N-1)\pi,\Xi)+I_-(\Delta t-(N+1)\pi,\Xi)\right],
  & \text{``near" $\gamma=\pi$} 
 \end{cases}
 \end{multline}
 for $N$ even, where $\Xi=\gamma$ ``near" $\gamma=0$ and $\Xi=\pi-\gamma$ ``near" $\gamma=\pi$. 
 For $N$ odd, merely:  (1) swap $I_{\pm}\to I_{\mp}$, and 
 (2) replace $N$ by $N-1$ if ``near" $\gamma=0$ or  $N$ by $N+1$ if ``near" $\gamma=\pi$ in (\ref{eq:G_+,N>1}).
 We can re-write:
 $\gamma^2-(\Delta t\pm N\pi-i\epsilon)^2=-\left[(\Delta t-i\epsilon)-(\mp N\pi-\gamma)\right]\left[(\Delta t-i\epsilon)-(\gamma\mp N\pi)\right]$.
 Note, however, that $I_{\pm}(T,\gamma)$ is regular at $X=\mp 1$.

 We then have that the singular behavior goes as
 \begin{equation}
 4\pi G_+(x,x')\sim 
 \frac{1}{2}\sqrt{\frac{\Xi}{\sin\Xi}}
 \begin{cases}
 +I_+(\Delta t,\gamma),  &0<\Delta t<\pi,\quad \gamma\sim 0
 \\
 -iI_+(\Delta t-\pi,\pi-\gamma),  &\pi<\Delta t<2\pi,\quad \gamma\sim \pi
 \\
 -I_+(\Delta t-2\pi,\gamma),  &2\pi<\Delta t<3\pi,\quad \gamma\sim 0
 \\
 +iI_+(\Delta t-3\pi,\pi-\gamma),  &3\pi<\Delta t<4\pi,\quad \gamma\sim \pi
 \end{cases}
 \end{equation}
 The 4-fold singularity structure arises clearly: a phase of $\pi/2$ is picked up every time the null geodesic
joining $x$ and $x'$ goes through a caustic ($\gamma=0$ or $\pi$).
 
 The expression for $G_+(x,x')$ is simplified by noting that, ``near" $\gamma=0$:
For $N$ even
\begin{multline}
 4\pi \left[G^N_+(x,x')+G^{N+1}_+(x,x')\right]\sim
 \\
  i^N\sqrt{\frac{\gamma}{\sin\gamma}}
\left[\frac{1}{\sqrt{\gamma^2-(\Delta t-N\pi-i\epsilon)^2}}+\frac{1}{\sqrt{\gamma^2-(\Delta t+N\pi-i\epsilon)^2}}\right],
\end{multline}
and for N odd
\begin{multline} 
4\pi \left[G^N_+(x,x')+G^{N+1}_+(x,x')\right]\\
\sim
 i^N\sqrt{\frac{\pi-\gamma}{\sin (\pi-\gamma)}}
 \Bigg[\frac{1}{\sqrt{(\pi-\gamma)^2-(\Delta t-N\pi-i\epsilon)^2}}\\+\frac{1}{\sqrt{(\pi-\gamma)^2-(\Delta t+N\pi-i\epsilon)^2}}\Bigg].
\end{multline}
 Similarly ``near" $\gamma=\pi$.
 The Poisson sum formula has yielded a sum over geodesic paths, labelled by the index $N$, and
 the large-order asymptotics for the Legendre functions have yielded the correct singularity structure near the null geodesics, allowing
 for the correct phase change at each caustic.



\bibliographystyle{apsrev2}
\lhead{}
\bibliography{thesis}

\begin{thebibliography}{188}
\expandafter\ifx\csname natexlab\endcsname\relax\def\natexlab#1{#1}\fi
\expandafter\ifx\csname bibnamefont\endcsname\relax
  \def\bibnamefont#1{#1}\fi
\expandafter\ifx\csname bibfnamefont\endcsname\relax
  \def\bibfnamefont#1{#1}\fi
\expandafter\ifx\csname citenamefont\endcsname\relax
  \def\citenamefont#1{#1}\fi
\expandafter\ifx\csname url\endcsname\relax
  \def\url#1{\texttt{#1}}\fi
\expandafter\ifx\csname urlprefix\endcsname\relax\def\urlprefix{URL }\fi
\providecommand{\bibinfo}[2]{#2}
\providecommand{\eprint}[2][]{\url{#2}}

\bibitem[{LIG()}]{LIGO}
\bibinfo{howpublished}{\url{http://www.ligo.caltech.edu}}.

\bibitem[{VIR()}]{VIRGO}
\bibinfo{howpublished}{\url{http://www.virgo.infn.it}}.

\bibitem[{GEO()}]{GEO}
\bibinfo{howpublished}{\url{http://geo600.aei.mpg.de}}.

\bibitem[{LIS()}]{LISA}
\bibinfo{howpublished}{\url{http://lisa.nasa.gov}}.

\bibitem[{\citenamefont{Bell}(2008)}]{Bell:2008}
\bibinfo{author}{\bibfnamefont{T.~E.} \bibnamefont{Bell}},
  \bibinfo{journal}{Nature} \textbf{\bibinfo{volume}{452}}, \bibinfo{pages}{18}
  (\bibinfo{year}{2008}).

\bibitem[{\citenamefont{Vallisneri}(2009)}]{Vallisneri:2009}
\bibinfo{author}{\bibfnamefont{M.}~\bibnamefont{Vallisneri}},
  \bibinfo{journal}{Class. Quantum Grav.} \textbf{\bibinfo{volume}{26}},
  \bibinfo{pages}{094024} (\bibinfo{year}{2009}), \eprint{arXiv:0812.0751}.

\bibitem[{\citenamefont{Pretorius}(2005)}]{Pretorius:2005}
\bibinfo{author}{\bibfnamefont{F.}~\bibnamefont{Pretorius}},
  \bibinfo{journal}{Phys. Rev. Lett.} \textbf{\bibinfo{volume}{95}},
  \bibinfo{pages}{121101} (\bibinfo{year}{2005}), \eprint{gr-qc/0507014}.

\bibitem[{\citenamefont{Gonzalez et~al.}(2008)\citenamefont{Gonzalez, Sperhake,
  and Bruegman}}]{Gonzalez-Sperhake-Bruegman-2008}
\bibinfo{author}{\bibfnamefont{J.~A.} \bibnamefont{Gonzalez}},
  \bibinfo{author}{\bibfnamefont{U.}~\bibnamefont{Sperhake}}, \bibnamefont{and}
  \bibinfo{author}{\bibfnamefont{B.}~\bibnamefont{Bruegman}}
  (\bibinfo{year}{2008}), \eprint{arXiv:0811.3952}.

\bibitem[{\citenamefont{Hughes et~al.}(2005)\citenamefont{Hughes, Drasco,
  Flanagan, and Franklin}}]{Hughes:Drasco:2005}
\bibinfo{author}{\bibfnamefont{S.~A.} \bibnamefont{Hughes}},
  \bibinfo{author}{\bibfnamefont{S.}~\bibnamefont{Drasco}},
  \bibinfo{author}{\bibfnamefont{E.~E.} \bibnamefont{Flanagan}},
  \bibnamefont{and} \bibinfo{author}{\bibfnamefont{J.}~\bibnamefont{Franklin}},
  \bibinfo{journal}{Phys. Rev. Lett.} \textbf{\bibinfo{volume}{94}},
  \bibinfo{pages}{221101} (\bibinfo{year}{2005}).

\bibitem[{\citenamefont{Gair et~al.}(2004)\citenamefont{Gair, Barack,
  Creighton, Cutler, Larson, Phinney, and Vallisneri}}]{Gair:2004}
\bibinfo{author}{\bibfnamefont{J.~R.} \bibnamefont{Gair}},
  \bibinfo{author}{\bibfnamefont{L.}~\bibnamefont{Barack}},
  \bibinfo{author}{\bibfnamefont{T.}~\bibnamefont{Creighton}},
  \bibinfo{author}{\bibfnamefont{C.}~\bibnamefont{Cutler}},
  \bibinfo{author}{\bibfnamefont{S.~L.} \bibnamefont{Larson}},
  \bibinfo{author}{\bibfnamefont{E.~S.} \bibnamefont{Phinney}},
  \bibnamefont{and}
  \bibinfo{author}{\bibfnamefont{M.}~\bibnamefont{Vallisneri}},
  \bibinfo{journal}{Classical and Quantum Gravity}
  \textbf{\bibinfo{volume}{21}}, \bibinfo{pages}{S1595} (\bibinfo{year}{2004}).

\bibitem[{\citenamefont{Detweiler and Whiting}(2003)}]{Detweiler-Whiting-2003}
\bibinfo{author}{\bibfnamefont{S.}~\bibnamefont{Detweiler}} \bibnamefont{and}
  \bibinfo{author}{\bibfnamefont{B.~F.} \bibnamefont{Whiting}},
  \bibinfo{journal}{Phys. Rev. D} \textbf{\bibinfo{volume}{67}},
  \bibinfo{pages}{024025} (\bibinfo{year}{2003}).

\bibitem[{\citenamefont{Dirac}(1938)}]{Dirac-1938}
\bibinfo{author}{\bibfnamefont{P.~A.~M.} \bibnamefont{Dirac}},
  \bibinfo{journal}{Proc. R. Soc. Lond. A} \textbf{\bibinfo{volume}{167}},
  \bibinfo{pages}{148} (\bibinfo{year}{1938}).

\bibitem[{\citenamefont{DeWitt and Brehme}(1960)}]{DeWitt:1960}
\bibinfo{author}{\bibfnamefont{B.~S.} \bibnamefont{DeWitt}} \bibnamefont{and}
  \bibinfo{author}{\bibfnamefont{R.~W.} \bibnamefont{Brehme}},
  \bibinfo{journal}{Ann. Phys.} \textbf{\bibinfo{volume}{9}},
  \bibinfo{pages}{220} (\bibinfo{year}{1960}).

\bibitem[{\citenamefont{Hobbs}(1968{\natexlab{a}})}]{Hobbs:1968a}
\bibinfo{author}{\bibfnamefont{J.~M.} \bibnamefont{Hobbs}},
  \bibinfo{journal}{Ann. Phys.} \textbf{\bibinfo{volume}{47}},
  \bibinfo{pages}{141} (\bibinfo{year}{1968}{\natexlab{a}}).

\bibitem[{\citenamefont{Quinn and Wald}(1997)}]{Quinn:Wald:1997}
\bibinfo{author}{\bibfnamefont{T.~C.} \bibnamefont{Quinn}} \bibnamefont{and}
  \bibinfo{author}{\bibfnamefont{R.~M.} \bibnamefont{Wald}},
  \bibinfo{journal}{Phys. Rev.} \textbf{\bibinfo{volume}{D56}},
  \bibinfo{pages}{3381} (\bibinfo{year}{1997}), \eprint{gr-qc/9610053}.

\bibitem[{\citenamefont{Gralla et~al.}(2009)\citenamefont{Gralla, Harte, and
  Wald}}]{Gralla:Harte:Wald:2009}
\bibinfo{author}{\bibfnamefont{S.~E.} \bibnamefont{Gralla}},
  \bibinfo{author}{\bibfnamefont{A.~I.} \bibnamefont{Harte}}, \bibnamefont{and}
  \bibinfo{author}{\bibfnamefont{R.~M.} \bibnamefont{Wald}},
  \bibinfo{journal}{Phys. Rev.} \textbf{\bibinfo{volume}{D80}},
  \bibinfo{pages}{024031} (\bibinfo{year}{2009}), \eprint{0905.2391}.

\bibitem[{\citenamefont{Mino et~al.}(1997)\citenamefont{Mino, Sasaki, and
  Tanaka}}]{Mino:Sasaki:Tanaka:1996}
\bibinfo{author}{\bibfnamefont{Y.}~\bibnamefont{Mino}},
  \bibinfo{author}{\bibfnamefont{M.}~\bibnamefont{Sasaki}}, \bibnamefont{and}
  \bibinfo{author}{\bibfnamefont{T.}~\bibnamefont{Tanaka}},
  \bibinfo{journal}{Phys. Rev.} \textbf{\bibinfo{volume}{D55}},
  \bibinfo{pages}{3457} (\bibinfo{year}{1997}), \eprint{gr-qc/9606018}.

\bibitem[{\citenamefont{Poisson}(2004)}]{Poisson:2003}
\bibinfo{author}{\bibfnamefont{E.}~\bibnamefont{Poisson}},
  \bibinfo{journal}{Living Rev. Relativity} \textbf{\bibinfo{volume}{7}},
  \bibinfo{pages}{6} (\bibinfo{year}{2004}), \eprint{gr-qc/0306052}.

\bibitem[{\citenamefont{Gralla and Wald}(2008)}]{Gralla:Wald:2008}
\bibinfo{author}{\bibfnamefont{S.~E.} \bibnamefont{Gralla}} \bibnamefont{and}
  \bibinfo{author}{\bibfnamefont{R.~M.} \bibnamefont{Wald}},
  \bibinfo{journal}{Class. Quantum Grav.} \textbf{\bibinfo{volume}{25}},
  \bibinfo{pages}{205009} (\bibinfo{year}{2008}), \eprint{arXiv:0806.3293}.

\bibitem[{\citenamefont{Quinn}(2000)}]{Quinn:2000}
\bibinfo{author}{\bibfnamefont{T.~C.} \bibnamefont{Quinn}},
  \bibinfo{journal}{Phys. Rev.} \textbf{\bibinfo{volume}{D62}},
  \bibinfo{pages}{064029} (\bibinfo{year}{2000}), \eprint{gr-qc/0005030}.

\bibitem[{\citenamefont{Detweiler}(2005)}]{Detweiler:2005}
\bibinfo{author}{\bibfnamefont{S.}~\bibnamefont{Detweiler}},
  \bibinfo{journal}{Class. Quantum Grav.} \textbf{\bibinfo{volume}{22}},
  \bibinfo{pages}{S681} (\bibinfo{year}{2005}), \eprint{gr-qc/0501004}.

\bibitem[{\citenamefont{Galley et~al.}(2006)\citenamefont{Galley, Hu, and
  Lin}}]{Galley:Hu:Lin:2006}
\bibinfo{author}{\bibfnamefont{C.~R.} \bibnamefont{Galley}},
  \bibinfo{author}{\bibfnamefont{B.~L.} \bibnamefont{Hu}}, \bibnamefont{and}
  \bibinfo{author}{\bibfnamefont{S.-Y.} \bibnamefont{Lin}},
  \bibinfo{journal}{Phys. Rev. D} \textbf{\bibinfo{volume}{74}},
  \bibinfo{pages}{024017} (\bibinfo{year}{2006}).

\bibitem[{\citenamefont{Harte}(2008)}]{Harte:2008}
\bibinfo{author}{\bibfnamefont{A.~I.} \bibnamefont{Harte}},
  \bibinfo{journal}{Class. Quantum Grav.} \textbf{\bibinfo{volume}{25}},
  \bibinfo{pages}{235020} (\bibinfo{year}{2008}).

\bibitem[{\citenamefont{Futamase et~al.}(2008)\citenamefont{Futamase, Hogan,
  and Itoh}}]{Futamase:Hogan:Itoh:2008}
\bibinfo{author}{\bibfnamefont{T.}~\bibnamefont{Futamase}},
  \bibinfo{author}{\bibfnamefont{P.~A.} \bibnamefont{Hogan}}, \bibnamefont{and}
  \bibinfo{author}{\bibfnamefont{Y.}~\bibnamefont{Itoh}},
  \bibinfo{journal}{Phys. Rev. D} \textbf{\bibinfo{volume}{78}},
  \bibinfo{pages}{104014} (\bibinfo{year}{2008}).

\bibitem[{\citenamefont{Flanagan and Hinderer}(2008)}]{Flanagan:Hinderer:2008}
\bibinfo{author}{\bibfnamefont{E.~E.} \bibnamefont{Flanagan}} \bibnamefont{and}
  \bibinfo{author}{\bibfnamefont{T.}~\bibnamefont{Hinderer}},
  \bibinfo{journal}{Phys. Rev. D} \textbf{\bibinfo{volume}{78}},
  \bibinfo{pages}{064028} (\bibinfo{year}{2008}).

\bibitem[{\citenamefont{Norton}(2009)}]{Norton:2009}
\bibinfo{author}{\bibfnamefont{A.~H.} \bibnamefont{Norton}},
  \bibinfo{journal}{Class. Quantum Grav.} \textbf{\bibinfo{volume}{26}}
  (\bibinfo{year}{2009}).

\bibitem[{\citenamefont{Ottewill and Wardell}(2008)}]{Ottewill:Wardell:2008}
\bibinfo{author}{\bibfnamefont{A.~C.} \bibnamefont{Ottewill}} \bibnamefont{and}
  \bibinfo{author}{\bibfnamefont{B.}~\bibnamefont{Wardell}},
  \bibinfo{journal}{Phys. Rev.} \textbf{\bibinfo{volume}{D77}},
  \bibinfo{pages}{104002} (\bibinfo{year}{2008}), \eprint{arXiv:0711.2469}.

\bibitem[{\citenamefont{Ottewill and Wardell}(2009)}]{Ottewill:Wardell:2009}
\bibinfo{author}{\bibfnamefont{A.~C.} \bibnamefont{Ottewill}} \bibnamefont{and}
  \bibinfo{author}{\bibfnamefont{B.}~\bibnamefont{Wardell}},
  \bibinfo{journal}{Phys. Rev.} \textbf{\bibinfo{volume}{D79}},
  \bibinfo{pages}{024031} (\bibinfo{year}{2009}), \eprint{arXiv:0810.1961}.

\bibitem[{\citenamefont{Casals et~al.}(2009{\natexlab{a}})\citenamefont{Casals,
  Dolan, Ottewill, and Wardell}}]{Casals:Dolan:Ottewill:Wardell:2009}
\bibinfo{author}{\bibfnamefont{M.}~\bibnamefont{Casals}},
  \bibinfo{author}{\bibfnamefont{S.~R.} \bibnamefont{Dolan}},
  \bibinfo{author}{\bibfnamefont{A.~C.} \bibnamefont{Ottewill}},
  \bibnamefont{and} \bibinfo{author}{\bibfnamefont{B.}~\bibnamefont{Wardell}},
  \bibinfo{journal}{Phys. Rev. D} \textbf{\bibinfo{volume}{79}},
  \bibinfo{pages}{124043} (\bibinfo{year}{2009}{\natexlab{a}}),
  \eprint{arXiv:0903.0395}.

\bibitem[{\citenamefont{Casals et~al.}(2009{\natexlab{b}})\citenamefont{Casals,
  Dolan, Ottewill, and Wardell}}]{QL}
\bibinfo{author}{\bibfnamefont{M.}~\bibnamefont{Casals}},
  \bibinfo{author}{\bibfnamefont{S.~R.} \bibnamefont{Dolan}},
  \bibinfo{author}{\bibfnamefont{A.~C.} \bibnamefont{Ottewill}},
  \bibnamefont{and} \bibinfo{author}{\bibfnamefont{B.}~\bibnamefont{Wardell}},
  \bibinfo{journal}{Phys. Rev. D} \textbf{\bibinfo{volume}{79}},
  \bibinfo{pages}{124044} (\bibinfo{year}{2009}{\natexlab{b}}),
  \eprint{arXiv:0903.5319}.

\bibitem[{\citenamefont{Ottewill and Wardell}()}]{Transport}
\bibinfo{author}{\bibfnamefont{A.~C.} \bibnamefont{Ottewill}} \bibnamefont{and}
  \bibinfo{author}{\bibfnamefont{B.}~\bibnamefont{Wardell}},
  \emph{\bibinfo{title}{A transport equation approach to calculations of
  {Green} functions and {HaMiDeW} coefficients}}, \eprint{arXiv:0906.0005}.

\bibitem[{\citenamefont{Misner et~al.}(1973)\citenamefont{Misner, Thorne, and
  Wheeler}}]{Misner:Thorne:Wheeler:1974}
\bibinfo{author}{\bibfnamefont{C.~W.} \bibnamefont{Misner}},
  \bibinfo{author}{\bibfnamefont{K.~S.} \bibnamefont{Thorne}},
  \bibnamefont{and} \bibinfo{author}{\bibfnamefont{J.~A.}
  \bibnamefont{Wheeler}}, \emph{\bibinfo{title}{{Gravitation}}}
  (\bibinfo{publisher}{{Freeman}}, \bibinfo{address}{San Francisco},
  \bibinfo{year}{1973}).

\bibitem[{\citenamefont{Avramidi}(2000)}]{Avramidi:2000}
\bibinfo{author}{\bibfnamefont{I.~G.} \bibnamefont{Avramidi}},
  \emph{\bibinfo{title}{Heat Kernel and Quantum Gravity}}
  (\bibinfo{publisher}{Springer, Berlin}, \bibinfo{year}{2000}).

\bibitem[{\citenamefont{Hadamard}(1923)}]{Hadamard}
\bibinfo{author}{\bibfnamefont{J.}~\bibnamefont{Hadamard}},
  \emph{\bibinfo{title}{Lectures on Cauchy's Problem in Linear Partial
  Differential Equations}} (\bibinfo{publisher}{Dover Publications},
  \bibinfo{address}{New York}, \bibinfo{year}{1923}).

\bibitem[{\citenamefont{Friedlander}(1975)}]{Friedlander}
\bibinfo{author}{\bibfnamefont{F.~G.} \bibnamefont{Friedlander}},
  \emph{\bibinfo{title}{{The Wave Equation on a Curved Space-time}}}
  (\bibinfo{publisher}{Cambridge University Press},
  \bibinfo{address}{Cambridge, England}, \bibinfo{year}{1975}).

\bibitem[{\citenamefont{D\'ecanini and Folacci}(2006)}]{Decanini:Folacci:2005a}
\bibinfo{author}{\bibfnamefont{Y.}~\bibnamefont{D\'ecanini}} \bibnamefont{and}
  \bibinfo{author}{\bibfnamefont{A.}~\bibnamefont{Folacci}},
  \bibinfo{journal}{Phys. Rev.} \textbf{\bibinfo{volume}{D73}},
  \bibinfo{pages}{044027} (\bibinfo{year}{2006}), \eprint{gr-qc/0511115}.

\bibitem[{\citenamefont{DeWitt}(1965)}]{DeWitt:1965}
\bibinfo{author}{\bibfnamefont{B.~S.} \bibnamefont{DeWitt}},
  \emph{\bibinfo{title}{{Dynamical theory of groups and fields}}}
  (\bibinfo{publisher}{{Gordon and Breach}}, \bibinfo{address}{New York},
  \bibinfo{year}{1965}).

\bibitem[{\citenamefont{Gibbons}(1979)}]{Gibbons}
\bibinfo{author}{\bibfnamefont{G.}~\bibnamefont{Gibbons}},
  \emph{\bibinfo{title}{Quantum Field Theory in Curved Spacetime}}
  (\bibinfo{publisher}{Cambridge University Press},
  \bibinfo{address}{Cambridge}, \bibinfo{year}{1979}),
  chap.~\bibinfo{chapter}{13}, pp. \bibinfo{pages}{639--679}.

\bibitem[{\citenamefont{Christensen}(1976)}]{Christensen:1976vb}
\bibinfo{author}{\bibfnamefont{S.~M.} \bibnamefont{Christensen}},
  \bibinfo{journal}{Phys. Rev.} \textbf{\bibinfo{volume}{D14}},
  \bibinfo{pages}{2490} (\bibinfo{year}{1976}).

\bibitem[{\citenamefont{Christensen}(1978)}]{Christensen:1978yd}
\bibinfo{author}{\bibfnamefont{S.~M.} \bibnamefont{Christensen}},
  \bibinfo{journal}{Phys. Rev.} \textbf{\bibinfo{volume}{D17}},
  \bibinfo{pages}{946} (\bibinfo{year}{1978}).

\bibitem[{\citenamefont{Christensen}(1995)}]{Christensen:1995}
\bibinfo{author}{\bibfnamefont{S.~M.} \bibnamefont{Christensen}},
  \emph{\bibinfo{title}{Computational Challenges in Heat Kernel Calculations}}
  (\bibinfo{publisher}{Texas A \& M University}, \bibinfo{year}{1995}), pp.
  \bibinfo{pages}{47--64}, Discourses in Mathematics and Its Applications.

\bibitem[{\citenamefont{Chandrasekhar}(1992)}]{Chandrasekhar}
\bibinfo{author}{\bibfnamefont{S.}~\bibnamefont{Chandrasekhar}},
  \emph{\bibinfo{title}{The Mathematical Theory of Black Holes}}
  (\bibinfo{publisher}{Oxford University Press}, \bibinfo{year}{1992}).

\bibitem[{\citenamefont{Hartle}(2003)}]{Hartle}
\bibinfo{author}{\bibfnamefont{J.~B.} \bibnamefont{Hartle}},
  \emph{\bibinfo{title}{Gravity}} (\bibinfo{publisher}{Addison Wesley},
  \bibinfo{year}{2003}).

\bibitem[{\citenamefont{Perlick}(2004)}]{Perlick}
\bibinfo{author}{\bibfnamefont{V.}~\bibnamefont{Perlick}},
  \bibinfo{journal}{Living Rev. Relativity} \textbf{\bibinfo{volume}{7}},
  \bibinfo{pages}{9} (\bibinfo{year}{2004}).

\bibitem[{\citenamefont{Carter}(1968)}]{Carter:1968}
\bibinfo{author}{\bibfnamefont{B.}~\bibnamefont{Carter}},
  \bibinfo{journal}{Phys. Rev.} \textbf{\bibinfo{volume}{174}},
  \bibinfo{pages}{1559} (\bibinfo{year}{1968}).

\bibitem[{\citenamefont{Boyer and Lindquist}(1967)}]{Boyer-Lindquist}
\bibinfo{author}{\bibfnamefont{C.~B.} \bibnamefont{Boyer}} \bibnamefont{and}
  \bibinfo{author}{\bibfnamefont{R.~W.} \bibnamefont{Lindquist}},
  \bibinfo{journal}{J. Math. Phys.} \textbf{\bibinfo{volume}{8}},
  \bibinfo{pages}{265} (\bibinfo{year}{1967}).

\bibitem[{\citenamefont{P\"oschl and Teller}(1933)}]{Poschl:Teller:1933}
\bibinfo{author}{\bibfnamefont{G.}~\bibnamefont{P\"oschl}} \bibnamefont{and}
  \bibinfo{author}{\bibfnamefont{E.}~\bibnamefont{Teller}},
  \bibinfo{journal}{Zeitschr. Phys.} \textbf{\bibinfo{volume}{83}},
  \bibinfo{pages}{143} (\bibinfo{year}{1933}).

\bibitem[{\citenamefont{Ferrari and Mashhoon}(1984)}]{Ferrari:Mashhoon:1984}
\bibinfo{author}{\bibfnamefont{V.}~\bibnamefont{Ferrari}} \bibnamefont{and}
  \bibinfo{author}{\bibfnamefont{B.}~\bibnamefont{Mashhoon}},
  \bibinfo{journal}{Phys. Rev. D} \textbf{\bibinfo{volume}{30}},
  \bibinfo{pages}{295} (\bibinfo{year}{1984}).

\bibitem[{\citenamefont{Berti and Cardoso}(2006)}]{Berti:Cardoso:2006}
\bibinfo{author}{\bibfnamefont{E.}~\bibnamefont{Berti}} \bibnamefont{and}
  \bibinfo{author}{\bibfnamefont{V.}~\bibnamefont{Cardoso}},
  \bibinfo{journal}{Phys. Rev. D} \textbf{\bibinfo{volume}{74}},
  \bibinfo{pages}{104020} (\bibinfo{year}{2006}).

\bibitem[{\citenamefont{Cardoso and Lemos}(2003)}]{Cardoso:Lemos:2003}
\bibinfo{author}{\bibfnamefont{V.}~\bibnamefont{Cardoso}} \bibnamefont{and}
  \bibinfo{author}{\bibfnamefont{J.~P.~S.} \bibnamefont{Lemos}},
  \bibinfo{journal}{Phys. Rev. D} \textbf{\bibinfo{volume}{67}},
  \bibinfo{pages}{084020} (\bibinfo{year}{2003}), \eprint{gr-qc/0301078}.

\bibitem[{\citenamefont{Zerbini and Vanzo}(2004)}]{Zerbini:Vanzo:2004}
\bibinfo{author}{\bibfnamefont{S.}~\bibnamefont{Zerbini}} \bibnamefont{and}
  \bibinfo{author}{\bibfnamefont{L.}~\bibnamefont{Vanzo}},
  \bibinfo{journal}{Phys. Rev. D} \textbf{\bibinfo{volume}{70}},
  \bibinfo{pages}{044030} (\bibinfo{year}{2004}), \eprint{hep-th/0402103}.

\bibitem[{\citenamefont{Nariai}(1950)}]{Nariai:1950}
\bibinfo{author}{\bibfnamefont{H.}~\bibnamefont{Nariai}},
  \bibinfo{journal}{Sci. Rep. Tohoku Univ.} \textbf{\bibinfo{volume}{34}},
  \bibinfo{pages}{160} (\bibinfo{year}{1950}).

\bibitem[{\citenamefont{Nariai}(1951)}]{Nariai:1951}
\bibinfo{author}{\bibfnamefont{H.}~\bibnamefont{Nariai}},
  \bibinfo{journal}{Sci. Rep. Tohoku Univ.} \textbf{\bibinfo{volume}{35}},
  \bibinfo{pages}{62} (\bibinfo{year}{1951}).

\bibitem[{\citenamefont{Ortaggio}(2002)}]{Ortaggio:2002}
\bibinfo{author}{\bibfnamefont{M.}~\bibnamefont{Ortaggio}},
  \bibinfo{journal}{Phys. Rev.} \textbf{\bibinfo{volume}{D65}},
  \bibinfo{pages}{084046} (\bibinfo{year}{2002}).

\bibitem[{\citenamefont{Hawking and Ellis}(1973)}]{Hawking:Ellis}
\bibinfo{author}{\bibfnamefont{S.~W.} \bibnamefont{Hawking}} \bibnamefont{and}
  \bibinfo{author}{\bibfnamefont{G.~F.~R.} \bibnamefont{Ellis}},
  \emph{\bibinfo{title}{The Large-Scale Structure of Spacetime}}
  (\bibinfo{publisher}{Cambridge University Press}, \bibinfo{year}{1973}).

\bibitem[{\citenamefont{Gibbons and Hawking}(1977)}]{Gibbons:Hawking:1977}
\bibinfo{author}{\bibfnamefont{G.}~\bibnamefont{Gibbons}} \bibnamefont{and}
  \bibinfo{author}{\bibfnamefont{S.}~\bibnamefont{Hawking}},
  \bibinfo{journal}{Phys. Rev.} \textbf{\bibinfo{volume}{D15}},
  \bibinfo{pages}{2738} (\bibinfo{year}{1977}).

\bibitem[{\citenamefont{Ginsparg and Perry}(1983)}]{Ginsparg:Perry:1983}
\bibinfo{author}{\bibfnamefont{P.}~\bibnamefont{Ginsparg}} \bibnamefont{and}
  \bibinfo{author}{\bibfnamefont{M.~J.} \bibnamefont{Perry}},
  \bibinfo{journal}{Nucl. Phys. B} \textbf{\bibinfo{volume}{222}},
  \bibinfo{pages}{245} (\bibinfo{year}{1983}).

\bibitem[{\citenamefont{Dias and Lemos}(2003)}]{Dias:Lemos:2003}
\bibinfo{author}{\bibfnamefont{O.~J.~C.} \bibnamefont{Dias}} \bibnamefont{and}
  \bibinfo{author}{\bibfnamefont{J.~P.~S.} \bibnamefont{Lemos}},
  \bibinfo{journal}{Phys. Rev. D} \textbf{\bibinfo{volume}{68}},
  \bibinfo{pages}{104010} (\bibinfo{year}{2003}).

\bibitem[{\citenamefont{Cardoso et~al.}(2004)\citenamefont{Cardoso, Dias, and
  Lemos}}]{Cardoso:Dias:Lemos:2004}
\bibinfo{author}{\bibfnamefont{V.}~\bibnamefont{Cardoso}},
  \bibinfo{author}{\bibfnamefont{O.~J.~C.} \bibnamefont{Dias}},
  \bibnamefont{and} \bibinfo{author}{\bibfnamefont{J.~P.~S.}
  \bibnamefont{Lemos}}, \bibinfo{journal}{Phys. Rev. D}
  \textbf{\bibinfo{volume}{70}}, \bibinfo{pages}{024002}
  (\bibinfo{year}{2004}), \eprint{hep-th/0401192}.

\bibitem[{\citenamefont{Bousso and Hawking}(1995)}]{Bousso:Hawking:1995}
\bibinfo{author}{\bibfnamefont{R.}~\bibnamefont{Bousso}} \bibnamefont{and}
  \bibinfo{author}{\bibfnamefont{S.~W.} \bibnamefont{Hawking}},
  \bibinfo{journal}{Phys. Rev. D} \textbf{\bibinfo{volume}{52}},
  \bibinfo{pages}{5659} (\bibinfo{year}{1995}).

\bibitem[{\citenamefont{Bousso and Hawking}(1996)}]{Bousso:Hawking:1996}
\bibinfo{author}{\bibfnamefont{R.}~\bibnamefont{Bousso}} \bibnamefont{and}
  \bibinfo{author}{\bibfnamefont{S.~W.} \bibnamefont{Hawking}},
  \bibinfo{journal}{Phys. Rev. D} \textbf{\bibinfo{volume}{54}},
  \bibinfo{pages}{6312} (\bibinfo{year}{1996}).

\bibitem[{\citenamefont{Casals et~al.}()\citenamefont{Casals, Dolan, Nolan,
  Ottewill, and Wardell}}]{Nolan:2009}
\bibinfo{author}{\bibfnamefont{M.}~\bibnamefont{Casals}},
  \bibinfo{author}{\bibfnamefont{S.~R.} \bibnamefont{Dolan}},
  \bibinfo{author}{\bibfnamefont{B.}~\bibnamefont{Nolan}},
  \bibinfo{author}{\bibfnamefont{A.~C.} \bibnamefont{Ottewill}},
  \bibnamefont{and} \bibinfo{author}{\bibfnamefont{B.}~\bibnamefont{Wardell}},
  \bibinfo{note}{in preparation}.

\bibitem[{\citenamefont{Allen and Jacobson}(1986)}]{Allen:Jac:86}
\bibinfo{author}{\bibfnamefont{B.}~\bibnamefont{Allen}} \bibnamefont{and}
  \bibinfo{author}{\bibfnamefont{T.}~\bibnamefont{Jacobson}},
  \bibinfo{journal}{Commun.Math.Phys.} \textbf{\bibinfo{volume}{103}},
  \bibinfo{pages}{669} (\bibinfo{year}{1986}).

\bibitem[{\citenamefont{Barack and Ori}(2000)}]{Barack:Ori:2000}
\bibinfo{author}{\bibfnamefont{L.}~\bibnamefont{Barack}} \bibnamefont{and}
  \bibinfo{author}{\bibfnamefont{A.}~\bibnamefont{Ori}},
  \bibinfo{journal}{Phys. Rev. D} \textbf{\bibinfo{volume}{61}},
  \bibinfo{pages}{061502} (\bibinfo{year}{2000}).

\bibitem[{\citenamefont{Barack}(2001)}]{Barack:2001}
\bibinfo{author}{\bibfnamefont{L.}~\bibnamefont{Barack}},
  \bibinfo{journal}{Phys. Rev. D} \textbf{\bibinfo{volume}{64}},
  \bibinfo{pages}{084021} (\bibinfo{year}{2001}).

\bibitem[{\citenamefont{Barack et~al.}(2002)\citenamefont{Barack, Mino, Nakano,
  Ori, and Sasaki}}]{Barack:Mino:Nakano:2002}
\bibinfo{author}{\bibfnamefont{L.}~\bibnamefont{Barack}},
  \bibinfo{author}{\bibfnamefont{Y.}~\bibnamefont{Mino}},
  \bibinfo{author}{\bibfnamefont{H.}~\bibnamefont{Nakano}},
  \bibinfo{author}{\bibfnamefont{A.}~\bibnamefont{Ori}}, \bibnamefont{and}
  \bibinfo{author}{\bibfnamefont{M.}~\bibnamefont{Sasaki}},
  \bibinfo{journal}{Phys. Rev. Lett.} \textbf{\bibinfo{volume}{88}},
  \bibinfo{pages}{091101} (\bibinfo{year}{2002}).

\bibitem[{\citenamefont{Barack and Sago}(2007)}]{Barack:Sago:2007}
\bibinfo{author}{\bibfnamefont{L.}~\bibnamefont{Barack}} \bibnamefont{and}
  \bibinfo{author}{\bibfnamefont{N.}~\bibnamefont{Sago}},
  \bibinfo{journal}{Phys. Rev.} \textbf{\bibinfo{volume}{D75}},
  \bibinfo{pages}{064021} (\bibinfo{year}{2007}), \eprint{gr-qc/0701069}.

\bibitem[{\citenamefont{Detweiler et~al.}(2003)\citenamefont{Detweiler,
  Messaritaki, and Whiting}}]{Detweiler:Messaritaki:Whiting:2002}
\bibinfo{author}{\bibfnamefont{S.}~\bibnamefont{Detweiler}},
  \bibinfo{author}{\bibfnamefont{E.}~\bibnamefont{Messaritaki}},
  \bibnamefont{and} \bibinfo{author}{\bibfnamefont{B.~F.}
  \bibnamefont{Whiting}}, \bibinfo{journal}{Phys. Rev.}
  \textbf{\bibinfo{volume}{D67}}, \bibinfo{pages}{104016}
  (\bibinfo{year}{2003}), \eprint{gr-qc/0205079}.

\bibitem[{\citenamefont{Vega and Detweiler}(2008)}]{Vega:Detweiler:2008}
\bibinfo{author}{\bibfnamefont{I.}~\bibnamefont{Vega}} \bibnamefont{and}
  \bibinfo{author}{\bibfnamefont{S.}~\bibnamefont{Detweiler}},
  \bibinfo{journal}{Phys. Rev. D} \textbf{\bibinfo{volume}{77}},
  \bibinfo{pages}{084008} (\bibinfo{year}{2008}).

\bibitem[{\citenamefont{Haas}(2007)}]{Haas:2007}
\bibinfo{author}{\bibfnamefont{R.}~\bibnamefont{Haas}}, \bibinfo{journal}{Phys.
  Rev.} \textbf{\bibinfo{volume}{D75}}, \bibinfo{pages}{124011}
  (\bibinfo{year}{2007}), \eprint{arXiv:0704.0797}.

\bibitem[{\citenamefont{Barack et~al.}(2007)\citenamefont{Barack, Golbourn, and
  Sago}}]{Barack:Golbourn:Sago:2007}
\bibinfo{author}{\bibfnamefont{L.}~\bibnamefont{Barack}},
  \bibinfo{author}{\bibfnamefont{D.~A.} \bibnamefont{Golbourn}},
  \bibnamefont{and} \bibinfo{author}{\bibfnamefont{N.}~\bibnamefont{Sago}},
  \bibinfo{journal}{Phys. Rev.} \textbf{\bibinfo{volume}{D76}},
  \bibinfo{pages}{124036} (\bibinfo{year}{2007}), \eprint{arXiv:0709.4588}.

\bibitem[{\citenamefont{Barack et~al.}(2008)\citenamefont{Barack, Ori, and
  Sago}}]{Barack:Ori:Sago:2008}
\bibinfo{author}{\bibfnamefont{L.}~\bibnamefont{Barack}},
  \bibinfo{author}{\bibfnamefont{A.}~\bibnamefont{Ori}}, \bibnamefont{and}
  \bibinfo{author}{\bibfnamefont{N.}~\bibnamefont{Sago}},
  \bibinfo{journal}{Phys. Rev. D} \textbf{\bibinfo{volume}{78}},
  \bibinfo{pages}{084021} (\bibinfo{year}{2008}), \eprint{arXiv:0808.2315}.

\bibitem[{\citenamefont{Sago et~al.}(2008)\citenamefont{Sago, Barack, and
  Detweiler}}]{Sago:Barack:Detweiler:2008}
\bibinfo{author}{\bibfnamefont{N.}~\bibnamefont{Sago}},
  \bibinfo{author}{\bibfnamefont{L.}~\bibnamefont{Barack}}, \bibnamefont{and}
  \bibinfo{author}{\bibfnamefont{S.}~\bibnamefont{Detweiler}},
  \bibinfo{journal}{Phys. Rev. D} \textbf{\bibinfo{volume}{78}},
  \bibinfo{pages}{124024} (\bibinfo{year}{2008}), \eprint{arXiv:0810.2530}.

\bibitem[{\citenamefont{Blanchet et~al.}(2009)\citenamefont{Blanchet,
  Grishchuk, and Schaefer}}]{Blanchet:Grishchuk:Schaefer:2009}
\bibinfo{author}{\bibfnamefont{L.}~\bibnamefont{Blanchet}},
  \bibinfo{author}{\bibfnamefont{L.~P.} \bibnamefont{Grishchuk}},
  \bibnamefont{and} \bibinfo{author}{\bibfnamefont{G.}~\bibnamefont{Schaefer}},
  in \emph{\bibinfo{booktitle}{Proceedings of the 11th Marcel Grossman Meeting
  on General Relativity}}, edited by
  \bibinfo{editor}{\bibfnamefont{H.}~\bibnamefont{Kleinert}} \bibnamefont{and}
  \bibinfo{editor}{\bibfnamefont{R.~T.} \bibnamefont{Jantzen}}
  (\bibinfo{publisher}{World Scientific}, \bibinfo{year}{2009}), p.
  \bibinfo{pages}{2475}.

\bibitem[{\citenamefont{Damour and Nagar}(2009)}]{Damour:Nagar:2009}
\bibinfo{author}{\bibfnamefont{T.}~\bibnamefont{Damour}} \bibnamefont{and}
  \bibinfo{author}{\bibfnamefont{A.}~\bibnamefont{Nagar}},
  \bibinfo{journal}{Phys. Rev. D} \textbf{\bibinfo{volume}{79}},
  \bibinfo{pages}{081503} (\bibinfo{year}{2009}), \eprint{arXiv:0902.0136}.

\bibitem[{\citenamefont{Detweiler}(2008)}]{Detweiler:2008}
\bibinfo{author}{\bibfnamefont{S.}~\bibnamefont{Detweiler}},
  \bibinfo{journal}{Phys. Rev. D} \textbf{\bibinfo{volume}{77}},
  \bibinfo{pages}{124026} (\bibinfo{year}{2008}), \eprint{arXiv:0804:3529}.

\bibitem[{\citenamefont{Wiseman}(2000)}]{Wiseman:2000}
\bibinfo{author}{\bibfnamefont{A.~G.} \bibnamefont{Wiseman}},
  \bibinfo{journal}{Phys. Rev.} \textbf{\bibinfo{volume}{D61}},
  \bibinfo{pages}{084014} (\bibinfo{year}{2000}), \eprint{gr-qc/0001025}.

\bibitem[{\citenamefont{Anderson et~al.}(2006)\citenamefont{Anderson,
  Eftekharzadeh, and Hu}}]{Anderson:Eftekharzadeh:Hu:2006}
\bibinfo{author}{\bibfnamefont{P.~R.} \bibnamefont{Anderson}},
  \bibinfo{author}{\bibfnamefont{A.}~\bibnamefont{Eftekharzadeh}},
  \bibnamefont{and} \bibinfo{author}{\bibfnamefont{B.~L.} \bibnamefont{Hu}},
  \bibinfo{journal}{Phys. Rev.} \textbf{\bibinfo{volume}{D73}},
  \bibinfo{pages}{064023} (\bibinfo{year}{2006}), \eprint{gr-qc/0507067}.

\bibitem[{\citenamefont{Anderson and Wiseman}(2005)}]{Anderson:Wiseman:2005}
\bibinfo{author}{\bibfnamefont{W.~G.} \bibnamefont{Anderson}} \bibnamefont{and}
  \bibinfo{author}{\bibfnamefont{A.~G.} \bibnamefont{Wiseman}},
  \bibinfo{journal}{Class. Quantum Grav.} \textbf{\bibinfo{volume}{22}},
  \bibinfo{pages}{S783} (\bibinfo{year}{2005}), \eprint{gr-qc/0506136}.

\bibitem[{\citenamefont{Anderson et~al.}(2005)\citenamefont{Anderson, Flanagan,
  and Ottewill}}]{Anderson:Flanagan:Ottewill:2004}
\bibinfo{author}{\bibfnamefont{W.~G.} \bibnamefont{Anderson}},
  \bibinfo{author}{\bibfnamefont{E.~E.} \bibnamefont{Flanagan}},
  \bibnamefont{and} \bibinfo{author}{\bibfnamefont{A.~C.}
  \bibnamefont{Ottewill}}, \bibinfo{journal}{Phys. Rev.}
  \textbf{\bibinfo{volume}{D71}}, \bibinfo{pages}{024036}
  (\bibinfo{year}{2005}), \eprint{gr-qc/0412009}.

\bibitem[{\citenamefont{Poisson and Wiseman}(1998)}]{Poisson:Wiseman:1998}
\bibinfo{author}{\bibfnamefont{E.}~\bibnamefont{Poisson}} \bibnamefont{and}
  \bibinfo{author}{\bibfnamefont{A.~G.} \bibnamefont{Wiseman}},
  \emph{\bibinfo{title}{Suggestion at the 1st capra ranch meeting on radiation
  reaction}} (\bibinfo{year}{1998}).

\bibitem[{\citenamefont{Leaver}(1986)}]{Leaver:1986}
\bibinfo{author}{\bibfnamefont{E.~W.} \bibnamefont{Leaver}},
  \bibinfo{journal}{Phys. Rev. D} \textbf{\bibinfo{volume}{34}},
  \bibinfo{pages}{384} (\bibinfo{year}{1986}).

\bibitem[{\citenamefont{Andersson}(1997)}]{Andersson:1997}
\bibinfo{author}{\bibfnamefont{N.}~\bibnamefont{Andersson}},
  \bibinfo{journal}{Phys. Rev. D} \textbf{\bibinfo{volume}{55}},
  \bibinfo{pages}{468} (\bibinfo{year}{1997}).

\bibitem[{\citenamefont{Smith and Will}(1980)}]{SW:1980}
\bibinfo{author}{\bibfnamefont{A.}~\bibnamefont{Smith}} \bibnamefont{and}
  \bibinfo{author}{\bibfnamefont{C.}~\bibnamefont{Will}},
  \bibinfo{journal}{Phys. Rev. D} \textbf{\bibinfo{volume}{22}},
  \bibinfo{pages}{1276} (\bibinfo{year}{1980}).

\bibitem[{\citenamefont{Rosenthal}(2004{\natexlab{a}})}]{Rosenthal:2004}
\bibinfo{author}{\bibfnamefont{E.}~\bibnamefont{Rosenthal}},
  \bibinfo{journal}{Phys. Rev.} \textbf{\bibinfo{volume}{D70}},
  \bibinfo{pages}{124016} (\bibinfo{year}{2004}{\natexlab{a}}),
  \eprint{gr-qc/0410022}.

\bibitem[{\citenamefont{Cho et~al.}(2007)\citenamefont{Cho, Tsokaros, and
  Wiseman}}]{CTW:2007}
\bibinfo{author}{\bibfnamefont{D.}~\bibnamefont{Cho}},
  \bibinfo{author}{\bibfnamefont{A.}~\bibnamefont{Tsokaros}}, \bibnamefont{and}
  \bibinfo{author}{\bibfnamefont{A.~G.} \bibnamefont{Wiseman}},
  \bibinfo{journal}{Class. Quantum Grav.} \textbf{\bibinfo{volume}{24}},
  \bibinfo{pages}{1035} (\bibinfo{year}{2007}).

\bibitem[{\citenamefont{Copson}(1928)}]{Copson:1928}
\bibinfo{author}{\bibfnamefont{E.}~\bibnamefont{Copson}},
  \bibinfo{journal}{Proc. R. Soc. London, A} \textbf{\bibinfo{volume}{118}},
  \bibinfo{pages}{184} (\bibinfo{year}{1928}).

\bibitem[{\citenamefont{Linet}(1976)}]{Linet:1976}
\bibinfo{author}{\bibfnamefont{B.}~\bibnamefont{Linet}}, \bibinfo{journal}{J.
  Phys. A: Math. Gen.} \textbf{\bibinfo{volume}{9}}, \bibinfo{pages}{1081}
  (\bibinfo{year}{1976}).

\bibitem[{\citenamefont{Zel'nikov and
  Frolov}(1982{\natexlab{a}})}]{Zelnikov:Frolov:1982a}
\bibinfo{author}{\bibfnamefont{A.}~\bibnamefont{Zel'nikov}} \bibnamefont{and}
  \bibinfo{author}{\bibfnamefont{V.}~\bibnamefont{Frolov}},
  \bibinfo{journal}{Zh. Eksp. Teor. Fiz.} \textbf{\bibinfo{volume}{82}},
  \bibinfo{pages}{321} (\bibinfo{year}{1982}{\natexlab{a}}).

\bibitem[{\citenamefont{Zel'nikov and
  Frolov}(1982{\natexlab{b}})}]{Zelnikov:Frolov:1982b}
\bibinfo{author}{\bibfnamefont{A.}~\bibnamefont{Zel'nikov}} \bibnamefont{and}
  \bibinfo{author}{\bibfnamefont{V.}~\bibnamefont{Frolov}},
  \bibinfo{journal}{Sov. Phys.-JETP} \textbf{\bibinfo{volume}{55}},
  \bibinfo{pages}{191} (\bibinfo{year}{1982}{\natexlab{b}}).

\bibitem[{\citenamefont{Rosenthal}(2004{\natexlab{b}})}]{Rosenthal:2003}
\bibinfo{author}{\bibfnamefont{E.}~\bibnamefont{Rosenthal}},
  \bibinfo{journal}{Phys. Rev.} \textbf{\bibinfo{volume}{D69}},
  \bibinfo{pages}{064035} (\bibinfo{year}{2004}{\natexlab{b}}),
  \eprint{gr-qc/0309103}.

\bibitem[{\citenamefont{Hobbs}(1968{\natexlab{b}})}]{Hobbs:1968b}
\bibinfo{author}{\bibfnamefont{J.}~\bibnamefont{Hobbs}}, \bibinfo{journal}{Ann.
  Phys.} \textbf{\bibinfo{volume}{47}}, \bibinfo{pages}{166}
  (\bibinfo{year}{1968}{\natexlab{b}}).

\bibitem[{\citenamefont{Bezerra and Khusnutdinov}(2009)}]{Bezerra:Khus}
\bibinfo{author}{\bibfnamefont{V.}~\bibnamefont{Bezerra}} \bibnamefont{and}
  \bibinfo{author}{\bibfnamefont{N.}~\bibnamefont{Khusnutdinov}},
  \bibinfo{journal}{Phys. Rev. D} \textbf{\bibinfo{volume}{79}},
  \bibinfo{pages}{064012} (\bibinfo{year}{2009}), \eprint{arXiv:0901.0480}.

\bibitem[{\citenamefont{Synge}(1960)}]{Synge}
\bibinfo{author}{\bibfnamefont{J.~L.} \bibnamefont{Synge}},
  \emph{\bibinfo{title}{Relativity: The General Theory}}
  (\bibinfo{publisher}{North-Holland}, \bibinfo{address}{Amsterdam},
  \bibinfo{year}{1960}), ISBN \bibinfo{isbn}{978-0720400663}.

\bibitem[{\citenamefont{Damour et~al.}(1998)\citenamefont{Damour, Iyer, and
  Sathyaprakash}}]{Damour:Iyer:Sathyaprakash}
\bibinfo{author}{\bibfnamefont{T.}~\bibnamefont{Damour}},
  \bibinfo{author}{\bibfnamefont{B.~R.} \bibnamefont{Iyer}}, \bibnamefont{and}
  \bibinfo{author}{\bibfnamefont{B.~S.} \bibnamefont{Sathyaprakash}},
  \bibinfo{journal}{Phys. Rev. D} \textbf{\bibinfo{volume}{57}},
  \bibinfo{pages}{885} (\bibinfo{year}{1998}).

\bibitem[{\citenamefont{Porter and Sathyaprakash}(2005)}]{Porter:Sathyaprakash}
\bibinfo{author}{\bibfnamefont{E.~K.} \bibnamefont{Porter}} \bibnamefont{and}
  \bibinfo{author}{\bibfnamefont{B.~S.} \bibnamefont{Sathyaprakash}},
  \bibinfo{journal}{Phys. Rev. D} \textbf{\bibinfo{volume}{71}},
  \bibinfo{pages}{024017} (\bibinfo{year}{2005}), \eprint{gr-qc/0406038}.

\bibitem[{\citenamefont{Anderson and Hu}(2004)}]{Anderson:2003}
\bibinfo{author}{\bibfnamefont{P.~R.} \bibnamefont{Anderson}} \bibnamefont{and}
  \bibinfo{author}{\bibfnamefont{B.~L.} \bibnamefont{Hu}},
  \bibinfo{journal}{Phys. Rev.} \textbf{\bibinfo{volume}{D69}},
  \bibinfo{pages}{064039} (\bibinfo{year}{2004}), \eprint{gr-qc/0308034}.

\bibitem[{\citenamefont{Howard}(1984)}]{Howard:1985}
\bibinfo{author}{\bibfnamefont{K.~W.} \bibnamefont{Howard}},
  \bibinfo{journal}{Phys. Rev.} \textbf{\bibinfo{volume}{D30}},
  \bibinfo{pages}{2532} (\bibinfo{year}{1984}).

\bibitem[{\citenamefont{Winstanley and Young}(2008)}]{Winstanley:2007}
\bibinfo{author}{\bibfnamefont{E.}~\bibnamefont{Winstanley}} \bibnamefont{and}
  \bibinfo{author}{\bibfnamefont{P.~M.} \bibnamefont{Young}},
  \bibinfo{journal}{Phys. Rev.} \textbf{\bibinfo{volume}{D77}},
  \bibinfo{pages}{024008} (\bibinfo{year}{2008}), \eprint{arXiv:0708.3820}.

\bibitem[{Had()}]{Hadamard-WKB-Code}
\bibinfo{howpublished}{\url{http://www.barrywardell.net/research/code/hadamard-wkb}}.

\bibitem[{\citenamefont{Watson}(1918)}]{Watson:1918}
\bibinfo{author}{\bibfnamefont{G.}~\bibnamefont{Watson}},
  \bibinfo{journal}{Proc. Roy. Soc. Lond. A} \textbf{\bibinfo{volume}{95}},
  \bibinfo{pages}{83} (\bibinfo{year}{1918}).

\bibitem[{\citenamefont{Ottewill et~al.}(2009)\citenamefont{Ottewill,
  Winstanley, and Young}}]{Ottewill:Winstanley:Young:2009}
\bibinfo{author}{\bibfnamefont{A.~C.} \bibnamefont{Ottewill}},
  \bibinfo{author}{\bibfnamefont{E.}~\bibnamefont{Winstanley}},
  \bibnamefont{and} \bibinfo{author}{\bibfnamefont{P.~M.} \bibnamefont{Young}}
  (\bibinfo{year}{2009}), \bibinfo{note}{in preparation}.

\bibitem[{\citenamefont{Gradshteyn and Ryzhik}(2007)}]{GradRyz}
\bibinfo{author}{\bibfnamefont{I.}~\bibnamefont{Gradshteyn}} \bibnamefont{and}
  \bibinfo{author}{\bibfnamefont{I.}~\bibnamefont{Ryzhik}},
  \emph{\bibinfo{title}{Table of Integrals, Series, and Products}}
  (\bibinfo{publisher}{Academic Press}, \bibinfo{address}{New York},
  \bibinfo{year}{2007}).

\bibitem[{\citenamefont{Anderson and Hu}(2007)}]{Anderson:2003:err1}
\bibinfo{author}{\bibfnamefont{P.~R.} \bibnamefont{Anderson}} \bibnamefont{and}
  \bibinfo{author}{\bibfnamefont{B.~L.} \bibnamefont{Hu}},
  \bibinfo{journal}{Phys. Rev.} \textbf{\bibinfo{volume}{D75}},
  \bibinfo{pages}{129901(E)} (\bibinfo{year}{2007}), \eprint{gr-qc/0308034}.

\bibitem[{\citenamefont{Anderson and Hu}(2008)}]{Anderson:2003:err2}
\bibinfo{author}{\bibfnamefont{P.~R.} \bibnamefont{Anderson}} \bibnamefont{and}
  \bibinfo{author}{\bibfnamefont{B.~L.} \bibnamefont{Hu}},
  \bibinfo{journal}{Phys. Rev.} \textbf{\bibinfo{volume}{D77}},
  \bibinfo{pages}{089901(E)} (\bibinfo{year}{2008}), \eprint{gr-qc/0308034}.

\bibitem[{\citenamefont{Morse and Feshbach}(1953)}]{M&F}
\bibinfo{author}{\bibfnamefont{P.}~\bibnamefont{Morse}} \bibnamefont{and}
  \bibinfo{author}{\bibfnamefont{H.}~\bibnamefont{Feshbach}},
  \emph{\bibinfo{title}{Methods of Theoretical Physics}}
  (\bibinfo{publisher}{McGraw-Hill}, \bibinfo{address}{New York},
  \bibinfo{year}{1953}).

\bibitem[{\citenamefont{Whittaker and Watson}(1927)}]{Whittaker:Watson}
\bibinfo{author}{\bibfnamefont{E.~T.} \bibnamefont{Whittaker}}
  \bibnamefont{and} \bibinfo{author}{\bibfnamefont{G.}~\bibnamefont{Watson}},
  \emph{\bibinfo{title}{A Course of Modern Analysis}}
  (\bibinfo{publisher}{Cambridge University Press}, \bibinfo{year}{1927}),
  \bibinfo{edition}{4th} ed.

\bibitem[{\citenamefont{Bender and Orszag}(1999)}]{Bender:Orszag}
\bibinfo{author}{\bibfnamefont{C.~M.} \bibnamefont{Bender}} \bibnamefont{and}
  \bibinfo{author}{\bibfnamefont{S.~A.} \bibnamefont{Orszag}},
  \emph{\bibinfo{title}{Advanced Mathematical Methods for Scientists and
  Engineers}} (\bibinfo{publisher}{Springer}, \bibinfo{address}{New York},
  \bibinfo{year}{1999}).

\bibitem[{\citenamefont{Press et~al.}(2007)\citenamefont{Press, Teukolsky,
  Betterling, and Flannery}}]{NumericalRecipes}
\bibinfo{author}{\bibfnamefont{W.~H.} \bibnamefont{Press}},
  \bibinfo{author}{\bibfnamefont{S.~A.} \bibnamefont{Teukolsky}},
  \bibinfo{author}{\bibfnamefont{W.~T.} \bibnamefont{Betterling}},
  \bibnamefont{and} \bibinfo{author}{\bibfnamefont{B.~P.}
  \bibnamefont{Flannery}}, \emph{\bibinfo{title}{Numerical Recipes}}
  (\bibinfo{publisher}{Cambridge University Press},
  \bibinfo{address}{Cambridge, England}, \bibinfo{year}{2007}),
  \bibinfo{edition}{3rd} ed.

\bibitem[{\citenamefont{Beyer}(1999)}]{Beyer:1999}
\bibinfo{author}{\bibfnamefont{H.~R.} \bibnamefont{Beyer}},
  \bibinfo{journal}{Commun. Math. Phys.} \textbf{\bibinfo{volume}{204}},
  \bibinfo{pages}{397} (\bibinfo{year}{1999}).

\bibitem[{\citenamefont{Kay et~al.}(1997)\citenamefont{Kay, Radzikowski, and
  Wald}}]{Kay:Radzikowski:Wald:1997}
\bibinfo{author}{\bibfnamefont{B.}~\bibnamefont{Kay}},
  \bibinfo{author}{\bibfnamefont{M.}~\bibnamefont{Radzikowski}},
  \bibnamefont{and} \bibinfo{author}{\bibfnamefont{R.}~\bibnamefont{Wald}},
  \bibinfo{journal}{Commun. Math. Phys.} \textbf{\bibinfo{volume}{183}},
  \bibinfo{pages}{533} (\bibinfo{year}{1997}).

\bibitem[{\citenamefont{Ori}(2008)}]{Ori1}
\bibinfo{author}{\bibfnamefont{A.}~\bibnamefont{Ori}} (\bibinfo{year}{2008}),
  \bibinfo{note}{{The four-fold structure of the singular part of the Green
  function beyond the caustics was found by Ori some time ago. He also
  conducted measurements in an analog acoustic system, which seem to verify the
  theoretical prediction. Private Communication (2008) and report (2009)
  available at \url{http://physics.technion.ac.il/~amos/acoustic.pdf}.}}

\bibitem[{\citenamefont{Vassilevich}(2003)}]{Vassilevich:2003}
\bibinfo{author}{\bibfnamefont{D.}~\bibnamefont{Vassilevich}},
  \bibinfo{journal}{Phys.Rept.} \textbf{\bibinfo{volume}{388}},
  \bibinfo{pages}{279} (\bibinfo{year}{2003}).

\bibitem[{\citenamefont{Gilkey}(1984)}]{Gilkey}
\bibinfo{author}{\bibfnamefont{P.}~\bibnamefont{Gilkey}},
  \emph{\bibinfo{title}{Invariance Theory, The Heat Equation, and the
  Atiyah-Singer Index Theorem}} (\bibinfo{publisher}{Publish or Perish, Inc.},
  \bibinfo{address}{Willmington, Delaware}, \bibinfo{year}{1984}).

\bibitem[{\citenamefont{Birrell and Davies}(1984)}]{Birrell:Davies}
\bibinfo{author}{\bibfnamefont{N.}~\bibnamefont{Birrell}} \bibnamefont{and}
  \bibinfo{author}{\bibfnamefont{P.}~\bibnamefont{Davies}},
  \emph{\bibinfo{title}{Quantum Fields in Curved Space}}
  (\bibinfo{publisher}{Cambridge University Press},
  \bibinfo{address}{Cambridge}, \bibinfo{year}{1984}).

\bibitem[{\citenamefont{Fulling et~al.}(1992)\citenamefont{Fulling, King,
  Wybourne, and Cummins}}]{Fulling:1992}
\bibinfo{author}{\bibfnamefont{S.~A.} \bibnamefont{Fulling}},
  \bibinfo{author}{\bibfnamefont{R.~C.} \bibnamefont{King}},
  \bibinfo{author}{\bibfnamefont{B.~G.} \bibnamefont{Wybourne}},
  \bibnamefont{and} \bibinfo{author}{\bibfnamefont{C.~J.}
  \bibnamefont{Cummins}}, \bibinfo{journal}{Class. Quantum Grav.}
  \textbf{\bibinfo{volume}{9}}, \bibinfo{pages}{1151} (\bibinfo{year}{1992}).

\bibitem[{\citenamefont{Avramidi}(1986)}]{Avramidi:1986}
\bibinfo{author}{\bibfnamefont{I.~G.} \bibnamefont{Avramidi}}, Ph.D. thesis,
  \emph{\bibinfo{title}{{Covariant methods for the calculation of the effective
  action in quantum field theory and investigation of higher-derivative quantum
  gravity. }}}, \bibinfo{school}{Moscow State University}
  (\bibinfo{year}{1986}), \eprint{hep-th/9510140}.

\bibitem[{Avr()}]{AvramidiCode}
\bibinfo{howpublished}{\url{http://www.barrywardell.net/research/code/covariantseries}}.

\bibitem[{Tra()}]{TransportCode}
\bibinfo{howpublished}{\url{http://www.barrywardell.net/research/code/transport}}.

\bibitem[{\citenamefont{Mart\'in-Garc\'ia}(2008)}]{xTensor}
\bibinfo{author}{\bibfnamefont{J.~M.} \bibnamefont{Mart\'in-Garc\'ia}},
  \bibinfo{journal}{Comp. Phys. Commun.} \textbf{\bibinfo{volume}{179}},
  \bibinfo{pages}{597} (\bibinfo{year}{2008}),
  \bibinfo{note}{\url{http://metric.iem.csic.es/Martin-Garcia/xAct/}}.

\bibitem[{\citenamefont{Bel'kov et~al.}(1996)\citenamefont{Bel'kov, Lanyov, and
  Schaale}}]{Belkov:1996}
\bibinfo{author}{\bibfnamefont{A.~A.} \bibnamefont{Bel'kov}},
  \bibinfo{author}{\bibfnamefont{A.~V.} \bibnamefont{Lanyov}},
  \bibnamefont{and} \bibinfo{author}{\bibfnamefont{A.}~\bibnamefont{Schaale}},
  \bibinfo{journal}{Comp. Phys. Commun.} \textbf{\bibinfo{volume}{95}},
  \bibinfo{pages}{123} (\bibinfo{year}{1996}).

\bibitem[{\citenamefont{Fulling}(1990)}]{Fulling:1990}
\bibinfo{author}{\bibfnamefont{S.~A.} \bibnamefont{Fulling}},
  \bibinfo{journal}{J. Symb. Comp.} \textbf{\bibinfo{volume}{9}},
  \bibinfo{pages}{73} (\bibinfo{year}{1990}).

\bibitem[{\citenamefont{Gusynin and Kornyak}(1994)}]{Gusynin:1994}
\bibinfo{author}{\bibfnamefont{V.~P.} \bibnamefont{Gusynin}} \bibnamefont{and}
  \bibinfo{author}{\bibfnamefont{V.~V.} \bibnamefont{Kornyak}},
  \bibinfo{journal}{J. Symb. Comp.} \textbf{\bibinfo{volume}{17}},
  \bibinfo{pages}{283} (\bibinfo{year}{1994}).

\bibitem[{\citenamefont{Bel'kov et~al.}(1993)\citenamefont{Bel'kov, Ebert,
  Lanyov, and Schaale}}]{Belkov:1993}
\bibinfo{author}{\bibfnamefont{A.~A.} \bibnamefont{Bel'kov}},
  \bibinfo{author}{\bibfnamefont{D.}~\bibnamefont{Ebert}},
  \bibinfo{author}{\bibfnamefont{A.~V.} \bibnamefont{Lanyov}},
  \bibnamefont{and} \bibinfo{author}{\bibfnamefont{A.}~\bibnamefont{Schaale}},
  \bibinfo{journal}{Int.J.Mod.Phys.} \textbf{\bibinfo{volume}{C4}},
  \bibinfo{pages}{775} (\bibinfo{year}{1993}).

\bibitem[{\citenamefont{Fulling}(1993)}]{Fulling:HeatKernel}
\bibinfo{author}{\bibfnamefont{S.~A.} \bibnamefont{Fulling}}, in
  \emph{\bibinfo{booktitle}{Proceedings of the Third International Colloquium
  on Differential Equations}}, edited by
  \bibinfo{editor}{\bibfnamefont{D.}~\bibnamefont{Bainov}} \bibnamefont{and}
  \bibinfo{editor}{\bibfnamefont{V.}~\bibnamefont{Covachev}}
  (\bibinfo{publisher}{VSP International Science Publishers},
  \bibinfo{year}{1993}), pp. \bibinfo{pages}{63--76}.

\bibitem[{\citenamefont{Gilkey}(1979)}]{Gilkey:1979}
\bibinfo{author}{\bibfnamefont{P.~B.} \bibnamefont{Gilkey}},
  \bibinfo{journal}{Compositio Mathematica} \textbf{\bibinfo{volume}{38}},
  \bibinfo{pages}{201} (\bibinfo{year}{1979}).

\bibitem[{\citenamefont{Gilkey}(1980)}]{Gilkey:1980}
\bibinfo{author}{\bibfnamefont{P.~B.} \bibnamefont{Gilkey}},
  \bibinfo{journal}{Duke Mathematical Journal} \textbf{\bibinfo{volume}{47}},
  \bibinfo{pages}{511} (\bibinfo{year}{1980}).

\bibitem[{\citenamefont{van~den Ven}(1998)}]{Ven:1998}
\bibinfo{author}{\bibfnamefont{A.~E.~M.} \bibnamefont{van~den Ven}},
  \bibinfo{journal}{Class. Quantum Grav.} \textbf{\bibinfo{volume}{15}},
  \bibinfo{pages}{2311} (\bibinfo{year}{1998}).

\bibitem[{\citenamefont{Anselmi and Benini}(2007)}]{Anselmi:2007}
\bibinfo{author}{\bibfnamefont{D.}~\bibnamefont{Anselmi}} \bibnamefont{and}
  \bibinfo{author}{\bibfnamefont{A.}~\bibnamefont{Benini}},
  \bibinfo{journal}{Journal of High Energy Physics}
  \textbf{\bibinfo{volume}{10}} (\bibinfo{year}{2007}).

\bibitem[{\citenamefont{Dowker and Kirsten}(1999)}]{Dowker:1999}
\bibinfo{author}{\bibfnamefont{J.~S.} \bibnamefont{Dowker}} \bibnamefont{and}
  \bibinfo{author}{\bibfnamefont{K.}~\bibnamefont{Kirsten}},
  \bibinfo{journal}{Class. Quantum Grav.} \textbf{\bibinfo{volume}{16}},
  \bibinfo{pages}{1917} (\bibinfo{year}{1999}), \eprint{hep-th/9806168}.

\bibitem[{\citenamefont{Gusynin and Kornyak}(1999)}]{Gusynin:1999}
\bibinfo{author}{\bibfnamefont{V.~P.} \bibnamefont{Gusynin}} \bibnamefont{and}
  \bibinfo{author}{\bibfnamefont{V.~V.} \bibnamefont{Kornyak}},
  \bibinfo{journal}{Fundamental and Applied Mathematics}
  \textbf{\bibinfo{volume}{5}}, \bibinfo{pages}{649} (\bibinfo{year}{1999}),
  \eprint{math/9909145}.

\bibitem[{\citenamefont{Salcedo}(2001)}]{Salcedo:2001}
\bibinfo{author}{\bibfnamefont{L.~L.} \bibnamefont{Salcedo}},
  \bibinfo{journal}{Eur. Phys. J.} \textbf{\bibinfo{volume}{C3}}
  (\bibinfo{year}{2001}), \eprint{hep-th/0107133}.

\bibitem[{\citenamefont{Salcedo}(2004)}]{Salcedo:2004}
\bibinfo{author}{\bibfnamefont{L.~L.} \bibnamefont{Salcedo}},
  \bibinfo{journal}{Eur. Phys. J.} \textbf{\bibinfo{volume}{C37}},
  \bibinfo{pages}{511} (\bibinfo{year}{2004}), \eprint{hep-th/0409140}.

\bibitem[{\citenamefont{Salcedo}(2007)}]{Salcedo:2007}
\bibinfo{author}{\bibfnamefont{L.~L.} \bibnamefont{Salcedo}},
  \bibinfo{journal}{Phys. Rev. D} \textbf{\bibinfo{volume}{76}},
  \bibinfo{pages}{44009} (\bibinfo{year}{2007}), \eprint{arXiv:0706.1875}.

\bibitem[{\citenamefont{Gayral et~al.}(2006)\citenamefont{Gayral, Iochum, and
  Vassilevich}}]{Gayral:2006}
\bibinfo{author}{\bibfnamefont{V.}~\bibnamefont{Gayral}},
  \bibinfo{author}{\bibfnamefont{B.}~\bibnamefont{Iochum}}, \bibnamefont{and}
  \bibinfo{author}{\bibfnamefont{D.~V.} \bibnamefont{Vassilevich}},
  \bibinfo{journal}{Commun. Math. Phys.} \textbf{\bibinfo{volume}{273}},
  \bibinfo{pages}{415} (\bibinfo{year}{2006}), \eprint{hep-th/0607078}.

\bibitem[{\citenamefont{Booth}(1998)}]{Booth:1998}
\bibinfo{author}{\bibfnamefont{M.~J.} \bibnamefont{Booth}}
  (\bibinfo{year}{1998}), \eprint{hep-th/9803113}.

\bibitem[{\citenamefont{Avramidi and
  Schimming}(1996)}]{Avramidi:Schimming:1996}
\bibinfo{author}{\bibfnamefont{I.~G.} \bibnamefont{Avramidi}} \bibnamefont{and}
  \bibinfo{author}{\bibfnamefont{R.}~\bibnamefont{Schimming}},
  \emph{\bibinfo{title}{Algorithms for the calculation of the heat kernel
  coefficients}} (\bibinfo{publisher}{Teubner-Texte zur Physik},
  \bibinfo{year}{1996}), vol.~\bibinfo{volume}{30}, pp.
  \bibinfo{pages}{150--162}, \eprint{hep-th/9510206}.

\bibitem[{\citenamefont{Barvinsky and Vilkovisky}(1985)}]{Barvinsky:1985}
\bibinfo{author}{\bibfnamefont{A.~O.} \bibnamefont{Barvinsky}}
  \bibnamefont{and} \bibinfo{author}{\bibfnamefont{G.~A.}
  \bibnamefont{Vilkovisky}}, \bibinfo{journal}{Phys. Repts.}
  \textbf{\bibinfo{volume}{119}}, \bibinfo{pages}{1 } (\bibinfo{year}{1985}).

\bibitem[{\citenamefont{Fliegner et~al.}(1998)\citenamefont{Fliegner, Haberl,
  Schmidt, and Schubert}}]{Fliegner:1998}
\bibinfo{author}{\bibfnamefont{D.}~\bibnamefont{Fliegner}},
  \bibinfo{author}{\bibfnamefont{P.}~\bibnamefont{Haberl}},
  \bibinfo{author}{\bibfnamefont{M.~G.} \bibnamefont{Schmidt}},
  \bibnamefont{and} \bibinfo{author}{\bibfnamefont{C.}~\bibnamefont{Schubert}},
  \bibinfo{journal}{Annals of Physics} \textbf{\bibinfo{volume}{264}},
  \bibinfo{pages}{51 } (\bibinfo{year}{1998}).

\bibitem[{\citenamefont{Galassi et~al.}(2009)}]{GSL}
\bibinfo{author}{\bibfnamefont{M.}~\bibnamefont{Galassi}} \bibnamefont{et~al.},
  \emph{\bibinfo{title}{GNU Scientific Library Reference Manual}},
  \bibinfo{edition}{3rd} ed. (\bibinfo{year}{2009}).

\bibitem[{GRT()}]{GRTensor}
\bibinfo{howpublished}{\url{http://www.grtensor.org}}.

\bibitem[{Map()}]{Maple}
\bibinfo{howpublished}{\url{http://www.maplesoft.com}}.

\bibitem[{\citenamefont{Frolov and Novikov}(1998)}]{Frolov}
\bibinfo{author}{\bibfnamefont{V.~P.} \bibnamefont{Frolov}} \bibnamefont{and}
  \bibinfo{author}{\bibfnamefont{I.~D.} \bibnamefont{Novikov}},
  \emph{\bibinfo{title}{Black Hole Physics: Basic Concepts and New
  Developments}} (\bibinfo{publisher}{Kluwer Academic Publishers},
  \bibinfo{year}{1998}).

\bibitem[{\citenamefont{Hartle and Wilkins}(1974)}]{Hartle:Wilkins:1974}
\bibinfo{author}{\bibfnamefont{J.~B.} \bibnamefont{Hartle}} \bibnamefont{and}
  \bibinfo{author}{\bibfnamefont{D.~C.} \bibnamefont{Wilkins}},
  \bibinfo{journal}{Commun. Math. Phys.} \textbf{\bibinfo{volume}{38}},
  \bibinfo{pages}{47} (\bibinfo{year}{1974}).

\bibitem[{\citenamefont{Leaver}(1985)}]{Leaver:1985}
\bibinfo{author}{\bibfnamefont{E.~W.} \bibnamefont{Leaver}},
  \bibinfo{journal}{Proc. Roy. Soc. Lond. A} \textbf{\bibinfo{volume}{402}},
  \bibinfo{pages}{285} (\bibinfo{year}{1985}).

\bibitem[{\citenamefont{Nollert}(1999)}]{Nollert:1999}
\bibinfo{author}{\bibfnamefont{H.-P.} \bibnamefont{Nollert}},
  \bibinfo{journal}{Class. Quantum Grav.} \textbf{\bibinfo{volume}{16}},
  \bibinfo{pages}{R159} (\bibinfo{year}{1999}).

\bibitem[{\citenamefont{Kokkotas and Schmidt}(1999)}]{Kokkotas:Schmidt:1999}
\bibinfo{author}{\bibfnamefont{K.~D.} \bibnamefont{Kokkotas}} \bibnamefont{and}
  \bibinfo{author}{\bibfnamefont{B.~G.} \bibnamefont{Schmidt}},
  \bibinfo{journal}{Living Rev. Relativity} \textbf{\bibinfo{volume}{2}},
  \bibinfo{pages}{2} (\bibinfo{year}{1999}).

\bibitem[{\citenamefont{Cardoso et~al.}(2009)\citenamefont{Cardoso, Miranda,
  Berti, Witek, and Zanchin}}]{Cardoso:Witek:2008}
\bibinfo{author}{\bibfnamefont{V.}~\bibnamefont{Cardoso}},
  \bibinfo{author}{\bibfnamefont{A.~S.} \bibnamefont{Miranda}},
  \bibinfo{author}{\bibfnamefont{E.}~\bibnamefont{Berti}},
  \bibinfo{author}{\bibfnamefont{H.}~\bibnamefont{Witek}}, \bibnamefont{and}
  \bibinfo{author}{\bibfnamefont{V.~T.} \bibnamefont{Zanchin}},
  \bibinfo{journal}{Phys. Rev. D} \textbf{\bibinfo{volume}{79}},
  \bibinfo{pages}{064016} (\bibinfo{year}{2009}), \eprint{arXiv:0812.1806}.

\bibitem[{\citenamefont{Iyer}(1987)}]{Iyer:1987}
\bibinfo{author}{\bibfnamefont{S.}~\bibnamefont{Iyer}}, \bibinfo{journal}{Phys.
  Rev.} \textbf{\bibinfo{volume}{D35}}, \bibinfo{pages}{3632}
  (\bibinfo{year}{1987}).

\bibitem[{\citenamefont{Cardoso et~al.}(2003)\citenamefont{Cardoso, Konoplya,
  and Lemos}}]{Konoplya:2003}
\bibinfo{author}{\bibfnamefont{V.}~\bibnamefont{Cardoso}},
  \bibinfo{author}{\bibfnamefont{R.}~\bibnamefont{Konoplya}}, \bibnamefont{and}
  \bibinfo{author}{\bibfnamefont{J.~P.~S.} \bibnamefont{Lemos}},
  \bibinfo{journal}{Phys. Rev.} \textbf{\bibinfo{volume}{D68}},
  \bibinfo{pages}{044024} (\bibinfo{year}{2003}), \eprint{gr-qc/0305037}.

\bibitem[{\citenamefont{Duistermaat and
  H\"ormander}(1972)}]{Duistermaat:Hormander:1972}
\bibinfo{author}{\bibfnamefont{J.}~\bibnamefont{Duistermaat}} \bibnamefont{and}
  \bibinfo{author}{\bibfnamefont{L.}~\bibnamefont{H\"ormander}},
  \bibinfo{journal}{Acta Mathematica} \textbf{\bibinfo{volume}{128}},
  \bibinfo{pages}{183} (\bibinfo{year}{1972}).

\bibitem[{\citenamefont{H{\"{o}}rmander}(1985)}]{Hormander:1985}
\bibinfo{author}{\bibfnamefont{L.}~\bibnamefont{H{\"{o}}rmander}},
  \emph{\bibinfo{title}{The Analysis of Linear Partial Differential Operators
  IV}} (\bibinfo{publisher}{Springer}, \bibinfo{address}{Berlin, Heidelberg,
  New York}, \bibinfo{year}{1985}).

\bibitem[{\citenamefont{D\'ecanini et~al.}(2003)\citenamefont{D\'ecanini,
  Folacci, and Jensen}}]{Decanini:Folacci:Jensen:2003}
\bibinfo{author}{\bibfnamefont{Y.}~\bibnamefont{D\'ecanini}},
  \bibinfo{author}{\bibfnamefont{A.}~\bibnamefont{Folacci}}, \bibnamefont{and}
  \bibinfo{author}{\bibfnamefont{B.}~\bibnamefont{Jensen}},
  \bibinfo{journal}{Phys. Rev. D} \textbf{\bibinfo{volume}{67}},
  \bibinfo{pages}{124017} (\bibinfo{year}{2003}).

\bibitem[{\citenamefont{Aki and Richards}(2002)}]{Aki:Richards}
\bibinfo{author}{\bibfnamefont{K.}~\bibnamefont{Aki}} \bibnamefont{and}
  \bibinfo{author}{\bibfnamefont{P.~G.} \bibnamefont{Richards}},
  \emph{\bibinfo{title}{Quantitative Seismology}}
  (\bibinfo{publisher}{University Science Books}, \bibinfo{year}{2002}).

\bibitem[{\citenamefont{Olver}(1974)}]{Olver:1974}
\bibinfo{author}{\bibfnamefont{F.~W.~J.} \bibnamefont{Olver}},
  \emph{\bibinfo{title}{Asymptotics and special functions}}
  (\bibinfo{publisher}{New York: Academic Press}, \bibinfo{year}{1974}).

\bibitem[{\citenamefont{D\'ecanini and Folacci}(2008)}]{Decanini:Folacci:2005b}
\bibinfo{author}{\bibfnamefont{Y.}~\bibnamefont{D\'ecanini}} \bibnamefont{and}
  \bibinfo{author}{\bibfnamefont{A.}~\bibnamefont{Folacci}},
  \bibinfo{journal}{Phys. Rev.} \textbf{\bibinfo{volume}{D78}},
  \bibinfo{pages}{044025} (\bibinfo{year}{2008}), \eprint{gr-qc/0512118}.

\bibitem[{\citenamefont{Van~Vleck}(1928)}]{VanVleck:1928}
\bibinfo{author}{\bibfnamefont{J.~H.} \bibnamefont{Van~Vleck}},
  \bibinfo{journal}{Proc. Nat. Acad. Sci.} \textbf{\bibinfo{volume}{14}},
  \bibinfo{pages}{178} (\bibinfo{year}{1928}).

\bibitem[{\citenamefont{Morette}(1951)}]{Morette:1951}
\bibinfo{author}{\bibfnamefont{C.}~\bibnamefont{Morette}},
  \bibinfo{journal}{Phys. Rev.} \textbf{\bibinfo{volume}{81}},
  \bibinfo{pages}{848} (\bibinfo{year}{1951}).

\bibitem[{\citenamefont{Visser}(1993)}]{Visser:1993}
\bibinfo{author}{\bibfnamefont{M.}~\bibnamefont{Visser}},
  \bibinfo{journal}{Phys. Rev. D} \textbf{\bibinfo{volume}{47}},
  \bibinfo{pages}{2395} (\bibinfo{year}{1993}).

\bibitem[{\citenamefont{Kravtsov}(1968)}]{Kravtsov:1968}
\bibinfo{author}{\bibfnamefont{Y.~A.} \bibnamefont{Kravtsov}},
  \bibinfo{journal}{Sov. Phys.-Acoust.} \textbf{\bibinfo{volume}{14}},
  \bibinfo{pages}{1} (\bibinfo{year}{1968}).

\bibitem[{\citenamefont{Arnold}(1990)}]{Arnold}
\bibinfo{author}{\bibfnamefont{V.~I.} \bibnamefont{Arnold}},
  \emph{\bibinfo{title}{Singularities of Caustics and Wave Fronts}}
  (\bibinfo{publisher}{Kluwer Academic Publishers}, \bibinfo{year}{1990}).

\bibitem[{\citenamefont{Berry and Mount}(1972)}]{B&M}
\bibinfo{author}{\bibfnamefont{M.}~\bibnamefont{Berry}} \bibnamefont{and}
  \bibinfo{author}{\bibfnamefont{K.}~\bibnamefont{Mount}},
  \bibinfo{journal}{Rept.\ Prog.\ Phys.} \textbf{\bibinfo{volume}{35}},
  \bibinfo{pages}{315} (\bibinfo{year}{1972}).

\bibitem[{\citenamefont{Maslov}(1965{\natexlab{a}})}]{Maslov'65}
\bibinfo{author}{\bibfnamefont{V.}~\bibnamefont{Maslov}},
  \emph{\bibinfo{title}{Theory of Perturbations and Asymptotic Methods [in
  Russian]}} (\bibinfo{publisher}{Izd-vo Moskov. Gos. Univ.},
  \bibinfo{address}{Moscow}, \bibinfo{year}{1965}{\natexlab{a}}),
  \bibinfo{note}{[French transl. (Dunod, Paris, 1972)]}.

\bibitem[{\citenamefont{Maslov}(1965{\natexlab{b}})}]{MaslovWKB}
\bibinfo{author}{\bibfnamefont{V.}~\bibnamefont{Maslov}},
  \bibinfo{journal}{``The WKB Method in the Multidimensional Case'', Appendix
  II to the Book: G.Heading, Introduction to the Phase Integral Method (WKB
  Method) (Wiley, New York) [Russian translation]} p. \bibinfo{pages}{177}
  (\bibinfo{year}{1965}{\natexlab{b}}).

\bibitem[{\citenamefont{Friedrich and Stewart}(1983)}]{Friedrich:Stewart:1983}
\bibinfo{author}{\bibfnamefont{H.}~\bibnamefont{Friedrich}} \bibnamefont{and}
  \bibinfo{author}{\bibfnamefont{J.~M.} \bibnamefont{Stewart}},
  \bibinfo{journal}{Proc. R. Soc. Lond. A} \textbf{\bibinfo{volume}{385}},
  \bibinfo{pages}{345} (\bibinfo{year}{1983}).

\bibitem[{\citenamefont{Ehlers and Newman}(2000)}]{Ehlers:Newman:2000}
\bibinfo{author}{\bibfnamefont{J.}~\bibnamefont{Ehlers}} \bibnamefont{and}
  \bibinfo{author}{\bibfnamefont{E.~T.} \bibnamefont{Newman}},
  \bibinfo{journal}{J. Math. Phys.} \textbf{\bibinfo{volume}{41}},
  \bibinfo{pages}{3344} (\bibinfo{year}{2000}).

\bibitem[{\citenamefont{Linet}(2005)}]{Linet:2005}
\bibinfo{author}{\bibfnamefont{B.}~\bibnamefont{Linet}}, \bibinfo{journal}{Gen.
  Rel. Grav.} \textbf{\bibinfo{volume}{37}}, \bibinfo{pages}{2145}
  (\bibinfo{year}{2005}), \eprint{gr-qc/0507072}.

\bibitem[{\citenamefont{Erdelyi et~al.}(1953)\citenamefont{Erdelyi, Magnus,
  Oberhettinger, and Tricomi}}]{Erdelyi:1953}
\bibinfo{author}{\bibfnamefont{A.}~\bibnamefont{Erdelyi}},
  \bibinfo{author}{\bibfnamefont{W.}~\bibnamefont{Magnus}},
  \bibinfo{author}{\bibfnamefont{F.}~\bibnamefont{Oberhettinger}},
  \bibnamefont{and} \bibinfo{author}{\bibfnamefont{F.}~\bibnamefont{Tricomi}},
  \emph{\bibinfo{title}{Higher Transcendental Functions Vol I}}
  (\bibinfo{publisher}{McGraw-Hill}, \bibinfo{address}{New York},
  \bibinfo{year}{1953}).

\bibitem[{\citenamefont{Candelas and Jensen}(1986)}]{Candelas:Jensen:1986}
\bibinfo{author}{\bibfnamefont{P.}~\bibnamefont{Candelas}} \bibnamefont{and}
  \bibinfo{author}{\bibfnamefont{B.}~\bibnamefont{Jensen}},
  \bibinfo{journal}{Phys. Rev. D} \textbf{\bibinfo{volume}{33}},
  \bibinfo{pages}{1596} (\bibinfo{year}{1986}).

\bibitem[{\citenamefont{Hardy}(1948)}]{Hardy}
\bibinfo{author}{\bibfnamefont{G.~H.} \bibnamefont{Hardy}},
  \emph{\bibinfo{title}{Divergent Series}} (\bibinfo{publisher}{Oxford
  University Press}, \bibinfo{address}{London}, \bibinfo{year}{1948}).

\bibitem[{\citenamefont{Ching et~al.}(1995)\citenamefont{Ching, Leung, Suen,
  and Young}}]{Ching:1995b}
\bibinfo{author}{\bibfnamefont{E.~S.~C.} \bibnamefont{Ching}},
  \bibinfo{author}{\bibfnamefont{P.~T.} \bibnamefont{Leung}},
  \bibinfo{author}{\bibfnamefont{W.~M.} \bibnamefont{Suen}}, \bibnamefont{and}
  \bibinfo{author}{\bibfnamefont{K.}~\bibnamefont{Young}},
  \bibinfo{journal}{Phys. Rev. D} \textbf{\bibinfo{volume}{52}},
  \bibinfo{pages}{2118} (\bibinfo{year}{1995}).

\bibitem[{\citenamefont{Barack and Golbourn}(2007)}]{Barack:Golbourn:2007}
\bibinfo{author}{\bibfnamefont{L.}~\bibnamefont{Barack}} \bibnamefont{and}
  \bibinfo{author}{\bibfnamefont{D.~A.} \bibnamefont{Golbourn}},
  \bibinfo{journal}{Phys. Rev.} \textbf{\bibinfo{volume}{D76}},
  \bibinfo{pages}{044020} (\bibinfo{year}{2007}), \eprint{arXiv:0705.3620}.

\bibitem[{\citenamefont{Phillips and Hu}(2003)}]{Phillips:Hu:2003}
\bibinfo{author}{\bibfnamefont{N.~G.} \bibnamefont{Phillips}} \bibnamefont{and}
  \bibinfo{author}{\bibfnamefont{B.~L.} \bibnamefont{Hu}},
  \bibinfo{journal}{Phys. Rev.} \textbf{\bibinfo{volume}{D67}},
  \bibinfo{pages}{104002} (\bibinfo{year}{2003}), \eprint{gr-qc/0209056}.

\bibitem[{\citenamefont{Mart\'in-Garc\'ia
  et~al.}(2007)\citenamefont{Mart\'in-Garc\'ia, Portugal, and
  Manssur}}]{Invar1}
\bibinfo{author}{\bibfnamefont{J.~M.} \bibnamefont{Mart\'in-Garc\'ia}},
  \bibinfo{author}{\bibfnamefont{R.}~\bibnamefont{Portugal}}, \bibnamefont{and}
  \bibinfo{author}{\bibfnamefont{L.}~\bibnamefont{Manssur}},
  \bibinfo{journal}{Comp. Phys. Commun.} \textbf{\bibinfo{volume}{177}},
  \bibinfo{pages}{640} (\bibinfo{year}{2007}), \eprint{arxiv:0704.1756}.

\bibitem[{\citenamefont{Mart\'in-Garc\'ia
  et~al.}(2008)\citenamefont{Mart\'in-Garc\'ia, Yllanes, and
  Portugal}}]{Invar2}
\bibinfo{author}{\bibfnamefont{J.~M.} \bibnamefont{Mart\'in-Garc\'ia}},
  \bibinfo{author}{\bibfnamefont{D.}~\bibnamefont{Yllanes}}, \bibnamefont{and}
  \bibinfo{author}{\bibfnamefont{R.}~\bibnamefont{Portugal}},
  \bibinfo{journal}{Comp. Phys. Commun.} \textbf{\bibinfo{volume}{179}},
  \bibinfo{pages}{586} (\bibinfo{year}{2008}), \eprint{arxiv:0802.1274}.

\bibitem[{\citenamefont{{Wolfram Research, Inc.}}(2008)}]{Mathematica}
\bibinfo{author}{\bibnamefont{{Wolfram Research, Inc.}}},
  \emph{\bibinfo{title}{Mathematica}} (\bibinfo{publisher}{Wolfram Research,
  Inc.}, \bibinfo{address}{Champaign, Illinois}, \bibinfo{year}{2008}),
  \bibinfo{edition}{{Version 7.0}} ed.

\bibitem[{\citenamefont{Price and Whiting}(2005)}]{Price:Whiting}
\bibinfo{author}{\bibfnamefont{L.~R.} \bibnamefont{Price}} \bibnamefont{and}
  \bibinfo{author}{\bibfnamefont{B.~F.} \bibnamefont{Whiting}},
  \bibinfo{journal}{Class. Quantum Grav.} \textbf{\bibinfo{volume}{22}},
  \bibinfo{pages}{S589} (\bibinfo{year}{2005}).

\bibitem[{\citenamefont{Geroch et~al.}(1973)\citenamefont{Geroch, Held, and
  Penrose}}]{GHP}
\bibinfo{author}{\bibfnamefont{R.}~\bibnamefont{Geroch}},
  \bibinfo{author}{\bibfnamefont{A.}~\bibnamefont{Held}}, \bibnamefont{and}
  \bibinfo{author}{\bibfnamefont{R.}~\bibnamefont{Penrose}},
  \bibinfo{journal}{J. Math. Phys.} \textbf{\bibinfo{volume}{14}},
  \bibinfo{pages}{874} (\bibinfo{year}{1973}).

\bibitem[{\citenamefont{Penrose and Rindler}(1984)}]{Penrose:Rindler:1}
\bibinfo{author}{\bibfnamefont{R.}~\bibnamefont{Penrose}} \bibnamefont{and}
  \bibinfo{author}{\bibfnamefont{W.}~\bibnamefont{Rindler}},
  \emph{\bibinfo{title}{Spinors and Space-time 1. Two Spinor Calculus and
  Relativistic Fields}} (\bibinfo{publisher}{Cambridge Monographs On
  Mathematical Physics}, \bibinfo{address}{Cambridge, UK},
  \bibinfo{year}{1984}).

\bibitem[{\citenamefont{Penrose and Rindler}(1986)}]{Penrose:Rindler:2}
\bibinfo{author}{\bibfnamefont{R.}~\bibnamefont{Penrose}} \bibnamefont{and}
  \bibinfo{author}{\bibfnamefont{W.}~\bibnamefont{Rindler}},
  \emph{\bibinfo{title}{Spinors and Space-time 2. Spinor and Twistor Methods in
  Space-time Geometry}} (\bibinfo{publisher}{Cambridge Monographs On
  Mathematical Physics}, \bibinfo{address}{Cambridge, UK},
  \bibinfo{year}{1986}).

\bibitem[{\citenamefont{Newman and Penrose}(1962)}]{NP}
\bibinfo{author}{\bibfnamefont{E.~T.} \bibnamefont{Newman}} \bibnamefont{and}
  \bibinfo{author}{\bibfnamefont{R.}~\bibnamefont{Penrose}},
  \bibinfo{journal}{J. Math. Phys.} \textbf{\bibinfo{volume}{3}},
  \bibinfo{pages}{566} (\bibinfo{year}{1962}).

\bibitem[{\citenamefont{Chisholm}(1973)}]{Chisholm:1973}
\bibinfo{author}{\bibfnamefont{J.~S.~R.} \bibnamefont{Chisholm}},
  \bibinfo{journal}{Mathematics of Computation} \textbf{\bibinfo{volume}{27}},
  \bibinfo{pages}{841} (\bibinfo{year}{1973}).

\bibitem[{\citenamefont{Chisholm and McEwan}(1974)}]{Chisholm:McEwan:1974}
\bibinfo{author}{\bibfnamefont{J.~S.~R.} \bibnamefont{Chisholm}}
  \bibnamefont{and} \bibinfo{author}{\bibfnamefont{J.}~\bibnamefont{McEwan}},
  \bibinfo{journal}{Proc. R. Soc. Lond. A} \textbf{\bibinfo{volume}{336}},
  \bibinfo{pages}{421} (\bibinfo{year}{1974}).

\bibitem[{\citenamefont{Stavroudis}(1972)}]{Stavroudis}
\bibinfo{author}{\bibfnamefont{O.~N.} \bibnamefont{Stavroudis}},
  \emph{\bibinfo{title}{Optics of Rays, Wavefronts and Caustics}}
  (\bibinfo{publisher}{Academic Press}, \bibinfo{year}{1972}).

\bibitem[{\citenamefont{Bozza}(2008)}]{Bozza:2008}
\bibinfo{author}{\bibfnamefont{V.}~\bibnamefont{Bozza}},
  \bibinfo{journal}{Phys. Rev.} \textbf{\bibinfo{volume}{D78}},
  \bibinfo{pages}{063014} (\bibinfo{year}{2008}), \eprint{arXiv:0806.4102}.

\bibitem[{\citenamefont{Brown and Ottewill}(1986)}]{Brown:Ottewill:1986}
\bibinfo{author}{\bibfnamefont{M.~R.} \bibnamefont{Brown}} \bibnamefont{and}
  \bibinfo{author}{\bibfnamefont{A.~C.} \bibnamefont{Ottewill}},
  \bibinfo{journal}{Phys. Rev.} \textbf{\bibinfo{volume}{D34}},
  \bibinfo{pages}{1776} (\bibinfo{year}{1986}).

\bibitem[{\citenamefont{G\"unther}(1988)}]{Gunther}
\bibinfo{author}{\bibfnamefont{P.}~\bibnamefont{G\"unther}},
  \emph{\bibinfo{title}{Huygens’ Principle and Hyperbolic Equations}}
  (\bibinfo{publisher}{Academic Press, London}, \bibinfo{year}{1988}).

\bibitem[{\citenamefont{Jones}(2001)}]{Jones'01}
\bibinfo{author}{\bibfnamefont{D.}~\bibnamefont{Jones}},
  \bibinfo{journal}{Math.\ Methods in the Applied Sciences}
  \textbf{\bibinfo{volume}{24}}, \bibinfo{pages}{369} (\bibinfo{year}{2001}).

\end{thebibliography}

\end{document}